\documentclass[12pt,lot, lof]{puthesis}
\newcommand{\proquestmode}{}

\usepackage{color}

\usepackage[caption=false]{subfig}
\usepackage{graphicx,booktabs,array}
\usepackage{multirow}
\DeclareGraphicsExtensions{.pdf,.png,.jpg,.eps}
\usepackage{epsfig}
\usepackage{epstopdf}
\usepackage{amssymb}
\usepackage{amsmath}
\usepackage{amsfonts}
\usepackage{mathrsfs}
\usepackage{amsthm}
\usepackage{float}
\usepackage{amsbsy}
\usepackage{bm}
\usepackage{grffile}
\usepackage{hyperref}
\hypersetup{
    colorlinks = true,
    citecolor = blue,
    urlcolor = blue,
    linkcolor = blue,
}
\usepackage{courier}
\usepackage{upgreek}
\usepackage{xspace}
\usepackage{xcolor}
\usepackage{calc} 
\usepackage{soul} 

\newcommand{\etal}{\textit{et al}.\xspace}
\newcommand{\ie}{\textit{i}.\textit{e}. }
\newcommand{\eg}{\textit{e}.\textit{g}. }
\newcommand{\etc}{\textit{etc.}\xspace} 

\newcommand{\Alfven}{Alfv\'{e}n\xspace}
\newcommand{\Alfvenic}{Alfv\'{e}nic\xspace}
\newcommand{\EGAE}{EP-GAE\xspace}
\newcommand{\Pade}{Pad\'{e}\xspace}

\newcommand{\va}{v_A}
\newcommand{\vs}{v_S}
\newcommand{\vb}{v_0}
\newcommand{\vc}{v_c}

\newcommand{\tor}{\phi}
\newcommand{\pol}{\theta}
\newcommand{\omegaci}{\omega_{ci}}

\newcommand{\omegacio}{\omega_{ci0}} 
\newcommand{\omegator}{\omega_\phi}
\newcommand{\omegapol}{\omega_\theta}
\newcommand{\omeganorm}{\omega / \omegaci}

\newcommand{\omegabar}{\bar{\omega}}
\newcommand{\omegace}{\omega_{ce}}
\newcommand{\omegape}{\omega_{pe}}

\newcommand{\omegaps}{\omega_{ps}}
\newcommand{\omegacs}{\omega_{cs}}
\newcommand{\ldebye}{\lambda_D}

\newcommand{\ktor}{k_\phi}

\newcommand{\db}{\delta B}
\newcommand{\de}{\delta E}
\newcommand{\dbpar}{\db_\parallel}
\newcommand{\depar}{\de_\parallel}
\newcommand{\deperp}{\de_\perp}
\newcommand{\dbperp}{\db_\perp}
\newcommand{\Epar}{E_\parallel}
\newcommand{\Eperp}{E_\perp}

\newcommand{\kpar}{k_\parallel}
\newcommand{\kperp}{k_\perp}
\newcommand{\kpars}{k_{\parallel,s}}

\newcommand{\kperpmin}{k_{\perp,\text{min}}}

\newcommand{\pphi}{p_\phi}
\newcommand{\pmin}{p_{\text{min}}}

\newcommand{\vinj}{\vb/\va}

\newcommand{\linj}{\lambda_0}
\newcommand{\dl}{\Delta\lambda}
\newcommand{\df}{\delta f}
\newcommand{\vcrit}{\vc/\vb}

\newcommand{\Jc}{\mathcal{J}}
\newcommand{\Jb}{J_\text{beam}}
\newcommand{\Jp}{J_\text{plasma}}
\newcommand{\Jnorm}{\Jb/\Jp}

\newcommand{\nbo}{n_b}
\newcommand{\neo}{n_e}
\newcommand{\nb}{\nbo / \neo}
\newcommand{\nv}{\nbo\vb / \neo\va}
\newcommand{\vpar}{v_\parallel}
\newcommand{\vpres}{v_{\parallel,\text{res}}}
\newcommand{\vpreslres}{v_{\parallel,\text{res},\lres}}
\newcommand{\vdrift}{v_{\text{Dr}}}
\newcommand{\vperp}{v_\perp}
\newcommand{\vperpz}{v_{\perp,0}}
\newcommand{\lres}{\ell}

\newcommand{\J}[2]{\mathscr{J}_{#1}^{#2}}
\newcommand{\Jlm}{\J{\lres}{m}}
\newcommand{\Jlg}{\J{\lres}{G}}
\newcommand{\Jlc}{\J{\lres}{C}}
\newcommand{\Jzm}{\J{0}{m}}
\newcommand{\Jzg}{\J{0}{G}}
\newcommand{\Jzc}{\J{0}{C}}

\newcommand{\Bres}{\eta}
\newcommand{\bres}{\Bres}

\newcommand{\bresl}{\bres_\lres}

\newcommand{\Jl}{J_\lres}
\newcommand{\Jlprime}{\deriv{\Jl}{\flr}}

\newcommand{\betae}{\beta_e}
\newcommand{\nperp}{n_\perp}
\newcommand{\npar}{n_\parallel}
\newcommand{\Nperp}{N_\perp}
\newcommand{\Npar}{N_\parallel}
\newcommand{\vtherme}{v_{th,e}}
\newcommand{\vthermi}{v_{th,i}}
\newcommand{\tep}{\text{EP}}
\newcommand{\vep}{v_\tep}

\newcommand{\W}{\mathcal{E}}

\newcommand{\Pb}{P_b}

\newcommand{\vE}{\vec{E}}
\newcommand{\vB}{\vec{B}}

\newcommand{\vv}{\vec{v}}
\newcommand{\vk}{\vec{k}}

\newcommand{\vkperp}{\vec{k}_\perp}
\newcommand{\vJ}{\vec{J}}
\newcommand{\vV}{\vec{V}}
\newcommand{\vx}{\vec{x}}

\newcommand{\xinj}{x_0}
\newcommand{\dx}{\Delta x}

\newcommand{\fb}{f_0}
\newcommand{\fl}{f_L} 
\newcommand{\fbeam}{f_b}
\newcommand{\omegacires}{\avg{\bar{\omega}_{ci}}}

\newcommand{\kratraw}{\kpar/\kperp}
\newcommand{\krat}{\abs{\kratraw}}
\newcommand{\rhob}{\rho_{\perp b}}

\newcommand{\zp}{\zeta}

\newcommand{\eqlab}[1]{\quad\text{#1}}
\newcommand{\CAElab}{\eqlab{CAE}}
\newcommand{\GAElab}{\eqlab{GAE}}
\newcommand{\Te}{T_e}
\newcommand{\Ti}{T_i}
\newcommand{\Veff}{V_\text{eff}}

\newcommand{\xm}{x_m}
\newcommand{\ftail}{f_\text{tail}}
\newcommand{\dv}{\Delta v}

\newcommand{\Ksym}{K}
\newcommand{\Kij}{\Ksym_{ij}}
\newcommand{\Kxx}{\Ksym_{11}}
\newcommand{\Kxy}{\Ksym_{12}}

\newcommand{\Kyx}{\Ksym_{21}}
\newcommand{\Kyy}{\Ksym_{22}}
\newcommand{\Kyz}{\Ksym_{23}}

\newcommand{\Kzy}{\Ksym_{32}}
\newcommand{\Kzz}{\Ksym_{33}}
\newcommand{\antiherm}{^{A}}
\newcommand{\herm}{^{H}}

\newcommand{\flr}{\xi}
\newcommand{\bvar}{A^{-1}}

\newcommand{\alphamin}{\alpha_\text{min}}
\newcommand{\alphamax}{\alpha_\text{max}}

\newcommand{\Rtan}{R_\text{tan}}
\newcommand{\Rzero}{R_0}

\newcommand{\Wwave}{\mathcal{W}}
\newcommand{\Pwave}{\mathcal{P}}
\newcommand{\gammadamp}{\gamma_{\text{damp}}}
\newcommand{\alphacrit}{\alpha_{\text{crit}}}
\newcommand{\EPGAE}{EP-GAE\xspace}
\newcommand{\pphipow}{\sigma}

\newcommand{\describe}[1]{\quad\quad\text{#1}}
\newcommand{\gammaa}{{\gamma_a}} 

\newcommand{\avg}[1]{\left\langle #1 \right\rangle}
\renewcommand{\vec}[1]{\bm{#1}}
\newcommand{\ten}[1]{\cdot 10^{#1}}
\newcommand{\abs}[1]{\left|#1\right|}
\renewcommand{\dot}{\cdot}
\newcommand{\cross}{\times}
\renewcommand{\div}{\nabla\dot}
\newcommand{\grad}{\nabla}
\newcommand{\curl}{\nabla\cross}
\newcommand{\defined}{\equiv}
\newcommand{\like}{\sim}
\newcommand{\conj}{{}^*}
\newcommand{\ord}[1]{\mathcal{O}\left(#1\right)}
\newcommand{\plusord}[1]{\, + \, \ord{#1}}

\newcommand{\approptoinn}[2]{\mathrel{\vcenter{
  \offinterlineskip\halign{\hfil$##$\cr
    #1\propto\cr\noalign{\kern2pt}#1\sim\cr\noalign{\kern-2pt}}}}}
\newcommand{\appropto}{\mathpalette\approptoinn\relax}

\newcommand{\tto}{\text{ to }}
\newcommand{\step}[1]{\theta\left(#1\right)}
\newcommand\numberthis{\addtocounter{equation}{1}\tag{\theequation}}
\renewcommand{\Re}[1]{\text{Re}\left[#1\right]}
\renewcommand{\Im}[1]{\text{Im}\left[#1\right]}

\newcommand{\pderiv}[2]{\frac{\partial #1}{\partial #2}}

\newcommand{\deriv}[2]{\frac{d #1}{d #2}}

\newcommand{\figref}[1]{Fig.\xspace\ref{#1}}
\renewcommand{\eqref}[1]{Eq.\xspace\ref{#1}}
\newcommand{\secref}[1]{Sec.\xspace\ref{#1}}
\newcommand{\citeref}[1]{Ref.\xspace\onlinecite{#1}}
\newcommand{\appref}[1]{Appendix\xspace\ref{#1}}
\newcommand{\tabref}[1]{Table\xspace\ref{#1}}
\newcommand{\chapref}[1]{Chapter\xspace\ref{#1}}

\newcommand{\code}[1]{\texttt{#1}\xspace}
\newcommand{\HYM}{\code{HYM}}
\newcommand{\TRANSP}{\code{TRANSP}}
\newcommand{\NUBEAM}{\code{NUBEAM}}
\newcommand{\NOVA}{\code{NOVA}}
\newcommand{\ORBIT}{\code{ORBIT}}
\newcommand{\CAETB}{\code{CAE3B}}

\newcommand{\Mathematica}{\code{Mathematica}}

\newcommand{\ViniciusFirst}{Vin\'{i}cius\xspace}
\newcommand{\StephaneFirst}{St\'{e}phane\xspace}

\newcommand{\Podesta}{Podest{\`a}}
\newcommand{\Fulop}{F{\"u}l{\"o}p}

\definecolor{darkgreen}{rgb}{0,0.5,0}

\usepackage[sort&compress,numbers,super]{natbib}
\usepackage{doi}%

\usepackage{appendix}
\usepackage{chngcntr}
\usepackage{etoolbox}
\AtBeginEnvironment{subappendices}{%
\chapter*{Appendices}
\addcontentsline{toc}{chapter}{Appendices}
\counterwithin{figure}{chapter}
\counterwithin{table}{chapter}
}

\newcommand{\onlinecite}[1]{\hspace{-1 ex} \nocite{#1}\citenum{#1}}

\makeatletter
\g@addto@macro\@floatboxreset{\centering}
\makeatother

\newcommand{\halfwidth}{.5\textwidth}
\newcommand{\midwidth}{.7\textwidth}
\newcommand{\fullwidth}{\textwidth}

\title{Theory and Simulations of Compressional and Global \Alfven Eigenmode Stability in Spherical Tokamaks}
\submitted{June 2020}  
\copyrightyear{2020}  
\author{Jeffrey Benjamin Lestz}
\adviser{Elena Belova and Nikolai Gorelenkov}  
\departmentprefix{Department of\hfill}  
\department{Astrophysical Sciences \\Program in Plasma Physics}

\usepackage{graphicx}

\usepackage{verbatim}

\usepackage{multirow}
\usepackage{longtable}

\usepackage{booktabs}

\ifdefined\printmode

\usepackage{url}

\else

\ifdefined\proquestmode

\usepackage{hyperref}
\hypersetup{bookmarksnumbered}

\makeatletter
\hypersetup{pdftitle=\@title,pdfauthor=\@author}
\makeatother

\else

\usepackage{hyperref}
\hypersetup{colorlinks,bookmarksnumbered}

\makeatletter
\hypersetup{pdftitle=\@title,pdfauthor=\@author}
\makeatother

\fi 
\fi 

\ifodd 0
\renewcommand{\maketitlepage}{}
\renewcommand*{\makecopyrightpage}{}
\renewcommand*{\makeabstract}{}

\else

\abstract{
Neutral-beam-driven, sub-cyclotron compressional (CAE) and global (GAE) \Alfven eigenmodes are routinely excited in spherical tokamaks such as NSTX(-U) and MAST, have been observed on the conventional aspect ratio tokamak DIII-D, and may be unstable in ITER burning plasmas. Their presence has been experimentally linked to the anomalous flattening of electron temperature profiles at high beam power in NSTX, potentially limiting fusion performance. A detailed understanding of CAE/GAE excitation, therefore, is vital to predicting (and ultimately controlling) their effects on plasma confinement. To this end, hybrid kinetic-MHD simulations, performed with the HYM code, are complemented with an analytic study of the linear stability properties of CAEs and GAEs. Perturbative, local analytic theory has been used to derive new instability conditions for CAEs/GAEs driven by realistic neutral beam distributions. A comprehensive set of simulations of NSTX-like plasmas has been performed for a wide range of beam parameters, providing a wealth of information on CAE and GAE stability in spherical tokamaks. This study is unique in that it uses a full orbit kinetic description of the beam ions in order to capture the Doppler-shifted cyclotron resonances which drive the modes. Linear simulations show that the excitation of CAEs vs GAEs has a complex dependence on the fast ion injection velocity and geometry, qualitatively described by the analytic theory developed in this thesis. Strong energetic particle modifications of GAEs are found in simulations, indicating the existence of a new type of high frequency energetic particle mode. A cross validation between the theoretical stability bounds, simulation results, and experimental measurements shows favorable agreement for both the unstable CAE and GAE spectra's dependence on fast ion parameters. The analytic results accurately explain the recent experimental discovery of GAE stabilization with small amounts of off-axis beam injection on NSTX-U and suggest new techniques for control of these instabilities in future experiments. 
}

\acknowledgements{
There are a huge number of people to credit for the existence of this thesis. First are my thesis advisors, Elena Belova and Nikolai Gorelenkov. I felt very fortunate to have advisors who were always available for discussions both on high level topics and also technical details in the research weeds. Together, you afforded me substantial freedom in determining my research priorities and approaching problems with my own style. Thank you for granting me the latitude to to tinker with subjects that I found alluring or even just personally amusing. I especially appreciated your counterweight to my perpetual impatience with my own research progress. It was a privilege to be mentored by such skilled physicists. I look forward to continued future collaboration. 

I am grateful to the NSTX-U collaboration for funding my thesis research and generously supporting my travel to many academic conferences and workshops. In particular, I thank Stan Kaye for approving the scope of this thesis, as well as signing a lot of paperwork for me. Officially this research was supported by the U.S. Department of Energy (NSTX contract DE-AC02-09CH11466) and computing resources at the National Energy Research Scientific Computing Center (NERSC). Thanks to everyone who pays their taxes so that I can do physics for a living. 

Beyond financial support, I received substantial assistance from many NSTX(-U) experimentalists over the years, both in collaboration on specific research projects and also in teaching me a vast amount of experimental miscellanea. Among these were Mario \Podesta, who also co-advised my first year ``experimental'' project, Eric Fredrickson, Neal Crocker, and Shawn Tang. Thank you for patiently entertaining my many inquiries and always being willing to help. Our interactions were consistently fruitful and enjoyable. I hope that we have the opportunity to continue in the future. 

Next, I must thank all of those involved in the approval of this thesis. In addition to my advisors, I deeply thank Amitava Bhattacharjee for consenting to read this thesis, and also for co-authoring my most referenced plasma physics textbook. I apologize for my verbosity. In addition, I thank Allan Reiman (who also acted as my faculty advisor), Roscoe White, and Eric Fredrickson for agreeing to serve on my thesis committee. Thank you all for your engagement and valuable contributions in the final stage of my PhD life cycle. A belated thanks to Ilya Dodin for agreeing to serve on the committee for my thesis proposal at the last minute a few years ago. 

To the rest of the faculty, thank you for providing us with comprehensive courses and acting as role models for us as scientists in training. I learned a lot from observing the types of questions and criticisms you raised and results that you regarded as significant in seminars and discussions. 

I was very fortunate to be assisted by three outstanding graduate program administrators during my tenure: Barbara Sarfaty, Beth Leman, and Dara Lewis. I appreciate that each of you took on so many projects beyond your official duties in order to improve the program for us. Collectively, you made navigating the hazards of graduate school boring and uneventful. I couldn't ask for anything more. 

My plasma physics research experience began prior to graduate school, advised by Roger Smith at the University of Washington and then here at PPPL by \StephaneFirst Ethier and Weixing Wang over a summer through the SULI program. I am grateful to Roger for taking a chance on me as a sophomore and teaching me to crawl in the research world. I thank \StephaneFirst and Weixing for making my summer research project enjoyable enough to make me want to come back for more. 

To all of my peers from the past six years, your presence consistently dulled the setbacks of graduate school, enhanced the occasional triumphs, and rendered the interim periods entertaining and enjoyable. I am grateful to Jonathan Shi, for providing truly inexhaustible technical, mathematical, and emotional support. What a privilege it has been to have you as a friend for the last decade. Thanks to my academic big brother, \ViniciusFirst Duarte. It was immensely helpful to have a companion to wander through the energetic particle wilderness with. I enjoyed my time in S219 with my long time office mate Jonathan Ng and his worthy successor Eric Palmerduca. Jonathan, thanks for your mentorship and for training me into a respectable ping pong player. Eric, thank you for allowing me to claim more than my share of the office spoils. 

I would be remiss not to recognize the fellow students in my program cohort: Ge Dong, Eugene Evans, Scott Keller, Denis St. Onge, and Qian Teng. It was a pleasure to march through courses, prelims, more courses, and finally generals together with such friendly and talented classmates. Each of you is more than capable of accomplishing the things that are important to you in life. 

I acknowledge all participants, enthusiastic and begrudging, for playing on the Tokabats softball team, which was one of the highlights of Princeton for me. A full list of players can be found in the official Tokastats spreadsheet, which I hope will be continued. For two years, our meager roster was absorbed by the Coprolites, who welcomed us onto their team, leading to a lot of great times and back to back championships. In addition to softball, a thriving ping pong scene emerged at PPPL during my tenure. I enjoyed facing off against all of the frequent players: Peter, Daniel, Lee, Jonathan, Lonathan, Noah, Vasily, Deepen, Mike, Charles, \ViniciusFirst, David, and many others. I still cherish my 2018 Bolgert Classic title. 

To the original Procter Hall/Butler/Roundabout crew -- Charles, Isabela, Ingrid, Kelsey, Eugene, Maria, Jacob, and Elijah -- thanks for the enjoyable meals, casual lounging, and lasting friendships thereafter. To the next generation cast of Eric, Suying, Oak, Laura, and Nick, thanks for tolerating my intermittent loitering in the party office and overall bolstering my time spent as a \emph{senior} graduate student. A second acknowledgment to the latter group plus Elijah, Mike, Ian, and Hongxuan for co-founding and sustaining Klub Tokamak. 

There are so many more students in our program whom I benefited tremendously from knowing, working with, and generally just being around. A non-exhaustive list of those not previously mentioned in roughly chronological order is: Brendan L., Tyler A., Josh B., Matt L., Dennis B., Seth D., Yuan S., Lee G., Peter J., Brian K., Jack M., Andy A., Alex G., Valentin S., Alec G., Ben I., Sierra J., Nick M., Eduardo R., Taz P., Alex L., Joe A., Kendra B., Nirbhav C., Steve M., and Tony Q. Each of you made the lab a more special place. 

Lastly, I acknowledge and am grateful for my family's immeasurable influence on this thesis. My parents instilled important values in me and entrusted me with the liberty to forge my own path in life. Thank you to mom and dad for all of the sacrifices you made in order to provide favorable initial conditions for this thesis. To my sister Loryn, thank you for taking the heat as the older child when we were younger, and for being a good influence as we matured into adults. I am grateful also for my grandfather Julian, who continues to serve as a role model to aspire towards. Without the unwavering love and support from my family, surely this thesis would not have been possible. 

}

\dedication{To Mom and Dad}

\fi  

\begin{document}

\makefrontmatter



\newcommand{\dirfigint}{ch-intro/figs}

\chapter{Introduction}
\label{ch:int:intro}

\section{A Brief Introduction to Fusion Energy} 
\label{sec:int:fusion}

Controlled thermonuclear fusion is currently being researched as a future source of sustainable commercial energy. Fusion is the most energy dense method of power generation aside from matter-antimatter annihilation. For example, roughly seven million kilograms of oil must be combusted to produce the same amount of energy as one kilogram of fuel for fusion. By comparison, roughly six kilograms of fuel would be required for the same energy output in a modern nuclear fission power plant. The benefits of fusion energy over existing methods of power production, as well as the outstanding scientific, technological, and engineering challenges have been written about extensively and will not be repeated here (see \citeref{FreidbergFusion} and references therein). 
Instead, a very brief, high level background will be given in order to place the work of this thesis into context for a less specialized audience and motivate its contribution towards this goal. 

\begin{figure}[tb]
\includegraphics[width = \midwidth]{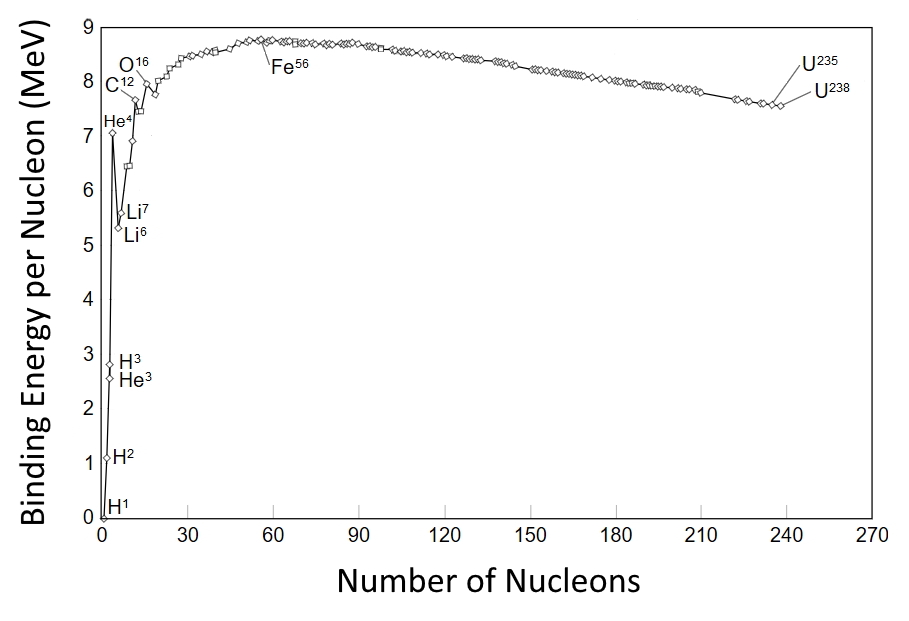}
\caption
[Nuclear binding energy curve.]
{Average nuclear binding energy per nucleon as a function of number of nucleons in the nucleus.
Reproduced from the \href{https://commons.wikimedia.org/wiki/File:Binding_energy_curve_-_common_isotopes.svg}{Wikimedia Commons} (public domain).}
\label{fig:int:bindcurve}
\end{figure}

Nuclear fusion occurs when two atomic nuclei are merged into a heavier nucleus. The difference in total mass of the initial vs final particles is released as energy in accordance with Einstein's relation $\Delta E = \Delta m c^2$. The nuclear binding energy curve in \figref{fig:int:bindcurve} describes how tightly bound each nucleus is. A nuclear reaction which starts at a state of lower binding energy and finishes with higher binding energy will release energy $(\Delta E > 0)$, while the opposite requires energy to occur $(\Delta E < 0)$. Since the nuclear binding energy curve peaks at an isotope of iron,  light nuclei, such as isotopes of hydrogen, helium, \etc release energy when fusion occurs, whereas all isotopes heavier than iron release energy during the opposite process: nuclear fission. 

Nuclear fusion occurs due to the strong nuclear force, which attracts nuclei to one another. However, this force only acts at extremely short ranges at the scale length of an atomic nucleus: $10^{-15}$ m. In order to induce fusion, two nuclei must be brought extremely close together. At larger distances, the force between charge neutral atomic nuclei is repulsive due to valence electron-electron forces (either electrostatic repulsion or chemical forces which lead to the formation of molecules but prevent overlap of the atomic nuclei). If the nuclei have instead been ionized, electrostatic repulsion exists due to the positively charged protons instead of the valence electrons. The general strategy to induce fusion is to heat ions to very high temperatures (greater than 10 million degrees Celsius) such that the the positively charged nuclei have a non-negligible chance of overcoming the Coulomb barrier and fusing. 

\begin{figure}[tb]
\includegraphics[width = 0.6\textwidth]{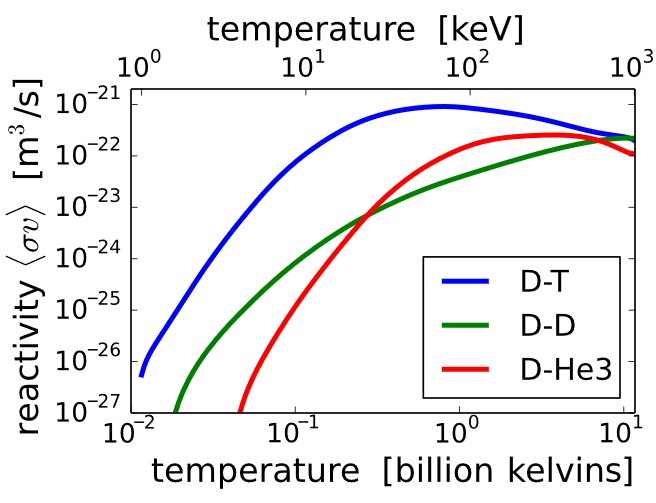}
\caption
[Fusion reactivity as a function of temperature.]
{Fusion reactivity as a function of temperature.
Reproduced from the \href{https://commons.wikimedia.org/wiki/File:Fusion_rxnrate.svg}{Wikimedia Commons}
under the \href{https://creativecommons.org/licenses/by/2.5/deed.en}{Creative Commons Attribution License}.}
\label{fig:int:reactivity}
\end{figure}

The likelihood that a fusion reaction will occur is characterized by the reactivity, which depends both on the ion temperature and specific pair of nuclear isotopes as shown in \figref{fig:int:reactivity} for a few common fusion reactions. Those most relevant to this thesis are 

\begin{subequations}
\begin{align}
\text{D}^2 + \text{D}^2 &\longrightarrow \text{He}^3 + \text{n} + 3.27 \text{ MeV} \,(50\%)\\ 
\text{D}^2 + \text{D}^2	&\longrightarrow \text{T}^3\hphantom{e} + \text{p} + 4.03 \text{ MeV} \,(50\%) \\ 
\text{D}^2 + \text{T}^3 &\longrightarrow \text{He}^4 + \text{n} + 17.6 \text{ MeV}
\label{eq:int:dt}
\end{align}
\end{subequations}

Here, D stands for deuterium, an isotope of hydrogen with one proton (p) and one neutron (n), and T represents tritium, an isotope of hydrogen with two neutrons. He$^3$ and He$^4$ are isotopes of helium with three and four total nucleons, respectively. He$^4$ is referred to as an $\alpha$ particle for historical reasons. Since DT fusion has the largest reactivity, it is the fuel of choice for future fusion reactors. However, tritium is radioactive and poses health risks if inhaled or ingested (for instance, through environmental contamination), so its use in modern fusion experiments is uncommon due to additional costs associated with safety procedures. Instead, pure deuterium fuel is prevalent in experiments, which is not radioactive and therefore poses no health risks.\footnote{Although DD reactions generate tritium as a fusion product, the rate of fusion in modern experiments is so low that the amount of tritium produced is negligible compared to the quantities needed for a DT reactor and consequently does not constitute a hazard.}

\begin{figure}[tb]
\includegraphics[width = 0.8\textwidth]{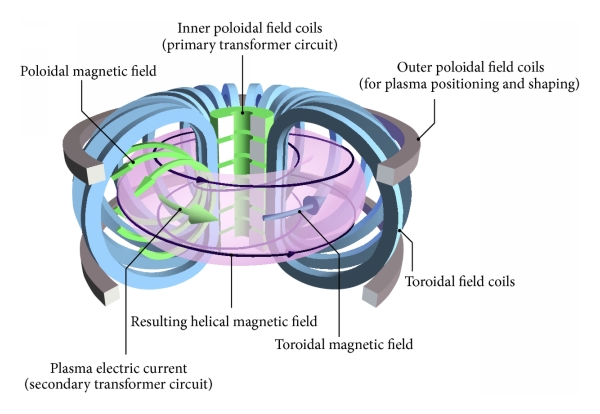}
\caption
[Schematic diagram of magnetic field generation for a tokamak.]
{Schematic diagram of magnetic field generation for a tokamak. Reproduced from \citeref{FigTokamak} 
under the \href{https://creativecommons.org/licenses/by/3.0/}{Creative Commons Attribution License}.}
\label{fig:int:tokamak}
\end{figure}

In addition to the temperature requirements, the plasma (ions and their stripped electrons) must also be confined to a certain volume so that the ions have a chance of colliding and initiating a fusion reaction. Stars such as the sun which are powered by fusion confine their plasma gravitationally -- this is not an option for terrestrial fusion reactors since it requires astronomical amounts of mass and space. While there are many methods of confinement, the one that has historically received the most research attention is toroidal magnetic confinement. In this approach, a strong helical magnetic field is established in toroidal geometry such that the charged particles will (ideally) spiral around the field lines indefinitely due to the Lorentz force. In a tokamak,\footnote{Tokamak is an English transliteration of a Russian acronym for ``toroidal chamber with axial magnetic field.''} the magnetic field is set up by a combination of external coils and the poloidal field generated by an inductively-driven toroidal current in the plasma. This configuration is shown in \figref{fig:int:tokamak}. Unfortunately, the plasma is not confined as well as this idealized scheme would imply. Several spontaneous mechanisms exist in reality which degrade the plasma's confinement, ranging from macroscopic instabilities to microscopic turbulent processes. The plasma also emits electromagnetic radiation due to the acceleration of charged particles. Consequently, the plasma will ``leak'' energy at a certain rate, cooling the plasma and reducing the rate of fusion reactions unless constant heating power is provided to maintain its temperature. 


Consider again the DT fusion reaction in \eqref{eq:int:dt}. Due to momentum conservation, 14.1 MeV of the energy released is carried by the neutron as kinetic energy and 3.5 MeV is carried by the $\alpha$ particle. Charge neutral, the neutron is unaffected by the magnetic field and rapidly leaves the confinement region. In a power plant, the neutrons will be caught in an exterior blanket, converting their kinetic energy into heat which can be used to generate electricity through conventional methods (\eg boil water to spin a turbine that drives an electric generator via induction). In contrast, the $\alpha$ particle is charged and can be confined by the magnetic field with the rest of the plasma. Ideally, the $\alpha$ particles will transfer their energy to the background fuel before leaving the system as cold ash, replenishing the plasma's stored energy as it is lost due to the mechanisms mentioned previously. Ignition occurs when the plasma heating by the fusion products balances the heat loss processes, without the need for auxiliary heating from external sources (to be discussed in \secref{sec:int:ep}). The goal for a future commercial fusion reactor is to reach ignition. 

A remaining milestone before ignition is ``scientific breakeven," where the fusion power generated is greater than the total heating power required to sustain the reaction. It is characterized by the gain factor $Q = P_\text{fusion}/P_\text{heat,ext}$. Hence, $Q > 1$ is the condition for the power generated by fusion to exceed the power required to heat the plasma to fusion temperatures (breakeven). Ignition corresponds to $Q = \infty$ since it requires no external heating -- the reaction is self-sustaining. To date, the highest achieved fusion gain was $Q = 0.62$ by the Joint European Torus (JET) in 1997 with DT fuel.\cite{Watkins1999} Shortly thereafter, the Japanese tokamak JT-60 reached plasma conditions with DD fuel that would correspond to $Q = 1.25$ if achieved with DT fuel.\cite{Fujita1999NF} A massive international collaboration is currently underway to construct the ITER tokamak in southern France, which is being designed to achieve a transient peak value of $Q = 10$ and sustain $Q = 5$ for several minutes. Its plasma volume will be ten times larger than the largest existing tokamak, JET. Construction is scheduled to finish in 2025, with initial experiments beginning soon thereafter, and high performance DT experiments commencing around 2035. 

\section{Plasma Heating, Energetic Particles, and Related Instabilities} 
\label{sec:int:ep}

In current experiments, the plasma is heated primarily by external sources due to the low amount of fusion power generated. Some heating is provided by the resistive dissipation of the inductively-driven plasma current. 
Additional heating is required to reach the high temperatures desired for fusion experiments. 
There are many methods to inject electromagnetic waves with external antennas in order to heat a resonant sub-population of the plasma. This resonant population subsequently heats the rest of the plasma via collisions. 

Another very common method which is intimately connected to the subject of this thesis is neutral beam injection (NBI). In this scheme, large linear devices external to the tokamak chamber electrostatically accelerate ions to high energies much larger than the desired temperature of the plasma. Just before being launched into the tokamak, the ions are sent through a volume of neutral gas in order to neutralize the energetic ions through charge exchange collisions, allowing the energetic particles to penetrate into the plasma. 

Once within the plasma, charge exchange will ionize most of the energetic particles near the plasma core (if they entered as ions instead of neutrals, they would be immediately confined to the edge by the magnetic field and therefore ineffectively heat the plasma). The energetic beam ions subsequently transfer their energy to the background plasma through collisions.\cite{WessonTokamak} Fusion born $\alpha$ particles heat the background plasma in a similar fashion. Fusion plasmas often have temperatures of the order $T \like 1 - 10$ keV,\footnote{Here and elsewhere, ``temperatures'' quoted as energies include implicit multiplication by the Boltzmann constant $k_B = 1.38\ten{-23}$ m$^2$ kg s$^{-2}$ K$^{-1}$, as is standard in the plasma physics community.} whereas NBI involves energies of $\W_{beam} \like 40 - 100$ keV in modern experiments. In ITER, 1 MeV beam ions and 3.5 MeV $\alpha$ particles will also be present. These ions satisfy the ordering $\vthermi \ll \vep \ll \vtherme$ where $\vthermi$ and $\vtherme$ are the background (thermal) ion and electron temperatures, respectively. Hence, they are referred to as ``energetic particles'' (EP) or ``fast ions'' in order to distinguish them from the Maxwellian background.\footnote{Runaway electrons can also become energetic particles in tokamaks, satisfying $\vtherme \ll \vep$, though their dynamics are somewhat different from those of fast ions and are not studied in this thesis.}

The steady state velocity distribution for a constant source of ions injected with velocity $\vb$ has previously been calculated analytically and is known as the slowing down distribution\cite{Gaffey1976JPP}

\begin{align}
f(v) &= \frac{\theta(\vb - v)}{v^3 + \vc^3}
\label{eq:int:slowingdown}
\end{align}

Here, $\theta(x)$ is the heaviside step function, defined piecewise as 1 for $x > 0$ and 0 otherwise. Hence it cuts off the distribution at velocities above the injection velocity. In the denominator, $\vc$ is the critical velocity, defined as $m_b \vc^2/2 = 14.8 A_b T_e / A_i^{2/3}$. Here, $A_i$ and $A_b$ are the atomic numbers of the thermal ions and neutral beam ions, respectively, and $m_b$ is the mass of the beam ions. When $\vb > \vc$, the injected ions will transfer most of their heat to the thermal electrons and the rest to the thermal ions, which then equilibrate through their own collisions. This is the usual case for neutral beam injection in modern experiments, and will also be true for energetic particles in ITER. The slowing down distribution is a valid description of fusion products (which are always born at a specific energy) and also NBI ions, with one caveat. Consider the case of deuterium NBI. Due to the neutralization process described in the beam generation, neutral molecular deuterium D$_2$ and D$_3$ also form in addition to neutral D atoms. Since these species all have the same energy (proportional to the beam voltage) but different masses, they enter the plasma at different energies. Hence, NBI distributions are more accurately described as a weighted sum over slowing down distributions with injection velocities $\vb$, $\vb/\sqrt{2}$, and $\vb/\sqrt{3}$. 


Because the beams are injected into the plasma at a specific location and with a specific orientation relative to the background magnetic field, the distribution function of their fast ions will be anisotropic in velocity space. The peak value of the pitch of the NBI distribution in the core can be estimated as $\vpar/v \approx \Rtan/\Rzero$, where $\Rtan$ is the value of the major radius where the beam line is tangent to the magnetic field (known as the tangency radius), and $\Rzero$ is the radius of the magnetic axis (unique distance where the poloidal magnetic field vanishes). Further discussion of suitable models of the distribution function will be discussed in \secref{sec:cyc:dispres} and \secref{sec:sim:model}. 

Although they comprise a minority of the plasma by density $(n_\text{EP} \ll n_e)$, the energetic particle pressure can be comparable to the thermal plasma pressure due to their large energy even in current devices. Therefore, they require special consideration in fusion plasmas because they introduce qualitatively different physical behavior from what is expected due to the background ions and electrons. Fundamentally, energetic particles must be treated as a kinetic species. Fast ion orbit widths and Larmor radii are much larger than those of the thermal species. 
Importantly, the typically non-Maxwellian fast ion distributions can be a source of free energy to destabilize plasma waves that would otherwise be stable. As will be discussed in detail in \secref{sec:cyc:derivation}, waves can be driven unstable or further stabilized by fast ions depending on the sign of gradients in the fast ion distribution. This fact was well known at the time of Mikhailovskii's 1975 review of instabilities in a non-Maxwellian plasma.\cite{Mikhailovskiiv6} Instabilities driven by fusion products were first identified by Kolesnichenko in 1967\cite{Kolesnichenko1967SAE} and treated specifically for shear \Alfven waves driven by neutral beam ions in a tokamak by Rosenbluth in 1975.\cite{Rosenbluth1975PRL}

In some cases, fast-ion-driven instabilities can have benign effects on plasma performance. In those circumstances, they need not be avoided and can even serve as effective diagnostics for background plasma properties,\cite{Fasoli2002PPCF} pellet injection,\cite{Sharapov2018NF,Oliver2019NF} or the fast ion distribution.\cite{McClements2015NF}
The interaction between energetic particles and plasma waves is most frequently studied within the context of fast ion transport. Fast ions can drive waves which consequently lead to fast ion redistribution or loss from the system, impacting their ability to heat the background plasma and potentially damaging the device.\cite{Heidbrink2020POP} 
Theoretically, judicious excitation of waves can also be used to more effectively transfer the energy of fusion products to the background plasma using the $\alpha$ channeling scheme.\cite{Fisch1992PRL,Fisch2000NF,Gorelenkov2010PPCF}
The motivation for the work of this thesis, to be described in more detail in \secref{sec:int:teflat}, is a situation where instabilities excited by fast ions can modify the electron temperature profile without any fast ion transport.\cite{Stutman2009PRL}
In general, the study of fast-ion-driven instabilities is concerned with predictability and control. In order to avoid their deleterious effects or harness advantageous ones, we must be able to reliably predict their excitation and self-consistent interaction with the fast ions and background plasma. 


Effective energy exchange (driving or damping the wave) between a wave and fast ions requires a resonant wave-particle interaction. Conceptually, resonance occurs when periodic motion of a particle is synchronized with the periodic fluctuation of a wave such that the particle experiences a net force due to the wave during each orbit. More precisely, this condition can be stated as $\oint \vv\dot\delta\vE \, dt \neq 0$, where the time integration is over a periodic orbit. Integration of an analogous expression over the full fast ion distribution yields the total energy exchange. In a tokamak, resonance occurs when the following general condition is satisfied: 

\begin{align}
\omega - n\avg{\omegator} - p\avg{\omegapol} = \lres\avg{\omegacs}
\label{eq:int:res}
\end{align}

Here, $\avg{\dots}$ denotes orbit averaging over time, $\omega$ is the wave frequency, $\avg{\omegator}$ is the average toroidal orbital frequency, $\avg{\omegapol}$ is the average poloidal orbital frequency, and $\avg{\omegacs}$ is the average cyclotron frequency of the species of interest: $\omegacs = \abs{q_s}B/m_s$. The frequencies must be related by integers $n$, $p$, and $\lres$. The toroidal mode number of the wave defines $n = \ktor \Rzero$, while $p$ can be different from the wave's poloidal mode number $m$ due to particle drift motion. The integer $\lres$ is the cyclotron coefficient, typically equal to zero for low frequency waves and non-zero for waves in the ion cyclotron frequency range. \eqref{eq:int:res} implies a ``slow instability,'' \eg with $\gamma \ll \omega_{\text{bounce}}$, since it describes a net wave-particle interaction over a full orbit (global resonance) as opposed to an instantaneous/transient resonant interaction which would describe a ``fast instability'' (local resonance). Further discussion of the local vs global resonance conditions can be found in \citeref{Gorelenkov1995POP}, \citeref{Belikov2003POP}, and \citeref{Zhang2015NF}. The resonance condition in \eqref{eq:int:res} can be equivalently written in terms of wave vectors and velocities instead of frequencies: 

\begin{align}
\omega - \avg{\kpar\vpar} - \avg{\kperp\vdrift} = \lres\avg{\omegaci}
\label{eq:int:reskpar}
\end{align} 

Here, $\kpar$ and $\kperp$ are projections of the wave vector $\vk$ parallel and perpendicular to the background magnetic field, while $\vpres$ and $\vdrift$ are the same projections for the fast ion guiding center velocity. These equivalent conditions are known as the general Doppler shifted cyclotron resonance condition. Note that here and for the rest of the paper, the ``Doppler shift" refers to the shift in the resonance due to a particle's parallel and drift motion, not the bulk rotation of the plasma. The Landau resonance corresponds to $\lres = 0$, the ``ordinary'' cyclotron resonance has $\lres = 1$, and the ``anomalous'' cyclotron resonance has $\lres = -1$. For sub-cyclotron frequencies, and in the usual case where $\abs{\avg{\kpar\vpar}} \gtrsim \abs{\avg{\kperp\vdrift}}$, counter-propagating modes $(\kpar < 0)$ can only satisfy the ordinary cyclotron resonance, while co-propagating modes can interact through the Landau or anomalous cyclotron resonances, depending on their frequency. In this work, co- and cntr-propagation are defined relative to the direction of plasma current and beam injection. 

With the understanding that fast ions can resonantly destabilize plasma waves, we will now introduce the specific types of waves that are the subject of this thesis. 

\section{Waves in a Uniform Magnetized Plasma} 
\label{sec:int:mhdwaves}

Only a small subset of the rich ecosystem of plasmas waves is relevant to this thesis. The rest are discussed in textbooks such as \citeref{StixWaves}. Of interest to this work are the waves which exist in magnetohydrodynamics (MHD). MHD is a model which describes the thermal plasma as a charged fluid. It is applicable when (1) the system is sufficiently collisional $(\lambda_\text{mfp} \ll L)$ such that the electrons and ions have Maxwellian distributions, (2) the plasma is macroscopically quasineutral $(\lambda_D \ll L)$, and (3) the plasma is non-relativistic such that wave phase velocities are small compared to the speed of light $(v_p = \omega/k \ll c)$. 

\subsection{Low Frequency MHD Waves}
\label{sec:int:lowfreqmhd}

First, consider the simple scenario of low frequency waves, where $\omega \ll \omegaci$. 
Under the MHD assumptions listed above, the ideal (zero resistivity) MHD system can be described by the following set of equations

\begin{align}
\curl \vE = -\pderiv{\vB}{t} &\describe{Faraday's Law} \label{eq:int:faraday}\\
\curl \vB = \mu_0\vJ &\describe{Ampere's Law} \label{eq:int:ampere}\\
\div \vB = 0 &\describe{Magnetic Laplace Equation} \label{eq:int:divb} \\ 
\pderiv{\rho}{t} + \div\left(\rho \vv\right) = 0 &\describe{Continuity Equation} \label{eq:int:cont} \\ 
\deriv{}{t}\left(\frac{P}{\rho^\gammaa}\right) = 0 &\describe{Adiabatic Equation of State} \label{eq:int:adiabatic} \\
\rho\deriv{\vv}{t} = \vJ \cross \vB - \grad P &\describe{Momentum Equation} \label{eq:int:momentum}\\
\vE + \vv \cross \vB = 0 &\describe{Ideal Ohm's Law} \label{eq:int:ohm}
\end{align}

Moreover, consider singly charged ions, such that $q_i = -q_e \defined e$, and also strong quasineutrality, so that $n_i = n_e \defined n$. The single fluids in the above equations are defined by: 

\begin{align}
\rho &= \sum_s m_s n_s \approx m_i n \\
\vv &= \frac{1}{\rho}\sum_s \rho_s \vv_s \approx \vv_i \\
\vJ &= \sum_s q_s n_s \vv_s = e n (\vv_i - \vv_e) \approx \vJ_e \\
P &= \sum_s m_s \int \left(\vv_s - \vv\right)\left(\vv_s - \vv\right)f_s d^3\vv 
\end{align}

\eqref{eq:int:faraday}, \eqref{eq:int:ampere}, and \eqref{eq:int:divb} are Maxwell's equations, with the assumption that the displacement current is small $(J \gg \varepsilon_0 \partial E/\partial t)$. Due to quasineutrality, Gauss's law is formally not included in the MHD system (see the rigorous discussion in \citeref{FreidbergMHD} and \citeref{GoedbloedMHDv1}). \eqref{eq:int:cont} describes particle conservation, \eqref{eq:int:adiabatic} imposes an adiabatic equation of state with adiabatic index $\gammaa$, \eqref{eq:int:momentum} is the momentum equation, and \eqref{eq:int:ohm} is the low frequency ideal Ohm's law -- it is only valid for waves where $\omega\ll\omegaci$. The system is arrived at by taking velocity moments of the Vlasov equation $d f_s/dt = 0$ for each species, summing over species, and imposing the MHD assumptions and electron-ion mass ordering $m_e \ll m_i$. 

For most of this thesis, we will consider the \emph{perturbative} effect of fast ions on MHD waves. A linearly growing or decaying wave oscillates with complex frequency $\omega + i\gamma$ like $e^{-i(\omega + i\gamma)t}$, where $\gamma > 0$ corresponds to wave growth and $\gamma < 0$ corresponds to decay. We assume that the kinetic effect of fast ions can be treated as a small perturbation on the MHD system, such that $\abs{\gamma} \ll \omega$ and the real part of the frequency $\omega$ is unchanged from its MHD description. When the wave amplitude grows large enough, nonlinear effects become relevant which end the exponential growth and eventually saturate the wave amplitude. This model has been successful in explaining experimental observations of fluctuations in the MHD frequency range. 

The nonperturbative regime can also exist in experiments, where the observed mode is not a solution of the MHD system, but rather its dispersion relation fundamentally depends on properties of the energetic particles. Such modes are referred to as energetic particle modes (EPMs) since they do not exist (stable or unstable) without the presence of energetic particles.\cite{Chen1994POP} A useful distinction between EPMs and other types of plasma waves is that ordinary plasma waves can be excited with an external antenna tuned to the mode frequency, whereas EPMs can not (unless energetic particles are also present). The fishbone instability was the first experimentally observed EPM,\cite{McGuire1983PRL} which was explained theoretically soon thereafter.\cite{Chen1984PRL} A more recent example is the energetic-particle-induced geodesic acoustic mode (EGAM).\cite{Fu2008PRL,Nazikian2008PRL}

A linearization procedure is used to solve for the MHD waves. Each quantity is decomposed into a steady state and fluctuating part, for example $\vB = \vB_0(\vx) + \delta\vB(\vx,t)$. The fluctuating component is assumed to be much smaller in magnitude than the background component, $\abs{\delta\vB} \ll \abs{\vB_0}$, such that only terms to leading order in fluctuating quantities are kept. The \emph{ansatz} $\delta\vB \like e^{-i\omega t}$ allows the replacement $\partial/\partial t \rightarrow -i\omega$. In a general inhomogeneous system with spatial dependence of the equilibrium quantities, a complicated, coupled second order vector system of partial differential equations will need to be solved in order to determine the eigenfrequency, eigenfunction solutions. 

It is instructive to start instead with a uniform slab model with constant background magnetic field, which can be easily solved analytically. The discussion of this scenario is based on the remarks presented in Ch. 6.5 of \citeref{GurnettMHD}. Due to the uniform background, spatial Fourier transforms can be taken to replace $\grad \rightarrow i\vk$. Then defining the $\hat{z}$ direction to be the direction of the background magnetic field, the fluctuations must obey the following relation which can be derived from the linearized MHD equations: 

\begin{align}
\omega^2\delta\vv = \va^2\left\{\vk\cross\left(\vk\cross\left[\delta\vv\cross\hat{z}\right]\right)\right\}\cross\hat{z} + \vs^2\vk\left(\vk\dot\delta\vv\right)
\label{eq:int:mhdwaves}
\end{align}

Above, $\va^2 = B_0^2 / \sqrt{\mu_0 \rho_0}$ is the \Alfven speed and $\vs^2 = \gammaa P_0 / \rho_0$ is the sound speed ($\gammaa$ is the adiabatic index from \eqref{eq:int:adiabatic}). Note that the ratio $\vs^2 / \va^2 \propto \beta \ll 1$ in tokamaks, where $\beta = 2 \mu_0 P_0 / B_0^2$ is the ratio of plasma pressure to magnetic pressure. \eqref{eq:int:mhdwaves} yields the following dispersion relation

\begin{align}
\left(\omega^2 - \kpar^2\va^2\right)\left(\omega^4 - \omega^2 k^2(\va^2 + \vs^2) + k^2\kpar^2\va^2\vs^2\right) = 0
\label{eq:int:mhddisp}
\end{align}

Here, $\kpar = \vk\dot\vB_0$ denotes the component of the wave vector that is parallel to the background magnetic field. The roots of this equation give the three MHD waves: 

\begin{align}
\omega = \abs{\kpar}\va &\describe{Shear \Alfven Wave} \label{eq:int:shear}\\
\omega = \frac{k}{2}\left[\va^2 + \vs^2 + \sqrt{(\va^2 - \vs^2)^2 + \frac{4\kperp^2\va^2\vs^2}{k^2}}\right] &\describe{Fast Magnetosonic Wave} \label{eq:int:fastwave}\\
\omega = \frac{k}{2}\left[\va^2 + \vs^2 - \sqrt{(\va^2 - \vs^2)^2 + \frac{4\kperp^2\va^2\vs^2}{k^2}}\right] &\describe{Slow Magnetosonic Wave} \label{eq:int:slowwave}
\end{align}

The shear \Alfven wave (SAW) is polarized with $\delta\vE$ aligned with $\vk_\perp$, while $\delta\vB\dot\vk = \delta\vv\dot\vk = \delta n = 0$. \ie the fluctuation is incompressible. Its group velocity points directly along the magnetic field lines. In fact, this wave is directly analogous to transverse waves on a string, where the parallel magnetic field pressure plays the role of the string tension, and the string mass density is replaced by the ion mass density. 

In general, the slow and fast magnetosonic waves behave as a combination of transverse electromagnetic waves and longitudinal sound waves. However, we are interested in the fusion-relevant case when $\beta \ll 1$, where the slow wave nearly vanishes and the fast wave dispersion relation simplifies significantly to $\omega \approx k\va$. This zero pressure limit of the magnetosonic waves will be referred to as the compressional \Alfven wave (CAW). It can propagate at any angle relative to the background magnetic field. In contrast to the shear wave, the fast wave is polarized such that $\delta\vE\dot\vk = 0$ and has finite $\delta n$, $\delta\vB\dot\vkperp$, and $\delta\vv\dot\vk$. 

\subsection{Two-Fluid Corrections}
\label{sec:int:subcycmhd}

For higher frequency waves approaching the ion cyclotron frequency, a different approach is needed. One way forward would be to generalize the Ohm's Law in \eqref{eq:int:ohm} to include the Hall term and electron pressure gradient $(\vJ \cross \vB - \grad P_e) / n e$ on the right hand side. However, since we are most interested in the low $\beta$ limit, we will instead start from the cold plasma dielectric tensor in order to capture the two-fluid corrections necessary at frequencies approaching the ion cyclotron scale. 

Maxwell's equations can be combined to give the homogeneous wave equation 

\begin{equation}
\abs{\epsilon_{ij} - n^2\left(\delta_{ij} - \frac{k_i k_j}{k^2}\right)} = 0 
\label{eq:int:fullwave}
\end{equation}

Here, $n = kc / \omega$ is the index of refraction ($c$ is the speed of light), and $\epsilon_{ij}(\omega,\vk) = \delta_{ij} + \sum_s \epsilon_{ij}^s$ is the dielectric tensor summed over the thermal electrons and ions. Defining $\vk = \kperp\hat{x} + \kpar\hat{z}$, its form is 

\begin{align}
\label{eq:int:dielectric}
\epsilon_{ij} &= \left(\begin{array}{ccc} 
S & -iD & 0 \\ 
iD & S & 0 \\ 
0 & 0 & P 
\end{array}\right) \qquad\text{where }\\ 
S &\defined 1 - \sum_s \frac{\omega_{ps}^2}{\omega^2 - \omega_{cs}^2} \approx \frac{c^2}{\va^2}\frac{1}{1 - \omegabar^2} = \frac{c^2}{\va^2}A \\ 
D &\defined \sum_s \frac{\omega_{cs}}{\omega}\frac{\omega_{ps}^2}{\omega^2 - \omega_{cs}^2} \approx -\frac{c^2}{\va^2}\frac{\omegabar}{1 - \omegabar^2} = -\frac{c^2}{\va^2}\omegabar A \\ 
P &\defined 1 - \sum_s \frac{\omegaps^2}{\omega^2} \approx -\frac{c^2}{\va^2}\frac{m_i}{m_e}\frac{1}{\omegabar^2} \\
A &\defined \frac{1}{1 - \omegabar^2}
\end{align}

The approximations for the $S$, $D$, and $P$ functions are valid when $\omega \ll \omegape,\abs{\omegace}$. 
For high conductivity conditions, $\Epar \ll \Eperp$, so only the perpendicular components of the dielectric tensor must be considered. Then the dispersion and polarization of the modes can be determined from 

\begin{align}
\begin{pmatrix}
A - \Npar^2 & i\omegabar A \\ 
-i\omegabar A & A - N^2
\end{pmatrix}
\begin{pmatrix}
E_x \\ E_y 
\end{pmatrix}
= 0 
\label{eq:int:twofluid}
\end{align}

Here, we have defined the \Alfven refractive index $N = k\va/\omega$. \eqref{eq:int:twofluid} determines the dispersion relation, which is 

\begin{align}
N^2 &= \frac{A G}{2 F^2}\left[1 \pm \sqrt{1 - \frac{4 F^2}{A G^2}}\right] 
\label{eq:int:twodisp}
\end{align}

The following shorthand has been introduced: $F^2 = \kpar^2/k^2$ and $G = 1 + F^2$. 
For $\omega < \omegaci$, the ``$+$'' solution of \eqref{eq:int:twodisp} is the shear wave and the ``$-$'' solution is the compressional wave. For $\omega > \omegaci$, the shear wave no longer propagates (though it can in a warm plasma model), and the ``$+$'' solution becomes the compressional wave instead. 

\begin{figure}[tb]
\subfloat[\label{fig:int:twofluid-disp}]{\includegraphics[width = \halfwidth]{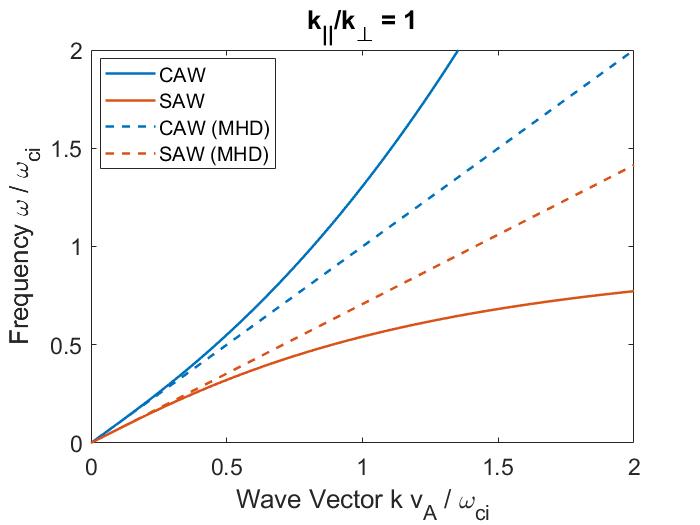}}
\subfloat[\label{fig:int:twofluid-pol}]{\includegraphics[width = \halfwidth]{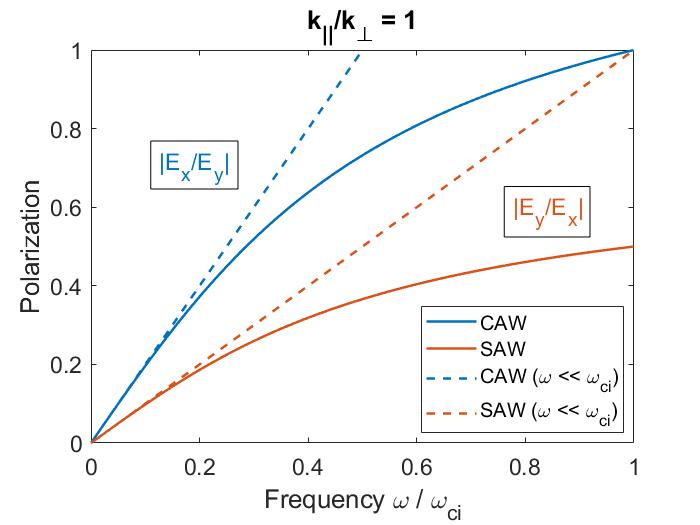}}
\caption
[Two fluid modification of MHD wave dispersion and polarization.]
{Two fluid modifications of the compressional and shear \Alfven waves for $\krat = 1$. (a) Normalized frequency $\omeganorm$ as a function of normalized wave vector $k\va/\omegaci$. Solid lines are the two-fluid dispersions -- blue for CAW and orange for SAW. Dashed lines are the MHD (single fluid) approximations. (b) Polarization as a function of $\omeganorm$. Blue solid line shows $\abs{E_x/E_y}$ for CAW. Orange solid line shows $\abs{E_y/E_x}$ for SAW. Dashed line shows the $\omeganorm \ll 1$ approximations.}
\label{fig:int:twofluid}
\end{figure}

The two-fluid corrections to the dispersion are shown in \figref{fig:int:twofluid-disp} for the case of $\krat = 1$. The modifications limit the SAW frequency at $\omega < \omegaci$, whereas the MHD dispersion is unbounded. The CAW frequency is increased relative to its single fluid form. The size of these corrections becomes larger for larger values of $\krat$ and can be important when interpreting experimental observations when $\omeganorm \ll 1$ is not satisfied.  

Importantly, the polarization of the two waves becomes mixed due to finite frequency effects. In the pure ideal MHD waves derived in \secref{sec:int:lowfreqmhd}, the CAW is polarized such that $\vE$ is in the $\hat{y}$ direction (perpendicular to $\vk$), and the SAW has $\vE$ in the $\hat{x}$ direction (aligned with $\vk_\perp$). This mixing is shown in \figref{fig:int:twofluid-pol}. From \eqref{eq:int:twofluid}, the polarization is given by $\abs{E_x/E_y} = (N^2 / A - 1)\omegaci/\omega$. To compare with the low frequency limit, the CAW polarization to leading order in $\omeganorm \ll 1$ is $\abs{E_x/E_y} \approx (\omeganorm)(1 + \kpar^2/\kperp^2)$ and the SAW in this same limit has $\abs{E_y/E_x} \approx (\omeganorm)(\kpar^2/\kperp^2)$. Hence, in the limit of $\kpar \gg \kperp$, both waves have $\abs{E_x} \approx \abs{E_y}$ even for $\omeganorm \ll 1$ -- a significant departure from the single fluid theory. The polarization affects how the wave interacts with fast ions on the Larmor radius scale, which will be discussed in more detail in \secref{sec:cyc:stabao} and \secref{sec:lan:stabao}. 

While the uniform slab model is a very crude approximation to a tokamak, it nonetheless provides intuition for the character of the waves that are present in more realistic conditions. 

\section{\Alfven Eigenmodes in a Tokamak}
\label{sec:int:tokamakwaves}


\begin{figure}[tb]
\includegraphics[width = \midwidth]{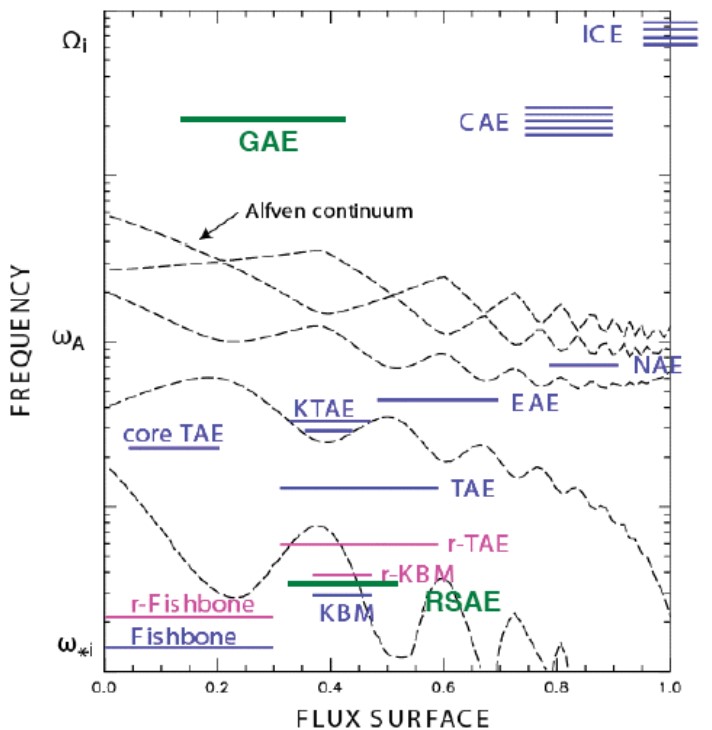}
\caption
[The zoo of \Alfven eigenmodes in a tokamak.]
{The zoo of \Alfven eigenmodes in a tokamak. The frequencies $\omega_{*i}$, $\omega_A$, and $\Omega_i$ are the ion diamagnetic freuqency, \Alfven frequency, and ion cyclotron frequency, respectively. Dashed curves indicate the \Alfven continuum. Adapted from \citeref{Heidbrink2002POP} by J.W. Van Dam. Reproduced with the permission of AIP Publishing.}
\label{fig:int:alfvenzoo}
\end{figure}

In toroidal geometry, the MHD waves discussed above become more complicated. There are still two main branches of the dispersion: the compressional and shear waves, but the eigenfrequencies and eigenfunctions (mode structures) will depend on details of the toroidal geometry and equilibrium plasma profiles. Tokamaks host a vibrant zoo of \Alfven eigenmodes, depicted in \figref{fig:int:alfvenzoo}, which may be driven unstable by phase space gradients in fast ions generated by heating in the ion cyclotron range of frequencies (ICRF), neutral beam injection, or fusion products. These waves may be excited at frequencies spanning many orders of magnitude: from EPMs and beta-induced \Alfven acoustic modes (BAAE)\cite{Gorelenkov2007PLA,Gorelenkov2007PPCF} at very low frequencies -- tens of kHz, and gap modes including toroidicity-induced \Alfven eigenmodes (TAEs)\cite{Cheng1985AP,Heidbrink2008POP} with $f \like \va/4\pi qR \approx 100$ kHz, to moderate frequency sub-cyclotron modes (CAEs, GAEs) $f \like f_{ci}/2 \approx 1$ MHz, and even above the cyclotron frequency (ion cyclotron emission, or ICE).\cite{Cottrell1988PRL,Gorelenkov2016NJP} 
A number of helpful reviews exist on the topics of energetic particles and fast-ion-driven instabilities in tokamaks and related devices which can provide much more detail than there is room for in this brief thesis introduction.\cite{Heidbrink1994NF,Wong1999PPCF,Heidbrink2008POP,Fasoli2007NF,Pinches2015POP,Breizman2011PPCF,Sharapov2013NF,Gorelenkov2014NF,McClements2017PPCF,Todo2019RMPP,Chen1995PS,Chen2007NF,Chen2016RMP}
There are also two textbooks\cite{CrossAE,CramerAE} devoted exclusively to the derivation of \Alfven waves in various theoretical frameworks. The two instabilities addressed in depth in this thesis will now be introduced in more detail. 

\subsection{Compressional \Alfven Eigenmodes (CAEs)}
\label{sec:int:cae}

Boundary conditions imposed by toroidal geometry discretize the spectrum of fast wave solutions, which are referred to as compressional \Alfven eigenmodes (CAEs) in a tokamak. In the zero pressure limit, the poloidal location of the mode can be approximated by a simplified 2D wave equation with an effective potential well\cite{Fredrickson2013POP} 

\begin{align}
\label{eq:int:Veff}
\left[\nabla_\perp^2 - \Veff(r,\theta)\right]\dbpar &= 0 \\ 
\text{where } \Veff(r,\theta) &= \kpar^2 - \frac{\omega^2}{\va^2}
\approx \left(\frac{n}{R}\right)^2 - \left(\frac{\omega}{\va}\right)^2
\end{align}

\begin{figure}[tb]
\includegraphics[width = 0.6\textwidth]{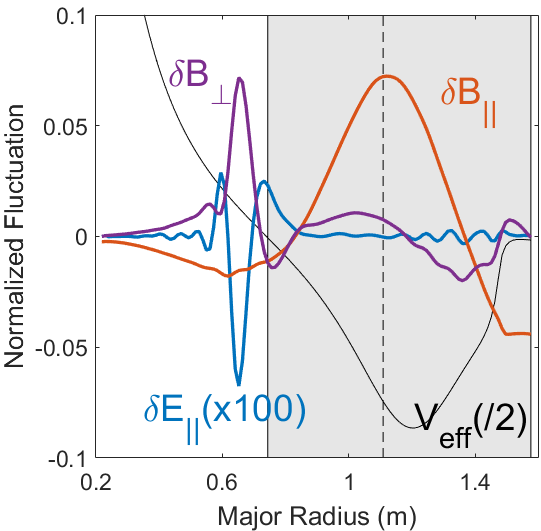}
\caption
[CAE confined within effective potential well in simulations]
{CAE mode structure calculated by a \HYM simulation. The effective potential is in black, with the shaded region marking $\Veff < 0$, where the wave can propagate. The dashed line marks the magnetic axis. The dominant $\dbpar$ fluctuation is localized in the well as expected. The coherent $\dbperp$ and $\depar$ structures near $R=0.60$ m are associated with the KAW that results from mode conversion at the \Alfven resonance location ($\Veff=0$).}
\label{fig:int:cae_ex}
\end{figure}

The wave can propagate in regions where $\Veff < 0$ and is evanescent for $\Veff >  0$, resulting in its radial and poloidal localization, as illustrated in \figref{fig:int:cae_ex}. More detailed expressions for the dispersion and effective potential including realistic toroidal effects have been derived by many authors,\cite{Mahajan1983bPF,Gorelenkova1998POP,Kolesnichenko1998NF,Gorelenkov2002POP,Gorelenkov2002NF,Smith2003POP,Gorelenkov2006NF} though the approximate dispersion and simplified wave equation are often sufficient for understanding qualitative features of CAEs in experiments and simulations. The spectral code \CAETB written by Smith\cite{Smith2009PPCF,Smith2017PPCF} has demonstrated the same behavior for the eigenmodes in both conventional and low aspect ratio conditions.

The spectrum of CAEs can be qualitatively explained by a heuristic dispersion which assumes discrete mode numbers\cite{Gorelenkov2006NF} $m$, $s$, and $n$, as well as a characteristic radial width $L_r$ of $\Veff$

\begin{equation}
\label{eq:int:CAEdiscrete}
\omega_{CAE}^2 \approx \left[\left(\frac{m}{r}\right)^2 + \left(\frac{\pi s}{L_r}\right)^2 + \left(\frac{n}{R}\right)^2\right]\va^2
\end{equation}

Another important feature of CAEs in tokamaks is their mode conversion to the kinetic \Alfven wave (KAW). The condition $\Veff = 0$ coincides with the \Alfven resonance location where the CAE frequency matches the local frequency of the shear \Alfven wave: $\omega_\text{CAE} = \kpar\va(R,Z)$. In ideal MHD, a logarithmic singularity would appear at this location.\cite{Hasegawa1976PF} However, when kinetic effects are considered, the singularity is replaced by a short scale fluctuation, the KAW, which has wavelength on the order $\kperp \like 1/(\rho^2 L_A)^{1/3}$ where $L_A$ is a a characteristic length scale of the $\va$ profile, and $\rho$ depends on the thermal and fast ion Larmor radii.\cite{Belova2017POP} 

Mode conversion and absorption of the compressional wave at the \Alfven resonance has been studied previously at the magnetopause\cite{Johnson1997GRL} and also in tokamak heating schemes such as ICRH\cite{Heikkinen1991NF} and \Alfven wave heating.\cite{Hasegawa1976PF} Heuristically, the coupling between the CAE and KAW is mediated by finite Larmor radius (FLR) effects. By including these effects to lowest order and also allowing for local shear \Alfven resonances, the simplified CAE dispersion is transformed into the following, from \citeref{Johnson1997GRL}


\begin{multline}
\label{eq:int:caekaw}
\kpar^2\va^2\rho^2\frac{\partial^4\phi}{\partial r^4} + 
\frac{\partial}{\partial r} r \left(\omega^2 - \kpar^2\va^2\right)\frac{\partial}{\partial r}\frac{\phi}{r} = \\
\left(\omega^2 - \kpar^2\va^2\right)\left[-\frac{1}{r^2}\frac{\partial^2}{\partial\theta^2} - \kpar^2 -
\frac{\omega^2}{v_{A0}^2}\frac{n(r)}{n_0}\left(1 + \epsilon\cos\theta\right)^2\right]\phi
\end{multline}

Above, $\rho$ is a scale length proportional to the thermal ion Larmor radius. In the existing literature, this equation is solved in two limits, far from the resonance ($\omega \gg \kpar\va$, the MHD scale) and close to the resonance ($\omega \approx \kpar\va$, the kinetic scale) and then asymptotically matched in order to determine the combined CAE-KAW solution. The former case reduces to the CAE dispersion, whereas the latter recovers a shear \Alfven wave. Hence it's clear that kinetic effects serve to couple the two fundamental MHD wave branches, with mode conversion occuring at the resonant location.

A representative example of CAE to KAW mode conversion present in a simulation is shown in \figref{fig:int:cae_ex}. Mode conversion can be an important energy channeling mechanism from the location of largest concentration of fast ions (which can drive CAEs unstable) to the \Alfven resonance (usually in the edge). It will be discussed in more detail in \secref{sec:int:teflat}. 

High frequency co-propagating and cntr-propagating CAEs have been observed in the spherical tokamaks NSTX(-U)\cite{Fredrickson2001PRL,Fredrickson2004POP,Fredrickson2013POP,Crocker2013NF,Fredrickson2019POP} and MAST.\cite{Appel2008PPCF,Sharapov2014PP,McClements2017PPCF} These instabilities are often excited in spherical tokamaks due to their low magnetic fields and large neutral beam power, which together generate a substantial population of super-\Alfvenic fast ions.\cite{McClements2017PPCF}
CAEs have also been implicated in observations of enhanced fast ion diffusion in TFTR which may be associated with alpha channeling.\cite{Fisch1992PRL,Gorelenkov2010PPCF} In addition, CAEs are the leading candidate to explain ICE observations in a variety of devices,\cite{Cottrell1988PRL,Cottrell1993NF,Fulop1997NF,Fulop1998NF,Gorelenkov2016NJP,Thome2019NF} which could serve as a passive diagnostic for the fast ion distribution function in future burning plasmas such as ITER.\cite{McClements2015NF} 

\subsection{Global \Alfven Eigenmodes (GAEs)}
\label{sec:int:gae}

The toroidal generalization of shear waves is somewhat more intricate than that of compressional waves. Consider the most simplified approximate dispersion $\omega \approx \kpar(r)\va(r)$. Taking the largest aspect ratio approximation of a cylinder, the parallel wave number can be expressed as $\kpar(r) \approx (n - m / q(r)) / R$, where $n$ is the toroidal mode number $(n\lambda_\phi = 2\pi R)$ and $m$ is the poloidal mode number $(m\lambda_\theta = 2\pi r)$. Consequently, $\kpar$ has radial dependence due to the dependence of the safety factor $q(r) = r B_\tor / R B_\pol$. Likewise, the \Alfven speed inherits radial dependence from the equilibrium magnetic field and density profiles. As a result, a \emph{continuum} of solutions exists for shear \Alfven waves in a tokamak\cite{Goedbloed1975PF} -- for a given frequency, the dispersion can be satisfied at each radius by a different value of $\kpar(r)$. These continuum modes suffer strong damping due to phase mixing since a wave packet would be rapidly sheared apart due to having different phase velocities at different radii.\cite{Heidbrink2008POP} Hence, continuum modes are rarely observed in experiments. 

In addition to the strongly damped continuum solutions, discrete eigenmodes exist with frequencies outside of the continuum. These fall into two categories: ``gap'' modes and ``extremum'' modes. In a torus, the two periodicity constraints couple different poloidal harmonics together, which can open up frequency gaps in the continuous spectrum of solutions, referred to as the \Alfven continuum. Within these gaps, discrete shear \Alfven eigenmodes exist\cite{Cheng1985AP,Heidbrink2008POP} which are not subject to the continuum damping, and therefore can be destabilized by fast ions. These gap modes, which include the toroidicity-induced \Alfven eigenmode (TAE), are the most commonly studied \Alfven wave in tokamaks since they can induce significant fast ion transport, jeopardizing our ability to efficiently heat the plasma. 

The final type of shear \Alfven wave existing in a tokamak is the ``extremum'' type mode. Like the gap modes, these are discrete eigenmode solutions which exist outside of the \Alfven continuum. Extremum type modes exist with frequency just below or above an extremum in the \Alfven continuum, radially localized near points where $d(\kpar(r)\va(r))/dr = 0$. These include the reverse-shear \Alfven eigenmode (RSAE), which achieves this condition when $q'(r) = 0$, as well as the global \Alfven eigenmode (GAE), which can occur due to any generic equilibrium profile variation leading to an extrema in the continuum. Unlike the gap modes, extremum type modes do not require poloidal coupling, and can exist in both cylindrical and toroidal plasmas. 

\begin{figure}[tb]
\includegraphics[width = 0.6\textwidth]{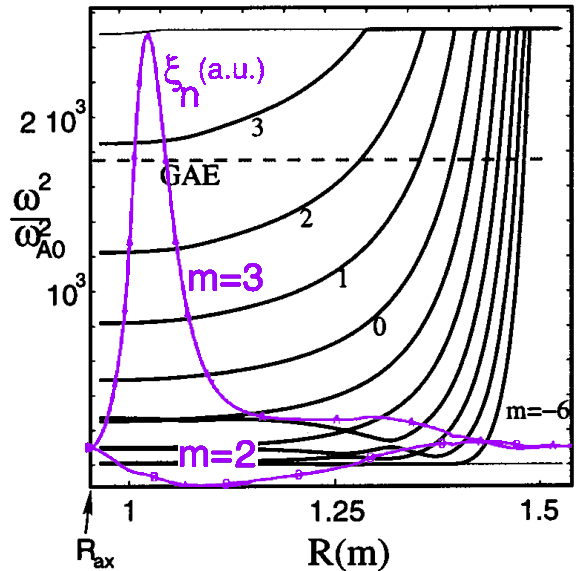}
\caption
[GAE solution in NSTX calculated by \NOVA.]
{GAE solution in NSTX calculated by \NOVA. In black is the $n = -3$ \Alfven continuum, including $m = -6 \tto 3$. Dashed line indicates the frequency of the GAE. In purple is the associated plasma displacement $\xi$ for the dominant poloidal harmonics $m = 3$ and $m = 2$, which peak near the radial location of the minimum of the continuum. Adapted from \citeref{Gorelenkov2004POP}. Reproduced with the permission of AIP Publishing.}
\label{fig:int:gae_ex}
\end{figure}

GAEs may be excited at a frequency slightly below a minimum of the \Alfven continuum, as illustrated in the \NOVA calculation in \figref{fig:int:gae_ex}. 
Their separation from the continuum arises from coupling to the magnetosonic mode, an equilibrium current density gradient, and inclusion of finite $\omeganorm$ effects.\cite{Appert1982PP,Mahajan1983PF,Mahajan1984PF,Li1987PF,DeAzevedo1991SP} ``Nonconventional" GAEs may also be excited above a local maxima in the continuum through similar mechanisms. \cite{Kolesnichenko2007POP} While their full dispersion can be quite complicated depending on how many realistic effects are maintained, the frequency is often close enough to the continuum that in practice it may be approximated by the slab MHD shear \Alfven dispersion 

\begin{equation}
\label{eq:int:GAEsimp}
\omega_\text{GAE} \leq \left[\kpar(r) \va(r)\right]_\text{min}
\end{equation}

GAEs were initially modeled numerically in cylindrical plasmas\cite{Ross1982PF,Appert1982PP} in order to explain resonant peaks in antenna loading observed in the TCA tokamak.\cite{DeChambrier1982POP} Further theoretical work found them to be stabilized by finite toroidicity effects\cite{Fu1989PF,VanDam1990FT} in the limit of $\omeganorm \ll 1$. More recently, counter-propagating GAEs excited by the ordinary Doppler-shifted cyclotron resonance have been the subject of experimental\cite{Fredrickson2006POP,Fredrickson2012NF,Crocker2013NF,Fredrickson2017PRL} and theoretical studies\cite{Gorelenkov2003NF,Gorelenkov2004POP,Kolesnichenko2006POP} due to their frequent excitation in spherical tokamaks. It may also be possible to excite co-propagating GAEs at high frequencies with very parallel neutral beam injection via the anomalous cyclotron resonance,\cite{Lestz2018POP,Lestz2020sim} though this awaits experimental confirmation.  

Cntr-propagating GAEs were commonly observed on the spherical tokamaks NSTX(-U)\cite{Gorelenkov2003NF,Gorelenkov2004POP,Fredrickson2012NF,Fredrickson2013POP,Crocker2013NF,Crocker2017NF,Fredrickson2018NF} and MAST.\cite{Appel2008PPCF,Sharapov2014PP,McClements2017PPCF} Dedicated experiments on the large aspect ratio tokamak DIII-D have also observed AE activity in this frequency range,\cite{Heidbrink2006NF,Tang2018APS,Crocker2018APS} allowing comparison between their excitation properties across these different configurations. Both CAEs and GAEs are prone to frequency chirping in NSTX, which can modify the characteristics of the fast ion transport (diffusive vs convective) and presents opportunities for validating nonlinear theories.\cite{Duarte2017NF} GAE chirping can trigger deleterious GAE ``avalanches'' -- sudden, broad spectrum, large amplitude bursts that can result in fast ion losses of up to $40\%$.\cite{Fredrickson2012NF}

\section{The National Spherical Torus Experiment (and Upgrade)}
\label{sec:int:nstx}

With the preceding physics background in mind, we will now review the specific experimental observations that motivate this work, and identify the problems that the thesis aims to address with a theoretical approach. 

\begin{figure}[tb]
\includegraphics[width = 0.6\textwidth]{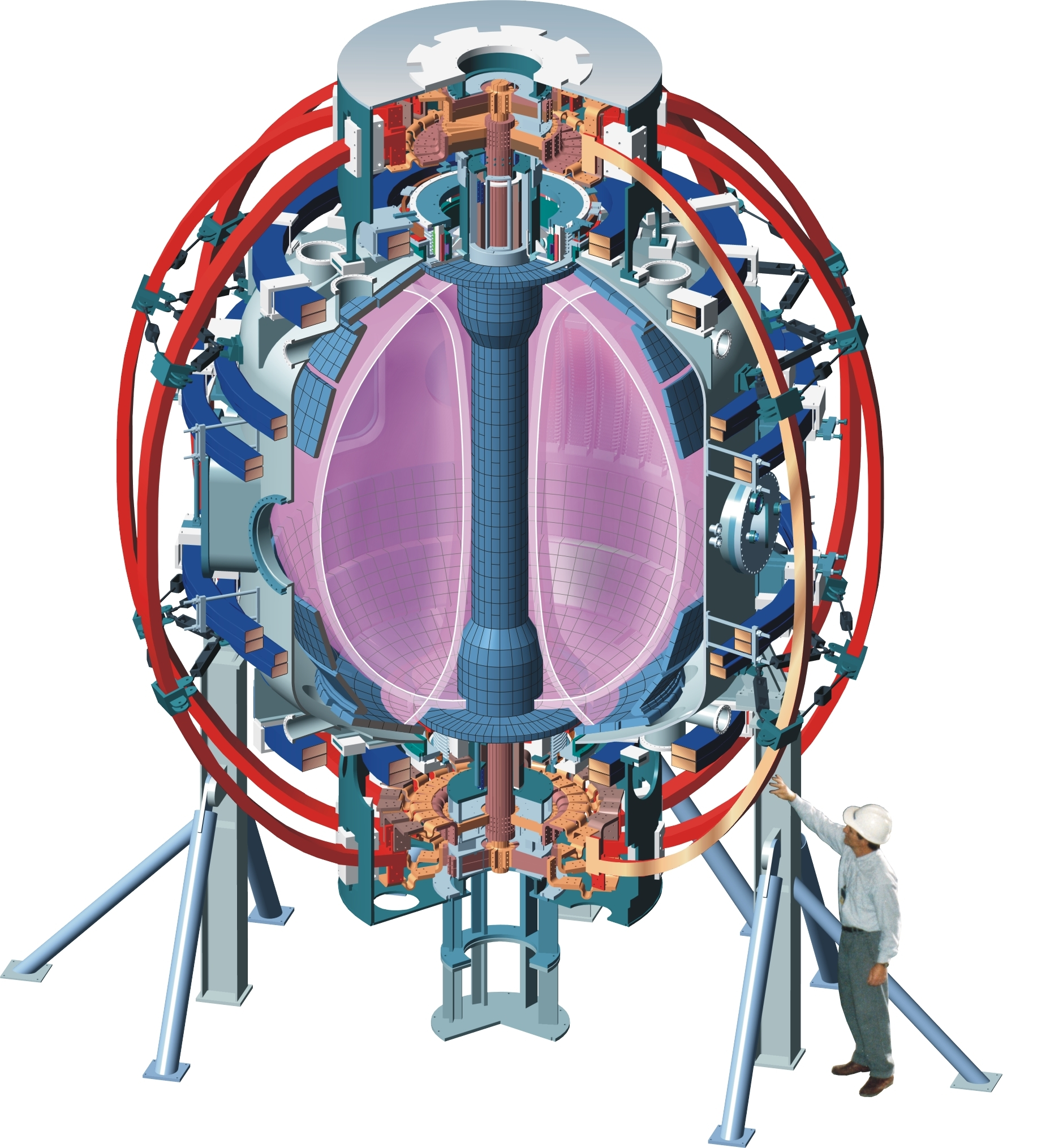}
\caption
[Cutaway of the National Spherical Torus Experiment (Upgrade).]
{Cutaway of the National Spherical Torus Experiment (Upgrade). 
Figure courtesy of the \href{https://nstx-u.pppl.gov/}{NSTX-U collaboration}.} 
\label{fig:int:nstx}
\end{figure}

The National Spherical Torus Experiment (NSTX) and its recent upgrade (NSTX-U),\cite{Menard2012NF} shown in \figref{fig:int:nstx}, are low aspect ratio $(R/a = 1.3)$ tokamaks, also known as a ``spherical tokamaks'' (STs). Spherical tokamaks are currently being researched as a potentially more cost-effective route to a fusion energy pilot plant\cite{Ono2015POP,Menard2011NF,Menard2016NF,Menard2019RSA} than conventional large tokamaks $(R/a = 2.5 - 4)$, since construction cost scales with device size. NSTX and the similar spherical tokamak MAST in the UK both found strong inverse confinement scaling with normalized electron collisionality $\nu_e^*$ ($\propto n_e / T_e^2$) -- indicating that plasma confinement may be more favorable at high temperatures in spherical tokamaks than in conventional ones.\cite{Kaye2007NF,Valovic2009NF,Valovic2011NF} This may be explained by magnetic field lines spending relatively more time in the ``good curvature'' region\cite{Rewoldt1996POP,Kinsey2007POP} and strong $\vE\cross\vB$ rotation shear.\cite{Roach2009PPCF,Guttenfelder2013NF,Guttenfelder2019NF} Spherical tokamaks are also capable of achieving high $\beta \approx 10 - 40\%$, compared to $\beta \approx 3 - 10\%$ for conventional aspect ratio tokamaks,\cite{Ono2015POP} allowing more efficient plasma confinement. 

NSTX has a major radius of 0.85 m, and a minor radius of 0.68 m. It achieves densities on the order of $10^{19}$ m$^{-3}$ and temperatures of approximately 1 keV. The upgrade from NSTX to NSTX-U included the installation of a new central magnetic and an additional neutral beam source. Consequently, the maximum toroidal field and plasma current were doubled from 0.5 T to 1 T and 1 MA to 2 MA, respectively, and discharge duration was increased from 1 s to 5 s. 

\begin{figure}[tb]
\includegraphics[width = 0.6\textwidth]{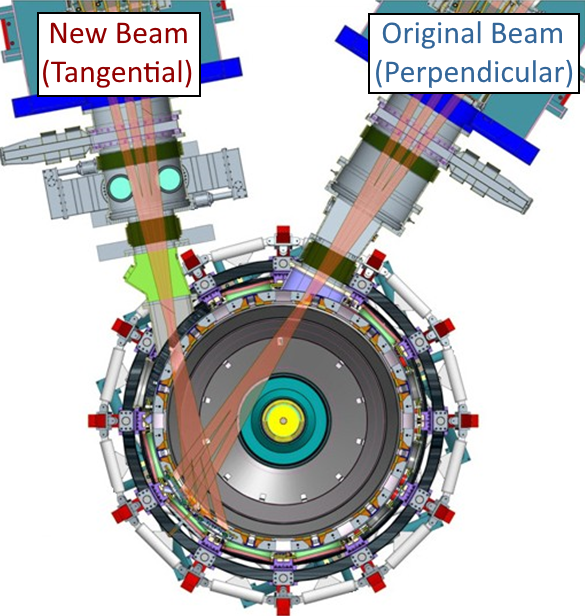}
\caption
[Overhead view of NSTX-U neutral beam geometry.]
{Overhead view of NSTX-U neutral beam geometry. 
Reproduced from \citeref{Menard2012NF} with permission from IAEA Publications.} 
\label{fig:int:beams}
\end{figure}

The NSTX-U deuterium neutral beams can operate at a maximum voltage of 90 kV, with total power doubled from 6 MW to 12 MW during the upgrade. Importantly, the new neutral beam source was installed in a different geometry from the original beam sources, as shown in \figref{fig:int:beams}. Each beam has three beam lines with slightly different injection angles. The original beam lines inject more radially, with tangency radii of $\Rtan = 0.5 \text{ m, } 0.6 \text{ m, and } 0.7 \text{ m}$, whereas the new beam sources lead to a more field-aligned (tangential) energetic particle population, due to tangency radii of $\Rtan = 1.1 \text{ m, } 1.2 \text{ m, and } 1.3 \text{ m}$ (the magnetic axis is located near $\Rzero \approx 1.05$ m due to the Shafranov shift). Combination of these beam sources provides substantial flexibility in tailoring the fast ion distribution in phase space in order to study fast-ion-driven instabilities. 

The primary diagnostics used to measure CAEs and GAEs on NSTX(-U) are magnetic Mirnov coils\cite{Fredrickson2013POP} and a reflectometer array.\cite{Crocker2011PPCF,Crocker2013NF,Crocker2017NF} The 10 Mirnov coils are toroidally spaced non-uniformly, with sufficient separation to detect $|n| \leq 18$, and detect frequencies up to 5 MHz ($\approx 1-2 f_{ci}$ in NSTX). They provide very precise measurements of the unstable mode frequencies and their time evolution throughout the discharge, as well as a measurement of the edge polarization of the fluctuations. The reflectometer array measures plasma displacements at 16 locations, bunched mostly in the edge/pedestal region, with a few points extending to near the magnetic axis. These internal measurements can be used to calculate radial mode structure of the modes across most of the plasma minor radius.\cite{Crocker2017NF} Both diagnostics can be used to measure the mode amplitudes. 

Previous comparison of the frequencies of the observed modes in the NSTX H-mode discharge \# 141398 and the most unstable modes in \HYM modeling show a close match in frequency for each toroidal harmonic.\cite{Belova2017POP} Comparison of the mode structures inferred from reflectometry measurements and present in \HYM simulations show qualitative similarities. Preliminary analysis of high-$k$ scattering measurements\cite{Smith2008RSI,Barchfield2018RSI} has shown possible signatures of CAE to KAW mode conversion in NSTX,\cite{Deng2017APS} though further analysis is needed for conclusive identification. 

NSTX-U provides an excellent laboratory for studying energetic-particle-driven instabilities. Due to its low magnetic field and high beam power, it is capable of operating across a wide range of $v_\tep/\va$ and $\beta_\tep/\beta_\text{tot}$, which are key parameters controlling the activity of these instabilities.\cite{Fredrickson2014NF} Moreover, dimensionless parameters for the fast ions in NSTX-U generated from neutral beam injection can be comparable to those of fusion alphas in large burning plasmas such as ITER, connecting these studies to future burning reactors. The focus of this thesis will be advancing the theoretical understanding of the global and compressional \Alfven eigenmode stability, motivated by two significant experimental observations anomalous electron temperature profile flattening in NSTX and robust GAE suppression with the off-axis beam sources in NTSX-U. 

\section{Electron Temperature Profile Flattening} 
\label{sec:int:teflat}

The primary motivation of this thesis is the anomalously flat electron temperature profile observed in NSTX during H-mode neutral-beam-heated discharges, which has been linked to the presence of high frequency \Alfven activity, identified as a mixture of CAEs and GAEs.\cite{Fredrickson2001PRL,Gorelenkov2003NF,Gorelenkov2004POP,Fredrickson2004POP,Fredrickson2013POP,Crocker2013NF,Crocker2017NF,Fredrickson2018NF,Fredrickson2019POP}
While the ion temperature matches predictions from the global transport codes \TRANSP,\cite{Goldston1982JCP} indicating neoclassical ion confinement, the electron temperature deviates from these descriptions at high beam power. As the beam power is increased, the electron temperature radially broadens while the central electron temperature stagnates or even decreases. The flattening occurs in both L-mode and H-mode discharges, most dramatically at higher beam power, presenting opportunity for its further investigation on NSTX-U with its doubled capacity of neutral beam heating to a maximum of 12 MW. The observed broadening of the electron temperature profiles contrasts with the expectation that increasing beam power will result in an increased on-axis temperature with negligible impact on its radial profile. Anomalously low electron temperature limits fusion performance (when not operating in a hot ion mode), and could imperil future spherical tokamak development. 

\begin{figure}[tb]
\includegraphics[width = 0.6\textwidth]{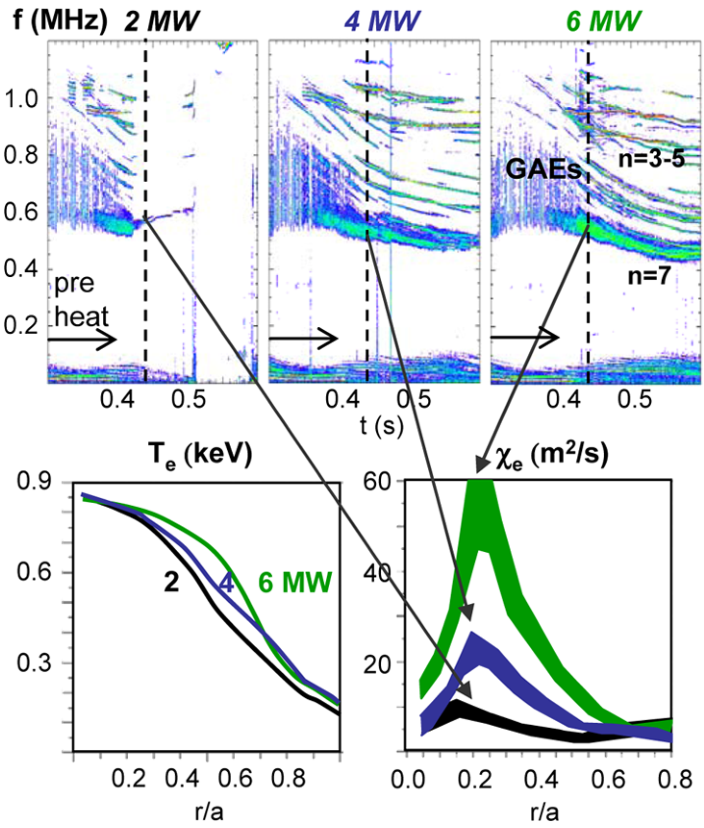}
\caption
[NSTX experiment demonstrating correlation between CAE/GAE activity and $\Te$ flattening]
{NSTX experiment demonstrating correlation between CAE/GAE activity and $\Te$ flattening, along with the inferred $\chi_e$. 
Reprinted figure with permission from \citeref{Stutman2009PRL}. Copyright 2009 by the American Physical Society.}
\label{fig:int:teflat}
\end{figure}

The inferred electron diffusion profile required to explain this observation is unusual and quite large in magnitude. Moreover, since the temperature profile is so flat, it is unlikely that microturbulence could be the source of this anomalous diffusion. Local gyrokinetic simulations of turbulence in the ``core flat region'' ($r/a \leq 0.15$ -- where gradients are absent) do not reproduce this level of anomalous transport.\cite{Guttenfelder2013NF,Ren2017NF} Conversely, high frequency \Alfven activity is often observed in discharges with the anomalous electron temperature flattening. This was demonstrated by Stutman \etal in an experiment where an H-mode plasma was reproduced with 2 MW, 4 MW, and 6 MW of injected beam power in successive discharges.\cite{Stutman2009PRL} As the beam power increased, the amplitude and number of distinct high frequency modes increased in conjunction with flattening of the temperature profile and an increase in the peak electron diffusivity by an order of magnitude, as depicted in \figref{fig:int:teflat}. In addition to dedicated experiments designed to investigate this phenomenon, a detailed database of shots with substantial \Alfvenic mode activity has been compiled by Fredrickson\cite{Fredrickson2014NF} and extended by Tang\cite{Tang2017TTF} to catalog details about CAEs and GAEs. Statistical analysis of this database further strengthens the link between between strong CAE/GAE activity and $\Te$ flattening. For these reasons, the CAEs and GAEs are suspected to be the primary cause of the unexplained electron energy transport. 

In response to observations of anomalously large ratios of $\Ti/\Te$ in early NSTX operations, it was also suggested that the simultaneous excitation of many CAEs could generate broadband turbulence, leading to efficient heating of the thermal ions through stochastic diffusion.\cite{Gates2001PRL,Kolesnychenko2005POP} In that case, further experimental analysis ruled this out as a viable mechanism for NSTX conditions based on the observed mode amplitudes.\cite{Fredrickson2002POP} Moreover, recent analytic studies by Kolesnichenko suggest that the CAE/GAE-induced energy transport could explain anomalously high confinement observed in certain JET discharges\cite{Kolesnichenko2018NF} –- introducing the possibility of harnessing these mechanisms for improved performance. Predictive capabilities and control of the $\Te$ profile with neutral beam heating is vital to achieving and exploring high performance plasmas in STs such as NSTX-U and future ST-FNSF designs. In order to probe the aforementioned favorable confinement scaling in regimes that can not be accessed by other devices, STs must be able to achieve their target $\Te$ profiles, which is jeopardized by their spontaneously flattening in the presence of CAEs and GAEs. The importance of this topic is emphasized in the NSTX-U 5 year plan:\cite{NSTXU5YP1418} ``The successful development and implementation of an energetic particle model for electron thermal transport is essential to achieve the high priority goal of $\Te$ and $\Ti$ profile predictions.'' 


Two theoretical mechanisms have been previously proposed to explain how the CAEs and GAEs could modify the electron temperature profile. The first involves the stochastization of electron orbits induced by the presence of many modes of sufficient amplitude. Test particle simulations using the guiding center particle code \ORBIT\cite{White1984PF} have shown that there is a sharp increase of two orders of magnitude in the electron diffusion coefficient if a threshold in the number of unstable GAEs is surpassed, provided that the mode amplitudes are sufficiently large.\cite{Gorelenkov2010NF} It is likely that a qualitatively similar effect occurs for sufficiently many CAEs, though this has not been confirmed. A strong dependence on the mode amplitude was also reported, as well as sensitivity to $\depar$ fluctuations. The second process is an energy-channeling mechanism where a core-localized CAE or GAE mode converts to a KAW at the \Alfven resonance location. Since the KAW efficiently damps on thermal electrons, this process modifies the effective beam energy deposition profile, redirecting neutral beam power from the core to the edge through the Poynting flux. This possibility has been studied analytically by Kolesnichenko \etal in the case of GAEs\cite{Kolesnichenko2010PRL,Kolesnichenko2010NF,Kolesnichenko2018POP,Kolesnichenko2018bPOP} and has been confirmed numerically for CAEs by Belova {\etal}\cite{Belova2015PRL,Belova2017POP} 

Both of these mechanisms -- energy channeling and orbit stochastization -- have been shown to have an effect in numerical calculations. However, there is a quantitative gap between the levels of effective transport they predict when mode amplitudes are scaled to experimental magnitudes and the amount necessary to explain the experimental $\Te$ profile flattening. While this phenomena is currently unique to NSTX, it could potentially be relevant to ITER where both the neutral beam ions and fusion alphas will be super-\Alfvenic and hence have the potential to excite CAEs/GAEs.\cite{Heidbrink2006NF} Solving this problem can be decomposed into two parts: 1) for a given plasma and beam scenario, which modes will be unstable? 2) for a given spectrum of modes, how will they affect the electron energy transport? My thesis focuses primarily on this first part, aiming to improve understanding of the CAE/GAE stability properties. 

\section{GAE Stabilization with Off-Axis Beams}
\label{sec:int:gaestab}

\begin{figure}[tb]
\includegraphics[width = \fullwidth]{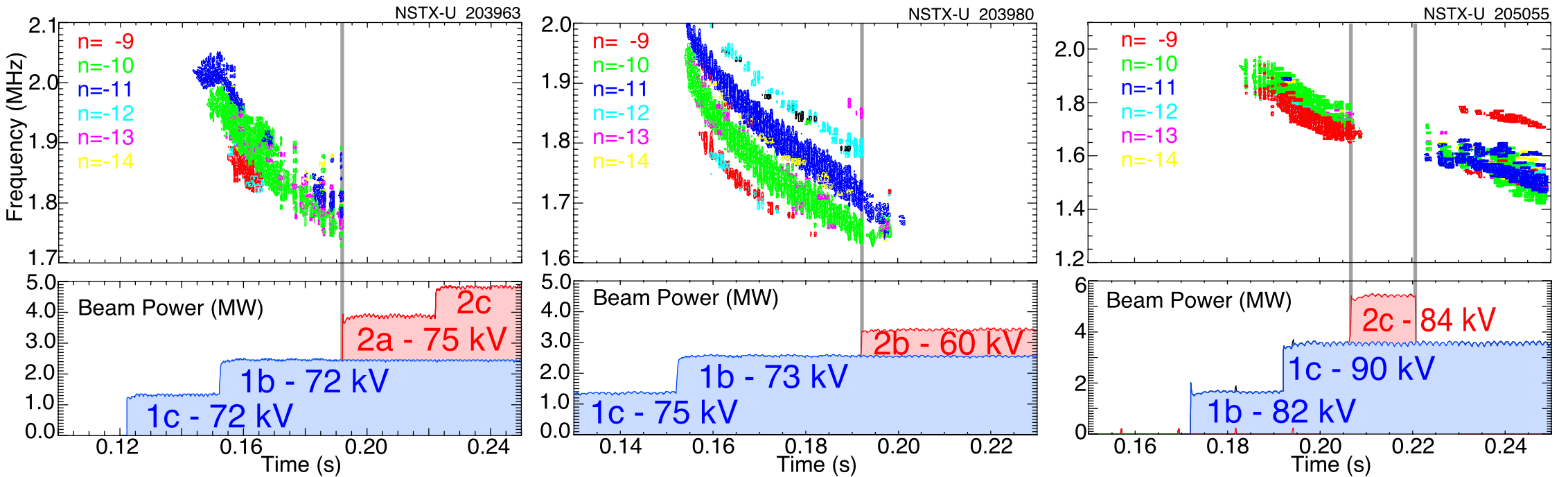}
\caption
[GAE suppression on NSTX-U due to new, off-axis beam sources.]
{GAE suppression on NSTX-U due to new, off-axis beam sources. Blue beams with labels 1b and 1c are the original NSTX sources, while red beams with labels 2a-c are the additional NSTX-U sources (see \figref{fig:int:beams}). 
Reproduced from \citeref{Fredrickson2018NF} with permission from IAEA Publications.}
\label{fig:int:gaesupp}
\end{figure}

The second experimental observation motivating the further theoretical development of CAE/GAE stability properties is the efficient suppression of GAEs with off-axis beam injection, recently discovered in NSTX-U. Early NSTX-U operations found that the original NSTX neutral beam sources excited a broad spectrum of high frequency \Alfven eigenmode activity (mostly GAEs), just as they did in NSTX. When additional beam power was supplied by the new, off-axis beam sources, all instabilities in this frequency range rapidly vanished.\cite{Fredrickson2017PRL} Three examples are shown in \figref{fig:int:gaesupp}, though this was an ubiquitous phenomenon present in at least 100 unique discharge time windows.\cite{Fredrickson2018NF} At first glance, this is a surprising finding. Typically, higher beam power will further destabilize \Alfven eigenmodes since it supplies the system with more energetic particles. Conversely, suppressing instabilities while increasing the plasma heating is the best case scenario. 

Subsequent \HYM modeling of an NSTX-U discharge reproduced both the excitation of GAEs with the original beam sources and also their complete stabilization with the addition of the new beam source.\cite{Belova2019POP} Further simulations found that the GAE suppression can be achieved with only 7\% of the total beam ions being supplied by the new beams, much lower than the 25\% of total beam ions supplied in the modeled experimental discharge. A more complete theoretical understanding of this very efficient stabilization mechanism could contribute to the development of additional phase space engineering techniques for control of fast ion instabilities.\cite{Fisch2000NF,Graves2012Nature,Bartolon2013PRL} In particular, theoretical advancements enabling the control of CAEs/GAEs will facilitate the investigation of their role in the anomalous electron energy transport. 

\section{Thesis Outline and Main Outcomes} 
\label{sec:int:outline}

The main goals of this thesis are to advance the theoretical understanding of CAE and GAE stability properties in application to the anomalous electron temperature profile flattening that they are associated with. Both analytic theory and numerical simulations are employed towards this goal. Each chapter in this thesis is written to be mostly self-contained, so there is an intentional degree of redundancy in some areas in order to remind readers of previous results and relevant background for each section. Appendices appear immediately after each chapter. The thesis is outlined as follows. 

In \chapref{ch:cyc:analytics-cyclotron}, analytic conditions for net fast ion drive are derived for beam-driven, sub-cyclotron CAEs and GAEs. Both co- and counter-propagating modes are investigated, driven by the ordinary and anomalous Doppler-shifted cyclotron resonance with fast ions. Whereas prior results were restricted to vanishingly narrow distributions in velocity space, broad parameter regimes are identified in this work which enable an analytic treatment for realistic fast ion distributions generated by neutral beam injection. The simple, approximate conditions derived in these regimes for beam distributions of realistic width compare well to the numerical evaluation of the full analytic expressions for fast ion drive. Moreover, previous results in the very narrow beam case are corrected and generalized to retain all terms in $\omeganorm$ and $\krat$, which are often assumed to be small parameters but can significantly modify the conditions of drive and damping when they are non-negligible. Favorable agreement is demonstrated between the approximate stability criterion, simulation results, and a large database of NSTX observations of cntr-GAEs. 

In \chapref{ch:lan:analytics-landau}, a similar analytic approach is taken for co-propagating CAEs and GAEs driven by the Landau resonance. Approximations applicable to realistic neutral beam distributions and mode characteristics observed in spherical tokamaks enable the derivation of marginal stability conditions for these modes. Such conditions successfully reproduce the stability boundaries found from numerical integration of the exact expression for local fast ion drive/damping. Coupling between the CAE and GAE branches of the dispersion due to finite $\omeganorm$ and $\krat$ is retained and found to be responsible for the existence of the GAE instability via this resonance. Encouraging agreement is demonstrated between the approximate stability criterion, simulation results, and a database of NSTX observations of co-CAEs. 

In \chapref{ch:sim:simulations}, a comprehensive numerical study is presented in order to investigate CAE/GAE stability properties for a wide range of beam parameters in realistic NSTX conditions. Linear simulations are performed with the hybrid MHD-kinetic initial value code \HYM in order to capture the general Doppler-shifted cyclotron resonance that drives the modes. The simulations reveal that unstable GAEs are more ubiquitous than unstable CAEs, consistent with experimental observations, as they are excited at lower beam energies and generally have larger growth rates. The local analytic theory derived in \chapref{ch:cyc:analytics-cyclotron} and \chapref{ch:lan:analytics-landau} is used to explain key features of the simulation results, including the preferential excitation of different modes based on beam injection geometry and the growth rate dependence on the beam injection velocity, critical velocity, and degree of velocity space anisotropy. Drive due to velocity space anisotropy is capable of explaining most trends theoretically, though it is found that gradients with respect to $\pphi$ can be responsible for a substantial fraction of the fast ion drive for co-propagating modes. The background damping rate is inferred from simulations and estimated analytically for relevant sources not present in the simulation model, indicating that co-CAEs are closer to marginal stability than modes driven by the cyclotron resonances. 

In \chapref{ch:epgae:epgae}, the numerical discovery of strong energetic particle modifications to GAEs in \HYM simulations of NSTX-like plasmas is presented and investigated. Key parameters defining the fast ion distribution function -- the injection velocity and injection geometry -- are varied in order to study their influence on the characteristics of the excited modes. It is found that the frequency of the most unstable mode changes significantly and continuously with beam parameters, in accordance with the Doppler-shifted cyclotron resonances which drive the modes, and depending most substantially on the injection velocity. This unexpected result is present for both counter-propagating GAEs, which are routinely excited in NSTX, and high frequency co-GAEs, which have not been previously studied. Large changes in frequency without clear corresponding changes in mode structure are signatures of an energetic particle mode, referred to here as an energetic-particle-modified GAE (\EGAE). Additional simulations conducted for a fixed MHD equilibrium demonstrate that the GAE frequency shift cannot be explained by the equilibrium changes due to energetic particle effects. 

Lastly, a summary of the main results and discussion of future research directions is given in \chapref{ch:conc:conc}.
\newcommand{\dirfigcyc}{ch-analytics-cyclotron/figs}

\chapter{Analytic Stability Boundaries for Interaction via Ordinary and Anomalous Cyclotron Resonances}
\label{ch:cyc:analytics-cyclotron}

\section{Introduction}
\label{sec:cyc:introduction}

The analysis of this chapter focuses on fast ions interacting with CAEs/GAEs through the ordinary or anomalous cyclotron resonances. Drive/damping due to the Landau resonance is treated in \chapref{ch:lan:analytics-landau}. General expressions for the growth rate of these instabilities were originally derived for mono-energetic beam\cite{Belikov1968ZHETF,Belikov1969JETP} and bi-Maxwellian\cite{Timofeevv5} distributions, as well as for an arbitrary distribution\cite{Mikhailovskiiv6} in a uniform plasma. These derivations were later extended and applied to NBI-driven CAEs/GAEs in various experimental conditions dating back to the TFTR era\cite{Gorelenkov1995POP,Gorelenkov1995NF} and continuing in more recent years with applications to JET\cite{Gorelenkov2002bNF} and NSTX.\cite{Gorelenkov2003NF,Kolesnichenko2006POP} The recent studies on NBI-driven modes had two key limitations. First, they did not correctly treat the cutoff at the injection energy, an approach suitable for shifted Maxwellians generated by heating in the ion cyclotron range of frequencies (ICRF), but not for slowing down distributions from NBI. Second, they assumed a delta function in pitch for tractability, which is unrealistic considering the more broad distributions present in experiments, as inferred from Monte Carlo codes such as the \NUBEAM\cite{Pankin2004CPC} module in \TRANSP.\cite{Goldston1982JCP} Prior studies also assume $\kpar \ll \kperp$ and $\omega \ll \omegaci$ as simplifying approximations, whereas the modes excited in spherical tokamaks such as NSTX may have frequencies approaching $\omega \lesssim \omegaci$ and $\kpar \like \kperp$.

Choice of parameter regimes to study has been informed by prior and ongoing numerical modeling of CAEs/GAEs with the 3D hybrid MHD-kinetic initial value code \HYM.\cite{Belova2017POP,Lestz2018POP,Belova2019POP} The simulation model couples a single fluid thermal plasma to a minority species of full orbit kinetic beam ions and also includes the contributions of the large beam current to the equilibrium self-consistently.\cite{Belova2003POP} 

The derivation presented in this chapter corrects and builds on prior work by providing a local expression for the fast ion drive due to an anisotropic beam-like distribution interacting via the ordinary and anomalous cyclotron resonances. The effect of finite injection energy of NBI distributions is included consistently, yielding a previously overlooked instability regime. Terms to all order in $\omeganorm$ and $\krat$ are kept for applicability to the entire possible spectrum of modes. As in previous works, full finite Larmor radius (FLR) terms are also retained. The analytic expression can be integrated numerically for any chosen parameters in order to determine if the full fast ion distribution is net driving or damping. More interestingly, it is found that when the beam is sufficiently wide in velocity space, such as realistic distributions resulting from NBI, the integral can be evaluated approximately in terms of elementary functions, yielding compact conditions for net fast ion drive/damping that depend only on a small set of parameters describing the fast ion and mode parameters. Such expressions grant new insights into the spectrum of CAEs and GAEs that may be excited by a given fast ion distribution, as well as providing intuition for interpreting experimental observations and simulation results. Since damping sources such as electron Landau and continuum damping are not addressed in this chapter, the net fast ion drive conditions derived here should be considered as necessary but not sufficient conditions for instability. 

The chapter is structured as follows. The dispersion relations, resonance condition, and model fast ion distribution function used in this chapter are described in \secref{sec:cyc:dispres}. In \secref{sec:cyc:derivation}, the local analytic expression for the CAE and GAE growth rates is adapted from \citeref{Mikhailovskiiv6} and applied to the fast ion distribution of interest. Approximations are applied to this expression in \secref{sec:cyc:approxstab} in order to derive useful instability criteria for the cases of a very narrow beam width in velocity space (\secref{sec:cyc:narrow}) and a beam with realistic width (\secref{sec:cyc:wide}) when FLR effects are small (\secref{sec:cyc:slow}) and large (\secref{sec:cyc:fast}). The derived conditions are also compared against the numerically calculated growth rates for realistic parameter values in \secref{sec:cyc:approxstab}. In \secref{sec:cyc:stabao}, the dependence of the fast ion drive/damping on the mode properties ($\omeganorm$ and $\krat$) is presented and compared against conclusions drawn from the approximate stability boundaries. A comparison of the approximate stability conditions against a database of cntr-GAE activity in NSTX and simulation results is shown in \secref{sec:cyc:expcomp}. Lastly, a summary of the main results and discussion of their significance is given in \secref{sec:cyc:summary}. The majority of the content of this chapter has been peer-reviewed and published in \citeref{Lestz2020p1}. 

\section{Dispersion, Resonance Condition, and Fast Ion Distribution} 
\label{sec:cyc:dispres}

One goal of this work is to extend previous derivations to include finite $\omeganorm$ and $\krat$ effects in the stability calculation, since experimental observations and modeling of NSTX suggests that these quantities may not always be small. Experimental observations often show CAEs with frequencies from $\omeganorm = 0.3$ to exceeding the cyclotron frequency. GAEs are observed with somewhat lower frequencies of $\omeganorm \approx 0.1 - 0.5$. While $\kperp$ can not be measured accurately on NSTX due to limited poloidal coil resolution, it can be calculated for the most unstable modes excited in simulations,\cite{Belova2017POP} which show that $\krat \approx 1$ is not uncommon, and can even reach $\krat > 3$ in some cases. This motivates using the full, unsimplified dispersion relations in uniform geometry when numerically calculating the growth rate, instead of using the common $\omeganorm\ll 1$ and $\krat \ll 1$ assumptions found in previous works. The more complicated eigenmode equations in nonuniform toroidal systems\cite{Appert1982PP,Mahajan1983PF,Mahajan1983bPF,Gorelenkov1995NF,Gorelenkov1995POP} have been derived in the past but are too complicated for our purposes. 

Define $\omegabar = \omega/\omegacio$, $N = k\va/\omega$, $A = (1 - \omegabar^2)^{-1}$, and also $F^2 = \kpar^2/k^2$, $G = 1 + F^2$. Here, $\omegacio$ is the on-axis ion cyclotron frequency. Then in uniform geometry, the local dispersion in the MHD limits of $E_\parallel \ll E_\perp$ and $\omega \ll \abs{\omegace},\omegape$ is readily given by\cite{Stix1975NF} 

\begin{equation}
N^2 = \frac{AG}{2F^2}\left[1 \pm \sqrt{1 - \frac{4F^2}{AG^2}}\right]
\label{eq:cyc:stixdisp}
\end{equation}

The ``$-$" solution corresponds to the compressional \Alfven wave (CAW), while the ``$+$" solution corresponds to the shear \Alfven wave (SAW). The coupled dispersion in \eqref{eq:cyc:stixdisp} will be used in the full analytic expression for fast ion drive. Notably, it can modify the polarization of the two modes, which in turn changes how the finite Larmor radius (FLR) effects from the fast ions contribute to the growth rate (see \eqref{eq:cyc:Jlm}). Its low frequency approximations are $\omega \approx k\va$ for CAWs and $\omega \approx \abs{\kpar}\va$ for SAWs. Throughout this chapter, CAW/CAE and SAW/GAE will be used interchangeably, where CAW and SAW formally refer to the solutions in a uniform slab, while CAE and GAE refer to their analogues in nonuniform and bounded geometries. Net energy transfer between a mode and the fast ions requires a sub-population of particles obeying the Doppler-shifted cyclotron resonance. 

\begin{align}
\omega - \avg{\kpar\vpar} - \avg{\kperp\vdrift} = \lres\avg{\omegaci}
\label{eq:cyc:rescon}
\end{align}

Here, $\avg{\dots}$ denotes poloidal orbit averaging and $\lres$ is an integer cyclotron resonance coefficient. Two resonances are studied in detail in this chapter for the sub-cyclotron modes: the $\lres = 1$ ordinary cyclotron resonance and $\lres = -1$ anomalous cyclotron resonance. Orbit averaging in \eqref{eq:cyc:rescon} is required to satisfy the global resonance condition, as opposed to the local resonance, which describes a net synchronization condition between the wave and particle on average over its orbit, even while not being in constant resonance at all points in time. This resonance condition is applicable so long as the growth rate of the mode is sufficiently smaller than the inverse particle transit time, which is satisfied by these modes according to \HYM simulations.
 
In this chapter, we will make the approximation of $\abs{\kperp\vdrift} \ll \abs{\kpar\vpar}$. Consequently, when $\omega < \omegaci$ and $\avg{\vpar} > 0$ (co-injection), \eqref{eq:cyc:rescon} can only be satisfied for $\lres = 1$ if $\kpar < 0$ (mode propagates counter to the fast ions). Likewise, $\lres = -1$ requires $\kpar > 0$, corresponding to co-propagation. Due to periodicity, the drift term can be approximated for passing particles\cite{Belikov2003POP} as $\avg{\kperp\vdrift} \approx s\avg{\vpar}/qR$ for integer $s$, though this term yields relatively small corrections due to the large values of $\abs{\kpar}$ relevant to these modes. In this approximation, the resonance condition can be rewritten as $\omega - \kpars\vpres = \lres\avg{\omegaci}$ with $\kpars = \kpar + s/q R$. Conversely, for trapped particles the drift term can be approximated as\cite{Belikov2004POP} $\avg{\kperp\vdrift} \approx s\omega_b$. \HYM simulations indicate that the $s = \pm 1$ sidebands are usually more relevant than larger $\abs{s}$.\cite{Belova2017POP} For quantitatively accurate growth rates, all sidebands should be summed over, as done in \citeref{Gorelenkov1995POP} in the limit of $\omega\gtrsim\omegaci\gg\omega_b$, and also in \citeref{Kolesnichenko2006POP}. Practically, these procedures require complicated non-local calculations which would preclude analytic progress except in extraordinarily special cases, contrary to the purpose of this chapter, which is to derive broadly applicable instability conditions. To this end, only the primary resonance $(s = 0)$ will be kept when deriving approximate stability boundaries in \secref{sec:cyc:approxstab}.  

Combination of the resonance condition with approximate dispersion relations can yield relations that will be useful later on. Introduce $\omegacires \defined \avg{\omegaci}/\omegacio$ as the average cyclotron frequency of the resonant particles, normalized to the on-axis cyclotron frequency $\omegacio$. This value is approximately 0.9, as inferred from inspection of the resonant particles in relevant \HYM simulations. Then defining $\vpres \defined \avg{\vpar} > 0$ (treating co-injected particles only) and rearranging \eqref{eq:cyc:rescon} gives

\begin{align}
\frac{\vpres}{\va} &= \abs{\frac{\omega}{\kpar\va}}\abs{1 - \frac{\lres\omegacires}{\omegabar}} 
\approx \left\{\begin{array}{ll}
\abs{1 - \frac{\lres\omegacires}{\omegabar}} & \GAElab \\ 
\sqrt{1 + \frac{\kperp^2}{\kpar^2}}\abs{1 - \frac{\lres\omegacires}{\omegabar}} & \CAElab
\end{array}\right.
\label{eq:cyc:vpres}
\end{align}

The stability calculation will be applied to a slowing down, beam-like background distribution of fast ions, motivated by theory and \NUBEAM modeling of NSTX discharges.\cite{Belova2017POP} In order to satisfy the steady state Vlasov equation, the distribution is written as a function of constants of motion $v = \sqrt{2\W/m_i}$ and $\lambda = \mu B_0 / \W$ in separable form: $\fb(v,\lambda)= C_f n_b f_1(v)f_2(\lambda)$, defined below

\begin{subequations}
\begin{align}
\label{eq:cyc:F1}
f_1(v) &= \frac{\ftail(v;v_0)}{v^3 + v_c^3} \\ 
\label{eq:cyc:F2}
f_2(\lambda) &= \exp\left(-\left(\lambda - \lambda_0\right)^2 / \Delta\lambda^2\right)
\end{align}
\label{eq:cyc:Fdistr}
\end{subequations}

The constant $C_f$ is for normalization.  
The first component $f_1(v)$ is a slowing down function in energy with a cutoff at the injection energy $v_0$ and a critical velocity $v_c$. The cutoff at $v = v_0$ is contained within $\ftail(v;v_0)$, which is in general a function which rapidly goes to zero for $v>v_0$. For ease of calculation, this is assumed to be a step function. The second component $f_2(\lambda)$ is a Gaussian distribution centered on some central value $\linj$ with width $\dl$. The variable $\lambda$ is a trapping parameter. To lowest order in $\mu \approx \mu_0$, it can be re-written as $\lambda = (\vperp^2/v^2)(\omegacio/\omegaci)$. Then, assuming a tokamak-like field $B \approx B_0/(1 + \epsilon\cos\theta)$ for $\epsilon = r/R$, passing particles will have $0 < \lambda < 1 - \epsilon$ and trapped particles will have $1 - \epsilon < \lambda < 1 + \epsilon$. Loosely, smaller $\lambda$ means the particle's velocity is more field aligned, such that $\lambda$ is a complementary variable to a particle's pitch $\vpar/v$. For analytic tractability, $\linj$ and $\dl$ are treated as constants in this model, ignoring any velocity dependence of these parameters which may be present, especially broadening in $\lambda$ at lower energies due to pitch angle scattering. The dependence on $\pphi$, is neglected in this study for simplicity, as it is expected to be less relevant for the high frequencies of interest for these modes. The model distribution does not include the two additional energy components that are present due to molecular deuterium production in the neutral beam source, as these have a quantitative but not qualitative impact on the analysis. Such effects can be recovered by summing over three beam distributions (with injection velocities $v_0$, $v_0/\sqrt{2}$, and $v_0/\sqrt{3}$) with appropriate weights. Comparison between the model distribution used in this study and those calculated with the Monte Carlo code \NUBEAM for NSTX and NSTX-U can be found in Fig. 5 of \citeref{Belova2003POP} and Fig. 4 of \citeref{Belova2019POP}, respectively.

The NSTX operating space spanned a range of normalized injection velocity $\vinj = 2 - 6$, depending on the beam voltage (typically $60 - 90$ keV at $2 - 6$ MW) and field strength ($0.25 - 0.50$ T) for each discharge. The beam injection geometry $\linj$ and width in velocity space $\dl$ are mostly determined by the neutral beam's geometry and collimation, yielding typical $\linj = 0.5 - 0.7$ and $\dl = 0.3$. For this study, $v_c = v_0/2$ is used as a characteristic value. The new beam line on NSTX-U has much more tangential injection, with $\linj \approx 0$, and also lower $\vinj = 1 - 3$ due to higher nominal field strength. A comparison between the model fast ion distribution used in this chapter (\eqref{eq:cyc:Fdistr}) and a \NUBEAM calculation for the well-studied H-mode discharge $\# 141398$, can be found in Fig. 5 of \citeref{Belova2003POP}.

\section{Fast Ion Drive for Anisotropic Beam Distribution in the Local Approximation} 
\label{sec:cyc:derivation}

In this section, the fast ion drive/damping is derived perturbatively in the local approximation for a two component plasma comprised of a cold bulk plasma and a minority hot ion kinetic population, and applied to the anisotropic beam distribution of interest. The formula presented here extends the results obtained in \citeref{Gorelenkov2003NF,Kolesnichenko2006POP}, which focused on $\omega \ll \omegaci$, $\kpar \ll \kperp$, and also did not study high frequency co-propagating  modes ($\lres = -1$ cyclotron resonance coefficient). In contrast, the following derivation is appropriate for all values of $\omeganorm$ and $\krat$, which is important since mode frequencies can be on the order $\omeganorm \like 0.5$ or larger, and in contrast to the common large tokamak assumption, $\krat$ can be of order unity, as inferred from simulations.\cite{Lestz2018POP} 

\subsection{Derivation}
\label{sec:cyc:subderiv}

The general dispersion is given by 

\begin{equation}
\abs{\epsilon_{ij} - n^2\left(\delta_{ij} - \frac{k_i k_j}{k^2}\right)} = 0 
\end{equation}

Here, $n = k c / \omega$ is the index of refraction, $\epsilon_{ij} = \delta_{ij} + \sum_s \epsilon_{ij}^s$ is the dielectric tensor. Without loss of generality, assume $\vB_0 = B_0\hat{z}$ and $\vk = \kpar\hat{z} + \kperp\hat{x}$. Then the dispersion is determined by 

\begin{equation}
\left(\begin{array}{cc} 
\epsilon_{11} - n_\parallel^2 & \epsilon_{12} \\ 
\epsilon_{21} & \epsilon_{22} - n^2
\end{array}\right)
\left(\begin{array}{cc}
E_x \\ E_y 
\end{array}\right) = 0
\end{equation}

The rest of the components are irrelevant in the MHD regime where $E_z \ll E_x,E_y$. For the cold bulk components,

\begin{equation}
\delta_{ij} + \epsilon_{ij}^{th,e} + \epsilon_{ij}^{th,i} = 
\left(\begin{array}{cc}
S & -iD \\ 
iD & S
\end{array}\right) 
\end{equation}

Above, $S = 1 - \sum_s \omegaps^2/(\omega^2 - \omegacs^2)$ and $D = \sum_s \omegacs\omegaps^2/(\omega(\omega^2 - \omegacs^2))$, where $\omegaps = \sqrt{n_s q_s^2/(m_s \epsilon_0)}$ and $\omegacs = q_s B_0 / m_s$ are the plasma frequency and signed cyclotron frequency for each species $s$. When $\omega\ll \omegape,\abs{\omegace}$, we can approximate $S \approx A c^2/\va^2$ and $D \approx -\omegabar A c^2/\va^2$, where as earlier $A = 1/(1 - \omegabar^2)$ and $\omegabar = \omega/\omegacio$. Setting $\Kij = \va^2 \epsilon_{ij}^b/c^2$ and also defining $y = \omega^2/(k^2\va^2) = N^{-2}$, the full dispersion is given by 

\begin{multline}
\left(y - F^2 \bvar - y \bvar \Kxx\right)\left(y - \bvar - y \bvar \Kyy\right) 
- y^2\left(\omegabar + \bvar\Kxy\right)^2 = 0 
\end{multline}

Neglecting the fast ion component (setting $\Kij = 0$) recovers the MHD dispersion in \eqref{eq:cyc:stixdisp}. Letting $\omega = \omega_0 + \omega_1$ with $\omega_1 \ll \omega_0$ and solving perturbatively to first order in $\Kij \like n_b/n_e \ll 1$ yields the growth rate as

\begin{align}
\frac{\omega_1}{\omega_0} = -\frac{y_0\left[\Kxx(y_0 - \bvar_0) -2\omegabar_0 y_0\abs{\Kxy} + (y_0 - F^2 \bvar_0)\Kyy\right]}{2\left(y_0^2 - F^2\right)}
\label{eq:cyc:omegapert}
\end{align}

As defined in \secref{sec:cyc:dispres}, $F^2 = \kpar^2/k^2$. All quantities with subscript $0$ are understood to be evaluated using $\omega = \omega_0$, \ie the unperturbed frequency given by \eqref{eq:cyc:stixdisp}. The tensor elements $\Kij$ can be calculated from Eq. A24 in \citeref{Mikhailovskiiv6}:

\begin{align}
\label{eq:cyc:ktens}
\Kij &= \frac{n_b}{n_e}\frac{\omegaci^2}{\omega}\int \vperp d\vperp d\vpar \sum_{\lres = -\infty}^\infty \frac{\vperp^2 g_{ij}^\lres(\flr)}{\omega - \kpar\vpar - \lres\omegaci}\hat{\pi}\fb \\ 
\label{eq:cyc:piparperp}
\text{where } \hat{\pi} &= \frac{1}{\vperp}\pderiv{}{\vperp} + \frac{\kpar}{\omega}\left(\pderiv{}{\vpar} - \frac{\vpar}{\vperp}\pderiv{}{\vperp}\right) \\ 
g_{ij}^\lres(\flr) &= \left(\begin{array}{cc}
\lres^2 \Jl^2/\flr^2 & i\lres \Jl^\prime \Jl/\flr \\ 
-i\lres \Jl^\prime \Jl/\flr & (\Jl^\prime)^2
\end{array}\right),\, \flr = \kperp\rhob
\end{align}

Above, $\rhob = \vperp/\omegaci$ is the Larmor radius of the fast ions, and the distribution is normalized such that $\int \fb\vperp d\vperp d\vpar = 1$. The finite Larmor radius (FLR) effects from the fast ions are contained in $g_{ij}^\lres(\flr)$, with $\Jl(\flr)$ denoting the $\lres^{th}$ order Bessel function of the first kind. In order to keep only the resonant contribution to the growth rate, we make the formal transformation $(\omega - \kpar\vpar - \lres\omegaci)^{-1} \rightarrow -i\pi\delta(\vpar - \vpreslres)/\abs{\kpar}$ with $\vpreslres = (\omega - \lres\omegaci)/\kpar$ the parallel velocity of the resonant fast ions. Then substituting \eqref{eq:cyc:ktens} into \eqref{eq:cyc:omegapert} and identifying the growth rate $\gamma = \text{Im}(\omega_1)$,  

\begin{align}
\label{eq:cyc:gammageneral}
&\frac{\gamma}{\omegaci} = \frac{\pi}{2}\frac{n_b}{n_e}\sum_\lres \abs{\frac{\vpreslres}{\omegabar - \lres}} 
\int d\vperp d\vpar \vperp^3 \delta(\vpar - \vpreslres)\hat{\pi}_\lres\fb \Jlm(\flr)  \\ 
&\text{where } \hat{\pi}_\lres = \frac{2}{v^2}\left[\left(\frac{\lres}{\omegabar} - x\right)\pderiv{}{x} + \frac{v}{2}\pderiv{}{v}\right] 
\end{align} 

The variable $x = \vperp^2/v^2 = \lambda\omegacires$ was introduced so that the gradients $\hat{\pi}\fb$ can be re-written in the natural coordinates of the distribution. Note that $\Jlm(\flr)$ is the ``FLR function" for cyclotron resonance $\lres$ and mode $m$ (= `$C$' for CAE and `$G$' for GAE), defined as 
\begin{align}
\Jlm(\flr) &\defined \frac{y_0}{y_0^2 - F^2}\left[\sqrt{y_0 - \bvar_0}\frac{\lres\Jl}{\flr} \mp \sqrt{y_0 - F^2 \bvar_0}\deriv{\Jl}{\flr}\right]^2
\label{eq:cyc:Jlm}
\end{align}

Above, the ``$-$" corresponds to CAEs and the ``$+$" for GAEs. Defining $\alpha = \krat$, the FLR parameter $\flr$ may also be re-written in the following form: 

\begin{align}
\label{eq:cyc:zsimp}
\flr &= \kperp\rhob \defined \zp \sqrt{\frac{x}{1-x}} \\ 
\zp &= \frac{\kperp\vpres}{\omegaci} = \frac{\abs{\omegabar - \lres\omegacires}}{\alpha}
\label{eq:cyc:zp}
\end{align}

The modulation parameter $\zp$ contains information about the mode characteristics and is a measure of how rapidly the integrand in \eqref{eq:cyc:gammageneral} is oscillating. The expression in \eqref{eq:cyc:zp} follows from the resonance condition in \eqref{eq:cyc:vpres}. The complicated form of $\Jlm(\flr)$ is due to coupling between the pure compressional and shear branches of the dispersion resulting from finite $\omeganorm$ and also modified by finite $\krat$, so it is worthwhile to highlight some of its properties. The FLR function $\Jlm(\flr)$ is non-negative for both modes when $\omeganorm < 1$. For CAEs, $y_0 \geq 1 \geq \bvar_0, F, F^2\bvar_0$ according to \eqref{eq:cyc:stixdisp}, so the square root arguments and leading factors are all positive. In contrast, for GAEs, $y_0 \leq \bvar_0, F, F^2\bvar_0$, so the arguments of the square roots as well as the leading factors are all negative, with signs canceling out. 

As a useful example, consider the limit of $\omeganorm \ll 1$. In that case, $y_0 = 1 + \omegabar^2\alpha^2 \plusord{\omegabar^4}$ for CAEs and $y_0 = F^2 - \omegabar^2\alpha^2 \plusord{\omegabar^4}$ for GAEs. Then $\Jlm(\flr)$ simplifies substantially to

\begin{subequations}
\label{eq:cyc:Jlmsmallappx}
\begin{align}
\label{eq:cyc:Jlmsmallappx-cae}
\lim_{\omegabar\rightarrow 0} \Jlc(\flr) &= \left(\Jlprime\right)^2
\CAElab \\ 
\label{eq:cyc:Jlmsmallappx-gae}
\lim_{\omegabar\rightarrow 0} \Jlg(\flr) &= \left\{\begin{array}{ll}
\left(\lres\Jl/\flr\right)^2 
& \lres \neq 0 \\ 
\left(\omegabar\alpha^2 J_1\right)^2 & \lres = 0 
\end{array}\right. 
\GAElab
\end{align}
\end{subequations} 

In another limit, where $0 < \omegabar < 1$ and $\alpha \gg 1$, the dispersion from \eqref{eq:cyc:stixdisp} reduces to $y_0 = 1 + \omegabar$ for CAEs and $y_0 = 1 - \omegabar$ for GAEs, simplifying the FLR function to 

\begin{subequations}
\label{eq:cyc:Jlmbigappx}
\begin{align}
\label{eq:cyc:Jlmbigappx-a}
\lim_{\alpha\rightarrow \infty} \Jlm(\flr) &= \frac{\left(1 \pm \omegabar\right)^2}{2\pm \omegabar}\left(\Jlprime \mp \frac{\lres\Jl}{\flr}\right)^2 \\ 
\label{eq:cyc:Jlmbigappx-cae}
\lim_{\alpha\rightarrow \infty} \Jlc(\flr) &= \frac{\left(1 + \omegabar\right)^2}{2 + \omegabar} J_{\lres+1}^2 \CAElab \\ 
\label{eq:cyc:Jlmbigappx-gae}
\lim_{\alpha\rightarrow \infty} \Jlg(\flr) &= \frac{\left(1 - \omegabar\right)^2}{2 - \omegabar} J_{\lres-1}^2 \GAElab
\end{align}
\end{subequations}

In \eqref{eq:cyc:Jlmbigappx-a}, the top signs are for CAEs, and the bottom signs for GAEs. The forms in \eqref{eq:cyc:Jlmsmallappx} match those used in \citeref{Gorelenkov2003NF,Kolesnichenko2006POP} in the same limit, and the limit of $\alpha \rightarrow 0$ of \eqref{eq:cyc:Jlm} reproduces the FLR function used in \citeref{Belikov2003POP,Belikov2004POP}. 

Since the distribution is written in terms of the variable $x$ instead of $\vperp$, it is useful to change variables after performing the trivial integration over $\vpar$ in \eqref{eq:cyc:gammageneral}. Using the definition $x = \vperp^2/v^2$ with differential relation $dx = 2\vperp(1-x)d\vperp/v^2$ gives 

\begin{align*}
\label{eq:cyc:gammageneralx}
&\frac{\gamma}{\omegaci} = \frac{\pi}{2} \frac{n_b}{n_e}\sum_\lres \abs{\frac{\vpreslres^3}{\omegabar - \lres}} 
\int \frac{x \Jlm(\flr)}{(1-x)^2} \left[\left(\frac{\lres}{\omegabar} - x\right)\pderiv{\fb}{x} + \frac{v}{2}\pderiv{\fb}{v}\right] dx \numberthis 
\end{align*} 

Lastly, \eqref{eq:cyc:gammageneralx} can be applied to the anisotropic beam distribution in \eqref{eq:cyc:Fdistr}. Defining $\bres_\lres = \vpreslres^2/v_0^2$ yields 

\begin{multline}
\frac{\gamma}{\omegaci} = -\frac{n_b}{n_e}\frac{\pi C_f v_0^3 }{v_c^3} \sum_\lres \frac{\bres_\lres^{3/2}}{\abs{\omegabar-\lres}} \times \\ 
\left\{ \int_0^{1-\bres_\lres} \frac{x \Jlm(\flr(x,\zp))}{(1-x)^2}\frac{e^{-(x-\xinj)^2/\dx^2}}{1 + \frac{v_0^3}{v_c^3}\left(\frac{\bres_\lres}{1-x}\right)^{3/2}} \right. 
\left[\frac{1}{\dx^2}\left(\frac{\lres}{\omegabar} - x\right)(x-\xinj) + \frac{3/4}{1 + \frac{v_c^3}{v_0^3}\left(\frac{1-x}{\bres_\lres}\right)^{3/2}}\right]dx   \\
\left.\vphantom{\left[\frac{3/4}{1 + \left(\frac{1-x}{4\bres_\lres}\right)^{3/2}}\right]}
+ \frac{\bres_\lres^{-1}-1}{2\left(1 + \frac{v_0^3}{v_c^3}\right)}e^{-(1 - \bres_\lres-\xinj)^2/\dx^2}\Jlm\left(\zp\sqrt{\bres_\lres^{-1}-1}\right)\right\}
\label{eq:cyc:gammabeam}
\end{multline}

The upper integration bound is a consequence of the finite injection energy since $\abs{\vpres} = v\sqrt{1 - x} < v_0 \sqrt{1 - x} \rightarrow x < 1 - \vpres^2/v_0^2$. All quantitative calculations in this chapter assume $v_c = v_0/2$ and $n_b/n_e = 5.3\%$, based on the conditions in the well-studied NSTX H-mode discharge $\# 141398$. The normalization constant is given by 

\begin{align}
C_f^{-1} &= \frac{1}{3}\ln\left(1 + \frac{v_0^3}{v_c^3}\right)\int_0^1 \frac{e^{-(x-\xinj)^2/\dx^2}}{\sqrt{1 - x}}dx 
\end{align}

This approach required two large assumptions in order to make the problem tractable. First, a local assumption was made in order to eliminate the spatial integrals, which require knowledge or detailed assumptions about the equilibrium profiles and mode structures, whereas we seek a simple criteria depending only on a few parameters ($\vinj, \linj, \omeganorm, \krat, \lres$) for broad comparison with experimental or simulation results. Hence, all equilibrium quantities in \eqref{eq:cyc:gammabeam} are understood to be taken at the peak of the mode structure, generally between the magnetic axis and mid-radius on the low-field side, where CAEs are localized due to a magnetic well and GAEs are localized due to a minimum in the \Alfven continuum. As a consequence, the accuracy of the drive/damping magnitude may be limited, however this approximation should not affect the sign of the expression, so it can still be used to distinguish net fast ion drive vs damping, which is the primary goal of this chapter. Second, the derivative with respect to $\pphi$ has been neglected in this derivation, which would be important for modes at lower frequencies (\eg for TAEs where it is the main source of drive) or fast ion distributions with very sharp spatial gradients, which is atypical for NBI. 

\subsection{Properties of Fast Ion Drive}
\label{sec:cyc:properties}

The expression in \eqref{eq:cyc:gammabeam} represents the local perturbative growth rate for CAE/GAEs in application to an anisotropic beam-like distribution of fast ions, keeping all terms from $\omeganorm$, $\krat$, and $\kperp\rhob$. The derivation presented in this section has some additional consequences worth highlighting. Observe that only the term in square brackets can change sign since the coefficient in front of the integral will always be negative, and the portions of the integrand not enclosed in square brackets are strictly nonnegative. Hence regions of the integrand where the term in brackets is negative are driving, and regions where these terms are positive are damping. 

Examining further, the second term in brackets and the term on the second line are due to $\partial\fb/\partial v$, which is always damping for the slowing down function. Both of these terms are negligible for $\lres \neq 0$, $\omeganorm < 1$ and $\dl < 1$, which is the case considered here. The first term in brackets is the fast ion drive/damping due to anisotropy $(\partial\fb/\partial\lambda)$, which usually dominates the $\partial\fb/\partial v$ terms except in a very narrow region where $\lambda \approx \linj$. Considering only fast ions with $\vpres > 0$, modes driven by the $\lres = -1$ resonance are destabilized by resonant particles with $\partial\fb/\partial\lambda < 0$ (equivalent to $\lambda > \lambda_0$ for our model distribution), whereas those interacting via the $\lres = 1$ resonance are driven by $\partial\fb/\partial\lambda > 0$ ($\lambda < \lambda_0$). This leads to a useful corollary to this expression without any further simplification: when $1 - \vpres^2/v_0^2 \leq \linj\omegacires$, the integrand does not change sign over the region of integration. Therefore, 

\begin{equation}
\label{eq:cyc:corollary}
1 - \vpres^2/v_0^2 \leq \linj\omegacires \rightarrow 
\left\{
\begin{array}{ll}
\gamma < 0 & \lres = -1 \\ 
\gamma > 0 & \lres = 1 
\end{array}
\right.
\end{equation}

For the single beam distribution in \eqref{eq:cyc:Fdistr}, if $1 - \vpres^2/v_0^2 \leq \linj\omegacires$, then modes driven by the $\lres = -1$ resonance (co-propagating) will be strictly damped by fast ions, while those driven by $\lres = 1$ (cntr-propagating) will exclusively be driven by fast ions. This represents a simple sufficient condition for net fast ion drive or damping when this relation between the mode properties ($\krat$ and $\omeganorm$, which determine $\vpres$ through the resonance condition) and fast ion distribution parameters ($v_0$ and $\linj$) is satisfied. 

Moreover, this condition reveals an instability regime unique to slowing down distributions generated by NBI with finite injection energy. This regime was not addressed in the initial studies, which considered either mono-energetic\cite{Belikov1968ZHETF} or bi-Maxwellian\cite{Timofeevv5} distributions for beam ions. Previous studies related to NBI-driven CAEs/GAEs\cite{Gorelenkov2003NF,Kolesnichenko2006POP} also overlooked this regime by implicitly assuming $\vpres \ll v_0$. Consequently, their results were used to interpret experimental observations in NSTX(-U)\cite{Fredrickson2004POP,Fredrickson2017PRL,Fredrickson2018NF} and DIII-D\cite{Heidbrink2006NF} in cases where they may not have been valid. In contrast, this new instability regime can more consistently explain the excitation and suppression of cntr-GAEs observed in NSTX-U,\cite{Fredrickson2017PRL,Belova2019POP} and also suggests that the properties of high frequency modes previously identified as CAEs in DIII-D\cite{Heidbrink2006NF} would in fact be more consistent with those of GAEs. 

Lastly, it is clear from the derivation and discussion in this section that $\lres = \pm 1$ instabilities can occur for any value of $\kperp\rhob$, depending on the parameters of the distribution $(\linj, \vinj)$ and the given mode properties $(\omeganorm,\krat)$. In contrast, in the previously studied regime where $\vpres \ll v_0$ and $\dl \ll 1$, net fast ion drive only occurs for specific ranges of $\kperp\rhob$ (provided that the resonance condition is satisfied).\cite{Gorelenkov2003NF} For further understanding of the relationships between the relevant parameters required for instability, analytic approximations or numerical methods must be employed. 

\section{Approximate Stability Criteria} 
\label{sec:cyc:approxstab}

The expression derived in \eqref{eq:cyc:gammabeam} can not be integrated analytically, and has complicated parametric dependencies on properties of the specific mode of interest: GAE vs CAE, $\krat$, $\omeganorm$, and the cyclotron coefficient $\lres$ as well as on properties of the fast ion distribution: $\vinj$, $\linj$, and $\dl$. For chosen values of these parameters, the net fast ion drive can be rapidly calculated via numerical integration. Whenever $1 - \vpres^2/v_0^2 \leq \linj\omegacires$, \eqref{eq:cyc:corollary} provides the sign of the drive/damping. When this inequality is not satisfied, there are also regimes where approximations can be made in order to gain insight into the stability properties analytically: one where the fast ion distribution is very narrow ($\dl \lesssim 0.10$) and one where it is moderately large $(\dl \gtrsim 0.20$). The former allows comparison with previous calculations,\cite{Gorelenkov2003NF,Kolesnichenko2006POP} while the latter includes the experimental regime where the distribution width in NSTX is typically $\dl \approx 0.30$. In this section, marginal stability criteria will be derived in these regimes. 

\subsection{Approximation of Very Narrow Beam}
\label{sec:cyc:narrow}

For the first regime, consider the approximation of a very narrow beam in velocity space. The purpose of this section is to determine when such an approximation can correctly capture the sign of the growth rate. For simplicity, also consider $\omeganorm \ll 1$ so that the anisotropy term dominates and also $\lres/\omegabar \gg x$. Then \eqref{eq:cyc:gammabeam} can be re-written as 

\begin{align}
\label{eq:cyc:gammasimp}
\frac{\gamma}{\omegaci} &\propto \int_0^{1-\bres} h(x)(x-\xinj)e^{-(x-\xinj)^2/\dx^2}dx \\ 
\text{where } h(x) &= -\frac{\lres C_f}{\dx^2} \frac{x}{(1-x)^2}\frac{\Jlm(\flr(x,\zp))}{1 + \frac{v_0^3}{v_c^3}\left(\frac{\bres}{1-x}\right)^{3/2}}
\end{align} 

If $\dx$ is very small, then the integral is dominated by a contribution in a narrow region $\xinj - \delta < x < \xinj + \delta$ where $\delta \approx 2\dx$. In this region, $h(x)$ can be approximated as a linear function, $h(x) \approx h(\xinj) +  (x-\xinj)h'(\xinj) \plusord{\dx^2}$. So long as $0 < \xinj - \delta$ and $\xinj + \delta < 1 - \bres$, this approximation can be applied:

\begin{align}
\frac{\gamma}{\omegaci} &\appropto h'(\xinj) \int_{\xinj-\delta}^{\xinj+\delta} (x - \xinj)^2e^{-(x-\xinj)^2/\dx^2}dx 
\label{eq:cyc:gammanarrow}
\end{align}

The integral is positive, so the sign of the growth rate is equal to the sign of $h'(\xinj)$. Note that this is the same instability regime as studied in previous papers on sub-cyclotron mode stability.\cite{Gorelenkov2003NF,Kolesnichenko2006POP} A comparison of the approximate narrow beam stability criteria to the unapproximated expression for cntr-GAEs with $\bres = 0.2$ is shown in \figref{fig:cyc:narrowcomp}. There, the dashed line shows the approximate analytic result \eqref{eq:cyc:gammanarrow} plotted as a function of $\xinj$ for $\dx = 0.04$ and different values of $\zp$. Values of $\xinj$ where $h'(\xinj) > 0$ indicate regions where the fast ions are net driving according to this assumption (shaded regions). For comparison, the full expression \eqref{eq:cyc:gammasimp} is integrated numerically for each value of $\xinj$ for varying $\dx = 0.04, 0.08, 0.16, 0.32$. This figure demonstrates where the narrow beam approximation correctly determines the sign of the fast ion drive, and how it depends on $\zp$. The curves for $\dx = 0.04$ and $\dx = 0.08$ have essentially the same roots as the analytic expression, whereas the zeros of $\dx = 0.16$ and $\dx = 0.32$ begin to drift away from the approximation or miss regions of instability entirely. The differences are most pronounced for larger values of $\zp$, since this causes the integrand to oscillate more rapidly. Hence, the approximate criteria in \eqref{eq:cyc:gammanarrow} is only reliable for $\dx \lesssim 0.10$, especially when $\zp \gg 1$, which is much more narrow than experimental fast ion distributions due to neutral beam injection which have $\dx \approx 0.30$ in NSTX. 

It is unsurprising that this type of approximation fails for realistically large values of $\dx$ since the width of the Gaussian spans nearly the entire integration region. Even for smaller $\dx$, the conclusion from \eqref{eq:cyc:gammanarrow} is restricted to situations when both $0 < \xinj - \delta$ and $\xinj + \delta < 1-\bres$ are satisfied. For instance, when $\bres = 0.2$ and $\dx = 0.1$, this expression is only strictly valid for $0.2 < \xinj < 0.6$. 


\newcommand{\figcycnarrowcomplen}{\halfwidth}
\begin{figure}[H]
\subfloat[]{\includegraphics[width = \figcycnarrowcomplen]{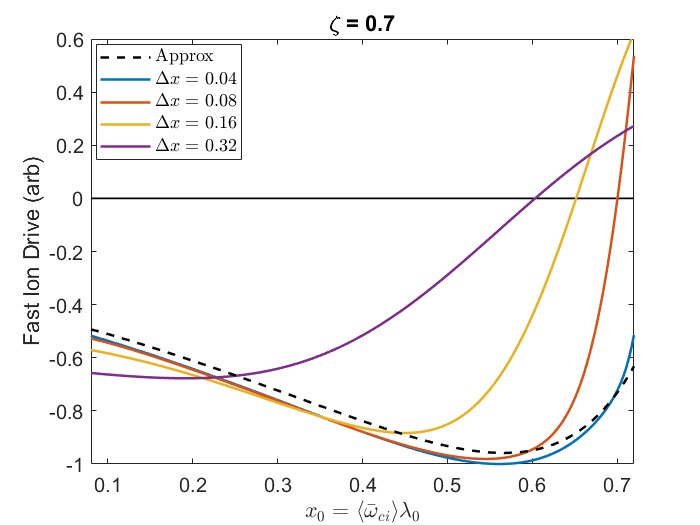}}
\subfloat[]{\includegraphics[width = \figcycnarrowcomplen]{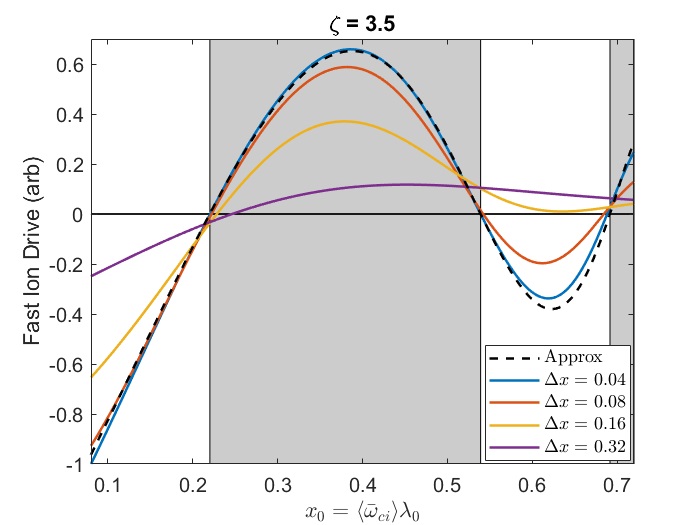}} \\
\subfloat[]{\includegraphics[width = \figcycnarrowcomplen]{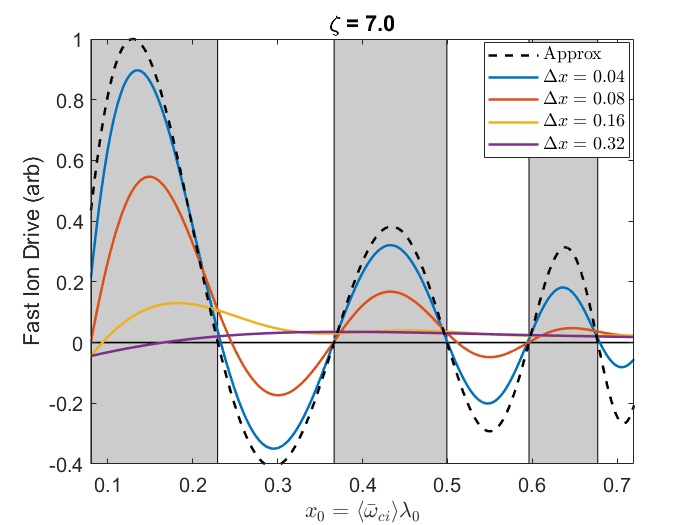}}
\caption
[Comparison of numerically integrated growth rate to narrow beam approximation for cntr-GAEs as a function of injection geometry.]
{Comparison of numerically integrated growth rate to narrow beam approximation for cntr-GAE with $\bres = 0.2$ as a function of the injection geometry $\xinj = \vperp^2/v^2$ of the beam distribution. Black dashed line shows the analytic approximation made in \eqref{eq:cyc:gammanarrow} for $\dx = 0.04$ and (a) $\zp = 0.7$, (b) $\zp = 3.5$, and (c) $\zp = 7.0$. Colored curves show numerical integration of \eqref{eq:cyc:gammasimp} for different values of $\dx$: blue $\dx = 0.04$, orange $\dx = 0.08$, gold $\dx = 0.16$, and purple $\dx = 0.32$. Shaded regions correspond to regions of drive according to the narrow beam approximation.}
\label{fig:cyc:narrowcomp}
\end{figure}


\subsection{Approximation of Realistically Wide Beam}
\label{sec:cyc:wide}

When the beam distribution instead has a non-negligible width in the trapping parameter $\lambda$, a complementary approach can be taken. For $\dx$ sufficiently large, one may approximate $d \exp(-(x-\xinj)^2/\dx^2)/dx \approx -2(x-\xinj)/\dx^2$. This is reasonable for $\xinj - \dx/\sqrt{2} < x < \xinj + \dx/\sqrt{2}$ since this linear approximation is accurate up to the local extrema in this function. When $\dx$ is large, this approximation region may cover nearly the entire region of integration. Throughout this section, $v_c = v_0/2$ will be taken as a representative figure, and the slowing down part of the distribution will be approximated as constant since it makes a small quantitative difference. Then \eqref{eq:cyc:gammabeam} may be well-approximated by

\begin{equation}
\gamma \appropto -\int_0^{1-\bres}\frac{x}{(1-x)^2}\Jlm(\flr)\left(\frac{\lres}{\omegabar} - x\right)\left(x - \xinj\right)dx
\label{eq:cyc:gammawide}
\end{equation}

This is still not possible to integrate directly because of the Bessel functions with complicated arguments in $\Jlm(\flr)$ since $\flr = \zp\sqrt{x/(1-x)}$. Substituting the values of $\omeganorm$ and $\krat$ from the most unstable modes in \HYM simulations into \eqref{eq:cyc:zp} shows that the majority of these modes have $\zp \approx 0.5 \tto 1$, with the largest values being $\zp \approx 3$. Since this parameter controls how rapidly $\Jlm(\flr)$ oscillates, we are motivated to consider two cases separately: the small ($\zp \ll 1$) and large ($\zp \gg 1$) FLR regimes. 

\subsubsection{Small FLR Regime \texorpdfstring{$(\zp \ll 1)$}{}}
\label{sec:cyc:slow}

For small $\zp$, the argument of the Bessel function will be small for most of the domain. For instance, $x = 1/(1 + \zp^2/\flr^2)$, so when $\zp = 0.5$, the small argument condition $\flr \ll 1$ is true for $x \ll 0.8$, which is the majority of the domain for $\bres$ not too small. 
The leading order approximation to $\Jlm(\flr)$ for $\lres = \pm 1$ and $\flr \ll 1$ is $c \plusord{\flr^2}$ with $c$ constant. For demonstration purposes, it will also be assumed that $\omegabar \ll 1$. This small correction is addressed in \appref{app:cyc:omegabar}. With this approximation, \eqref{eq:cyc:gammawide} can be simplified and then integrated exactly as 

\begin{align}
\gamma &\appropto -\lres\int_0^{1-\Bres}\frac{x(x-\xinj)}{(1-x)^2}dx
\end{align}

Solving for the marginal stability condition $\gamma = 0$ yields 

\begin{align}
\label{eq:cyc:wideslowGAEgamraw}
\xinj &= \frac{1 - \bres^2 + 2\bres\log\bres}{1 - \bres + \bres\log\bres} 
\approx 1 - \bres^{2/3} \\ 
\Rightarrow v_0 &= \frac{\vpres}{\left(1 - \xinj\right)^{3/4}}
\label{eq:cyc:wideslowGAEgam}
\end{align}

The serendipitous approximation is better than $1\%$ accurate everywhere. It is arrived at by noticing that \eqref{eq:cyc:wideslowGAEgamraw} is a smooth, convex, monotonically decreasing function on $(0,1) \rightarrow (0,1)$, which suggests an \emph{ansatz} of the form $f(x) = 1 - x^p$ for $0 < p < 1$. The choice of $p = 2/3$ is made in order to match the value of the derivative at the $x = 1$ boundary, which coincidentally also matches the second derivative there. At the the $x = 0$ boundary, the limit of the derivatives of both the function and approximation is $-\infty$ for odd derivatives and $+\infty$ for even derivatives. The success of this approximation technique for inverting the marginal stability condition motivates its repeated use in other cases studied in this chapter.

This stability condition depends implicitly on the mode parameters $\omegabar \defined \omeganorm$ and $\alpha \defined \krat$ through the dependence of $\vpres$, as in \eqref{eq:cyc:vpres}. The cases of $\lres = \pm 1$ have the same stability boundary, with an overall sign difference.  Hence, when $\zp \ll 1$, the cntr-propagating $\lres = +1$ CAEs/GAEs are destabilized by fast ion distributions with $v_0 < \vpres/(1 - \xinj)^{3/4}$ and the co-propagating $\lres = -1$ CAEs/GAEs have net fast ion drive when $v_0 > \vpres/(1 - \xinj)^{3/4}$. 

\begin{figure}[tb]
\includegraphics[width = \midwidth]{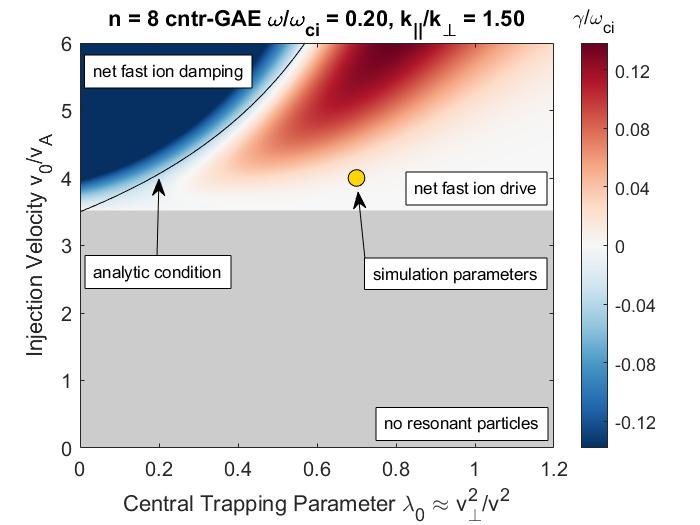}
\caption
[cntr-GAE growth rate dependence on beam injection geometry $\linj$ and velocity $\vinj$ in the wide beam, small FLR regime ($\zp \ll 1$).]
{Numerical integration of full growth rate expression \eqref{eq:cyc:gammabeam} as a function of fast ion distribution parameters $\vinj$ and $\linj$ with $\dx = 0.30$ for a cntr-GAE with properties inferred from \HYM simulations: $\omeganorm = 0.20$ and $\krat = 1.50$, implying $\zp = 0.47$. Red indicates net fast ion drive, blue indicates net fast ion damping, and gray indicates beam parameters with insufficient energy to satisfy the resonance condition. Black curve shows approximate stability condition derived in \eqref{eq:cyc:wideslowGAEgam}.}
\label{fig:cyc:wideslowGAEfig}
\end{figure} 

It is prudent to compare this approximate analytic condition against the numerical evaluation of \eqref{eq:cyc:gammabeam} for a characteristic mode. This is done in \figref{fig:cyc:wideslowGAEfig}, where the full expression for fast ion drive of $\lres = +1$ GAE is integrated numerically for a beam distribution with $\dl = 0.30$ (estimated experimental value) and a range of values of $\linj$ and $\vinj$. A representative $n = 8$ cntr-GAE is chosen from \HYM simulations which had $\omeganorm = 0.20$ and $\krat = 1.50$, implying a value of $\zp = 0.47$. The color indicates the sign of the growth rate: red is positive (net fast ion drive), blue is negative (net fast ion damping), while gray is used for beam parameters with insufficient energy to satisfy the resonance condition. The analytic instability condition derived in \eqref{eq:cyc:wideslowGAEgam} is shown as the black curve, demonstrating a remarkably good approximation to the full numerical calculation.

Similarly good agreement between the approximation and numerical calculation shown in \figref{fig:cyc:wideslowGAEfig} holds even up to $\zp \lesssim 2$ since $\flr = \zp\sqrt{x/(1-x)} \lesssim 1$ is typically still obeyed for most of the integration region in that case, so long as $\bres$ is not too small. Since $\zp = \abs{\omegabar - \lres}/\alpha$ (\eqref{eq:cyc:zp}), typically values of $\alpha \gtrsim 0.5$ lead to validity of this regime. When $\zp$ becomes too large, the lowest order Bessel function expansion of $\Jlm(\flr)$ employed in this section is no longer valid over enough of the integration domain for the result to be accurate. For values of $\omegabar$ and $\alpha$ which lead to $\zp \gg 2$, the asymptotic form of the Bessel functions must be used instead to find different stability boundaries, which are derived in \secref{sec:cyc:fast}. The ``wide beam'' approximate stability conditions remain a good approximation to the numerical calculation for about $0.20 < \dx < 0.80$. If $\dx$ is smaller than this minimum value, the wide beam approximation begins to break down, while $\dx$ larger than the maximum value is where the damping due to the neglected $\partial\fb/\partial v$ term begins to become more important and lead to a nontrivial correction.  

\subsubsection{Large FLR Regime \texorpdfstring{$(\zp \gg 1)$}{}}
\label{sec:cyc:fast}

Another limit can be explored, that is of the wide beam and rapidly oscillating integrand regime, namely $\zp \gg 1$. This limit is applicable when very large FLR effects dominate most of the region of integration. Based on the most unstable modes found in the \HYM simulations, this is not the most common regime for NSTX-like plasmas, but it can occur and is treated for completeness and comparison to the slowly oscillating results. 

This approximation allows the use of the asymptotic form of the Bessel functions: $J_n(\flr) \like \sqrt{2/\pi \flr}\cos\left(\flr - (2n+1)\pi/4\right) \plusord{\flr^{-3/2}}$, which is very accurate for $\flr > 2$. Note also that $\zp \gg 1$ implies $\alpha \ll 1$ since $\zp = \abs{\lres - \omegabar}/\alpha < 2/\alpha$ for $\abs{\lres} \leq 1 $. 
Since $\alpha \ll 1$, the FLR functions for $\lres = \pm 1$ are well-approximated by $\J{\pm1}{G} \like J_1^2(\flr)/\flr^2 \like 
(1 - \sin(2\flr))/ \flr^3$ for GAEs and $\J{\pm 1}{C}(\flr) \like J_0^2(\flr) \like (1 - \sin(2\flr))/ \flr$ for CAEs. Considering first the case of the $\lres = \pm 1$ GAEs, the relevant integral is 

\begin{align}
\gamma &\appropto -\lres\int_0^{1-\bres} \frac{dx}{\sqrt{x(1-x)}}\left[1 - \sin\left(2\zp\sqrt{\frac{x}{1-x}}\right)\right](x-\xinj) \label{eq:cyc:boappx}\\ 
&= -\lres\int_0^{1-\bres} \frac{(x-\xinj)}{\sqrt{x(1-x)}}dx \\ 
&= -\sqrt{\bres(1-\bres)} + (1 - 2\xinj)\arccos\sqrt{\bres}
\label{eq:cyc:highpsiGAE}
\end{align}

The first line is \eqref{eq:cyc:gammawide} using the asymptotic expansion of the Bessel functions, then the second line is obtained using the stationary phase approximation for rapidly oscillating integrands.\cite{BenderOrszagStationaryPhase} Specifically, the Riemann-Lebesgue lemma\cite{BenderOrszagStationaryPhase} guarantees that $\int_a^b f(t) e^{ixt}dt \rightarrow 0$ for $x\rightarrow \infty$ with integrable $\abs{f(t)}$, which is clear with the substitution of $t = 2\sqrt{x/(1-x)}$ in \eqref{eq:cyc:boappx}. Then as before, the marginal stability condition can be found and inverted after an approximation procedure: 

\begin{align}
\label{eq:cyc:highpsiGAEboundpre}
\xinj &= \frac{1}{2}\left(1 - \frac{\sqrt{\bres(1-\bres)}}{\arccos\sqrt{\bres}}\right) \approx \frac{1}{2}\left(1 - \bres^{2/3}\right) \\ 
\Rightarrow v_0 &= \frac{\vpres}{\left(1 - 2\xinj\right)^{3/4}}
\label{eq:cyc:highpsiGAEbound}
\end{align}

The approximation above is found with the same procedure as described for \eqref{eq:cyc:wideslowGAEgamraw}, and has a maximum relative error of $3\%$. Interestingly, this condition is similar to the one derived for $\zp \ll 1$ except that $(1 - \xinj)$ has been replaced by $(1 - 2\xinj)$. This condition describes the boundary for $\lres = \pm 1$ GAEs, with $v_0 > \vpres/(1 - 2\xinj)^{3/4}$ indicating net fast ion drive for $\lres = -1$ co-GAEs and net fast ion damping for $\lres = +1$ cntr-GAEs. 

\begin{figure}[tb]
\includegraphics[width = \midwidth]{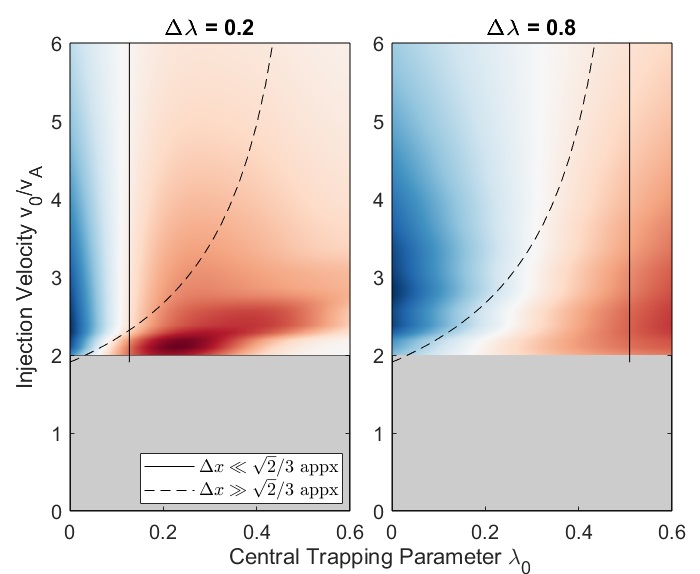}
\caption
[cntr-GAE growth rate dependence on beam injection geometry $\linj$ and velocity $\vinj$ in the wide beam, large FLR regime ($\zp \gg 1$).]
{Comparison of approximations for marginal fast ion drive for cntr-GAEs with $\zp \gg 1$ and $\dx \lesssim \sqrt{2}/3$ vs $\dx \gtrsim \sqrt{2}/3$. Left is the former (with $\dx = 0.20$) and right is the latter (with $\dx = 0.80$). Both use $\omeganorm = 0.3$ and $\krat = 0.07$ so that $\zp = 8.6$, and also $\omegacires = 0.9$. Red indicates net fast ion drive, while blue indicates net fast ion damping, and gray indicates beam parameters with insufficient energy to satisfy the resonance condition. The vertical line is the approximate marginal stability boundary of $\xinj = \dx/\sqrt{2}$, valid when $\dx \lesssim \sqrt{2}/3$ for $\zp \gg 1$. The dashed curve is the approximate marginal stability boundary of $\vinj = \vpres/(1 - 2\linj\omegacires)^{3/4}$, valid when $\dx \gtrsim \sqrt{2}/3$ for $\zp \gg 1$.}
\label{fig:cyc:widefastGAEfig}
\end{figure} 

When compared to the exact numerical calculation in this regime, \eqref{eq:cyc:highpsiGAEbound} captures the qualitative feature that the stability boundary occurs at much lower $\xinj$ than in the low $\zp$ regime. However, the quantitative agreement is not as good unless $\dx \approx 0.6$. For smaller values of $\dx$, the approximations become poor for large $x \gtrsim \xinj + \dx \sqrt{2}$ where the Gaussian decay would tend to dominate the diverging term $1/\sqrt{1-x}$ at $x\rightarrow 1$. This can be seen in \figref{fig:cyc:widefastGAEfig} where the marginal stability boundary approaches a vertical asymptote. To capture this behavior, the wide beam approximation can still be used, but with the integration running from $x = 0 \tto a = \xinj + \dx\sqrt{2}$ instead of $x = 0 \tto 1 - \bres$ to replicate the decay expected beyond this region. Then, the fast ion drive is approximately 

\begin{align}
%
\gamma &\appropto \lres\int_0^{a} \frac{(x-\xinj)}{\sqrt{x(1-x)}}dx \\ 
&= -\sqrt{a(1 - a)} + (1 - 2\xinj)\arcsin\sqrt{a} \\ 
\Rightarrow \xinj &= \frac{1}{2}\left[1 - \frac{\sqrt{a(1-a)}}{\arcsin\sqrt{a}}\right] 
\approx \frac{1}{2}\left[1 - (1 - \xinj - \dx\sqrt{2})^{2/3}\right] 
\label{eq:cyc:highpsiGAElim}
\end{align}

The approximation in the last line has a maximum global error of $3\%$. If $\xinj + \dx\sqrt{2}$ is close to 1, then the term in round braces is small, and the limit of $\xinj \rightarrow 1/2$ is recovered from \eqref{eq:cyc:highpsiGAEbound}. Hence, the other case of interest is when $\xinj + \dx\sqrt{2}$ is small, in which case a linear approximation admits a solution for \eqref{eq:cyc:highpsiGAElim} of $\xinj = \dx/\sqrt{2}$, which gives much better agreement with the numerically calculated boundary shown in \figref{fig:cyc:widefastGAEfig}. Hence, \eqref{eq:cyc:highpsiGAEbound} is applicable for $\dx \gtrsim \sqrt{2}/3$, whereas $\xinj = \dx/\sqrt{2}$ gives the limiting boundary for smaller $\dx$. 

A similar procedure can be used to approximate the marginal stability boundaries for CAEs, however it is rare for CAEs to be excited with $\zp \gg 1$ for the parameters studied here. This is because the CAE dispersion combined with the resonance condition yields $\zp \approx \omegabar\vpres/\va$ for $\zp \gg 1$, which can not be very large for $\vinj < 6$ considering $\vpres \like v_0/2$ is common, as is $\omeganorm \like 1/2$. The case is different for GAEs since their dispersion yields a parallel resonant velocity that is independent of $\alpha$, such that $\zp$ can be made arbitrarily large by choosing $\alpha$ sufficiently small without constraining the size of $\vpres/\va$. The case of $\zp \gg 1$ for CAEs with $\lres = \pm 1$ is treated in \appref{app:cyc:fastcae}. 

\subsection{Summary of Necessary Conditions for Net Fast Ion Drive}

\newcommand{\vroomcyc}{\vphantom{{\Huge text}}}
\newcommand{\hroomcyc}{\hspace{1ex}}
\newcommand{\cheadcyc}[1]{\multicolumn{1}{c}{#1}}
\begin{table*}[t]\centering
\renewcommand\arraystretch{1.5}
\begin{tabular}{c}
\begin{tabular}[t]{l l l}
\hline\hline 
\multicolumn{3}{c}{CAE fast ion drive conditions} \\ \hline\hline
 & \cheadcyc{$\lres = +1$ (cntr)} 
 & \cheadcyc{$\lres=-1$ (co)} \\ 
\hline
$\zp \lesssim 2$ \hroomcyc\hroomcyc & $v_0 < \dfrac{\vpres}{(1 - \xinj)^{3/4}}$ \hroomcyc\hroomcyc & $v_0 > \dfrac{\vpres}{(1 - \xinj)^{3/4}}$ \hroomcyc\vroomcyc \\
$\zp \gg 2$ & $v_0 < \dfrac{\vpres}{(1 - \xinj)^{5/6}}$ & $v_0 > \dfrac{\vpres}{(1 - \xinj)^{5/6}}$ \vroomcyc \vspace{1ex}\\ 
\hline\hline
\end{tabular}
\\
\begin{tabular}[t]{l l l l}
\hline\hline 
\multicolumn{4}{c}{GAE fast ion drive conditions} \\ \hline\hline
 & & \cheadcyc{$\lres=+1$ (cntr)} & \cheadcyc{$\lres=-1$ (co)} \\ 
\hline
$\zp \lesssim 2$ & & $v_0 < \dfrac{\vpres}{(1 - \xinj)^{3/4}}$ \hroomcyc & $v_0 > \dfrac{\vpres}{(1 - \xinj)^{3/4}}$ \vroomcyc \\
\multirow{2}{3em}{$\zp \gg 2$} & $\dx \lesssim \sqrt{2}/3$ \hroomcyc & $\xinj > \dx/\sqrt{2}$ & $\xinj < \dx/\sqrt{2}$ \vroomcyc \\
& $\dx \gtrsim \sqrt{2}/3$ & $v_0 < \dfrac{\vpres}{(1 - 2\xinj)^{3/4}}$ & $v_0 > \dfrac{\vpres}{(1 - 2\xinj)^{3/4}}$ \vroomcyc \vspace{1ex} \\ 
\hline\hline
\end{tabular}
\end{tabular}

\caption
[Approximate net fast ion drive conditions for GAEs and CAEs driven by $\lres = \pm 1$ resonances in the wide beam approximation.]
{Approximate net fast ion drive conditions for GAEs and CAEs driven by $\lres = \pm 1$ resonances in the wide beam approximation, valid for $0.2 < \dx < 0.8$ where $\dx = \dl\omegacires$ characterizes the velocity anisotropy of the beam. The quantity $\zp = \kperp\vpres/\omegaci$ is the ``modulation parameter" (see \eqref{eq:cyc:zp}) and $x_0 = \lambda_0\omegacires = \vperpz^2/v_0^2$.}
\label{tab:cyc:appxcons}
\end{table*}

For clarity, it is worthwhile to summarize all of the conditions for net fast ion drive derived in this section and remind the reader of their respective ranges of validity. When $1 - \vpres^2/v_0^2 \leq \linj\omegacires$ is satisfied, $\lres = -1$ modes will be net damped by fast ions, while those interacting via the $\lres = 1$ resonance will be net driven. All other results address the scenarios when this inequality is not satisfied, which is the parameter regime considered by previous authors.\cite{Gorelenkov2003NF,Kolesnichenko2006POP} When $\dl$ is sufficiently small $(\dl \lesssim 0.10)$, the narrow beam approximation can be made, which yields \eqref{eq:cyc:gammanarrow} and implies that net drive vs damping depends on the sign of $h'(\xinj)$. When $\dl$ is sufficiently large $(0.20 \lesssim \dl \lesssim 0.80)$, the wide beam approximation is justified. This includes the nominal NSTX case of $\dl \approx 0.3$. For most of the unstable modes in \HYM simulations, $\zp \lesssim 2$ is also valid, which facilitates the results obtained in the case of a wide beam with small FLR effects. The complementary limit of $\zp \gg 2$ is also tractable when the beam is sufficiently wide, though this is not the typical case in NSTX conditions, except for some low $n$ cntr-GAEs. All conditions for the cases involving wide beams are organized in \tabref{tab:cyc:appxcons}. 

\section{Preferential excitation as a function of mode parameters} 
\label{sec:cyc:stabao}

For fixed beam parameters, the theory can determine which parts of the spectrum may be excited -- complementary to the previous figures which addressed how the excitation conditions depend on the two beam parameters for given mode properties. Such an examination can also illustrate the importance of coupling between the compressional and shear branches due to finite frequency effects on the most unstable parts of the spectra. All fast ion distributions in this section will be assumed to have $\dl = 0.3$ and $\omegacires = 0.9$ for the resonant ions. 

\subsection{GAE Stability}
\label{sec:cyc:GAEao}

Consider first the GAEs. As a consequence of the approximate dispersion $\omega \approx \abs{\kpar}\va$, the necessary condition $\vpres < v_0$ for resonant interaction, and the net fast ion drive condition derived in \eqref{eq:cyc:wideslowGAEgam},
the region in $(\omegabar,\alpha)$ space corresponding to net fast ion drive in the typical case of $\zp\lesssim 1$ is nearly independent of $\alpha$. For counter-propagating modes with $\lres = 1$,   

\begin{align}
\frac{\omegacires}{\vinj + 1} < 
\left(\frac{\omega}{\omegaci}\right)_{\lres = 1}^{GAE} < 
\frac{\omegacires}{\vinj\left(1 - \linj\omegacires\right)^{3/4} + 1}
\label{eq:cyc:GAEmfreqrange}
\end{align}

Hence, the theory predicts a relatively small band of unstable frequencies. Larger $\vinj$ decreases both boundaries, leading to a range of unstable frequencies of about $(\omega_\text{max} - \omega_\text{min})/\omegaci \approx 10 - 20\%$. 

For co-propagating GAEs driven by $\lres = -1$, there is instead a lower bound on the unstable frequencies: 

\begin{align}
\frac{\omegacires}{\vinj(1-\linj\omegacires)^{3/4}-1} < \left(\frac{\omega}{\omegaci}\right)_{\lres = -1}^{GAE} < 1
\label{eq:cyc:GAEpfreqrange}
\end{align}

\begin{figure}[tb]
\subfloat[]{\includegraphics[width = \halfwidth]{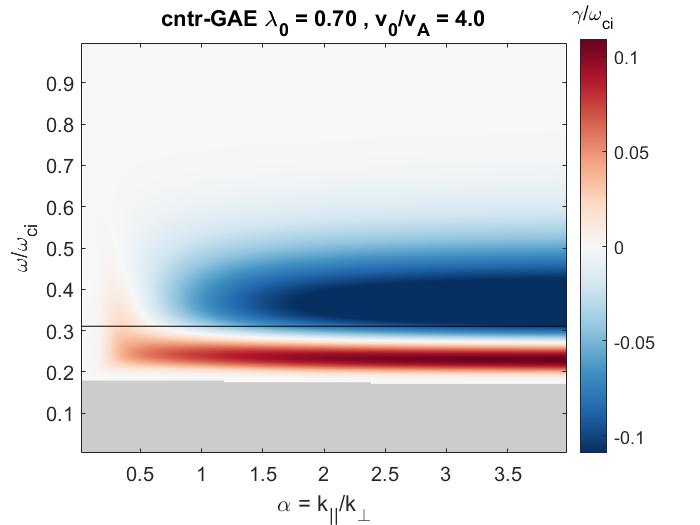} \label{fig:cyc:GAEao-cntr}}
\subfloat[]{\includegraphics[width = \halfwidth]{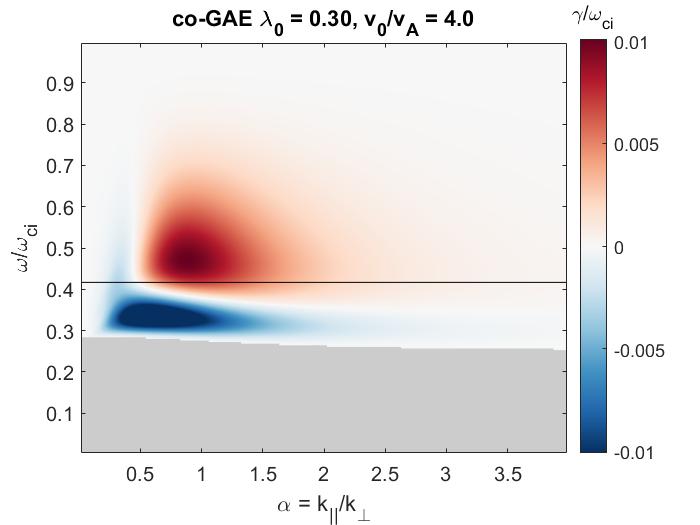} \label{fig:cyc:GAEao-co}}
\caption
[GAE growth rate dependence on mode parameters: normalized frequency $\omegabar = \omeganorm$ and wave vector direction $\alpha = \krat$.]
{Numerically calculated fast ion drive/damping for GAEs as a function of $\omegabar = \omeganorm$ and $\alpha = \krat$, when driven by a beam distribution with (a) $\linj = 0.7$ for cntr-GAEs and (b) $\linj = 0.3$ for co-GAEs. Also, $\vinj = 4.0$, $\dl = 0.3$, and assuming $\omegacires \approx 0.9$. Red corresponds to net fast ion drive, blue to damping, and gray to regions excluded by the resonance condition. Black line is the marginal frequency for fast ion drive predicted by the approximate analytic conditions in \eqref{eq:cyc:GAEmfreqrange} and \eqref{eq:cyc:GAEpfreqrange}.}
\label{fig:cyc:GAEao}
\end{figure}

These conditions can be compared against the net fast ion drive calculated from \eqref{eq:cyc:gammabeam} as a function of $\omeganorm$ and $\krat$ for a distribution with $\vinj = 4$. Injection geometries $\linj = 0.7$ (somewhat radial) and $\linj = 0.3$ (somewhat tangential) are used for cntr- and co-GAEs, respectively. The calculation is shown in \figref{fig:cyc:GAEao}. The simple analytic conditions are reasonably close to the true marginal stability on these figures. Further improved agreement could be achieved by substituting the full coupled dispersions from \eqref{eq:cyc:stixdisp} into the formula for $\vpres$ in \eqref{eq:cyc:vpres}, though the resulting boundaries would be implicit. The deviation from the analytic line on the figure at very low $\alpha$ is due to the inapplicability of the assumption $\zp \ll 1$ which was used to derive the approximate boundary, since very low $\alpha$ implies very large $\zp$ according to \eqref{eq:cyc:zp}, which has a different instability condition, as discussed in \secref{sec:cyc:fast}. 

The variation of the growth rate as a function of $\alpha$ is due to coupling between the shear and compressional branches, as well as FLR effects, contained within \eqref{eq:cyc:stixdisp} and \eqref{eq:cyc:Jlm}. For large $\alpha \gg 1$, the FLR functions in \eqref{eq:cyc:Jlmbigappx-gae} are valid, and as discussed previously, $\alpha \rightarrow \infty$ is equivalent to $\flr \rightarrow 0$. For the cntr-GAEs, $\J{1}{G} \propto J_0^2$, which peaks at $\flr = 0$, thus explaining why the growth rate in \figref{fig:cyc:GAEao-cntr} increases monotonically with $\alpha$ for the cntr-GAE, and eventually saturating. In contrast, the co-GAEs have $\J{-1}{G} \propto J_2^2$ in this limit, which vanishes for $\flr \rightarrow 0$. When coupling with the compressional branch is not taken into account, the co-GAE would also have its growth rate strictly increasing with $\alpha$ since it would have the same FLR function as the cntr-GAE. 

Conversely, $\alpha \rightarrow 0$ implies $\flr \rightarrow \infty$, where all Bessel functions of the first kind $J_\lres(\flr)$ decay to zero, such that the net drive vanishes for small $\alpha$. For the co-GAE, the growth rate decreasing at both large and small $\alpha$ results in a local maximum in the growth rate at $\alpha \like 1$. When the coupling is neglected, the maximum co-GAE growth rate is increased by a factor of 4 relative to when coupling is included (in addition to being shifted from $\alpha \like 1$ to $\alpha \rightarrow \infty$), whereas the cntr-GAE growth rate is hardly affected.

\subsection{CAE Stability}
\label{sec:cyc:CAEao}

The cntr-CAEs also have a band of unstable frequencies, though this band also depends on $\alpha$. The analogous inequalities using the approximate $\omega \approx k\va$ are

\begin{align}
\frac{\omegacires}{\frac{\abs{\kpar}v_0}{k\va} + 1} < 
\left(\frac{\omega}{\omegaci}\right)_{\lres = 1}^{CAE} < 
\frac{\omegacires}{\frac{\abs{\kpar}v_0}{k\va}\left(1 - \linj\omegacires\right)^{3/4} + 1}
\label{eq:cyc:CAEmfreqrange}
\end{align}

\begin{figure}[tb]
\subfloat[]{\includegraphics[width = \halfwidth]{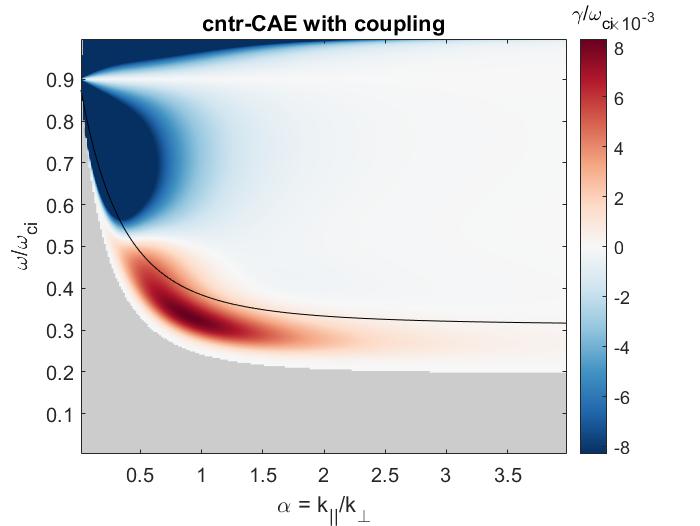}\label{fig:cyc:CAEao-full}}
\subfloat[]{\includegraphics[width = \halfwidth]{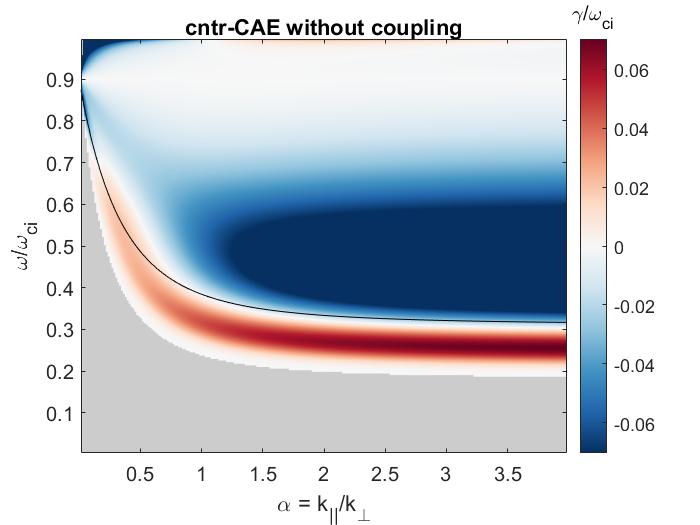}\label{fig:cyc:CAEao-simp}}
\caption
[cntr-CAE growth rate dependence on mode parameters: normalized frequency $\omegabar = \omeganorm$ and wave vector direction $\alpha = \krat$, with and without coupling to the shear branch.]
{Numerically calculated fast ion drive/damping for cntr-CAEs as a function of $\omegabar = \omeganorm$ and $\alpha = \krat$ when coupling to the shear branch is (a) included and (b) neglected. In both calculations, the modes are driven by a beam distribution with $\linj = 0.7$, $\vinj = 4.0$, $\dl = 0.3$, and assuming $\omegacires \approx 0.9$. Red corresponds to net fast ion drive, blue to damping, and gray to regions excluded by the resonance condition. Black line is the marginal frequency for fast ion drive predicted by the approximate analytic condition in \eqref{eq:cyc:CAEmfreqrange}.}
\label{fig:cyc:CAEao}
\end{figure}

The comparison between the full numerical calculation of fast ion drive as a function of $\omegabar,\alpha$ for cntr-CAEs against this approximate boundary is shown in \figref{fig:cyc:CAEao}, both when coupling to the shear branch is (a) included and (b) neglected. The agreement between the approximate condition and the numerical marginal stability is quite reasonable in both cases. These two calculations are shown in order to highlight the importance of including this coupling, which comes from finite $\omeganorm$ and FLR effects. Consider first the simpler case when no coupling is present. Then the growth rate increases monotonically with $\alpha$ like it did for the cntr-GAE. The difference between \eqref{eq:cyc:CAEmfreqrange} for the cntr-CAEs and \eqref{eq:cyc:GAEmfreqrange} for the cntr-GAEs is the additional factor of $\abs{\kpar}/k = \alpha/\sqrt{1 + \alpha^2}$ for the CAEs, which tends to one for large $\alpha$, where \figref{fig:cyc:CAEao-simp} and \figref{fig:cyc:GAEao-cntr} agree with similar growth rates. 

As was the case with the co-GAEs, the effect of coupling between the two branches is also significant for the cntr-CAEs, and for similar reasons. When coupling is included, \eqref{eq:cyc:Jlmbigappx-cae} shows that when $\alpha \gg 1$ for cntr-CAEs, $\J{1}{C} \propto J_2^2$, which goes to zero for small $\flr$. In the approximation of no coupling, instead $\J{1}{C} \propto J_0^2$, which is maximized at $\flr = 0$, just as $\J{1}{G}$ is, explaining the agreement between \figref{fig:cyc:CAEao-simp} and \figref{fig:cyc:GAEao-cntr} at large $\alpha$. As with the GAEs, the CAE growth rates go to zero for $\alpha \rightarrow 0$ since this is the $\flr \rightarrow \infty$ limit of the Bessel functions, where they decay. Hence, the cntr-CAE has a maximum in its growth rate near $\alpha \like 1$ just as the co-GAE did in \secref{sec:cyc:GAEao}. Likewise, the inclusion of coupling reduces the maximum cntr-CAE growth rate by almost an order of magnitude for the beam parameters used in \figref{fig:cyc:CAEao}. It is worth pointing out that the cntr-GAE growth rates are larger than those for the cntr-CAEs at nearly every set of mode and beam parameters, possibly explaining why the GAEs were more frequently observed in NSTX experiments. This may also explain why initial value simulations of NSTX with the \HYM code finds unstable cntr-GAEs but not cntr-CAEs.\cite{Belova2017POP} 

The analysis of this section shows that coupling between the two branches (due to two-fluid effects in this model) is important in determining the growth rate of the cntr-CAEs and co-GAEs via their influence on the FLR effects from the fast ions. Hence, a two-fluid description of the thermal plasma (such as Hall-MHD) may be important in order to accurately model cntr-CAEs and co-GAEs. 

\section{Experimental Comparison} 
\label{sec:cyc:expcomp}

\begin{figure}[tb] 
\includegraphics[width = \midwidth]{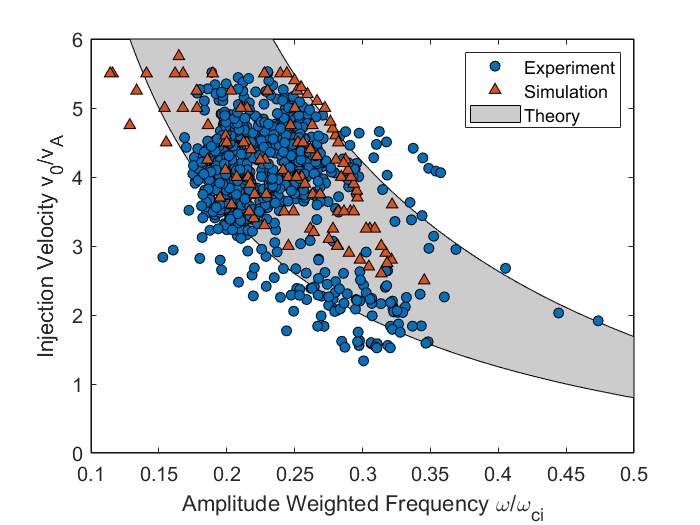}
\caption
[Comparison of approximate GAE instability conditions to simulation results and experimental observations.]
{Comparison between theory, simulations, and experiment. (a) Blue circles represent amplitude-weighted quantities from 50 ms time windows of NSTX discharges identified as having mostly cntr-GAE activity. Red triangles show cntr-GAEs excited in \HYM simulations. Theory predicts net fast ion drive in the shaded region between the two curves, as in \eqref{eq:cyc:GAEmfreqrange}.}
\label{fig:cyc:allcomp}
\end{figure}

An experimental database of CAE and GAE activity in NSTX has previously been compiled and analyzed.\cite{Tang2017TTF} This database includes approximately 200 NSTX discharges, separated into over 1000 individual 50 ms analysis windows. For each time slice, fluctuation power-weighted averages of mode quantities were calculated. The simplified instability conditions derived here relating the beam injection parameters to the mode parameters depends only on $\linj,\vinj,\omegabar$ for GAEs, which are relatively well-known and measured quantities. Hence, a comparison can be made between the marginal fast ion drive conditions and the experimental observations, shown in \figref{fig:cyc:allcomp}. This comparison assumes that the $\zp \lesssim 1$ regime (which described the most unstable modes in \HYM simulations) is valid for the experimental modes. 

The blue circles are amplitude-weighted observations in discharges with \Alfvenic activity determined to be predominantly GAE-like. Specifically, the selected time slices satisfy $-10 \leq \avg{n} \leq -4$, $\avg{f} > 200$ kHz, $T_e > 500$ eV, and $\Pb > 1$ MW. These properties were found to correlate with GAE-like modes dominating the spectrum from inspection of the database. 

The red triangles represent unstable cntr-GAEs from 3D hybrid MHD-kinetic \HYM simulations with $\linj = 0.5 - 0.9$, covering the typical range for NSTX NBI distributions. The theory developed in this chapter predicts net fast ion drive in the shaded region between the two curves. Further analysis of the linear simulation results shown on \figref{fig:cyc:allcomp} is described in \chapref{ch:sim:simulations}. The simulations used equilibrium profiles from the well-studied H-mode discharge $\#141398$,\cite{Fredrickson2013POP,Crocker2013NF,Crocker2017NF,Belova2017POP} and fast ion distributions with the same $(\lambda,v)$ dependence studied in this chapter, and given in \eqref{eq:cyc:Fdistr}. The peak fast ion density in all cases is $n_b/n_e = 5.3\%$, matching its experimental value in the model discharge. 

The theoretically predicted unstable region according to \eqref{eq:cyc:GAEmfreqrange} lies in the shaded region between the two curves, which was calculated with $\omegacires = 0.9$, motivated by the mean value of the resonant fast ions in \HYM simulations across a wide range of simulation parameters, and also $\linj \approx 0.7$ as a characteristic value of the NSTX beam geometry. There is strong agreement, especially considering the variety of assumptions required to derive the simplified stability boundaries. When evaluating the instability bounds for the specific values of $\linj$, $\vinj$, and $\omeganorm$ for each data point shown in the figure, $82\%$ of the experimental points are calculated to be theoretically unstable, and $94\%$ of the simulation points. 

An analogous comparison would be more difficult to perform for the other modes discussed in this chapter. First, co-propagating GAEs have not yet been observed in experiments since their excitation requires much smaller $\linj$ than was possible on NSTX. If they are observed in future NSTX-U experiments, as they could be in low field scenarios with the new, more tangential beam sources, a comparison could be made. Moreover, there appear to be fewer discharges dominated by cntr-CAEs than cntr-GAEs, hence requiring time-intensive inspection of many discharges in order to confidently identify cntr-CAE modes for comparison. The cntr-CAE instability boundaries (given in \eqref{eq:cyc:CAEmfreqrange}) also depend on both $\omeganorm$ and $\krat$, increasing the parameter space of the comparison. Nonetheless, these would be interesting avenues for further cross-validation.   

\section{Summary and Discussion}
\label{sec:cyc:summary}

The fast ion drive/damping for compressional (CAE) and global (GAE) \Alfven eigenmodes has been investigated analytically for a model slowing down, beam-like fast ion distribution in 2D velocity space, such as distributions generated by neutral beam injection in NSTX. Growth rate expressions previously derived by Gorelenkov\cite{Gorelenkov2003NF} and Kolesnichenko\cite{Kolesnichenko2006POP} were generalized to retain all terms in 
$\flr = \kperp\rhob$, $\alpha = \krat$ and $\omegabar = \omeganorm$ 
for sub-cyclotron modes in the local approximation driven by the Doppler-shifted ordinary $(\lres = 1)$ and anomalous $(\lres = -1)$ cyclotron resonances. This general expression for fast ion drive was evaluated numerically to determine the dependence of the fast ion drive/damping on key distribution parameters (injection velocity $\vinj$ and injection geometry $\linj = \mu B_0/\W$) and mode parameters (normalized frequency $\omeganorm$ and direction of propagation $\krat$) for each mode type and resonance. Retaining finite $\omeganorm$ and $\krat$, a source of coupling between the shear and compressional branches, was found to be responsible for significantly modifying the cntr-CAE and co-GAE growth rate dependence on $\krat$. 

The derived growth rate led to an immediate corollary: when $1 - \vpres^2/v_0^2 \leq \linj\omegacires$, cntr-propagating modes are strictly driven by fast ions while co-propagating modes are strictly damped. This condition occurs due to a finite beam injection energy, and it uncovers a new instability regime that was not considered in previous studies except recently in \citeref{Belova2019POP}, which were valid only in the $\vpres \ll v_0$ limit. 
Recall that $\omegacires \defined \avg{\omegaci}/\omegacio$ is the orbit-averaged cyclotron frequency of the resonant particles, normalized to the on-axis cyclotron frequency.

For cases where $1 - \vpres^2/v_0^2 \leq \linj\omegacires$ is not satisfied, approximate methods were employed to derive conditions necessary for net fast ion drive. Previous analytic conditions were also limited to delta functions in $\lambda$, which are a poor approximation for fast ions generated by NBI. In this chapter, broad parameter regimes were identified which allow for tractable integration, leading to the first compact net fast ion drive conditions as a function of fast ion and mode parameters which properly integrate over the full beam-like distribution. For the narrow beam case discussed in \secref{sec:cyc:narrow}, the sign of the growth rate depends on a function of $\linj$ only, similar to the instability regime studied previously.\cite{Gorelenkov2003NF,Kolesnichenko2006POP} Numerical integration showed that this result was only reliable for beams much more narrow $(\dl \lesssim 0.1)$ than those in experiments $(\dl \approx 0.3)$, underscoring the limitations of past results. In particular, those previous studies identified $\kperp\rhob > 1$ and $\kperp\rhob > 2$ as the most unstable parameters for cntr-CAE and cntr-GAE instabilities, respectively, whereas this work demonstrates that these instabilities may be excited for any value of $\kperp\rhob$, with $\kperp\rhob \lesssim 1$ instabilities perhaps more common for NSTX conditions. 

The approximation of a sufficiently wide beam $(\dl \gtrsim 0.2)$ in conjunction with a small or large FLR assumption allowed the derivation of very simple conditions for net fast ion drive, summarized in \tabref{tab:cyc:appxcons}. These expressions depend on the fast ion injection velocity $\vinj$, injection geometry $\linj$, and mode properties $\omeganorm$, $\krat$ which determine $\vpres$ along with the cyclotron resonance coefficient $\lres$. It is found that the wide beam, small FLR assumption is valid over a wide enough range of parameters ($\zp = \kperp\vpres/\omegaci \lesssim 2$) that it encompasses the typical conditions for NSTX fast ions and properties of the most unstable CAEs/GAEs inferred from experiments and simulations. 

Comparison between full numerical evaluation of the exact analytic expression and the approximate stability boundaries demonstrate excellent agreement within the ranges of applicability. These regimes include fast ion parameters motivated by \TRANSP/\NUBEAM modeling of NSTX beam profiles, as well as properties $(\omeganorm,\krat)$ of the most unstable modes excited in hybrid simulations with the \HYM code. In addition to providing insight into an individual mode's growth rate as a function of fast ion parameters, the new instability conditions also yield information about the properties of the unstable modes for a fixed beam distribution. Namely, cntr-propagating GAEs are unstable for a specific range of frequencies (as a function of beam parameters) nearly independent of $\krat$, whereas cntr-CAEs are more sensitive to $\krat$. This condition for cntr-GAEs compares well against NSTX data across many discharges, providing support for the theoretical underpinnings of the growth rate calculation, as well as the series of mathematical approximations made to arrive at these compact marginal stability conditions. 

The approximate conditions for net fast ion drive were only made possible by a series of simplifications, which should be kept in mind when applying these results. Integration over space and $\pphi$ were neglected, restricting the analysis to 2D phase space. Moreover, the derived stability boundaries do not include damping on the background plasma, such that net fast ion drive as calculated in this chapter is a necessary but not sufficient condition for overall instability. Including the electron Landau damping rate and the continuum/radiative damping due to interaction with the \Alfven continuum is an area for future work.

The results derived here can be applied in the future to help interpret experimental results and improve physics understanding of first principles simulations. Ideally, they can be used to guide expectations about the spectrum of unstable modes that will be generated by a specific neutral beam configuration. For instance, if a specific mode is driven unstable by an initial beam distribution, these expressions show where additional neutral beam power may be added that would act to stabilize this mode, or drive it further unstable, if desired. This enables systematic analysis and prediction of scenarios like those of the cntr-GAE stabilization observed in NSTX-U.\cite{Fredrickson2017PRL,Fredrickson2018NF,Belova2019POP}  

\begin{subappendices}

\section{Remarks on Serendipitous Approximations}
\label{app:cyc:approx}

In this appendix, I will share some additional remarks on the serendipitous mathematical approximations that many of the key results of this chapter relied upon. The canonical example is \eqref{eq:cyc:wideslowGAEgamraw}, reprinted below: 

\begin{align}
f(x) &= \frac{1 - x^2 + 2 x \log x}{1 - x + x \log x} 
\approx 1 - x^{2/3}
\label{eq:cyc:wideslowGAEgamrawAPP}
\end{align}

\begin{figure}[htb]
\subfloat[\label{fig:cyc:approx23a}]{\includegraphics[width = \halfwidth]{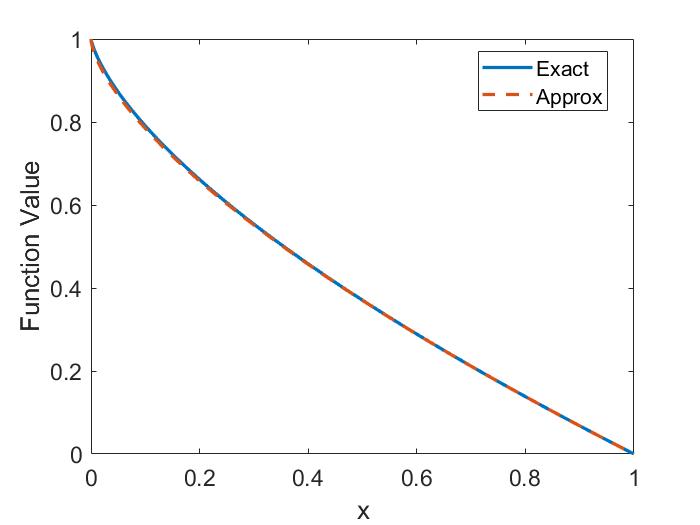}} 
\subfloat[\label{fig:cyc:approx23e}]{\includegraphics[width = \halfwidth]{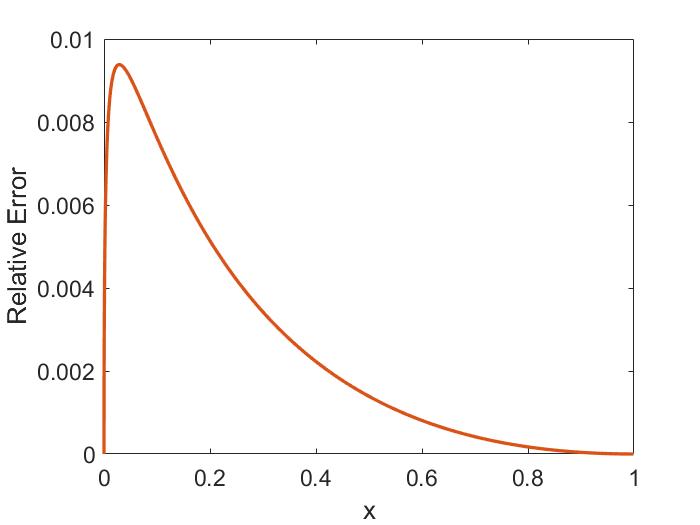}}
\caption
[Comparison of approximation in \eqref{eq:cyc:wideslowGAEgamraw} to the exact function.]
{Comparison of approximation in \eqref{eq:cyc:wideslowGAEgamraw} to the exact function (identical to \eqref{eq:cyc:wideslowGAEgamrawAPP}). Left: blue curve shows exact function, dashed orange curve shows approximation. Right: relative error of the approximation.}
\label{fig:cyc:approx23}
\end{figure}

The relative error between the exact function and the approximate form is shown in \figref{fig:cyc:approx23}, with a maximum error of $1\%$. This approximation was sought so that the expression could be inverted to find an expression for $x$ in terms of $f(x)$. Without approximation, no such expression can be written in terms of elementary functions. Since we are interested in the entire range $0 \leq x \leq 1$, a small parameter expansion assuming $x \ll 1$ can not be used. Hence, a \emph{global} approximation must be employed in order to ensure accuracy everywhere, not just at one end point of this range. To achieve this, we will choose a suitable functional form with qualitative properties similar to the exact function and then enforce quantitative conditions by setting values of free parameters. 

Asserting a specific functional form may appear to introduce an artificial degree of arbitrariness. However, this is really no different from how more familiar approximations are made in general. Consider a local series expansion. When expanding to $n^{th}$ order, a functional form of $a_0 + a_1 (x-x_a) + \dots + a_n (x - x_a)^n$ is assumed, and then the $\left\{a_i\right\}$ are chosen to match the value of the function and the first $n-1$ derivative values at the location $x = x_a$. This type of approximation is excellent near $x = x_a$ but poor far from it. 

Consider also a \Pade approximation, where a function is approximated by a rational function $P(x) / Q(x)$ where both $P(x)$ and $Q(x)$ are series truncated after the desired number of terms. This type of approximation can achieve global accuracy since constraints can be imposed to match the original function behavior at multiple locations. For instance, a \Pade approximation often used in plasma physics\cite{Kuvshinov1994PPCF} is $f(x) = 1 - e^{-x} I_0(x) \approx x / (1 + x)$, where $I_0(x)$ is the zero-order modified Bessel function of the first kind. This has the correct behavior for $x \ll 1$ of $f(x) \approx x$ and also the correct asymptotic behavior when $x \gg 1$ of $f(x) \like 1$. At intermediate values of $x$, the approximation is correct to within $10\%$, making this a valid global approximation. 

While our problem will use neither a local series expansion nor a global \Pade approximation, our approach will be fundamentally the same. We will assume a certain functional form and then choose conditions to ensure that certain important properties of the original function are preserved by the approximation. Namely, we will mostly use the following general functional form (and modest variations)

\begin{align}
f(x) = (1 - x^p)^q
\label{eq:cyc:appform}
\end{align} 

This form is motivated by noticing that the unapproximated function in \eqref{eq:cyc:wideslowGAEgamrawAPP} is smooth, convex, and monotonically decreasing on $(0,1) \rightarrow (0,1)$. Both the original function and the approximate form have $f(0) = 1$, $f(1) = 0$, and $f^{(n)}(x) \rightarrow (-1)^{n}\infty$ as $x \rightarrow 0$ for $0 < p < 1$ and $q > 0$. For positive integer $q$, the first $q-1$ derivatives vanish at $x = 1$. Then, the value of $p$ is chosen so that the approximation matches the value of the first non-vanishing derivative at the endpoint $x = 1$. In this way, the behavior of the function at both end points is matched, which is why the error converges to zero at those points in \figref{fig:cyc:approx23e}. Many other functions encountered in this work that we seek to approximate share these features. 

So it is very clear why the approximation is successful near both endpoints. What is less clear is why it is still good far away from either one -- at $x = 0.5$, for instance. Although not proven, this is likely a consequence of the smoothness and monotonicity of the function. Intuitively it makes sense that if a well-behaved function is approximated in a way that captures its value and derivatives at two distant points, then it will probably be fairly accurate between those points as well. This feature is what makes the approximations ``serendipitous.'' This approximation procedure consistently worked well for many different types of functions. Hence, it is unclear which specific properties of those functions are essential to its efficacy. The approximations used in this chapter will be catalogued in this appendix, with accompanying figures to demonstrate their accuracy. 

The approximation in \eqref{eq:cyc:highpsiGAEboundpre}: 

\begin{align}
f(x) &= \frac{1}{2}\left(1 - \frac{\sqrt{x(1-x)}}{\arccos\sqrt{x}}\right) \approx \frac{1}{2}\left(1 - x^{2/3}\right) 
\label{eq:cyc:highpsiGAEboundAPP}
\end{align}

\begin{figure}[htb]
\subfloat[\label{fig:cyc:approx23halfa}]{\includegraphics[width = \halfwidth]{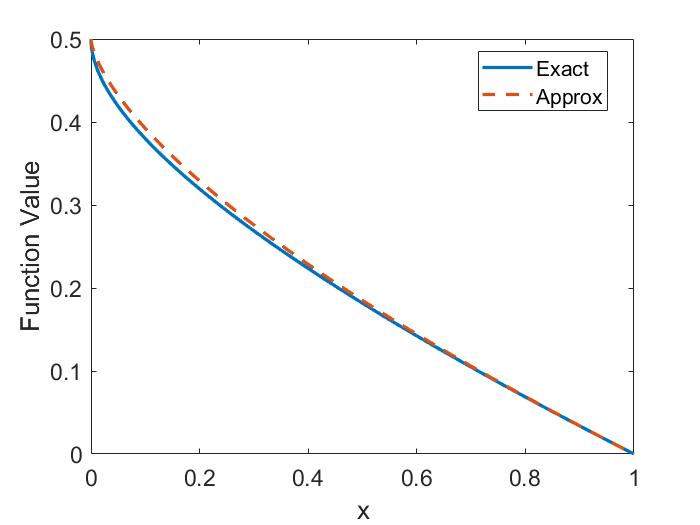}} 
\subfloat[\label{fig:cyc:approx23halfe}]{\includegraphics[width = \halfwidth]{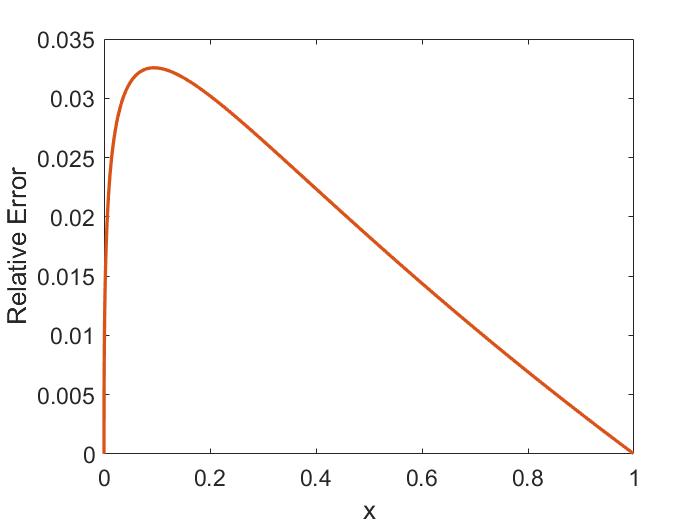}}
\caption
[Comparison of approximation in \eqref{eq:cyc:highpsiGAEboundpre} to the exact function.]
{Comparison of approximation in \eqref{eq:cyc:highpsiGAEboundpre} to the exact function (identical to \eqref{eq:cyc:highpsiGAEboundAPP}). Left: blue curve shows exact function, dashed orange curve shows approximation. Right: relative error of the approximation.}
\label{fig:cyc:approx23half}
\end{figure}

The approximation in \eqref{eq:cyc:highpsiGAElim}: 

\begin{align}
f(x) &= \frac{1}{2}\left[1 - \frac{\sqrt{x(1-x)}}{\arcsin\sqrt{x}}\right] 
\approx \frac{1}{2}\left[1 - (1 - x)^{2/3}\right] 
\label{eq:cyc:highpsiGAElimAPP}
\end{align}

\begin{figure}[H]
\subfloat[\label{fig:cyc:approx23sina}]{\includegraphics[width = \halfwidth]{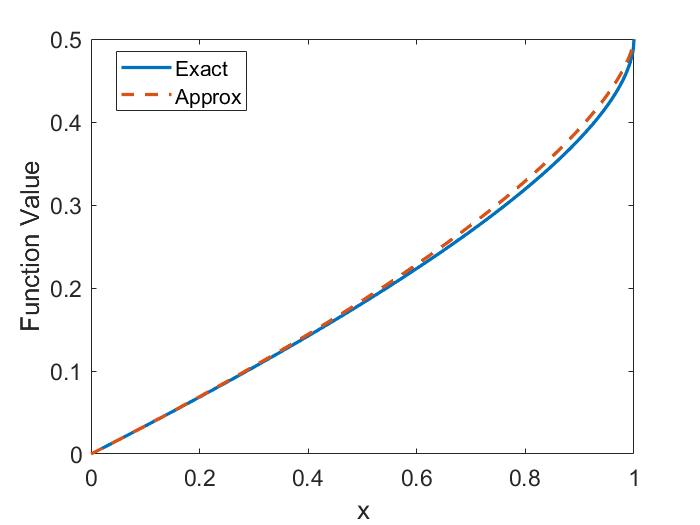}} 
\subfloat[\label{fig:cyc:approx23sine}]{\includegraphics[width = \halfwidth]{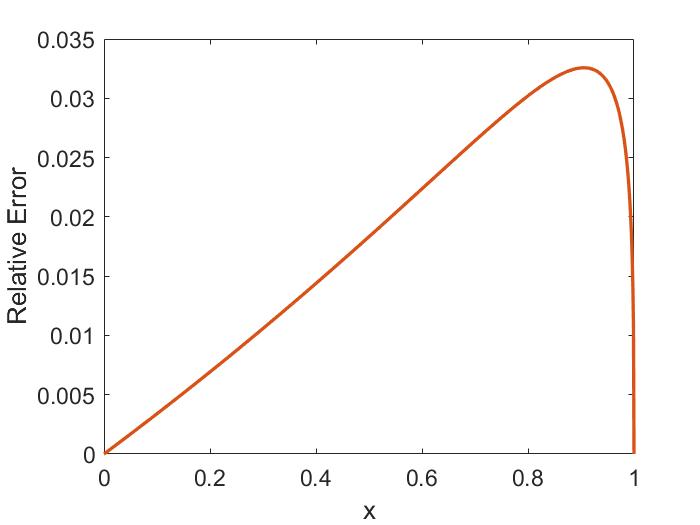}}
\caption
[Comparison of approximation in \eqref{eq:cyc:highpsiGAElim} to the exact function.]
{Comparison of approximation in \eqref{eq:cyc:highpsiGAElim} to the exact function (identical to \eqref{eq:cyc:highpsiGAElimAPP}). Left: blue curve shows exact function, dashed orange curve shows approximation. Right: relative error of the approximation.}
\label{fig:cyc:approx23sin}
\end{figure}

There are also two related approximations which were found during the course of this work, but were not applied in this thesis. They are documented here for completeness. Curiously, there does not appear to be a generalization of these formulas. Consider

\begin{align}
\label{eq:cyc:japprox2}
f(x) &= 1 - x + x \log x 
\approx \left(1 - x^{1/2^{1/2}}\right)^2 \\
\label{eq:cyc:japprox3}
f(x) &= 1 - x^2 + 2 x \log x 
\approx \left(1 - x^{1/3^{1/3}}\right)^3
\end{align}

\begin{figure}[H]
\subfloat[]{\includegraphics[width = \halfwidth]{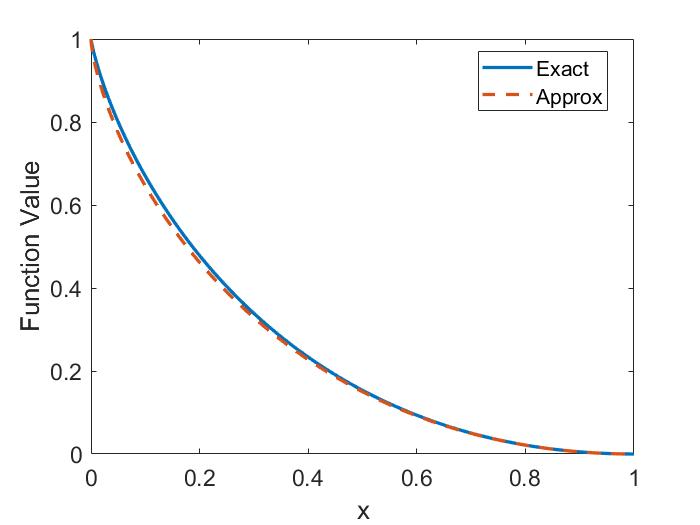}} 
\subfloat[]{\includegraphics[width = \halfwidth]{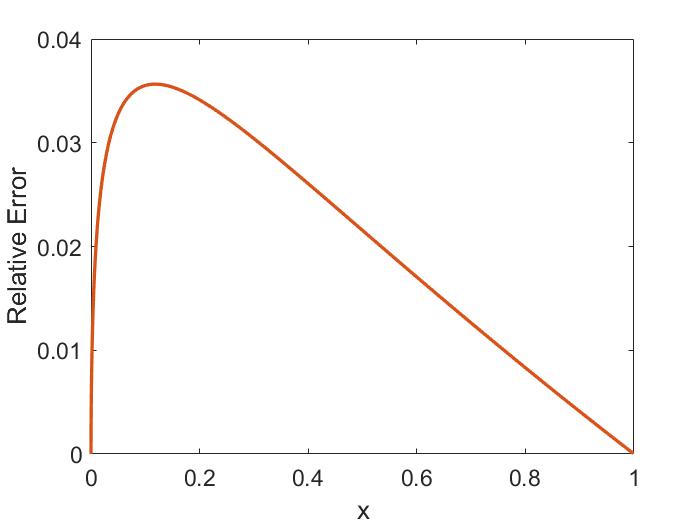}}
\caption
[Comparison of approximation in \eqref{eq:cyc:japprox2} to the exact function.]
{Comparison of approximation in \eqref{eq:cyc:japprox2} to the exact function). Left: blue curve shows exact function, dashed orange curve shows approximation. Right: relative error of the approximation.}
\label{fig:cyc:approxj2}
\end{figure}

\begin{figure}[H]
\subfloat[]{\includegraphics[width = \halfwidth]{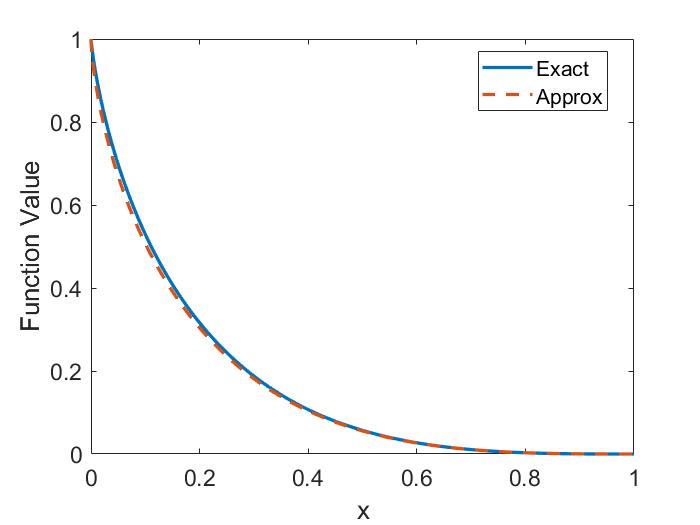}} 
\subfloat[]{\includegraphics[width = \halfwidth]{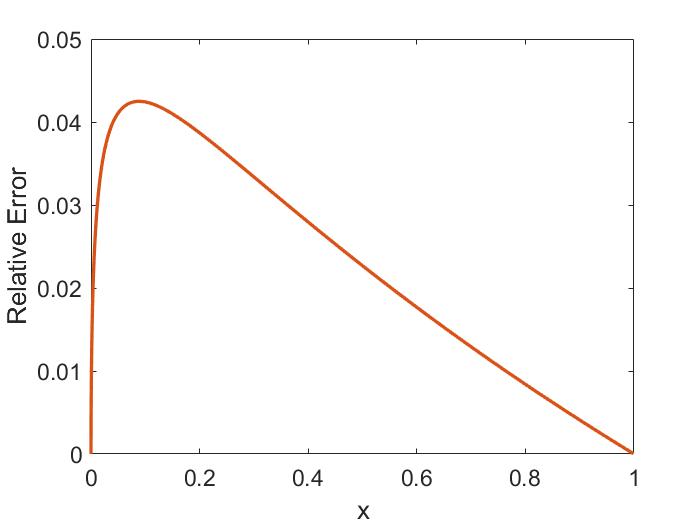}}
\caption
[Comparison of approximation in \eqref{eq:cyc:japprox3} to the exact function.]
{Comparison of approximation in \eqref{eq:cyc:japprox3} to the exact function. Left: blue curve shows exact function, dashed orange curve shows approximation. Right: relative error of the approximation.}
\label{fig:cyc:approxj3}
\end{figure}

\section{Correction for Finite Frequency in Small FLR Regime \texorpdfstring{($\zp\ll 1$)}{}}
\label{app:cyc:omegabar}

Here, the correction due to finite $\omegabar$ for the wide beam, low $\zp$ approximation is addressed. This term was neglected in \secref{sec:cyc:wide}. Including this term, the integral of interest is 

\begin{align}
\gamma &\appropto -\int_0^{1-\bres} \frac{x(x - \xinj)}{(1-x)^2}\left(\frac{\lres}{\omegabar} - x\right)dx = 0 \\ 
\Rightarrow \xinj &= \frac{\lres f(\bres) + \omegabar g(\bres)/2}{\lres h(\bres) - \omegabar f(\bres)} \\ 
f(\bres) &= 1 - \bres^2 + 2\bres\log\bres \\ 
g(\bres) &= -2 - 3\bres + 6\bres^2 - \bres^3 - 6\bres\log\bres \\ 
h(\bres) &= 1 - \bres + \bres\log\bres 
\end{align}

This function can be approximated to leading order in $\omegabar < 1$, and will take advantage of the known approximation from earlier $f(\bres)/h(\bres) \approx 1 - \bres^{2/3}$. 

\begin{align}
\label{eq:cyc:appb}
\xinj &= \frac{f(\bres)}{h(\bres)} + \frac{\omegabar}{\lres}\left[\left(\frac{f(\bres)}{h(\bres)}\right)^2 + \frac{g(\bres)}{2h(\bres)}\right] \\ 
&\approx 1-\bres^{2/3} - \frac{\omegabar}{8\lres}\bres^{2/3}\left(1-\bres^{2/3}\right)^2
\end{align}

The second term in the second line is the approximation to the function in brackets. Again using $\omegabar$ as a small parameter, assume a solution of the form $\bres = \bres_0 + \omegabar\bres_1$ where $\bres_0 = (1-\xinj)^{3/2}$. Then the leading order correction in $\omegabar$ to the $\omegabar\rightarrow 0$ solution found in \secref{sec:cyc:wide} is 

\begin{align}
v_0 = \frac{\vpres}{\left(1 - \xinj\right)^{3/4}}\left(1 + \frac{3\omegabar\xinj^2}{32\lres}\right)
\end{align}

\begin{figure}[H]
\subfloat[]{\includegraphics[width = \halfwidth]{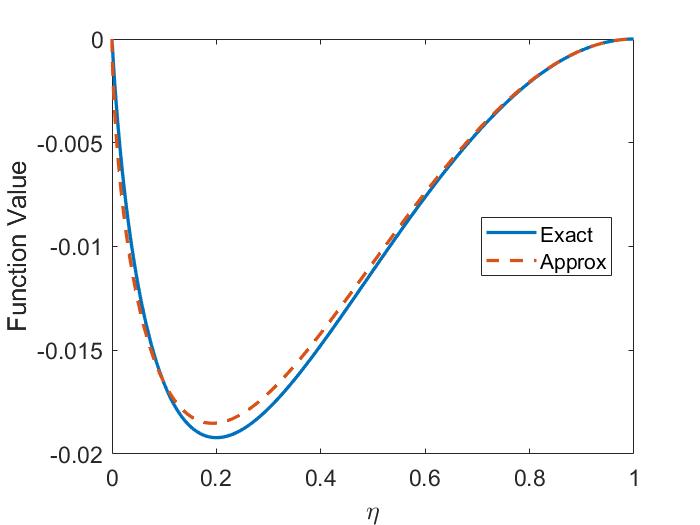}} 
\subfloat[]{\includegraphics[width = \halfwidth]{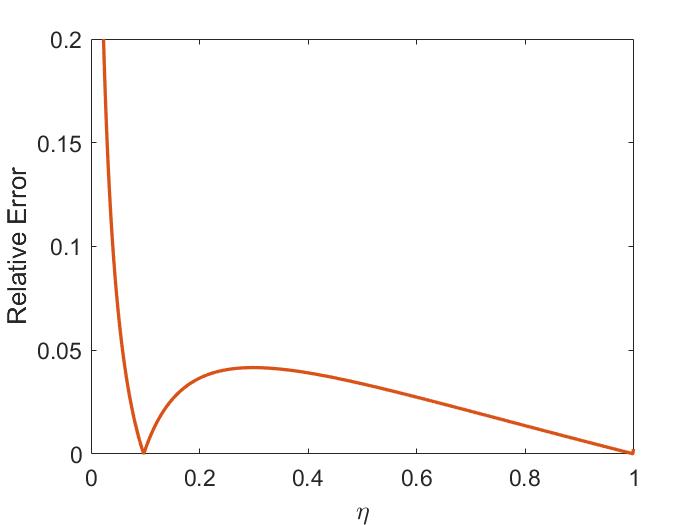}}
\caption
[Comparison of approximation in \eqref{eq:cyc:appb} to the exact function.]
{Comparison of approximation in \eqref{eq:cyc:appb} to the exact function. Left: blue curve shows exact function, dashed orange curve shows approximation. Right: relative error of the approximation.}
\label{fig:cyc:appb}
\end{figure}

The accuracy of the approximation of the term in square brackets in \eqref{eq:cyc:appb} is shown in \figref{fig:cyc:appb}. The relative error is large for $x\rightarrow 0$, but this is tolerable since the term in brackets is a correction to the leading order term, and this is where the correction vanishes as well. For $0.06 < x < 1$, the relative error is less than $5\%$. 

\section{Large FLR Regime for CAEs \texorpdfstring{$(\zp \gg 1)$}{}}
\label{app:cyc:fastcae}

Using the large $\zp \gg 1$ (equivalently small $\alpha \ll 1$) expansion for CAEs with $\lres = \pm 1$ gives  $\J{\pm 1}{C}(\flr) \like J_0^2(\flr) \like (1 - \sin(2\flr))/ \flr$. As in \secref{sec:cyc:fast}, the rapidly varying $\sin(2\flr)$ will average to zero in the integral, leaving 

\begin{align}
\gamma &\appropto -\lres\int_0^{1-\bres} \frac{\sqrt{x}(x-\xinj)}{(1-x)^{3/2}}dx 
\label{eq:cyc:highpsiCAE}
\end{align}

Integrating and finding the marginal stability condition $\gamma = 0$ results in 

\begin{align}
\label{eq:cyc:highpsiCAEcrit}
\xinj &= \frac{8\sqrt{\bres^{-1}-1} + 4\sqrt{\bres(1-\bres)} - 3\pi - 6\arctan\left(\frac{1-2\bres}{2\sqrt{\bres(1-\bres)}}\right)}
{8\left(\sqrt{\bres^{-1}-1} - \arccos\sqrt{\bres}\right)}  
\approx 1 - \bres^{3/5} \\ 
\Rightarrow v_0 &= \frac{\vpres}{(1 - \xinj)^{5/6}}
\label{eq:cyc:highpsiCAEcon}
\end{align}

\begin{figure}[H]
\subfloat[]{\includegraphics[width = \halfwidth]{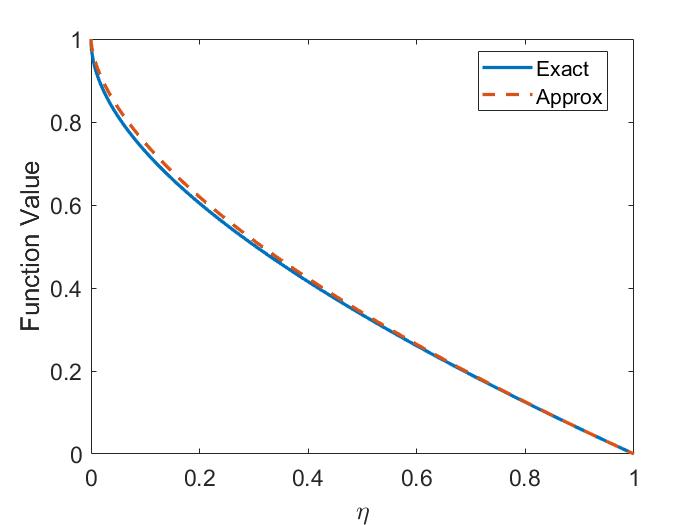}} 
\subfloat[]{\includegraphics[width = \halfwidth]{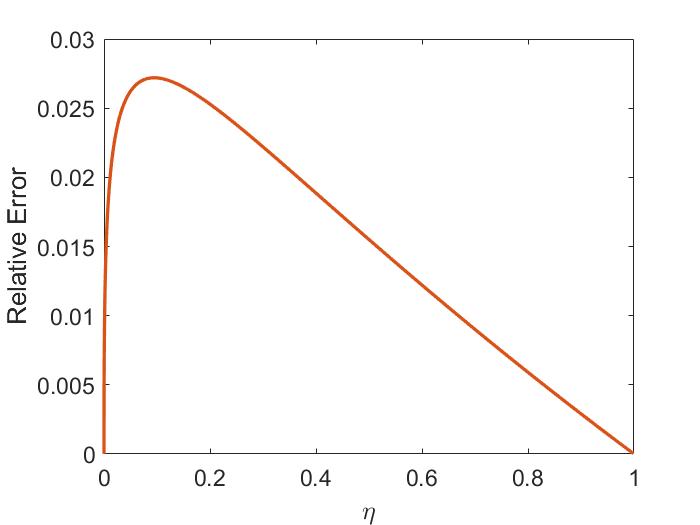}}
\caption
[Comparison of approximation in \eqref{eq:cyc:highpsiCAEcrit} to the exact function.]
{Comparison of approximation in \eqref{eq:cyc:highpsiCAEcrit} to the exact function. Left: blue curve shows exact function, dashed orange curve shows approximation. Right: relative error of the approximation.}
\label{fig:cyc:appc}
\end{figure}

The approximation in \eqref{eq:cyc:highpsiCAEcrit} has a maximum global error of $3\%$, as shown in \figref{fig:cyc:appc}. The instability condition for cntr-propagating modes $(\lres = 1)$ is $v_0 < \vpres/(1 - \xinj)^{5/6}$, while the co-propagating modes $(\lres = -1)$ are driven for $v_0 > \vpres/(1 - \xinj)^{5/6}$.

\end{subappendices}

\newcommand{\dirfiglan}{ch-analytics-landau/figs}

\chapter{Analytic Stability Boundaries for Interaction via Landau Resonance}
\label{ch:lan:analytics-landau}

\section{Introduction}
\label{sec:lan:introduction}

The analysis of this chapter focuses on fast ions interacting with CAEs/GAEs through the Landau (\ie non-cyclotron) resonance. Drive/damping due to the ordinary and anomalous cyclotron resonances is treated in \chapref{ch:cyc:analytics-cyclotron}. The stability of CAEs driven by neutral beam injection (NBI) due to the Landau resonance has previously been studied by Belikov\cite{Belikov2003POP,Belikov2004POP} in application to NSTX. In those works, a delta function distribution in pitch was assumed for the fast ions, which is an unrealistic model for the typically broad distributions inferred from experimental modeling. Previous works also assumed $\kpar \ll \kperp$ and $\omega \ll \omegaci$, whereas experimental observations and simulations demonstrate that $\kpar \like \kperp$ and $\omega \like \omegaci/2$ are common. Here, prior work is extended to provide a local expression for the fast ion drive due to an anisotropic beam-like distribution through the Landau resonance. Terms to all order in $\omeganorm$ and $\krat$ are kept for applicability to the entire possible spectrum of modes. In particular, finite $\omeganorm$ and $\krat$ introduces coupling between the compressional and shear branches of the dispersion which enables the GAE to be excited through this resonance. Full finite Larmor radius (FLR) terms are also retained, similar to prior studies. As in \chapref{ch:cyc:analytics-cyclotron} for the cyclotron resonances, experimentally relevant regimes have been identified where approximate stability boundaries can be derived. Since other damping sources have not been included in this chapter, the derived conditions for net fast ion should be treated as necessary but insufficient conditions for instability. The choice of specific parameter regimes has been aided by initial value simulations of CAEs with the 3D hybrid MHD-kinetic code \HYM.\cite{Belova2017POP} The simulation model couples a single fluid thermal plasma to a minority species of full orbit kinetic beam ions and also includes the contributions of the large beam current to the equilibrium self-consistently.\cite{Belova2003POP} 

The chapter is structured as follows. The fast ion drive for CAEs/GAEs from the Landau resonance is derived analytically in the local approximation in \secref{sec:lan:derivation}, based on the framework in \citeref{Mikhailovskiiv6} and applied to a parametrized neutral beam distribution. Approximations are made to this expression in \secref{sec:lan:approxstab} in order to derive marginal stability conditions in the limits of very narrow (\secref{sec:lan:narrow}) and realistically broad (\secref{sec:lan:wide}) fast ion distributions. Within \secref{sec:lan:wide}, the limits of small and large FLR effects are treated separately in \secref{sec:lan:slow} and \secref{sec:lan:fast}, respectively, and the dependence of the drive/damping on fast ion parameters for fixed mode properties is discussed and compared to the approximate analytic conditions. A complementary discussion of the fast ion drive/damping as a function of the mode properties for fixed fast ion parameters is presented in \secref{sec:lan:stabao}, including the role of compressional-shear mode coupling in setting the stability boundaries. A comparison of the approximate stability conditions against a database of co-CAE activity in NSTX and simulations results is shown in \secref{sec:lan:expcomp}. Lastly, a summary of the main results and discussion of their significance is given in \secref{sec:lan:summary}. The majority of the content of this chapter has been peer-reviewed and published in \citeref{Lestz2020p2}. 

\section{Fast Ion Drive for Anisotropic Beam Distribution in the Local Approximation for the Landau Resonance} 
\label{sec:lan:derivation}

As in \chapref{ch:cyc:analytics-cyclotron}, we note that due to the large frequencies of these modes in experiments: $\omeganorm = 0.3 \tto 1$ and $\krat$ often order unity in simulations, it is worthwhile to consider the dispersion relation for the shear and compressional branches including coupling due to thermal plasma two-fluid effects. Additional coupling can be induced by spatial gradients present in realistic experimental profiles, which is not included in our analysis. 

\subsection{Starting Equations}

Define $\omegabar = \omega/\omegacio$, $N = k\va/\omega$, $A = (1 - \omegabar^2)^{-1}$, and also $F^2 = \kpar^2/k^2$, $G = 1 + F^2$. Here, $\omegacio$ is the on-axis ion cyclotron frequency. Then in uniform geometry, the local dispersion in the MHD limits of $E_\parallel \ll E_\perp$ and $\omega \ll \abs{\omegace},\omegape$ is\cite{Stix1975NF} 

\begin{equation}
N^2 = \frac{AG}{2F^2}\left[1 \pm \sqrt{1 - \frac{4F^2}{AG^2}}\right]
\label{eq:lan:stixdisp}
\end{equation}

The ``$-$" solution corresponds to the compressional \Alfven wave (CAW), while the ``$+$" solution corresponds to the shear \Alfven wave (SAW). The coupled dispersion can modify the polarization of the two modes relative to the uncoupled approximation. In \secref{sec:cyc:stabao}, it was shown that the growth rates for the cyclotron resonance-driven cntr-CAEs and co-GAEs have local maxima with respect to $\krat$, whereas they increase monotonically when this coupling is neglected. The low frequency approximation of \eqref{eq:lan:stixdisp} is $\omega \approx k\va$ for CAEs and $\omega \approx \abs{\kpar}\va$ for GAEs, which can simplify analytic results when valid. The Landau resonance describes a wave-particle interaction satisfying the following relation

\begin{align}
\omega - \avg{\kpar\vpar} - \avg{\kperp\vdrift} = 0
\label{eq:lan:rescon}
\end{align}

Above, $\avg{\dots}$ denotes poloidal orbit averaging appropriate for the ``global'' resonance (see further discussion in \secref{sec:cyc:dispres} and also \citeref{Belikov2003POP}). An equivalent way of writing this resonance condition is $\omega - n \omega_\phi - p\omega_\theta = 0$ for integers $n,p$ where $\omega_\phi$ and $\omega_\theta$ are the orbit-averaged toroidal and poloidal particle frequencies, respectively. Satisfaction of the global resonance condition in this form has previously been demonstrated in \HYM simulations.\cite{Belova2017POP} As in \secref{sec:cyc:dispres}, we consider the approximation of $\abs{\kperp\vdrift} \ll \abs{\kpar\vpar}$, focusing on the primary resonance and neglecting sidebands. Hence, all modes satisfying this resonance with particles with $\vpres \defined \avg{\vpar} > 0$ must be co-propagating with $\kpar > 0$. For the local calculations in this section, we will approximate \eqref{eq:lan:rescon} as $\omega - \kpar\vpres = 0$. 

The stability calculation will be applied to the same model fast ion distribution as in \secref{sec:cyc:dispres}, motivated by theory and \NUBEAM modeling of NSTX discharges,\cite{Belova2017POP} written as a function of $v = \sqrt{2\W/m_i}$ and $\lambda = \mu B_0 / \W$ in separable form: $\fb(v,\lambda)= C_f n_b f_1(v)f_2(\lambda)$, defined below

\begin{subequations}
\begin{align}
\label{eq:lan:F1}
f_1(v) &= \frac{\ftail(v;v_0)}{v^3 + v_c^3} \\ 
\label{eq:lan:F2}
f_2(\lambda) &= \exp\left(-\left(\lambda - \lambda_0\right)^2 / \Delta\lambda^2\right)
\end{align}
\label{eq:lan:Fdistr}
\end{subequations}

The constant $C_f$ is for normalization. The first component $f_1(v)$ is a slowing down function in energy with a cutoff at the injection energy $v_0$ and a critical velocity $v_c$, with $\ftail(v;v_0)$ a step function. The second component $f_2(\lambda)$ is a Gaussian distribution in $\lambda$. To lowest order in $\mu \approx \mu_0$, it can be re-written as $\lambda = (\vperp^2/v^2)(\omegacio/\omegaci)$. The distribution in the final velocity component, $\pphi$, is neglected in this study for simplicity, as it is expected to be less relevant for the high frequencies of interest for these modes. NSTX typically operated with $\vinj = 2 - 6$ and $\linj = 0.5 - 0.7$ with $\dl = 0.3$. Early operations of NSTX-U had $\vinj < 3$, featuring an additional beam line with $\linj \approx 0$. For this study, $v_c = v_0/2$ is used as a characteristic value. Comparison between the model distribution used in this study and those calculated with the Monte Carlo code \NUBEAM for NSTX and NSTX-U can be found in Fig. 5 of \citeref{Belova2003POP} and Fig. 4 of \citeref{Belova2019POP}, respectively.

In \secref{sec:cyc:derivation}, the fast ion drive/damping was derived perturbatively in the local approximation for a two component plasma comprised of a cold bulk plasma and a minority hot ion population. Restricting \eqref{eq:cyc:gammageneralx} to the $\lres = 0$ Landau resonance and applying to the model distribution gives 

\begin{multline}
\frac{\gamma}{\omegaci} = -\frac{n_b}{n_e}\frac{\pi C_f v_0^3 \bres^{3/2}}{v_c^3\omegabar} 
\times \\
\left\{ \int_0^{1-\bres} \frac{x \Jzm(\flr(x,\zp))}{(1-x)^2}\frac{e^{-(x-\xinj)^2/\dx^2}}{1 + \frac{v_0^3}{v_c^3}\left(\frac{\bres}{1-x}\right)^{3/2}} 
\left[-\frac{x(x-\xinj)}{\dx^2} + \frac{3/4}{1 + \frac{v_c^3}{v_0^3}\left(\frac{1-x}{\bres}\right)^{3/2}}\right]dx  \right. \\  
\left.\vphantom{\left[\frac{3/4}{1 + \left(\frac{1-x}{4\bres}\right)^{3/2}}\right]}
+ \frac{\bres^{-1}-1}{2\left(1 + \frac{v_0^3}{v_c^3}\right)}e^{-(1 - \bres-\xinj)^2/\dx^2}\Jzm\left(\zp\sqrt{\bres^{-1}-1}\right)\right\}
\label{eq:lan:gammabeam}
\end{multline}

All notation is the same as defined in \secref{sec:cyc:derivation}. Briefly, the integration variable is $x = \vperp^2/v^2 = \lambda\omegacires$ where $\omegacires \defined \avg{\omegaci}/\omegacio$ is the orbit-averaged cyclotron frequency of the resonant particles, normalized to the on-axis cyclotron frequency. Similarly, $\xinj = \linj\omegacires$ and $\dx = \dl\omegacires$. The resonant parallel energy fraction is $\bres = \vpres^2/v_0^2$. The beam ion density is given as $n_b$, with the electron density as $n_e$. The first term in square brackets is the contribution from $\partial\fb/\partial\lambda$ (anisotropy) while the second term in brackets and also the term outside the integral come from $\partial\fb/\partial v$ (slowing down). \eqref{eq:lan:gammabeam} is valid for arbitrary $\omeganorm$ and $\krat$, generalizing results published in \citeref{Belikov2003POP,Belikov2004POP} for the co-CAE driven by the Landau resonance, which were restricted to $\omega \ll \omegaci$ and $\kpar \ll \kperp$. This generalization is contained mostly in the FLR effects, within the function $\Jlm(\flr)$, defined for arbitrary $\lres$ in \eqref{eq:cyc:Jlm}, which simplifies for $\lres = 0$ to 

\begin{align}
\label{eq:lan:Jlm}
\Jzm(\flr) &= \frac{N^{-2}\left(N^{-2} - F^2(1 - \omegabar^2)\right)}{N^{-4} - F^2}J_1^2(\flr) \\ 
\text{where } \label{eq:lan:zsimp}
\flr &= \kperp\rhob \defined \zp \sqrt{\frac{x}{1-x}} \\ 
\text{and }
\zp &= \frac{\kperp\vpres}{\omegaci} = \frac{\omegabar}{\alpha}
\label{eq:lan:zp}
\end{align}

Above, $\rhob = \vperp/\omegaci$ is the Larmor radius of the fast ions, $\flr$ is the FLR parameter, and $\zp$ is the modulation parameter describing how rapidly the integrand of \eqref{eq:lan:gammabeam} oscillates, which depends on the mode parameters $\omegabar = \omeganorm$ and $\alpha \defined \krat$. The $m$ in $\Jzm(\flr)$ denotes the mode dispersion (= `$C$' for CAE and `$G$' for GAE), as contained within $N$ (given in \eqref{eq:lan:stixdisp} for CAEs using the ``$-$" solution and GAEs using the ``$+$'' solution). As argued in \secref{sec:cyc:derivation} $\Jzm(\flr) \geq 0$ for both modes. In the limit of $\omeganorm \ll 1$, 

\begin{subequations}
\label{eq:lan:Jlmsmallappx}
\begin{align}
\label{eq:lan:Jlmsmallappx-cae}
\lim_{\omegabar\rightarrow 0} \Jzc(\flr) &= J_1^2(\flr)
\CAElab \\ 
\label{eq:lan:Jlmsmallappx-gae}
\lim_{\omegabar\rightarrow 0} \Jzg(\flr) &= 
\omegabar^2\alpha^4 J_1^2(\flr)
\GAElab
\end{align}
\end{subequations}

Hence, GAEs may only interact with fast ions via the Landau resonance when finite $\omeganorm$ and $\krat$ are considered. In another limit, where $0 < \omegabar < 1$ and $\alpha \gg 1$, FLR function reduces to 

\begin{align}
\lim_{\alpha\rightarrow \infty} \Jzm(\flr) &= \frac{\left(1 \pm \omegabar\right)^2}{2\pm \omegabar}J_1^2(\flr) 
\label{eq:lan:Jlmbigappx}
\end{align}

In \eqref{eq:lan:Jlmbigappx}, the top signs are for CAEs, and the bottom signs for GAEs. The expression in \eqref{eq:lan:gammabeam} represents the local perturbative growth rate for CAE/GAEs in application to an anisotropic beam-like distribution of fast ions, keeping all terms from $\omeganorm$, $\krat$, and $\kperp\rhob$. The derivative with respect to $\pphi$ has been omitted, as it is expected to less relevant for the high frequency modes studied here. Moreover, the local approximation ignores spatial profile dependencies, sacrificing accuracy in the magnitude of the growth/damping rate in favor of deriving more transparent instability conditions. 

\subsection{Properties of Fast Ion Drive}

Notice that only regions of the integrand where the term in brackets is negative are driving. For modes interacting via the Landau resonance, this requires $\partial\fb/\partial\lambda < 0$, equivalent to $\lambda > \linj$ for a distribution peaked at $\linj$. Unlike the cyclotron resonance-driven modes analyzed in \chapref{ch:cyc:analytics-cyclotron}, the damping from $\partial \fb/\partial v$ (second term in square brackets) can be comparable to the drive/damping from velocity space anisotropy over a nontrivial fraction of the integration region. Consequently, an immediate stability condition can be extracted. 

When $1 - \vpres^2/v_0^2 \leq \linj\omegacires$, the integrand is non-negative over the region of integration, such that $\gamma < 0$. As a corollary, when $1 - \vpres^2/v_0^2 \leq \linj\omegacires$, modes interacting through the Landau resonance are strictly stable. For CAEs, $\vpres$ depends on $\krat$, and hence this relation provides information about the allowed mode properties driven by a given distribution of fast ions. 

As mentioned above, the co-GAE instability due to the Landau resonance is possible only when coupling to the compressional branch is considered. Neglecting coupling, its FLR function $\J{0}{G}$ would be identically zero according to \eqref{eq:lan:Jlm} in the limit of $N^{-2} = F^2 (1 - \omegabar^2)$ exactly. However, even when considering the coupling, its growth rate is much smaller compared to the co-CAE due to the additional factor of $\omegabar^2\alpha^4$ in \eqref{eq:lan:Jlmsmallappx-gae}, which is typically small for $\omegabar < 1$ and $\alpha \like 1$. Consequently, the co-GAE will be at most weakly unstable due to this resonance, and perhaps stabilized entirely by electron Landau or continuum damping.\cite{Fu1989PF} In contrast, co-CAEs have less barriers to excitation, consistent with their measurement in NSTX\cite{Fredrickson2001PRL,Fredrickson2013POP} and MAST,\cite{Appel2008PPCF,Sharapov2014PP} and also their appearance in \HYM modeling of NSTX.\cite{Belova2017POP} Both instabilities require finite $\kperp\rhob$ for excitation, since their FLR functions are $\J{0}{m} \propto J_1^2(\flr) \rightarrow 0$ for $\flr \rightarrow 0$. 

The expression for growth rate in \eqref{eq:lan:gammabeam} also demonstrates that instability can occur for any value of $\kperp\rhob > 0$, depending on the parameters of the fast ion distribution. This extends the region of instability found for co-CAEs driven by passing particles in \citeref{Belikov2003POP}, which asserted that $\sqrt{\linj}(\omeganorm)(\vinj) < 2$ was necessary for instability, due to the additional assumption of a delta function distribution in $\lambda$. Similarly, the conclusions from the same authors in \citeref{Belikov2004POP} regarding co-CAE stabilization by trapped particles, while qualitatively consistent with the findings here, are likewise limited to the case of a vanishingly narrow distribution in $\lambda$. For further understanding of the relationships between the relevant parameters required for instability, analytic approximations or numerical methods must be employed. 

\newcommand{\figlannarrowcomplen}{\halfwidth}
\begin{figure}[tb]
\subfloat[]{\includegraphics[width = \figlannarrowcomplen]{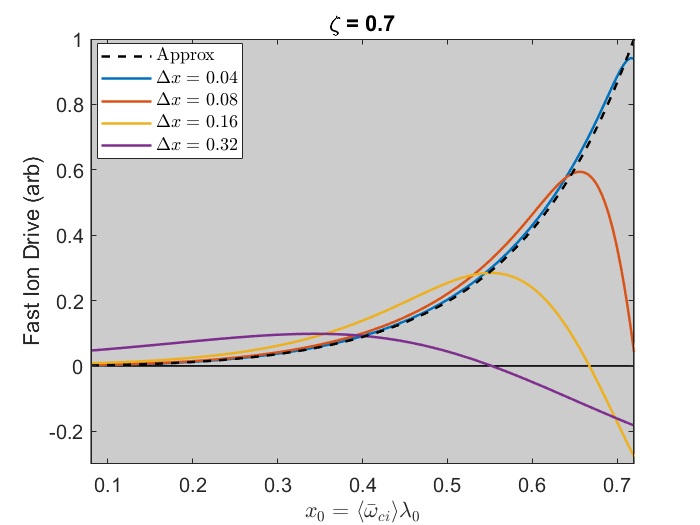}}
\subfloat[]{\includegraphics[width = \figlannarrowcomplen]{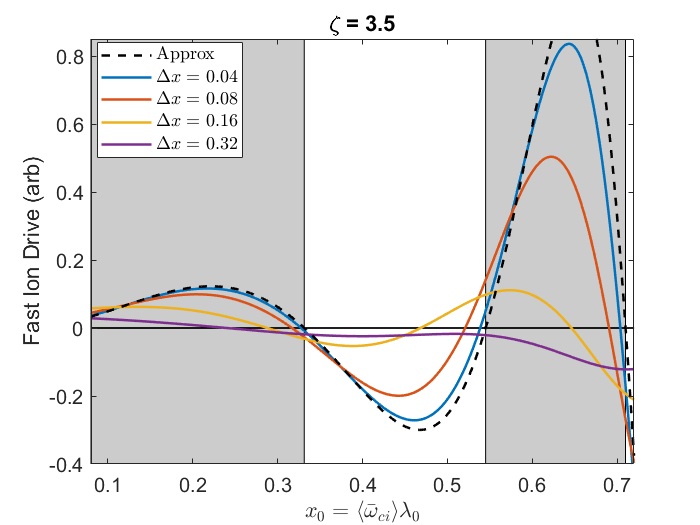}} \\
\subfloat[]{\includegraphics[width = \figlannarrowcomplen]{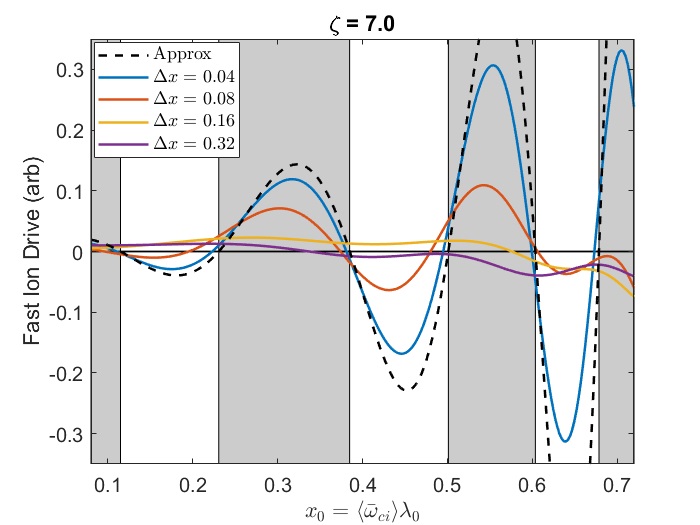}}
\caption
[Comparison of numerically integrated growth rate to narrow beam approximation for co-CAEs as a function of injection geometry.]
{Comparison of numerically integrated growth rate to narrow beam approximation for co-CAEs/GAEs with $\bres = 0.2$ as a function of the injection geometry $\xinj = \vperp^2/v^2$ of the beam distribution. Black dashed line shows the analytic approximation made in \eqref{eq:lan:gammasimp} for $\dx = 0.04$ and (a) $\zp = 0.7$, (b) $\zp = 3.5$, and (c) $\zp = 7.0$. Colored curves show numerical integration of \eqref{eq:lan:gammabeam} for different values of $\dx$: blue $\dx = 0.04$, orange $\dx = 0.08$, gold $\dx = 0.16$, and purple $\dx = 0.32$. Shaded regions correspond to regions of drive according to the narrow beam approximation.}
\label{fig:lan:narrowcomp}
\end{figure}

\section{Approximate Stability Criteria} 
\label{sec:lan:approxstab}

The expression in \eqref{eq:lan:gammabeam} can not be integrated analytically, and has complicated parametric dependencies on properties of the specific mode of interest: GAE vs CAE, $\krat$, $\omeganorm$, as well as on properties of the fast ion distribution: $\vinj$, $\linj$, and $\dl$. For chosen values of these parameters, the net fast ion drive can be rapidly calculated via numerical integration. Whenever $1 - \vpres^2/v_0^2 \leq \linj\omegacires$, both modes are damped via the Landau resonance provided that the fast ion distribution is monotonically decreasing in energy (\eg slowing down) and has a single peak in $\lambda$ at $\lambda = \linj$, such as the beam distribution given in \eqref{eq:lan:Fdistr}. There are also regimes where approximations can be made in order to gain insight into the stability properties analytically: one where the fast ion distribution is very narrow ($\dl \lesssim 0.10$) and one where it is moderately large $(\dl \gtrsim 0.20$). The former allows comparison with previous calculations,\cite{Belikov2003POP,Belikov2004POP} while the latter includes the experimental regime where the distribution width in NSTX is typically $\dl \approx 0.30$. In this section, marginal stability criteria will be derived in these regimes and compared to the numerical evaluation of \eqref{eq:lan:gammabeam}, using $v_c = v_0/2$ and $n_b/n_e = 5.3\%$, based on the conditions in the well-studied NSTX H-mode discharge $\# 141398$. 

\subsection{Approximation of Very Narrow Beam}
\label{sec:lan:narrow}

For the first regime, consider the approximation of a very narrow beam in velocity space. The purpose of this section is to determine when such an approximation can correctly capture the sign of the growth rate. Hence, consider $\dx \ll 1$ such that only a small region $\xinj - \delta < x < \xinj + \delta$ contributes to the integral, where $\delta \approx 2\dx$. So long as $0 < \xinj - \delta$ and $\xinj + \delta < 1 - \bres$, two linear approximations can be made such that to leading order in $\dx$, \eqref{eq:lan:gammabeam} is approximately 

\begin{align}
\label{eq:lan:gammasimp}
\frac{\gamma}{\omegaci} &\propto C_f \dx\sqrt{\pi}\left[2 h_1'(\xinj) - 3 h_2(\xinj)\right] \\
\text{where } h_1(x) &= \frac{x^2}{(1-x)^2}\frac{\Jzm(\flr(x,\zp))}{1 + \frac{v_0^3}{v_c^3}\left(\frac{\bres}{1-x}\right)^{3/2}} \\ 
\text{and } h_2(x) &= \frac{h_1(x)}{x}\frac{1}{1 + \frac{v_c^3}{v_0^3}\left(\frac{1-x}{\bres}\right)^{3/2}}
\end{align} 

The above expressions apply equally to CAEs and GAEs. Whereas for the cyclotron resonances discussed in \secref{sec:cyc:narrow}, the narrow beam approximation yielded a growth rate with sign depending only on the sign of a single function, for the Landau resonance, a second function must be kept to include the non-negligible contribution from $\partial\fb/\partial v$. A comparison of the approximate narrow beam stability criteria to the exact expression with $\bres = 0.2$ is shown in \figref{fig:lan:narrowcomp}. There, the dashed line shows the approximate analytic result \eqref{eq:lan:gammasimp} plotted as a function of $\xinj$ for $\dx = 0.04$ and different values of $\zp$. Values of $\xinj$ where $\gamma > 0$ according to \eqref{eq:lan:gammasimp} indicate regions where the fast ions are net driving according to this assumption (shaded regions). For comparison, the full expression \eqref{eq:lan:gammabeam} is integrated numerically for each value of $\xinj$ for varying $\dx = 0.04, 0.08, 0.16, 0.32$. 
This figure demonstrates where the narrow beam approximation correctly determines the sign of the fast ion drive, and how it depends on $\zp$. As in \secref{sec:cyc:narrow} for cntr-GAEs driven by the ordinary cyclotron resonance, it is demonstrated that $\dx \approx 0.1$ gives an acceptable (albeit strained) agreement between the approximation and numerically integrated expression. For any larger values (such as $\dx = 0.16$ and $\dx = 0.32$ shown), the approximation no longer captures the correct sign of the growth rate as a function of $\xinj$, with more pronounced disagreement occurring at larger values of $\xinj$. Moreover, it is clear that larger $\zp$ leads to more distinct regions of net drive and damping, leading to more areas where the approximate formula may incorrectly predict stability or instability. 

\subsection{Approximation of Realistically Wide Beam}
\label{sec:lan:wide}

For sufficiently wide beam distributions (such as those generated with NBI in NSTX with $\dx \approx 0.3$), one may approximate $d \exp(-(x-\xinj)^2/\dx^2)/dx \approx -2(x-\xinj)/\dx^2$. This linear approximation is appropriate for $\xinj - \dx/\sqrt{2} < x < \xinj + \dx/\sqrt{2}$. When this range extends over a large fraction of the integration region, it can be used to provide very accurate marginal stability conditions. Throughout this section, $v_c = v_0/2$ will be taken as a representative figure, and the slowing down part of the distribution will be approximated as constant since it makes a small quantitative difference. However, this approximation alone is insufficient to evaluate \eqref{eq:lan:gammabeam} in terms of elementary functions, as the Bessel functions with complicated arguments remain intractable. 

For the cyclotron resonances analyzed in \secref{sec:cyc:wide}, the fast ion damping due to $\partial\fb/\partial v$ could be neglected since it was smaller than the drive/damping due to $\partial\fb/\partial \lambda$ in that case by a factor of $\omegabar\dx^2 \ll 1$ except in a very small region near $x = \xinj$. For modes driven by the Landau resonance, it can compete with the drive/damping from anisotropy over a wider range of the integration region. Hence, the contributions from $\partial\fb/\partial v$ must be kept in this section, leading to somewhat more complicated instability boundaries than those derived in \secref{sec:cyc:wide}.

Substituting the values of $\omeganorm$ and $\krat$ from the most unstable modes in \HYM simulations into \eqref{eq:lan:zp} shows that the majority of these modes have $\zp \like \ord{1}$. Since this parameter controls how rapidly $\Jlm(\flr)$ oscillates, we are motivated to consider two cases separately: $\zp \ll 1$ (small FLR, more common) and $\zp \gg 1$ (large FLR, uncommon for $\omega < \omegaci$). 

\subsubsection{Small FLR Regime \texorpdfstring{$(\zp \ll 1)$}{}}
\label{sec:lan:slow}

Consider first the case of small FLR effects. For small argument, $\Jlm(\flr) \propto J_1^2(\flr) \approx \flr^2/4 \plusord{\flr^4}$. Then the simplified integral to consider is 

\begin{multline}
\gamma \appropto \int_0^{1-\bres} \frac{x^3(x-\xinj)}{(1-x)^3}dx  
- \frac{3\dx^2}{4}\int_0^{1-\bres} \frac{x^2}{(1-x)^3}\frac{dx}{1 + \left(\frac{1-x}{4\bres}\right)^{3/2}} \\
- \frac{\dx^2}{2}\left(\bres^{-1}-1\right)^2 e^{-(1-\bres-\xinj)^2/\dx^2}
\label{eq:lan:CAEdampgam}
\end{multline}

As a reminder, $\eta = \vpres^2/v_0^2$ such that the upper bound of the integration describes a cutoff in the distribution function at the finite injection velocity $v_0$. The integrals can be evaluated exactly and well-approximated. Solving for the marginal stability condition $\gamma = 0$, neglecting the third term for now, yields 

\begin{align}
\label{eq:lan:CAEdampgamsym}
\xinj &= \frac{g_0(\bres) -\dx^2 g_1(\bres)}{g_2(\bres)} \\ 
\label{eq:lan:CAExcrit}
 &\approx 1 - \bres^{4/5} - \frac{2\dx^2}{3(1 - \bres^{4/5})} \\
\label{eq:lan:wideslowCAEgam}
\Rightarrow v_0 &= \frac{\vpres}{\left[1 - \frac{1}{2}\left(\xinj + \sqrt{\xinj^2 + \frac{8\dx^2}{3}}\right)\right]^{5/8}}
\end{align}

The exact forms of $g_0(\bres)/g_2(\bres)$ and $g_1(\bres)/g_2(\bres)$ are given in \appref{app:lan:damprefsmall}. The first function can be excellently approximated by $g_0(\bres)/g_2(\bres) \approx 1 - \bres^{4/5}$, with a maximum relative error of less than $1\%$. The second function, $g_1(\bres)/g_2(\bres)$ is substantially more complicated. Noting its singularity as $\bres\rightarrow 1$, and considering that the goal is to find a closed form for $\bres$ as a function of $\xinj$, an \emph{ansatz} of the form $c/(1 - \bres^{4/5})$ is chosen, with $c = 2/3$ giving a maximum relative error of $15\%$, and usually half that. With this approximation, the marginal stability condition could be derived. 

When $\dx$ is small, \eqref{eq:lan:wideslowCAEgam} would reduce to $v_0 = \vpres/(1 - \xinj)^{5/8}$, similar in form to the marginal stability condition found in \secref{sec:cyc:slow} for cyclotron resonance-driven modes with $\zp \ll 1$, except with a power of $5/8$ instead of $3/4$ due to the different $\Jlm(\flr)$ functions. Note that $\gamma < 0$ for \emph{all} values of $\xinj,v_0$ when $\dx^2 > 5/3$ according to \eqref{eq:lan:CAExcrit}. This condition represents the beam width necessary to balance the maximum anisotropy drive with the slowing down damping. While it indicates a theoretical avenue for stabilizing all CAEs/GAEs driven by the Landau resonance, it is unlikely to be useful in practice since it requires a nearly uniform distribution in $\lambda$, which would not allow sufficient flexibility in the current profile that is desirable for other plasma performance objectives. 

The third term in \eqref{eq:lan:CAEdampgam} was neglected because its inclusion would prevent an algebraic solution for $\xinj$ at marginal stability. However, it can be comparable in magnitude to the second term in the integration, and can be included in an \emph{ad-hoc} fashion by solving for its effect at $\xinj = 0$, and multiplying the full result by this factor. We will also apply a rational function approximation to the Gaussian dependence, so that at $\xinj = 0$, the marginal stability condition for $\bres$ is 

\begin{multline}
\label{eq:lan:bdampcrit}
\int_0^{1-\bres} \frac{x^4}{(1-x)^3}dx 
- \frac{3\dx^2}{4}\int_0^{1-\bres} \frac{x^2}{(1-x)^3}\frac{dx}{1 + \left(\frac{1-x}{4\bres}\right)^{3/2}} \\ 
- \frac{\dx^4}{2}\frac{\left(\bres^{-1}-1\right)^2}{\dx^2 + (1-\bres)^2} = 0
\end{multline}

This expression yields a quadratic formula for $\dx^2$, given in \appref{app:lan:damprefsmall}, which can be approximated to within $10\%$ globally and inverted to yield 

\begin{align}
\bres \approx (1 - \dx^{4/5})^{5/4}
\label{eq:lan:slowdxapprox}
\end{align}


Hence, the instability condition resulting from matching the correction due to the third term in \eqref{eq:lan:CAEdampgam} at $\xinj = 0$ is

\begin{align}
v_0 &> \frac{\vpres}{\left[1 - \frac{1}{2}\left(\xinj + \sqrt{\xinj^2 + \frac{8\dx^2}{3}}\right)\right]^{5/8}}\left(\frac{1 - \dx\sqrt{2/3}}{1-\dx^{4/5}}\right)^{5/8}
\label{eq:lan:wideslowCAEgamtail}
\end{align}

\begin{figure}[H]
\subfloat[\label{fig:lan:wideslowCAE}]{\includegraphics[width = 0.6\textwidth]{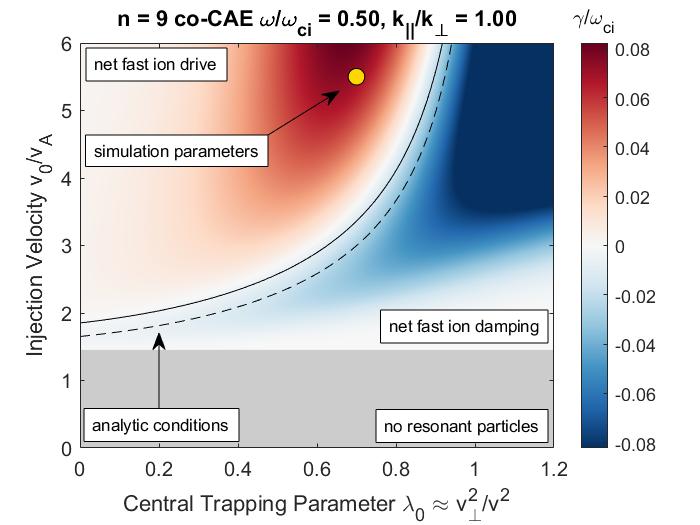}} \\
\vspace*{-0.5cm}
\subfloat[\label{fig:lan:wideslowGAE}]{\includegraphics[width = 0.6\textwidth]{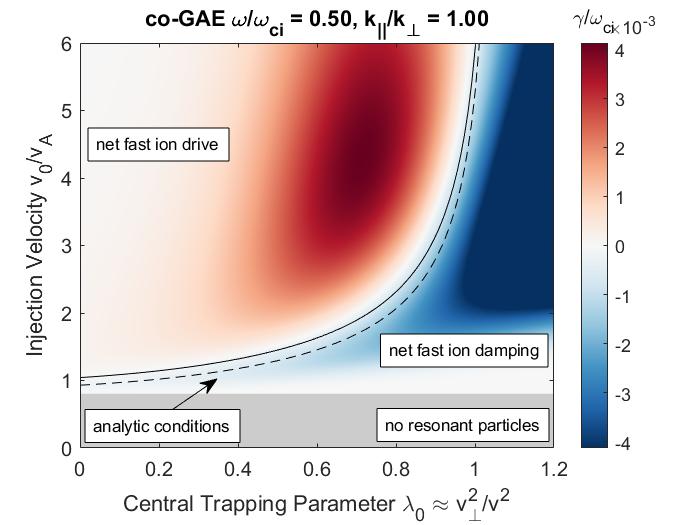}} 
\caption
[co-CAE and co-GAE growth rate dependence on beam injection geometry $\linj$ and velocity $\vinj$ in the wide beam, small FLR regime ($\zp \ll 1$).]
{Numerical integration of full growth rate expression \eqref{eq:lan:gammabeam} as a function of fast ion distribution parameters $\vinj$ and $\linj$ with $\dx = 0.30$ for a Landau-resonance driven (a) co-CAE and (b) co-GAE in the small FLR regime ($\zp \ll 1$) with properties inferred from \HYM simulations: $\omeganorm = 0.5$ and $\krat = 1$, implying $\zp = 0.5$. Red indicates net fast ion drive, blue indicates net fast ion damping, and gray indicates beam parameters with insufficient energy to satisfy the resonance condition. Dashed curve shows approximate stability condition excluding damping from the tail, derived in \eqref{eq:lan:wideslowCAEgam}. Solid curve shows approximate stability condition including damping from the tail, derived in \eqref{eq:lan:wideslowCAEgamtail}.}
\label{fig:lan:wideslowfig}
\end{figure}

This marginal stability condition encompasses both the CAEs and GAEs, since the only difference is that the GAE drive/damping has a reduced magnitude, as described by \eqref{eq:lan:Jlmsmallappx} and \eqref{eq:lan:Jlmbigappx} when $\omegabar \ll 1$ and $\alpha \gg 1$, respectively. The condition derived in \eqref{eq:lan:wideslowCAEgamtail} can also be compared against the full numerically integrated expression in 2D beam parameter space for a typical case, shown in \figref{fig:lan:wideslowCAE}. There, an $n = 9$ co-CAE driven by the Landau resonance in \HYM simulations has been chosen, using mode parameters of $\omegabar \defined \omeganorm = 0.5$ and $\alpha \defined \krat = 1$, implying $\zp = 0.5$ and a distribution with $\dx = 0.30$. There, the solid curve includes the contribution from the tail of the distribution (\eqref{eq:lan:wideslowCAEgamtail}), while the dashed curve neglects this contribution (\eqref{eq:lan:wideslowCAEgam}). The former better tracks the numerically computed stability boundary. Note also that the boundary is shifted upwards due to the damping from including the velocity derivative terms.  

Compared to the cntr-propagating modes driven by the ordinary cyclotron resonance, co-CAEs driven by the Landau resonance require relatively large $\vinj$ for excitation. To see this, consider \eqref{eq:lan:wideslowCAEgamtail} and substitute $\vpres/\va \approx k/\abs{\kpar}$, which follows from the approximate dispersion $\omega \approx k\va$ for CAEs. Then the minimum $\vinj$ for instability occurs at $\xinj = 0$, such that $\vinj > \abs{k/\kpar}/(1 - \dx^{4/5})^{5/8}$. This expression in turn is minimized for $\abs{k/\kpar} \rightarrow 1$, which for $\dx = 0.3$ yields $\vinj > 1.3$ as a strict lower bound for this instability. With more realistic perpendicular beam injection, such as the original NSTX beam with $\linj \approx 0.7$, the requirement increases to $\vinj > 2.9$ in the same limit of $\abs{k/\kpar} \rightarrow 1$, and even larger at $\vinj > 4.1$ for common values of $\kpar/\kperp \approx 1$. 

In contrast, cyclotron resonance-driven cntr-GAE excitation features no such constraints, as modes can in principle be excited even for $\vinj < 1$ so long as the frequency is sufficiently large to satisfy the resonance condition in \eqref{eq:lan:rescon}. The same is true for cntr-CAEs, with the caveat that $\krat$ must be sufficiently large as well ($\krat \approx 1$ usually sufficient).  
These considerations can explain both simulation results and experimental observations. In \HYM simulations of NSTX for a given set of plasma profiles, co-CAEs are found to require $\vinj \gtrsim 4.5$, whereas cntr-GAEs are excited for a wider range of $\vinj$. In NSTX experiments, counter-propagating modes were more commonly observed than co-CAEs, with the latter appearing only very rarely in NSTX-U experiments which typically operated at much lower $\vinj \lesssim 2$ due to the increased toroidal field strength relative to NSTX. 

A similar comparison can be made for co-GAEs, using the same mode parameters of $\omeganorm = 0.5$ and $\krat = 1$, shown in \figref{fig:lan:wideslowGAE}. Due to the difference in dispersion relation, the co-GAE can sustain a resonant interaction with a fast ion distribution with smaller $\vinj$ than the co-CAE can. The peak growth rate for co-GAEs with these parameters is reduced by an order of magnitude relative to the co-CAE, as expected based on the factor $\omegabar^2\alpha^4$ in front of its FLR function in \eqref{eq:lan:Jlmsmallappx-gae}. Although the co-GAE growth rate peaks at lower $\vinj$ in this example, even at its absolute peak, the co-CAE growth rate is larger. This may explain why co-GAEs driven via the Landau resonance were not observed in NSTX experiments. Furthermore, such modes would have been even more difficult to excite in \HYM simulations, as their drive is strongly enhanced by coupling to the compressional mode, and this coupling is under-estimated in the \HYM model due to the absence of thermal plasma two-fluid effects (see \citeref{Belova2017POP} for a detailed description of the simulation model). 

\subsubsection{Large FLR Regime \texorpdfstring{$(\zp \gg 1)$}{}}
\label{sec:lan:fast}

The complementary limit, of large FLR effects, or rapidly oscillating integrand regime due to $\zp \gg 1$ can also be explored. Based on the most unstable modes found in the \HYM simulations, this is not the most common regime for NSTX-like plasmas, but it can occur and is treated for completeness and comparison to the slowly oscillating (small FLR) results. 

This approximation allows the use of the asymptotic form of the Bessel functions: $J_n(\flr) \like \sqrt{2/\pi \flr}\cos\left(\flr - (2n+1)\pi/4\right) \plusord{\flr^{-3/2}}$, which is very accurate for $\flr > 2$. Note also that $\zp \gg 1$ implies $\alpha \ll 1$ since $\zp =\omegabar/\alpha < 1/\alpha$. For both CAEs and GAEs, the FLR function has asymptotic behavior $\J{0}{m}(\flr) \like J_0^2(\flr) \like (1 - \sin(2\flr))/\flr$, where the rapidly varying $\sin(2\flr)$ component will average out in the integrand by the Riemann-Lebesgue Lemma\cite{BenderOrszagStationaryPhase} (see \secref{sec:cyc:fast} for further description of this procedure). Then the simplified integral to consider is 

\begin{multline}
\label{eq:lan:dampgamfast}
\gamma \appropto \int_0^{1-\bres} \frac{x^{3/2}(x-\xinj)}{(1-x)^{3/2}}dx 
- \frac{3\dx^2}{4}\int_0^{1-\bres}\frac{\sqrt{x}}{(1-x)^{3/2}}\frac{dx}{1 + \left(\frac{1-x}{4\bres}\right)^{3/2}} \\ 
- \frac{\dx^2}{2}\sqrt{\bres^{-1}-1}e^{-(1-\bres-\xinj)^2/\dx^2}
\end{multline}

Following the same method as in the small FLR regime, first find the marginal stability condition $\gamma = 0$ while neglecting the third term: 



\begin{align}
\label{eq:lan:dampfastx0notail}
\xinj &= \frac{h_0(\bres) - \dx^2 h_1(\bres)}{h_2(\bres)} \\ 
&\approx 1-\bres^{5/7} - \dx^2 \left[-\frac{2\left(1 - \bres^{5/7}\right)}{5} + \frac{4}{5\left(1 - \bres^{5/7}\right)}\right] \\
\Rightarrow v_0 &= \frac{\vpres}{\left[1 - \frac{\xinj + \sqrt{\xinj^2 + 16\dx^2\left(1 + 2\dx^2/5\right)/5}}{2\left(1 + 2\dx^2/5\right)}\right]^{7/10}}
\label{eq:lan:dampfastvcrit}
\end{align}

The first part of the approximation ($h_0(\bres)/h_2(\bres)$) is accurate to within $4\%$, while the second part ($h_1(\bres)/h_2(\bres)$) has a maximum relative error of $17\%$, with the error reducing to less than $5\%$ for $\bres > 0.01$. Note that this expression is different from Eq. 26 in \citeref{Lestz2020p2} where it was first published due to an error discovered in that version. The discrepancy creates only a small quantitative difference, but results in a substantial change to the symbolic expression. Further details can be found in \appref{app:lan:dampreflarge}. 

Now consider comparing \eqref{eq:lan:dampfastvcrit} to the analogous instability condition for the same resonance when $\zp \ll 1$ (\eqref{eq:lan:wideslowCAEgam}). When $\dx = 0$, the $\zp \gg 1$ condition is somewhat more restrictive due to the different exponents, and for finite $\dx$, the correction due to the slowing down part of the $\partial \fb/\partial v$ term is also larger than it is when $\zp \ll 1$, as in \secref{sec:lan:slow}.  

The contribution from the third term in \eqref{eq:lan:dampgamfast} will be treated in the same fashion as in \secref{sec:lan:slow}. Hence, consider solving \eqref{eq:lan:dampgamfast} for marginal stability setting $\xinj = 0$ and approximating $\exp(-x^2) \approx 1/(1 + x^2)$. 

\begin{multline}
\label{eq:lan:bdampcritlarge}
\int_0^{1-\bres} \frac{x^{5/2}}{(1-x)^{3/2}}dx 
- \frac{3\dx^2}{4}\int_0^{1-\bres} \frac{\sqrt{x}}{(1-x)^{3/2}}\frac{dx}{1 + \left(\frac{1-x}{4\bres}\right)^{3/2}} \\ 
- \frac{\dx^4}{2}\frac{\sqrt{\bres^{-1}-1}}{\dx^2 + (1-\bres)^2} = 0
\end{multline}

Then $\dx^2$ can be isolated from a quadratic formula, approximated and inverted as shown in \appref{app:lan:dampreflarge}. This procedure gives the following condition for marginal stability at $\xinj = 0$, accurate to within $15\%$

\begin{align}
\bres &\approx \left(1 - (5/7)^{2/5}\dx^{4/5}\right)^{7/5}
\label{eq:lan:fastdxapprox}
\end{align}

This can be combined with \eqref{eq:lan:dampfastvcrit} to determine the modification to the instability condition required to match the solution at $\xinj = 0$


\begin{align}
v_0 &> \frac{\vpres}{\left[1 - \frac{\xinj + \sqrt{\xinj^2 + 16\dx^2\left(1 + 2\dx^2/5\right)/5}}{2\left(1 + 2\dx^2/5\right)}\right]^{7/10}}
\left(\frac{1 - \frac{2}{\sqrt{5}\sqrt{1 + 2\dx^2/5}}\dx}{1 - (5/7)^{2/5}\dx^{4/5}}\right)^{7/10}
\label{eq:lan:widefastCAEgamtail}
\end{align}

\begin{figure}[H]
\subfloat[\label{fig:lan:widefastCAE}]{\includegraphics[width = 0.6\textwidth]{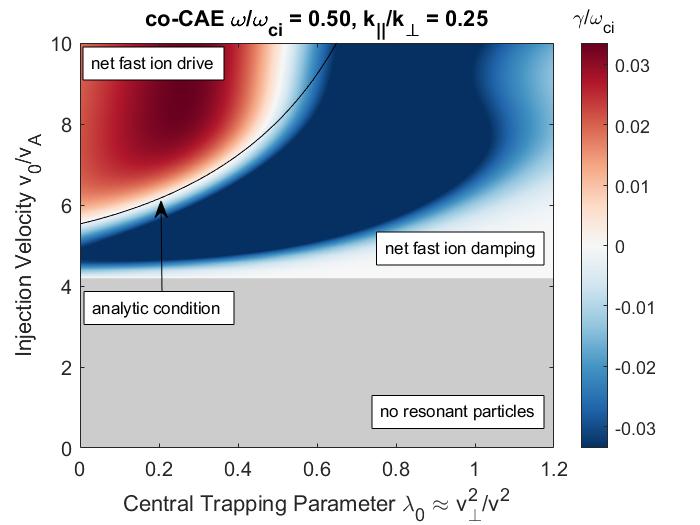}} \\
\vspace*{-0.5cm}
\subfloat[\label{fig:lan:widefastGAE}]{\includegraphics[width = 0.6\textwidth]{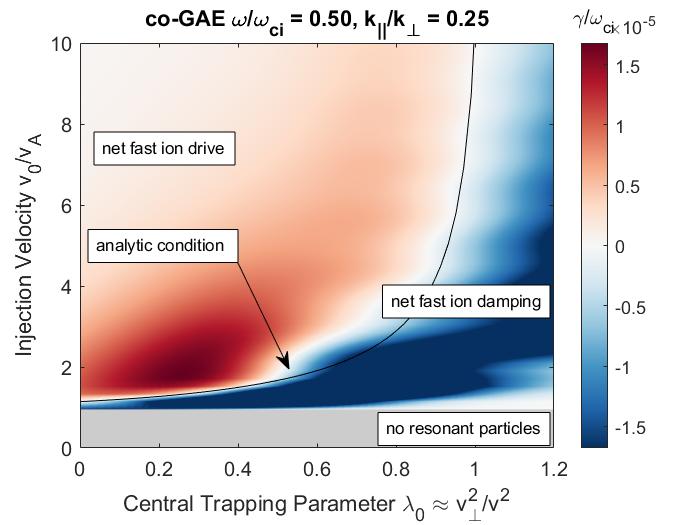}} 
\caption
[co-CAE and co-GAE growth rate dependence on beam injection geometry $\linj$ and velocity $\vinj$ in the wide beam, large FLR regime ($\zp \gg 1$).]
{Numerical integration of full growth rate expression \eqref{eq:lan:gammabeam} as a function of fast ion distribution parameters $\vinj$ and $\linj$ with $\dx = 0.30$ for a Landau resonance-driven (a) co-CAE and (b) co-GAE in the large FLR regime ($\zp \gg 1$): $\omeganorm = 0.5$ and $\krat = 0.25$, implying $\zp = 2$. Red indicates net fast ion drive, blue indicates net fast ion damping, and gray indicates beam parameters with insufficient energy to satisfy the resonance condition. Solid curve shows approximate stability condition including damping from the tail, derived in \eqref{eq:lan:widefastCAEgamtail}.}
\label{fig:lan:widefastfig}
\end{figure}

This $\zp \gg 1$ marginal stability bound is compared to the numerically evaluated fast ion drive/damping in \figref{fig:lan:widefastfig} for a co-CAE and co-GAE with $\omegabar \defined \omeganorm = 0.5$ and $\alpha \defined \krat = 0.25$ such that $\zp = 2$. While $\zp = 2$ is only marginally within the $\zp \gg 1$ regime, the agreement is still acceptable. Note that in the figures, a maximum value of $\vinj = 10$ is shown, which far exceeds the NSTX range of $\vinj < 6$. This is because the CAE dispersion combined with the resonance condition yields $\zp \approx \omegabar\vpres/\va$ for $\zp \gg 1$, which can not be very large for $\vinj < 6$ considering $\vpres \like v_0/2$ is common, as is $\omeganorm \like 0.5$. The case is different for GAEs since their dispersion yields a parallel resonant velocity that is independent of $\alpha$, such that $\zp$ can be made arbitrarily large by choosing $\alpha$ sufficiently small without constraining the size of $\vpres/\va$. This explains why the co-CAE in the figure has no wave particle interaction when $\vinj < 4$, while an interaction with the co-GAE becomes possible near $\vinj \approx 1$. Although the co-GAE can in principle be driven by fast ions for more accessible values of $\vinj$, note that the growth rate is vastly reduced due to the factor of $\krat^4 \lll 1$. Thus, one would expect that the minuscule magnitude of fast ion drive for the co-GAE shown in \figref{fig:lan:widefastGAE} would be far outweighed by damping on the background plasma. For these reasons, the $\zp \gg 1$ regime is less relevant to modern experimental conditions than the $\zp \ll 1$ regime, except possibly for CAEs with $\omega > \omegaci$ which can be excited at more reasonable values of $\vinj$ (to be addressed in a future work).

\subsection{Summary of Necessary Conditions for Net Fast Ion Drive}

Here, we briefly summarize the different stability boundaries derived up to this point, along with their ranges of validity. When $1 - \vpres^2/v_0^2 \leq \linj\omegacires$ is satisfied, Landau resonance-driven co-propagating CAEs/GAEs will be net damped by fast ions. All other results address the scenarios when this inequality is not satisfied. When $\dl$ is sufficiently small $(\dl \lesssim 0.10)$, the narrow beam approximation can be made, which yields \eqref{eq:lan:gammasimp}, where the sign of the growth rate depends on $\xinj$ and can be evaluated without further integration. When $\dl$ is sufficiently large $(0.20 \lesssim \dl \lesssim 0.80)$, the wide beam approximation is justified. This includes the nominal NSTX case of $\dl \approx 0.30$. For most of the unstable modes in \HYM simulations, $\zp \lesssim 2$ is also valid, which enables the results contained in the case of a wide beam and slowly oscillating integrand. The complementary limit of $\zp \gg 2$ is also tractable when the beam is sufficiently wide, though this is not the typical case for CAEs and GAEs interacting with fast ions through the Landau resonance. All conditions for the cases involving wide beams are organized in \tabref{tab:lan:appxcons}. 

\newcommand{\vroomlan}{\vphantom{{\Huge text}}}
\newcommand{\hroomlan}{\hspace{1ex}}
\newcommand{\cheadlan}[1]{\multicolumn{1}{c}{#1}}
\begin{table*}\centering
\renewcommand\arraystretch{1.5}
\begin{tabular}{l l l}
\hline\hline 
\multicolumn{2}{c}{CAE/GAE fast ion drive conditions (Landau resonance)} \\ \hline\hline 
 & \cheadlan{} 
 & \cheadlan{}\vspace{-3ex} \\ 
$\zp \lesssim 2$ \hroomlan\hroomlan & $v_0 > \dfrac{\vpres}{\left[1 - \frac{1}{2}\left(\xinj + \sqrt{\xinj^2 + 8\dx^2/3}\right)\right]^{5/8}} \left(\dfrac{1 - \dx\sqrt{2/3}}{1-\dx^{4/5}}\right)^{5/8}$ \hroomlan\vroomlan \\
$\zp \gg 2$ & $v_0 > \dfrac{\vpres}{\left[1 - \frac{\xinj + \sqrt{\xinj^2 + 16\dx^2\left(1 + 2\dx^2/5\right)/5}}{2\left(1 + 2\dx^2/5\right)}\right]^{7/10}}
\left(\dfrac{1 - \frac{2}{\sqrt{5}\sqrt{1 + 2\dx^2/5}}\dx}{1 - (5/7)^{2/5}\dx^{4/5}}\right)^{7/10}$ \hroomlan\vroomlan 
\vphantom{{\Huge text}}\\
\hline\hline
\end{tabular}

\caption
[Approximate net fast ion drive conditions for GAEs and CAEs driven by the Landau resonance in the wide beam approximation.]
{Approximate net fast ion drive conditions for GAEs and CAEs driven by the Landau resonance in the wide beam approximation, valid for $0.2 < \dx < 0.8$ where $\dx = \dl\omegacires$ characterizes the velocity anisotropy of the beam. The quantity $\zp = \kperp\vpres/\omegaci$ is the ``modulation parameter" (see \eqref{eq:lan:zp}) and $x_0 = \lambda_0\omegacires = \vperpz^2/v_0^2$.}
\label{tab:lan:appxcons}
\end{table*}

\section{Preferential Excitation as a Function of Mode Parameters} 
\label{sec:lan:stabao}

For fixed beam parameters, the theory can determine which parts of the spectrum may be excited -- complementary to the previous figures which addressed how the excitation conditions depend on the two beam parameters for given mode properties. Such an examination can also illustrate the importance of coupling between the compressional and shear branches due to finite frequency effects on the most unstable parts of the spectra. All fast ion distributions in this section will be assumed to have $\dl = 0.3$ and $\omegacires = 0.9$ for the resonant ions. For the modes driven by the Landau resonance studied here in the small FLR regime, the instability conditions can be written generally as 

\newcommand{\cterm}{d}
\begin{align}
\cterm^2 &> \vpres^2(\omeganorm,\krat)/\va^2 \label{eq:lan:cvpres} \\ 
\text{where } \cterm &= \frac{v_0}{\va}\left[1 - \frac{1}{2}\left(\xinj + \sqrt{\xinj^2 + \frac{8\dx^2}{3}}\right)\right]^{5/8} 
\left(\frac{1-\dx^{4/5}}{1 - \dx\sqrt{2/3}}\right)^{5/8} 
\label{eq:lan:dvalue}
\end{align}

In the large FLR regime, $\cterm$ can be replaced by the analogous quantity from \eqref{eq:lan:widefastCAEgamtail}, though analysis in this section will focus on the more experimentally relevant small FLR regime. 
Determining the unstable regions of the spectrum as a function of $\omegabar \defined \omeganorm$ and $\alpha \defined \krat$ therefore relies on the dependence of $\vpres$ on these quantities. This dependence can be well approximated as 

\begin{align}
\vpres^{CAE}/\va &\approx \sqrt{\frac{1}{\alpha^2} + 1 + \omegabar} \label{eq:lan:vprescae}\\ 
\vpres^{GAE}/\va &\approx \sqrt{1 - \omegabar^{\frac{2 + \alpha^2}{1 + \alpha^2}}} \label{eq:lan:vpresgae}
\end{align}

These expressions have a maximum relative error of $3\%$ and $6\%$ respectively for $0 < \omeganorm < 1$ and all values of $\krat$. Using these expressions, the approximate stability conditions become 

\begin{align}
0 &< \left(\frac{\omega}{\omegaci}\right)^{CAE} < \cterm^2 - \left(1 + \frac{1}{\alpha^2}\right) \label{eq:lan:caezao}\\ 
\left(1 - \cterm^2\right)^{\frac{1 + \alpha^2}{2 + \alpha^2}} &< \left(\frac{\omega}{\omegaci}\right)^{GAE} \label{eq:lan:gaezao} < 1
\end{align}

\begin{figure}[tb]
\subfloat[\label{fig:lan:caezao30}]{\includegraphics[width = \halfwidth]{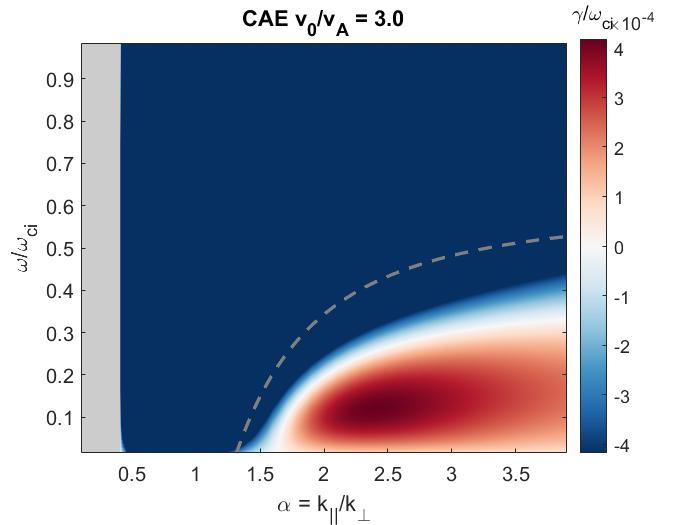}}
\subfloat[\label{fig:lan:caezao35}]{\includegraphics[width = \halfwidth]{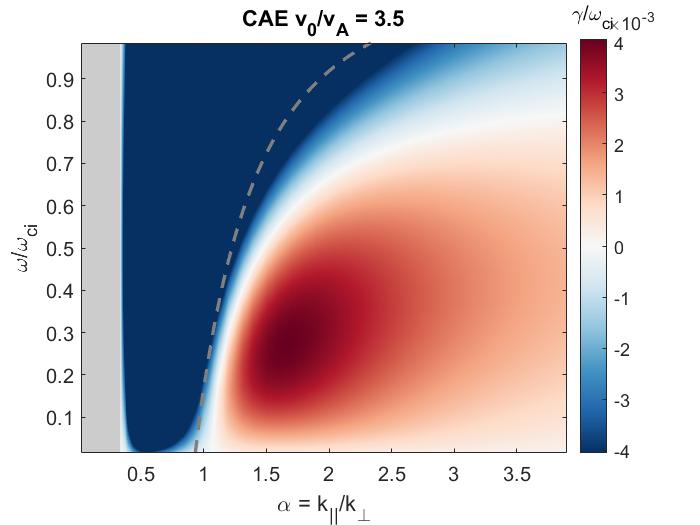}} \\
\subfloat[\label{fig:lan:caezao40}]{\includegraphics[width = \halfwidth]{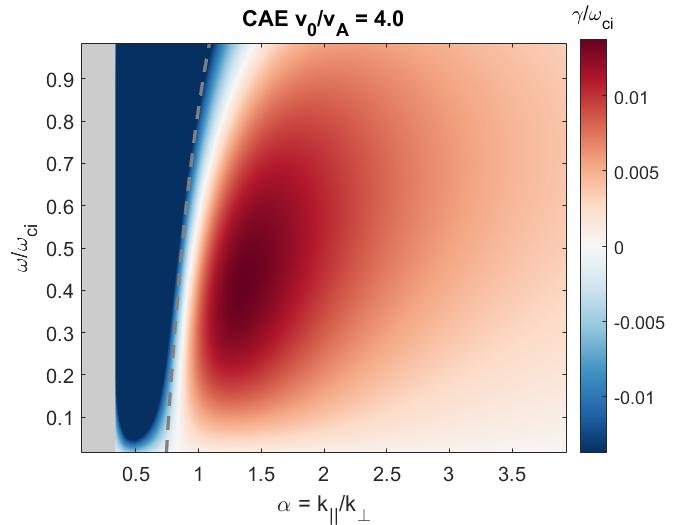}}
\caption
[co-CAE growth rate dependence on mode parameters: normalized frequency $\omegabar = \omeganorm$ and wave vector direction $\alpha = \krat$.]
{Numerically calculated fast ion drive/damping for Landau resonance-driven co-CAEs for (a) $\vinj = 3.0$, (b) $\vinj = 3.5$, and (c) $\vinj = 4.0$ as a function of $\omegabar = \omeganorm$ and $\alpha = \krat$, when driven by a beam distribution with $\linj = 0.7$, $\dl = 0.3$, and assuming $\omegacires \approx 0.9$. Red corresponds to net fast ion drive, blue to damping, and gray to regions excluded by the resonance condition. Gray curves indicate approximate marginal stability conditions.}
\label{fig:lan:caezao}
\end{figure}

Consider first the case of the CAEs. A comparison between these boundaries and the numerically integrated expression for growth rate is shown in \figref{fig:lan:caezao}. There, a fast ion distribution with $\linj = 0.7$ is assumed, similar to NSTX conditions, and the calculation is shown for different values of $\vinj$.  Note that there is a minimum value of $\alpha$ below which all frequencies are stable. This follows from \eqref{eq:lan:caezao} when $\cterm^2 < 1 + 1/\alpha^2$. For small values of $\vinj$, only small values of $\omeganorm$ can be driven by the fast ions, even though the resonance condition is satisfied for all frequencies. For larger values of $\vinj$, the frequency dependence of this boundary becomes very weak, with the boundary converging simply to $\alpha > \alphamin$. Note that if coupling to the shear mode were neglected, $\vpres$ for the CAEs would be independent of $\alpha$, which would remove the frequency dependence of the marginal stability boundary even in the case of small $\vinj$. The dashed gray curves plot \eqref{eq:lan:caezao}, demonstrating qualitative agreement with the numerically evaluated expression. The quantitative disagreement is mostly inherited from the inaccuracy of the \emph{ad-hoc} correction for the damping coming from the tail of the distribution, which used a factor to match the solution at $\linj = 0$, leading to larger errors at larger $\linj$ such as $\linj = 0.7$ used for these plots. 

\begin{figure}[tb]
\subfloat[\label{fig:lan:gaezao15}]{\includegraphics[width = \halfwidth]{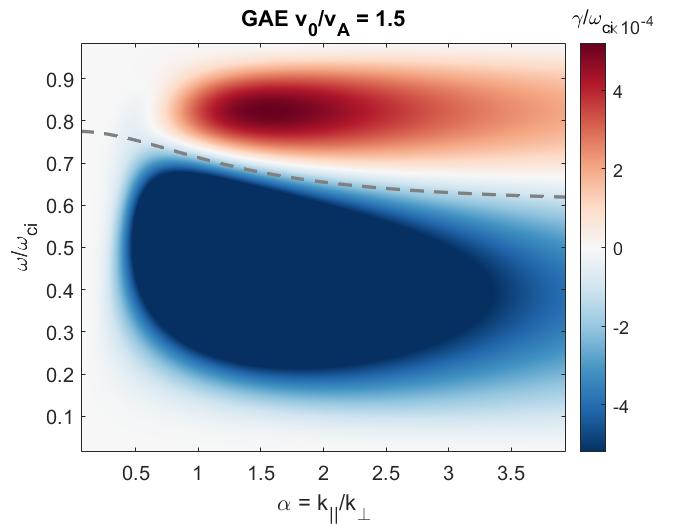}}
\subfloat[\label{fig:lan:gaezao20}]{\includegraphics[width = \halfwidth]{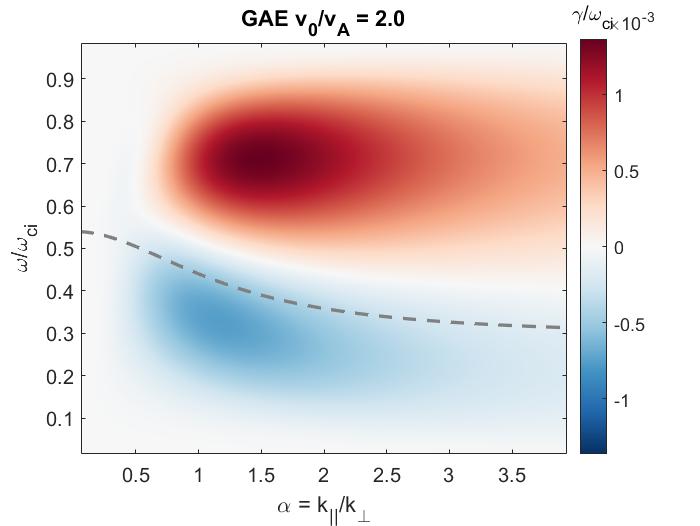}} \\
\subfloat[\label{fig:lan:gaezao25}]{\includegraphics[width = \halfwidth]{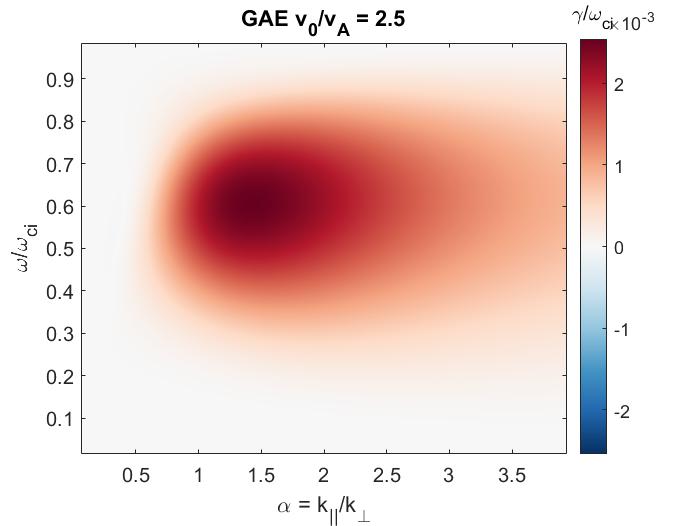}}
\caption
[co-GAE growth rate dependence on mode parameters: normalized frequency $\omegabar = \omeganorm$ and wave vector direction $\alpha = \krat$.]
{Numerically calculated fast ion drive/damping for Landau resonance-driven co-GAEs for (a) $\vinj = 1.5$, (b) $\vinj = 2.0$, and (c) $\vinj = 2.5$ as a function of $\omegabar = \omeganorm$ and $\alpha = \krat$, when driven by a beam distribution with $\linj = 0.7$, $\dl = 0.3$, and assuming $\omegacires \approx 0.9$. Red corresponds to net fast ion drive, blue to damping, and gray to regions excluded by the resonance condition. Gray curves indicate approximate marginal stability conditions.}
\label{fig:lan:gaezao}
\end{figure}

Considering now the GAEs, not only is their drive only made possible due to coupling to the compressional branch, as discussed in \secref{sec:lan:derivation}, but the unstable spectrum can also only be described when considering the coupled dispersion relation. Suppose instead that the simplified dispersion were used. Then $\vpres/\va \approx 1$ would be true for the GAEs, implying $\cterm^2 > 1$ for instability, which is completely independent of $\omeganorm$ and $\krat$. However, \figref{fig:lan:gaezao} clearly shows a minimum frequency for instability when $\vinj$ is not too large. This results from coupling to the compressional branch, which results in the modification to $\vpres$ included in \eqref{eq:lan:vpresgae}. The dashed curves compare the approximate instability conditions to the numerically integrated growth rate, showing that this correction is qualitatively captured. Again, there is some quantitative mismatch between the analytic condition and the true marginal stability boundary due to the less accurate treatment of damping from the tail. Moreover, it is worth pointing out that unlike the co-CAEs, as $\vinj$ is increased for the co-GAEs, it becomes possible to destabilize modes with \emph{smaller} frequencies. 

Note that for sufficiently large values of $\vinj$ (determined by $\cterm^2 > 1$), the GAEs can be strictly driven for all values of $\omeganorm$ and $\krat$. Such an example is shown in \figref{fig:lan:gaezao25}. However, the drive can become extremely small for regions of this parameter space far from the most favorable parameters, where modes will be stabilized by any damping mechanisms (thermal plasma, continuum) not considered here. The peak growth rate occurs near $\alpha \approx 1.5$ and $\omeganorm \approx 0.6$ in this case. This can be qualitatively understood from the form of the FLR function. For very small $\alpha$, the coefficient $\alpha^4$ in \eqref{eq:lan:Jlmsmallappx-gae} substantially decreases the growth rate. In contrast, at large $\alpha$, the coefficient in front of the Bessel function can be order unity, however the argument $\flr = \zp\sqrt{x/(1-x)}$ becomes small since $\zp = \omegabar/\alpha$, and hence $J_1^2(\omegabar/\alpha) \propto 1/\alpha^2$ for $\alpha \gg 1$. The local maximum in frequency can be understood similarly, as at low frequency, there is a coefficient $\omegabar^2$ in front of the Bessel function, and also the Bessel function will expand as $\omegabar^2$. For the limit of $\omeganorm \rightarrow 1$, the coefficient in \eqref{eq:lan:Jlmsmallappx-gae} vanishes for the GAEs. 

No special weight should be assigned to the values of $\vinj$ used in \figref{fig:lan:caezao} and \figref{fig:lan:gaezao} in relation to the shapes of the stability boundaries in general, since these conditions also depend on $\linj$. They are relevant to NSTX since the value used in the figure, $\linj = 0.7$, is characteristic of the neutral beam geometry used for that experiment. For instance, for a different value of $\linj$, the co-GAEs would become unstable for all frequencies (\eg \figref{fig:lan:gaezao25}) at some other value of $\vinj$. Likewise, the co-CAE boundary will also converge to $\alpha > \alphamin$ for a value of $\vinj$ depending on $\linj$. 

\section{Experimental Comparison} 
\label{sec:lan:expcomp}

Co-CAEs were studied in depth in NSTX in many discharges in \citeref{Fredrickson2013POP} and manually analyzed to determine the toroidal mode number and frequency of each observed eigenmode (in contrast to the database of cntr-GAEs discussed in \secref{sec:cyc:expcomp}, which was more massive and therefore relied on spectrum-averaged quantities calculated via automated analysis). Co-CAEs can be unambiguously distinguished\cite{Fredrickson2013POP,Appel2008PPCF,Sharapov2014PP} from cntr-GAEs due to the direction of propagation and the absence of other modes in the high frequency range studied ($\omeganorm \gtrsim 0.5$). From a simplified 2D dispersion solver, these high $\abs{n} (> 10)$ modes were inferred to be localized in a potential well near the low field side edge, typically with low $\abs{m} \lesssim 2$. It is worth noting that these high frequency co-CAEs were mostly observed when a low frequency $n = 1$ kink mode was present, though the source of their nonlinear interaction is not precisely known.\cite{Fredrickson2013POP}

Whereas the cntr-GAE stability condition in \eqref{eq:cyc:GAEmfreqrange} yielded lower and upper bounds on the unstable range of frequencies for a given $(\linj,\vinj)$, the marginal stability condition for co-CAEs (given in \eqref{eq:lan:caezao}) instead yields a lower bound on the allowed value of $\alpha \defined \krat$ in the low coupling limit of $\omeganorm \ll 1$, which is usually more restrictive than the lower bound on $\krat$ resulting from the requirement $\vpres < v_0$. Hence, one of these lower bounds will always be redundant. An upper bound on $\krat$ can be derived heuristically, considering that the CAEs are trapped in a local effective potential well\cite{Gorelenkova1998POP,Kolesnichenko1998NF,Smith2003POP,Smith2009PPCF,Smith2017PPCF} 
of characteristic width $\Delta R\approx R_0/2$. To satisfy this constraint, an integer number of half wavelengths must fit within the potential well, such that $k_{R,\text{min}} = \pi/\Delta R$. Similarly, poloidal symmetry requires $k_{\theta,\text{min}} = m/a$ for integer $m$. Hence, $\kperpmin \approx (2\pi/R_0)\sqrt{1 + (R_0/2\pi a)^2} \approx 2\pi/R_0$. Moreover, $\kpar \approx k_\phi = n/R_0$ is a reasonable approximation for the observed high $\abs{n}$, low $\abs{m}$ modes. Hence, $\krat_\text{max} = \kpar / \kperpmin \approx n/2\pi$. 

Although $\kperp$ is not a reliably measured experimental quantity, it can be inferred from the measured frequency and toroidal mode number using the approximate dispersion $\omega \approx k\va$, such that $\krat = 1/\sqrt{\omega^2 R_0^2 / n^2 \va^2 - 1}$. Within this local framework, $\va$ is evaluated near the plasma edge, where the mode exists, to calculate $\krat$. 

\begin{figure}[tb]
\includegraphics[width = \midwidth]{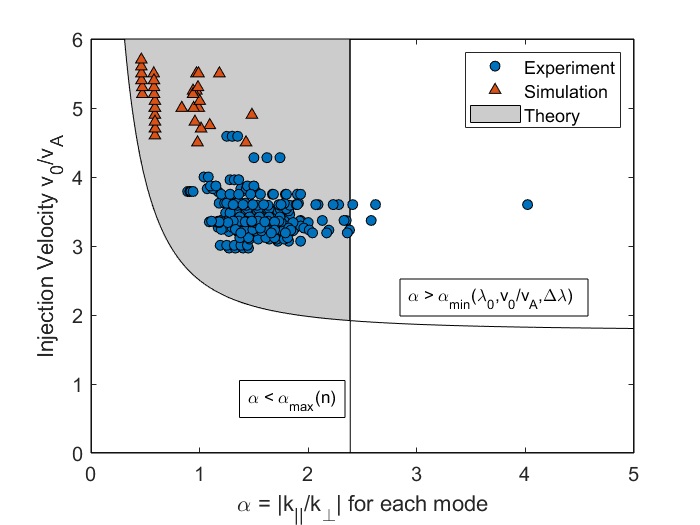}
\caption
[Comparison of approximate CAE instability conditions to simulation results and experimental observations.]
{Comparison between theory, simulations, and experiment. Blue circles represent individual co-CAE modes from NSTX discharges. Red triangles show co-CAEs excited in \HYM simulations with $\linj = 0.7$. Theory predicts net fast ion drive in the shaded region between the two curves.}
\label{fig:lan:allcomp-co}
\end{figure}

The comparison of these bounds with the experimental observations (blue circles) and simulation results (red triangles) is shown in \figref{fig:lan:allcomp-co}. The curve is calculated from \eqref{eq:lan:caezao} using $\linj = 0.65$, which was the average value for the studied discharges. Also, $N = 1$ was chosen for consistency with the formula used to calculate the wavenumber from the measured frequency. The straight line represents the heuristic upper bound on $\krat$ using $n = 15$, which was the maximum value in the experimental database. Hence, our theory predicts net fast ion drive in the shaded region between the curve and vertical line. Only simulations with $\linj = 0.7$ are shown in order to remain close to the average value of $\linj = 0.65$ for the experimental conditions shown. All points have $\krat > \alphamin$, in agreement with \eqref{eq:lan:caezao}, and most of the points are also consistent with $\krat < \alphamax$. When calculating these boundaries for the specific properties of each mode, it is found that all of the simulation points fall within the allowed range, while $82\%$ of the experimental co-CAEs agree with theory. The outliers with $\krat > \alphamax$ could be due to either a wider potential well width in those discharges or slight errors in the $n$ number identification due to limited toroidal resolution. Such an experimental comparison can not be made with co-GAEs at this time, as none were identified in NSTX, likely due to their reduced growth rate relative to the co-CAEs. 

Further analysis of the linear simulation results shown on \figref{fig:lan:allcomp-co} is described in \chapref{ch:sim:simulations}. The simulation set up and properties of the modes can be found in \citeref{Lestz2018POP}. The simulations used equilibrium profiles from the well-studied H-mode discharge $\#141398$,\cite{Fredrickson2013POP,Crocker2013NF,Crocker2017NF,Belova2017POP} and fast ion distributions with the same $(\lambda,v)$ dependence studied in this work, and given in \eqref{eq:lan:Fdistr}. The $\pphi$ dependence was fit from \TRANSP to a power law, as described in \citeref{Belova2017POP}. The peak fast ion density in all cases is $n_b/n_e = 5.3\%$, matching its experimental value in the model discharge. 

\section{Summary and Discussion}
\label{sec:lan:summary}

The fast ion drive/damping for compressional (CAE) and global (GAE) \Alfven eigenmodes due to the Landau resonance has been investigated analytically for a model slowing down, beam-like fast ion distribution, such as those generated by neutral beam injection in NSTX. The local growth rate includes contributions to all orders in $\alpha = \krat$ and $\omegabar = \omeganorm$, addressing parameter regimes that were not treated by previous work studying this instability.\cite{Belikov2003POP,Belikov2004POP} Retaining finite $\omeganorm$ and $\krat$ was demonstrated to be important for capturing the coupling between the shear and compressional branches (present due to two-fluid effects in our model), which was in turn vital to the existence of the co-GAE instability. The full FLR dependence was also kept in this derivation, as in previous work. The dependence of the fast ion drive was studied as a function of four key parameters: the beam injection velocity $\vinj$, the beam injection geometry $\linj = \mu B_0 / \W$, the mode frequency $\omeganorm$, and the direction of the wave vector $\krat$. It was shown that CAEs require relatively large $\vinj$ in order to have an appreciable growth rate, explaining why they were observed much less frequently in NSTX-U than NSTX. Moreover, the growth rate of the GAE carries an additional small coefficient of $(\omeganorm)^2\krat^4$ relative to the CAE, suggesting why these are rarely observed. 

Without further approximation, the derived growth rate led to an immediate corollary: when $1 - \vpres^2/v_0^2 \leq \linj\omegacires$, only damping occurs from the Landau resonance. For cases where this condition is not satisfied, approximate conditions for net fast ion drive were derived by making experimentally relevant approximations. Previous analytic conditions\cite{Belikov2003POP,Belikov2004POP} for net fast ion drive of CAEs driven by the Landau resonance were limited to delta functions in $\lambda$, which are a poor approximation for fast ions generated by NBI. In contrast, the instability conditions derived here result from integrating over the full beam-like distribution with finite width in velocity space. It was found in \secref{sec:lan:narrow} that the approximation of a narrow beam was only valid when $\dl \lesssim 0.1$, much smaller than the experimental value of $\dl \approx 0.3$. Consequently, our more general derivation allows for instability at any value of $\flr = \kperp\rhob$, whereas prior work concluded a limited range. 

The approximation of a sufficiently wide beam in conjunction with a small or large FLR assumption yielded an integral in the growth rate expression which could be evaluated exactly and led to useful conditions for net fast ion drive, listed in \tabref{tab:lan:appxcons}. In particular, the condition for a wide beam and small FLR effects ($\zp = \kperp\vpres/\omegaci \lesssim 2$) is typically applicable to NSTX conditions, as determined from observations and simulations of these modes. 

Comparison between the numerical integration of the analytic expression for growth rate and the approximate stability boundaries indicates strong agreement within the broad parameter regimes that they apply. Since these stability conditions depend on both fast ion parameters ($\linj, \vinj$) and mode parameters $(\omeganorm,\krat)$, they can provide information both about how a specific mode's stability depends on the properties of the fast ions, as well as the properties of the modes that may be driven unstable by a specific beam distribution. Namely, co-propagating CAEs are unstable for sufficiently large $\krat$, nearly independent of frequency when $\vinj$ is sufficiently large. In contrast, when $\vinj$ is not too large, GAEs can only be excited at high frequencies. The approximate condition for CAE stability was compared against NSTX data from many discharges, yielding greater than $80\%$ agreement, demonstrating the utility of these results in interpreting observations and guiding future experiments. One area of ongoing work is the application of this theory to predict ways to stabilize co-propagating modes with the addition of a second beam source, complementary to the cntr-GAE suppression observed in NSTX-U\cite{Fredrickson2017PRL} and reproduced numerically\cite{Belova2019POP,Kaye2019NF} with small amounts of power in the new, off-axis beam sources. 

It is worth reminding one final time of the simplifications used in deriving these results. Contributions from the gradient in $\pphi$ were not analyzed, though this is not expected to be a substantial correction based on past simulations.\cite{Belova2017POP} The calculation was also local, not accounting for spatial profiles or mode structures. Consequently, the magnitude of the drive/damping shown in figures should not be considered absolute, but rather relative. Lastly, the net drive conditions do not include sources of damping coming from the background plasma, so they should be interpreted as necessary but not sufficient conditions for instability. Careful analysis of these damping sources and their dependence on all of the parameters studied here (including kinetic contributions from the large fast ion current) is left for future work. 

\begin{subappendices}

\section{Details of Approximations for Small FLR Regime \texorpdfstring{$(\zp \ll 1)$}{}} 
\label{app:lan:damprefsmall}

Details from the calculations in \secref{sec:lan:slow} are listed here for reference. The full form of \eqref{eq:lan:CAEdampgamsym} is 

\begin{subequations}
\begin{align}
\label{eq:smallform}
\xinj &= \frac{g_0(\bres) -\dx^2 g_1(\bres)}{g_2(\bres)} \\ 
g_0(\bres) &= \frac{1 - 8\bres + 8\bres^3 - \bres^4 - 12\bres^2\log\bres}{2\bres^2} \\ 
g_1(\bres) &= \frac{A + B + C + D}{64\bres^2} \\ 
A &= 12\left(1 + \sqrt{\bres} - 8\bres + 6\bres^2\right) \\ 
B &= 2\sqrt{3}(1 + 8\bres)\arctan\left(\frac{\bres^{-1/2}-1}{\sqrt{3}}\right) \\ 
C &= \log\left[3\left(\frac{(1 - 2\sqrt{b})^2 + 2\sqrt{b}}{(1 + 2\sqrt{b})^2}\right)\right] \\
D &= 8\bres\left(2\bres\log\left[\frac{81}{\bres^3(8 + \bres^{3/2})^2}\right] 
+ \log\left[\frac{1}{3} + \frac{2\sqrt{\bres}}{(1 - 2\sqrt{\bres})^2 + 2\sqrt{\bres}}\right]\right) \\ 
g_2(\bres) &= \frac{1 - 6\bres + 3\bres^2 + 2\bres^3 - 6\bres^2\log\bres}{2\bres^2}
\end{align}
\end{subequations}

To perform the integral represented by $g_1(\bres)$ in \Mathematica, one must make the substitution $u = (1 - x)/4\bres$. Note also that the above has some differences from Eq. A1 printed in \citeref{Lestz2020p2}. That equation had some mistakes which have been rectified here. Namely, the argument of the $\arctan$ in term $B$ was incorrect and an overall factor of $8\bres$ was missing from term $D$. However, those mistakes were purely typographical, so they did not affect any of the subsequent approximations or results. The following approximations were used in \secref{sec:lan:slow} in order to substantially simplify the preceding expression:  

\begin{align}
\label{eq:lan:appg0}
 \frac{g_0(\bres)}{g_2(\bres)} &\approx 1 - \bres^{4/5} \\ 
\label{eq:lan:appg1}
 \frac{g_1(\bres)}{g_2(\bres)} &\approx  \frac{2}{3(1 - \bres^{4/5})}
\end{align}

\begin{figure}[H]
\subfloat[]{\includegraphics[width = \halfwidth]{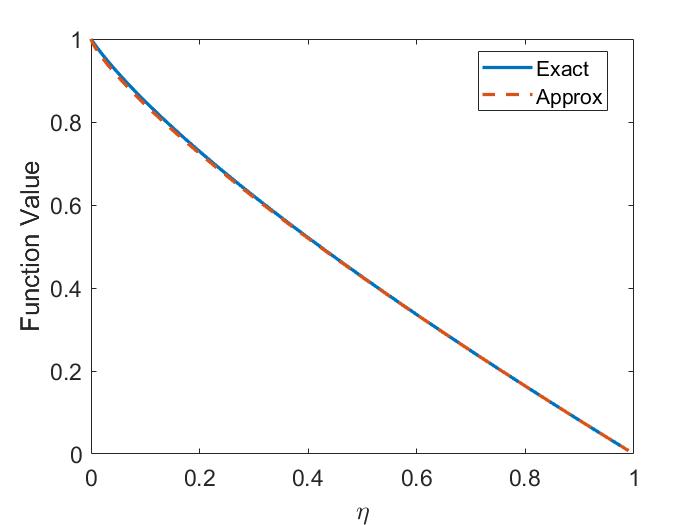}} 
\subfloat[]{\includegraphics[width = \halfwidth]{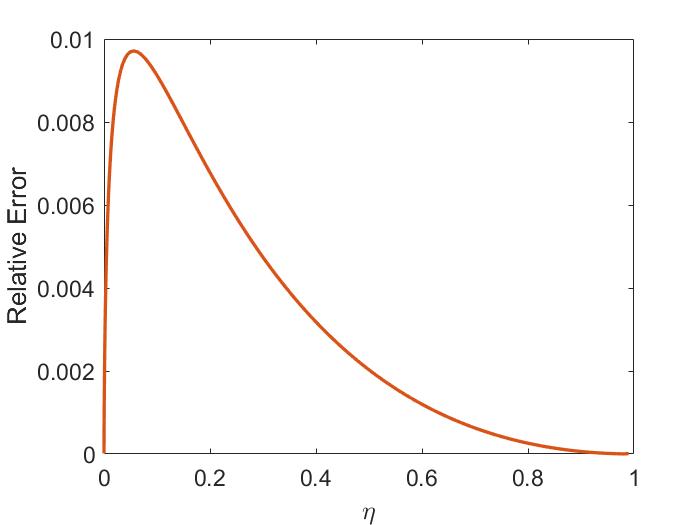}}
\caption
[Comparison of approximation for $g_0(\bres)/g_2(\bres)$ in \eqref{eq:lan:CAExcrit} to the exact function.]
{Comparison of approximation for $g_0(\bres)/g_2(\bres)$ in \eqref{eq:lan:CAExcrit} to the exact function (identical to \eqref{eq:lan:appg0}). Left: blue curve shows exact function, dashed orange curve shows approximation. Right: relative error of the approximation.}
\label{fig:lan:appg0}
\end{figure}

\begin{figure}[H]
\subfloat[]{\includegraphics[width = \halfwidth]{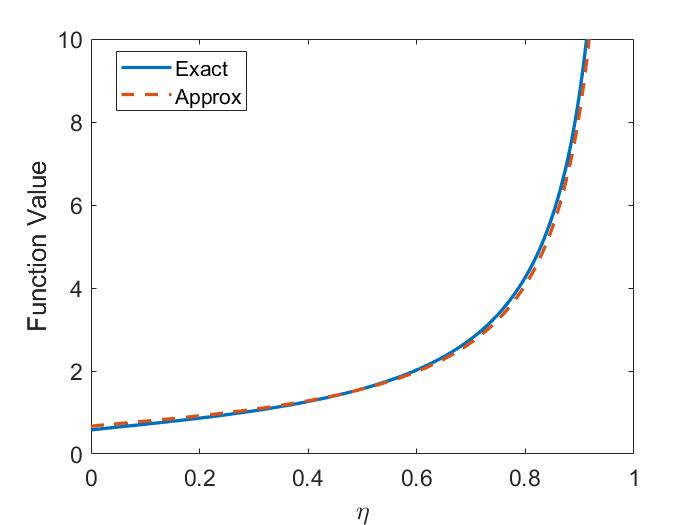}} 
\subfloat[]{\includegraphics[width = \halfwidth]{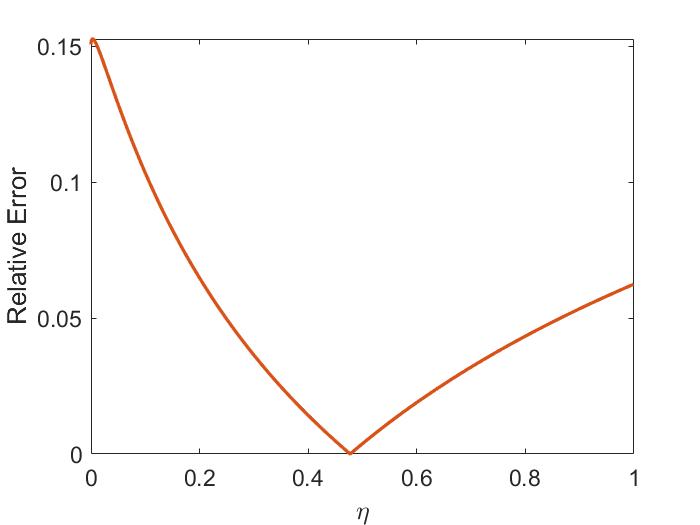}}
\caption
[Comparison of approximation for $g_1(\bres)/g_2(\bres)$ in \eqref{eq:lan:CAExcrit} to the exact function.]
{Comparison of approximation for $g_1(\bres)/g_2(\bres)$ in \eqref{eq:lan:CAExcrit} to the exact function (identical to \eqref{eq:lan:appg1}). Left: blue curve shows exact function, dashed orange curve shows approximation. Right: relative error of the approximation.}
\label{fig:lan:appg1}
\end{figure}

The accuracy of these approximations is shown in \figref{fig:lan:appg0} and \figref{fig:lan:appg1}.
While $g_0(\bres)/g_2(\bres)$ was approximated using the procedure described in \appref{app:cyc:approx} by matching end behavior of the exact function, a different method was needed for $g_1(\bres)/g_2(\bres)$. Since our goal is to invert the expression in \eqref{eq:smallform} to isolate $\bres$ in terms of $\xinj$, the form of $g_1(\bres)/g_2(\bres)$ is chosen as $c g_2(\bres)/ g_0(\bres)$ in order to allow this. The parameter $c$ is then determined through nonlinear least squares fitting to minimize the error between the approximation for $g_1(\bres)/g_2(\bres)$ and its exact form. 

The solution of \eqref{eq:lan:bdampcrit} is a quadratic formula for $\dx^2$, given by 


\begin{subequations}
\begin{align}
\dx^2 &= \frac{-B - \sqrt{B^2 - 4AC}}{2A} \\ 
\text{where } A &= -g_1(\bres) - (\bres^{-1} - 1)^2/2 \\ 
B &= g_0(\bres) - (1-\bres)^2 g_1(\bres) \\ 
C &= (1-\bres)^2 g_0(\bres) 
\end{align}
\end{subequations}

In \eqref{eq:lan:slowdxapprox}, this solution is approximated by

\begin{align}
\dx^2 \approx \left(1 - \bres^{4/5}\right)^{5/2}
\label{eq:lan:slowdxapproxAPP}
\end{align}

\begin{figure}[H]
\subfloat[]{\includegraphics[width = \halfwidth]{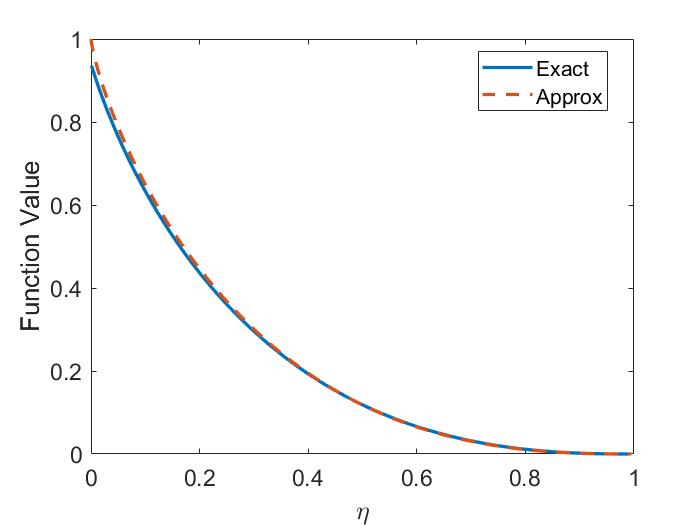}} 
\subfloat[]{\includegraphics[width = \halfwidth]{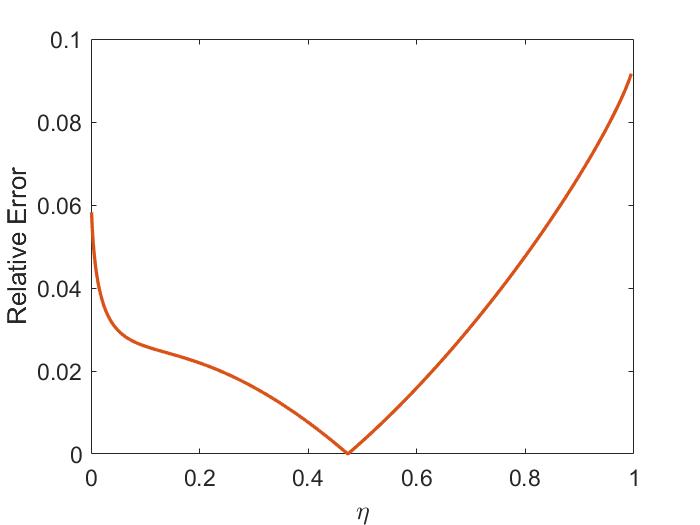}}
\caption
[Comparison of approximation in \eqref{eq:lan:slowdxapprox} to the exact function.]
{Comparison of approximation in \eqref{eq:lan:slowdxapprox} to the exact function (identical to \eqref{eq:lan:slowdxapproxAPP}). Left: blue curve shows exact function, dashed orange curve shows approximation. Right: relative error of the approximation.}
\label{fig:lan:appdxslow}
\end{figure}

This approximation was arrived at by nonlinear least squares optimization on the function $\dx^2 = \left(1 - \bres^{c_1}\right)^{c_2}$. The accuracy of this approximation is shown in \figref{fig:lan:appdxslow}

\section{Details of Approximations for Large FLR Regime \texorpdfstring{$(\zp \gg 1)$}{}} 
\label{app:lan:dampreflarge}

Details from the calculations in \secref{sec:lan:fast} are listed here for reference. The full form of \eqref{eq:lan:dampfastx0notail} is


\begin{align}
\xinj &= \frac{h_0(\bres) - \dx^2 h_1(\bres)}{h_2(\bres)} \\ 
h_0(\bres) &= \left[8 + 9\bres - 2\bres^2)\sqrt{\bres^{-1} - 1} - 15\arccos\sqrt{\bres}\right]/4 \\ 
h_1(\bres) &= \frac{3}{4}\int_0^{1-\bres}\frac{\sqrt{x}}{(1-x)^{3/2}}\frac{dx}{1 + \left(\frac{1-x}{4\bres}\right)^{3/2}} \\
h_2(\bres) &= (2 + \bres)\sqrt{\bres^{-1}-1}-3\arccos\sqrt{\bres}
\end{align}

\begin{figure}[H]
\subfloat[]{\includegraphics[width = \halfwidth]{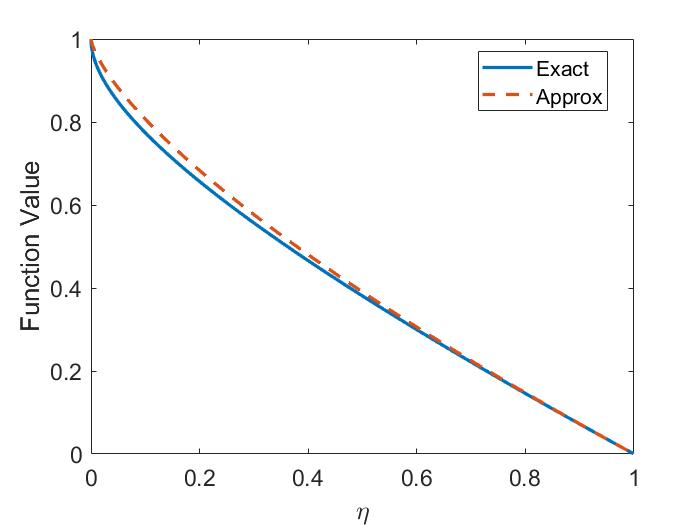}} 
\subfloat[]{\includegraphics[width = \halfwidth]{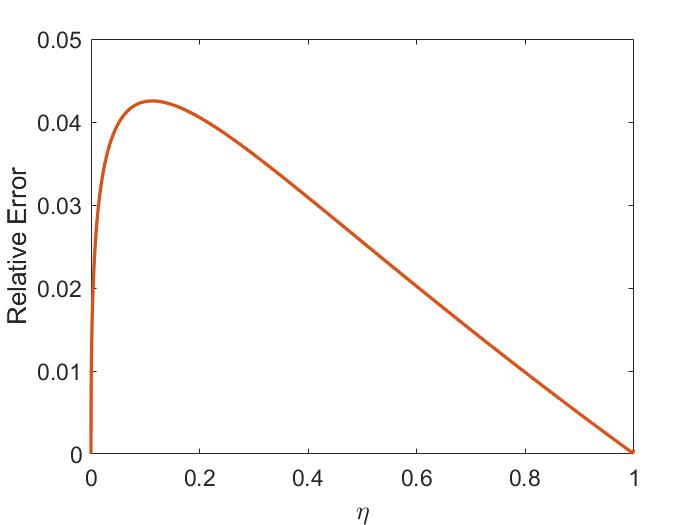}}
\caption
[Comparison of approximation for $h_0(\bres)/h_2(\bres)$ in \eqref{eq:lan:dampfastx0notail} to the exact function.]
{Comparison of approximation for $h_0(\bres)/h_2(\bres)$ in \eqref{eq:lan:dampfastx0notail} to the exact function (identical to \eqref{eq:lan:h0app}). Left: blue curve shows exact function, dashed orange curve shows approximation. Right: relative error of the approximation.}
\label{fig:lan:apph0}
\end{figure}

\begin{figure}[H]
\subfloat[]{\includegraphics[width = \halfwidth]{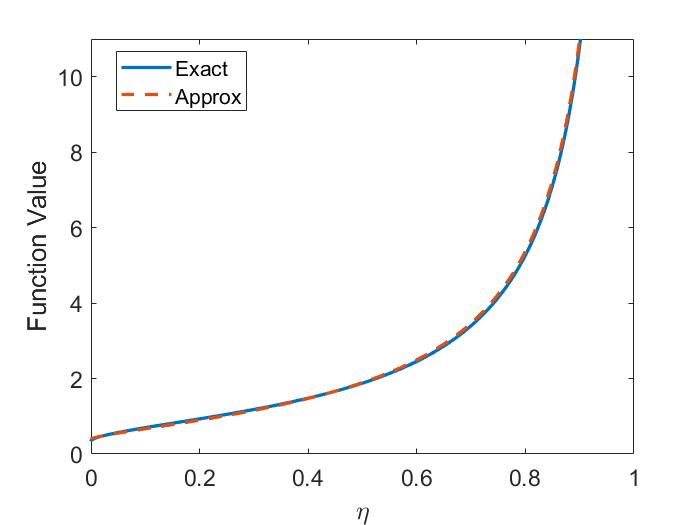}} 
\subfloat[]{\includegraphics[width = \halfwidth]{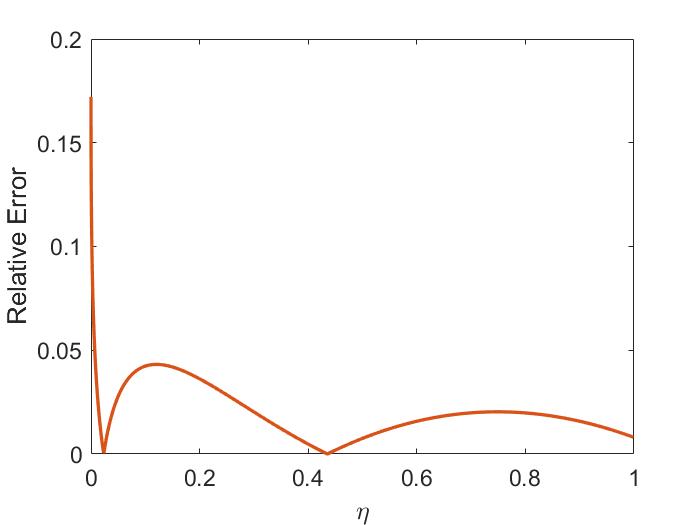}}
\caption
[Comparison of approximation for $h_1(\bres)/h_2(\bres)$ in \eqref{eq:lan:dampfastx0notail} to the exact function.]
{Comparison of approximation for $h_1(\bres)/h_2(\bres)$ in \eqref{eq:lan:dampfastx0notail} to the exact function (identical to \eqref{eq:lan:h1app}). Left: blue curve shows exact function, dashed orange curve shows approximation. Right: relative error of the approximation.}
\label{fig:lan:apph1}
\end{figure}

The integral for $h_1(\bres)$ given above can be evaluated by \Mathematica by making the substitution $u = (1 - x)/4\bres$ 
, but the result is horrendously long and not useful to print here. Importantly, a mistake was made in Eq. 23 of \citeref{Lestz2020p1} -- the expression printed there for $h_1(\bres)$ is the integral only of $-(3/4)\int_0^{1-\bres} \sqrt{x} / ( 1 - x)^{3/2} dx$, which is missing a complicated term in the integrand (the sign change is not an error, but rather a change in convention for consistency). Correcting this error leads to a slight quantitative change in results which can be incorporated by updating the approximations used here as well for the remainder of \secref{sec:lan:fast}. 
The following approximations are made to substantially simplify the above, and the accuracy of these approximations is shown in \figref{fig:lan:apph0} and \figref{fig:lan:apph1}. 

\begin{align}
\label{eq:lan:h0app}
\frac{h_0(\bres)}{h_2(\bres)} &\approx 1 - \bres^{5/7} \\ 
\label{eq:lan:h1app}
\frac{h_1(\bres)}{h_2(\bres)} &\approx -\frac{2\left(1-\bres^{5/7}\right)}{5} +\frac{4}{5\left(1 - \bres^{5/7}\right)}
\end{align}

The solution of \eqref{eq:lan:bdampcritlarge} is a quadratic formula for $\dx^2$, given by 

\begin{subequations}
\begin{align}
\dx^2 &= \frac{-B - \sqrt{B^2 - 4AC}}{2A} \\ 
\text{where } A &= -h_1(\bres) - \sqrt{\bres^{-1} - 1}/2 \\ 
B &= h_0(\bres) - (1-\bres)^2 h_1(\bres) \\ 
C &= (1-\bres)^2 h_0(\bres)
\end{align}
\end{subequations}

In \eqref{eq:lan:fastdxapprox}, this solution is approximated by

\begin{align}
\dx^2 \approx \frac{7}{5}\left(1 - \bres^{5/7}\right)^{5/2}
\label{eq:lan:fastdxapproxAPP}
\end{align}

\begin{figure}[H]
\subfloat[]{\includegraphics[width = \halfwidth]{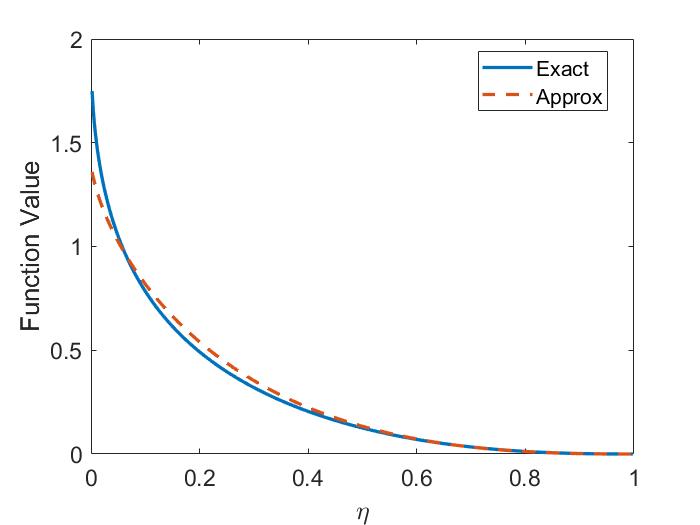}} 
\subfloat[]{\includegraphics[width = \halfwidth]{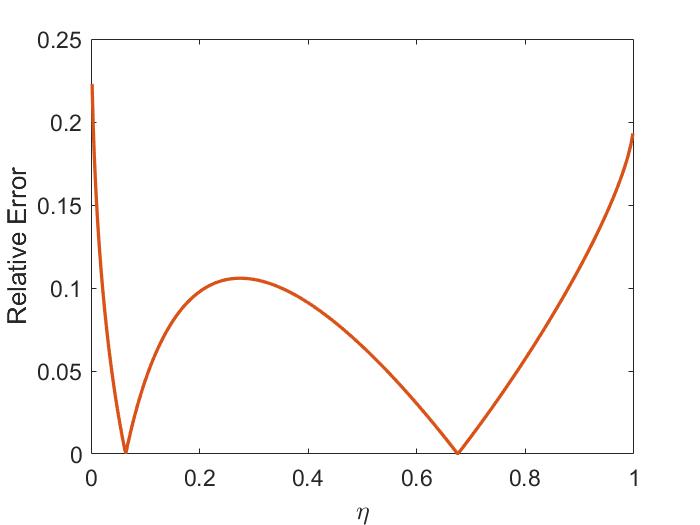}}
\caption
[Comparison of approximation in \eqref{eq:lan:fastdxapprox} to the exact function.]
{Comparison of approximation in \eqref{eq:lan:fastdxapprox} to the exact function (identical to \eqref{eq:lan:fastdxapproxAPP}). Left: blue curve shows exact function, dashed orange curve shows approximation. Right: relative error of the approximation.}
\label{fig:lan:appdxfast}
\end{figure}

This approximation was arrived at by nonlinear least squares optimization on the function $\dx^2 = c_1 \left(1 - \bres^{c_2}\right)^{c_3}$. The accuracy of this approximation is shown in \figref{fig:lan:appdxfast}. Due to the complexity of the function $h_1(\bres)$, this approximation was less accurate than others made in this thesis. 

\end{subappendices}

\newcommand{\dirfigsim}{ch-simulations/figs}

\chapter{Hybrid Simulations of Sub-Cyclotron \Alfven Eigenmode Stability}
\label{ch:sim:simulations}

\section{Introduction}
\label{sec:sim:Intro}

The purpose of this study is to investigate how the linear stability properties of CAEs and GAEs depend on key fast ion properties using hybrid kinetic-MHD simulations of realistic NSTX conditions. These instabilities have been modeled by Belova \etal for specific discharges in NSTX,\cite{Belova2015PRL,Belova2017POP} NSTX-U,\cite{Belova2019POP} and DIII-D,\cite{Belova2020IAEA} but there is still much to explore regarding the details of their excitation. 

The simulation model is described in \secref{sec:sim:model}. In \secref{sec:sim:id}, an overview of the basic properties of the three different types of mode studied -- co-CAEs, cntr-GAEs, and co-GAEs -- is given. \secref{sec:sim:stabres} contains the bulk of the stability results. The most heavily investigated parameters were the beam injection geometry, characterized by the parameter $\linj = \mu B_0/\W$, and the normalized beam injection velocity $\vinj$. The results of a comprehensive parameter scan of these quantities is presented in \secref{sec:sim:simres}. 

The analytic theory derived earlier is briefly reviewed in \secref{sec:sim:theory} and then used to interpret and explain the simulation results. Predictions from the analytic calculations are compared against the dependence of the growth rate of the most unstable mode on $\linj$ and $\vinj$. Previously derived approximate stability boundaries are found to be in reasonable agreement with the simulation results. Key differences in the properties of the different types of unstable modes are also well-described by this theory. 

In \secref{sec:sim:vcdl}, a brief examination of the dependence of the growth rate on two additional beam parameters -- the normalized critical velocity $\vcrit$ and velocity space anisotropy, as characterized by $\dl$ -- is presented. In \secref{sec:sim:pphi}, the effect of gradients in the fast ion distribution with respect to $\pphi$ is considered and found to resolve some discrepancies. Lastly, \secref{sec:sim:damp} addresses the level of CAE/GAE damping on the background plasma, both as inferred from simulations, and also calculated for the electron damping not included in the simulation model. A discussion of the main results is given in \secref{sec:sim:summary}. The majority of the content of this chapter is currently being prepared for submission to Physics of Plasmas.\cite{Lestz2020sim}

\section{Simulation Scheme}
\label{sec:sim:model}

The simulations in this work are conducted using the hybrid MHD-kinetic initial value code \HYM in toroidal geometry.\cite{Belova1997JCP,Belova2017POP} In this code, the thermal electrons and ions are modeled as a single resistive, viscous MHD fluid. The minority energetic beam ions $(\nbo \ll \neo)$ are treated kinetically with a full-orbit $\delta f$ scheme. When studying high frequency modes with $\omega \lesssim \omegaci$, resolving the fast ion gyromotion is crucial to capturing the general Doppler-shifted cyclotron resonance that drives the modes (see \eqref{eq:sim:reskpar}). The two species are coupled together using current coupling in the momentum equation below

\begin{align}
\rho \frac{d \vV}{d t} = -\nabla P + (\vJ - \vJ_b) \cross \vB - e n_b (\vE - \eta \delta\vJ) + \upmu \Delta \vV
\label{eq:sim:momentum}
\end{align}

Here, $\rho, \vec{V}, P$ are the thermal plasma mass density, fluid velocity, and pressure, respectively. The magnetic field can be decomposed into an equilibrium and perturbed part $\vB = \vB_0 + \delta\vB$, while the electric field $\vE$ has no equilibrium component. The beam density and current are $n_b$ and $\vec{J_b}$. The total plasma current is determined by $\mu_0\vec{J} = \curl \vec{B}$ while $\mu_0\delta\vec{J} = \curl \delta\vec{B}$ is the perturbed current. Non-ideal MHD physics are introduced through the viscosity coefficient $\upmu$ and resistivity $\eta$. \eqref{eq:sim:momentum} results from summing over thermal ion and electron momentum equations, taking advantage of $m_e \ll m_i$, and enforcing quasineutrality $n_e = n_b + n_i$. In addition to \eqref{eq:sim:momentum}, the thermal plasma evolves according to the following set of fluid equations 

\begin{subequations}
\begin{align}
\label{eq:sim:fluids1}
\vec{E} &= - \vec{V} \cross \vec{B} + \eta \delta\vec{J} \\
\label{eq:sim:fluids2}
\frac{\partial \vec{B}}{\partial t} &= - \curl \vec{E} \\ 
\label{eq:sim:fluids3}
\frac{\partial \rho}{\partial t} &= - \div \left( \rho \vec{V} \right) \\ 
\label{eq:sim:fluids4}
\frac{d}{dt}&\left( \frac{P}{\rho^\gamma} \right) = 0 
\end{align}
\label{eq:sim:fluids}
\end{subequations}

In fully nonlinear simulations, the pressure equation includes Ohmic and viscous heating in order to conserve the system's energy (see Eq. 2 of \citeref{Belova2017POP}). 
These effects are neglected in the linearized simulations presented here, reducing to the adiabatic equation of state in \eqref{eq:sim:fluids4} with the adiabatic index $\gamma = 5/3$. Note that the terms $\vE - \eta\delta\vJ$ appear in \eqref{eq:sim:momentum}, \eqref{eq:sim:fluids1}, and \eqref{eq:sim:parts2} due to collisional drag between the thermal plasma and beam ions. 

Field quantities are evolved on a cyclindrical grid, with particle quantities stored on a Cartesian grid sharing the $Z$ axis with the fluid grid. For simulations of $n < 8$, a field grid of $N_Z\times N_R\times N_\phi = 120 \times 120 \times 64$ is used. For larger $n$, the resolution is refined to $120 \times 96 \times 128$. The particle grid has $N_Z \times N_X \times N_Y = 120 \times 51 \times 51$, with at least $500,000$ particles used in each simulation. Convergence studies were conducted on the grid resolution and number of simulation particles, which can lead to slight variations in growth rate but no change in frequency or mode structure. 

The fast ion distribution is decomposed into an equilibrium and perturbed part, $\fbeam = \fb + \df$. Each numerical particle has a weight $w = \df / \fl$ where $\fl$ is a function of integrals of motion used for particle loading $(d\fl/dt = 0)$. The $\df$ particles representing the fast ions evolve according to the following equations of motion

\begin{subequations}
\begin{align}
\label{eq:sim:parts1}
\frac{d\vec{x}}{dt} &= \vec{v} \\
\label{eq:sim:parts2}
\frac{d\vec{v}}{dt} &= \frac{q_i}{m_i}\left(\vec{E} - \eta \delta\vec{J} + \vec{v} \cross \vec{B}\right) \\ 
\label{eq:sim:dwdt}
\frac{dw}{dt} &= -\left(\frac{\fbeam}{\fl} - w\right)\frac{d \ln \fb}{dt}
\end{align}
\label{eq:sim:parts}
\end{subequations}

Particle weights are used to calculate the perturbed beam density $\delta n_b$ and current $\delta \vJ_b$ which appear in \eqref{eq:sim:momentum}. The $\delta f$ scheme has two advantages: the reduction of numerical noise and intrinsic identification of resonant particles, which are those with largest weights at the end of the simulation. 

The equilibrium fast ion distribution function is written as a function of the constants of motion $\W$, $\lambda$, and $\pphi$. The first, $\W = \frac{1}{2}m_i v^2$, is the particle's kinetic energy. Next, $\lambda = \mu B_0 / \W$ is a trapping parameter, where first order corrections in $\rho_{EP}/L_B$ to the magnetic moment $\mu$ are kept for improved conservation in the simulations.\cite{Belova2003POP} This correction is more relevant in low aspect ratio devices since the fast ion Larmor radius can be a significant fraction of the minor radius, leading to nontrivial variation in $B_0$ during a gyro-orbit. To lowest order in $\mu \approx \mu_0$, one may re-write $\lambda \approx (\vperp^2/v^2)(\omegacio/\omegaci)$ such that in a tokamak, passing particles will have $0 < \lambda < 1 - r/R$ and trapped particles will have $1 - r/R < \lambda < 1 + r/R$. Hence, $\lambda$ is a complementary variable to a particle's pitch $\vpar/v$. Lastly, $\pphi = -q_i \psi + m_i R v_\phi$ is the canonical toroidal angular momentum, conserved due to the axisymmetric equilibria used in these simulations. Here, $\psi$ is the poloidal magnetic flux and $\psi_0$ is its on-axis value. A separable form of the beam distribution is assumed:\cite{Belova2003POP} $\fb (v,\lambda,\pphi) = C_f f_1 (v) f_2 (\lambda) f_3 (\pphi,v)$

\begin{subequations}
\begin{align}
\label{eq:sim:F1}
f_1(v) &= \frac{1}{v^3 + v_c^3} \quad \text{ for } v < \vb \\
\label{eq:sim:F2}
f_2(\lambda) &= \exp\left(-\left(\lambda - \linj\right)^2 / \dl^2\right) \\ 
\label{eq:sim:F3}
f_3\left(\pphi,v\right) &= \left(\frac{\pphi - \pmin}{m_i R_0 v - q_i \psi_0 - \pmin}\right)^\pphipow \text{ for } \pphi > \pmin
\end{align}
\label{eq:sim:F0}
\end{subequations}

The energy dependence, $f_1(v)$, is a slowing down function with injection velocity $\vb$ and critical velocity\cite{Gaffey1976JPP} $\vc$ (defined later in \eqref{eq:vcrit}). For $v > \vb$, $f_1(v) = \exp(-(v-\vb)^2/\dv^2)/(v^3 + \vc^3)$ is used to model a smooth, rapid decay near the injection energy with $\dv = 0.1 \vb$. A beam-like distribution in $\lambda$ is used for $f_2(\lambda)$, centered around $\linj$ with constant width $\dl$. Characteristic profiles of beam density calculated by the global transport code \TRANSP\cite{Goldston1982JCP} and Monte Carlo fast ion module \NUBEAM\cite{Pankin2004CPC} motivate the ad-hoc form of $f_3(\pphi,v)$. A prompt-loss boundary condition at the last closed flux surface is imposed by requiring $\pphi > \pmin = -0.1\psi_0$. Lastly, $C_f$ is a normalization constant determined numerically to match the peak value of $\nb$ to its desired value. Values for the parameters appearing in the distribution are set in order to match the distribution in \NUBEAM from the target discharge. In this case, this yields $\vc = \vb/2$, $\vinj = 4.9$, $\linj = 0.7$, $\dl = 0.3$, $\pphipow = 6$, and $n_b/n_e = 5.3\%$. The distribution function in \eqref{eq:sim:F0} is a qualitative fit that is sufficiently realistic in order to extract growth rate dependencies on the fast ion parameters. 

Importantly, \HYM includes the energetic particles self-consistently when solving for the equilibrium. Inclusion of the fast ions results in a modified Grad-Shafranov equation, which leads to pressure anisotropy, an increased Shafranov shift, and more peaked current profiles.\cite{Belova2003POP} Although the beam density is small, $\nbo \ll \neo$, the current carried by the beam can nevertheless be comparable to the thermal plasma current due to the large fast ion energy. 

This study is concerned with linear stability. Nonlinear simulations of CAEs\cite{Belova2017POP} and GAEs\cite{Belova2019POP} have also been conducted with the \HYM code. Since the modes in this simulation do not saturate, the most linearly unstable mode will dominate and obscure all other modes. In order to study all of the eigenmodes, each toroidal mode number $n$ is simulated separately by eliminating all but one toroidal harmonic systematically throughout the simulation.

In this study, approximately 600 simulations are performed to scan over the normalized injection velocity $\vinj = 2 - 6$ and injection geometry $\linj = 0.1 - 0.9$ of the beam ion distribution. The operating range for $\vinj$ in NSTX is approximately the same as the scanned range, while $\linj$ is restricted to approximately $0.5-0.7$ in experiment. The new, off-axis NSTX-U beam sources\cite{Gerhardt2012NF} have much more tangential injection, with $\linj \approx 0$. 

\begin{figure}[tb]
{\includegraphics[width = \midwidth]{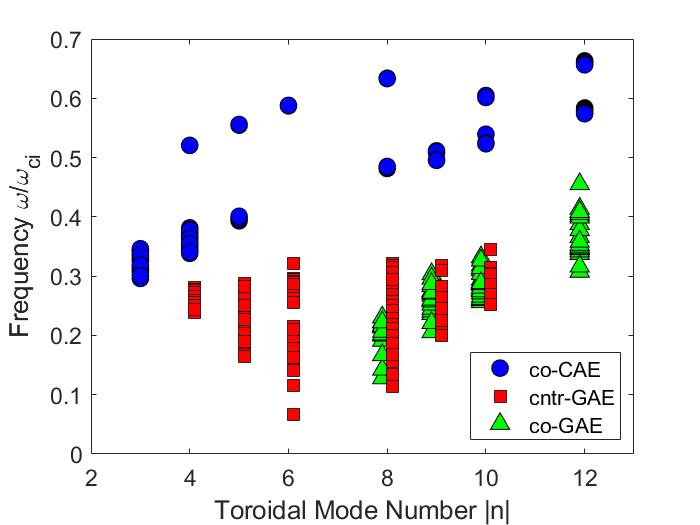}}
\caption
[Frequency of each type of mode as a function of toroidal mode number in simulations.]
{Frequency of each type of mode as a function of toroidal mode number in simulations. Simulations shown use beam distributions with $\linj = 0.1 - 0.9$ and $\vinj = 2.5 - 6.0$.}
\label{fig:sim:frequency_vs_n}
\end{figure}

\begin{figure}[tb]
\includegraphics[width = \midwidth]{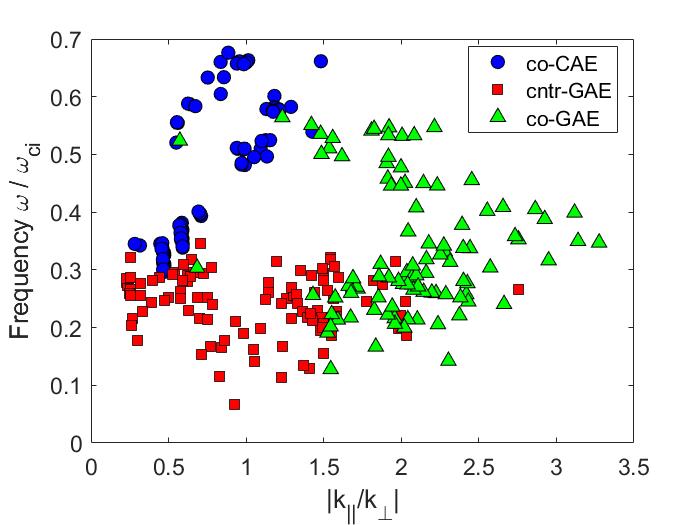}
\caption
[Frequency and wave vector directions calculated from unstable modes in simulations.]
{Frequency and wave vector directions calculated from unstable modes in simulations. Simulations shown use beam distributions with $\linj = 0.1 - 0.9$ and $\vinj = 2.5 - 6.0$.}
\label{fig:sim:ao_prop}
\end{figure}

The bulk plasma equilibrium properties in these simulations are based on the well-diagnosed NSTX H-mode discharge 141398, which had $n_e = 6.7\ten{19}$ m$^{-3}$ and $B_0 = 0.325$ T on axis, with 6 MW of 90 keV beams corresponding to $\vinj = 4.9$, centered on $\linj \approx 0.7$. The on-axis ion cyclotron frequency was $f_{ci} = 2.5$ MHz. This plasma was chosen as the nominal scenario to study due to its rich spectrum of high frequency modes and pre-existing experimental analysis.\cite{Crocker2013NF,Fredrickson2013POP}

A perfectly conducting boundary condition ($\dbperp = \depar = 0$, $\dbpar,\deperp$ allowed to be finite) is imposed at the bounding surfaces of the simulation volume. Projected onto the poloidal plane, this boundary has a box geometry. Since the shape of this boundary is not identical to the irregular shape of the NSTX(-U) vessel, the location of the boundary condition imposed in the simulations is different than it is in experiments. Previous simulations have shown that a smaller distance between the last closed flux surface and the bounding box can decrease the growth rate, so this discrepancy could lead to quantitative differences. However, this is not a concern for the goals of this study since it should not affect the trends in the growth rate with fast ion parameters. 

\section{Identification of Modes and Structures}
\label{sec:sim:id} 

Three distinct types of modes are found in this simulation study: co-propagating CAEs, counter-propagating GAEs, and co-propagating GAEs. \figref{fig:sim:frequency_vs_n} summarizes the frequency range of each of these modes for the simulated toroidal mode numbers, while \figref{fig:sim:ao_prop} shows the ranges of $\omeganorm$ and $\alpha \defined \krat$ of each mode. Note that while $\kpar$ is often inferred from the large tokamak expression $\kpar = (n - m / q) / R$, this relation was not applied here due to the low aspect ratio of NSTX and ambiguous poloidal mode numbers in simulations. Instead, $\kpar$ was determined by taking a Fourier transform along equilibrium field lines traced on the flux surface with largest RMS fluctuation magnitude in $\dbpar$ for CAEs and a component of $\dbperp$ for GAEs. Meanwhile, peaks in the spatial Fourier transforms in $Z$, $R$, and $\phi$ give $k = \sqrt{k_R^2 + k_Z^2 + k_\phi^2}$, which can be used to infer $\kperp = \sqrt{k^2 - \kpar^2}$. This section will detail the characteristics of each simulated mode and contrast their properties. Distinguishing between CAEs and GAEs is notoriously difficult\cite{Crocker2013NF} in experiments due to similar frequency ranges and mixed polarization of the modes at the outboard side where magnetic coils are available for measurements. Hence, the wealth of information provided by simulations is valuable in guiding future experimental analysis. 

Fast ions can interact with the modes through the general Doppler shifted cyclotron resonance condition: 

\begin{align}
\omega - \avg{\kpar\vpar} - \avg{\kperp\vdrift} = \lres\avg{\omegaci}
\label{eq:sim:reskpar}
\end{align} 

Here, $\avg{\dots}$ denotes orbit-averaging. The cyclotron resonance coefficient $\lres = -1,0,1$ for the frequency range studied here, though a resonance may exist for any integer value. The Landau resonance corresponds to $\lres = 0$, the ``ordinary'' cyclotron resonance has $\lres = 1$, and the ``anomalous'' cyclotron resonance has $\lres = -1$. Note that for sub-cyclotron frequencies, and in the usual case where $\abs{\avg{\kpar\vpar}} \gtrsim \abs{\avg{\kperp\vdrift}}$, counter-propagating modes $(\kpar < 0)$ can only satisfy the ordinary cyclotron resonance, while co-propagating modes can interact through the Landau or anomalous cyclotron resonances, depending on their frequency. \eqref{eq:sim:reskpar} can be equivalently written in terms of orbital frequencies: 

\begin{align}
\omega - n\avg{\omegator} - p\avg{\omegapol} = \lres\avg{\omegaci}
\label{eq:sim:resomega}
\end{align}

In this expression, $\omegator$ and $\omegapol$ are the toroidal and poloidal orbital frequencies, respectively, and $p$ is an integer. During simulations, $\avg{\omegator}$, $\avg{\omegapol}$, and $\avg{\omegaci}$ are numerically computed for each fast ion, which enables determination of $p$ for the resonant particles, and hence identification of the dominant resonance in each simulation. 

\subsection{Co-Propagating CAEs} 
\label{sec:sim:idcae}

Our simulations find co-propagating CAEs for $n \geq 3$ with $0.28\omegaci < \omega < 0.70\omegaci$, marked by blue circles on \figref{fig:sim:frequency_vs_n}. For each toroidal mode number, the unstable CAE frequencies are larger than those of GAEs. Moreover, \figref{fig:sim:ao_prop} shows that the CAEs have $\alpha\defined\abs{\krat} = 0.3 - 1.5$, with larger $\krat$ corresponding to the modes with higher $n$ numbers. As will also be the case with the GAEs, it is important to note that $\krat$ can range from small to order unity, violating the $\kpar\ll\kperp$ assumption that is often made in previous theoretical work for large aspect ratio tokamaks. This motivated in part the reexamination of the instability conditions for CAEs and GAEs in \chapref{ch:cyc:analytics-cyclotron} and \chapref{ch:lan:analytics-landau}, which will be used to interpret the simulation results presented here. 

\begin{figure}[tb]
\includegraphics[width = \midwidth]{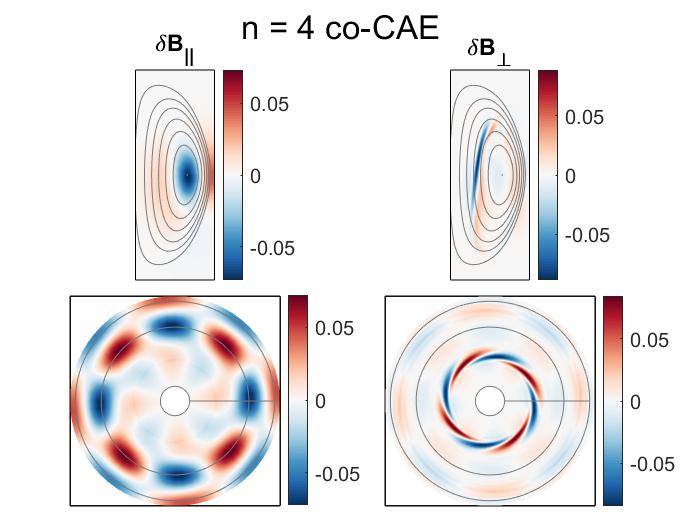}
\caption
[Mode structure of an $n = 4$ co-CAE.]
{Mode structure of an $n = 4$ co-CAE driven by a fast ion distribution parameterized by $\vinj = 5.0$, $\linj = 0.7$. Top row is a poloidal cut, where gray lines are $\psi$ contours. Bottom row is a toroidal cut at the midplane, where gray lines represent the high field side, low field side, and magnetic axis. The first column is $\dbpar$, while the second is one of the orthogonal components of $\dbperp$.}
\label{fig:sim:CAEn4l0.7v5.0}
\end{figure}

This group of modes was heuristically identified as CAEs since they have magnetic fluctuations dominated by the compressional component near the core, where $\dbpar/\dbperp \gg 1$. In the simulations, $\dbpar$ for the low to moderate $n$ modes ($n < 10$) usually peaks on axis with low poloidal mode number $(m = 0 - 2)$. A typical example is shown in \figref{fig:sim:CAEn4l0.7v5.0}, which shows an $n = 4$ co-propagating CAE with beam distribution parameterized by $\vinj = 5.0$ and $\linj = 0.7$ -- the parameters most closely matching the conditions of the NSTX discharge which this set of simulations are modeling. The top left figure shows a poloidal cross section of $\dbpar$ taken at a toroidal angle where $\dbpar$ is near its maximum. The core-localization of these modes agrees with previous nonlinear HYM simulations,\cite{Belova2017POP} contrasting with the analytic studies of CAEs under large aspect ratio assumptions, which predict localization near the edge.\cite{Smith2003POP} 

These modes also exhibit a substantial fluctuation in $\dbpar$ on the low field side beyond the last closed flux surface, which is also a generic feature of counter-propagating GAE (see \figref{fig:sim:GAEMn6l0.7v5.0}). This has complicated previous efforts to delineate between high frequency AEs in experiment.\cite{Crocker2013NF} In fact, the high frequency AEs observed in NSTX were initially identified as CAEs before further analysis revealed that GAEs were also present. While $\dbperp$ is very small in the core for linear simulations dominated by a single CAE, there can still be large $\dbperp$ closer to the edge. This structure is most prominent on the high field side, as shown in \figref{fig:sim:CAEn4l0.7v5.0}, but is also visible to a lesser degree on the low field side. The feature has been previously identified as a kinetic \Alfven wave (KAW),\cite{Belova2015PRL,Belova2017POP} located at the \Alfven resonance location $\omega = \kpar \va(R,Z)$ where CAEs undergo mode conversion. The KAW appears whenever a CAE is unstable in these simulations. 

\begin{figure}[tb]
\includegraphics[width = \midwidth]{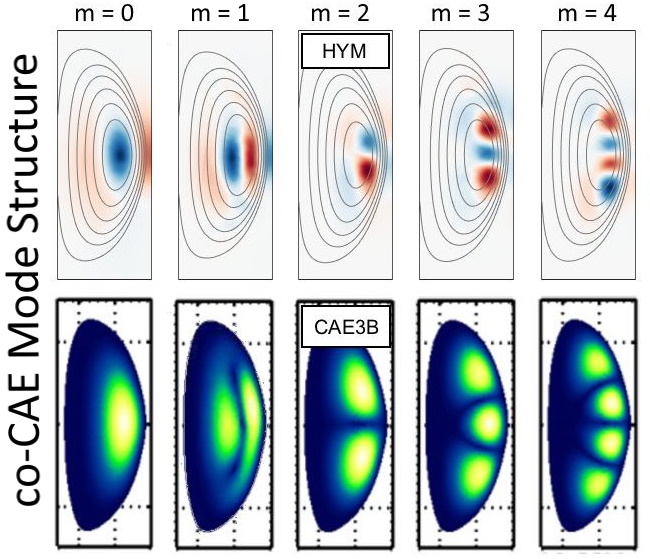} 
\caption
[Comparison of CAE mode structures between the initial value code \HYM and the spectral code \CAETB.]
{(a) Representative examples of poloidal mode structure ($\dbpar$) of co-CAEs from HYM simulations, labeled with qualitative poloidal mode numbers. From left to right, these modes have the following toroidal mode number and were simulated with fast ion populations parameterized by normalized injection velocity $\vinj$ and central pitch $\linj$: (1) $n=4, \linj = 0.7, \vinj = 5.0$, (2) $n = 6, \linj = 0.3, \vinj = 5.5$, (3) $n = 12, \linj = 0.9, \vinj = 5.1$, (4) $n =  10, \linj = 0.7, \vinj = 5.0$, (5) $n = 12, \linj = 0.7, \vinj = 5.1$. (b) Representative examples of poloidal mode structure of CAEs from the spectral code \CAETB, labeled with mode numbers analogous to those in (a). All of these modes are $n = -3$ cntr-CAEs, solved with an equilibrium based on NSTX discharge 130335.}
\label{fig:sim:CAE_HYM_struct}
\end{figure}

\begin{figure}[tb]
\includegraphics[width = \midwidth]{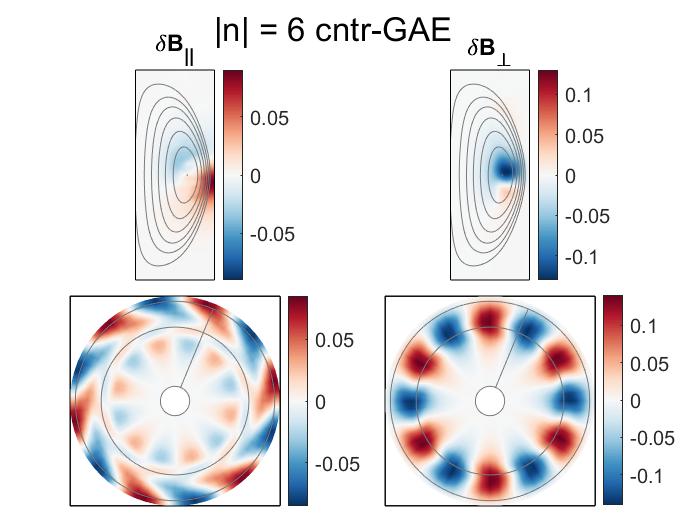}
\caption
[Mode structure of an $n = -6$ cntr-GAE.]
{Mode structure of a $n = -6$ cntr-GAE driven by a fast ion distribution parameterized by $\vinj = 5.0$, $\linj = 0.9$. Top row is a poloidal cut, where gray lines are $\psi$ contours. Bottom row is a toroidal cut at the midplane, where gray lines represent the high field side, low field side, and magnetic axis. The first column is $\dbpar$, while the second is one of the orthogonal components of $\dbperp$.}
\label{fig:sim:GAEMn6l0.7v5.0}
\end{figure}

The modes with higher $n$ are localized closer to the edge on the low field side  and have somewhat higher poloidal mode number ($m = 2 - 4$). The full range of poloidal mode structures of CAEs observed in linear HYM simulations is shown in \figref{fig:sim:CAE_HYM_struct}. The first two modes, with $n = 4$ and $n = 6$, are more concentrated in the core, whereas the last three modes ($n = 12, 10, \text{ and } 12$, respectively) are more localized near the edge. 

The labeling of these modes with poloidal mode numbers is somewhat arbitrary, as the modes do not lend themselves well to description with a single poloidal harmonic $m$ along the $\theta$ direction, as the structure can peak on axis or have structures that are poorly aligned with $\psi$ contours. This dilemma is also discussed in \citeref{Smith2009PPCF}, where the spectral code \CAETB is used to solve for CAEs at low aspect ratio. Moreover, the CAE has a standing wave structure, so multiple signs of $m$ are present, broadening the spectrum of $\kpar$ and $\kperp$ values. 

Regardless of the poloidal mode numbers ascribed to them, the CAE mode structures from the HYM simulations presented here qualitatively match the CAE eigenmodes found by the \CAETB eigensolver for a separate NSTX discharge 130335.\cite{Geiser2016APS} The similarity between the HYM and \CAETB mode structures provides further evidence that the instabilities seen in HYM and heuristically identified as CAE due to large $\dbpar$ in the core are indeed CAE solutions. Comparison against the CAE dispersion relation further supports the classification of these modes. In a uniform plasma, the magnetosonic dispersion including finite $\beta$ may be written 

\begin{equation}
\omega_{CAE}^2 = \frac{k^2 \va^2}{2}\left[1 + u^2 + \sqrt{\left(1 - u^2\right)^2 + \left(2u\kperp/k\right)^2}\right]
\label{eq:sim:caebeta}
\end{equation}

Here, $u\defined \vs/\va = \sqrt{2\gamma\beta}$ where $\gamma = 5/3$ is the adiabatic index and $\beta = 8\pi P / B^2$. The finite $\beta$ corrections are important because the large pressure can make $\vs \like \va$ in the vicinity of the mode due to the beam contribution. In lieu of well-defined poloidal mode numbers, the mode structures from simulations are Fourier transformed to determined $k_R$ and $k_Z$. The toroidal wave number $k_\phi$ is determined via Fourier transform along the field lines. Fair agreement is found between \eqref{eq:sim:caebeta} calculated with the inferred wave vectors and the mode frequencies in the simulations. 

The co-propagating CAEs are driven by the Landau resonance: $\omega - n\avg{\omegator} - p\avg{\omegapol} = 0$, where $\avg{\omegator}$ and $\avg{\omegapol}$ are the orbit-averaged toroidal and poloidal frequencies, respectively, of the resonant fast ions. The toroidal mode number is $n$ (positive for co-propagation, negative for cntr-propagation), and $p$ is an integer. This condition has been verified in previous \HYM simulations\cite{Belova2017POP} by examining the fast ions with largest weights, which reveal the resonant particles due to \eqref{eq:sim:dwdt}. Such examinations generally show that only a small number of resonances (\ie integers $p$) contribute nontrivially to each unstable mode -- a primary resonance $p$ and often two sub-dominant sidebands with $p\pm 1$. 

\begin{figure}[tb]
\includegraphics[width = \midwidth]{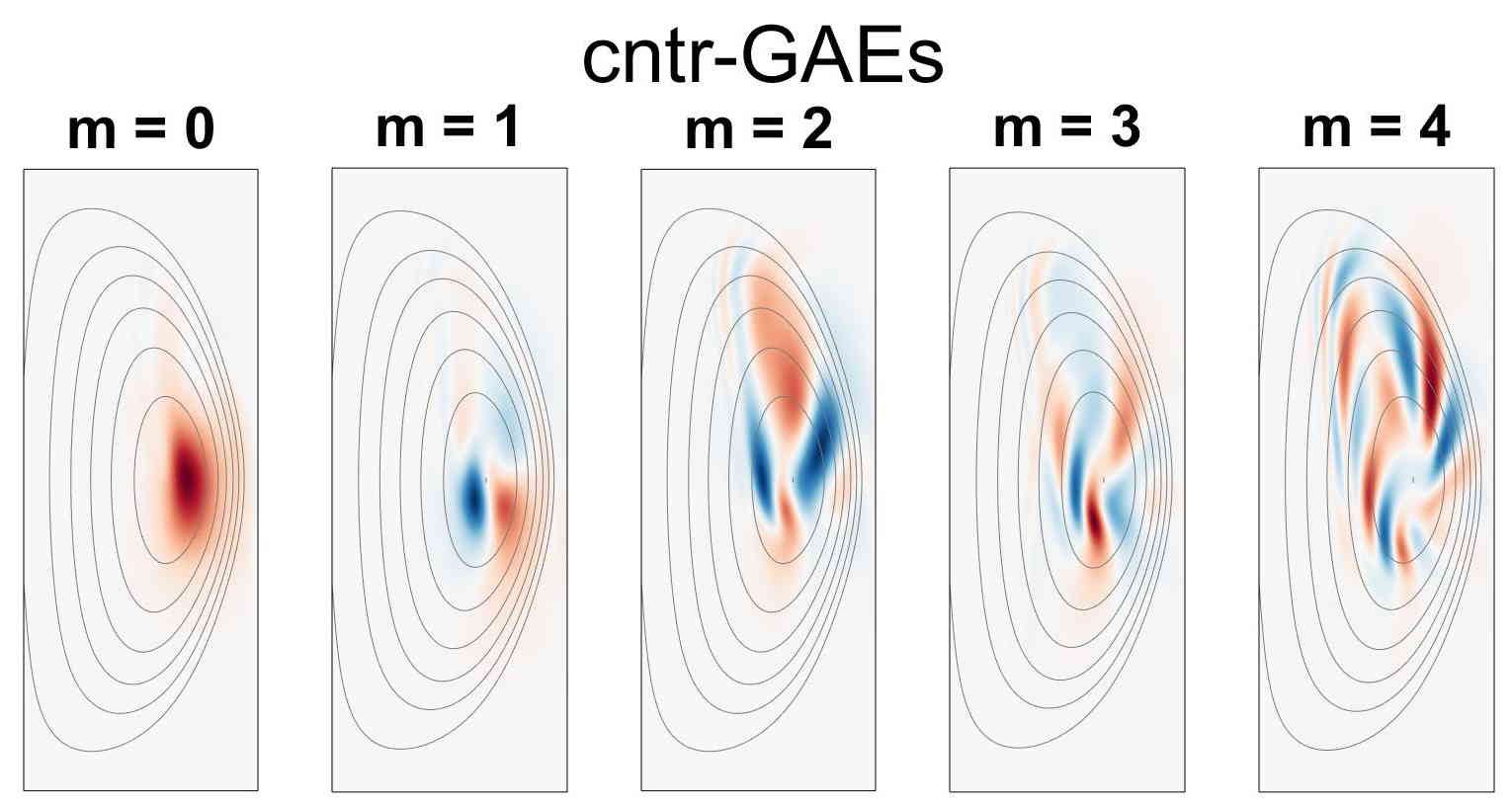}
\caption
[Mode structures of cntr-GAEs with different dominant poloidal harmonics.]
{Example mode structures of GAEs with different dominant poloidal harmonics. From left to right, $m = 0$ ($n = 8, \linj = 0.9, \vinj = 4.5$), $m = 1$, ($n = 6, \linj = 0.7, \vinj = 4.5$), $m  = 2$ ($n = 6, \linj = 0.9, \vinj = 4.5$), $m = 3$ ($n = 5, \linj = 0.7, \vinj = 4.75$), and $m = 4$ ($n = 4, \linj = 0.9, \vinj = 5.0$).}
\label{fig:sim:GAEMcomp}
\end{figure}

\subsection{Counter-Propagating GAEs}
\label{sec:sim:idgaem}

The global \Alfven eigenmodes in these simulations appeared in two flavors: co- and counter-propagating relative to the direction of the plasma current. The counter-propagating modes were excited for $\abs{n} = 4 - 10$ in the frequency range $0.05\omegaci < \omega < 0.35\omegaci$, and are indicated as red squares on \figref{fig:sim:frequency_vs_n}. They exhibit a very wide range of wave vector directions, with unstable modes having $\krat = 0.2 - 3$ in the simulations.  

The cntr-GAEs were distinguished from CAEs due to their dominant shear polarization, \eg $\dbperp \gg \dbpar$ for peak amplitudes, and their dispersion which generally scaled with the shear \Alfven dispersion $\omega \propto \kpar\va$. Counter-GAEs were routinely observed in NSTX\cite{Crocker2013NF} and NSTX-U\cite{Fredrickson2018NF} experiments with these basic characteristics, and previous comparisons between \HYM simulations and experimental measurements have revealed close agreement between the frequency of the most unstable counter-GAE for each $n$.\cite{Belova2019POP} All counter-propagating modes which appeared in these simulations were determined to be GAEs, even though cntr-CAEs have also been observed in NSTX experiments.\cite{Crocker2013NF} This may be attributed to the initial value nature of these simulations, which are dominated by the most unstable mode. As discussed in \secref{sec:cyc:stabao}, cntr-CAEs and cntr-GAEs are driven unstable by the same fast ion parameters, but the growth rate for the cntr-GAEs is typically larger. 

Similar to the CAEs, the counter-GAEs did not feature poloidal structure with well-defined mode numbers. Using an effective mode number $m$ loosely corresponding to the number of full wavelengths in the azimuthal direction, the GAEs ranged from $m = 0 - 5$, with lower $\abs{n}$ modes typically having larger $m = 3-5$ and modes with $\abs{n} > 7$ preferring smaller poloidal mode numbers of $m = 0 - 2$. Some representative examples are shown in \figref{fig:sim:GAEMcomp}. The counter-GAE mode structure is typically more complex than that of the co-CAEs, likely due to the close proximity of the GAEs to the \Alfven continuum, which introduces shorter scale fluctuations on a kinetic scale that modulates the slower varying MHD structure. Comparing the mode structures in \figref{fig:sim:CAE_HYM_struct} and \figref{fig:sim:GAEMcomp}, one can see that while the CAEs are trapped in a potential well on the low field side,\cite{Mahajan1983bPF,Gorelenkova1998POP,Kolesnichenko1998NF,Smith2003POP,Gorelenkov2006NF} the GAEs can access all poloidal angles. 

The most interesting property of the GAEs in these simulations is how significantly the beam distribution influences the frequency of the most unstable mode. This discovery will be described in detail in \chapref{ch:epgae:epgae}, and will be briefly summarized here. It was found that almost uniformly for each toroidal mode number, the frequency of the most unstable GAE was proportional to the normalized injection velocity of the fast ion distribution. 
In most cases, the mode structure of the most unstable mode remains relatively stable despite these large changes in frequency -- yielding small quantitative changes in the mode's width, elongation, or radial location, but not changing mode numbers. This behavior is in contrast to what is seen for the CAEs, where the frequency of the most unstable mode is nearly unchanged for large intervals of $\vinj$, then switching to a new frequency at some critical value, which also coincided with the appearance of a new poloidal harmonic.

Due to their sub-cyclotron frequency and $n < 0$, the cntr-GAEs can only interact with fast ions through the ordinary Doppler-shifted cyclotron resonance. It may be written as $\omega - n\avg{\omegator} - p\avg{\omegapol} = \avg{\omegaci}$ where $\avg{\omegaci}$ is the orbit-averaged cyclotron frequency of resonant particles. 

\subsection{Co-Propagating GAEs} 
\label{sec:sim:idgaep}

In addition to the counter-GAEs, high frequency co-propagating GAEs were found to be unstable in simulations for certain beam parameters, namely small $\linj$ and large $\vinj$, with frequencies $0.15\omegaci < \omega < 0.60$ across $n = 8 - 12$. Almost uniformly these modes have $m = 0$ or 1. Due to the large $n$ values and small $m$, co-GAEs tend to have large $\krat > 1$ in simulations. The co-GAEs may simultaneously resonate with the high energy fast ions through the ``anomalous'' Doppler-shifted cyclotron resonance, $\omega - n\avg{\omegator} - p\avg{\omegapol} = -\avg{\omegaci}$, as well as with fast ions with $\vpar\like\va$ through the Landau resonance $\omega - n\avg{\omegator} - p\avg{\omegapol} = 0$, as shown on \figref{fig:sim:GAEPres}. In the simulations, most of the drive comes from the high energy fast ions, though the absent two-fluid effects may make the Landau resonance of comparable importance in experiments. Just as with the cntr-GAEs, the high frequency co-GAEs behave more like energetic particle modes than MHD eigenmodes, exhibiting large changes in frequency in proportion to changes in $\vinj$, without any significant changes to the mode structure. 

\begin{figure}[tb]
\includegraphics[width = \midwidth]{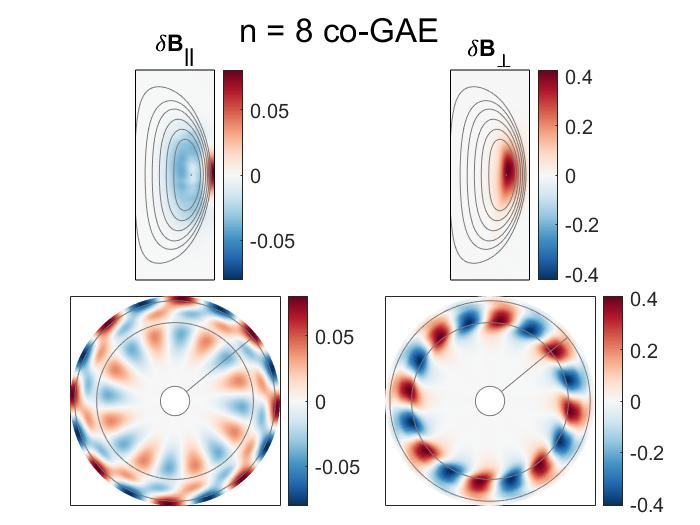}
\caption
[Mode structure of an $n = 8$ co-GAE.]
{Mode structure of an $n = 8$ co-GAE driven by a fast ion distribution parameterized by $\vinj = 5.3$, $\linj = 0.1$. Top row is a poloidal cut, where gray lines are $\psi$ contours. Bottom row is a toroidal cut at the midplane, where gray lines represent the high field side, low field side, and magnetic axis. The first column is $\dbpar$, while the second is one of the orthogonal components of $\dbperp$.}
\label{fig:sim:GAEPn8l0.1v5.3}
\end{figure}

Whereas cntr-GAEs are frequently observed experimentally and have been the subject of recent theoretical studies, co-GAEs are less commonly discussed. The early development of GAE theory did involve co-propagating modes, but these were restricted to very low frequencies and consequently only considered the $\lres = 0$ Landau resonance.\cite{Ross1982PF,Appert1982PP,DeChambrier1982POP,Fu1989PF,VanDam1990FT} This type of low frequency co-GAE driven by the Landau resonance was not found to be the most unstable mode for any set of fast ion parameters in \HYM simulations. As explored in \chapref{ch:lan:analytics-landau}, this may be due to the fact that they have a similar instability condition to the co-CAEs, but with growth rate reduced by a typically small factor $(\omega/\omegaci)^2$. High frequency co-GAEs were never observed in NSTX, nor have they been documented in other devices. As detailed in \secref{sec:sim:theory}, this is at least partly because they are most unstable for very tangential injection with $\linj \rightarrow 0$ (necessary to satisfy the $\lres = -1$ resonance), which is far from the operational constraints of the beam lines available on NSTX. However, the additional neutral beam installed on NSTX-U allows for much more tangential injection,\cite{Gerhardt2012NF} which should be able to excite high frequency co-GAEs for discharges with sufficiently large $\vinj$ (experimentally achievable with a low magnetic field). 

\begin{figure}[tb]
\includegraphics[width = \midwidth]{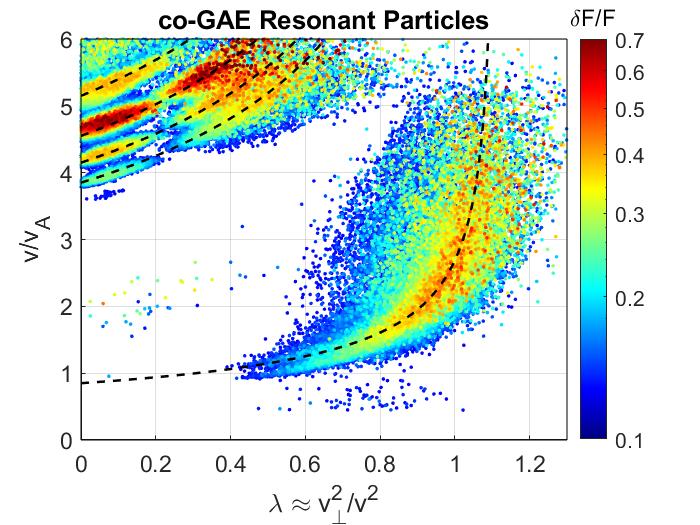}
\caption
[Resonant particles for an $n = 9$ co-GAE.]
{Resonant particles for an $n = 9$ co-GAE driven by beam ions with $\linj = 0.3$ and $\vinj = 5.1$. Dashed lines indicate contours of constant $\vpres = 0.85, 3.85, 4.15, 4.55, 5.15$. The color scale indicates the normalized particle weights $\delta f/\fbeam$ on a log scale.}
\label{fig:sim:GAEPres}
\end{figure}

\section{Stability Results} 
\label{sec:sim:stabres}

\subsection{Simulation Results}
\label{sec:sim:simres}

\begin{figure}[htb]
\includegraphics[width = \fullwidth]{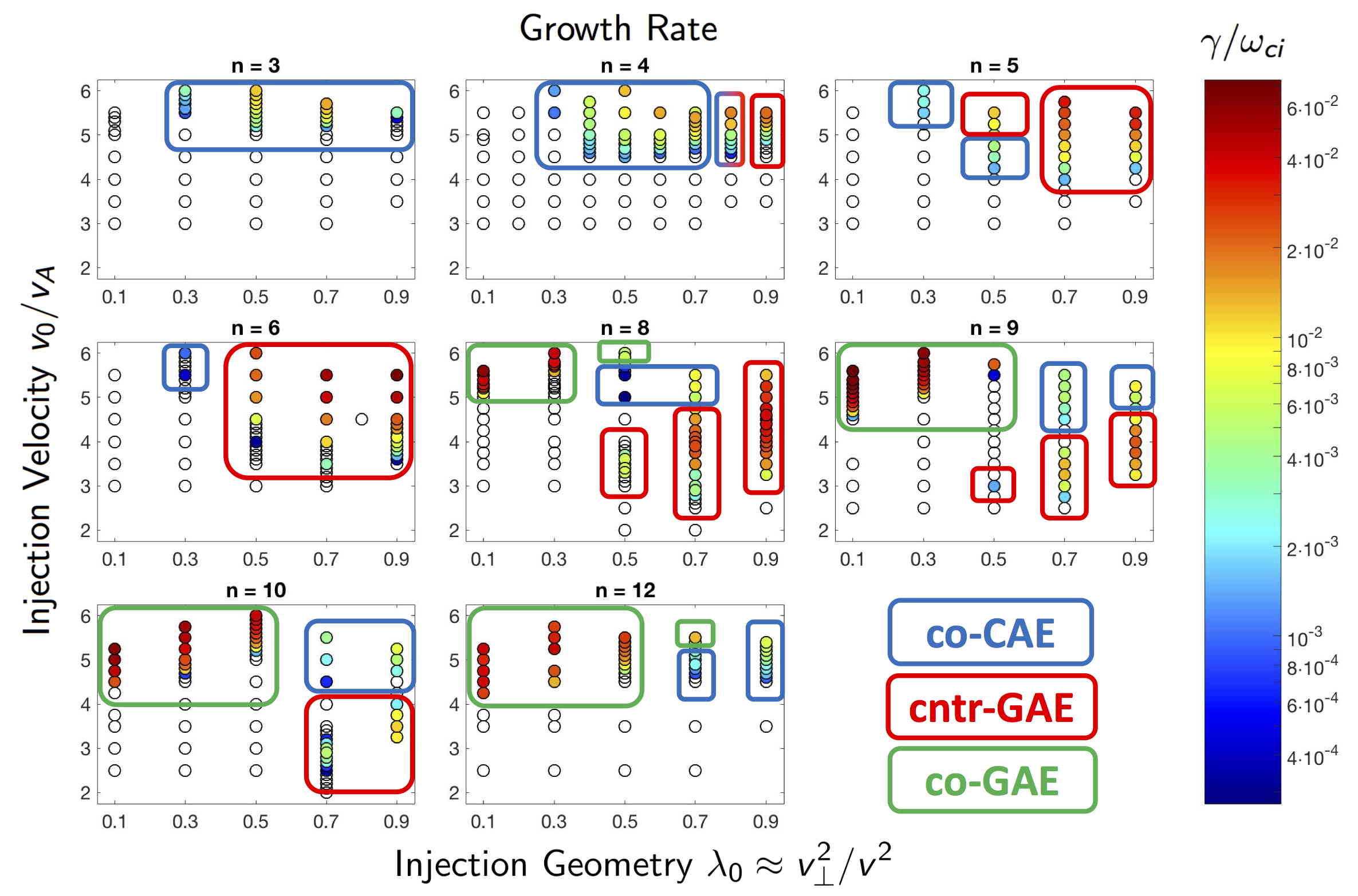}
\caption
[Linear growth rates of CAE/GAE modes as a function of beam injection geometry $\linj$ and injection velocity $\vinj$ for all simulated toroidal harmonics.]
{Linear growth rates of CAE/GAE modes. Each subplot represents a single toroidal harmonic $(|n|)$, showing growth rate of the most unstable mode for each distribution parameterized by $(\linj,\vinj)$. Individual white data points represent simulations where no unstable mode was found. Colored circles indicate the magnitude of the growth rate. Data points enclosed by blue boxes are co-CAEs, red boxes are cntr-GAEs, and green boxes are co-GAEs.}
\label{fig:sim:stab_all}
\end{figure} 

This section presents the linear stability trends from simulations and their interpretation with analytic theory. Unstable CAE/GAE modes are found for a variety of beam parameters, and all simulated toroidal mode numbers $|n| = 3 - 12$. In general, co-GAEs have the largest growth rate, followed by cntr-GAEs, and then co-CAEs. However, co-GAEs are only the most unstable mode at the periphery of realistic NSTX parameters, which may help explain why they have not been discussed in the experimental literature. For realistic NSTX beam geometry $(\linj = 0.5 - 0.7)$, cntr-GAEs usually have the largest growth rate. 

The summary of results from a large set of simulations is shown in \figref{fig:sim:stab_all}, where each subplot corresponds to simulations restricted to a single toroidal harmonic $\abs{n}$. Within each plot, each circle represents a separate simulation with the fast ion distribution in \eqref{eq:sim:F0} parameterized by injection geometry $\linj \approx \vperp^2/v^2$ and injection geometry $\vinj$. The white circles indicate simulations where all modes were stable, while the color scale indicates the growth rate of the most unstable mode. As a reminder, these linear initial value simulations are dominated by the most unstable mode in each simulation. Consequently, the presence of a specific type of unstable mode in a given simulation does not necessarily imply that other modes are stable, rather that if they are unstable they must have smaller growth rate. 

The normalized growth rates span three orders of magnitude in the simulations, ranging from $\gamma/\omegaci = 10^{-4}$ to $10^{-1}$. 
When instead normalized to the mode frequency, growth rates for CAEs are in the range $\gamma/\omega = 0.005 - 0.05$, while most of the GAEs have $\gamma/\omega = 0.01 - 0.2$, with a few more unstable cases having $\gamma/\omega$ up to 0.4. While there is reason to believe that GAE growth rates \emph{are} typically larger than those of CAEs, the nearly order of magnitude difference seen in these simulations is primarily due to the different resonant interactions, as will be explained shortly. The colored boxes on \figref{fig:sim:stab_all} indicate the most unstable type of mode in each simulation: co-CAE, cntr-GAE, or co-GAE. The direction of propagation for each mode is identified from the relative phase of the fluctuations at three closely spaced toroidal points. The determination of CAE vs GAE is made from the dominant polarization of the mode and comparison with the appropriate dispersion relation, as discussed in \secref{sec:sim:id}. 

\begin{figure}[tb]
{\includegraphics[width = \midwidth]{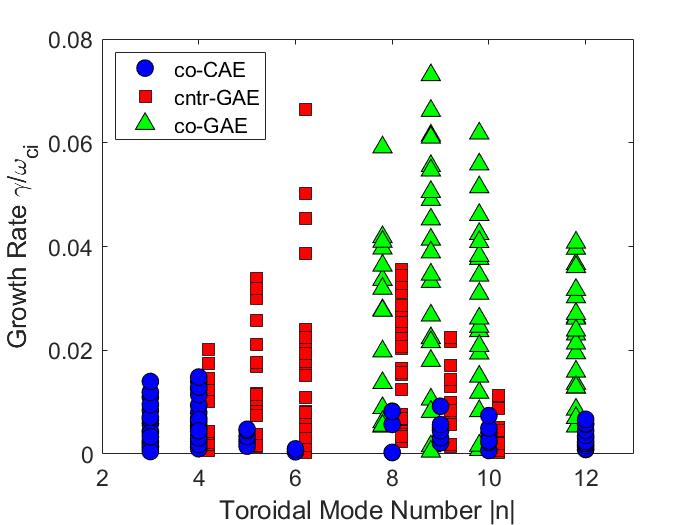}}
\caption
[Growth rate of each type of mode as a function of toroidal mode number in simulations.]
{Growth rate of each type of mode as a function of toroidal mode number in simulations. Simulations shown use beam distributions with $\linj = 0.1 - 0.9$ and $\vinj = 2.5 - 6.0$.}
\label{fig:sim:gamma_vs_n}
\end{figure}

\figref{fig:sim:gamma_vs_n} summarizes how the growth rate depends on toroidal mode number for each type of mode. In simulations restricted to $\abs{n} = 1$ and $\abs{n} = 2$, the only unstable modes had $\omeganorm < 0.05$, with high $m$ numbers and mixed polarization. Hence these modes are qualitatively different from the CAEs/GAEs being studied, and their further investigation is left to future work. At low $\abs{n} = 3 - 4$, co-CAEs were the most common mode. For each value of $\abs{n} =  3 - 12$, there existed at least one set of beam parameters $(\linj,\vinj)$ such that a co-CAE was the most unstable mode, with the growth rate maximized for $n = 4$. In contrast, the GAE growth rates peaked at moderately large values of $\abs{n} = 6$ and $\abs{n} = 9$ for cntr-GAEs and co-GAEs, respectively. Note that the co-GAEs are excited only at large $n$ since the $\lres = -1$ resonance requires a large Doppler shift $\kpar\vpres \approx n\vpres/R > \omegaci$. 

The simulation results will now be compared with the analytic theory of beam-driven, sub-cyclotron CAE/GAE stability developed earlier in this thesis. In \chapref{ch:cyc:analytics-cyclotron}, a local expression for the fast ion drive due to an anisotropic neutral beam-like distribution is derived. Terms to all orders in $\omeganorm$, $\krat$, and $\kperp\rhob$ are kept for applicability to the entire possible spectrum of modes. Analysis was restricted to 2D velocity space in order to avoid making assumptions about equilibrium profiles, mode structures, particle orbits, \etc. In particular, this excludes drive/damping due to gradients in $\pphi$, which is addressed in \secref{sec:sim:pphi}. Moreover, the calculation does not include bulk damping sources, hence it is most reliable when applied far from marginal stability. To supplement this analysis, quantification of the magnitude of damping present in \HYM simulations as well as an estimate of the thermal electron damping rate (absent in the simulation model) will be presented in \secref{sec:sim:damp}. In a nutshell, the velocity space drive (or damping) due to fast ions is a weighted integral over gradients of the fast ion distribution:

\begin{align}
\gamma \appropto \int d\lambda h(\lambda)\left[\left(\frac{\lres\omegaci}{\omega} - \lambda\right)\pderiv{}{\lambda} + \frac{v}{2}\pderiv{}{v}\right]\fb
\label{eq:sim:gamma_dfdxdv}
\end{align} 

Here $\lres$ is the cyclotron coefficient in the general resonance condition $\omega - n\avg{\omegator} - p\avg{\omegapol} = \lres\avg{\omegaci}$ and $h(\lambda)$ is a complicated positive function that weights the integrand. The full expression for the fast ion drive for the model neutral beam distribution given in \eqref{eq:sim:F0} can be found in \eqref{eq:sim:gammabeam}. We can gain a qualitative understanding of the simulation results by relying on \eqref{eq:sim:gamma_dfdxdv}. For the model distribution, the $\partial\fb/\partial v$ term is always negative (damping) since the slowing down function decreases monotonically. Velocity space anisotropy $(\partial\fb/\partial\lambda)$ can provide either drive or damping depending on its sign. For $\lres \neq 0$ resonances and the experimental value of $\dl \approx 0.3$, the $\partial\fb/\partial v$ contribution is much smaller than that from $\partial\fb/\partial\lambda$, though they can be comparable when $\lres = 0$.

Now it is clear why the co-CAE growth rates are typically smaller than those of the GAEs. As discussed previously, the co-CAEs are driven by the $\lres = 0$ resonance, while the co-GAEs and cntr-GAEs are driven by $\lres = \pm 1$, which leads to the large factor $\omegaci/\omega$ multiplying their growth rates. Cntr-CAEs have also been observed in experiments,\cite{Crocker2013NF,Sharapov2014PP} which should also have relatively large growth rates due to this factor for the $\lres = 1$ resonance, but they are not seen in \HYM simulations. As discussed in \secref{sec:cyc:stabao}, local theory predicts the linear growth rate of cntr-CAEs to be slightly smaller than that of cntr-GAEs driven by the same fast ion distribution, so these subdominant modes would not appear in linear initial value simulations. 

One might also ask why the co-GAEs are driven by the $\lres = -1$ resonance, but the co-CAEs are not, since it would enhance their growth rate. Due to the difference in dispersion, fast ions must have a larger parallel velocity to resonate with CAEs than GAEs. For GAEs, the $\lres = -1$ resonance requires $\vpres/\va \approx (1 + \omegaci/\omega)$, while for CAEs, $\vpres/\va \approx \abs{k/\kpar}(1 + \omegaci/\omega)$. For the ranges of $\abs{k/\kpar}$ and $\omegaci/\omega$ inferred from modes in the simulations, this resonance would require fast ion velocities above the beam injection velocity, hence the condition can not be satisfied and co-CAEs are restricted to the $\lres = 0$ resonance with typically smaller growth rate. 

\begin{figure}[tb]
\includegraphics[width = \midwidth]{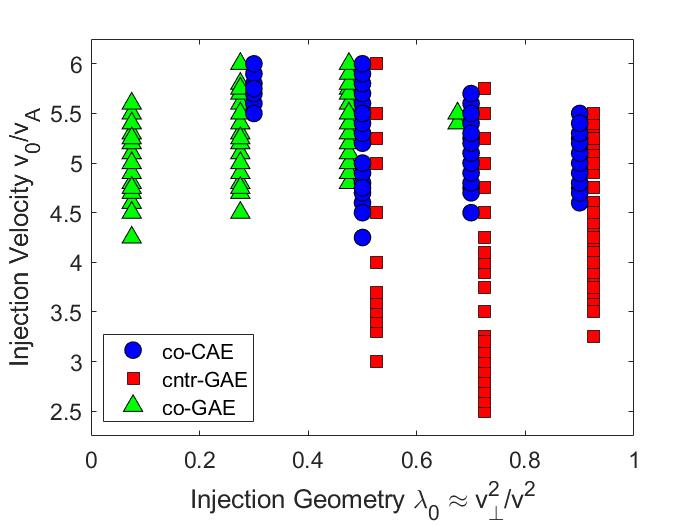}
\caption{Existence of unstable mode type as a function of the beam injection geometry $\linj$ and velocity $\vinj$.}
\label{fig:sim:lv_prop}
\end{figure}

In order to compare the stability properties of each mode against one another as a function of beam parameters, the results contained in \figref{fig:sim:stab_all} have been condensed into \figref{fig:sim:lv_prop}, which includes all simulated toroidal harmonics. Clearly, the cntr-GAEs prefer large $\linj$ whereas the co-GAEs prefer small $\linj$. This is reasonable for a distribution peaked in $\lambda$ based on \eqref{eq:sim:gamma_dfdxdv}. cntr-GAEs interacting with beams ions through the $\lres = 1$ resonance are driven by $\partial\fb/\partial\lambda > 0$, while co-GAEs are driven by regions of phase space with $\partial\fb/\partial\lambda < 0$ due to the $\lres = -1$ resonance. Hence when the distribution peaks at large $\linj \rightarrow 1$, a larger region of phase space contributes to drive the cntr-GAEs (and damp the co-GAEs). Conversely when $\linj \rightarrow 0$, more resonant fast ions contribute to driving the co-GAEs (and damp cntr-GAEs). A more quantitative comparison will be shown in \secref{sec:sim:theory} to elaborate on this intuition.  

The co-CAE dependence on $\linj$ is somewhat more subtle. From \eqref{eq:sim:gamma_dfdxdv}, one would expect that small $\linj$ favors their excitation, just as in the previous argument given for co-GAEs. Yet, \figref{fig:sim:lv_prop} shows unstable CAEs across a wide range $0.3 \leq \linj \leq 0.9$. Although the fast ions do provide drive for co-CAEs for $\linj = 0.1$ according to theory, the predicted growth rate is too small to overcome the background damping in simulations. Analytic theory predicts that the drive from fast ions peaks at some intermediate injection geometry $\linj \approx 0.5$, which is similar to what occurs in simulations which include some damping on the background plasma. 

With respect to the injection energy, the cntr-GAEs are unstable at significantly lower beam voltages than the CAEs. Whereas no CAE is found to be unstable for $\vinj < 4$, unstable cntr-GAEs were found with $\vinj > 2.5$ for a given set of plasma profiles based on a specific NSTX H-mode discharge. This is consistent with NSTX experiments, where cntr-GAEs were more routinely observed and in a wider array of operating parameters. For co-CAEs driven by the $\lres = 0$ resonance, the fast ion damping from $\partial\fb/\partial v$ competes more closely with the drive from anisotropy $\partial\fb/\partial\lambda$ than when $\lres \neq 0$, leading to a larger $\vinj$ for instability. As discussed in \secref{sec:lan:slow}, net drive for co-CAEs driven by beams with $\linj = 0.7$ and $\dl = 0.3$ requires $\vinj > 4.1$, similar to what is found in \HYM simulations. The co-GAEs also require relatively large $\vinj$ for excitation, due to the requirement of a sufficiently large Doppler shift $\kpar\vpres$ in order to satisfy the strong $\lres = -1$ resonance which drives them. Note that $\vinj \gtrsim 2.5$ and $\vinj \gtrsim 4$ should not be regarded as universal conditions necessary for cntr-GAE and co-GAE excitation, respectively. CAE/GAE excitation also depends on equilibrium profiles, which determine the background damping rate as well as the spectrum of eigenmodes. For instance, cntr-GAEs were routinely excited in early operation of NSTX-U\cite{Fredrickson2018NF} with $\vinj \approx 1 - 2$, while co-CAEs were rarely observed. In addition, cntr-propagating, sub-cyclotron modes have also been observed in dedicated low field DIII-D discharges\cite{Heidbrink2006NF,Tang2019APS} with $\vinj \approx 1$, in agreement with corresponding \HYM simulations.\cite{Belova2020IAEA} 

\subsection{Comparison with Local Analytic Theory}
\label{sec:sim:theory}

\subsubsection{Review of Theory}
\label{sec:sim:theoryrev}

It is worthwhile to study how well the local analytic theory can capture the stability properties determined by the realistic hybrid simulations. In general, the fast ion drive has a complicated dependence on the beam parameters $\linj$, $\vinj$, $\dl$, and $\vcrit$, the mode parameters $\omeganorm$ and $\krat$, and the specific resonance $\lres$ driving the mode. To compare with simulations, \eqref{eq:cyc:gammabeam} has been modified slightly in order to incorporate the exact form of the tail of the fast ion distribution above the injection velocity used in simulations. The expression is given below. 

\begin{multline}
\frac{\gamma}{\omegaci} = -\frac{n_b}{n_e}\frac{\pi C_f v_0^3 }{v_c^3} \sum_\lres \frac{\bresl^{3/2}}{\abs{\omegabar-\lres}}
\left\{ 
\vphantom{\frac{e^{-(x-\xinj)^2/\dx^2}e^{-\step{\sqrt{\frac{\bresl}{1-x}}-1}\left(\sqrt{\frac{\bresl}{1-x}}-1\right)^2\vb^2/\dv^2}}{1 + \frac{v_0^3}{v_c^3}\left(\frac{\bresl}{1-x}\right)^{3/2}}}
\int_0^1 \frac{x \Jlm(\flr(x,\zp))}{(1-x)^2} \times \right. 
\\
 \left. \left[\frac{1}{\dx^2}\left(\frac{\lres}{\omegabar} - x\right)(x-\xinj) + \frac{3/4}{1 + \frac{v_c^3}{v_0^3}\left(\frac{1-x}{\bres_\lres}\right)^{3/2}}\right] 
\frac{e^{-(x-\xinj)^2/\dx^2}e^{-\step{\sqrt{\frac{\bresl}{1-x}}-1}\left(\sqrt{\frac{\bresl}{1-x}}-1\right)^2\vb^2/\dv^2}}{1 + \frac{v_0^3}{v_c^3}\left(\frac{\bresl}{1-x}\right)^{3/2}} dx \right\} 
\label{eq:sim:gammabeam}
\end{multline}

As a reminder, $x \defined \vperp^2/v^2 = \lambda\omegacires$ where $\omegacires$ is the orbit-averaged cyclotron frequency of resonant particles normalized to the on-axis cyclotron frequency $\omegacio$. Similarly, $\xinj = \linj\omegacires$ and $\dx = \dl\omegacires$. It is found that resonant particles typically have $\omegacires \approx 0.9$ in simulations, and $\dl = 0.3$ is the default beam width. Also, $\bresl = \vpres^2/\vb^2$, $\step{x}$ is a step function with argument $v - \vb$ such that the distribution decays when $v > \vb$ with characteristic speed $\dv = 0.1\vb$ in simulations. The second line is the fast ion distribution $\fb(x,v(x,\bresl))$ using the relation $v = \vpres / \sqrt{1 - x} = \vb\sqrt{\bresl/(1-x)}$. The finite Larmor radius terms are contained within the function $\Jlm(\flr)$, where the lower index $\lres$ is the cyclotron resonance coefficient and $m$ denotes different functions for the mode types -- $m = ``C''$ for CAEs and $m = ``G''$ for GAEs. Its definition is 

\begin{align}
\Jlm(\flr) &\defined \frac{y}{y^2 - F^2}\left[\sqrt{y - \bvar}\frac{\lres\Jl}{\flr} \mp \sqrt{y - F^2 \bvar}\deriv{\Jl}{\flr}\right]^2
\label{eq:sim:Jlm}
\end{align}

The ``$-$'' solution is for CAEs and the ``$+$'' solution is for GAEs. As in \chapref{ch:cyc:analytics-cyclotron}, $F^2 = \kpar^2/k^2$ and $A = 1/(1 - \omegabar^2)$. Its argument is $\flr = \kperp\rhob = \zp\sqrt{x/(1 -x)}$ and $\zp = \kperp\vpres/\omegaci$. Using the resonance condition, $\zp$ can be re-written as $\zp = \abs{\omegabar - \lres\omegacires}/\alpha$ with $\omegabar \defined \omeganorm$ and $\alpha \defined \krat$. Lastly, $y\defined N^{-2} = \omega^2/k^2\va^2$, and defining $G = 1 + F^2$, the cold plasma two-fluid dispersion is 

\begin{equation}
N^2 = \frac{AG}{2F^2}\left[1 \pm \sqrt{1 - \frac{4F^2}{AG^2}}\right]
\label{eq:sim:twodisp}
\end{equation}

For $0 < \omegabar < 1$, the ``$-$" solution corresponds to the CAE, while the ``$+$" solution corresponds to GAE. For comparison with simulations which do not include two-fluid effects, we will make the approximations $N^2 \approx 1$ for CAEs and $N^2 \approx 1/F^2$ for GAEs, which give $\Jlc(\flr) \approx \left(J_\lres'(\flr)\right)^2$ for CAEs and $\Jlg(\flr) \approx \left(\lres J_\lres/\flr\right)^2$. A discussion of the impact of these two-fluid effects on the growth rate can be found in \secref{sec:cyc:stabao} and \secref{sec:lan:stabao}, but we restrict our attention to the single fluid limit since our aim is interpretation of the simulations. 

\subsubsection{Maximum Growth Rate Dependence on Beam Injection Geometry and Velocity}
\label{sec:sim:gammamaxlv}

\begin{figure}[htb]
\subfloat[]{\includegraphics[width = \halfwidth]{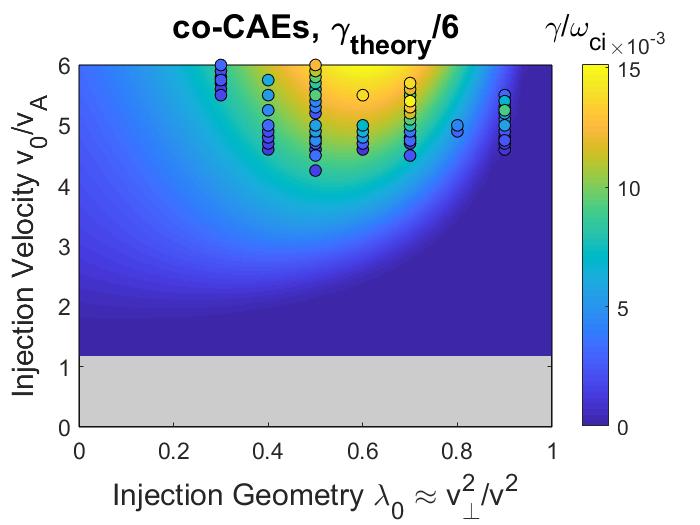}}
\subfloat[]{\includegraphics[width = \halfwidth]{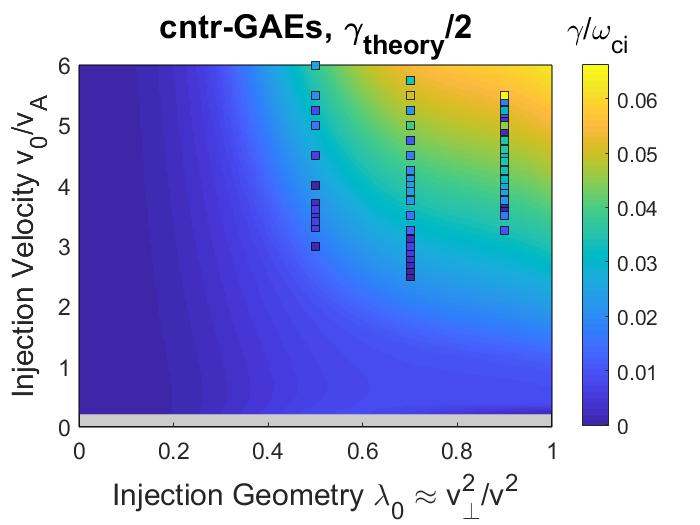}} \\
\subfloat[]{\includegraphics[width = \halfwidth]{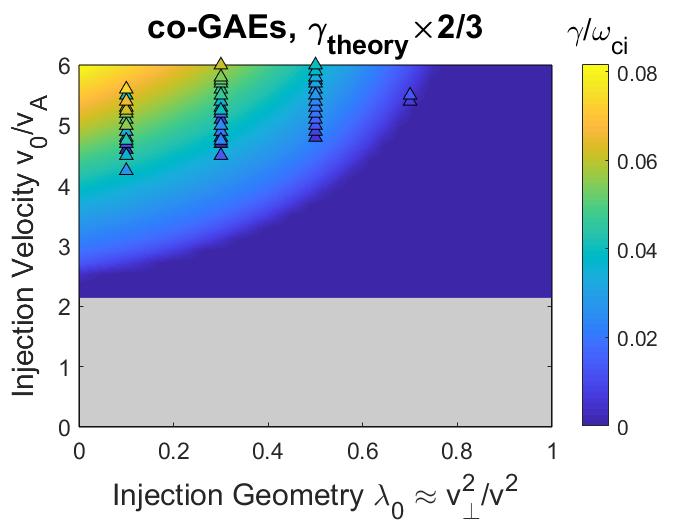}}
\caption
[Comparison of growth rate of most unstable mode calculated from analytic theory and \HYM simulations as a function of injection geometry $\linj$ and velocity $\vinj$.]
{Growth rate of most unstable mode as a function of injection geometry $\linj$ and velocity $\vinj$, calculated from analytic theory (background color) and \HYM simulations (individual points). From left to right: (a) co-CAEs, (b) cntr-GAEs, and (c) co-GAEs. In order to more clearly show trends, the analytically calculated growth rate has been re-scaled as indicated.}
\label{fig:sim:lv_comp}
\end{figure}

As a first comparison, we will examine the dependence of the growth rate on the beam injection geometry $\linj$ and velocity $\vinj$. Since the growth rate is sensitive to the specific mode properties (frequency $\omeganorm$ and wave vector direction $\alpha = \krat$) whereas only the mode with the largest growth rate survives in the simulations, it makes sense to use \eqref{eq:sim:gammabeam} to calculate the growth rate of the most unstable mode across all values of $0 < \omeganorm < 1$ and $\krat$. Such a comparison is shown in \figref{fig:sim:lv_comp}, with separate subplots for co-CAEs, cntr-GAEs, and co-GAEs. In these figures, the background color gives the growth rate calculated analytically, while the points show the growth rates of unstable modes in separate simulations. 

The analytic growth rate is typically larger than that from simulations, so their magnitudes have been re-scaled as indicated in the figures so that a comparison of trends can be made. There are two reasons why the analytic growth rates are larger than those from simulations for two reasons. First, the simulations include drive/damping from fast ions and also damping of the mode on the single fluid resistive background plasma. As will be discussed in \secref{sec:sim:damp}, this damping can be of order $20\% - 60\%$ of the fast ion drive, while it is not present at all in the analytic calculation which perturbatively considers the contribution from fast ions alone. Second, the analytic theory being used is a \emph{local} approximation which does not take into account equilibrium profiles of the background plasma, fast ion density, or spatial dependence of the mode structure. The value of $\nb$ that we use in \eqref{eq:sim:gammabeam} is its peak value ($5.3\%$ for the simulations in this section), leading to a calculated growth rate larger than it would be if spatial dependencies were taken into account. Consequently, the value of comparing this calculation with the simulation results lies in the trends with beam parameters, not the absolute values of the growth rate. 

We see that for each type of mode, there is reasonable agreement between the regions of instability predicted by analytic theory and the beam parameters of unstable modes. For co-CAEs, theory predicts the largest growth rate near moderate values of $\linj \approx 0.5$ and large $\vinj$, which are also the beam parameters preferred by unstable modes in simulations. According to theory, cntr-GAEs generally become more unstable for larger values of $\linj$ (more perpendicularly injected beams), while co-GAEs are most unstable for small $\linj$ (very tangential injection). This is exactly the trend seen in simulations, where the unstable cntr-GAEs generally have largest growth rate for  $\linj = 0.7 - 0.9$ and the co-GAEs are most unstable for $\linj = 0.1$, the smallest simulated value. Hence, numerical evaluation of the analytic expression for fast ion drive confirms many of the qualitative arguments given when interpreting \figref{fig:sim:lv_comp} through the lens of \eqref{eq:sim:gamma_dfdxdv}. For all types of modes, larger beam velocity leads to a larger maximum growth rate, though that growth rate may correspond to different mode parameters $\omeganorm$ and $\krat$ than the most unstable mode for a smaller beam velocity. 

Lastly, note that the trends from the calculation shown in \figref{fig:sim:lv_comp} rely on the aforementioned search over all mode parameters for the most unstable mode. For instance, when the same calculation is done for cntr-GAEs with specific values of $\omeganorm$ and $\krat$, the growth rate is no longer a strictly increasing function of $\vinj$ and $\linj$, but rather it can peak and then decrease, as in \figref{fig:cyc:wideslowGAEfig}. This occurs when the beam parameters are varied in such a way that not as many particles resonate with the particular mode of interest. An example of such behavior is given in the $\abs{n} = 8$ subplot of \figref{fig:sim:stab_all}, as the cntr-GAE growth rate for injection geometry $\linj = 0.7$ first increases, peaks, and then decreases, with the cntr-GAE eventually being replaced by a more unstable co-CAE. Since only the $\abs{n} = 8$ toroidal harmonic is kept in that simulation, the range of mode properties is also restricted.  

\subsubsection{Approximate Stability Boundaries}
\label{sec:sim:stabapprox}

In \chapref{ch:cyc:analytics-cyclotron} and \chapref{ch:lan:analytics-landau}, approximations relevant to the simulate conditions were used in order to derive simple necessary conditions for net fast ion drive. These conditions will be compared with the simulation results in order to assess the capability of the local analytic theory to reproduce simulation results. As discussed in \secref{sec:cyc:narrow} and \secref{sec:lan:narrow}, these approximations can be made when $0.2 \lesssim \dl \lesssim 0.8$ (the NSTX(-U) value is $\dl \approx 0.3$) and either $\zp \defined \kperp\vpres/\omegaci \lesssim 2$ or $\zp \gg 2$. Note that this constraint is related to a condition on the finite Larmor radius (FLR) of the fast ions, 
since $\kperp\rhob = \zp\sqrt{x / (1 - x)}$. 
Since the resonant fast ions have identical $\vpar = \vpres$ but varying energies, they also have different values of $\kperp\rhob$. Hence $\zp \lesssim 2$ implies that FLR effects are generally small, and $\zp \gg 2$ implies they are generally large, but neither are as strict as $\kperp\rhob \ll 1$ or $\kperp\rhob \gg 1$. Notice that this parameter can be re-written using the resonance condition in the form $\omega - \kpar\vpres = \lres\omegacires$ as $\zp = \abs{\omegabar - \lres\omegacires}/\alpha$, allowing the calculation of $\zp$ from mode properties alone. 

Using the data shown in \figref{fig:sim:ao_prop}, one can calculate that $\zp = 0.5 - 1$ for the co-CAEs and $\zp = 0.5 - 1.5$ for the co-GAEs, so all modes fall within the $\zp \lesssim 2$ regime. Meanwhile, $\zp = 0.5 - 3$ for cntr-GAEs, though the most unstable ones do have $\zp < 2$. Hence the approximate instability conditions derived in \secref{sec:cyc:slow} for $\lres = \pm 1$ resonances and and \secref{sec:lan:slow} for the $\lres = 0$ resonance are applicable to the unstable modes in NSTX simulations presented here. Using these approximations, and recalling the earlier definition $\xinj = \linj\omegacires$, the following condition is found for GAE instability due to beam anisotropy 

\begin{align}
\label{eq:sim:cntrGAE}
\vb &< \frac{\vpres}{(1 - \linj\omegacires)^{3/4}} \eqlab{cntr-GAEs} \\ 
\label{eq:sim:coGAE}
\vb &> \frac{\vpres}{(1 - \linj\omegacires)^{3/4}} \eqlab{co-GAEs} 
\end{align}

Meanwhile, for co-CAEs, the $\partial\fb/\partial v$ terms must also be taken into account, which leads to a more complicated condition

\begin{align}
\label{eq:sim:coCAE}
\vb &> \vpres\left[\frac{1 - \dx\sqrt{2/3}}{\left[1 - \frac{1}{2}\left(\xinj + \sqrt{\xinj^2 + 8\dx^2/3}\right)\right]
\left[1 - \dx^{4/5}\right]}\right]^{5/8}
\end{align}

Note that in each of these equations, $\vpres$ implicitly depends on $\omeganorm$ and $\krat$ through the resonance condition written as $\omega - \kpar\vpres = \lres\omegacires$ and the appropriate dispersion relation for each mode. Hence, these conditions place constraints on the beam parameters ($\linj$, $\vinj$, as well as $\dl$ for co-CAEs) and mode properties for a mode to be driven unstable. Just as in our previous qualitative analysis of \eqref{eq:sim:gamma_dfdxdv}, \eqref{eq:sim:cntrGAE} and \eqref{eq:sim:coGAE} demonstrate that the cntr-GAEs and co-GAEs have complementary instability conditions resulting from the opposite sign of $\lres$ in the resonance driving each mode. Consequently, co-GAE excitation is favored when $\linj$ is small, while cntr-GAEs prefer large $\linj$. 

\begin{figure}[H]
\subfloat[\label{fig:sim:quant_gae}]{\includegraphics[width = \midwidth]{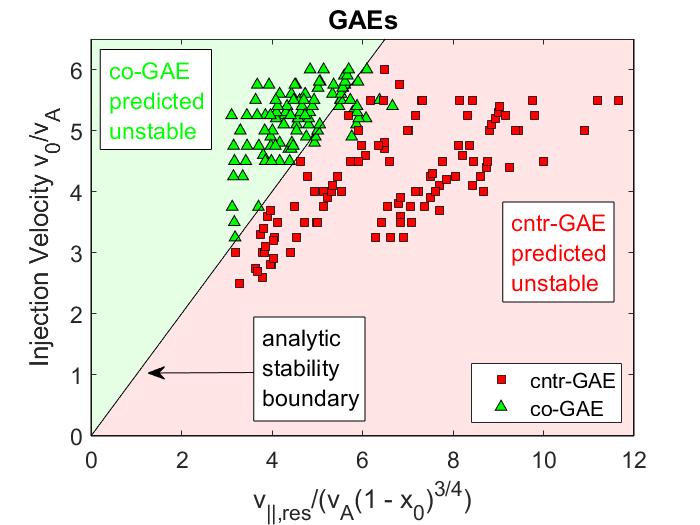}} \\
\subfloat[\label{fig:sim:quant_cae}]{\includegraphics[width = \midwidth]{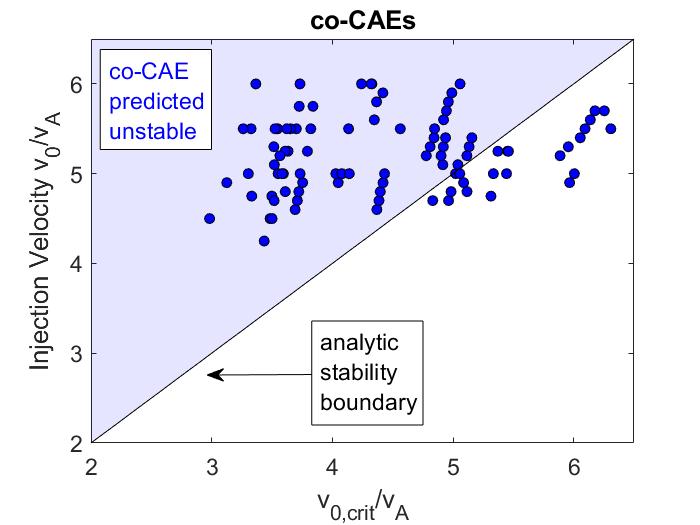}} 
\caption
[Comparison of unstable modes in simulations against approximate analytic stability boundaries.]
{Quantitative comparison of unstable modes in simulations against analytic predictions for (a) GAEs and (b) CAEs. The shaded areas indicate regions of instability predicted by the approximate conditions given in \eqref{eq:sim:cntrGAE}, \eqref{eq:sim:coGAE}, and \eqref{eq:sim:coCAE} for cntr-GAEs, co-GAEs, and co-CAEs, respectively. The points are unstable modes of the labeled types from simulations. In (b), $v_{0,\text{crit}}$ is defined as the right hand side of \eqref{eq:sim:coCAE}}
\label{fig:sim:quant}
\end{figure}

A comparison between these conditions and the simulation results can be used to verify theory and also understand the simulation results. Such a comparison is shown in \figref{fig:sim:quant}, where $\vpres = (\omega - \lres\omegacires)/\kpar$ is determined from the resonance condition with $\kpar$ calculated from the mode structure in each simulation as described at the beginning of \secref{sec:sim:theory}. 

Each point represents an unstable mode from an individual simulation, with the beam injection velocity used in the simulation plotted against the marginal value of $\vinj$ necessary for instability, as determined by \eqref{eq:sim:cntrGAE}, \eqref{eq:sim:coCAE}, and \eqref{eq:sim:coCAE}. The shaded regions -- green for co-GAEs, red for cntr-GAEs, and blue for co-CAEs -- indicate the regions of instability predicted by the theoretical conditions. For the GAEs, there is quite good agreement between theory and simulations, with only a few of the co-GAEs appearing far from the predicted boundary and all of the cntr-GAEs falling in the predicted region. The agreement for CAEs is not as strong, though the majority of unstable modes are still in the theoretically predicted region. 

There are three reasons why this might be the case. First, recall that the local theory which led to \eqref{eq:sim:cntrGAE}, \eqref{eq:sim:coGAE}, and \eqref{eq:sim:coCAE} neglected the $\pphi$ gradient contribution to the fast ion drive/damping. As will be discussed briefly in \secref{sec:sim:pphi}, this gives a contribution proportional to $n\partial\fb/\partial\pphi$. Hence it provides additional drive for co-CAEs and co-GAEs, which would extend the region of instability. Second, as described in \secref{sec:sim:model}, the ``tail'' of the fast ion distribution in the simulation is modeled with a rapid Gaussian decay above the injection energy whereas the theory assumes a delta function for analytic tractability. These particles in the tail can similarly provide additional drive for co-propagating modes because they provide more resonant particles with $\lambda > \linj$ (corresponding to $\partial\fb/\partial\lambda < 0$). Numerical calculation of the analytic growth rate replacing the delta function tail with the Gaussian decay present in simulations improves the agreement in \figref{fig:sim:quant_cae}. Lastly, the approximate determination of $\vpres \approx \omega/\kpar$ may be inadequate, since it assumes that sideband resonances are sub-dominant. However, examination of resonant particles in simulations indicates that although $p = 0$ is often the dominant resonance for co-CAEs, there can be several other sidebands present as well. Since much of the disagreement between theory and simulation occurs for beams with larger values of $\linj\gtrsim 0.5$, it may be that trapped particles are playing a more important role. To improve the quantitative accuracy of the approximate instability condition for co-CAEs, the theory should be extended to properly weight and sum over all sideband resonances. Altogether, this comparison suggest that features of nonlocal theory are needed to improve accuracy for the co-CAEs, since they grow more slowly than the GAEs. 

\subsubsection{Properties of Unstable Modes} 
\label{sec:sim:stabao}
\begin{figure}[tb]
\subfloat[\label{fig:sim:ao_comp_caez}]{\includegraphics[width = \halfwidth]{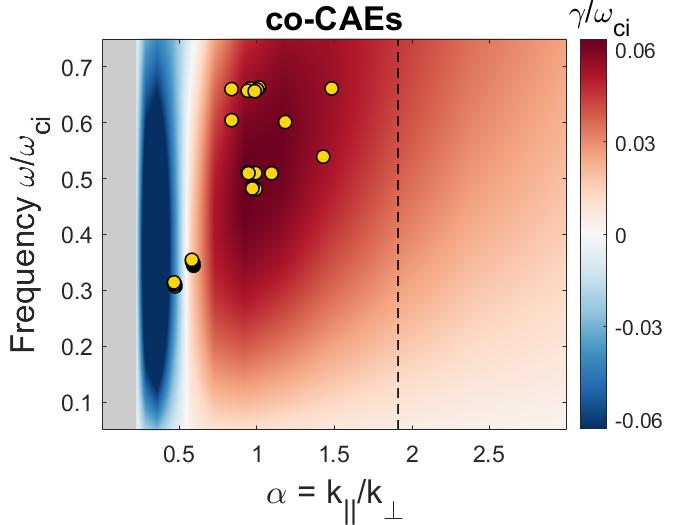}}
\subfloat[\label{fig:sim:ao_comp_gaem}]{\includegraphics[width = \halfwidth]{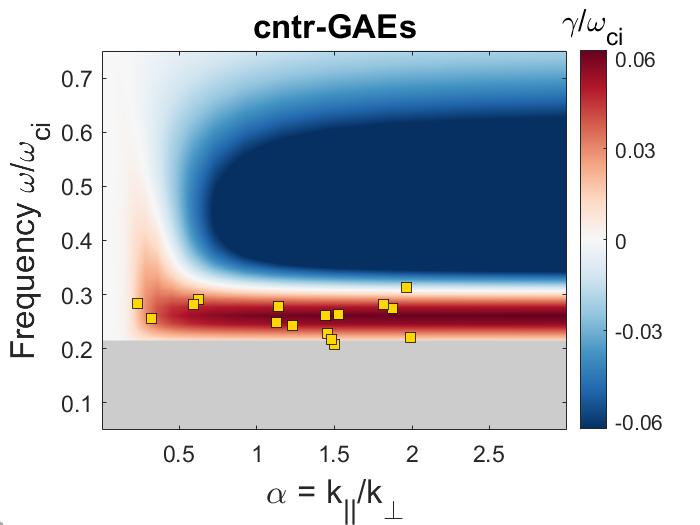}} \\
\subfloat[\label{fig:sim:ao_comp_gaep}]{\includegraphics[width = \halfwidth]{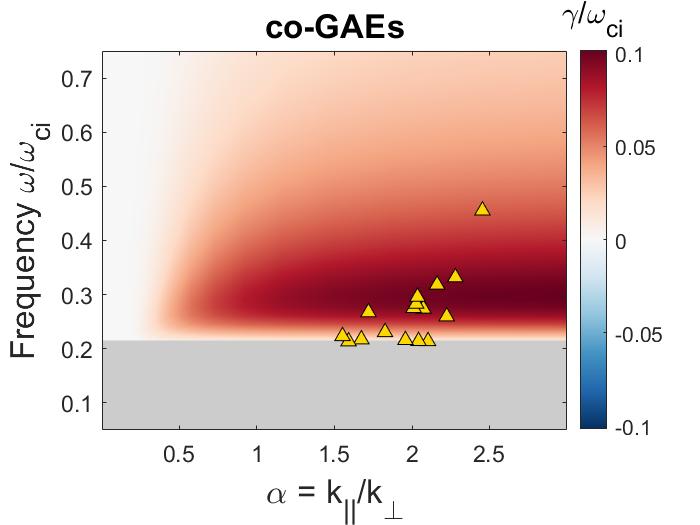}}
\caption
[Analytically calculated growth rate as a function of frequency $(\omeganorm)$ and wave vector direction $(\alpha = \krat)$ compared to the properties of unstable modes excited in simulations.]
{Analytically calculated growth rate (background color) as a function of frequency $(\omeganorm)$ and wave vector direction $(\alpha = \krat)$ compared to the properties of unstable modes excited in simulations (individual points). (a) co-CAEs driven by a fast ion distribution parameterized by $\linj = 0.7$ and $4.5 \leq \vinj \leq 5.5$, (b) cntr-GAEs driven by $\linj = 0.7$ and $3.0 \leq \vinj \leq 4.0$, and (c) co-GAEs driven by $\linj = 0.1$ and $5.0 \leq \vinj \leq 6.0$.}
\label{fig:sim:ao_comp}
\end{figure}

Another approach for comparison between simulation and theory is to instead fix the beam parameters and consider how the growth rate depends on the properties of each mode, namely its frequency and direction of wave vector. Such analysis can be useful in the interpretation of experimental observations in a given discharge. This comparison is made in \figref{fig:sim:ao_comp}. In these figures, the background color is the analytically computed growth rate, with darker red indicating larger predicted growth rate, and blue indicating regions with negative growth rate (damped by fast ions). Gray regions indicate where no resonance is possible, occurring when $\vb < \vpres$. The gold points represent unstable modes in \HYM simulations for beam distributions with fixed values of $\linj$ and $\vinj$ as specified on the plots for each type of mode. A small range of $\vinj$ around a central value is included in each case in order to include enough examples to show a trend. 

For co-CAEs, a minimum value of $\alpha$ is needed in order to resonate with the mode at all, and an even larger value is needed in order to have net fast ion drive. This is reasonable since combination of the $\lres = 0$ resonance condition and approximate CAE dispersion $\omega \approx k\va$ yields $\vpres/\va = \sqrt{1 + \kperp^2/\kpar^2}$. For a resonance to exist, $\vb > \vpres$ must be satisfied, which requires sufficiently large $\alpha = \krat$, as shown on \figref{fig:sim:ao_comp_caez}. Meanwhile, the approximate instability condition written in \eqref{eq:sim:coCAE} is of the form $\vb > \vpres g(\xinj,\dx)$ where $0 < g(\xinj,\dx) < 1$ depends on the beam injection geometry and weakly on the degree of anisotropy. Hence an even larger value of $\krat$ is necessary for the modes to be unstable than simply for the resonance to be satisfied. On the other hand, a maximum value of $\krat$ for the existence of co-CAEs was derived heuristically in \secref{sec:lan:expcomp} based on the condition that the CAE is trapped within a magnetic well on the low field side, which gives $\krat \lesssim n / 2\pi$, shown as a dashed line on \figref{fig:sim:ao_comp_caez}. The agreement between simulations and theory in the figure is close but imperfect. While the bulk of the unstable modes from simulations are well within the unstable region and even cluster around the mode properties with maximum growth rate, there are two sets of modes very close to the stability boundary, some even in the blue region where theory predicts that they should be stable. A likely explanation is the lack of $\pphi$ gradient in the theory calculation, which is present in simulations and should provide additional drive for the co-propagating modes $(n > 0)$.

The same comparison is very successful for both cntr- and co-GAEs. \figref{fig:sim:ao_comp_gaem} and \figref{fig:sim:ao_comp_gaep} illustrate a key difference between the unstable spectrum of GAEs vs CAEs. While instability for co-CAEs requires a sufficiently large value of $\krat$, the GAEs require sufficiently large \emph{frequency} in order to satisfy the resonance condition. This can be understood as a consequence of the distinct dispersion relations for CAEs vs GAEs. For GAEs, $\omega \approx \abs{\kpar}\va$, so $\vpres/\va \approx (\omegaci/\omega) - \lres$. Consequently, a resonance is possible when $\vb > \vpres$, which requires sufficiently large $\omeganorm$. For cntr-GAEs, the approximate condition for instability given in \eqref{eq:sim:cntrGAE} takes the form $\vb < \vpres h(\xinj)$, where $0 < h(\xinj) < 1$. Consequently, larger frequencies can violate this condition, so a band of unstable frequencies results. Nearly all of the simulation points fall within this band, with the few that fall just within the gray ``no resonance possible'' region are due to having $\vinj$ slightly larger than the value it is being computed for in the figure (or due to sideband resonances). 

In the case of co-GAEs, there is no maximum frequency for unstable modes since the inequality in their approximate instability condition is flipped relative to that for the cntr-GAEs. Hence, both the resonance condition and the instability condition provide upper bounds on $\vpres$, or equivalently lower bounds on $\omeganorm$. 
This is similar to the co-CAEs, which had two distinct lower bounds on $\alpha$ coming from satisfying the resonance condition and instability conditions. 
For the beam parameters shown in \figref{fig:sim:ao_comp_gaep}, the lower bound on $\omeganorm$ coming from the instability condition is less restrictive than that coming from the resonance condition, so there are no frequencies where a resonance is possible but the mode is stable. However this is not always the case. For smaller values of $\vinj$ (for example $\vinj = 4.0$, not shown), such a band of stable frequencies exist that are damped by the fast ions. Consistency is found between the properties of unstable co-GAEs in simulations and those predicted by analytic theory, as all of the simulation points appear above the minimum frequency required for resonance, and with values of the wave vector direction $\krat$ near the region of largest growth rate. 

To summarize, analytic theory predicts that instability for co-CAEs occurs for modes with $\krat$ above a threshold value (specific value depending on beam parameters). A maximum value of $\krat$ for CAEs can also be determined heuristically. Meanwhile, co-GAEs are unstable for frequencies that are sufficiently large. Lastly, the most unstable cntr-GAEs are predicted to occur within a specific range of frequencies. The unstable modes in \HYM simulations exhibit these properties in the majority of cases, with outliers likely explained by known limitations of the local analytic theory. 

\subsection{Dependence on Critical Velocity and Beam Anisotropy} 
\label{sec:sim:vcdl}

Whereas the majority of the simulation parameter scan performed for this study focused on varying the beam injection geometry and velocity, there are other parameter dependencies that can be understood even with a less comprehensive set of simulations. Namely, the ratio of the critical velocity to the beam injection velocity $\vcrit$ controls the steepness of the distribution with respect to velocity, while the width of the beam in velocity space $\dl$ controls the level of anisotropy. 

\begin{figure}[tb]
\includegraphics[width = \midwidth]{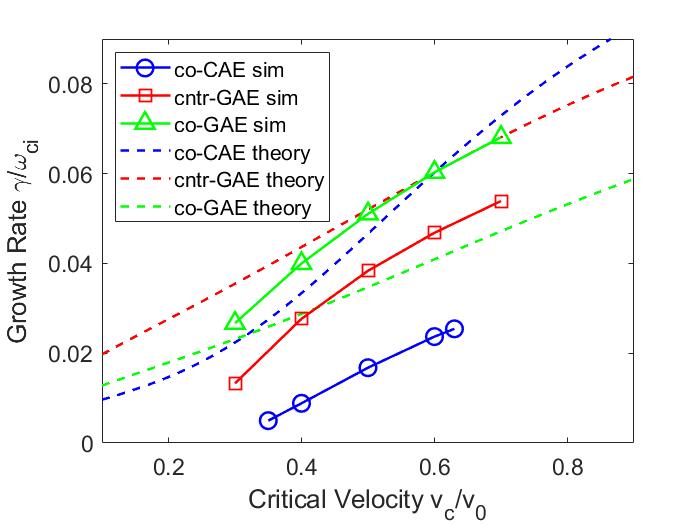}
\caption
[Growth rate as a function of the normalized critical velocity $\vcrit$ from simulations and calculated by analytic theory.]
{Growth rate as a function of the normalized critical velocity $\vcrit$ from simulations (points connected by solid lines) and calculated by analytic theory (dashed lines). Blue curves are for $n = 4$ co-CAEs, with $\linj = 0.7$, $\vinj = 5.5$, $\dl = 0.3$. Red curves are for $n = -6$ cntr-GAEs, with $\linj = 0.7$, $\vinj = 5.0$, $\dl = 0.3$. Green curves are for $n = 9$ co-GAEs, with $\linj = 0.1$, $\vinj = 5.0$, $\dl = 0.3$.}
\label{fig:sim:vc_scan}
\end{figure}

The dependence of the growth rate on $\vcrit$ is shown in \figref{fig:sim:vc_scan} as determined by simulations and also calculated by the analytic expression previously discussed. It is clear that larger values of $\vcrit$ tend to make all modes more unstable -- whether they are CAEs or GAEs and co- or cntr-propagating. There are two key factors leading to this effect. First, increasing $\vcrit$ while keeping the total number of particles fixed (\ie properly normalized) leads to a larger number of high energy resonant particles relative to those with low energy, thus providing more energy to drive the mode. Second, larger $\vcrit$ corresponds to smaller magnitude of $\partial\fb/\partial v$, so the fast ion ``damping'' from this term is also decreased. In \figref{fig:sim:vc_scan}, the range of simulated values of $\vcrit$ is constrained by the equilibrium solver. Note that the slopes of the analytic curves are quite similar to those of the simulation results, indicating similar dependence on $\vcrit$, even without quantitative agreement for all of the reasons previously discussed. 

Recall the definition of the critical velocity:\cite{Gaffey1976JPP} 

\begin{align}
m_b \vc^2 = 14.8 A_b T_e / A_i^{2/3}
\label{eq:vcrit}
\end{align}

Here, $A_i$ and $A_b$ are the atomic numbers of the thermal ions and neutral beam ions, respectively, and $m_b$ is the mass of the beam ions. Hence, a lower temperature plasma (with fixed beam voltage) should tend to render CAEs and GAEs somewhat more stable. Extrapolating to ITER-like parameters with $T_e \approx 20$ keV with 1 MeV Deuterium beams being injected into a DT plasma implies $\vcrit \approx 0.57$, slightly larger than the approximate NSTX value of $\vcrit \approx 0.5$. 

\begin{figure}[tb]
\includegraphics[width = \midwidth]{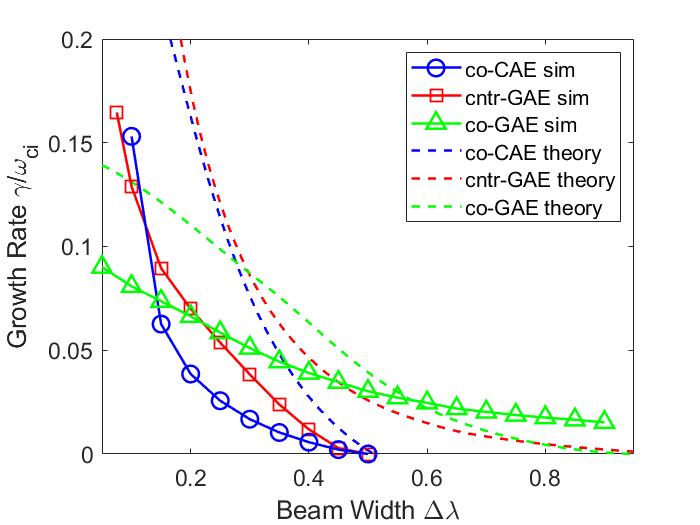}
\caption
[Growth rate as a function of the beam width in velocity space $\dl$ from simulations and calculated by analytic theory.]
{Growth rate as a function of the beam width in velocity space $\dl$ from simulations (points connected by solid lines) and calculated by analytic theory (dashed lines). Blue curves are for $n = 4$ co-CAEs, with $\linj = 0.7$, $\vinj = 5.5$, $\vcrit = 0.5$. Red curves are for $n = -6$ cntr-GAEs, with $\linj = 0.7$, $\vinj = 5.0$, $\vcrit = 0.5$. Green curves are for $n = 9$ co-GAEs, with $\linj = 0.1$, $\vinj = 5.0$, $\vcrit = 0.5$.}
\label{fig:sim:dl_scan}
\end{figure}

Consider now the growth rate's dependence on $\dl$, which is shown in \figref{fig:sim:dl_scan}. The simulations and analytic calculations agree that all types of modes become more unstable as $\dl$ is decreased, consistent with the theoretical understanding that the dominant source of drive is beam anisotropy. Again, we find that there is qualitative agreement between the analytically calculated dependence on $\dl$ and that determined directly from simulations. To make the comparison shown in \figref{fig:sim:dl_scan}, the normalized frequency $\omeganorm$ and wave vector direction $\krat$ were calculated from simulation results, and then a small range around these values was used to find the maximum analytic growth rate as $\dl$ was varied. 

Theoretically, the scaling with $\dl$ can be understood in two different regimes: the experimental regime where $\dl$ is relatively large $(\dl \gtrsim 0.2)$, and then the limiting case of $\dl \ll 1$. In \appref{app:sim:dl}, it is demonstrated that $\gamma \propto 1/\dl^2$ when $\dl \gtrsim 0.2$, and $\gamma \propto 1/\dl$ in the limit of $\dl \rightarrow 0$. 

The scaling of the simulation results is approximately $1/\dl$ for the cntr-GAEs and co-GAEs, with a $1/\dl^2$ trend found for the co-CAEs. Interpretation of the simulation results is complicated by the presence of nontrivial damping in the simulations, which becomes more relevant as $\dl$ is increased. Hence it is consistent that the simulations would mostly capture the $\dx \ll 1$ scaling and have additional complications for the $\dx \gtrsim 0.2$ regime. It's unclear why the co-CAEs from the simulations exhibit the stronger $1/\dl^2$ dependence overall. It's also important to keep in mind that the analytic theory that we are comparing with is \emph{perturbative} in the sense that it assumes $\gamma \ll \omega$. Hence, any calculations predicting $\gamma \like \omega$ are explicitly unreliable. This places a lower bound on the value of $\dl$ that can be used to calculate the growth rate, as even $\dl \approx 0.2$ is giving $\gamma/\omegaci \approx 0.2$, corresponding to $\gamma/\omega \approx 0.6$ for sub-cyclotron frequencies $\omeganorm \approx 0.3$. The simulation model has no such restriction, but are still constrained to sufficiently large values of $\dl$ in order to satisfy constraints on the equilibrium for the modeled NSTX discharge. Whereas the examined co-CAE and cntr-GAE are stabilized for $\dl = 0.5$, the co-GAE remains unstable even for the largest simulated value of $\dl = 0.9$, which corresponds to extremely weak anisotropy. To explain this result, the previously neglected contribution from gradients with respect to $\pphi$ must be considered, as discussed in the next section. 

\subsection{Effect of \texorpdfstring{$\pphi$}{pphi} Gradients}
\label{sec:sim:pphi}

Most of the analysis so far has focused on fast ion drive resulting from velocity space anistropy or gradients in energy present in the fast ion distribution since these are the terms present in the local analytic theory derived by Mikhailovskii\cite{Mikhailovskiiv6} and subsequently adapted for the analysis of sub-cyclotron modes in this work. However, a more complete treatment would also include a contribution from gradients with respect to $\pphi$. In this section, we will discuss the qualitative effect of this term that is absent in our theoretical analysis, and discuss its role in resolving certain disagreements between the self consistent simulation results and predictions from local analytic theory. 

A general form of the growth rate is adapted from \citeref{Kaufman1972PF} in \appref{app:sim:pphi}. The relevant result is that 

\begin{align}
\gamma &\propto \int d\Gamma \left[\left(\frac{\lres}{\omegabar} - \lambda\right)\pderiv{\fb}{\lambda} + \W\pderiv{\fb}{\W} + \frac{n}{\omegabar}\frac{\W}{\omegaci}\pderiv{\fb}{\pphi}\right]
\label{eq:sim:gammapphi}
\end{align}

Here, $d\Gamma$ is the differential volume of phase space and $\W = m_i v^2 / 2$, whereas elsewhere in the thesis $\W$ may not carry units of mass. The first two terms in brackets are the gradients present in the local theory derived in \secref{sec:cyc:subderiv}, while the third term results from plasma non-uniformity. 

An important consequence of \eqref{eq:sim:gammapphi} is that the contribution to the growth rate from the $\partial\fb/\partial\pphi$ term depends on the sign of $n$. Hence, for non-hollow fast ion distributions ($\partial\fb/\partial\pphi > 0$ everywhere), it has a destabilizing effect for co-propagating modes $(n > 0)$ and a stabilizing effect for cntr-propagating modes $(n < 0)$. This consequence has been previously noted in the literature, and used to explain transitions between co- and cntr-propagation of toroidal \Alfven eigenmodes (TAEs) in TFTR\cite{Wong1999PLA} and NSTX-U.\cite{Podesta2018NF} This term is expected to be most relevant for large values of $\abs{n}$. 

The relative importance of this term in the conditions simulated here has been estimated in two ways. First, the $\pphipow$ parameter in \eqref{eq:sim:F3} was varied in order to make the fast ion distribution more or less peaked, thus affecting the magnitude of the gradients in this variable. Recalling that $\pphipow = 6$ was used to most closely match the experimental distribution reconstruction from \TRANSP, it was found that varying $\pphipow = 4 - 7$ changed the growth rate by at most $10\%$, demonstrating that it is usually sub-dominant to the effect of anisotropy. Varying $\pphipow$ further beyond this range was incompatible with the boundary conditions imposed on the self-consistent equilibrium solution. 

\begin{figure}[tb]
\includegraphics[width = \midwidth]{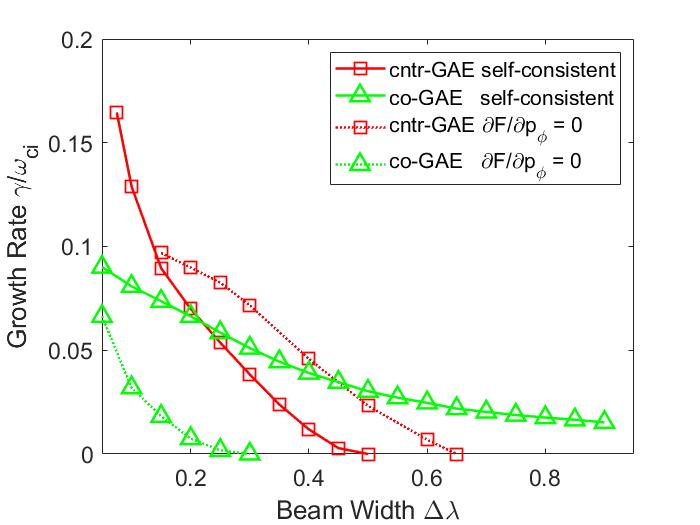}
\caption
[Growth rate as a function of the beam width in velocity space $\dl$ from self-consistent simulations and those excluding the effect of $\partial\fb/\partial\pphi$.]
{Growth rate as a function of the beam width in velocity space $\dl$ from self-consistent simulations (solid lines, reproduced from \figref{fig:sim:dl_scan}) and those excluding the effect of $\partial\fb/\partial\pphi$ (dotted lines). Red curves are for cntr-GAEs, with $\linj = 0.7$, $\vinj = 5.0$, $\vcrit = 0.5$ and restricted to $n = -6$. Green curves are for co-GAEs, with $\linj = 0.1$, $\vinj = 5.0$, $\vcrit = 0.5$, and restricted to $n = 9$.}
\label{fig:sim:nopphi}
\end{figure}

Second, a set of non-self-consistent simulations was run where the $\pphi$ derivative was neglected from the time evolution equation for the particle weights (see \eqref{eq:sim:dwdt}). In essence, this ``turns off'' the effect of $\partial\fb/\partial\pphi$ on the instability. Simulations were conducted with and without this term for cntr- and co-GAEs while varying $\dl$ in order to determine its influence on the growth rate, shown in \figref{fig:sim:nopphi}. The solid curves are the same self-consistent simulation results as shown in \figref{fig:sim:dl_scan}, while the dotted curves are ones where $\partial\fb/\partial\pphi = 0$ is imposed. It is immediately apparent that the gradient in $\pphi$ has a destabilizing effect for co-GAEs and stabilizing effect for cntr-GAEs, just as predicted by \eqref{eq:sim:gammapphi}. Moreover, removing the contribution from $\partial\fb/\partial\pphi$ leads the co-GAE to be stabilized for $\dl > 0.3$, whereas its destabilizing effect supports the instability in self consistent simulations even when $\dl = 0.9$. It is reasonable that this effect might be strong for the co-GAEs which typically have large $\abs{n}$, necessary to generate a sufficiently large Doppler shift $\kpar\vpres \approx n\vpres/R$ to satisfy the anomalous Doppler-shifted cyclotron resonance condition $(\lres = -1)$. 

When the same type of parameter scan was conducted for the co-CAEs shown in \figref{fig:sim:dl_scan}, the modes were either stable or replaced by a more unstable cntr-GAE. This demonstrates that the gradient in $\pphi$ is crucial to scenarios where the co-CAE is the most unstable modes -- it further destabilizes co-CAEs and damps cntr-GAEs that might otherwise have larger growth rate. 

Recall also \figref{fig:sim:quant_cae}, which showed relatively poor agreement between the approximate analytic instability conditions for co-CAEs and simulation results. This discrepancy can be explained by the destabilizing effect of $\partial\fb/\partial\pphi$ on co-propagating modes, which is present in simulations but not in the local analytic theory used to derive the approximate stability boundaries. Hence, the instability condition shown on the plot underestimates fast ion drive for co-CAEs, sometimes enough to incorrectly predict a mode to be stable when it may actually be unstable. A similar but less significant disagreement was found for co-GAEs, as shown on \figref{fig:sim:quant_gae}, which can be understood in the same way. However, the co-GAE's larger growth rates in general (discussed in \secref{sec:sim:simres}) make this correction less likely to make the difference between stability and instability. 

\subsection{Background Damping}
\label{sec:sim:damp}

\begin{figure}[tb]
\includegraphics[width = \midwidth]{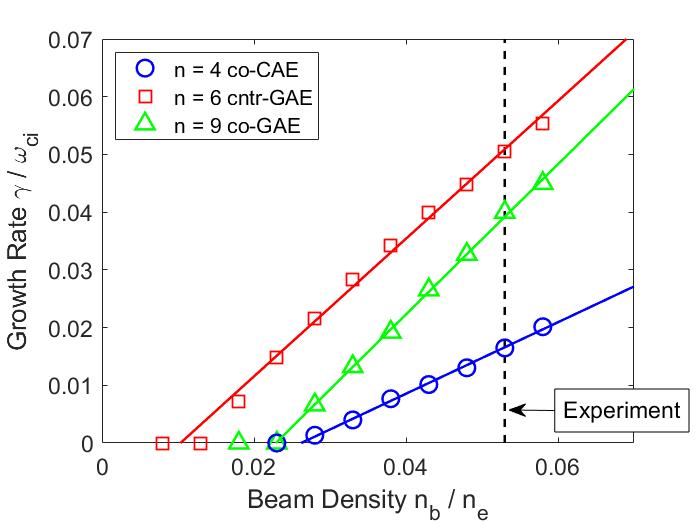}
\caption
[Linear growth rates of representative cases of CAE/GAE modes as a function of beam density.]
{Linear growth rates of representative cases of CAE/GAE modes as a function of beam density. Points are growth rates measured from simulations, whereas the lines are linear fits. Growth rate of zero indicates a simulation with no unstable mode. All simulations used $\vinj = 5.5$. Blue circles are an $n = 4$ co-CAE driven by NBI with $\linj = 0.7$, red squares are an $n = -6$ cntr-GAE for $\linj = 0.7$, and green triangles are an $n = 9$ co-GAE with $\linj = 0.3$ fast ions.}
\label{fig:sim:growth_damping}
\end{figure}

In addition to the drive/damping that comes from the fast ions, the modes can also be damped due to interactions with the thermal plasma. Since the thermal plasma is modeled as a fluid, the simulation will necessarily lack damping due to kinetic effects, such as Landau damping and corrections to continuum damping due to kinetic thermal ions. The total damping rate present in the simulation for a specific mode can be determined by varying the beam density fraction. This is shown in \figref{fig:sim:growth_damping} for one example of a co-CAE, cntr-GAE, and co-GAE. For each mode, there is a critical beam ion density $n_c/n_e$, below which the mode is stable. Above the critical density, the growth rate is proportional to density, as is expected in the perturbative regime where $\abs{\gamma} \ll \omega$. Hence, the relationship $\gamma_\text{net} = \gamma_0(n_b - n_c)/n_e$ is implied, allowing the inference of the damping rate $\gamma_0 n_c/n_e$. These critical densities imply thermal damping rates of $\gammadamp/\omegaci = 0.02 - 0.05$, corresponding to $20 - 60\%$ of the fast ion drive for the case with the nominal experimental fast ion density of $n_b/n_e = 0.053$. 

Given this relatively large damping rate, it is natural to consider its primary source. The resistivity and viscosity in the simulations were varied to determine their influence on the damping. It was found that the damping rate was not very sensitive to either of these quantities. Changing the viscosity by an order of magnitude changed the total growth rate by a few percent, and changing the resistivity had an even smaller effect. Numerical damping could also be present in the simulations, though previous convergence studies of the growth rate for CAEs rules this out as a major source of damping. 

For CAEs, interaction with the \Alfven continuum has been previously identified as the likely dominant damping source, since mode conversion to a kinetic \Alfven wave near the \Alfven resonance location is apparent in the simulations.\cite{Belova2015PRL,Belova2017POP} For the GAEs, the robustness of the damping rate to the viscosity and resistivity also suggest that the primary damping source may be continuum damping, since continuum damping is known to be independent of the independent of the details of the specific damping mechanism.\cite{Tataronis1975JPP,Rosenbluth1992PRL,Rosenbluth1992PFB,Berk1992PFB} However, unlike the CAEs, coupling to the continuum is not always obvious in the mode structure. As investigated previously, this may be due to the intrinsic non-perturbative nature of the GAEs -- they may fundamentally be energetic particle modes excited in the continuum, in which case their interaction with the continuum would be near the center of the mode instead of at its periphery, consequently obscuring the interaction. A more definitive identification of the GAE damping as due to its interaction with the continuum would require the calculation of the \Alfven continuum including the kinetic effects of thermal and fast ions, which is an avenue for further theoretical development. 

It is worthwhile to estimate the magnitude of the absent kinetic thermal damping and compare it to the sources present in the simulations. The thermal ion damping can be neglected because only a very small sub-population will have sufficient energy to resonate with the mode. However, a large number of thermal electrons can interact with the mode. The total electron damping rate in a uniform plasma has been derived in \appref{app:sim:damping} for $\omega\ll\abs{\omegace},\omegape$, generalizing a derivation published by Stix in \citeref{Stix1975NF} which was restricted to $\kpar\ll\kperp$. In contrast, the modes in the simulations have $\krat \leq 3$, violating that condition.  

The general damping rate is given in \eqref{eq:sim:dampfull}, including Landau damping, transit-time magnetic pumping, and their cross term. The standard fast wave Landau damping rate\cite{Stix1975NF} can be obtained in the limit $\kpar\ll\kperp$: 

\begin{align}
\lim_{\alpha\ll 1} \frac{\gammadamp^{\text{CAE}}}{\omega} = -\frac{\betae\sqrt{\pi} y e^{-y^2}}{2}
\end{align}

Here, $y = \omega/\kpar\vtherme$, $\betae = 8\pi P_e / B^2$, and $\alpha = \krat$. 
In the opposing limit of $\kpar \gg \kperp$, \eqref{eq:sim:dampfull} gives 

\begin{align}
\lim_{\alpha\gg 1} \frac{\gammadamp}{\omega} = \frac{\betae\sqrt{\pi} y e^{-y^2}}{2\alpha^2}\frac{1 \pm 2\omegabar + 2\omegabar^2}{(2\pm\omegabar)(1\pm\omegabar)}
\end{align}

Above, the ``$+$'' solution corresponds to compressional modes and the ``$-$'' solution is for shear modes. Hence for both modes, electron damping scales like $\gammadamp \propto \kperp^2/\kpar^2$, reducing damping for modes with larger $\krat$. The general CAE damping rate is mostly sensitive to $\krat$, depending very weakly on $\omeganorm$. The maximum CAE damping rate occurs with a sharp peak at $y = 1/\sqrt{2}$, corresponding to $\alphacrit = \sqrt{2 m_e / m_i \betae} \ll 1$, hence the maximum damping rate is $\gammadamp^{\text{CAE}}/\omega \leq \sqrt{\pi/8e}\betae = 0.38\betae$. However, most CAEs from the simulations have $\alpha\gg\alphacrit$, reducing the expected electron damping rate. For shear modes, the maximum growth rate typically occurs at some $\alpha \like \ord{1}$. Unlike the compressional modes, the damping rate depends strongly on both $\alpha$ and $\omega$. Numerical evaluation of \eqref{eq:sim:dampfull} shows that $\gammadamp^{\text{GAE}}/\omega \leq 0.0021$ for all values of $\betae < 1$. 

To estimate the importance of the electron damping for the modes studied in the \HYM simulations, we can evaluate the electron damping rate expression in \eqref{eq:sim:dampfull} numerically without any approximation, using $\omegabar$ and $\alpha$ for each mode from the simulation, and $\betae = 8\%$ on-axis for each. 
For the GAEs, this exercise shows that the absent electron damping is relatively insignificant -- at most $10\%$ of the net growth rate in the simulation, and in most cases less than $0.1 - 1\%$. In contrast, the continuum damping present in the simulation is approximately $50\%$ of the net growth rate, so the hybrid model in \HYM which ignores kinetic electron effects is well-justified for calculations of GAE stability.

\begin{figure}[tb]
\includegraphics[width = \midwidth]{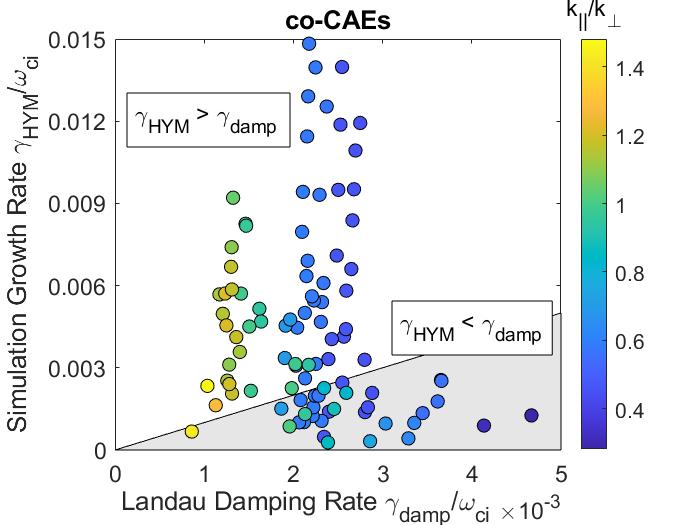}
\caption
[Comparison of co-CAE growth rates from \HYM simulations and analytically calculated electron Landau damping rates.]
{Comparison of co-CAE growth rates from \HYM simulations and analytically calculated electron Landau damping rates. Shaded region indicates where electron damping (absent in simulation) would stabilize the mode. Color shows the value of $\alpha = \krat$ for each mode, calculated from the mode structure in each simulation.}
\label{fig:sim:landau_comp}
\end{figure}  

For the CAEs, the electron damping can be more important, as shown in \figref{fig:sim:landau_comp} where the net drive of each CAE in the simulation is compared to the analytically calculated electron damping rate. The shaded region shows where the electron damping exceeds the net drive in the simulations. This indicates that in some cases the electron damping missing from the hybrid model could be large enough to stabilize a mode that is marginally unstable in the simulations. 

While this section considered the collisionless electron damping in a uniform plasma, further improvements could be made in future work by considering the effects of trapped particles, non-uniform geometry, and collisions. As discussed in \citeref{Gorelenkov2002NF}, the inclusion of trapped particles modifies the damping rate by a factor $0 < y M / \sqrt{1 + y^2 M^2} < 1$ with $M = (1 - \chi_s^2)/\chi_s$ where $\chi_s = \vpar/v$ evaluated at the trapped passing boundary. Since $\chi_s \like r/R$ scales with inverse aspect ratio, the damping rate is reduced in low aspect ratio devices such as NSTX(-U). Work has been done by Grishanov and co-authors\cite{Grishanov2001PPCF,Grishanov2002POP,Grishanov2003PPCF} to develop the theory of electron Landau damping in the context of realistic low aspect ratio geometry, though further work must be done to include transit-time magnetic pumping and cross terms. Lastly, the effect of collisional trapped electron damping has previously been considered for TAEs, which typically enhances the damping rate.\cite{Gorelenkov1992PS} However, these results can not presently be applied outright to GAEs due to the different frequency range, nor to CAEs due to differences in dispersion and polarization. 

\section{Summary and Discussion}
\label{sec:sim:summary}

Hybrid initial value simulations of NSTX-like plasmas were performed in order to investigate the influence of fast ion parameters on the stability and spectral properties of sub-cyclotron compressional (CAE) and global (GAE) \Alfven eigenmodes. The simulations coupled a single fluid thermal plasma to full orbit $\df$ fast ions in order to capture the general Doppler-shifted cyclotron resonance condition which drives the modes. The model NBI distribution included several parameters that were studied: injection geometry $\linj$, normalized injection velocity $\vinj$, normalized critical velocity $\vcrit$, and the degree of velocity space anisotropy $\dl$.  

Depending on the fast ion parameters, the simulations demonstrated unstable co-propagating CAEs, co-propagating GAEs, and cntr-propagating GAEs across many toroidal harmonics $\abs{n} = 3 - 12$ in the frequency range $\omeganorm = 0.05 - 0.70$ with normalized growth rates of $\gamma/\omegaci = 10^{-4} - 10^{-2}$. Modes were identified based on 
comparison with approximate dispersion relations
and inspection of the mode polarization of the fluctuation in the core.
All modes are seen to have strongly mixed polarization in the outboard edge, complicating comparison with experiments which detect the modes primarily with Mirnov coils at the edge. Internal measurements from reflectometry are also available, which can provide information about radial mode structure.\cite{Crocker2011PPCF}

The toroidal direction of propagation of the mode determines the sign of $\kpar$, which was then used to verify the specific resonance driving each mode by inspecting the resonant particles, as determined by those with largest weights. It was confirmed that the co-CAEs are driven by the Landau resonance ($\lres = 0$), cntr-GAEs are driven by the ordinary cyclotron resonance $(\lres = 1)$, and co-GAEs can interact with fast ions both through the anomalous cyclotron ($\lres = -1$) and Landau resonances. 

While both co- and cntr-propagating CAEs have been identified in NSTX experiments,\cite{Fredrickson2013POP} the cntr-CAEs never appear as the most unstable mode in \HYM simulations. This can be explained by the initial value nature of the simulations, which become dominated exclusively by the most unstable mode at long times. Local analytic theory predicts that the beam distributions capable of destabilizing cntr-CAEs can usually also excite a cntr-GAE with larger growth rate, making the cntr-CAE sub-dominant in most cases. 

Both co- and counter-propagating GAEs were unstable in the simulation. The cntr-GAEs have similar frequencies and toroidal harmonics as those observed experimentally in the model discharge for these simulations.\cite{Belova2017POP} Meanwhile, the co-GAEs are higher frequency and have not previously been observed in NSTX, likely due to beam geometry constraints. The co-GAEs should be excitable with the new off-axis beam sources installed on NSTX-U, given a discharge with sufficiently large $\vinj$. 

In simulations, the GAEs behaved more like energetic particle modes than perturbative MHD modes, in that the frequency of the most unstable mode was proportional to the beam ion injection velocity, without corresponding changes in poloidal mode numbers. In contrast, the co-CAEs appeared to be conventional MHD eigenmodes with different frequencies corresponding to distinct eigenstructures, and their poloidal mode structures from \HYM simulations were qualitatively consistent with the spectral MHD code \CAETB. 

Moreover, it was found that a large fraction of unstable modes violated the common large aspect ratio assumption of $\kpar \ll \kperp$, instead having $\krat = 0.5 - 3$. Hence, previous theoretical results which utilized this assumption (and also $\omeganorm \ll 1$) may not be reliably applied, motivating the generalization of local analytic expressions for the growth rate to include these factors. 


The simulations revealed that cntr-propagating GAEs can be excited at a significantly lower $\vinj$ than the co-propagating CAEs and GAEs, which was explained in terms of the different resonance conditions that govern the modes. In terms of beam geometry, it was found that the cntr-GAEs prefer perpendicular injection $(\linj \rightarrow 1)$, the co-GAEs prefer tangential injection $(\linj \rightarrow 0)$, and co-CAEs are most unstable for a moderate value of $\linj \approx 0.5$. Combination of NSTX-U's new tangential beam source with its original more radial one should provide sufficient flexibility to test this growth rate dependency. 

In addition, the GAEs typically have larger growth rates than the co-CAEs by about an order of magnitude. These simulation results are consistent with NSTX observations, where GAEs are more commonly observed. Due to the increased nominal on-axis magnetic field on NSTX-U, this result indicates that typical NSTX-U conditions should favor unstable GAEs over CAEs even more heavily than in NSTX. Early NSTX-U discharges appear to corroborate this conclusion, though confirmation awaits more extensive operations. The maximum growth rate of each type of mode in the simulations increases with increased $\vinj$. 

The simulation results were compared against calculations based on local analytic theory. For typical NSTX NBI parameters, the gradient due to anisotropy dominates, explaining why the different types of modes become most unstable for different ranges of the injection geometry $\linj$. 
%
Compact, approximate instability conditions derived from the local analytic theory were compared against the simulation results. Good agreement was found for the cntr-GAEs and co-GAEs in terms of the beam parameters necessary to drive the modes. The agreement in this area was not as good for co-CAEs, though this is consistent with a more general theory including gradients in the fast ion distribution due to $\pphi$. Inclusion of this term provides an additional source of drive for co-propagating modes (such as the CAEs), while it damps those that cntr-propagate. 


The growth rate dependence on the normalized critical velocity $\vcrit$ and beam anistropy $\dl$ determined in simulations could also be explained by theory. For all modes, increasing the parameter $\vcrit$ led to larger growth rates in simulations. 
Thus the modes might be expected to be more unstable at higher plasma temperatures in proportion to $\sqrt{T_e/\W_\text{beam}}$. 

Larger fast ion anisotropy in velocity space (smaller $\dl$) increases the growth rate in simulations. Theory predicts that the fast ion drive should scale between $\gamma \like 1/\dl$ and $\gamma \like 1/\dl^2$, depending on the size of $\dl$. The scaling inferred from simulations does fall within this range, though not exactly where predicted. Interestingly, it is found that the large $n$ co-GAEs receive substantial drive from $\partial\fb/\partial\pphi$, allowing them to remain unstable when there is much weaker anisotropy than required to drive the cntr-GAEs and co-CAEs. It is also determined that co-CAEs in the experimental range of parameters require drive both from $\lambda$ and $\pphi$ gradients in order to overcome the background damping. The $\pphi$ gradient has a stabilizing effect on cntr-GAEs, though smaller in magnitude than the typical drive from anisotropy. 

Lastly, an assessment of the background damping was made. In simulations, background damping rates of $20 - 60\%$ of the net drive was found for the beam parameters most closely matching the modeled experimental conditions. This damping has been attributed to radiative and continuum damping, as the CAE coupling to the kinetic \Alfven wave (KAW) is clearly visible in the mode structure and Poynting flux.\cite{Belova2015PRL,Belova2017POP} For GAEs the source is less certain, but past theoretical work has concluded that continuum damping is the primary mechanism.\cite{Gorelenkov2003NF} The electron damping absent in the \HYM model has also been estimated analytically, generalizing the well-known expression to arbitrary $\krat$ and $0 < \omeganorm < 1$. It was found that the electron damping was negligible relative to the fast ion drive of GAEs in all cases considered, but could be important for the co-CAEs with smaller growth rates, potentially suppressing some of the marginally unstable co-CAEs found in simulations. 

Together, the large set of simulations combined with their mostly successful theoretical interpretation within a simple theoretical framework helps explain how the spectrum of unstable CAEs and GAEs is influenced by the properties of the fast ion distribution. The  information gathered about the properties of the modes can help guide the notoriously difficult task of distinguishing CAEs from GAEs in experimental observations.\cite{Crocker2013NF} 
With the large beam parameter space available on NSTX-U, the results presented here can be used to guide future experiments to perhaps isolate the effects of CAEs vs GAEs on the enhanced transport. 

A detailed simulation study of multi-beam effects on CAE/GAE stability is a compelling next step, especially when considering the very robust stabilization of GAEs found with the off-axis, tangential beam sources on NSTX-U.\cite{Fredrickson2017PRL,Fredrickson2018NF,Belova2019POP} While this work focused on the influence of fast ion parameters, the stability properties also depend on the equilibrium profiles, which could be explored in future work. Development of a complete nonlocal analytic theory including both fast ion drive and relevant background damping sources would be the next step forward for the local theory used here which works well qualitatively but does not give reliable values for the total growth rate. Ideally, the combination of the present work with these additional steps could enable the development of a reduced model for predicting the most unstable CAE and GAE modes in a given discharge scenario, en route to a more complete understanding of their influence on the electron temperature profiles.

\begin{subappendices}

\section{Growth Rate Scaling with Anisotropy}
\label{app:sim:dl}

The purpose of this derivation is to determine the overall scaling of the growth rate with the beam anisotropy parameter $\dl$, so overall factors multiplying the growth rate will not be accounted for. Consider the contribution from anisotropy alone in \eqref{eq:sim:gammabeam} in the small FLR limit (such that $\J{\pm 1}{G}(\flr) \approx c - \ord{\flr}^2$ with $0 < c < 1/4$ a constant depending on $\omeganorm$ and $\krat$). Similar arguments can be made for the large FLR limit, which does not affect the result. Then recalling the definitions $x = \lambda\omegacires = \vperp^2/v^2$ and $\dx = \dl\omegacires$,  

\begin{align}
\gamma \appropto \frac{1}{C_f(\dx)\dx^2}\int_0^{\xm} \frac{x(x-\xinj)}{(1-x)^2}e^{-(x-\xinj)^2/\dx^2} dx
\end{align}

Here, the upper limit of integration $\xm = 1 - \vpres^2/\vb^2$ is approximated as $\xm \approx \xinj$, approximately corresponding to the condition for largest growth rate. 
%
%
For large $\dx$ such that $\xinj - \dx\sqrt{2} \gtrsim 0$, the Gaussian dependence on $x$ in \eqref{eq:sim:F2} is very weak and can be approximated by a constant, which also removes the $\dx$ dependence from the normalization constant $C_f$. Then we have 

\begin{align}
\gamma \appropto \frac{1}{\dx^2}\int_0^{\xinj} \frac{x(x-\xinj)}{(1-x)^2} dx \like \frac{1}{\dx^2}
\end{align}

Conversely, consider very small $\dx$ where the distribution is so narrow that only the behavior of the integrand very close to $x \approx \xinj$ is relevant. Then the normalization with respect to $\dx$ can be approximated as

\begin{align}
C_f^{-1} = \int_0^1 \frac{e^{-(x-\xinj)^2/\dx^2}}{\sqrt{1-x}}dx \approx \frac{\dx\sqrt{\pi}}{\sqrt{1 - \xinj}}
\end{align}

Subsequent Taylor expansion of the rest of the integrand gives $x(x-\xinj)/(1 - x)^2 \approx \xinj(x-\xinj)/(1 - \xinj)^2$ permits integration:

\begin{align}
\gamma &\appropto \frac{1}{\dx^3}\frac{\xinj}{(1 - \xinj)^2}\int_0^{\xinj} (x - \xinj)e^{-(x-\xinj)^2/\dx^2} dx \\ 
&= \frac{1}{\dx}\frac{\xinj}{2(1 - \xinj)^2}\left(-1 + e^{-\xinj^2/\dx^2}\right) \like \frac{1}{\dx}
\end{align}

Numerical evaluation of the unapproximated analytic expression in \eqref{eq:sim:gammabeam} confirms that the growth rate scales as $1/\dx^2$ for $\dx \gtrsim 0.2$, transitioning to a different asymptotic scaling of $1/\dx$ in the limit of $\dx \ll 1$. Analogous arguments to those given above can also be made for the $\lres = 0$ resonance (relevant for CAEs), which result in the same scalings. So long as $\omegacires$ is not sensitive to $\dl$ (confirmed by simulations), the scaling for $\dx$ is equivalent to the scaling for $\dl$. 

\section{Growth Rate Correction due to Gradients in \texorpdfstring{$\pphi$}{pphi}}
\label{app:sim:pphi}

A general form of the growth rate is given by Kaufman in Eq. 37 of \citeref{Kaufman1972PF} in terms of action angle coordinates as 

\begin{align}
\gamma &\propto \int d\Gamma \left(\vec{\lres}\dot\grad_{\vec{J}}\right) \fb
\label{eq:sim:kaufstart}
\end{align}

Here, $d\Gamma$ is the differential volume of phase space, $\vec{\lres} = \left(\lres, n, \lres_P\right)$ is the vector of integers for the resonance condition, and $\vec{J} = \left(\mu, \pphi, J_P\right)$ is the vector of actions. $J_P$ is a constant motion defined as an integral over poloidal motion (see Eq. 13 of \citeref{Kaufman1972PF}). There are additional terms present in the integrand, but we will ignore those for now since the aim of this section is to obtain qualitative understanding of the effect of the gradient in $\pphi$. The chain rule can be used to transform \eqref{eq:sim:kaufstart} into the variables $(\lambda,\pphi,\W)$: 

\begin{align}
\label{eq:sim:kaufpphia}
\gamma &\appropto \int d\Gamma \left[ \lres \pderiv{\fb}{\mu} + n\pderiv{\fb}{\pphi} + \left(\lres\pderiv{\W}{\mu} + n\pderiv{\W}{\pphi} + \lres_P\pderiv{\W}{J_P}\right)\pderiv{\fb}{\W} \right] \\ 
\label{eq:sim:kaufpphib}
&= \int d\Gamma \left[ \lres \pderiv{\fb}{\mu} + n\pderiv{\fb}{\pphi} + \omega\pderiv{\fb}{\W} \right] \\ 
&= \int d\Gamma  \frac{\omega}{\W}\left[\left(\frac{\lres}{\omegabar} - \lambda\right)\pderiv{\fb}{\lambda} + \W\pderiv{\fb}{\W} + \frac{n}{\omegabar}\frac{\W}{\omegaci}\pderiv{\fb}{\pphi}\right]
\label{eq:sim:kaufpphi}
\end{align}

An alternative (and equivalent) form of the resonance condition was used to simplify from \eqref{eq:sim:kaufpphia} to \eqref{eq:sim:kaufpphib}: $\omega = \lres(\partial\W/\partial\mu) + n(\partial\W/\partial\pphi) + \lres_P(\partial\W/\partial J_P)$, as in Eq. 30 in \citeref{Kaufman1972PF}. The form of \eqref{eq:sim:kaufpphi} is consistent with a similar expression found in \citeref{Wong1999PLA}, \citeref{Coppi1986PF}, \citeref{Gorelenkov1995POP}, and \citeref{Kolesnichenko2006POP}, which use an approximation $m_i \vpar \ll e Z_i \psi'(r)$ in order to re-write $\partial\fb/\partial\pphi \approx - [q / (\omegaci m_i r)](\partial\fb/\partial r)$. Note that \citeref{Kolesnichenko2006POP} uses an opposite convention for the sign of $n$ from what is used in other works (including this one), leading to a relative sign difference. 

\section{Electron Damping Beyond \texorpdfstring{$\kpar \ll \kperp$}{kpar << kperp}}
\label{app:sim:damping}

In this appendix, the electron damping rate for a uniform plasma is generalized from \citeref{Stix1975NF} to include the cases where $\kpar \ll \kperp$ is not satisfied. Consequently, this rate will include Landau damping, transit-time magnetic pumping, as well as their cross term. The normalized damping rate is given by $\gammadamp/\omega = -\Pwave/\omega \Wwave$, where $\Pwave$ is the power density transferred to the particles from the wave, and $\Wwave$ is the wave energy density. To ensure accuracy, the complete two-fluid dispersion instead of the approximate forms $\omega \approx k\va$ (CAEs) and $\omega \approx \kpar\va$ (GAEs) will be used. In a uniform geometry with $B_0$ oriented in the $\hat{z}$ direction, and $\kperp$ in the $\hat{x}$ direction, the cold plasma dispersion is determined by 

\begin{align}
\begin{pmatrix}
\Kxx - \npar^2 & \Kxy & \npar\nperp \\ 
\Kyx & \Kyy - n^2 & 0  \\ 
\npar\nperp & 0 & \Kzz - \nperp^2
\end{pmatrix}
\begin{pmatrix}
E_x \\ E_y \\ E_z 
\end{pmatrix}
= 0
\label{eq:sim:wavetens}
\end{align}

Above, $n = k c / \omega$ is the index of refraction. The directions are defined such that $\vec{k} = \kperp\hat{x} + \kpar\hat{z}$. $\Kij$ are the usual cold plasma dielectric tensor elements.\cite{StixWaves} As explained by Stix, the low frequency, high conductivity limit gives the MHD condition $\Epar \ll \Eperp$. Again taking the low frequency limit ($\omega \ll \omegape,\abs{\omegace}$), defining $\omegabar = \omega/\omegaci$, we have 

\begin{subequations}
\begin{align}
\label{eq:sim:tens-S}
\Kxx &= \Kyy = S \approx \frac{c^2}{\va^2}A \\ 
\label{eq:sim:tens-D}
\Kxy &= -\Kyx = -i D \approx i\omegabar\frac{c^2}{\va^2}A \\ 
\label{eq:sim:tens-A}
\text{where } A &\defined \frac{1}{1 - \omegabar^2}
\end{align}
\label{eq:sim:wavetens-cold}
\end{subequations}

Define the \Alfven refractive index $N = k\va/\omega$, $F^2 = \kpar^2/k^2$, and $G = 1 + F^2$. Then the two-fluid coupled dispersion is readily found by neglecting $E_z$: 

\begin{align}
N^2 &= \frac{A G}{2 F^2}\left[1 \pm \sqrt{1 - \frac{4 F^2}{A G^2}}\right]
\label{eq:sim:stixdisp}
\end{align}

For $\omega < \omegaci$, the ``$+$'' solution corresponds to the shear wave, and ``$-$'' solution to the compressional wave. For $\omega > \omegaci$, the shear wave does not propagate, and the ``$+$'' solution corresponds to the compressional wave. The polarization will be needed to compute $\Pwave$ and $\Wwave$. The second line of \eqref{eq:sim:wavetens} gives 

\begin{equation}
H \defined i\frac{E_x}{E_y} = -\frac{1}{\omegabar}\left(\frac{N^2}{A} - 1\right)
\label{eq:sim:Hexp}
\end{equation}

While $\Kzy$ and $\Kzz$ were not needed to calculate the cold dispersion, their finite temperature forms are needed to accurately capture the $\Epar$ effects. In the limit of $\omega^2/\kpar^2\vtherme^2 \ll 1$, which is the regime studied here, the relevant finite temperature tensor elements are 

\begin{align}
\Kzy &\approx \frac{i\kperp\omegape^2}{\kpar\omega\abs{\omegace}} = \frac{i}{\alpha\omegabar}\frac{c^2}{\va^2} \\ 
\Kzz &\approx \frac{1}{\kpar^2\ldebye^2} = \frac{2}{\Npar^2\omegabar^2\betae}\frac{c^2}{\va^2}
\end{align}

Note $\omegape^2 = 4\pi n e^2 / m_e$ is the electron plasma frequency, $\ldebye^{-1/2} = 4\pi n e^2/T_e$ is the Debye Length, $\vtherme = \sqrt{2 T_e/m_e}$ defines the thermal electron velocity, and $\betae = 8\pi n_e T_e / B^2$ is the electron pressure normalized to the magnetic pressure. As elsewhere in the thesis, $\alpha = \kpar/\kperp$. Note the following relations which are useful for simplifying expressions above and below: $\omegaci\vtherme^2 = \betae\abs{\omegace}\va^2$, $\omegape^2\va^2 = c^2\omegaci\abs{\omegace}$, and $2\ldebye\omegape = \vtherme^2$. The first and third lines of \eqref{eq:sim:wavetens}, modified to include the hot forms of $\Kzy$ and $\Kzz$, implies 

\begin{align}
J &\defined \frac{E_z}{E_y} = \frac{-\npar\nperp \Kxy + (\Kxx - \npar^2)\Kzy}{\npar^2\nperp^2 - (\Kxx - \npar^2)(\Kzz - \npar^2)} \\ 
&= i\left[\frac{-\omegabar A \Nperp\Npar + (A - \Npar^2)/\alpha\omegabar}{\Nperp^2\Npar^2 - (A - \Npar^2)\left(\frac{2}{\beta\omegabar^2\Npar^2} - \Nperp^2\right)}\right]
\label{eq:sim:Jexp}
\end{align}

The absorbed power density $\Pwave$ is given by 

\begin{align}
\Pwave &= -\frac{i\omega}{16\pi}\vE\conj\dot\Ksym\dot\vE + \text{ c.c.} \\
&= -\frac{i\omega}{8\pi}\Re{\vE\conj\dot\Ksym\antiherm\dot\vE} \\ 
&= -\frac{i\omega\abs{E_y}^2}{8\pi}\left[\Kyy\antiherm + 2 i \Kyz\antiherm\Im{J} + \Kzz\antiherm\abs{J}^2\right] 
\label{eq:sim:pwave-raw}
\end{align}

Above, $\Ksym$ is the hot dielectric tensor and $\Ksym\antiherm$ is its anti-Hermitian part. We also define $\vE$ such that the real wave field $\vE_0 = \Re{\vE e^{i(\vk\dot\vec{x} - \omega t)}}$. Only the nonzero terms for the $\omega - \kpar\vpar = 0$ resonance are kept since $\omega\ll\abs{\omegace}$. Contributions from higher resonances will be smaller by a factor of $\exp(-\omegace^2/\kpar^2\vtherme^2) \ll 1$. The anti-Hermitian tensor elements are given in Eq. 12 in \citeref{Stix1975NF} and then simplified as 

\newcommand{\Gstix}{Q}
\begin{align}
\Kyy\antiherm &= \frac{i\kperp^2\vtherme^2}{\omega\abs{\omegace}}\Gstix
= i\omegabar\Nperp^2\betae \Gstix \\ 
\Kyz\antiherm &= \frac{\Gstix}{\alpha} \\ 
\Kzz\antiherm &= \frac{2i\omega\abs{\omegace}}{\kpar^2\vtherme^2}\Gstix
= \frac{2i \Gstix}{\Npar^2 \omegabar\betae} \\ 
\text{where }\Gstix &= \frac{\sqrt{\pi}\omegape^2 e^{-y^2}}{\abs{\omegace}\kpar\vtherme} 
= \sqrt{\pi}\frac{c^2}{\va^2}\frac{y}{\omegabar}e^{-y^2} \\ 
\text{and }y &= \frac{\omega}{\kpar\vtherme} 
\end{align}

Substitution into \eqref{eq:sim:pwave-raw} yields 

\begin{align}
\Pwave &= -\frac{\omega\abs{E_y}^2}{4\sqrt{\pi}}\frac{c^2}{\va^2\omegabar}y e^{-y^2}
\left(\frac{\betae\omegabar\Nperp^2}{2} + \frac{\abs{J}^2}{\betae\omegabar\Npar^2} + \frac{\Im{J}}{\alpha}\right)
\label{eq:sim:Pexp}
\end{align}

We also need the wave energy density $\Wwave$ since $\gammadamp = -\Pwave/\Wwave$. It is defined as 

\begin{align}
\Wwave &= \frac{1}{16\pi}\left[\abs{B}^2 + \vE\conj\dot\pderiv{\left(\omega\Ksym\herm\right)}{\omega}\dot\vE\right]
\end{align}

Above, $\Ksym\herm$ is the Hermitian part of the cold dielectric tensor written in \eqref{eq:sim:wavetens-cold}. For the magnetic field part, use Faraday's Law $c\curl\vE = -\partial\vB/\partial t$. Then this may be evaluated as 

\begin{multline}
\Wwave = \frac{\abs{E_y}^2}{16\pi}\frac{c^2}{\va^2}\left\{N^2(1 + H^2 F^2) 
\vphantom{\frac{A^2}{2}\left[(1+H)^2(1-\omegabar)^2 + (1-H)^2(1+\omegabar)^2\right]}
+ \frac{A^2}{2}\left[(1+H)^2(1-\omegabar)^2 + (1-H)^2(1+\omegabar)^2\right]\right\}
\label{eq:sim:Wexp}
\end{multline}

Combination of \eqref{eq:sim:Pexp} and \eqref{eq:sim:Wexp}, along with the definitions of $N$ in \eqref{eq:sim:stixdisp}, $H$ in \eqref{eq:sim:Hexp}, and $J$ in \eqref{eq:sim:Jexp}, gives the total damping rate below

\begin{multline}
\frac{\gammadamp}{\omega} = -\frac{4\sqrt{\pi}}{\omegabar}y e^{-y^2} 
\left(\frac{\betae\omegabar\Nperp^2/2 + \abs{J}^2/(\betae\omegabar\Npar^2) + \Im{J}/\alpha}
{N^2(1 + H^2 F^2) + A^2\left[(1+H)^2(1-\omegabar)^2 + (1-H)^2(1+\omegabar)^2\right]/2}\right)
\label{eq:sim:dampfull}
\end{multline}

This is a general expression for the total electron damping rate for compressional and shear \Alfven waves when $y \ll 1$ and $\omega \ll \abs{\omegace},\omegape$. It depends on the mode type (compressional vs shear dispersion), frequency $(\omegabar = \omeganorm)$, wave vector direction $(\alpha = \krat)$, and electron pressure $(\betae = 8\pi P_e / B^2)$. 

In order to recover the standard fast wave Landau damping rate in the limit of $\kpar\ll\kperp$, approximate $\Npar^2 \ll \Nperp^2 \approx 1$. Then it follows that $H \approx \omegabar$ and also $J \approx -2i\Npar\Nperp\omegabar\betae$ such that

\begin{align}
\lim_{\alpha\ll 1} \frac{\gammadamp^{\text{CAE}}}{\omega} = -\frac{\betae\sqrt{\pi} y e^{-y^2}}{2}
\end{align}

This is the familiar formula from \citeref{Stix1975NF}. 
The damping rate can also be simplified in the complementary limit of $\kpar \gg \kperp$. In this limit, one can approximate $\Npar^2 \approx 1/(1 \pm \omegabar)$, where the ``$+$'' solution corresponds to CAEs and the ``$-$'' solution is for GAEs. Consequently, $H \approx \pm 1$ and $J \approx -i\omegabar\betae/(2\alpha(1 \pm \omegabar)^2$). Then we find

\begin{align}
\lim_{\alpha\gg 1} \frac{\gammadamp}{\omega} = \frac{\betae\sqrt{\pi} y e^{-y^2}}{2\alpha^2}\frac{1 \pm 2\omegabar + 2\omegabar^2}{(2\pm\omegabar)(1\pm\omegabar)}
\end{align}

\end{subappendices}

\newcommand{\dirfigepgae}{ch-epgae/figs}

\chapter{Energetic-Particle-Modified Global \Alfven Eigenmodes}
\label{ch:epgae:epgae}

\section{Introduction}
\label{ch:epgae:intro}

Linear 3D hybrid simulations presented here demonstrate that the high frequency shear \Alfven waves excited in NSTX conditions can be strongly nonperturbative -- a fact that has not been recognized before. Consequently, this mode could be considered an energetic particle mode, or an energetic-particle-modified global \Alfven eigenmode (\EGAE). This is primarily concluded due to large changes in the frequency of the most unstable mode in proportion to the maximum energetic particle velocity without clear corresponding changes in the mode structure or location tracking the minimum of the \Alfven continuum. 
This behavior is pervasive for both co- and counter-propagating modes for all examined toroidal mode numbers, $\abs{n} = 4 - 12$. If the resonant value of $\vpar$ is proportional to the injection velocity $\vb$, then the large frequency changes can be qualitatively explained by the resonance condition. The most unstable mode frequency is determined to a large degree by features of the energetic particle population, in addition to properties of the thermal plasma -- a key signature of energetic particle modes (EPM).\cite{Heidbrink2008POP} 
These may be the first example of EPM-type fluctuations that are excited at a significant fraction of the ion cyclotron frequency, typically $\omeganorm \approx 0.1 - 0.5$. The goal of this chapter is to study the properties of unstable {\EGAE}s in simulations, in order to guide future theoretical studies of these modes and enable experimental tests of their distinguishing features. 

This chapter is organized as follows. The primary simulation results which this chapter seeks to explain are detailed in \secref{sec:epgae:freq}. The relative importance of changes to the equilibrium versus changes to the fast ions in accounting for this effect is investigated in \secref{sec:epgae:eqps}. The poloidal mode structure of the excited modes is shown for a range of EP energies in \secref{sec:epgae:struct}, and the frequency of the most unstable mode for a wide variety of beam parameters is compared against the shear \Alfven dispersion relation. Lastly, the characteristics of the resonant particles are examined in \secref{sec:epgae:res} as a function of the injection energy in order to clarify the role that the resonant wave-particle interaction plays in setting the frequency of the most unstable mode. A summary of the key results and discussion of implications for NSTX-U is given in \secref{sec:epgae:discussion}. The majority of the content of this chapter has been peer-reviewed and published in \citeref{Lestz2018POP}.

\section{Frequency Dependence on Fast Ion Parameters}
\label{sec:epgae:freq}

This section describes results obtained from the self-consistent hybrid simulations, \eg those with an equilibrium that self-consistently includes fast ion effects.\cite{Belova2003POP} Since linear initial value simulations are conducted, only the mode with the largest growth rate can be seen. Consequently, the results in this section represent the properties of the most unstable mode in each simulation. A filter for a single toroidal harmonic is imposed on the simulation so that many distinct eigenmodes can be studied independently. The simulations are conducted with the \HYM code, described in \secref{sec:sim:model}, using the same model fast ion distribution function as given in \eqref{eq:sim:F0}. 

Each of the simulations is based on the conditions of the well-analyzed NSTX H-mode discharge 141398,\cite{Fredrickson2013POP,Crocker2013NF} which has nominal experimental beam parameters of $\nb = 0.053$ and $\vinj = 4.9$, while $\linj = 0.7, \dl = 0.3, \vc = \vb/2,$ and $\pphipow = 6$ are  chosen to reproduce the beam ion distribution function calculated by \NUBEAM. In ordinary NSTX operations, $\vinj = 3 - 6$ and $\linj \approx 0.5 - 0.7$, while early NSTX-U experiments had $\vinj = 1 - 3$ with $\linj \approx 0$ from the new tangential beam sources in addition to the original perpendicular sources. In this set of simulations, the normalized injection velocity $\vinj$ and the injection geometry $\linj$ of the energetic particle distribution are varied in order to explore their effect on characteristics of the excited sub-cyclotron modes. 

Generally, unstable modes in the simulations are identified as GAEs instead of CAEs when $\dbperp \gg \dbpar$ near the plasma core. This identification is supported by previous cross validation between experiment, \HYM, and the \NOVA eigenmode solver.\cite{Gorelenkov2003NF,Gorelenkov2004POP,Crocker2013NF,Crocker2017IAEA} The modes identified as GAEs have linear growth rates ranging from $\gamma/\omegaci = 0.1 - 5\%$, with most around $1\%$ or less. Normalized instead to the mode frequency yields $\gamma/\omega = 1 - 20 \%$, with a few percent typical. 

Unexpectedly, the frequency of the most unstable GAE for a single toroidal harmonic changes significantly as the energetic particle distribution is changed from one simulation to the next. The change in frequency is not usually accompanied by significant changes in the mode structure. Most notably, varying the injection velocity by a factor of two results in a factor of two change in the mode frequency. Since these modes are a non-negligible fraction of the cyclotron frequency ($\omeganorm \approx 0.1 - 0.5$), this can represent a dramatic change in frequency of hundreds of kilohertz. As GAEs are expected to have frequencies slightly below a minimum of the \Alfven continuum, such large changes in frequency with beam parameters clashes with their orthodox MHD description. In contrast, CAEs excited in similar simulations do not exhibit this same strong frequency dependence on fast ion parameters. Instead, the frequency of the most unstable CAE is nearly constant except for jumps in frequency at specific values of $\vinj$, which are also accompanied by a clear change in poloidal mode number.

\begin{figure}[tb]
\includegraphics[width = \midwidth]{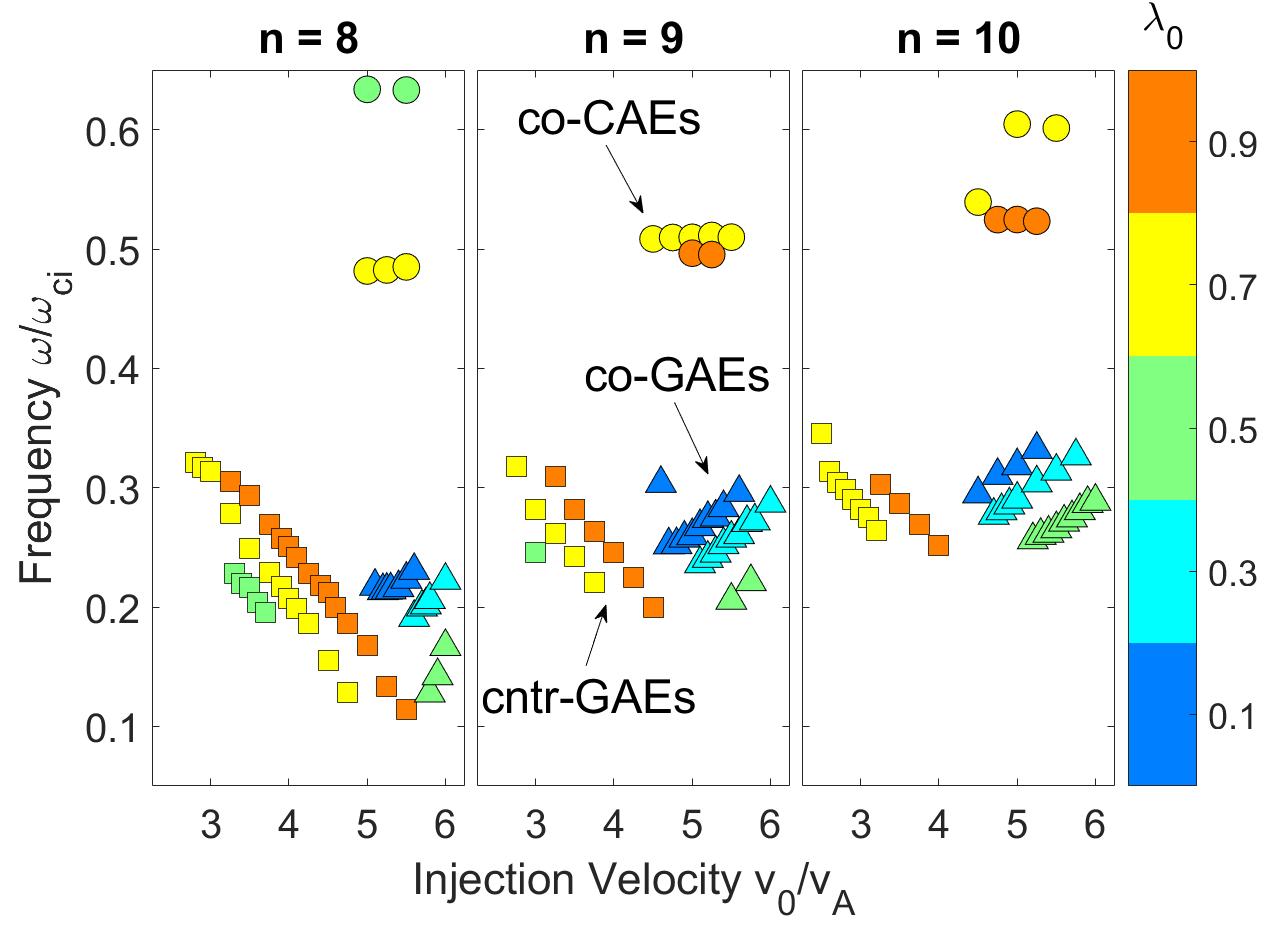}
\caption
[Frequency of the most unstable modes for $\abs{n} = 8 - 10$ as a function of normalized injection velocity $\vinj$.]
{Frequency of the most unstable modes for $\abs{n} = 8 - 10$ as a function of normalized injection velocity $\vinj$. Each plot shows modes for a single toroidal mode number $\abs{n}$, with cntr-GAEs marked by squares, co-GAEs marked by triangles, and co-CAEs by circles. Color denotes the injection geometry $\linj$ of the fast ion distribution in each individual simulation.}
\label{fig:epgae:all_shifts}
\end{figure}

For sufficiently large beam injection velocities, GAEs propagating both with and against the direction of plasma current/beam injection are found to be unstable in the simulations. Analysis of the wave-particle interactions shows that co-GAEs and cntr-GAEs are driven by the Doppler-shifted cyclotron resonance with $\lres=-1$ and $\lres=1$, respectively. Counter-propagating GAEs are commonly observed in NSTX discharges while the co-propagating GAEs are yet to be detected. This is primarily due to geometric constraints of the neutral beam sources, since the co-GAEs are typically excited in the simulations when the energetic particle population has very low values of $\linj \lesssim 0.5$, whereas the typical regime for NSTX is $\lambda \approx 0.5 - 0.7$. The additional beam sources on NSTX-U are more tangential and thus different beam mixtures could potentially excite modes propagating in either direction in future experiments, given sufficiently large $\vinj$.

For cntr-GAEs, the frequency of the most unstable mode decreases as injection velocity increases, whereas it increases for co-GAEs. \figref{fig:epgae:all_shifts} shows how the frequency changes with the normalized injection velocity $\vinj$ for each toroidal mode number $\abs{n} = 8 - 10$, where both co- and counter-propagating GAEs are excited in this set of simulations. Each point on the figure represents an individual simulation conducted with the energetic particle distribution from \eqref{eq:sim:F0} parameterized by values of $\left( \vinj, \linj \right)$ in a 2D beam ion parameter scan. For each distribution, the equilibrium is re-calculated to self-consistently capture the EP effects on the thermal plasma profiles. It is clear that the frequency of the most unstable mode in each simulation depends linearly on the injection velocity, except for some outliers near marginal stability. The injection geometry $\linj$ of the distribution also impacts the frequency, though this effect is not as pronounced. Especially noteworthy is the continuous nature of the change in frequency with injection velocity. 

Even at the smallest investigated increments of $\Delta\vinj = 0.1$, the change in frequency remains proportional to the change in injection velocity. This suggests the existence of either a continuum of modes which are being excited or very densely packed discrete eigenmodes. In the case of discrete eigenfrequencies, one would expect to see a discontinuous ``staircase" pattern in the frequency of the most unstable mode as a function of the injection velocity; a single discrete eigenmode with constant frequency would be the most unstable for some range of $\vinj$, with a jump to a new frequency when a different discrete mode becomes more unstable for the next velocity range. However, this is not what is observed, at least to the resolution of $\Delta\vinj = 0.1$. Overall, GAEs propagating with or against the plasma current exhibit a change in frequency proportional to the change in the normalized injection velocity of the energetic particles. The direction of this change matches the sign of $\kpar$, implicating the Doppler shift in the resonance condition as the likely explanation.

\begin{figure}[tb]
\includegraphics[width = \fullwidth]{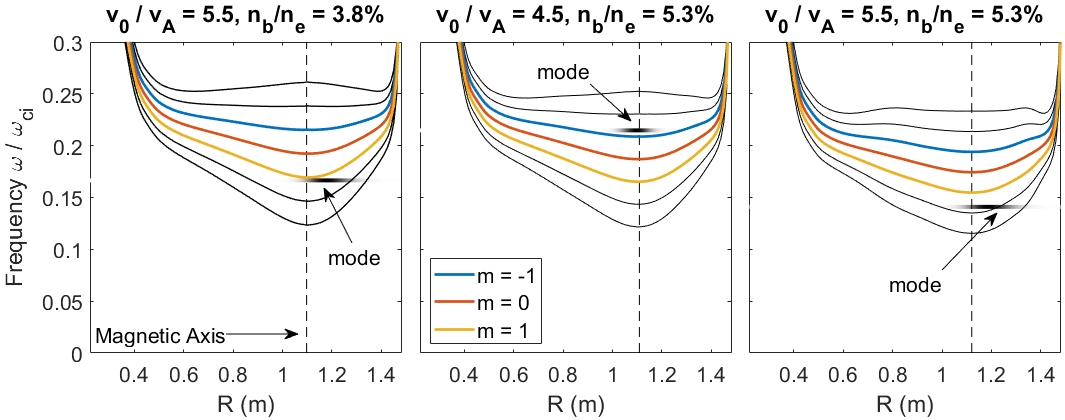}
\caption
[\Alfven continuum for $\abs{n} = 6$, including poloidal harmonics with $\abs{m} \leq 3$, for the self consistent equilibrium with different beam parameters.]
{\Alfven continuum for $\abs{n} = 6$, including poloidal harmonics with $\abs{m} \leq 3$, for the self consistent equilibrium with different beam parameters -- left: $\vinj = 5.5$, $\nb = 3.8\%$, center: $\vinj = 4.5, \nb = 5.3\%$, right: $\vinj = 5.5, \nb = 5.3\%$. The thick horizontal lines mark the frequency and location of the mode excited in each simulation, where the darkness is proportional to the average amplitude of $\dbperp$. The $m = -1,0,1$ branches of the continuum are labeled for reference.}
\label{fig:epgae:continuum}
\end{figure}

Moreover, these modes are global eigenmodes in the sense that the fluctuations oscillate at the same frequency at all points in space, and that the mode structure is converged at long times (once the mode has grown long enough to dominate the initial random perturbations). Comparing the location of these modes relative to the \Alfven continuum can also help elucidate the character of these modes. Since these modes have been identified as GAEs in previous experimental and numerical analysis, one would expect them to be radially localized near a local minimum of the continuum with frequency near that value. For example, previous \HYM simulations of a separate NSTX discharge with smaller $\nb$ demonstrated excitation of a GAE with the expected characteristics, in particular with a frequency just below a minimum of the \Alfven continuum.\cite{Gorelenkov2003NF,Gorelenkov2004POP} If instead the modes substantially intersect the continuum, strong continuum damping would make their excitation unlikely, or suggest that they may not be shear \Alfven eigenmodes at all. The continuum is calculated using the $q(r)$ and $n(r)$ profiles from the self-consistently calculated equilibrium for three separate cases, and shown in \figref{fig:epgae:continuum}. The left-most case has $\vinj = 5.5, \,nb = 3.8\%$, and the mode peaks quite close to an on-axis minimum of the continuum. In the middle figure, $\vinj = 4.5,\,\nb = 5.3\%$, and the GAE actually occurs above the minimum, but nonetheless avoids intersecting the continuum due to its limited radial extent. The right-most case is $\vinj = 5.5,\, \nb = 5.3\%$, and moderately overlaps the continuum. These examples demonstrate that as the relative fast ion pressure becomes larger, either through increased density or energy, the modes can depart from their textbook description. 

A limitation of this analysis is that kinetic corrections to the MHD continuum could become important for an accurate comparison in this regime. For instance, Kuvshinov has shown that in a single fluid Hall MHD model, the kinetic corrections to the shear \Alfven dispersion due to finite Larmor radius effects is $n_b \kperp^2\rho_\perp^2 / n_e (1 + \kperp^2\rho_\perp^2)$, which is equivalent to a Pad{\'e} approximation to the full ion-kinetic response.\cite{Kuvshinov1994PPCF} Near peak beam density, $n_b/n_e$ can approach $20\%$ in these simulations, and large fast ion energies can yield $\kperp\rho_\perp \approx 2$, which yields a roughly $15\%$ correction from this term. Developing a model of the continuous spectrum including fast ions self-consistently would make this comparison more definitive, but is beyond the scope of this work, as it represents a quite substantial enterprise itself. 

\section{Equilibrium vs Fast Ion Effects}
\label{sec:epgae:eqps}

The purpose of this section is to determine numerically if these large changes in frequency (as large as $20 - 50\%$, or $100 - 500$ kHz) can be explained by energetic particle effects, or if they can be interpreted some other way. Since the preceding results were from simulations which included EP effects self-consistently in the equilibrium, one possible explanation is that increasing the beam energy is modifying the equilibrium (and \Alfven continuum), indirectly changing the characteristic GAE frequency. While $\nb$  is small (of order $5\%$) in these simulations, the fast ion current can be comparable to the thermal plasma current due to large beam energies. Previous work has demonstrated the substantial effects that the beam contribution can have on the equilibrium.\cite{Belova2003POP} Moreover, there is recent work showing that the inclusion of alpha particles can significantly deform the \Alfven continuum.\cite{Slaby2016POP} It is important to investigate if these changes in frequency can be attributed to changes in the self-consistent equilibrium or changes in the fast particles driving the mode, independent of the equilibrium. The latter would be typical of nonperturbative energetic particle modes while the former would fit well with an MHD description of GAEs. 

\subsection{Equilibrium Effects}
\label{sec:epgae:eq}

In order to distinguish between these competing interpretations, these simulations were first reproduced at decreased EP density, since this decreases the ratio of the beam current to thermal plasma current, which is the key parameter controlling the impact of EP effects on the equilibrium profiles. These additional simulations are conducted for representative examples of both counter- and co-propagating GAEs. In the former case, an $n = 6$ mode driven by a beam distribution parameterized by $\vinj = 5.5, \linj = 0.7$ is studied, and for the latter, an $n = 9$ mode driven by a $\vinj = 5.5, \linj = 0.3$ distribution is selected. By varying $\nb$ with fixed $\vinj$ and combining with the previous simulation results which were conducted for constant $\nb$ and varying $\vinj$, the frequencies can be plotted against $\Jc \equiv \nv \propto \Jnorm$. If the frequency depends on this parameter in the same way in both sets of simulations, then it can be concluded that the large changes in frequency of the GAEs seen in the simulation are due to the EP-related changes to the equilibrium. 

\begin{figure}[tb]
\subfloat[\label{fig:epgae:sc_mhd_cntr}]{\includegraphics[width = \halfwidth]{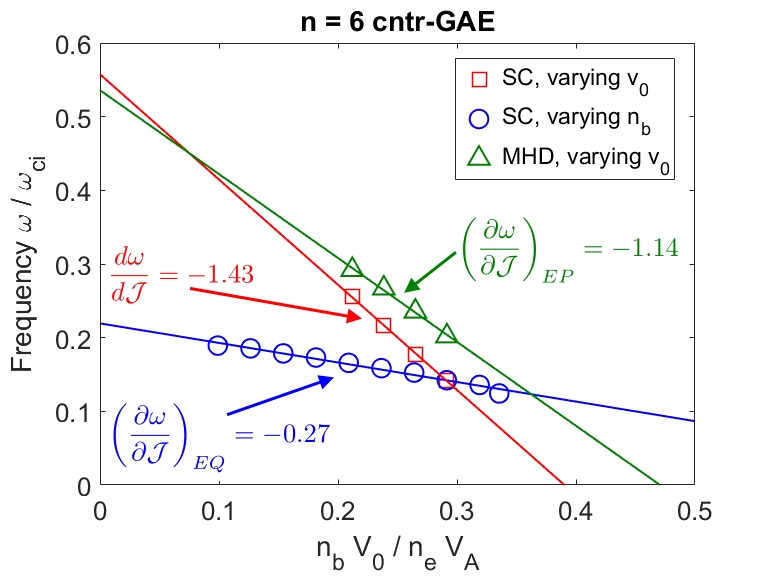}} 
\subfloat[\label{fig:epgae:sc_mhd_co}]{\includegraphics[width = \halfwidth]{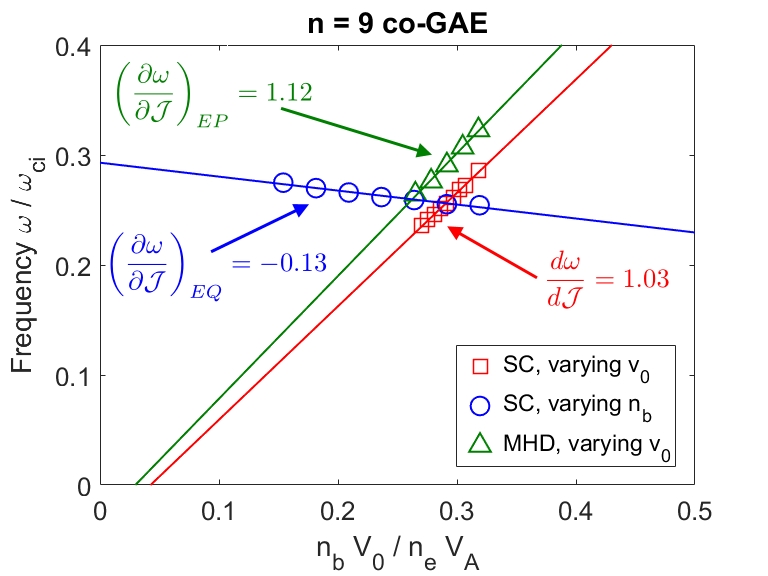}}
\caption
[Frequency changes of modes as $\Jc = \nv \propto \Jnorm$ is varied under different conditions.]
{Frequency changes of modes as $\Jc = \nv \propto \Jnorm$ is varied under different conditions. (Red) Equilibrium includes EP self-consistently (``SC"); injection velocity $\vinj$ is varied while beam density $\nb$ is constant. (Blue) SC equilibrium; $\nb$ is varied, $\vinj$ is constant. (Green) Equilibrium determined without EP contributions (``MHD-only"); $\vinj$ is varied, $\nb$ is fixed. (a) counter-propagating $n = 6$ mode. (b) co-propagating $n = 9$ mode.}
\label{fig:epgae:sc_mhd}
\end{figure}

The results of this comparison are shown in \figref{fig:epgae:sc_mhd_cntr} for the cntr-GAE modes and \figref{fig:epgae:sc_mhd_co} for the co-GAEs. The red squares are simulations with fixed beam density and differing injection velocity (same conditions as those shown in \figref{fig:epgae:all_shifts}) whereas the blue circles show simulations where the EP distributions share a single value of $\vinj$ and have varying $\nb$. For both co- and cntr-GAEs, increasing beam density results in a modest decrease in mode frequency. This likely reflects changes in the equilibrium, and is supported by work done by Slaby \etal which found that the continuum frequencies are decreased in the presence of increased alpha particle pressure.\cite{Slaby2016POP} Also apparent in this comparison is that the mode has a different stability threshold in $\Jc$ depending on if $\Jc$ is decreased through $\nb$ or $\vinj$, as the mode can still exist for small $\Jc$ provided that $\vinj$ is sufficiently large. The mode frequency exhibits a linear dependence on EP density, with the slope for the two modes studied differing by a factor of two. The change in frequency due to this effect is less than $20\%$ of the magnitude of the change due to changing beam energy at constant beam density. Moreover, it has the opposite sign of that seen in the first set of simulations for the co-GAEs, which increase in frequency as $\vinj$ increases. These results demonstrate that changes to the equilibrium, proportional to $\Jnorm$, are not the primary cause of the large changes in frequency.

\subsection{Fast Ion Effects}
\label{sec:epgae:ps}

Since the previous results suggest that the frequency changes can not be an equilibrium effect alone, the direct effects of the energetic particles should be isolated from the changes in the equilibrium. To do this, complementary simulations are conducted where the equilibrium is no longer calculated self-consistently to include the beam contribution. Instead, the equilibrium is solved for considering only the effects of the thermal plasma. This ``MHD-only" equilibrium is calculated with the same total current as the self-consistent one, and the plasma pressure is set to be comparable to the total thermal and beam pressure. These simulations will serve as a definitive test of the effects of the different energetic particle parameters on the excited mode frequency and structure for a single, fixed equilibrium. 

The simulations are repeated for the same $n = 6$ counter- and $n = 9$ co-propagating GAEs as introduced in \secref{sec:epgae:freq}. The results correspond to the green triangles on \figref{fig:epgae:sc_mhd}. The simulations with the fixed ``MHD-only" equilibrium and changing beam velocity reproduce the trend and approximate magnitude of the frequency shifts observed in simulations with the self-consistent equilibria (labeled ``SC" on the figure) for both the $n=6$ cntr-GAEs and $n=9$ co-GAEs. In order to distinguish between the various frequency dependencies, the following conventions are adopted for the different types of simulations conducted. $d\omega/d\Jc$ is the slope of the most unstable mode frequency with respect to $\Jc$ for simulations conducted with self-consistent equilibria and varying $\vinj$, which are the red squares on \figref{fig:epgae:sc_mhd}. These simulations represent the total frequency dependence on $\Jc$ since the changes to $\vinj$ alter both the equilibrium profiles and the location of resonant particles in phase space (detailed in \secref{sec:epgae:res}). Changes in frequency in simulations with self-consistent equilibria with varying $\nb$ only, the blue circles, are purely due to changes in the equilibrium, so that slope is labeled as $\left(\partial\omega/\partial\Jc\right)_\text{EQ}$. Varying $\vinj$ for a fixed MHD-only equilibrium is a pure energetic particle effect on the frequency, associated with $\left(\partial\omega/\partial\Jc\right)_\text{EP}$ and shown as the green triangles. The effects on the GAE frequencies due to equilibrium and energetic particle effects appear to be nearly linear, succinctly stated in \eqref{eq:epgae:dflin}, which is accurate to within $5\%$ for the two cases studied in \figref{fig:epgae:sc_mhd}. This further supports that there are two independent factors determining the GAE frequency, and that the nonperturbative energetic particle influence on the mode dominates over the effects due to EP-induced changes to the equilibrium. 

\begin{equation}
\label{eq:epgae:dflin}
\frac{d \omega}{d \Jc} \approx 
\left(\frac{\partial\omega}{\partial \Jc}\right)_\text{EQ} + 
\left(\frac{\partial\omega}{\partial \Jc}\right)_\text{EP} = 
\neo\va
\left[
\frac{1}{\vb}\frac{\partial\omega}{\partial \nbo} + 
\frac{1}{\nbo}\frac{\partial\omega}{\partial\vb}
\right]
\end{equation}

For completeness, a final set of ``MHD-only" simulations were conducted where the beam energy is fixed and the beam density is varied. The changes in frequency due to varying this parameter are much smaller than any other, though they imply a negative partial derivative for both types of modes, similar to the SC EQ effect. This effect is labeled NR for non-resonant since it results from changes to the energetic particles, but not how they resonantly interact with the mode. It can be attributed to the small change of the continuum frequencies due to the change in total density when $\nbo$ is changed. For small $\nb$, this can be estimated as $\left.\partial\omega/\partial\Jc\right|_{\vinj} = -(\va/2v_0)\left(\kpar B_0 / \sqrt{n_e}\right)$ which evaluates to a slope of approximately $-0.02$ for the cntr-GAE case and $-0.03$ for the co-GAE case, which are of the right magnitude to explain the effect shown in the figure, and also very close to the less than $5\%$ discrepancy in \eqref{eq:epgae:dflin}. The relative magnitudes of these different effects are summarized in \eqref{eq:epgae:dftot}.  

\begin{equation}
\label{eq:epgae:dftot}
\Delta\omega \approx 
\left(\Delta\omega\right)_\text{EP} \gg 
\left(\Delta\omega\right)_\text{EQ} \gg 
\left(\Delta\omega\right)_\text{NR}
\end{equation} 

\section{Mode Structure and Dispersion}
\label{sec:epgae:struct}

In order to determine if these are ideal MHD eigenmodes or strongly energetic-particle-modified modes such as EPMs, inspection of the mode structure is necessary. If MHD modes, one would expect that changes in frequency would be associated with some qualitative change in mode structure, such as the presence of different poloidal or radial harmonics, marking a new eigenmode. Conversely, in a nonperturbative energetic particle regime, the mode structure can be preserved even as the frequency changes significantly, such as in the theory and observation of chirping modes\cite{Fredrickson2006POP,Podesta2012NF,Duarte2017POP} or in the case of fishbones.\cite{McGuire1983PRL,Chen1984PRL} In these simulations of GAEs, the mode structure is frequently qualitatively unaffected by the large changes in frequency which accompany changes in the normalized EP beam energy. Quantitative changes are typically subtle, including slight changes in radial location, mode width, or elongation. A key difference between chirping modes, fishbones, and the GAEs studied here is that the first two fundamentally involve nonlinear physics, whereas the latter is a linear mode with nonperturbative EP modifications.

\begin{figure}[t]
\subfloat[Poloidal structure at a single toroidal angle, slice taken at angle shown by radial line in (b).]{\includegraphics[width = \fullwidth]{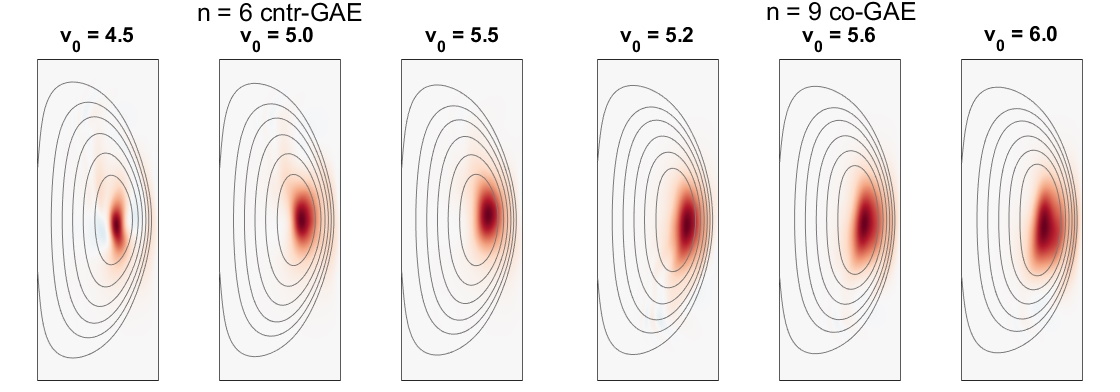}} \newline
\subfloat[Toroidal structure at midplane. Circles indicate the the last closed flux surface and magnetic axis.]{\includegraphics[width = \fullwidth]{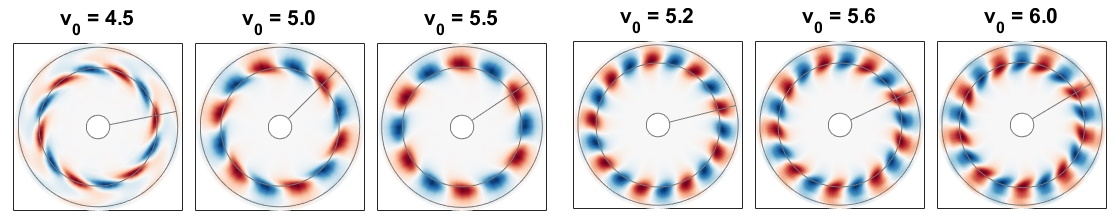}} \newline
\subfloat[Fourier amplitude of generalized poloidal harmonics along the $\vartheta = \nabla\psi \cross \nabla\phi$ direction, summed over all toroidal angles.]{\includegraphics[width = \fullwidth]{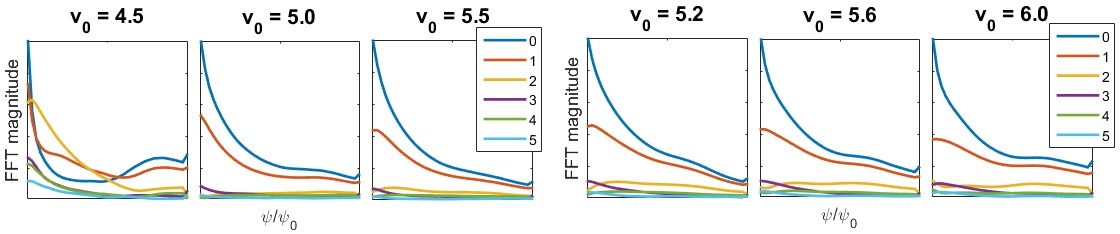}} \newline 
\caption
[Mode structure of cntr-GAEs and co-GAEs in self-consistent simulations as $\vinj$ is varied.]
{Mode structure of GAEs in self-consistent simulations. Left column: $n = 6$ cntr-GAE excited by EP with $\linj = 0.7$ and $\vinj = 4.5, 5.0, 5.5$ in self-consistent simulations, with frequencies $\omeganorm = 0.214, 0.178, 0.141$. Right column: $n = 9$ co-GAE excited by EP with $\linj = 0.3$ and $\vinj = 5.2, 5.6, 6.0$ in self-consistent simulations, with frequencies $\omeganorm = 0.239, 0.264, 0.289$. The fluctuation shown is $\dbperp$ in the $\nabla R \cross \vec{B_0}$ direction.}
\label{fig:epgae:struct}
\end{figure}

This endeavor is complicated by the fact that the GAEs, the counter-propagating modes especially, may interact with the continuum and excite a kinetic \Alfven wave, inferred through the presence of a well-localized $\depar$ fluctuation on the high field side and coincident short-scale modulation of the $\dbperp$ mode structure near this region. The coupling of the KAW with the compressional mode in \HYM simulations was studied in depth in a recent publication,\cite{Belova2017POP} which identified key signatures of the KAW in the simulation which can also be leveraged in the case of the GAEs. Some of the more dramatic changes in mode structure can be attributed to gradual suppression or excitation of KAW features, which has dominant $\dbperp$ polarization just as the GAEs do. This can be subjectively distinguished from the GAE mode structure since the KAW has a characteristic ``tilted" structure near the \Alfven resonance location whereas the GAE is usually concentrated between the axis and mid-radius, often towards the low-field side. The left column of \figref{fig:epgae:struct} shows how the mode structure evolves as a function of $\vinj$ for the $n = 6$ cntr-GAE in fully self-consistent simulations. Visually, the structure could be assigned a poloidal mode number of $m = 0$ or $m = 1/2$ since it has a single peak. Fourier decomposition in the generalized poloidal direction $(\vartheta = \nabla\psi \cross \nabla\phi)$ yields the same answer, some mix of $m = 0$ and $m = 1$. From $\vinj = 4.5$ (first column) to $\vinj = 5.5$ (last column), the frequency changes by $34\%$, or about 175 kHz, yet no new poloidal or radial harmonic emerges. Qualitatively, the structure becomes broader as $\vinj$ increases, and also gradually shifts towards the low field side, as can be seen in the midplane slices. 

For co-GAEs, there is even less change. Generally, the co-GAE mode structure is more broad radially and more elongated than the cntr-GAE structure. The poloidal structure of the co-GAEs looks very similar when excited by energetic particles with $\vinj = 5.2 - 6.0$, as shown in the right column of \figref{fig:epgae:struct}. Again, Fourier decomposition yields $m = 0 - 1$, matching visual intuition, and remaining unchanged as $\vinj$ is varied. For the case shown, the frequency changes by more than $20\%$, equivalent to 150 kHz. In contrast to the cntr-GAE, the co-GAEs migrate slightly towards the high field side for larger EP energies. Similar to the cntr-GAEs, this constancy of the mode structure despite large changes in frequency would be very atypical of MHD eigenmodes. Since these modes are $m = 0$ or 1 with $n = 9$, the approximation $\kpar \approx k_\phi = n/R$ is justified. Hence, this change in mode location to lower $R$ tends to increase $\kpar$. Furthermore, $\va$ has its minimum near the magnetic axis, so the local \Alfven speed can also change due to shifts in the mode location. It is then possible that a change in mode location could occur such that the frequency changes while conserving $\omega \approx \kpar \va$ without changing the mode numbers. However, this would necessarily move the mode away from an extremum in the \Alfven continuum (if it were originally near one when excited by lower $\vinj$), leading it to intersect the \Alfven continuum, which typically results in strong damping. This is essentially what was observed in the ``MHD-only" simulations and shown in \figref{fig:epgae:continuum}. 

\begin{figure}[tb]
\subfloat[\label{fig:epgae:dispersion_cntr}]{\includegraphics[width = \halfwidth]{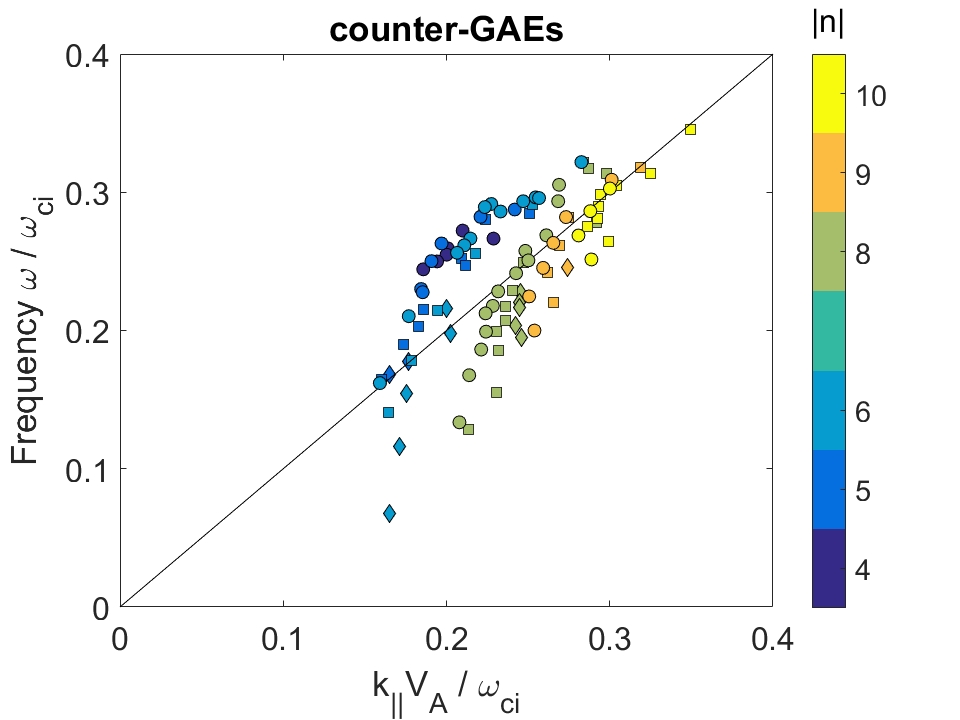}} 
\subfloat[\label{fig:epgae:dispersion_co}]{\includegraphics[width = \halfwidth]{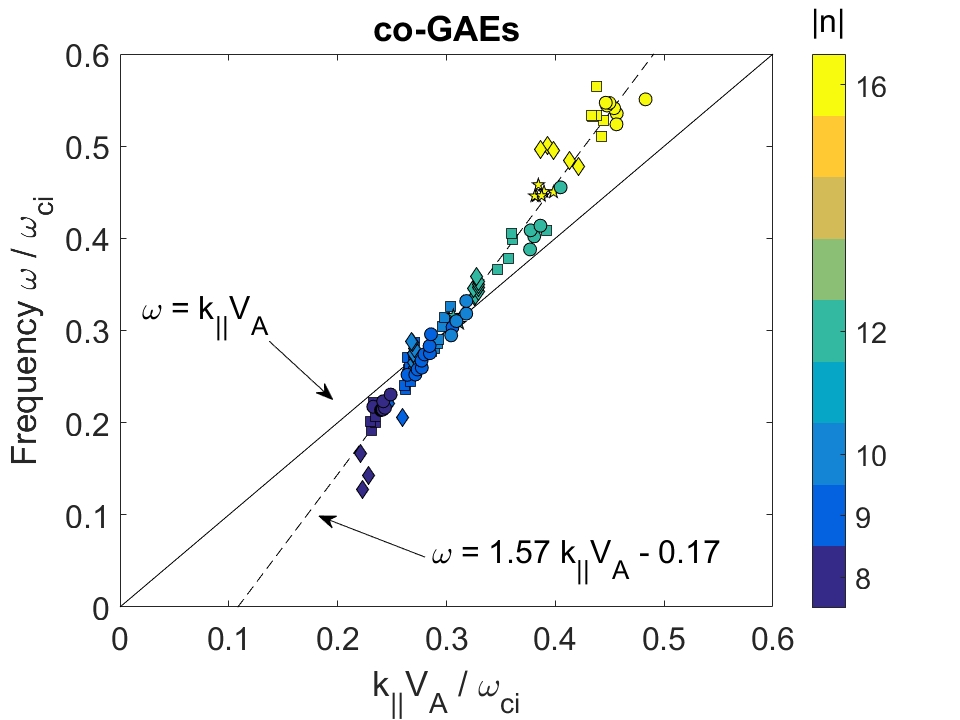}}
\caption
[Comparison of mode frequencies to shear \Alfven dispersion.]
{Comparison of mode frequencies to shear \Alfven dispersion with $\kpar$ and $\va$ evaluated at the peak mode location. Solid line indicates $\omega = \kpar\va$, dashed line indicates linear fit to simulation data. Color: toroidal mode number of the simulated mode. (a) cntr-GAEs. (b) co-GAEs.}
\label{fig:epgae:dispersion}
\end{figure}

Since counter-propagating \Alfven eigenmodes with shear polarization have typically been identified as perturbative GAEs in NSTX plasmas both experimentally\cite{Gorelenkov2003NF,Crocker2013NF} and in simulations,\cite{Gorelenkov2004POP} it is necessary to determine if their frequencies lie close to the shear \Alfven dispersion, $\omega_A = \kpar\va$, or if they deviate significantly due to the large frequency changes with beam parameters. While perturbative GAEs should have frequencies shifted somewhat below the \Alfven frequency, the difference should be small, \eg $\lesssim 10\%$ and often much less.\cite{Appert1982PP} For accuracy, the dispersion relation should be evaluated at the mode location. Calculating $\va$ at the mode location is only nontrivial due to the mode structure being broad, though this is easily solved by defining the mode location to be the $\db^2$ weighted average of $R$. The parallel wave number is less well defined. In a large aspect ratio tokamak, it is accurately represented by the familiar formula $\kpar = (n - m/q)/R$. However, this is only valid for $\epsilon = r/R \ll 1$ and requires $m$ to be well defined. In contrast, these simulations are carried out at the low aspect ratio of NSTX, where $\epsilon \approx 3/4$, and there is often no clear poloidal harmonic present in the mode structure, as discussed in section \ref{sec:epgae:struct}. For high $n$ numbers, the approximation $\kpar \approx k_\phi$ becomes more reliable since typically $nq > m$ for the modes excited in the simulations. However, this is a poor approximation for the cntr-GAEs which may have, for instance, $n = 4$ and $m = 2 - 4$. As an alternative, the most literal interpretation of $\kpar$ is used, that is the peak in the Fourier spectrum of the fluctuation when projected onto the background field lines near the mode location with a field-line following code. This method is sufficient to determine if the mode frequencies are at least ``near" the shear \Alfven frequency, as in \figref{fig:epgae:dispersion}. 

For both counter- and co-propagating modes, there is a clear correlation between the frequency of the modes and the shear \Alfven dispersion, as expected for GAEs. However, the cntr-GAEs show significant deviation from this relation for low $\abs{n}$ modes, while the co-GAEs show a steeper than expected slope. The co-GAEs are well fit by the relation $\omega = 1.57\kpar\va - 0.17$. The deviations from the shear \Alfven dispersion are not explained at this time. A complete explanation likely requires modification of the GAE dispersion to include beam contributions to the eigenequation nonperturbatively, as well as coupling to the compressional mode. In order to remain consistent with the simulation results, the modification must at least include a term proportional to $\kpar\vb$. One route to pursue would be to build upon the theory developed by Berk \etal for reverse-shear \Alfven eigenmodes (RSAE) which employs energetic particle effects to localize the eigenmode near local extrema in the \Alfven continuum.\cite{Berk2001PRL,Breizman2003PP} In particular, Eq. 5 of \citeref{Berk2001PRL} includes terms proportional to $\avg{n_h}$ and $\kpar\avg{J_{\parallel h}}$ which could help explain the results in \figref{fig:epgae:sc_mhd}. The derivation of an accurate dispersion for the \EGAE is left for future work. 

\section{Resonant Particles}
\label{sec:epgae:res} 

Ultimately, the resonance condition is determined to be responsible for key properties of these modes. Investigation of the properties of the resonant particles identified in the simulation with explanations supported by analytic theory can shed light on the origins of the unusual features of these modes. 

\subsection{Influence of Resonance Condition}
\label{sec:epgae:resa}

Since a $\df$ scheme is employed, the particle weights can reveal information about resonant particles. The weights will evolve according to \eqref{eq:sim:dwdt}. Hence weights with large magnitudes correspond to regions of phase space with large changes in the distribution function, \eg particles which interact strongly with the waves. Particles can resonate with the wave through the general Doppler-shifted cyclotron resonance, reproduced below

\begin{equation}
\label{eq:epgae:res_one}
\omega - \avg{\kpar\vpar} - \avg{\kperp\vdrift} \approx \omega - \avg{\kpar\vpar} = \lres\avg{\omegaci}
\end{equation} 

\begin{figure}[tb]
\subfloat[\label{fig:epgae:resonance_cntr}]{\includegraphics[width = \halfwidth]{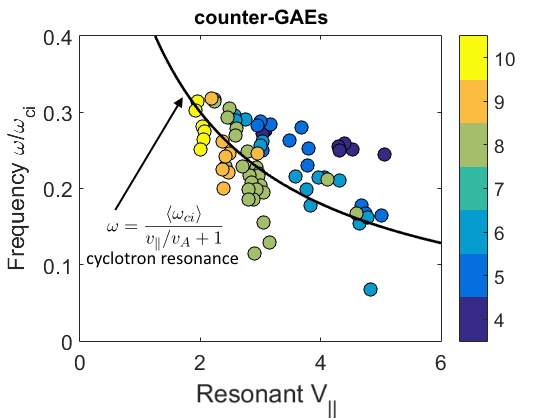}} 
\subfloat[\label{fig:epgae:resonance_co}]{\includegraphics[width = \halfwidth]{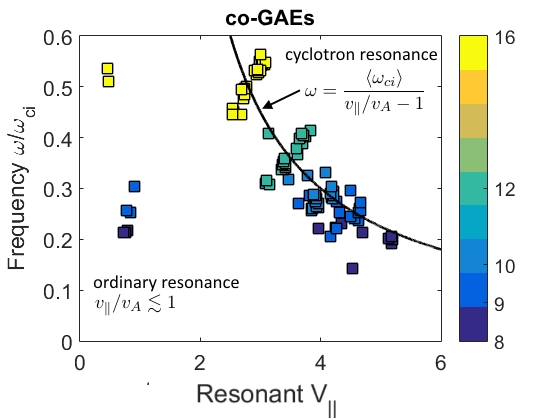}}
\caption
[Frequency and approximate $\vpar$ of resonant particles in simulations.]
{Frequency and approximate $\vpar$ of resonant particles in simulations. Solid line is the expression from \eqref{eq:epgae:res_vpar} required by the dispersion and resonance condition, assuming $s = 0$. Frequency and velocity normalized by on-axis values of $\omegaci$ and $\va$. (a) cntr-GAEs. (b) co-GAEs.}
\label{fig:epgae:resonance}
\end{figure}

On the right hand side of \eqref{eq:epgae:res_one}, the drift term $\kperp\vdrift$ is being neglected because it is usually sub-dominant in the simulated conditions, as discussed earlier in this thesis. For modes satisfying \eqref{eq:epgae:res_one} with $0 < \omega < \omegaci$ and $\vpar > 0$, counter propagation $(\kpar < 0$) implies $\lres > 0$, and co-propagation implies $\lres \leq 0$. While the $\lres = 0$ resonance is present in the some of the simulations for the co-GAEs, it is usually subdominant to $\lres = -1$ (visible in \figref{fig:epgae:resonance_co}). Consequently, attention is restricted to the cases where $\lres = \pm 1$, which also leads to the correspondence $\lres = -\text{sign}\,\kpar$. Combining \eqref{eq:epgae:res_one} with the presumed shear \Alfven dispersion, an expression can be written for the frequency of the excited mode as a function of the resonant $\vpar$ of the EP driving it unstable: 

\begin{equation}
\label{eq:epgae:res_vpar}
\omega = \frac{\avg{\omegaci}}{\lres + \avg{\vpar}/\va} \quad \text{for } \lres = \pm 1 = - \text{sign}\,\kpar
\end{equation}

Although $\vpar$ is not a constant of motion, it can be represented to lowest order in $\mu$ for each particle as 

\begin{equation}
\label{eq:epgae:vpar}
\vpar \approx v \sqrt{1 - \frac{\omegaci}{\omega_{ci0}}\lambda}
\end{equation}

\figref{fig:epgae:resonance} shows the parallel velocity (approximated by \eqref{eq:epgae:vpar}) of the EP with the largest weights, plotted against the frequency of the most unstable mode in each simulation. The relation between the mode frequency and parallel velocity of the most resonant particles generally adheres to \eqref{eq:epgae:res_vpar}, shown on the figures as the solid line. For co-GAEs, the condition is essentially obeyed, with some deviation due to a combination of drift term corrections and errors in the approximate expression for the resonant value of $\vpar$. In general, \eqref{eq:epgae:res_vpar} suggests that the frequency of the excited mode is inversely proportional to the parallel velocity of the resonant particles. While for co-GAEs the opposite trend is seen for fixed $n$ -- frequency increases with parallel velocity instead of decreases -- this is anticipated by the resonance condition. Since $\kpar \propto n$, the Doppler shift will increase with $\vpar$ at constant $n$. For cntr-GAEs, the mode frequencies still cluster near the curve representing \eqref{eq:epgae:res_vpar}, though there is substantial spread inherited from the deviations from the shear \Alfven dispersion due to ambiguous $\kpar$ as discussed in section \ref{sec:epgae:struct}. 

\begin{figure}[tb]
\subfloat[\label{fig:epgae:res_conts_lv}]{\includegraphics[width = \halfwidth]{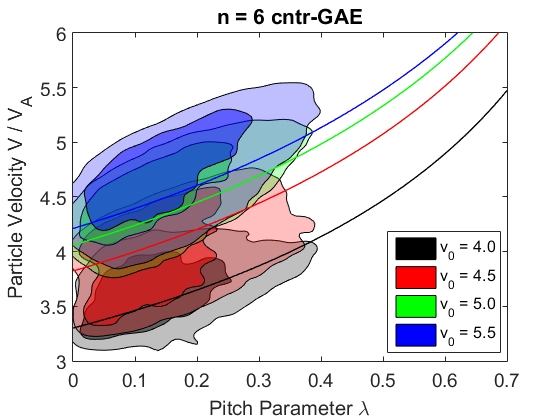}} 
\subfloat[\label{fig:epgae:res_conts_oc}]{\includegraphics[width = \halfwidth]{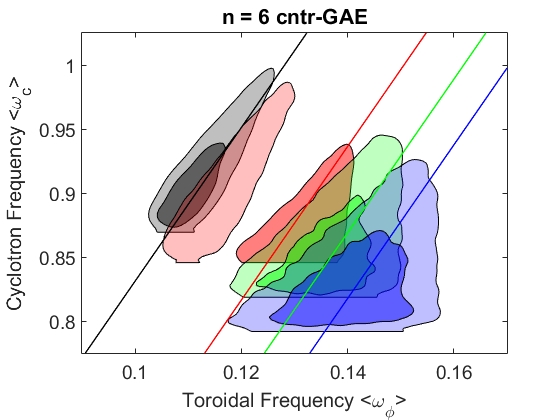}}
\caption
[Resonant particles in $(\lambda,v/\va)$ and $(\avg{\omegator},\omegacires)$ space for $n = -6$ cntr-GAEs for varying $\vinj$.]
{Resonant particles for $n = -6$ cntr-GAE excited by $\vinj = 4.0 - 5.5$. (a) Shaded contours show the location of the resonant particles in pitch-velocity $(\lambda,v)$ constant of motion space. Curves are contours of constant $\vpar$ determined by a $w$-weighted average of $\vpar$ over all resonant particles. (b) orbit-averaged toroidal and cyclotron frequencies of resonant particles. Solid lines show the resonance condition for each mode, averaging $\omegapol$ over all resonant particles and using the dominant $p$ in \eqref{eq:int:res} for each mode.}
\label{fig:epgae:res_conts}
\end{figure}

The mode frequency's sensitivity to the location of fast ions in phase space is reminiscent of energetic particle modes where the EPM frequency tracks typical particle orbit frequencies. Although the cyclotron and orbital frequencies are not constants of motion, a unique value of each can be calculated for each $\df$ particle as an orbit-averaged value. On \figref{fig:epgae:res_conts}, the shaded contours show the characteristic frequencies of the resonant particles in each simulation, where the resonant particles are defined as those with weights in the top $5\%$ at the end of the simulation. As the injection velocity increases, the resonant particles migrate to larger toroidal frequencies and smaller cyclotron frequencies. The lines imposed on the plot of toroidal vs cyclotron frequency represent the relation expected by the Doppler shifted cyclotron resonance written in terms of particle frequencies (see \eqref{eq:int:res}). The resonant particles in each simulation cluster around these lines, showing that the frequency of the most unstable mode is being set by the location of the resonant particles in this phase space. In other words, the mode frequency adapts to the energetic particle attributes in order to satisfy the resonance condition. It is also helpful to examine where the resonant particles exist in the constant-of-motion space, $(v,\lambda,\pphi)$, which are the natural variables for the distribution function. This is shown in \figref{fig:epgae:res_conts_lv}. The resonant particles move towards higher energy as those regions become accessible with the larger injection velocity. For each distribution, a curve representing constant $\vpar$ is shown, with value determined by averaging over all resonant particles. Each shaded contour roughly tracks this line of constant $\vpar$, with value increasing with increasing $\vinj$. 

Overall, \figref{fig:epgae:res_conts} demonstrates a clear linear relation between the energetic particle parameters and the frequency of the excited mode, a hallmark quality of energetic particle modes.\cite{Todo2006POP} This finding contradicts the conventional ``beam-driven MHD mode" paradigm where the energetic particles provide drive but otherwise do not affect the excited MHD mode. On the one hand, a resonant wave-particle interaction is necessary to drive the mode unstable, in which case it is natural that the frequency of the mode matches the combined orbital and cyclotron motion of the resonant particles. However, it is quite remarkable that the frequency of the mode is changing without clear changes in the mode structure. If this were a perturbative MHD mode, then one would expect that the changes in frequency would correspond to changes in mode structure, \ie poloidal or radial mode numbers. Alternatively, if only a single, specific eigenmode were being excited, then its frequency should not change as the energetic particle population does -- the mode would simply pick out the same resonant particles as $\vinj$ is increased. In view of these findings, this mode, formerly identified as a GAE from ideal MHD theory must be strongly altered by nonperturbative energetic particle effects, and thus could be considered as an energetic particle mode. This is different from the energetic particle modes commonly observed in experiments and discussed in the literature (fishbone, EGAM, etc) typically have much lower frequencies, on the order of orbital frequencies.\cite{Heidbrink2008POP} To our knowledge, this is the first evidence of an EPM that is driven by a cyclotron resonance and with a frequency that can be an appreciable fraction of the cyclotron frequency.

\subsection{Relationship Between Injection and Resonant Velocities}
\label{sec:epgae:resb}

The key takeaway is that if the resonant value of $\vpar$ is proportional to the injection velocity $\vb$, then the large frequency changes of these GAEs are qualitatively explained by the resonance condition. This is plausible based on the local analytic theory developed in \chapref{ch:cyc:analytics-cyclotron}. Recall that for the cyclotron resonances ($\lres \neq 0$) which drive the GAEs in simulations, the growth rate is dominated by the contribution from anisotropy, such that 

\begin{align}
\gamma \appropto \lres \int_0^{\xm} h(x) \pderiv{\fb}{x} dx 
\label{eq:epgae:gammaqual}
\end{align}

As a reminder, $x = \vperp^2 / v^2 \approx \lambda \omegacires/\omegacio$ and $h(x)$ is a complicated non-negative function of $x$ that weights the integrand, including the FLR terms. The upper limit of integration $\xm = 1 - \vpres^2/\vb^2$ is a consequence of the finite beam injection energy since $\vpres = v\sqrt{1 - x} < \vb\sqrt{1 - x} \rightarrow x < 1 - \vpres^2/\vb^2$. Due to the singly peaked fast ion distribution in $x$ used in this study, the integrand has the sign of $\lres$ for $x < \xinj$ (since $\partial\fb/\partial x > 0$ there) and the opposite sign for $x > \xinj$. 
%
%
As discussed in \secref{sec:cyc:properties}, a sufficient condition for fast ion drive for cntr-GAEs (driven by $\lres = 1$) is therefore $\xm \leq \xinj$. Similarly, a necessary condition for instability for co-GAEs (driven primarily by $\lres = -1$) is $\xm \geq \xinj$. Although these conditions do not necessarily maximize the growth rate to give the most unstable mode, the condition $\xm \approx \xinj$ is usually pretty close to achieving this. Since $\xm = 1 - \vpres^2/\vb^2$, these instability conditions impose constraints on the GAE frequency (related to $\vpres$ through the resonance condition) as a function of $\vb$.  In other words, theory predicts that the frequency of the most unstable mode changes as $\vb$ changes since there is a preferred value of $\vpres/\vb$ that maximizes the growth rate. 

\begin{figure}[tb]
\subfloat[\label{fig:epgae:integral_integrand}]{\includegraphics[width = \halfwidth]{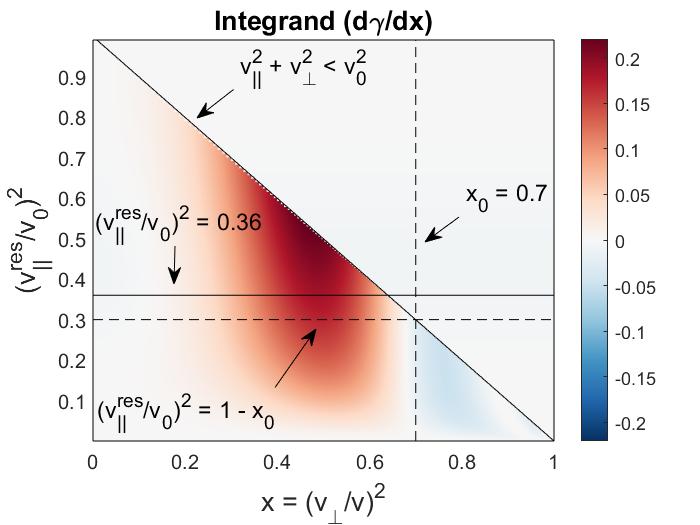}} 
\subfloat[\label{fig:epgae:integral_gamma}]{\includegraphics[width = \halfwidth]{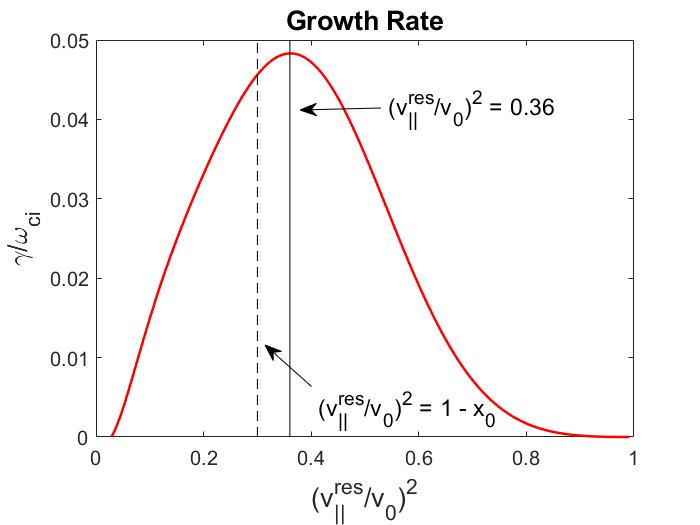}}
\caption
[Integrand of growth rate integral $(d\gamma/dx)$ as a function of $\vperp^2/v^2$ and $\vpres^2/\vb^2$. Growth rate as a function of $\vpres^2/\vb^2$.]
{(a) Integrand of growth rate integral $(d\gamma/dx)$, from \eqref{eq:cyc:gammabeam}, for a cntr-GAE with $\xinj = 0.7$. $\alpha = \kpar/\kperp = 0.5$ and $\vinj = 5.0$ are chosen as typical values. Vertical dashed line shows the central value $\xinj = 0.7$, horizontal dashed line shows the sufficient condition for net drive at $\vpres^2/\vb^2 = 1 - \xinj$, and the solid line shows the value $\vpres^2/\vb^2 = 0.36$ that maximizes the growth rate. (b) Integral of plot (a) with respect to $x$, showing growth rate as a function of $\vpres^2/\vb^2$. Vertical lines match horizontal lines on plot (a).}
\label{fig:epgae:integral}
\end{figure}

To supplement the preceding qualitative argument regarding the condition for marginal drive from the fast ions, the full expression for the growth rate given in \eqref{eq:cyc:gammabeam} can be evaluated numerically to determine how the maximum growth rate depends on the three independent parameters $\vinj$, $\vpres^2/\vb^2$, and $\alpha = \krat$ for a cntr-GAE $(\lres = +1)$ with $\linj = 0.7$. The sum of Bessel functions embedded in the FLR function $\Jlm(\flr)$ (see \eqref{eq:cyc:Jlm}) is the main obstacle to gaining intuition about the growth rate's dependencies by inspection or calculus. The parameter $\alpha$ enters through this Bessel term, since the FLR argument can be rewritten as

\begin{equation}
\label{eq:epgae:ztransform} 
\flr = \kperp\rhob = \frac{\kperp\vperp}{\omegaci} = 
\frac{\kperp}{\abs{\kpar}}\frac{\abs{\kpar}\va}{\omegaci}\frac{\vperp}{\va} = 
\frac{1}{\alpha}\frac{\omega}{\omegaci}\frac{\vb}{\va}\sqrt{\frac{(\vpres^2/\vb^2) x}{1-x}}
\end{equation}

Above we have used $\omega \approx \abs{\kpar}\va$ for illustration purposes, though the full dispersion can be substituted for the term $\abs{\kpar}\va/\omega = \Npar$ by using \eqref{eq:int:twodisp}. Note also that $\omega$ is not an independent parameter since it is determined by the three chosen parameters ($\vinj,$ $\vpres^2/\vb^2$, and $\krat$) by the resonance condition in \eqref{eq:int:reskpar}. 


The integrand with $\vinj = 5.0, \krat = 0.5, \xinj = 0.7$ is shown in \figref{fig:epgae:integral_integrand}, revealing complicated dependence on both the integration variable $x$ and the parameter $\vpres^2/\vb^2$. Generally, decreasing $\krat$ makes the details of the integrand even more intricate, as more zeros of $\Jlm(\flr)$ become contained within the integration region. Visualized this way it is clear why the sufficient condition for net drive from the energetic particles exists: at sufficiently large $\vpres^2/\vb^2$, the upper integration bound excludes the regions of velocity phase space which damp the wave. 
\figref{fig:epgae:integral_gamma} shows the growth rate's dependence on $\vpres^2/\vb^2$ for these specific values of $\vinj$ and $\krat$, demonstrating a local maximum exceeding the sufficient threshold for net drive at $\vpres^2/\vb^2 = 1 - \xinj$ (dashed line). This optimal value of $\vpres^2/\vb^2$ is also marked on \figref{fig:epgae:integral_integrand} with the solid line near $\vpres^2/\vb^2 = 0.36$. 

\begin{figure}[tb]
\subfloat[\label{fig:epgae:gamma_scan_vb}]{\includegraphics[width = \halfwidth]{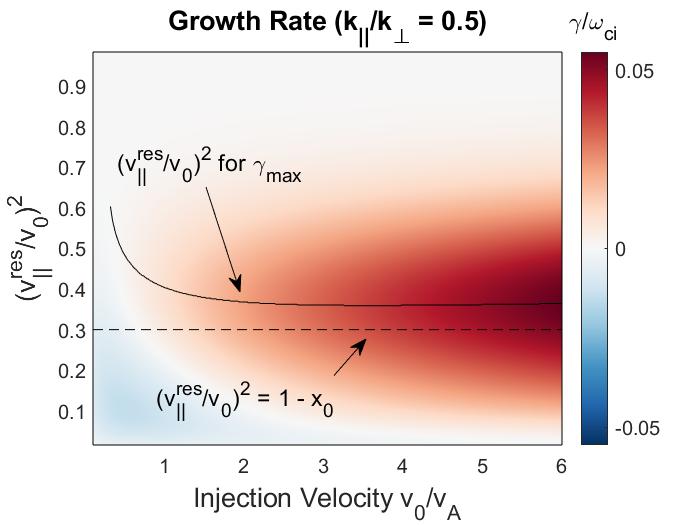}}
\subfloat[\label{fig:epgae:gamma_scan_ab}]{\includegraphics[width = \halfwidth]{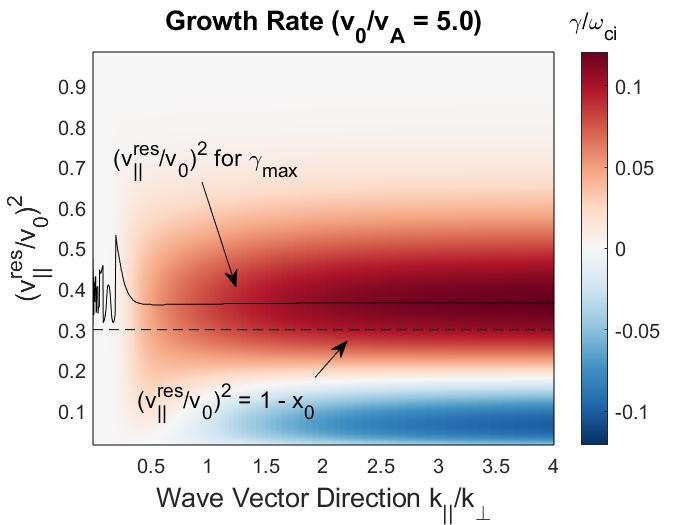}}
\caption
[Growth rate as a function of $(\vinj,\vpres^2/\vb^2)$ and $(\krat,\vpres^2/\vb^2)$.]
{(a): Growth rate as a function of $\vinj$ and $\vpres^2/\vb^2$ for $\kpar/\kperp = 0.5$. Dashed horizontal line is the sufficient condition for drive $\vpres^2/\vb^2 > 1 - \xinj$. The solid curve shows the value of $\vpres^2/\vb^2$ which maximizes the growth rate as a function of $\vinj$. 
(b): Growth rate as a function of $\krat$ and $\vpres^2/\vb^2$ for $\vinj = 5.0$. Solid curve and dashed line have the same definition as in (a).}
\label{fig:epgae:gamma_scan}
\end{figure}

Numerical integration can be performed over a range of values of $\vinj$, $\vpres^2/\vb^2$, and $\krat$ in order to determine if the growth rate prefers changing the frequency of the mode as $\vinj$ is varied, which would explain the simulation results. These scans are shown in \figref{fig:epgae:gamma_scan}. Note that on \figref{fig:epgae:gamma_scan_ab}, the rapid oscillation of the optimal value of $\vpres^2/\vb^2$ in the limit of $\krat \ll 1$ is due to rapid oscillation of the FLR terms in the integrand, corresponding to $\zp \gg 1$ in \secref{sec:cyc:fast} and \secref{sec:lan:fast}, referred to as the ``Bessel regime'' in the appendix of \citeref{Belova2019POP}. For $\vinj \gtrsim 2.5$ and $\krat \gtrsim 0.4$, there is a clear preference for $\vpres^2/\vb^2 \approx 0.36$ in order to maximize the growth rate, as the optimal value of $\vpres^2/\vb^2$ is within $1\%$ of this value in this range of parameters, which also encompasses the properties of the simulated modes. This calculation implies that the energetic particle drive is maximized for a mode resonantly excited by a sub-population of fast ions with parallel velocity at a specific fraction of the injection velocity, explaining the connection between the injection velocity and resonant parallel velocity. Then, the frequency dependence due to the resonance condition becomes 

\begin{equation}
\label{eq:epgae:res_exp}
\omega = \lres\omegaci + \kpar v_0\sqrt{\Bres} = \lres\left[\omegaci - \abs{\kpar} v_0\sqrt{\Bres}\right]
\end{equation}

Above we have used the fact that $\lres = - \text{sign}\,\kpar$. In the case of cntr-GAEs $(\lres = +1)$, the Doppler shift is less than the cyclotron frequency, and so the preferred mode frequency decreases linearly as a function of $\vb$. Conversely, co-GAEs excited by the $\lres = -1$ resonance have a Doppler shift exceeding the cyclotron frequency, so the frequency of the most unstable mode will increase linearly with increasing $\vb$. While this result reproduces the frequency trend of the most unstable modes from the simulations, the calculation is limited by not including the sources of bulk plasma damping. It is fair to assume that the thermal damping will affect each mode similarly, and hence, the maximum growth rate argument could remain valid. However, the amount of continuum damping each mode is subject to could vary substantially depending on quantitative details of the mode structure and differences in the self-consistent equilibria generated by fast ion populations of different injection velocities. Simplified analytic calculations have been performed in order to understand the numerical results, and they do not include the effects of continuum damping. Nonetheless, the presence of this frequency dependence both in simulations with signs of coupling to the continuum (via the appearance of short scale structures near the ideal \Alfven resonance location) as well as in those where they are absent indicates that the impact of continuum damping may not be crucial to developing a qualitative understanding of this phenomenon. The determination of the most unstable mode based on maximizing drive from the fast ions may be suitable to describe the robust numerical results. 

\section{Summary and Discussion}
\label{sec:epgae:discussion} 

Hybrid simulations have been conducted to study how the properties of high frequency shear \Alfven eigenmodes depend on parameters of the energetic particle distribution in NSTX-like low aspect ratio conditions. In simulations that solve for the equilibrium with self-consistent inclusion of energetic particle effects, it is found that the frequency of the most unstable GAE changes significantly with the energetic particle parameters. The frequency changes most significantly with the normalized injection velocity $\vinj$, which shows a clear linear relation. With increasing injection velocity, counter-propagating modes have a decrease in frequency, while co-propagating modes increase in frequency. The linear dependence and sign of the change are consistent with the Doppler-shifted cyclotron resonance condition. 

However, there are no clear concurrent changes in mode structure that would indicate that these frequencies correspond to distinct eigenmodes, especially for the co-GAEs. Moreover, the frequencies change continuously as a function of the injection velocity, not in a discrete stair-stepping pattern one would expect if different discrete eigenmodes were being excited. In contrast, the frequencies of  compressional modes excited in the simulations are largely unaffected by the fast ions, and modes with distinct frequencies have different poloidal mode numbers. 

At fixed injection energy, the frequency of both co- and counter-propagating modes decrease as the normalized EP density $\nb$ is increased, though the frequency change is an order of magnitude less than that caused by changing the injection energy. Although there was some difficulty in determining a reliable value of $\kpar$ for these modes due to low aspect ratio and poorly defined $m$ numbers, the modes do roughly obey the shear \Alfven dispersion relation $\omega \approx \left[\kpar(r)\va(r)\right]_{r=r_0}$, evaluated at the mode location, to within $10 - 20\%$. Lastly, the substantial changes in frequency persist even when the energetic particles are ignored in the equilibrium solver, implying that the change in frequency directly due to changes in the energetic particle population is much larger than the indirect change in frequency due to changes in the equilibrium from fast particle contributions. 

Put together, these results call into question the description of these modes as the global \Alfven eigenmodes described by ideal MHD theory. Since GAEs are shear \Alfven MHD modes, in order to be weakly damped they must have frequencies just below a minimum of the \Alfven continuum. Large frequency shifts with changing beam parameters can displace the modes from being localized near these extrema, and lead them to intersect the continuum where they woudl be expected to suffer strong damping. The energetic particles are clearly exerting a nonperturbative effect on the modes since the eigenfrequency is changing without clear corresponding changes in the mode structure that would indicate excitation of a different eigenmode. Instead, these results could be interpreted as defining a high frequency energetic particle mode, regarded here as an energetic-particle-modified global \Alfven eigenmode (\EGAE). For excitation, the mode must be resonant with a sub-population of energetic particles with a specific value of $\vpar$. As the injection velocity is increased, new values of $\vpar$ become accessible. It was shown that the drive from the fast ions is maximized for a resonant parallel velocity at a specific fraction of the injection velocity, given the same degree of anisotropy. As the resonant value of $\vpar$ changes, both $\omega$ and $\kpar$ must also change according to the resonance condition and the approximate dispersion. An energetic particle mode defined by a continuum of $\kpar$ values to choose from as the injection velocity is varied is consistent with these findings. This is unusual since energetic particle modes typically have much lower frequencies which track the characteristic energetic particle orbital frequency.\cite{Todo2006POP} In contrast, the modes excited in these simulations can be an appreciable fraction of the cyclotron frequency, $\omega \approx 0.1 - 0.5\omegaci$ for the range of toroidal harmonics $\abs{n} = 4 - 12$, and have frequencies which track a combination of the energetic particle orbital and cyclotron frequencies. 

There have been previous studies showing an MHD mode's eigenfrequency changing in proportion to energetic particle velocity. One is the so-called ``resonant toroidicity-induced \Alfven eigenmode" (RTAE), which is characterized by the mode frequency decreasing in order to remain in resonance with fast particles as $T_{EP}/T_i$ decreases.\cite{Cheng1995NF} Cheng \etal remark that this trend can lead the RTAE to have a frequency much below the characteristic TAE gap frequency that it is associated with, just as the GAEs in these simulation results can be significantly displaced from the minimum in the \Alfven continuum. In addition, previous hybrid gyrokinetic simulations have demonstrated a transition from TAE to a lower frequency kinetic ballooning modes (KBM) as the maximum energetic particle energy is increased.\cite{Santoro1996PP} During this transition, the frequency of the KBM changes in proportion to the energetic particle velocity, similar to the results presented here. 

Although the exact dispersion of the \EGAE has not yet been determined, it is clear that it is fundamentally affected by the energetic particles nonperturbatively, leading to a departure from its previous perturbative MHD description. In addition to the interest to basic plasma physics of the discovery of a high frequency energetic particle mode with frequencies tracking the combined orbital and cyclotron motion, there are also potential implications for NSTX-U which should be explored in the future. The simulations presented here show that the nonperturbative regime for these modes was routinely accessed in NSTX operating conditions. The basic picture of an energetic beam driving an MHD mode of the thermal plasma without modifying its attributes breaks down in conditions where $\Jb$ is comparable to $\Jp$. Even with the nominal factor of two increase in toroidal field in NSTX-U which will tend to decrease $\vinj$, these modes may still be unstable due to the increase in beam power,\cite{Gerhardt2012NF} though early operations indicate they can be suppressed with the addition of off-axis injection.\cite{Fredrickson2017PRL}

NSTX experiments have established a robust link between sub-cyclotron \Alfven modes and anomalous electron temperature flattening.\cite{Stutman2009PRL,Ren2017NF} Both of the existing theoretical mechanisms proposed to explain how \Alfvenic modes could generate this anomalous heat diffusivity have previously assumed that they are accurately described as perturbative ideal MHD GAEs.\cite{Gorelenkov2004POP,Kolesnichenko2010PRL,Gorelenkov2010NF} Since it has now been shown that there can be quite substantial nonperturbative corrections to this description, the polarization and mode structure of these modes may be quite different from those assumed by these previous analyses. In particular, Gorelenkov \etal investigated how several overlapping GAEs could collectively stochasticize electron orbits and enhance the radial diffusion. Nonperturbative modifications of the mode characteristics could alter the thresholds in number of overlapping modes and mode amplitudes required to generate the level of diffusion necessary to explain the experimental observations. While compressional modes have received more attention for their potential to channel energy away from the core to the edge through mode conversion to kinetic \Alfven waves,\cite{Belova2015PRL,Belova2017POP} GAEs also couple to KAWs in principle\cite{Kolesnichenko2010PRL,Kolesnichenko2010NF} and may also contribute. At least in the case of GAE-KAW mode conversion, the simulation results presented here suggest that nonperturbative inclusion of the energetic particles should be further explored for a more accurate description of that coupling in application to energy channeling in fusion conditions. Examining the impact of these corrections on previous quantitative predictions of anomalous electron heat transport will be the subject of future work. 

Prospects for future experimental verification of the EP-GAE are promising, as its defining characteristics should be observable in suitably designed experiments on NSTX-U. Analysis without such dedicated experiments may prove challenging since it is necessary to separate the changes in mode frequency due to the change in beam energy (the nonperturbative effect) from the changes in the equilibrium (MHD effect). The preferred approach would be to reproduce a discharge multiple times with different beam voltages for each shot so that the time evolution of the equilibrium profiles can be factored out of the observed change in frequency, such as the experiments conducted in \citeref{Fredrickson2002POP}. Measurement of the change in frequency due to this effect could be further complicated by chirping, which sometimes occurs for the high frequency \Alfvenic modes in NSTX. Fortunately, existing analysis shows that this usually takes the form of symmetric chirping (as opposed to monotonic frequency sweeping) about the linear mode frequency.\cite{Fredrickson2006POP} In this case, the frequency dependence on $\vinj$ should still be detectable. In addition to the signature change in frequency in proportion to the injection velocity, the gradual shift of the counter-propagating mode further towards the low field side with increasing beam energy as discussed in \secref{sec:epgae:struct} may be observable with reflectometer measurements.\cite{Crocker2011PPCF,Crocker2013NF}

\chapter{Summary and Future Work}
\label{ch:conc:conc}

\section{Summary}
\label{ch:conc:summary}

In this chapter, a very brief summary of the work discussed in this thesis will be given. The goals of this thesis were to develop further understanding of the stability properties of neutral beam-driven, sub-cyclotron compressional (CAE) and global (GAE) \Alfven eigenmodes. As discussed in the introduction, the presence of these instabilities has been experimentally linked to anomalous electron temperature profile flattening in NSTX. Consequently, techniques for their avoidance or control are needed in order to achieve desired temperature profiles. A theoretical approach was taken, leveraging both analytic theory and state-of-the-art numerical simulations. 

In \chapref{ch:cyc:analytics-cyclotron}, analytic theory was used to derive the perturbative fast ion drive for CAEs and GAEs in the local approximation. This derivation extended previous work by including terms to all orders in $\omeganorm$ (in the regime of $\omega \ll \omegape,\abs{\omegace}$) and $\krat$, which lead to coupling between the CAE and GAE that affects the finite Larmor radius (FLR) terms. The general expression was then applied to a model neutral beam distribution for the fast ions, which led to the discovery of a new instability regime that had not been considered by previous authors in their studies of shifted Maxwellians. Experimentally relevant physical approximations were made in order to evaluate the complicated integral expressions for the growth rate analytically. It was determined that the excitation of the modes is dominated by the contribution from velocity space anisotropy of the fast ion distribution. Finally, excellent ``global'' mathematical approximations were found which enabled the derivation of simple, compact marginal stability conditions $(\gamma = 0)$ in many useful cases. 

Several compromises were made in the analytic theory since it was necessary to sacrifice some amount of physics in order to preserve analytic tractability and allow for broadly applicable conclusions to be made. These limitations are enumerated here for reference. (1) The theory is only valid for the perturbative regime of $\gamma \ll \omega$. (2) Local theory is used, ignoring spatial variation of plasma profiles. (3) Sideband resonances are not considered, aside from their influence on $\vpres$. (4) The background damping rates are not calculated, so $\gamma > 0$ corresponds to a necessary (but not sufficient) condition for for instability, making the results most useful far from marginal stability. 

First, the derived instability conditions were analyzed in \chapref{ch:cyc:analytics-cyclotron} in detail for the case of the ordinary and anomalous cyclotron resonances which drive counter-propagating modes and high frequency co-propagating modes (those with fast ion Doppler shift larger exceeding the ion cyclotron frequency). It was found that the approximate stability boundaries faithfully reproduced those calculated numerically from the full (unapproximated) analytic expression for the growth rate, both as a function of the fast ion parameters (injection velocity $\vinj$ and injection geometry $\linj$) and the mode parameters (frequency $\omeganorm$ and wave vector direction $\krat$). The approximate marginal stability conditions implied a constraint on the frequency of unstable cntr-GAEs as a function of the injection velocity. Comparison with a large experimental database and set of simulations found strong agreement between the theoretical predictions, simulation results, and experimental observations, lending confidence to the theoretical approach taken. 

In \chapref{ch:lan:analytics-landau}, the local analytic theory was similarly applied to the Landau (non-cyclotron) resonance for the case of co-propagating CAEs and GAEs. Analysis of this resonance was somewhat more involved than the cyclotron resonances because of increased competition between contributions from velocity space anisotropy and fast ion damping from the slowing down function in determining the net drive. Again, a combination of physical and mathematical function approximation yielded approximate stability conditions which depended on only a small set of fast ion and mode parameters. Favorable agreement was found between these approximate expressions and the numerically computed analytic marginal stability boundaries. In addition, it was demonstrated that in the approximation of $\depar = 0$, GAEs can only be driven by this resonance due to coupling to the compressional mode,\footnote{Finite $\depar$ introduces sound waves into the system, which can similarly couple to the shear waves and allow them to be driven by the Landau resonance.} included in this work via two-fluid effects. In the case of CAEs, the derived instability conditions predicted an unstable range of $\krat$, which was then compared against simulations and experimental measurements from NSTX. This endeavor again found encouraging consistency between the properties of unstable modes in experiments, simulations, and the analytic theory. 

With the theoretical framework established and well-studied, the work of this thesis turned to analyzing and interpreting a comprehensive simulation study of CAE and GAE stability in realistic NSTX conditions. These simulations were performed with the hybrid kinetic-MHD code \HYM in toroidal geometry. The simulation model couples full-orbit kinetic beam ions to a thermal background MHD plasma in order to efficiently simulate the excitation of these instabilities. 

The work in \chapref{ch:sim:simulations} focused primarily on understanding the preferential excitation of different modes and their growth rate dependencies on the beam injection geometry and velocity. It was found that modes driven by the cyclotron resonances were more unstable than those driven by the Landau resonance, consistent with the theory developed in earlier chapters. As predicted by theory, high frequency co-GAEs driven by very tangential beam injection with large $\vinj$ were found to be unstable in simulations, whereas unstable cntr-GAEs required more perpendicular injection. The co-CAEs preferred moderately perpendicular injection and large $\vinj$ in order to simultaneously satisfy the resonance condition and dispersion relation. The growth rate of the most unstable mode of each type increased with larger normalized injection velocity $\vinj$. All modes were found to become more unstable with larger critical velocity fraction $\vcrit$ and velocity space anisotropy, with similar trends to those calculated from analytic theory. Moreover, it was found that spatial gradients in the fast ion distribution, neglected in the local analytic theory developed in this thesis, were responsible for a large fraction of the drive of the high $n \geq 8$ co-GAEs in simulations. A local expression of thermal electron damping was generalized to include finite $\omeganorm$ and $\krat$ effects. Application of the derived damping rate found a trivial effect for GAEs, but one that could stabilize some of the marginally unstable CAEs found in \HYM simulations, providing context for the applicability of the \HYM model to study CAE/GAE excitation. 

Lastly, \chapref{ch:epgae:epgae} presented a numerical investigation of the energetic particle modifications to the unstable GAEs found in \HYM simulations. It was found that the frequency of the most unstable GAE in each simulation changed in proportion to the beam injection velocity -- frequency decreasing with larger $\vinj$ for cntr-GAEs and increasing for co-GAEs. However, the mode structure of the mode did not change significantly during these large frequency modifications, suggesting that the fast ions could be changing the dispersion relation non-perturbatively. In contrast, no such behavior was exhibited by the unstable CAEs, which had distinct frequencies associated with distinct eigenstructures. Simulations were performed with equilibria that self-consistently included the effects of fast ions and also those that excluded them, leading to the conclusion that changes to the equilibrium profiles due to different fast ion parameters resulted in sub-dominant effect in setting the most unstable mode frequency. Instead, the properties of these modes in simulations were interpreted as a new type of high frequency energetic particle mode (EPM) driven by the Doppler shifted cyclotron resonances. 

\section{Future Work}
\label{ch:conc:future}

There are many avenues for continuation of this work that could be explored in the future. As described in the introduction, solving the electron temperature flattening problem has two parts: (1) understanding the excitation of the CAEs/GAEs and (2) understanding the effect of the modes on the background plasma. Presently, neither of the two proposed theoretical mechanisms can generate enough transport to reproduce the observed temperature profiles when using experimentally observed mode spectra as inputs to the theories. Therefore, further theoretical development of the interaction of CAEs/GAEs with the background plasma is the highest priority item towards the solution of this problem. The work in this thesis made significant progress on part (1), leaving part (2) ripe for subsequent study. In that direction, each of the two proposed mechanisms of temperature profile flattening have clear directions for continued work. 

For energy channeling via mode conversion to kinetic \Alfven waves (KAWs), prior simulations have quantified the amount of energy transported from CAEs to KAWs numerically\cite{Belova2015PRL,Belova2017POP} in the \HYM model of kinetic fast ions coupled to a thermal background plasma. However, kinetic effects from thermal ions could also be relevant, so it would be interesting to perform simulations with fully kinetic ions (beam and thermal ions treated on equal footing) in order to examine their impact on the energy channeling. Such a hybrid model is one of several models built into the \HYM code, and has previously been used in studies of field-reversed configurations (FRCs).\cite{Belova2000POP} Energy channeling due to GAE to KAW conversion has previously been considered by Kolesnichenko and collaborators,\cite{Kolesnichenko2010PRL,Kolesnichenko2010NF} but has not been studied numerically in toroidal simulations. This could be done by analysis of the energy flux in simulations with GAEs, such as those presented in \chapref{ch:sim:simulations} and \chapref{ch:epgae:epgae}, in order to determine if this is a relevant energy transport channel. 

The second mechanism of electron temperature profile flattening was orbit stochastization due to the presence of sufficiently many GAEs. This process was initially demonstrated to significantly enhance the electron energy transport in the guiding center code \ORBIT, using idealized GAE mode structures.\cite{Gorelenkov2010NF} Those studies also found sensitive dependence of the transport to $\depar$ fluctuations, imposed at levels to account for FLR effects. While preliminary studies were done to similarly include the effect of CAEs,\cite{Tritz2012APS} more work needs to be done in this area to quantify their contribution to this process. There are two additional sources of $\depar$ that were not considered in the initial \ORBIT studies. Unlike MHD modes, the KAW resulting from mode conversion has large $\depar$ (comparable to $\db$), which could cause additional transport. Second, the energetic particle modifications of GAEs studied in \chapref{ch:epgae:epgae} appear to also lead to somewhat larger $\depar$ fluctuations than would be expected by their orthodox MHD description. Hence, it would be interesting to perform electron test particle simulations using realistic CAE/GAE/KAW mode structures produced by self-consistent \HYM simulations, since this could determine if there are any important corrections to the original studies in \citeref{Gorelenkov2010NF} due to the use of more realistic eigenmodes including non-ideal MHD effects. 

There are also several areas for further theoretical work that are more direct extensions of the projects contained in this thesis. As demonstrated in \secref{sec:sim:pphi}, an important limitation of the analytic theory used in this thesis is the local assumption which excludes radial plasma profiles from the analysis, importantly neglecting $\partial\fb/\partial\pphi$ from contributing to the fast ion drive/damping. Previous authors have studied CAE/GAE instability with a global treatment, though that approach requires additional assumptions to make analytic progress since the resulting system is more complicated. In the case of the work by Gorelenkov \etal,\cite{Gorelenkov1995POP,Gorelenkov1995NF} these studies were focused primarily on understanding ion cyclotron emission (ICE) observations with $\omega \gtrsim \omegaci$. It could be useful to revisit those works in order to adapt them to the beam-driven, sub-cyclotron regime studied here. As a complementary approach, it would be interesting to extend the local theory applied here to the high frequency $\omega\gtrsim\omegaci$ regime in order to study how the two approaches differ when making predictions relative to ICE. Perhaps there are some features of ICE that could be more simply understood with the less involved local theory if plasma inhomogeneity is not crucial to its excitation in some cases. 

To extend the simulation study of CAE/GAE linear stability properties presented in \chapref{ch:sim:simulations}, it would be beneficial to model several additional NSTX and NSTX-U discharges in order to elucidate the influence of the equilibrium plasma profiles on the CAE/GAE instability properties. For instance, the steepness of the effective potential well and the safety factor shear should affect the damping due to interaction with the continuum for CAEs and GAEs, respectively. Work is currently being done by Belova\cite{Belova2020IAEA} to model GAEs in the conventional aspect ratio DIII-D. Since DIII-D has several different beam sources with an array of injection geometries, it would be an excellent candidate for validating the results of the simulation study presented here, especially the preferential excitation of CAEs vs GAEs and direction of propagation due to beam geometry. Such a study could also shed light on similarities and differences between CAE/GAE excitation in low vs high aspect ratio devices in order to extrapolate to other machines. Perhaps most importantly in this area, modeling of ITER plasma scenarios should be done in order to determine if the super-\Alfvenic neutral beam ions or fusion products will destabilize these modes. Back of the envelope estimates suggest that counter-propagating modes could be unstable since ITER fast ions may have similar dimensionless parameters as NSTX-U discharges, where cntr-GAEs were commonly observed. Determining how unstable CAEs/GAEs will be in ITER would be the first in forecasting if they will be expected to induce enhanced electron temperature transport like they did in NSTX. 

Next, the energetic particle modifications to GAEs discovered in \HYM simulations also point to new directions for research. While the numerical results were interpreted as indicating the existence of an energetic particle mode, this conclusion would be most definitively confirmed by deriving the dispersion relation. To do this, one would need to study the combined beam-thermal plasma system nonperturbatively in order to allow the eigenfrequencies to be influenced by the fast ion properties. A related area of theory that would benefit from further development is the kinetic modification of the \Alfven continuum by beam ion populations. Some work has been done by Kuvshinov\cite{Kuvshinov1994PPCF} in studying the ``ion-kinetic regime,'' but this approach would need to be adapted to account for the non-Maxwellian fast ion population. Numerically, Slaby\cite{Slaby2016POP} has studied the non-perturbative effect of a fusion $\alpha$ population on the continuum in the TAE frequency range, but again, the effects due to a characteristic neutral beam distribution could be somewhat different. An overarching question to answer is under what conditions due GAEs behave more like perturbative MHD modes vs. ones with strong energetic particle modifications. Potential strategies for detecting key signatures of the \EPGAE are outlined at the end of \secref{sec:epgae:discussion}. 

The combined analytic and numerical results contained in this thesis indicate potential techniques that could be used to control the CAE and GAE instabilities, similar to the GAE suppression that was observed in NSTX-U with the addition of the off-axis beam source. As discussed in \citeref{Belova2019POP}, that result can be explained by the theory presented in \chapref{ch:cyc:analytics-cyclotron}. The original NSTX beam sources inject fast ions with relatively large trapping parameter $\lambda$, generating a large region of phase space with $\partial\fb/\partial\lambda > 0$, which drives the cntr-GAEs unstable. The more tangentially injected NSTX-U beam sources have very small $\lambda$, introducing fast ion damping from $\partial\fb/\partial\lambda < 0$. Preliminary simulations indicate that co-CAEs can also be suppressed by tangential injection, possibly due to enhancement of the radiative damping due to fast ion-induced changes to the \Alfven continuum.\cite{Lestz2019APS} Further work to pin down this stabilization mechanism inferred from numerical work would indicate how it can be used to stabilize CAEs in experiments. 

\newcommand{\vpresone}{\vpres^{(1)}}
\newcommand{\vprestwo}{\vpres^{(2)}}
Moreover, \eqref{eq:cyc:gammageneralx} suggests additional stabilization techniques that have yet to be explored. Due to the coefficient $\lres$ multiplying the $\partial\fb/\partial\lambda$ term, it could be possible to stabilize a mode driven by the $\lres = 1$ resonance by injecting a second population of fast ions that satisfy the $\lres = -1$ resonance. This could be achieved with a counter-injected beam with $\vpar < 0$ such that the two resonances could be satisfied by different groups of fast ions which would then make oppositely signed contributions to the growth rate for the same sign of $\partial\fb/\partial\lambda$.\footnote{Assuming cntr-propagating modes ($\kpar < 0$) and $\omega \like \omegaci/2$, the two opposing resonances could be simultaneously satisfied as (1) $\omega - \kpar\vpresone = \avg{\omegaci}$ and (2) $\omega - \kpar\vprestwo = -\avg{\omegaci}$, provided that $\vprestwo = \vpresone - 2\avg{\omegaci}/\abs{\kpar}$, and assuming $\vpresone > 0$. A similar construction can be made with the signs of $\vpresone$ and $\vprestwo$ exchanged.} 
Such a scheme could in principle be tested on DIII-D, which does have both co- and cntr-injecting beam sources. A separate strategy would be to instead co-inject a second beam at a lower beam voltage than the initial driving beam, which should suppress co-CAEs according to the $(\linj,\vinj)$ stability diagrams shown in \chapref{ch:lan:analytics-landau} (for instance, see \figref{fig:lan:wideslowCAE} and consider adding a beam at smaller $\vinj$ in the blue region where $\gamma < 0$). Lastly, lengthening the tail of the fast ion distribution above the neutral beam injection velocity could also contribute to stabilizing cntr-GAEs since these particles could be tailored to have a stabilizing $\partial\fb/\partial\lambda < 0$. For instance, high harmonic fast wave heating (HHFW) has the potential to generate such high energy tails in the fast ion distribution. Self-consistent modeling capabilities of the interaction between fast ions and HHFW are currently under development.\cite{Bertelli2019NF} Fascinatingly, robust suppression of fast-ion-driven instabilities, including GAEs, with HHFW was previously observed in NSTX and is yet to be fully understood.\cite{Fredrickson2014NF} 

Overall, future extension of the results of this thesis on CAE/GAE instability behavior should be oriented in the direction of developing experimental tools for control of these instabilities in order to enable investigation into their role in the anomalous thermal electron energy transport and to aid in the identification of \Alfvenic transport mechanisms. \hfill\qedsymbol

\singlespacing

\cleardoublepage
\ifdefined\phantomsection
  \phantomsection  
\else
\fi
\addcontentsline{toc}{chapter}{Bibliography}

\bibliography{all_bib}

\begin{thebibliography}{180}%
\makeatletter
\providecommand \@ifxundefined [1]{%
 \@ifx{#1\undefined}
}%
\providecommand \@ifnum [1]{%
 \ifnum #1\expandafter \@firstoftwo
 \else \expandafter \@secondoftwo
 \fi
}%
\providecommand \@ifx [1]{%
 \ifx #1\expandafter \@firstoftwo
 \else \expandafter \@secondoftwo
 \fi
}%
\providecommand \natexlab [1]{#1}%
\providecommand \enquote  [1]{``#1''}%
\providecommand \bibnamefont  [1]{#1}%
\providecommand \bibfnamefont [1]{#1}%
\providecommand \citenamefont [1]{#1}%
\providecommand \href@noop [0]{\@secondoftwo}%
\providecommand \href [0]{\begingroup \@sanitize@url \@href}%
\providecommand \@href[1]{\@@startlink{#1}\@@href}%
\providecommand \@@href[1]{\endgroup#1\@@endlink}%
\providecommand \@sanitize@url [0]{\catcode `\\12\catcode `\$12\catcode
  `\&12\catcode `\#12\catcode `\^12\catcode `\_12\catcode `\%12\relax}%
\providecommand \@@startlink[1]{}%
\providecommand \@@endlink[0]{}%
\providecommand \url  [0]{\begingroup\@sanitize@url \@url }%
\providecommand \@url [1]{\endgroup\@href {#1}{\urlprefix }}%
\providecommand \urlprefix  [0]{URL }%
\providecommand \Eprint [0]{\href }%
\providecommand \doibase [0]{http://dx.doi.org/}%
\providecommand \selectlanguage [0]{\@gobble}%
\providecommand \bibinfo  [0]{\@secondoftwo}%
\providecommand \bibfield  [0]{\@secondoftwo}%
\providecommand \translation [1]{[#1]}%
\providecommand \BibitemOpen [0]{}%
\providecommand \bibitemStop [0]{}%
\providecommand \bibitemNoStop [0]{.\EOS\space}%
\providecommand \EOS [0]{\spacefactor3000\relax}%
\providecommand \BibitemShut  [1]{\csname bibitem#1\endcsname}%
\let\auto@bib@innerbib\@empty
\bibitem [{\citenamefont {Friedberg}(2007)}]{FreidbergFusion}%
  \BibitemOpen
  \bibfield  {author} {\bibinfo {author} {\bibfnamefont {J.~P.}\ \bibnamefont
  {Friedberg}},\ }\href@noop {} {\emph {\bibinfo {title} {Plasma Physics and
  Fusion Energy}}}\ (\bibinfo  {publisher} {Cambridge University Press},\
  \bibinfo {year} {2007})\BibitemShut {NoStop}%
\bibitem [{\citenamefont {Li}\ \emph {et~al.}(2014)\citenamefont {Li},
  \citenamefont {Jiang}, \citenamefont {Ren},\ and\ \citenamefont
  {Xu}}]{FigTokamak}%
  \BibitemOpen
  \bibfield  {author} {\bibinfo {author} {\bibfnamefont {S.}~\bibnamefont
  {Li}}, \bibinfo {author} {\bibfnamefont {H.}~\bibnamefont {Jiang}}, \bibinfo
  {author} {\bibfnamefont {Z.}~\bibnamefont {Ren}}, \ and\ \bibinfo {author}
  {\bibfnamefont {C.}~\bibnamefont {Xu}},\ }\href {\doibase
  10.1155/2014/940965} {\bibfield  {journal} {\bibinfo  {journal} {Abstract and
  Applied Analysis}\ }\textbf {\bibinfo {volume} {2014}},\ \bibinfo {pages}
  {940965} (\bibinfo {year} {2014})}\BibitemShut {NoStop}%
\bibitem [{\citenamefont {{The JET Team (prepared by M.L
  Watkins)}}(1999)}]{Watkins1999}%
  \BibitemOpen
  \bibfield  {author} {\bibinfo {author} {\bibnamefont {{The JET Team (prepared
  by M.L Watkins)}}},\ }\href {\doibase 10.1088/0029-5515/39/9y/302} {\bibfield
   {journal} {\bibinfo  {journal} {Nuclear Fusion}\ }\textbf {\bibinfo {volume}
  {39}},\ \bibinfo {pages} {1227} (\bibinfo {year} {1999})}\BibitemShut
  {NoStop}%
\bibitem [{\citenamefont {Fujita}\ \emph {et~al.}(1999)\citenamefont {Fujita},
  \citenamefont {Kamada}, \citenamefont {Ishida}, \citenamefont {Neyatani},
  \citenamefont {Oikawa}, \citenamefont {Ide}, \citenamefont {Takeji},
  \citenamefont {Koide}, \citenamefont {Isayama}, \citenamefont {Fukuda},
  \citenamefont {Hatae}, \citenamefont {Ishii}, \citenamefont {Ozeki},
  \citenamefont {Shirai},\ and\ \citenamefont {the JT-60~Team}}]{Fujita1999NF}%
  \BibitemOpen
  \bibfield  {author} {\bibinfo {author} {\bibfnamefont {T.}~\bibnamefont
  {Fujita}}, \bibinfo {author} {\bibfnamefont {Y.}~\bibnamefont {Kamada}},
  \bibinfo {author} {\bibfnamefont {S.}~\bibnamefont {Ishida}}, \bibinfo
  {author} {\bibfnamefont {Y.}~\bibnamefont {Neyatani}}, \bibinfo {author}
  {\bibfnamefont {T.}~\bibnamefont {Oikawa}}, \bibinfo {author} {\bibfnamefont
  {S.}~\bibnamefont {Ide}}, \bibinfo {author} {\bibfnamefont {S.}~\bibnamefont
  {Takeji}}, \bibinfo {author} {\bibfnamefont {Y.}~\bibnamefont {Koide}},
  \bibinfo {author} {\bibfnamefont {A.}~\bibnamefont {Isayama}}, \bibinfo
  {author} {\bibfnamefont {T.}~\bibnamefont {Fukuda}}, \bibinfo {author}
  {\bibfnamefont {T.}~\bibnamefont {Hatae}}, \bibinfo {author} {\bibfnamefont
  {Y.}~\bibnamefont {Ishii}}, \bibinfo {author} {\bibfnamefont
  {T.}~\bibnamefont {Ozeki}}, \bibinfo {author} {\bibfnamefont
  {H.}~\bibnamefont {Shirai}}, \ and\ \bibinfo {author} {\bibnamefont {the
  JT-60~Team}},\ }\href {\doibase 10.1088/0029-5515/39/11y/302} {\bibfield
  {journal} {\bibinfo  {journal} {Nuclear Fusion}\ }\textbf {\bibinfo {volume}
  {39}},\ \bibinfo {pages} {1627} (\bibinfo {year} {1999})}\BibitemShut
  {NoStop}%
\bibitem [{\citenamefont {Wesson}(2004)}]{WessonTokamak}%
  \BibitemOpen
  \bibfield  {author} {\bibinfo {author} {\bibfnamefont {J.}~\bibnamefont
  {Wesson}},\ }\href@noop {} {\emph {\bibinfo {title} {Tokamaks}}},\ \bibinfo
  {edition} {3rd}\ ed.\ (\bibinfo  {publisher} {Oxford University Press},\
  \bibinfo {year} {2004})\BibitemShut {NoStop}%
\bibitem [{\citenamefont {Gaffey}(1976)}]{Gaffey1976JPP}%
  \BibitemOpen
  \bibfield  {author} {\bibinfo {author} {\bibfnamefont {J.~D.}\ \bibnamefont
  {Gaffey}},\ }\href {\doibase 10.1017/S0022377800020134} {\bibfield  {journal}
  {\bibinfo  {journal} {Journal of Plasma Physics}\ }\textbf {\bibinfo {volume}
  {16}},\ \bibinfo {pages} {149–169} (\bibinfo {year} {1976})}\BibitemShut
  {NoStop}%
\bibitem [{\citenamefont {Mikhailovskii}(1975)}]{Mikhailovskiiv6}%
  \BibitemOpen
  \bibfield  {author} {\bibinfo {author} {\bibfnamefont {A.~B.}\ \bibnamefont
  {Mikhailovskii}},\ }in\ \href@noop {} {\emph {\bibinfo {booktitle} {Reviews
  of Plasma Physics}}},\ Vol.~\bibinfo {volume} {6},\ \bibinfo {editor} {edited
  by\ \bibinfo {editor} {\bibfnamefont {M.~A.}\ \bibnamefont {Leontovich}}}\
  (\bibinfo  {publisher} {Consultants Bureau},\ \bibinfo {address} {New York},\
  \bibinfo {year} {1975})\ pp.\ \bibinfo {pages} {77 -- 159}\BibitemShut
  {NoStop}%
\bibitem [{\citenamefont {Kolesnichenko}\ and\ \citenamefont
  {Oraevskii}(1967)}]{Kolesnichenko1967SAE}%
  \BibitemOpen
  \bibfield  {author} {\bibinfo {author} {\bibfnamefont {Y.}~\bibnamefont
  {Kolesnichenko}}\ and\ \bibinfo {author} {\bibfnamefont {V.~N.}\ \bibnamefont
  {Oraevskii}},\ }\href {https://doi.org/10.1007/BF01120459} {\bibfield
  {journal} {\bibinfo  {journal} {Soviet Atomic Energy}\ }\textbf {\bibinfo
  {volume} {23}},\ \bibinfo {pages} {1028} (\bibinfo {year}
  {1967})}\BibitemShut {NoStop}%
\bibitem [{\citenamefont {Rosenbluth}\ and\ \citenamefont
  {Rutherford}(1975)}]{Rosenbluth1975PRL}%
  \BibitemOpen
  \bibfield  {author} {\bibinfo {author} {\bibfnamefont {M.~N.}\ \bibnamefont
  {Rosenbluth}}\ and\ \bibinfo {author} {\bibfnamefont {P.~H.}\ \bibnamefont
  {Rutherford}},\ }\href {\doibase 10.1103/PhysRevLett.34.1428} {\bibfield
  {journal} {\bibinfo  {journal} {Phys. Rev. Lett.}\ }\textbf {\bibinfo
  {volume} {34}},\ \bibinfo {pages} {1428} (\bibinfo {year}
  {1975})}\BibitemShut {NoStop}%
\bibitem [{\citenamefont {Fasoli}\ \emph {et~al.}(2002)\citenamefont {Fasoli},
  \citenamefont {Testa}, \citenamefont {Sharapov}, \citenamefont {Berk},
  \citenamefont {Breizman}, \citenamefont {Gondhalekar}, \citenamefont
  {Heeter}, \citenamefont {Mantsinen},\ and\ \citenamefont {contributors to~the
  EFDA-JET~Workprogramme}}]{Fasoli2002PPCF}%
  \BibitemOpen
  \bibfield  {author} {\bibinfo {author} {\bibfnamefont {A.}~\bibnamefont
  {Fasoli}}, \bibinfo {author} {\bibfnamefont {D.}~\bibnamefont {Testa}},
  \bibinfo {author} {\bibfnamefont {S.}~\bibnamefont {Sharapov}}, \bibinfo
  {author} {\bibfnamefont {H.~L.}\ \bibnamefont {Berk}}, \bibinfo {author}
  {\bibfnamefont {B.}~\bibnamefont {Breizman}}, \bibinfo {author}
  {\bibfnamefont {A.}~\bibnamefont {Gondhalekar}}, \bibinfo {author}
  {\bibfnamefont {R.~F.}\ \bibnamefont {Heeter}}, \bibinfo {author}
  {\bibfnamefont {M.}~\bibnamefont {Mantsinen}}, \ and\ \bibinfo {author}
  {\bibnamefont {contributors to~the EFDA-JET~Workprogramme}},\ }\href
  {http://stacks.iop.org/0741-3335/44/i=12B/a=312} {\bibfield  {journal}
  {\bibinfo  {journal} {Plasma Physics and Controlled Fusion}\ }\textbf
  {\bibinfo {volume} {44}},\ \bibinfo {pages} {B159} (\bibinfo {year}
  {2002})}\BibitemShut {NoStop}%
\bibitem [{\citenamefont {Sharapov}\ \emph {et~al.}(2018)\citenamefont
  {Sharapov}, \citenamefont {Oliver}, \citenamefont {Breizman}, \citenamefont
  {Fitzgerald},\ and\ \citenamefont {and}}]{Sharapov2018NF}%
  \BibitemOpen
  \bibfield  {author} {\bibinfo {author} {\bibfnamefont {S.~E.}\ \bibnamefont
  {Sharapov}}, \bibinfo {author} {\bibfnamefont {H.~J.}\ \bibnamefont
  {Oliver}}, \bibinfo {author} {\bibfnamefont {B.~N.}\ \bibnamefont
  {Breizman}}, \bibinfo {author} {\bibfnamefont {M.}~\bibnamefont
  {Fitzgerald}}, \ and\ \bibinfo {author} {\bibfnamefont {L.~G.}\ \bibnamefont
  {and}},\ }\href {\doibase 10.1088/1741-4326/aabb67} {\bibfield  {journal}
  {\bibinfo  {journal} {Nuclear Fusion}\ }\textbf {\bibinfo {volume} {58}},\
  \bibinfo {pages} {082008} (\bibinfo {year} {2018})}\BibitemShut {NoStop}%
\bibitem [{\citenamefont {Oliver}\ \emph {et~al.}(2019)\citenamefont {Oliver},
  \citenamefont {Sharapov}, \citenamefont {Breizman}, \citenamefont
  {Fontanilla}, \citenamefont {Spong},\ and\ \citenamefont
  {and}}]{Oliver2019NF}%
  \BibitemOpen
  \bibfield  {author} {\bibinfo {author} {\bibfnamefont {H.~J.}\ \bibnamefont
  {Oliver}}, \bibinfo {author} {\bibfnamefont {S.~E.}\ \bibnamefont
  {Sharapov}}, \bibinfo {author} {\bibfnamefont {B.~N.}\ \bibnamefont
  {Breizman}}, \bibinfo {author} {\bibfnamefont {A.~K.}\ \bibnamefont
  {Fontanilla}}, \bibinfo {author} {\bibfnamefont {D.~A.}\ \bibnamefont
  {Spong}}, \ and\ \bibinfo {author} {\bibfnamefont {D.~T.}\ \bibnamefont
  {and}},\ }\href {\doibase 10.1088/1741-4326/ab382b} {\bibfield  {journal}
  {\bibinfo  {journal} {Nuclear Fusion}\ }\textbf {\bibinfo {volume} {59}},\
  \bibinfo {pages} {106031} (\bibinfo {year} {2019})}\BibitemShut {NoStop}%
\bibitem [{\citenamefont {McClements}\ \emph {et~al.}(2015)\citenamefont
  {McClements}, \citenamefont {D'Inca}, \citenamefont {Dendy}, \citenamefont
  {Carbajal}, \citenamefont {Chapman}, \citenamefont {Cook}, \citenamefont
  {Harvey}, \citenamefont {Heidbrink},\ and\ \citenamefont
  {Pinches}}]{McClements2015NF}%
  \BibitemOpen
  \bibfield  {author} {\bibinfo {author} {\bibfnamefont {K.~G.}\ \bibnamefont
  {McClements}}, \bibinfo {author} {\bibfnamefont {R.}~\bibnamefont {D'Inca}},
  \bibinfo {author} {\bibfnamefont {R.~O.}\ \bibnamefont {Dendy}}, \bibinfo
  {author} {\bibfnamefont {L.}~\bibnamefont {Carbajal}}, \bibinfo {author}
  {\bibfnamefont {S.~C.}\ \bibnamefont {Chapman}}, \bibinfo {author}
  {\bibfnamefont {J.~W.}\ \bibnamefont {Cook}}, \bibinfo {author}
  {\bibfnamefont {R.~W.}\ \bibnamefont {Harvey}}, \bibinfo {author}
  {\bibfnamefont {W.~W.}\ \bibnamefont {Heidbrink}}, \ and\ \bibinfo {author}
  {\bibfnamefont {S.~D.}\ \bibnamefont {Pinches}},\ }\href
  {http://stacks.iop.org/0029-5515/55/i=4/a=043013} {\bibfield  {journal}
  {\bibinfo  {journal} {Nuclear Fusion}\ }\textbf {\bibinfo {volume} {55}},\
  \bibinfo {pages} {043013} (\bibinfo {year} {2015})}\BibitemShut {NoStop}%
\bibitem [{\citenamefont {Heidbrink}\ and\ \citenamefont
  {White}(2020)}]{Heidbrink2020POP}%
  \BibitemOpen
  \bibfield  {author} {\bibinfo {author} {\bibfnamefont {W.~W.}\ \bibnamefont
  {Heidbrink}}\ and\ \bibinfo {author} {\bibfnamefont {R.~B.}\ \bibnamefont
  {White}},\ }\href {\doibase 10.1063/1.5136237} {\bibfield  {journal}
  {\bibinfo  {journal} {Physics of Plasmas}\ }\textbf {\bibinfo {volume}
  {27}},\ \bibinfo {pages} {030901} (\bibinfo {year} {2020})}\BibitemShut
  {NoStop}%
\bibitem [{\citenamefont {Fisch}\ and\ \citenamefont
  {Rax}(1992)}]{Fisch1992PRL}%
  \BibitemOpen
  \bibfield  {author} {\bibinfo {author} {\bibfnamefont {N.~J.}\ \bibnamefont
  {Fisch}}\ and\ \bibinfo {author} {\bibfnamefont {J.-M.}\ \bibnamefont
  {Rax}},\ }\href {\doibase 10.1103/PhysRevLett.69.612} {\bibfield  {journal}
  {\bibinfo  {journal} {Phys. Rev. Lett.}\ }\textbf {\bibinfo {volume} {69}},\
  \bibinfo {pages} {612} (\bibinfo {year} {1992})}\BibitemShut {NoStop}%
\bibitem [{\citenamefont {Fisch}(2000)}]{Fisch2000NF}%
  \BibitemOpen
  \bibfield  {author} {\bibinfo {author} {\bibfnamefont {N.~J.}\ \bibnamefont
  {Fisch}},\ }\href {http://stacks.iop.org/0029-5515/40/i=6/a=307} {\bibfield
  {journal} {\bibinfo  {journal} {Nuclear Fusion}\ }\textbf {\bibinfo {volume}
  {40}},\ \bibinfo {pages} {1095} (\bibinfo {year} {2000})}\BibitemShut
  {NoStop}%
\bibitem [{\citenamefont {Gorelenkov}\ \emph
  {et~al.}(2010{\natexlab{a}})\citenamefont {Gorelenkov}, \citenamefont
  {Fisch},\ and\ \citenamefont {Fredrickson}}]{Gorelenkov2010PPCF}%
  \BibitemOpen
  \bibfield  {author} {\bibinfo {author} {\bibfnamefont {N.~N.}\ \bibnamefont
  {Gorelenkov}}, \bibinfo {author} {\bibfnamefont {N.~J.}\ \bibnamefont
  {Fisch}}, \ and\ \bibinfo {author} {\bibfnamefont {E.}~\bibnamefont
  {Fredrickson}},\ }\href {http://stacks.iop.org/0741-3335/52/i=5/a=055014}
  {\bibfield  {journal} {\bibinfo  {journal} {Plasma Physics and Controlled
  Fusion}\ }\textbf {\bibinfo {volume} {52}},\ \bibinfo {pages} {055014}
  (\bibinfo {year} {2010}{\natexlab{a}})}\BibitemShut {NoStop}%
\bibitem [{\citenamefont {Stutman}\ \emph {et~al.}(2009)\citenamefont
  {Stutman}, \citenamefont {Delgado-Aparicio}, \citenamefont {Gorelenkov},
  \citenamefont {Finkenthal}, \citenamefont {Fredrickson}, \citenamefont
  {Kaye}, \citenamefont {Mazzucato},\ and\ \citenamefont
  {Tritz}}]{Stutman2009PRL}%
  \BibitemOpen
  \bibfield  {author} {\bibinfo {author} {\bibfnamefont {D.}~\bibnamefont
  {Stutman}}, \bibinfo {author} {\bibfnamefont {L.}~\bibnamefont
  {Delgado-Aparicio}}, \bibinfo {author} {\bibfnamefont {N.}~\bibnamefont
  {Gorelenkov}}, \bibinfo {author} {\bibfnamefont {M.}~\bibnamefont
  {Finkenthal}}, \bibinfo {author} {\bibfnamefont {E.}~\bibnamefont
  {Fredrickson}}, \bibinfo {author} {\bibfnamefont {S.}~\bibnamefont {Kaye}},
  \bibinfo {author} {\bibfnamefont {E.}~\bibnamefont {Mazzucato}}, \ and\
  \bibinfo {author} {\bibfnamefont {K.}~\bibnamefont {Tritz}},\ }\href
  {\doibase 10.1103/PhysRevLett.102.115002} {\bibfield  {journal} {\bibinfo
  {journal} {Phys. Rev. Lett.}\ }\textbf {\bibinfo {volume} {102}},\ \bibinfo
  {pages} {115002} (\bibinfo {year} {2009})}\BibitemShut {NoStop}%
\bibitem [{\citenamefont {Gorelenkov}\ and\ \citenamefont
  {Cheng}(1995{\natexlab{a}})}]{Gorelenkov1995POP}%
  \BibitemOpen
  \bibfield  {author} {\bibinfo {author} {\bibfnamefont {N.~N.}\ \bibnamefont
  {Gorelenkov}}\ and\ \bibinfo {author} {\bibfnamefont {C.~Z.}\ \bibnamefont
  {Cheng}},\ }\href {\doibase 10.1063/1.871281} {\bibfield  {journal} {\bibinfo
   {journal} {Physics of Plasmas}\ }\textbf {\bibinfo {volume} {2}},\ \bibinfo
  {pages} {1961} (\bibinfo {year} {1995}{\natexlab{a}})}\BibitemShut {NoStop}%
\bibitem [{\citenamefont {Belikov}\ \emph {et~al.}(2003)\citenamefont
  {Belikov}, \citenamefont {Kolesnichenko},\ and\ \citenamefont
  {White}}]{Belikov2003POP}%
  \BibitemOpen
  \bibfield  {author} {\bibinfo {author} {\bibfnamefont {V.~S.}\ \bibnamefont
  {Belikov}}, \bibinfo {author} {\bibfnamefont {Y.~I.}\ \bibnamefont
  {Kolesnichenko}}, \ and\ \bibinfo {author} {\bibfnamefont {R.~B.}\
  \bibnamefont {White}},\ }\href {\doibase http://dx.doi.org/10.1063/1.1625375}
  {\bibfield  {journal} {\bibinfo  {journal} {Physics of Plasmas}\ }\textbf
  {\bibinfo {volume} {10}},\ \bibinfo {pages} {4771} (\bibinfo {year}
  {2003})}\BibitemShut {NoStop}%
\bibitem [{\citenamefont {Zhang}\ \emph {et~al.}(2015)\citenamefont {Zhang},
  \citenamefont {Fu}, \citenamefont {White},\ and\ \citenamefont
  {Wang}}]{Zhang2015NF}%
  \BibitemOpen
  \bibfield  {author} {\bibinfo {author} {\bibfnamefont {R.~B.}\ \bibnamefont
  {Zhang}}, \bibinfo {author} {\bibfnamefont {G.~Y.}\ \bibnamefont {Fu}},
  \bibinfo {author} {\bibfnamefont {R.~B.}\ \bibnamefont {White}}, \ and\
  \bibinfo {author} {\bibfnamefont {X.~G.}\ \bibnamefont {Wang}},\ }\href
  {\doibase 10.1088/0029-5515/55/12/122002} {\bibfield  {journal} {\bibinfo
  {journal} {Nuclear Fusion}\ }\textbf {\bibinfo {volume} {55}},\ \bibinfo
  {pages} {122002} (\bibinfo {year} {2015})}\BibitemShut {NoStop}%
\bibitem [{\citenamefont {Stix}(1992)}]{StixWaves}%
  \BibitemOpen
  \bibfield  {author} {\bibinfo {author} {\bibfnamefont {T.~H.}\ \bibnamefont
  {Stix}},\ }\href@noop {} {\emph {\bibinfo {title} {Waves in Plasmas}}}\
  (\bibinfo  {publisher} {American Institute of Physics},\ \bibinfo {year}
  {1992})\BibitemShut {NoStop}%
\bibitem [{\citenamefont {Friedberg}(2014)}]{FreidbergMHD}%
  \BibitemOpen
  \bibfield  {author} {\bibinfo {author} {\bibfnamefont {J.~P.}\ \bibnamefont
  {Friedberg}},\ }\href@noop {} {\emph {\bibinfo {title} {Ideal
  Magnetohydrodynamics}}}\ (\bibinfo  {publisher} {Cambridge University
  Press},\ \bibinfo {year} {2014})\BibitemShut {NoStop}%
\bibitem [{\citenamefont {Goedbloed}\ and\ \citenamefont
  {Poedts}(2004)}]{GoedbloedMHDv1}%
  \BibitemOpen
  \bibfield  {author} {\bibinfo {author} {\bibfnamefont {J.~P.}\ \bibnamefont
  {Goedbloed}}\ and\ \bibinfo {author} {\bibfnamefont {S.}~\bibnamefont
  {Poedts}},\ }\href@noop {} {\emph {\bibinfo {title} {Principles of
  Magnetohydrodynamics}}}\ (\bibinfo  {publisher} {Cambridge University
  Press},\ \bibinfo {year} {2004})\BibitemShut {NoStop}%
\bibitem [{\citenamefont {Chen}(1994)}]{Chen1994POP}%
  \BibitemOpen
  \bibfield  {author} {\bibinfo {author} {\bibfnamefont {L.}~\bibnamefont
  {Chen}},\ }\href {\doibase 10.1063/1.870702} {\bibfield  {journal} {\bibinfo
  {journal} {Physics of Plasmas}\ }\textbf {\bibinfo {volume} {1}},\ \bibinfo
  {pages} {1519} (\bibinfo {year} {1994})}\BibitemShut {NoStop}%
\bibitem [{\citenamefont {McGuire}\ \emph {et~al.}(1983)\citenamefont
  {McGuire}, \citenamefont {Goldston}, \citenamefont {Bell}, \citenamefont
  {Bitter}, \citenamefont {Bol}, \citenamefont {Brau}, \citenamefont
  {Buchenauer}, \citenamefont {Crowley}, \citenamefont {Davis}, \citenamefont
  {Dylla}, \citenamefont {Eubank}, \citenamefont {Fishman}, \citenamefont
  {Fonck}, \citenamefont {Grek}, \citenamefont {Grimm}, \citenamefont
  {Hawryluk}, \citenamefont {Hsuan}, \citenamefont {Hulse}, \citenamefont
  {Izzo}, \citenamefont {Kaita}, \citenamefont {Kaye}, \citenamefont {Kugel},
  \citenamefont {Johnson}, \citenamefont {Manickam}, \citenamefont {Manos},
  \citenamefont {Mansfield}, \citenamefont {Mazzucato}, \citenamefont {McCann},
  \citenamefont {McCune}, \citenamefont {Monticello}, \citenamefont {Motley},
  \citenamefont {Mueller}, \citenamefont {Oasa}, \citenamefont {Okabayashi},
  \citenamefont {Owens}, \citenamefont {Park}, \citenamefont {Reusch},
  \citenamefont {Sauthoff}, \citenamefont {Schmidt}, \citenamefont {Sesnic},
  \citenamefont {Strachan}, \citenamefont {Surko}, \citenamefont {Slusher},
  \citenamefont {Takahashi}, \citenamefont {Tenney}, \citenamefont {Thomas},
  \citenamefont {Towner}, \citenamefont {Valley},\ and\ \citenamefont
  {White}}]{McGuire1983PRL}%
  \BibitemOpen
  \bibfield  {author} {\bibinfo {author} {\bibfnamefont {K.}~\bibnamefont
  {McGuire}}, \bibinfo {author} {\bibfnamefont {R.}~\bibnamefont {Goldston}},
  \bibinfo {author} {\bibfnamefont {M.}~\bibnamefont {Bell}}, \bibinfo {author}
  {\bibfnamefont {M.}~\bibnamefont {Bitter}}, \bibinfo {author} {\bibfnamefont
  {K.}~\bibnamefont {Bol}}, \bibinfo {author} {\bibfnamefont {K.}~\bibnamefont
  {Brau}}, \bibinfo {author} {\bibfnamefont {D.}~\bibnamefont {Buchenauer}},
  \bibinfo {author} {\bibfnamefont {T.}~\bibnamefont {Crowley}}, \bibinfo
  {author} {\bibfnamefont {S.}~\bibnamefont {Davis}}, \bibinfo {author}
  {\bibfnamefont {F.}~\bibnamefont {Dylla}}, \bibinfo {author} {\bibfnamefont
  {H.}~\bibnamefont {Eubank}}, \bibinfo {author} {\bibfnamefont
  {H.}~\bibnamefont {Fishman}}, \bibinfo {author} {\bibfnamefont
  {R.}~\bibnamefont {Fonck}}, \bibinfo {author} {\bibfnamefont
  {B.}~\bibnamefont {Grek}}, \bibinfo {author} {\bibfnamefont {R.}~\bibnamefont
  {Grimm}}, \bibinfo {author} {\bibfnamefont {R.}~\bibnamefont {Hawryluk}},
  \bibinfo {author} {\bibfnamefont {H.}~\bibnamefont {Hsuan}}, \bibinfo
  {author} {\bibfnamefont {R.}~\bibnamefont {Hulse}}, \bibinfo {author}
  {\bibfnamefont {R.}~\bibnamefont {Izzo}}, \bibinfo {author} {\bibfnamefont
  {R.}~\bibnamefont {Kaita}}, \bibinfo {author} {\bibfnamefont
  {S.}~\bibnamefont {Kaye}}, \bibinfo {author} {\bibfnamefont {H.}~\bibnamefont
  {Kugel}}, \bibinfo {author} {\bibfnamefont {D.}~\bibnamefont {Johnson}},
  \bibinfo {author} {\bibfnamefont {J.}~\bibnamefont {Manickam}}, \bibinfo
  {author} {\bibfnamefont {D.}~\bibnamefont {Manos}}, \bibinfo {author}
  {\bibfnamefont {D.}~\bibnamefont {Mansfield}}, \bibinfo {author}
  {\bibfnamefont {E.}~\bibnamefont {Mazzucato}}, \bibinfo {author}
  {\bibfnamefont {R.}~\bibnamefont {McCann}}, \bibinfo {author} {\bibfnamefont
  {D.}~\bibnamefont {McCune}}, \bibinfo {author} {\bibfnamefont
  {D.}~\bibnamefont {Monticello}}, \bibinfo {author} {\bibfnamefont
  {R.}~\bibnamefont {Motley}}, \bibinfo {author} {\bibfnamefont
  {D.}~\bibnamefont {Mueller}}, \bibinfo {author} {\bibfnamefont
  {K.}~\bibnamefont {Oasa}}, \bibinfo {author} {\bibfnamefont {M.}~\bibnamefont
  {Okabayashi}}, \bibinfo {author} {\bibfnamefont {K.}~\bibnamefont {Owens}},
  \bibinfo {author} {\bibfnamefont {W.}~\bibnamefont {Park}}, \bibinfo {author}
  {\bibfnamefont {M.}~\bibnamefont {Reusch}}, \bibinfo {author} {\bibfnamefont
  {N.}~\bibnamefont {Sauthoff}}, \bibinfo {author} {\bibfnamefont
  {G.}~\bibnamefont {Schmidt}}, \bibinfo {author} {\bibfnamefont
  {S.}~\bibnamefont {Sesnic}}, \bibinfo {author} {\bibfnamefont
  {J.}~\bibnamefont {Strachan}}, \bibinfo {author} {\bibfnamefont
  {C.}~\bibnamefont {Surko}}, \bibinfo {author} {\bibfnamefont
  {R.}~\bibnamefont {Slusher}}, \bibinfo {author} {\bibfnamefont
  {H.}~\bibnamefont {Takahashi}}, \bibinfo {author} {\bibfnamefont
  {F.}~\bibnamefont {Tenney}}, \bibinfo {author} {\bibfnamefont
  {P.}~\bibnamefont {Thomas}}, \bibinfo {author} {\bibfnamefont
  {H.}~\bibnamefont {Towner}}, \bibinfo {author} {\bibfnamefont
  {J.}~\bibnamefont {Valley}}, \ and\ \bibinfo {author} {\bibfnamefont
  {R.}~\bibnamefont {White}},\ }\href {\doibase 10.1103/PhysRevLett.50.891}
  {\bibfield  {journal} {\bibinfo  {journal} {Phys. Rev. Lett.}\ }\textbf
  {\bibinfo {volume} {50}},\ \bibinfo {pages} {891} (\bibinfo {year}
  {1983})}\BibitemShut {NoStop}%
\bibitem [{\citenamefont {Chen}\ \emph {et~al.}(1984)\citenamefont {Chen},
  \citenamefont {White},\ and\ \citenamefont {Rosenbluth}}]{Chen1984PRL}%
  \BibitemOpen
  \bibfield  {author} {\bibinfo {author} {\bibfnamefont {L.}~\bibnamefont
  {Chen}}, \bibinfo {author} {\bibfnamefont {R.~B.}\ \bibnamefont {White}}, \
  and\ \bibinfo {author} {\bibfnamefont {M.~N.}\ \bibnamefont {Rosenbluth}},\
  }\href {\doibase 10.1103/PhysRevLett.52.1122} {\bibfield  {journal} {\bibinfo
   {journal} {Phys. Rev. Lett.}\ }\textbf {\bibinfo {volume} {52}},\ \bibinfo
  {pages} {1122} (\bibinfo {year} {1984})}\BibitemShut {NoStop}%
\bibitem [{\citenamefont {Fu}(2008)}]{Fu2008PRL}%
  \BibitemOpen
  \bibfield  {author} {\bibinfo {author} {\bibfnamefont {G.~Y.}\ \bibnamefont
  {Fu}},\ }\href {\doibase 10.1103/PhysRevLett.101.185002} {\bibfield
  {journal} {\bibinfo  {journal} {Phys. Rev. Lett.}\ }\textbf {\bibinfo
  {volume} {101}},\ \bibinfo {pages} {185002} (\bibinfo {year}
  {2008})}\BibitemShut {NoStop}%
\bibitem [{\citenamefont {Nazikian}\ \emph {et~al.}(2008)\citenamefont
  {Nazikian}, \citenamefont {Fu}, \citenamefont {Austin}, \citenamefont {Berk},
  \citenamefont {Budny}, \citenamefont {Gorelenkov}, \citenamefont {Heidbrink},
  \citenamefont {Holcomb}, \citenamefont {Kramer}, \citenamefont {McKee},
  \citenamefont {Makowski}, \citenamefont {Solomon}, \citenamefont {Shafer},
  \citenamefont {Strait},\ and\ \citenamefont {Zeeland}}]{Nazikian2008PRL}%
  \BibitemOpen
  \bibfield  {author} {\bibinfo {author} {\bibfnamefont {R.}~\bibnamefont
  {Nazikian}}, \bibinfo {author} {\bibfnamefont {G.~Y.}\ \bibnamefont {Fu}},
  \bibinfo {author} {\bibfnamefont {M.~E.}\ \bibnamefont {Austin}}, \bibinfo
  {author} {\bibfnamefont {H.~L.}\ \bibnamefont {Berk}}, \bibinfo {author}
  {\bibfnamefont {R.~V.}\ \bibnamefont {Budny}}, \bibinfo {author}
  {\bibfnamefont {N.~N.}\ \bibnamefont {Gorelenkov}}, \bibinfo {author}
  {\bibfnamefont {W.~W.}\ \bibnamefont {Heidbrink}}, \bibinfo {author}
  {\bibfnamefont {C.~T.}\ \bibnamefont {Holcomb}}, \bibinfo {author}
  {\bibfnamefont {G.~J.}\ \bibnamefont {Kramer}}, \bibinfo {author}
  {\bibfnamefont {G.~R.}\ \bibnamefont {McKee}}, \bibinfo {author}
  {\bibfnamefont {M.~A.}\ \bibnamefont {Makowski}}, \bibinfo {author}
  {\bibfnamefont {W.~M.}\ \bibnamefont {Solomon}}, \bibinfo {author}
  {\bibfnamefont {M.}~\bibnamefont {Shafer}}, \bibinfo {author} {\bibfnamefont
  {E.~J.}\ \bibnamefont {Strait}}, \ and\ \bibinfo {author} {\bibfnamefont
  {M.~A.~V.}\ \bibnamefont {Zeeland}},\ }\href {\doibase
  10.1103/PhysRevLett.101.185001} {\bibfield  {journal} {\bibinfo  {journal}
  {Phys. Rev. Lett.}\ }\textbf {\bibinfo {volume} {101}},\ \bibinfo {pages}
  {185001} (\bibinfo {year} {2008})}\BibitemShut {NoStop}%
\bibitem [{\citenamefont {Gurnett}\ and\ \citenamefont
  {Bhattacharjee}(2005)}]{GurnettMHD}%
  \BibitemOpen
  \bibfield  {author} {\bibinfo {author} {\bibfnamefont {D.~A.}\ \bibnamefont
  {Gurnett}}\ and\ \bibinfo {author} {\bibfnamefont {A.}~\bibnamefont
  {Bhattacharjee}},\ }\href@noop {} {\emph {\bibinfo {title} {Introduction to
  Plasma Physics with Space and Laboratory Applications}}}\ (\bibinfo
  {publisher} {Cambridge University Press},\ \bibinfo {year}
  {2005})\BibitemShut {NoStop}%
\bibitem [{\citenamefont {Heidbrink}(2002)}]{Heidbrink2002POP}%
  \BibitemOpen
  \bibfield  {author} {\bibinfo {author} {\bibfnamefont {W.~W.}\ \bibnamefont
  {Heidbrink}},\ }\href {\doibase 10.1063/1.1461383} {\bibfield  {journal}
  {\bibinfo  {journal} {Physics of Plasmas}\ }\textbf {\bibinfo {volume} {9}},\
  \bibinfo {pages} {2113} (\bibinfo {year} {2002})}\BibitemShut {NoStop}%
\bibitem [{\citenamefont {Gorelenkov}\ \emph
  {et~al.}(2007{\natexlab{a}})\citenamefont {Gorelenkov}, \citenamefont {Berk},
  \citenamefont {Fredrickson}, \citenamefont {Sharapov},\ and\ \citenamefont
  {{JET EFDA Contributors}}}]{Gorelenkov2007PLA}%
  \BibitemOpen
  \bibfield  {author} {\bibinfo {author} {\bibfnamefont {N.~N.}\ \bibnamefont
  {Gorelenkov}}, \bibinfo {author} {\bibfnamefont {H.~L.}\ \bibnamefont
  {Berk}}, \bibinfo {author} {\bibfnamefont {E.}~\bibnamefont {Fredrickson}},
  \bibinfo {author} {\bibfnamefont {S.~E.}\ \bibnamefont {Sharapov}}, \ and\
  \bibinfo {author} {\bibnamefont {{JET EFDA Contributors}}},\ }\href {\doibase
  http://dx.doi.org/10.1016/j.physleta.2007.05.113} {\bibfield  {journal}
  {\bibinfo  {journal} {Physics Letters A}\ }\textbf {\bibinfo {volume}
  {370}},\ \bibinfo {pages} {70 } (\bibinfo {year}
  {2007}{\natexlab{a}})}\BibitemShut {NoStop}%
\bibitem [{\citenamefont {Gorelenkov}\ \emph
  {et~al.}(2007{\natexlab{b}})\citenamefont {Gorelenkov}, \citenamefont {Berk},
  \citenamefont {Crocker}, \citenamefont {Fredrickson}, \citenamefont {Kaye},
  \citenamefont {Kubota}, \citenamefont {Park}, \citenamefont {Peebles},
  \citenamefont {Sabbagh}, \citenamefont {Sharapov}, \citenamefont {Stutman},
  \citenamefont {Tritz}, \citenamefont {Levinton}, \citenamefont {Yuh},
  \citenamefont {the NSTX~Team},\ and\ \citenamefont {{JET EFDA
  Contributors}}}]{Gorelenkov2007PPCF}%
  \BibitemOpen
  \bibfield  {author} {\bibinfo {author} {\bibfnamefont {N.~N.}\ \bibnamefont
  {Gorelenkov}}, \bibinfo {author} {\bibfnamefont {H.~L.}\ \bibnamefont
  {Berk}}, \bibinfo {author} {\bibfnamefont {N.~A.}\ \bibnamefont {Crocker}},
  \bibinfo {author} {\bibfnamefont {E.~D.}\ \bibnamefont {Fredrickson}},
  \bibinfo {author} {\bibfnamefont {S.}~\bibnamefont {Kaye}}, \bibinfo {author}
  {\bibfnamefont {S.}~\bibnamefont {Kubota}}, \bibinfo {author} {\bibfnamefont
  {H.}~\bibnamefont {Park}}, \bibinfo {author} {\bibfnamefont {W.}~\bibnamefont
  {Peebles}}, \bibinfo {author} {\bibfnamefont {S.~A.}\ \bibnamefont
  {Sabbagh}}, \bibinfo {author} {\bibfnamefont {S.~E.}\ \bibnamefont
  {Sharapov}}, \bibinfo {author} {\bibfnamefont {D.}~\bibnamefont {Stutman}},
  \bibinfo {author} {\bibfnamefont {K.}~\bibnamefont {Tritz}}, \bibinfo
  {author} {\bibfnamefont {F.~M.}\ \bibnamefont {Levinton}}, \bibinfo {author}
  {\bibfnamefont {H.}~\bibnamefont {Yuh}}, \bibinfo {author} {\bibnamefont {the
  NSTX~Team}}, \ and\ \bibinfo {author} {\bibnamefont {{JET EFDA
  Contributors}}},\ }\href {http://stacks.iop.org/0741-3335/49/i=12B/a=S34}
  {\bibfield  {journal} {\bibinfo  {journal} {Plasma Physics and Controlled
  Fusion}\ }\textbf {\bibinfo {volume} {49}},\ \bibinfo {pages} {B371}
  (\bibinfo {year} {2007}{\natexlab{b}})}\BibitemShut {NoStop}%
\bibitem [{\citenamefont {Cheng}\ \emph {et~al.}(1985)\citenamefont {Cheng},
  \citenamefont {Chen},\ and\ \citenamefont {Chance}}]{Cheng1985AP}%
  \BibitemOpen
  \bibfield  {author} {\bibinfo {author} {\bibfnamefont {C.}~\bibnamefont
  {Cheng}}, \bibinfo {author} {\bibfnamefont {L.}~\bibnamefont {Chen}}, \ and\
  \bibinfo {author} {\bibfnamefont {M.}~\bibnamefont {Chance}},\ }\href
  {\doibase http://dx.doi.org/10.1016/0003-4916(85)90335-5} {\bibfield
  {journal} {\bibinfo  {journal} {Annals of Physics}\ }\textbf {\bibinfo
  {volume} {161}},\ \bibinfo {pages} {21 } (\bibinfo {year}
  {1985})}\BibitemShut {NoStop}%
\bibitem [{\citenamefont {Heidbrink}(2008)}]{Heidbrink2008POP}%
  \BibitemOpen
  \bibfield  {author} {\bibinfo {author} {\bibfnamefont {W.~W.}\ \bibnamefont
  {Heidbrink}},\ }\href {\doibase 10.1063/1.2838239} {\bibfield  {journal}
  {\bibinfo  {journal} {Physics of Plasmas}\ }\textbf {\bibinfo {volume}
  {15}},\ \bibinfo {pages} {055501} (\bibinfo {year} {2008})}\BibitemShut
  {NoStop}%
\bibitem [{\citenamefont {Cottrell}\ and\ \citenamefont
  {Dendy}(1988)}]{Cottrell1988PRL}%
  \BibitemOpen
  \bibfield  {author} {\bibinfo {author} {\bibfnamefont {G.~A.}\ \bibnamefont
  {Cottrell}}\ and\ \bibinfo {author} {\bibfnamefont {R.~O.}\ \bibnamefont
  {Dendy}},\ }\href {\doibase 10.1103/PhysRevLett.60.33} {\bibfield  {journal}
  {\bibinfo  {journal} {Phys. Rev. Lett.}\ }\textbf {\bibinfo {volume} {60}},\
  \bibinfo {pages} {33} (\bibinfo {year} {1988})}\BibitemShut {NoStop}%
\bibitem [{\citenamefont {Gorelenkov}(2016)}]{Gorelenkov2016NJP}%
  \BibitemOpen
  \bibfield  {author} {\bibinfo {author} {\bibfnamefont {N.~N.}\ \bibnamefont
  {Gorelenkov}},\ }\href {http://stacks.iop.org/1367-2630/18/i=10/a=105010}
  {\bibfield  {journal} {\bibinfo  {journal} {New Journal of Physics}\ }\textbf
  {\bibinfo {volume} {18}},\ \bibinfo {pages} {105010} (\bibinfo {year}
  {2016})}\BibitemShut {NoStop}%
\bibitem [{\citenamefont {Heidbrink}\ and\ \citenamefont
  {Sadler}(1994)}]{Heidbrink1994NF}%
  \BibitemOpen
  \bibfield  {author} {\bibinfo {author} {\bibfnamefont {W.}~\bibnamefont
  {Heidbrink}}\ and\ \bibinfo {author} {\bibfnamefont {G.}~\bibnamefont
  {Sadler}},\ }\href {\doibase 10.1088/0029-5515/34/4/i07} {\bibfield
  {journal} {\bibinfo  {journal} {Nuclear Fusion}\ }\textbf {\bibinfo {volume}
  {34}},\ \bibinfo {pages} {535} (\bibinfo {year} {1994})}\BibitemShut
  {NoStop}%
\bibitem [{\citenamefont {Wong}(1999)}]{Wong1999PPCF}%
  \BibitemOpen
  \bibfield  {author} {\bibinfo {author} {\bibfnamefont {K.-L.}\ \bibnamefont
  {Wong}},\ }\href {\doibase 10.1088/0741-3335/41/1/001} {\bibfield  {journal}
  {\bibinfo  {journal} {Plasma Physics and Controlled Fusion}\ }\textbf
  {\bibinfo {volume} {41}},\ \bibinfo {pages} {R1} (\bibinfo {year}
  {1999})}\BibitemShut {NoStop}%
\bibitem [{\citenamefont {Fasoli}\ \emph {et~al.}(2007)\citenamefont {Fasoli},
  \citenamefont {Gormenzano}, \citenamefont {Berk}, \citenamefont {Breizman},
  \citenamefont {Briguglio}, \citenamefont {Darrow}, \citenamefont
  {Gorelenkov}, \citenamefont {Heidbrink}, \citenamefont {Jaun}, \citenamefont
  {Konovalov}, \citenamefont {Nazikian}, \citenamefont {Noterdaeme},
  \citenamefont {Sharapov}, \citenamefont {Shinohara}, \citenamefont {Testa},
  \citenamefont {Tobita}, \citenamefont {Todo}, \citenamefont {Vlad},\ and\
  \citenamefont {Zonca}}]{Fasoli2007NF}%
  \BibitemOpen
  \bibfield  {author} {\bibinfo {author} {\bibfnamefont {A.}~\bibnamefont
  {Fasoli}}, \bibinfo {author} {\bibfnamefont {C.}~\bibnamefont {Gormenzano}},
  \bibinfo {author} {\bibfnamefont {H.~L.}\ \bibnamefont {Berk}}, \bibinfo
  {author} {\bibfnamefont {B.}~\bibnamefont {Breizman}}, \bibinfo {author}
  {\bibfnamefont {S.}~\bibnamefont {Briguglio}}, \bibinfo {author}
  {\bibfnamefont {D.~S.}\ \bibnamefont {Darrow}}, \bibinfo {author}
  {\bibfnamefont {N.}~\bibnamefont {Gorelenkov}}, \bibinfo {author}
  {\bibfnamefont {W.~W.}\ \bibnamefont {Heidbrink}}, \bibinfo {author}
  {\bibfnamefont {A.}~\bibnamefont {Jaun}}, \bibinfo {author} {\bibfnamefont
  {S.~V.}\ \bibnamefont {Konovalov}}, \bibinfo {author} {\bibfnamefont
  {R.}~\bibnamefont {Nazikian}}, \bibinfo {author} {\bibfnamefont {J.-M.}\
  \bibnamefont {Noterdaeme}}, \bibinfo {author} {\bibfnamefont
  {S.}~\bibnamefont {Sharapov}}, \bibinfo {author} {\bibfnamefont
  {K.}~\bibnamefont {Shinohara}}, \bibinfo {author} {\bibfnamefont
  {D.}~\bibnamefont {Testa}}, \bibinfo {author} {\bibfnamefont
  {K.}~\bibnamefont {Tobita}}, \bibinfo {author} {\bibfnamefont
  {Y.}~\bibnamefont {Todo}}, \bibinfo {author} {\bibfnamefont {G.}~\bibnamefont
  {Vlad}}, \ and\ \bibinfo {author} {\bibfnamefont {F.}~\bibnamefont {Zonca}},\
  }\href {http://stacks.iop.org/0029-5515/47/i=6/a=S05} {\bibfield  {journal}
  {\bibinfo  {journal} {Nuclear Fusion}\ }\textbf {\bibinfo {volume} {47}},\
  \bibinfo {pages} {S264} (\bibinfo {year} {2007})}\BibitemShut {NoStop}%
\bibitem [{\citenamefont {Pinches}\ \emph {et~al.}(2015)\citenamefont
  {Pinches}, \citenamefont {Chapman}, \citenamefont {Lauber}, \citenamefont
  {Oliver}, \citenamefont {Sharapov}, \citenamefont {Shinohara},\ and\
  \citenamefont {Tani}}]{Pinches2015POP}%
  \BibitemOpen
  \bibfield  {author} {\bibinfo {author} {\bibfnamefont {S.~D.}\ \bibnamefont
  {Pinches}}, \bibinfo {author} {\bibfnamefont {I.~T.}\ \bibnamefont
  {Chapman}}, \bibinfo {author} {\bibfnamefont {P.~W.}\ \bibnamefont {Lauber}},
  \bibinfo {author} {\bibfnamefont {H.~J.~C.}\ \bibnamefont {Oliver}}, \bibinfo
  {author} {\bibfnamefont {S.~E.}\ \bibnamefont {Sharapov}}, \bibinfo {author}
  {\bibfnamefont {K.}~\bibnamefont {Shinohara}}, \ and\ \bibinfo {author}
  {\bibfnamefont {K.}~\bibnamefont {Tani}},\ }\href {\doibase
  10.1063/1.4908551} {\bibfield  {journal} {\bibinfo  {journal} {Physics of
  Plasmas}\ }\textbf {\bibinfo {volume} {22}},\ \bibinfo {pages} {021807}
  (\bibinfo {year} {2015})}\BibitemShut {NoStop}%
\bibitem [{\citenamefont {Breizman}\ and\ \citenamefont
  {Sharapov}(2011)}]{Breizman2011PPCF}%
  \BibitemOpen
  \bibfield  {author} {\bibinfo {author} {\bibfnamefont {B.~N.}\ \bibnamefont
  {Breizman}}\ and\ \bibinfo {author} {\bibfnamefont {S.~E.}\ \bibnamefont
  {Sharapov}},\ }\href {\doibase 10.1088/0741-3335/53/5/054001} {\bibfield
  {journal} {\bibinfo  {journal} {Plasma Physics and Controlled Fusion}\
  }\textbf {\bibinfo {volume} {53}},\ \bibinfo {pages} {054001} (\bibinfo
  {year} {2011})}\BibitemShut {NoStop}%
\bibitem [{\citenamefont {Sharapov}\ \emph {et~al.}(2013)\citenamefont
  {Sharapov}, \citenamefont {Alper}, \citenamefont {Berk}, \citenamefont
  {Borba}, \citenamefont {Breizman}, \citenamefont {Challis}, \citenamefont
  {Classen}, \citenamefont {Edlund}, \citenamefont {Eriksson}, \citenamefont
  {Fasoli}, \citenamefont {Fredrickson}, \citenamefont {Fu}, \citenamefont
  {Garcia-Munoz}, \citenamefont {Gassner}, \citenamefont {Ghantous},
  \citenamefont {Goloborodko}, \citenamefont {Gorelenkov}, \citenamefont
  {Gryaznevich}, \citenamefont {Hacquin}, \citenamefont {Heidbrink},
  \citenamefont {Hellesen}, \citenamefont {Kiptily}, \citenamefont {Kramer},
  \citenamefont {Lauber}, \citenamefont {Lilley}, \citenamefont {Lisak},
  \citenamefont {Nabais}, \citenamefont {Nazikian}, \citenamefont {Nyqvist},
  \citenamefont {Osakabe}, \citenamefont {von Thun}, \citenamefont {Pinches},
  \citenamefont {Podesta}, \citenamefont {Porkolab}, \citenamefont {Shinohara},
  \citenamefont {Schoepf}, \citenamefont {Todo}, \citenamefont {Toi},
  \citenamefont {Zeeland}, \citenamefont {Voitsekhovich}, \citenamefont
  {White},\ and\ \citenamefont {Yavorskij}}]{Sharapov2013NF}%
  \BibitemOpen
  \bibfield  {author} {\bibinfo {author} {\bibfnamefont {S.~E.}\ \bibnamefont
  {Sharapov}}, \bibinfo {author} {\bibfnamefont {B.}~\bibnamefont {Alper}},
  \bibinfo {author} {\bibfnamefont {H.~L.}\ \bibnamefont {Berk}}, \bibinfo
  {author} {\bibfnamefont {D.~N.}\ \bibnamefont {Borba}}, \bibinfo {author}
  {\bibfnamefont {B.~N.}\ \bibnamefont {Breizman}}, \bibinfo {author}
  {\bibfnamefont {C.~D.}\ \bibnamefont {Challis}}, \bibinfo {author}
  {\bibfnamefont {I.~G.}\ \bibnamefont {Classen}}, \bibinfo {author}
  {\bibfnamefont {E.~M.}\ \bibnamefont {Edlund}}, \bibinfo {author}
  {\bibfnamefont {J.}~\bibnamefont {Eriksson}}, \bibinfo {author}
  {\bibfnamefont {A.}~\bibnamefont {Fasoli}}, \bibinfo {author} {\bibfnamefont
  {E.~D.}\ \bibnamefont {Fredrickson}}, \bibinfo {author} {\bibfnamefont
  {G.~Y.}\ \bibnamefont {Fu}}, \bibinfo {author} {\bibfnamefont
  {M.}~\bibnamefont {Garcia-Munoz}}, \bibinfo {author} {\bibfnamefont
  {T.}~\bibnamefont {Gassner}}, \bibinfo {author} {\bibfnamefont
  {K.}~\bibnamefont {Ghantous}}, \bibinfo {author} {\bibfnamefont
  {V.}~\bibnamefont {Goloborodko}}, \bibinfo {author} {\bibfnamefont {N.~N.}\
  \bibnamefont {Gorelenkov}}, \bibinfo {author} {\bibfnamefont {M.~P.}\
  \bibnamefont {Gryaznevich}}, \bibinfo {author} {\bibfnamefont
  {S.}~\bibnamefont {Hacquin}}, \bibinfo {author} {\bibfnamefont {W.~W.}\
  \bibnamefont {Heidbrink}}, \bibinfo {author} {\bibfnamefont {C.}~\bibnamefont
  {Hellesen}}, \bibinfo {author} {\bibfnamefont {V.~G.}\ \bibnamefont
  {Kiptily}}, \bibinfo {author} {\bibfnamefont {G.~J.}\ \bibnamefont {Kramer}},
  \bibinfo {author} {\bibfnamefont {P.}~\bibnamefont {Lauber}}, \bibinfo
  {author} {\bibfnamefont {M.~K.}\ \bibnamefont {Lilley}}, \bibinfo {author}
  {\bibfnamefont {M.}~\bibnamefont {Lisak}}, \bibinfo {author} {\bibfnamefont
  {F.}~\bibnamefont {Nabais}}, \bibinfo {author} {\bibfnamefont
  {R.}~\bibnamefont {Nazikian}}, \bibinfo {author} {\bibfnamefont
  {R.}~\bibnamefont {Nyqvist}}, \bibinfo {author} {\bibfnamefont
  {M.}~\bibnamefont {Osakabe}}, \bibinfo {author} {\bibfnamefont {C.~P.}\
  \bibnamefont {von Thun}}, \bibinfo {author} {\bibfnamefont {S.~D.}\
  \bibnamefont {Pinches}}, \bibinfo {author} {\bibfnamefont {M.}~\bibnamefont
  {Podesta}}, \bibinfo {author} {\bibfnamefont {M.}~\bibnamefont {Porkolab}},
  \bibinfo {author} {\bibfnamefont {K.}~\bibnamefont {Shinohara}}, \bibinfo
  {author} {\bibfnamefont {K.}~\bibnamefont {Schoepf}}, \bibinfo {author}
  {\bibfnamefont {Y.}~\bibnamefont {Todo}}, \bibinfo {author} {\bibfnamefont
  {K.}~\bibnamefont {Toi}}, \bibinfo {author} {\bibfnamefont {M.~A.~V.}\
  \bibnamefont {Zeeland}}, \bibinfo {author} {\bibfnamefont {I.}~\bibnamefont
  {Voitsekhovich}}, \bibinfo {author} {\bibfnamefont {R.~B.}\ \bibnamefont
  {White}}, \ and\ \bibinfo {author} {\bibfnamefont {V.}~\bibnamefont
  {Yavorskij}},\ }\href {\doibase 10.1088/0029-5515/53/10/104022} {\bibfield
  {journal} {\bibinfo  {journal} {Nuclear Fusion}\ }\textbf {\bibinfo {volume}
  {53}},\ \bibinfo {pages} {104022} (\bibinfo {year} {2013})}\BibitemShut
  {NoStop}%
\bibitem [{\citenamefont {Gorelenkov}\ \emph {et~al.}(2014)\citenamefont
  {Gorelenkov}, \citenamefont {Pinches},\ and\ \citenamefont
  {Toi}}]{Gorelenkov2014NF}%
  \BibitemOpen
  \bibfield  {author} {\bibinfo {author} {\bibfnamefont {N.~N.}\ \bibnamefont
  {Gorelenkov}}, \bibinfo {author} {\bibfnamefont {S.~D.}\ \bibnamefont
  {Pinches}}, \ and\ \bibinfo {author} {\bibfnamefont {K.}~\bibnamefont
  {Toi}},\ }\href {http://stacks.iop.org/0029-5515/54/i=12/a=125001} {\bibfield
   {journal} {\bibinfo  {journal} {Nuclear Fusion}\ }\textbf {\bibinfo {volume}
  {54}},\ \bibinfo {pages} {125001} (\bibinfo {year} {2014})}\BibitemShut
  {NoStop}%
\bibitem [{\citenamefont {McClements}\ and\ \citenamefont
  {Fredrickson}(2017)}]{McClements2017PPCF}%
  \BibitemOpen
  \bibfield  {author} {\bibinfo {author} {\bibfnamefont {K.~G.}\ \bibnamefont
  {McClements}}\ and\ \bibinfo {author} {\bibfnamefont {E.~D.}\ \bibnamefont
  {Fredrickson}},\ }\href {http://stacks.iop.org/0741-3335/59/i=5/a=053001}
  {\bibfield  {journal} {\bibinfo  {journal} {Plasma Physics and Controlled
  Fusion}\ }\textbf {\bibinfo {volume} {59}},\ \bibinfo {pages} {053001}
  (\bibinfo {year} {2017})}\BibitemShut {NoStop}%
\bibitem [{\citenamefont {Todo}(2018)}]{Todo2019RMPP}%
  \BibitemOpen
  \bibfield  {author} {\bibinfo {author} {\bibfnamefont {Y.}~\bibnamefont
  {Todo}},\ }\href {\doibase 10.1007/s41614-018-0022-9} {\bibfield  {journal}
  {\bibinfo  {journal} {Reviews of Modern Plasma Physics}\ }\textbf {\bibinfo
  {volume} {3}},\ \bibinfo {pages} {1} (\bibinfo {year} {2018})}\BibitemShut
  {NoStop}%
\bibitem [{\citenamefont {Chen}\ and\ \citenamefont
  {Zonca}(1995)}]{Chen1995PS}%
  \BibitemOpen
  \bibfield  {author} {\bibinfo {author} {\bibfnamefont {L.}~\bibnamefont
  {Chen}}\ and\ \bibinfo {author} {\bibfnamefont {F.}~\bibnamefont {Zonca}},\
  }\href {\doibase 10.1088/0031-8949/1995/t60/011} {\bibfield  {journal}
  {\bibinfo  {journal} {Physica Scripta}\ }\textbf {\bibinfo {volume} {T60}},\
  \bibinfo {pages} {81} (\bibinfo {year} {1995})}\BibitemShut {NoStop}%
\bibitem [{\citenamefont {Chen}\ and\ \citenamefont
  {Zonca}(2007)}]{Chen2007NF}%
  \BibitemOpen
  \bibfield  {author} {\bibinfo {author} {\bibfnamefont {L.}~\bibnamefont
  {Chen}}\ and\ \bibinfo {author} {\bibfnamefont {F.}~\bibnamefont {Zonca}},\
  }\href {\doibase 10.1088/0029-5515/47/10/s20} {\bibfield  {journal} {\bibinfo
   {journal} {Nuclear Fusion}\ }\textbf {\bibinfo {volume} {47}},\ \bibinfo
  {pages} {S727} (\bibinfo {year} {2007})}\BibitemShut {NoStop}%
\bibitem [{\citenamefont {Chen}\ and\ \citenamefont
  {Zonca}(2016)}]{Chen2016RMP}%
  \BibitemOpen
  \bibfield  {author} {\bibinfo {author} {\bibfnamefont {L.}~\bibnamefont
  {Chen}}\ and\ \bibinfo {author} {\bibfnamefont {F.}~\bibnamefont {Zonca}},\
  }\href {\doibase 10.1103/RevModPhys.88.015008} {\bibfield  {journal}
  {\bibinfo  {journal} {Rev. Mod. Phys.}\ }\textbf {\bibinfo {volume} {88}},\
  \bibinfo {pages} {015008} (\bibinfo {year} {2016})}\BibitemShut {NoStop}%
\bibitem [{\citenamefont {Cross}(1988)}]{CrossAE}%
  \BibitemOpen
  \bibfield  {author} {\bibinfo {author} {\bibfnamefont {R.}~\bibnamefont
  {Cross}},\ }\href@noop {} {\emph {\bibinfo {title} {An Introduction to
  \Alfven Waves}}}\ (\bibinfo  {publisher} {CRC Press},\ \bibinfo {year}
  {1988})\BibitemShut {NoStop}%
\bibitem [{\citenamefont {Cramer}(2001)}]{CramerAE}%
  \BibitemOpen
  \bibfield  {author} {\bibinfo {author} {\bibfnamefont {N.~F.}\ \bibnamefont
  {Cramer}},\ }\href@noop {} {\emph {\bibinfo {title} {The Physics of \Alfven
  Waves}}}\ (\bibinfo  {publisher} {Wiley-VCH},\ \bibinfo {year}
  {2001})\BibitemShut {NoStop}%
\bibitem [{\citenamefont {Fredrickson}\ \emph {et~al.}(2013)\citenamefont
  {Fredrickson}, \citenamefont {Gorelenkov}, \citenamefont {Podesta},
  \citenamefont {Bortolon}, \citenamefont {Crocker}, \citenamefont {Gerhardt},
  \citenamefont {Bell}, \citenamefont {Diallo}, \citenamefont {LeBlanc},
  \citenamefont {Levinton},\ and\ \citenamefont {Yuh}}]{Fredrickson2013POP}%
  \BibitemOpen
  \bibfield  {author} {\bibinfo {author} {\bibfnamefont {E.~D.}\ \bibnamefont
  {Fredrickson}}, \bibinfo {author} {\bibfnamefont {N.~N.}\ \bibnamefont
  {Gorelenkov}}, \bibinfo {author} {\bibfnamefont {M.}~\bibnamefont {Podesta}},
  \bibinfo {author} {\bibfnamefont {A.}~\bibnamefont {Bortolon}}, \bibinfo
  {author} {\bibfnamefont {N.~A.}\ \bibnamefont {Crocker}}, \bibinfo {author}
  {\bibfnamefont {S.~P.}\ \bibnamefont {Gerhardt}}, \bibinfo {author}
  {\bibfnamefont {R.~E.}\ \bibnamefont {Bell}}, \bibinfo {author}
  {\bibfnamefont {A.}~\bibnamefont {Diallo}}, \bibinfo {author} {\bibfnamefont
  {B.}~\bibnamefont {LeBlanc}}, \bibinfo {author} {\bibfnamefont {F.~M.}\
  \bibnamefont {Levinton}}, \ and\ \bibinfo {author} {\bibfnamefont
  {H.}~\bibnamefont {Yuh}},\ }\href {\doibase 10.1063/1.4801663} {\bibfield
  {journal} {\bibinfo  {journal} {Physics of Plasmas}\ }\textbf {\bibinfo
  {volume} {20}},\ \bibinfo {pages} {042112} (\bibinfo {year}
  {2013})}\BibitemShut {NoStop}%
\bibitem [{\citenamefont {Mahajan}\ and\ \citenamefont
  {Ross}(1983)}]{Mahajan1983bPF}%
  \BibitemOpen
  \bibfield  {author} {\bibinfo {author} {\bibfnamefont {S.~M.}\ \bibnamefont
  {Mahajan}}\ and\ \bibinfo {author} {\bibfnamefont {D.~W.}\ \bibnamefont
  {Ross}},\ }\href {\doibase 10.1063/1.864445} {\bibfield  {journal} {\bibinfo
  {journal} {The Physics of Fluids}\ }\textbf {\bibinfo {volume} {26}},\
  \bibinfo {pages} {2561} (\bibinfo {year} {1983})}\BibitemShut {NoStop}%
\bibitem [{\citenamefont {Gorelenkova}\ and\ \citenamefont
  {Gorelenkov}(1998)}]{Gorelenkova1998POP}%
  \BibitemOpen
  \bibfield  {author} {\bibinfo {author} {\bibfnamefont {M.~V.}\ \bibnamefont
  {Gorelenkova}}\ and\ \bibinfo {author} {\bibfnamefont {N.~N.}\ \bibnamefont
  {Gorelenkov}},\ }\href {\doibase 10.1063/1.873133} {\bibfield  {journal}
  {\bibinfo  {journal} {Physics of Plasmas}\ }\textbf {\bibinfo {volume} {5}},\
  \bibinfo {pages} {4104} (\bibinfo {year} {1998})}\BibitemShut {NoStop}%
\bibitem [{\citenamefont {Kolesnichenko}\ \emph {et~al.}(1998)\citenamefont
  {Kolesnichenko}, \citenamefont {Fülöp}, \citenamefont {Lisak},\ and\
  \citenamefont {Anderson}}]{Kolesnichenko1998NF}%
  \BibitemOpen
  \bibfield  {author} {\bibinfo {author} {\bibfnamefont {Y.}~\bibnamefont
  {Kolesnichenko}}, \bibinfo {author} {\bibfnamefont {T.}~\bibnamefont
  {Fülöp}}, \bibinfo {author} {\bibfnamefont {M.}~\bibnamefont {Lisak}}, \
  and\ \bibinfo {author} {\bibfnamefont {D.}~\bibnamefont {Anderson}},\ }\href
  {http://stacks.iop.org/0029-5515/38/i=12/a=311} {\bibfield  {journal}
  {\bibinfo  {journal} {Nuclear Fusion}\ }\textbf {\bibinfo {volume} {38}},\
  \bibinfo {pages} {1871} (\bibinfo {year} {1998})}\BibitemShut {NoStop}%
\bibitem [{\citenamefont {Gorelenkov}\ \emph
  {et~al.}(2002{\natexlab{a}})\citenamefont {Gorelenkov}, \citenamefont
  {Cheng},\ and\ \citenamefont {Fredrickson}}]{Gorelenkov2002POP}%
  \BibitemOpen
  \bibfield  {author} {\bibinfo {author} {\bibfnamefont {N.~N.}\ \bibnamefont
  {Gorelenkov}}, \bibinfo {author} {\bibfnamefont {C.~Z.}\ \bibnamefont
  {Cheng}}, \ and\ \bibinfo {author} {\bibfnamefont {E.}~\bibnamefont
  {Fredrickson}},\ }\href {\doibase 10.1063/1.1492803} {\bibfield  {journal}
  {\bibinfo  {journal} {Physics of Plasmas}\ }\textbf {\bibinfo {volume} {9}},\
  \bibinfo {pages} {3483} (\bibinfo {year} {2002}{\natexlab{a}})}\BibitemShut
  {NoStop}%
\bibitem [{\citenamefont {Gorelenkov}\ \emph
  {et~al.}(2002{\natexlab{b}})\citenamefont {Gorelenkov}, \citenamefont
  {Cheng}, \citenamefont {Fredrickson}, \citenamefont {Belova}, \citenamefont
  {Gates}, \citenamefont {Kaye}, \citenamefont {Kramer}, \citenamefont
  {Nazikian},\ and\ \citenamefont {White}}]{Gorelenkov2002NF}%
  \BibitemOpen
  \bibfield  {author} {\bibinfo {author} {\bibfnamefont {N.~N.}\ \bibnamefont
  {Gorelenkov}}, \bibinfo {author} {\bibfnamefont {C.~Z.}\ \bibnamefont
  {Cheng}}, \bibinfo {author} {\bibfnamefont {E.}~\bibnamefont {Fredrickson}},
  \bibinfo {author} {\bibfnamefont {E.}~\bibnamefont {Belova}}, \bibinfo
  {author} {\bibfnamefont {D.}~\bibnamefont {Gates}}, \bibinfo {author}
  {\bibfnamefont {S.}~\bibnamefont {Kaye}}, \bibinfo {author} {\bibfnamefont
  {G.~J.}\ \bibnamefont {Kramer}}, \bibinfo {author} {\bibfnamefont
  {R.}~\bibnamefont {Nazikian}}, \ and\ \bibinfo {author} {\bibfnamefont
  {R.}~\bibnamefont {White}},\ }\href
  {http://stacks.iop.org/0029-5515/42/i=8/a=306} {\bibfield  {journal}
  {\bibinfo  {journal} {Nuclear Fusion}\ }\textbf {\bibinfo {volume} {42}},\
  \bibinfo {pages} {977} (\bibinfo {year} {2002}{\natexlab{b}})}\BibitemShut
  {NoStop}%
\bibitem [{\citenamefont {Smith}\ \emph {et~al.}(2003)\citenamefont {Smith},
  \citenamefont {\Fulop}, \citenamefont {Lisak},\ and\ \citenamefont
  {Anderson}}]{Smith2003POP}%
  \BibitemOpen
  \bibfield  {author} {\bibinfo {author} {\bibfnamefont {H.}~\bibnamefont
  {Smith}}, \bibinfo {author} {\bibfnamefont {T.}~\bibnamefont {\Fulop}},
  \bibinfo {author} {\bibfnamefont {M.}~\bibnamefont {Lisak}}, \ and\ \bibinfo
  {author} {\bibfnamefont {D.}~\bibnamefont {Anderson}},\ }\href {\doibase
  10.1063/1.1566441} {\bibfield  {journal} {\bibinfo  {journal} {Physics of
  Plasmas}\ }\textbf {\bibinfo {volume} {10}},\ \bibinfo {pages} {1437}
  (\bibinfo {year} {2003})}\BibitemShut {NoStop}%
\bibitem [{\citenamefont {Gorelenkov}\ \emph {et~al.}(2006)\citenamefont
  {Gorelenkov}, \citenamefont {Fredrickson}, \citenamefont {Heidbrink},
  \citenamefont {Crocker}, \citenamefont {Kubota},\ and\ \citenamefont
  {Peebles}}]{Gorelenkov2006NF}%
  \BibitemOpen
  \bibfield  {author} {\bibinfo {author} {\bibfnamefont {N.~N.}\ \bibnamefont
  {Gorelenkov}}, \bibinfo {author} {\bibfnamefont {E.~D.}\ \bibnamefont
  {Fredrickson}}, \bibinfo {author} {\bibfnamefont {W.~W.}\ \bibnamefont
  {Heidbrink}}, \bibinfo {author} {\bibfnamefont {N.~A.}\ \bibnamefont
  {Crocker}}, \bibinfo {author} {\bibfnamefont {S.}~\bibnamefont {Kubota}}, \
  and\ \bibinfo {author} {\bibfnamefont {W.~A.}\ \bibnamefont {Peebles}},\
  }\href {http://stacks.iop.org/0029-5515/46/i=10/a=S10} {\bibfield  {journal}
  {\bibinfo  {journal} {Nuclear Fusion}\ }\textbf {\bibinfo {volume} {46}},\
  \bibinfo {pages} {S933} (\bibinfo {year} {2006})}\BibitemShut {NoStop}%
\bibitem [{\citenamefont {Smith}\ and\ \citenamefont
  {Verwichte}(2009)}]{Smith2009PPCF}%
  \BibitemOpen
  \bibfield  {author} {\bibinfo {author} {\bibfnamefont {H.~M.}\ \bibnamefont
  {Smith}}\ and\ \bibinfo {author} {\bibfnamefont {E.}~\bibnamefont
  {Verwichte}},\ }\href {http://stacks.iop.org/0741-3335/51/i=7/a=075001}
  {\bibfield  {journal} {\bibinfo  {journal} {Plasma Physics and Controlled
  Fusion}\ }\textbf {\bibinfo {volume} {51}},\ \bibinfo {pages} {075001}
  (\bibinfo {year} {2009})}\BibitemShut {NoStop}%
\bibitem [{\citenamefont {Smith}\ and\ \citenamefont
  {Fredrickson}(2017)}]{Smith2017PPCF}%
  \BibitemOpen
  \bibfield  {author} {\bibinfo {author} {\bibfnamefont {H.~M.}\ \bibnamefont
  {Smith}}\ and\ \bibinfo {author} {\bibfnamefont {E.~D.}\ \bibnamefont
  {Fredrickson}},\ }\href {http://stacks.iop.org/0741-3335/59/i=3/a=035007}
  {\bibfield  {journal} {\bibinfo  {journal} {Plasma Physics and Controlled
  Fusion}\ }\textbf {\bibinfo {volume} {59}},\ \bibinfo {pages} {035007}
  (\bibinfo {year} {2017})}\BibitemShut {NoStop}%
\bibitem [{\citenamefont {Hasegawa}\ and\ \citenamefont
  {Chen}(1976)}]{Hasegawa1976PF}%
  \BibitemOpen
  \bibfield  {author} {\bibinfo {author} {\bibfnamefont {A.}~\bibnamefont
  {Hasegawa}}\ and\ \bibinfo {author} {\bibfnamefont {L.}~\bibnamefont
  {Chen}},\ }\href {\doibase 10.1063/1.861427} {\bibfield  {journal} {\bibinfo
  {journal} {The Physics of Fluids}\ }\textbf {\bibinfo {volume} {19}},\
  \bibinfo {pages} {1924} (\bibinfo {year} {1976})}\BibitemShut {NoStop}%
\bibitem [{\citenamefont {Belova}\ \emph {et~al.}(2017)\citenamefont {Belova},
  \citenamefont {Gorelenkov}, \citenamefont {Crocker}, \citenamefont {Lestz},
  \citenamefont {Fredrickson}, \citenamefont {Tang},\ and\ \citenamefont
  {Tritz}}]{Belova2017POP}%
  \BibitemOpen
  \bibfield  {author} {\bibinfo {author} {\bibfnamefont {E.~V.}\ \bibnamefont
  {Belova}}, \bibinfo {author} {\bibfnamefont {N.~N.}\ \bibnamefont
  {Gorelenkov}}, \bibinfo {author} {\bibfnamefont {N.~A.}\ \bibnamefont
  {Crocker}}, \bibinfo {author} {\bibfnamefont {J.~B.}\ \bibnamefont {Lestz}},
  \bibinfo {author} {\bibfnamefont {E.~D.}\ \bibnamefont {Fredrickson}},
  \bibinfo {author} {\bibfnamefont {S.}~\bibnamefont {Tang}}, \ and\ \bibinfo
  {author} {\bibfnamefont {K.}~\bibnamefont {Tritz}},\ }\href {\doibase
  10.1063/1.4979278} {\bibfield  {journal} {\bibinfo  {journal} {Physics of
  Plasmas}\ }\textbf {\bibinfo {volume} {24}},\ \bibinfo {pages} {042505}
  (\bibinfo {year} {2017})}\BibitemShut {NoStop}%
\bibitem [{\citenamefont {Johnson}\ and\ \citenamefont
  {Cheng}(1997)}]{Johnson1997GRL}%
  \BibitemOpen
  \bibfield  {author} {\bibinfo {author} {\bibfnamefont {J.~R.}\ \bibnamefont
  {Johnson}}\ and\ \bibinfo {author} {\bibfnamefont {C.~Z.}\ \bibnamefont
  {Cheng}},\ }\href {\doibase 10.1029/97GL01333} {\bibfield  {journal}
  {\bibinfo  {journal} {Geophysical Research Letters}\ }\textbf {\bibinfo
  {volume} {24}},\ \bibinfo {pages} {1423} (\bibinfo {year}
  {1997})}\BibitemShut {NoStop}%
\bibitem [{\citenamefont {Heikkinen}\ \emph {et~al.}(1991)\citenamefont
  {Heikkinen}, \citenamefont {Hellsten},\ and\ \citenamefont
  {Alava}}]{Heikkinen1991NF}%
  \BibitemOpen
  \bibfield  {author} {\bibinfo {author} {\bibfnamefont {J.~A.}\ \bibnamefont
  {Heikkinen}}, \bibinfo {author} {\bibfnamefont {T.}~\bibnamefont {Hellsten}},
  \ and\ \bibinfo {author} {\bibfnamefont {M.~J.}\ \bibnamefont {Alava}},\
  }\href {http://stacks.iop.org/0029-5515/31/i=3/a=002} {\bibfield  {journal}
  {\bibinfo  {journal} {Nuclear Fusion}\ }\textbf {\bibinfo {volume} {31}},\
  \bibinfo {pages} {417} (\bibinfo {year} {1991})}\BibitemShut {NoStop}%
\bibitem [{\citenamefont {Fredrickson}\ \emph {et~al.}(2001)\citenamefont
  {Fredrickson}, \citenamefont {Gorelenkov}, \citenamefont {Cheng},
  \citenamefont {Bell}, \citenamefont {Darrow}, \citenamefont {Johnson},
  \citenamefont {Kaye}, \citenamefont {LeBlanc}, \citenamefont {Menard},
  \citenamefont {Kubota},\ and\ \citenamefont {Peebles}}]{Fredrickson2001PRL}%
  \BibitemOpen
  \bibfield  {author} {\bibinfo {author} {\bibfnamefont {E.~D.}\ \bibnamefont
  {Fredrickson}}, \bibinfo {author} {\bibfnamefont {N.}~\bibnamefont
  {Gorelenkov}}, \bibinfo {author} {\bibfnamefont {C.~Z.}\ \bibnamefont
  {Cheng}}, \bibinfo {author} {\bibfnamefont {R.}~\bibnamefont {Bell}},
  \bibinfo {author} {\bibfnamefont {D.}~\bibnamefont {Darrow}}, \bibinfo
  {author} {\bibfnamefont {D.}~\bibnamefont {Johnson}}, \bibinfo {author}
  {\bibfnamefont {S.}~\bibnamefont {Kaye}}, \bibinfo {author} {\bibfnamefont
  {B.}~\bibnamefont {LeBlanc}}, \bibinfo {author} {\bibfnamefont
  {J.}~\bibnamefont {Menard}}, \bibinfo {author} {\bibfnamefont
  {S.}~\bibnamefont {Kubota}}, \ and\ \bibinfo {author} {\bibfnamefont
  {W.}~\bibnamefont {Peebles}},\ }\href {\doibase
  10.1103/PhysRevLett.87.145001} {\bibfield  {journal} {\bibinfo  {journal}
  {Phys. Rev. Lett.}\ }\textbf {\bibinfo {volume} {87}},\ \bibinfo {pages}
  {145001} (\bibinfo {year} {2001})}\BibitemShut {NoStop}%
\bibitem [{\citenamefont {Fredrickson}\ \emph {et~al.}(2004)\citenamefont
  {Fredrickson}, \citenamefont {Gorelenkov},\ and\ \citenamefont
  {Menard}}]{Fredrickson2004POP}%
  \BibitemOpen
  \bibfield  {author} {\bibinfo {author} {\bibfnamefont {E.~D.}\ \bibnamefont
  {Fredrickson}}, \bibinfo {author} {\bibfnamefont {N.~N.}\ \bibnamefont
  {Gorelenkov}}, \ and\ \bibinfo {author} {\bibfnamefont {J.}~\bibnamefont
  {Menard}},\ }\href {\doibase 10.1063/1.1760094} {\bibfield  {journal}
  {\bibinfo  {journal} {Physics of Plasmas}\ }\textbf {\bibinfo {volume}
  {11}},\ \bibinfo {pages} {3653} (\bibinfo {year} {2004})}\BibitemShut
  {NoStop}%
\bibitem [{\citenamefont {Crocker}\ \emph {et~al.}(2013)\citenamefont
  {Crocker}, \citenamefont {Fredrickson}, \citenamefont {Gorelenkov},
  \citenamefont {Peebles}, \citenamefont {Kubota}, \citenamefont {Bell},
  \citenamefont {Diallo}, \citenamefont {LeBlanc}, \citenamefont {Menard},
  \citenamefont {\Podesta}, \citenamefont {Tritz},\ and\ \citenamefont
  {Yuh}}]{Crocker2013NF}%
  \BibitemOpen
  \bibfield  {author} {\bibinfo {author} {\bibfnamefont {N.~A.}\ \bibnamefont
  {Crocker}}, \bibinfo {author} {\bibfnamefont {E.~D.}\ \bibnamefont
  {Fredrickson}}, \bibinfo {author} {\bibfnamefont {N.~N.}\ \bibnamefont
  {Gorelenkov}}, \bibinfo {author} {\bibfnamefont {W.~A.}\ \bibnamefont
  {Peebles}}, \bibinfo {author} {\bibfnamefont {S.}~\bibnamefont {Kubota}},
  \bibinfo {author} {\bibfnamefont {R.~E.}\ \bibnamefont {Bell}}, \bibinfo
  {author} {\bibfnamefont {A.}~\bibnamefont {Diallo}}, \bibinfo {author}
  {\bibfnamefont {B.~P.}\ \bibnamefont {LeBlanc}}, \bibinfo {author}
  {\bibfnamefont {J.~E.}\ \bibnamefont {Menard}}, \bibinfo {author}
  {\bibfnamefont {M.}~\bibnamefont {\Podesta}}, \bibinfo {author}
  {\bibfnamefont {K.}~\bibnamefont {Tritz}}, \ and\ \bibinfo {author}
  {\bibfnamefont {H.}~\bibnamefont {Yuh}},\ }\href
  {http://stacks.iop.org/0029-5515/53/i=4/a=043017} {\bibfield  {journal}
  {\bibinfo  {journal} {Nuclear Fusion}\ }\textbf {\bibinfo {volume} {53}},\
  \bibinfo {pages} {043017} (\bibinfo {year} {2013})}\BibitemShut {NoStop}%
\bibitem [{\citenamefont {Fredrickson}\ \emph {et~al.}(2019)\citenamefont
  {Fredrickson}, \citenamefont {Gorelenkov}, \citenamefont {Bell},
  \citenamefont {Diallo}, \citenamefont {LeBlanc},\ and\ \citenamefont
  {Podestà}}]{Fredrickson2019POP}%
  \BibitemOpen
  \bibfield  {author} {\bibinfo {author} {\bibfnamefont {E.~D.}\ \bibnamefont
  {Fredrickson}}, \bibinfo {author} {\bibfnamefont {N.~N.}\ \bibnamefont
  {Gorelenkov}}, \bibinfo {author} {\bibfnamefont {R.~E.}\ \bibnamefont
  {Bell}}, \bibinfo {author} {\bibfnamefont {A.}~\bibnamefont {Diallo}},
  \bibinfo {author} {\bibfnamefont {B.~P.}\ \bibnamefont {LeBlanc}}, \ and\
  \bibinfo {author} {\bibfnamefont {M.}~\bibnamefont {Podestà}},\ }\href
  {\doibase 10.1063/1.5081047} {\bibfield  {journal} {\bibinfo  {journal}
  {Physics of Plasmas}\ }\textbf {\bibinfo {volume} {26}},\ \bibinfo {pages}
  {032111} (\bibinfo {year} {2019})}\BibitemShut {NoStop}%
\bibitem [{\citenamefont {Appel}\ \emph {et~al.}(2008)\citenamefont {Appel},
  \citenamefont {\Fulop}, \citenamefont {Hole}, \citenamefont {Smith},
  \citenamefont {Pinches}, \citenamefont {Vann},\ and\ \citenamefont {the
  MAST~Team}}]{Appel2008PPCF}%
  \BibitemOpen
  \bibfield  {author} {\bibinfo {author} {\bibfnamefont {L.~C.}\ \bibnamefont
  {Appel}}, \bibinfo {author} {\bibfnamefont {T.}~\bibnamefont {\Fulop}},
  \bibinfo {author} {\bibfnamefont {M.~J.}\ \bibnamefont {Hole}}, \bibinfo
  {author} {\bibfnamefont {H.~M.}\ \bibnamefont {Smith}}, \bibinfo {author}
  {\bibfnamefont {S.~D.}\ \bibnamefont {Pinches}}, \bibinfo {author}
  {\bibfnamefont {R.~G.~L.}\ \bibnamefont {Vann}}, \ and\ \bibinfo {author}
  {\bibnamefont {the MAST~Team}},\ }\href
  {http://stacks.iop.org/0741-3335/50/i=11/a=115011} {\bibfield  {journal}
  {\bibinfo  {journal} {Plasma Physics and Controlled Fusion}\ }\textbf
  {\bibinfo {volume} {50}},\ \bibinfo {pages} {115011} (\bibinfo {year}
  {2008})}\BibitemShut {NoStop}%
\bibitem [{\citenamefont {Sharapov}\ \emph {et~al.}(2014)\citenamefont
  {Sharapov}, \citenamefont {Lilley}, \citenamefont {Akers}, \citenamefont
  {Ayed}, \citenamefont {Cecconello}, \citenamefont {Cook}, \citenamefont
  {Cunningham},\ and\ \citenamefont {Verwichte}}]{Sharapov2014PP}%
  \BibitemOpen
  \bibfield  {author} {\bibinfo {author} {\bibfnamefont {S.~E.}\ \bibnamefont
  {Sharapov}}, \bibinfo {author} {\bibfnamefont {M.~K.}\ \bibnamefont
  {Lilley}}, \bibinfo {author} {\bibfnamefont {R.}~\bibnamefont {Akers}},
  \bibinfo {author} {\bibfnamefont {N.~B.}\ \bibnamefont {Ayed}}, \bibinfo
  {author} {\bibfnamefont {M.}~\bibnamefont {Cecconello}}, \bibinfo {author}
  {\bibfnamefont {J.~W.~S.}\ \bibnamefont {Cook}}, \bibinfo {author}
  {\bibfnamefont {G.}~\bibnamefont {Cunningham}}, \ and\ \bibinfo {author}
  {\bibfnamefont {E.}~\bibnamefont {Verwichte}},\ }\href {\doibase
  10.1063/1.4891322} {\bibfield  {journal} {\bibinfo  {journal} {Physics of
  Plasmas}\ }\textbf {\bibinfo {volume} {21}},\ \bibinfo {pages} {082501}
  (\bibinfo {year} {2014})}\BibitemShut {NoStop}%
\bibitem [{\citenamefont {Cottrell}\ \emph {et~al.}(1993)\citenamefont
  {Cottrell}, \citenamefont {Bhatnagar}, \citenamefont {Costa}, \citenamefont
  {Dendy}, \citenamefont {Jacquinot}, \citenamefont {McClements}, \citenamefont
  {McCune}, \citenamefont {Nave}, \citenamefont {Smeulders},\ and\
  \citenamefont {Start}}]{Cottrell1993NF}%
  \BibitemOpen
  \bibfield  {author} {\bibinfo {author} {\bibfnamefont {G.~A.}\ \bibnamefont
  {Cottrell}}, \bibinfo {author} {\bibfnamefont {V.~P.}\ \bibnamefont
  {Bhatnagar}}, \bibinfo {author} {\bibfnamefont {O.~D.}\ \bibnamefont
  {Costa}}, \bibinfo {author} {\bibfnamefont {R.~O.}\ \bibnamefont {Dendy}},
  \bibinfo {author} {\bibfnamefont {J.}~\bibnamefont {Jacquinot}}, \bibinfo
  {author} {\bibfnamefont {K.~G.}\ \bibnamefont {McClements}}, \bibinfo
  {author} {\bibfnamefont {D.~C.}\ \bibnamefont {McCune}}, \bibinfo {author}
  {\bibfnamefont {M.~F.}\ \bibnamefont {Nave}}, \bibinfo {author}
  {\bibfnamefont {P.}~\bibnamefont {Smeulders}}, \ and\ \bibinfo {author}
  {\bibfnamefont {D.~F.}\ \bibnamefont {Start}},\ }\href
  {http://stacks.iop.org/0029-5515/33/i=9/a=I10} {\bibfield  {journal}
  {\bibinfo  {journal} {Nuclear Fusion}\ }\textbf {\bibinfo {volume} {33}},\
  \bibinfo {pages} {1365} (\bibinfo {year} {1993})}\BibitemShut {NoStop}%
\bibitem [{\citenamefont {\Fulop}\ \emph {et~al.}(1997)\citenamefont {\Fulop},
  \citenamefont {Kolesnichenko}, \citenamefont {Lisak},\ and\ \citenamefont
  {Anderson}}]{Fulop1997NF}%
  \BibitemOpen
  \bibfield  {author} {\bibinfo {author} {\bibfnamefont {T.}~\bibnamefont
  {\Fulop}}, \bibinfo {author} {\bibfnamefont {Y.}~\bibnamefont
  {Kolesnichenko}}, \bibinfo {author} {\bibfnamefont {M.}~\bibnamefont
  {Lisak}}, \ and\ \bibinfo {author} {\bibfnamefont {D.}~\bibnamefont
  {Anderson}},\ }\href {http://stacks.iop.org/0029-5515/37/i=9/a=I08}
  {\bibfield  {journal} {\bibinfo  {journal} {Nuclear Fusion}\ }\textbf
  {\bibinfo {volume} {37}},\ \bibinfo {pages} {1281} (\bibinfo {year}
  {1997})}\BibitemShut {NoStop}%
\bibitem [{\citenamefont {\Fulop}\ and\ \citenamefont
  {Lisak}(1998)}]{Fulop1998NF}%
  \BibitemOpen
  \bibfield  {author} {\bibinfo {author} {\bibfnamefont {T.}~\bibnamefont
  {\Fulop}}\ and\ \bibinfo {author} {\bibfnamefont {M.}~\bibnamefont {Lisak}},\
  }\href {http://stacks.iop.org/0029-5515/38/i=5/a=309} {\bibfield  {journal}
  {\bibinfo  {journal} {Nuclear Fusion}\ }\textbf {\bibinfo {volume} {38}},\
  \bibinfo {pages} {761} (\bibinfo {year} {1998})}\BibitemShut {NoStop}%
\bibitem [{\citenamefont {Thome}\ \emph {et~al.}(2019)\citenamefont {Thome},
  \citenamefont {Pace}, \citenamefont {Pinsker}, \citenamefont {Zeeland},
  \citenamefont {Heidbrink},\ and\ \citenamefont {Austin}}]{Thome2019NF}%
  \BibitemOpen
  \bibfield  {author} {\bibinfo {author} {\bibfnamefont {K.~E.}\ \bibnamefont
  {Thome}}, \bibinfo {author} {\bibfnamefont {D.~C.}\ \bibnamefont {Pace}},
  \bibinfo {author} {\bibfnamefont {R.~I.}\ \bibnamefont {Pinsker}}, \bibinfo
  {author} {\bibfnamefont {M.~A.~V.}\ \bibnamefont {Zeeland}}, \bibinfo
  {author} {\bibfnamefont {W.~W.}\ \bibnamefont {Heidbrink}}, \ and\ \bibinfo
  {author} {\bibfnamefont {M.~E.}\ \bibnamefont {Austin}},\ }\href {\doibase
  10.1088/1741-4326/ab20e7} {\bibfield  {journal} {\bibinfo  {journal} {Nuclear
  Fusion}\ }\textbf {\bibinfo {volume} {59}},\ \bibinfo {pages} {086011}
  (\bibinfo {year} {2019})}\BibitemShut {NoStop}%
\bibitem [{\citenamefont {Goedbloed}(1975)}]{Goedbloed1975PF}%
  \BibitemOpen
  \bibfield  {author} {\bibinfo {author} {\bibfnamefont {J.~P.}\ \bibnamefont
  {Goedbloed}},\ }\href {\doibase 10.1063/1.861012} {\bibfield  {journal}
  {\bibinfo  {journal} {The Physics of Fluids}\ }\textbf {\bibinfo {volume}
  {18}},\ \bibinfo {pages} {1258} (\bibinfo {year} {1975})}\BibitemShut
  {NoStop}%
\bibitem [{\citenamefont {Appert}\ \emph {et~al.}(1982)\citenamefont {Appert},
  \citenamefont {Gruber}, \citenamefont {Troyuon},\ and\ \citenamefont
  {Vaclavik}}]{Appert1982PP}%
  \BibitemOpen
  \bibfield  {author} {\bibinfo {author} {\bibfnamefont {K.}~\bibnamefont
  {Appert}}, \bibinfo {author} {\bibfnamefont {R.}~\bibnamefont {Gruber}},
  \bibinfo {author} {\bibfnamefont {F.}~\bibnamefont {Troyuon}}, \ and\
  \bibinfo {author} {\bibfnamefont {J.}~\bibnamefont {Vaclavik}},\ }\href
  {http://stacks.iop.org/0032-1028/24/i=9/a=010} {\bibfield  {journal}
  {\bibinfo  {journal} {Plasma Physics}\ }\textbf {\bibinfo {volume} {24}},\
  \bibinfo {pages} {1147} (\bibinfo {year} {1982})}\BibitemShut {NoStop}%
\bibitem [{\citenamefont {Mahajan}\ \emph {et~al.}(1983)\citenamefont
  {Mahajan}, \citenamefont {Ross},\ and\ \citenamefont {Chen}}]{Mahajan1983PF}%
  \BibitemOpen
  \bibfield  {author} {\bibinfo {author} {\bibfnamefont {S.~M.}\ \bibnamefont
  {Mahajan}}, \bibinfo {author} {\bibfnamefont {D.~W.}\ \bibnamefont {Ross}}, \
  and\ \bibinfo {author} {\bibfnamefont {G.}~\bibnamefont {Chen}},\ }\href
  {\doibase 10.1063/1.864404} {\bibfield  {journal} {\bibinfo  {journal} {The
  Physics of Fluids}\ }\textbf {\bibinfo {volume} {26}},\ \bibinfo {pages}
  {2195} (\bibinfo {year} {1983})}\BibitemShut {NoStop}%
\bibitem [{\citenamefont {Mahajan}(1984)}]{Mahajan1984PF}%
  \BibitemOpen
  \bibfield  {author} {\bibinfo {author} {\bibfnamefont {S.~M.}\ \bibnamefont
  {Mahajan}},\ }\href {\doibase 10.1063/1.864878} {\bibfield  {journal}
  {\bibinfo  {journal} {The Physics of Fluids}\ }\textbf {\bibinfo {volume}
  {27}},\ \bibinfo {pages} {2238} (\bibinfo {year} {1984})}\BibitemShut
  {NoStop}%
\bibitem [{\citenamefont {Li}\ \emph {et~al.}(1987)\citenamefont {Li},
  \citenamefont {Mahajan},\ and\ \citenamefont {Ross}}]{Li1987PF}%
  \BibitemOpen
  \bibfield  {author} {\bibinfo {author} {\bibfnamefont {Y.~M.}\ \bibnamefont
  {Li}}, \bibinfo {author} {\bibfnamefont {S.~M.}\ \bibnamefont {Mahajan}}, \
  and\ \bibinfo {author} {\bibfnamefont {D.~W.}\ \bibnamefont {Ross}},\ }\href
  {\doibase 10.1063/1.866260} {\bibfield  {journal} {\bibinfo  {journal} {The
  Physics of Fluids}\ }\textbf {\bibinfo {volume} {30}},\ \bibinfo {pages}
  {1466} (\bibinfo {year} {1987})}\BibitemShut {NoStop}%
\bibitem [{\citenamefont {De~Azevedo}\ \emph {et~al.}(1991)\citenamefont
  {De~Azevedo}, \citenamefont {De~Assis}, \citenamefont {Shigueoka},\ and\
  \citenamefont {Sakanaka}}]{DeAzevedo1991SP}%
  \BibitemOpen
  \bibfield  {author} {\bibinfo {author} {\bibfnamefont {C.~A.}\ \bibnamefont
  {De~Azevedo}}, \bibinfo {author} {\bibfnamefont {A.~S.}\ \bibnamefont
  {De~Assis}}, \bibinfo {author} {\bibfnamefont {H.}~\bibnamefont {Shigueoka}},
  \ and\ \bibinfo {author} {\bibfnamefont {P.~H.}\ \bibnamefont {Sakanaka}},\
  }\href {https://doi.org/10.1007/BF00151748} {\bibfield  {journal} {\bibinfo
  {journal} {Solar Physics}\ }\textbf {\bibinfo {volume} {131}},\ \bibinfo
  {pages} {119} (\bibinfo {year} {1991})}\BibitemShut {NoStop}%
\bibitem [{\citenamefont {Kolesnichenko}\ \emph {et~al.}(2007)\citenamefont
  {Kolesnichenko}, \citenamefont {Lutsenko}, \citenamefont {Weller},
  \citenamefont {Werner}, \citenamefont {Yakovenko}, \citenamefont {Geiger},\
  and\ \citenamefont {Fesenyuk}}]{Kolesnichenko2007POP}%
  \BibitemOpen
  \bibfield  {author} {\bibinfo {author} {\bibfnamefont {Y.~I.}\ \bibnamefont
  {Kolesnichenko}}, \bibinfo {author} {\bibfnamefont {V.~V.}\ \bibnamefont
  {Lutsenko}}, \bibinfo {author} {\bibfnamefont {A.}~\bibnamefont {Weller}},
  \bibinfo {author} {\bibfnamefont {A.}~\bibnamefont {Werner}}, \bibinfo
  {author} {\bibfnamefont {Y.~V.}\ \bibnamefont {Yakovenko}}, \bibinfo {author}
  {\bibfnamefont {J.}~\bibnamefont {Geiger}}, \ and\ \bibinfo {author}
  {\bibfnamefont {O.~P.}\ \bibnamefont {Fesenyuk}},\ }\href {\doibase
  10.1063/1.2789558} {\bibfield  {journal} {\bibinfo  {journal} {Physics of
  Plasmas}\ }\textbf {\bibinfo {volume} {14}},\ \bibinfo {pages} {102504}
  (\bibinfo {year} {2007})}\BibitemShut {NoStop}%
\bibitem [{\citenamefont {Ross}\ \emph {et~al.}(1982)\citenamefont {Ross},
  \citenamefont {Chen},\ and\ \citenamefont {Mahajan}}]{Ross1982PF}%
  \BibitemOpen
  \bibfield  {author} {\bibinfo {author} {\bibfnamefont {D.~W.}\ \bibnamefont
  {Ross}}, \bibinfo {author} {\bibfnamefont {G.~L.}\ \bibnamefont {Chen}}, \
  and\ \bibinfo {author} {\bibfnamefont {S.~M.}\ \bibnamefont {Mahajan}},\
  }\href {\doibase 10.1063/1.863789} {\bibfield  {journal} {\bibinfo  {journal}
  {The Physics of Fluids}\ }\textbf {\bibinfo {volume} {25}},\ \bibinfo {pages}
  {652} (\bibinfo {year} {1982})}\BibitemShut {NoStop}%
\bibitem [{\citenamefont {de~Chambrier}\ \emph {et~al.}(1982)\citenamefont
  {de~Chambrier}, \citenamefont {Cheetham}, \citenamefont {Heym}, \citenamefont
  {Hofmann}, \citenamefont {Joye}, \citenamefont {Keller}, \citenamefont
  {Lietti}, \citenamefont {Lister},\ and\ \citenamefont
  {Pochelon}}]{DeChambrier1982POP}%
  \BibitemOpen
  \bibfield  {author} {\bibinfo {author} {\bibfnamefont {A.}~\bibnamefont
  {de~Chambrier}}, \bibinfo {author} {\bibfnamefont {A.~D.}\ \bibnamefont
  {Cheetham}}, \bibinfo {author} {\bibfnamefont {A.}~\bibnamefont {Heym}},
  \bibinfo {author} {\bibfnamefont {F.}~\bibnamefont {Hofmann}}, \bibinfo
  {author} {\bibfnamefont {B.}~\bibnamefont {Joye}}, \bibinfo {author}
  {\bibfnamefont {R.}~\bibnamefont {Keller}}, \bibinfo {author} {\bibfnamefont
  {A.}~\bibnamefont {Lietti}}, \bibinfo {author} {\bibfnamefont {J.~B.}\
  \bibnamefont {Lister}}, \ and\ \bibinfo {author} {\bibfnamefont
  {A.}~\bibnamefont {Pochelon}},\ }\href {\doibase 10.1088/0032-1028/24/8/003}
  {\bibfield  {journal} {\bibinfo  {journal} {Plasma Physics}\ }\textbf
  {\bibinfo {volume} {24}},\ \bibinfo {pages} {893} (\bibinfo {year}
  {1982})}\BibitemShut {NoStop}%
\bibitem [{\citenamefont {Fu}\ and\ \citenamefont {Dam}(1989)}]{Fu1989PF}%
  \BibitemOpen
  \bibfield  {author} {\bibinfo {author} {\bibfnamefont {G.~Y.}\ \bibnamefont
  {Fu}}\ and\ \bibinfo {author} {\bibfnamefont {J.~W.~V.}\ \bibnamefont
  {Dam}},\ }\href {\doibase 10.1063/1.859175} {\bibfield  {journal} {\bibinfo
  {journal} {Physics of Fluids B: Plasma Physics}\ }\textbf {\bibinfo {volume}
  {1}},\ \bibinfo {pages} {2404} (\bibinfo {year} {1989})}\BibitemShut
  {NoStop}%
\bibitem [{\citenamefont {Dam}\ \emph {et~al.}(1990)\citenamefont {Dam},
  \citenamefont {Fu},\ and\ \citenamefont {Cheng}}]{VanDam1990FT}%
  \BibitemOpen
  \bibfield  {author} {\bibinfo {author} {\bibfnamefont {J.~W.~V.}\
  \bibnamefont {Dam}}, \bibinfo {author} {\bibfnamefont {G.~Y.}\ \bibnamefont
  {Fu}}, \ and\ \bibinfo {author} {\bibfnamefont {C.~Z.}\ \bibnamefont
  {Cheng}},\ }\href {\doibase 10.13182/FST90-A29282} {\bibfield  {journal}
  {\bibinfo  {journal} {Fusion Technology}\ }\textbf {\bibinfo {volume} {18}},\
  \bibinfo {pages} {461} (\bibinfo {year} {1990})}\BibitemShut {NoStop}%
\bibitem [{\citenamefont {Fredrickson}\ \emph {et~al.}(2006)\citenamefont
  {Fredrickson}, \citenamefont {Bell}, \citenamefont {Darrow}, \citenamefont
  {Fu}, \citenamefont {Gorelenkov}, \citenamefont {LeBlanc}, \citenamefont
  {Medley}, \citenamefont {Menard}, \citenamefont {Park}, \citenamefont
  {Roquemore}, \citenamefont {Heidbrink}, \citenamefont {Sabbagh},
  \citenamefont {Stutman}, \citenamefont {Tritz}, \citenamefont {Crocker},
  \citenamefont {Kubota}, \citenamefont {Peebles}, \citenamefont {Lee},\ and\
  \citenamefont {Levinton}}]{Fredrickson2006POP}%
  \BibitemOpen
  \bibfield  {author} {\bibinfo {author} {\bibfnamefont {E.~D.}\ \bibnamefont
  {Fredrickson}}, \bibinfo {author} {\bibfnamefont {R.~E.}\ \bibnamefont
  {Bell}}, \bibinfo {author} {\bibfnamefont {D.~S.}\ \bibnamefont {Darrow}},
  \bibinfo {author} {\bibfnamefont {G.~Y.}\ \bibnamefont {Fu}}, \bibinfo
  {author} {\bibfnamefont {N.~N.}\ \bibnamefont {Gorelenkov}}, \bibinfo
  {author} {\bibfnamefont {B.~P.}\ \bibnamefont {LeBlanc}}, \bibinfo {author}
  {\bibfnamefont {S.~S.}\ \bibnamefont {Medley}}, \bibinfo {author}
  {\bibfnamefont {J.~E.}\ \bibnamefont {Menard}}, \bibinfo {author}
  {\bibfnamefont {H.}~\bibnamefont {Park}}, \bibinfo {author} {\bibfnamefont
  {A.~L.}\ \bibnamefont {Roquemore}}, \bibinfo {author} {\bibfnamefont {W.~W.}\
  \bibnamefont {Heidbrink}}, \bibinfo {author} {\bibfnamefont {S.~A.}\
  \bibnamefont {Sabbagh}}, \bibinfo {author} {\bibfnamefont {D.}~\bibnamefont
  {Stutman}}, \bibinfo {author} {\bibfnamefont {K.}~\bibnamefont {Tritz}},
  \bibinfo {author} {\bibfnamefont {N.~A.}\ \bibnamefont {Crocker}}, \bibinfo
  {author} {\bibfnamefont {S.}~\bibnamefont {Kubota}}, \bibinfo {author}
  {\bibfnamefont {W.}~\bibnamefont {Peebles}}, \bibinfo {author} {\bibfnamefont
  {K.~C.}\ \bibnamefont {Lee}}, \ and\ \bibinfo {author} {\bibfnamefont
  {F.~M.}\ \bibnamefont {Levinton}},\ }\href {\doibase 10.1063/1.2178788}
  {\bibfield  {journal} {\bibinfo  {journal} {Physics of Plasmas}\ }\textbf
  {\bibinfo {volume} {13}},\ \bibinfo {pages} {056109} (\bibinfo {year}
  {2006})}\BibitemShut {NoStop}%
\bibitem [{\citenamefont {Fredrickson}\ \emph {et~al.}(2012)\citenamefont
  {Fredrickson}, \citenamefont {Gorelenkov}, \citenamefont {Belova},
  \citenamefont {Crocker}, \citenamefont {Kubota}, \citenamefont {Kramer},
  \citenamefont {LeBlanc}, \citenamefont {Bell}, \citenamefont {\Podesta},
  \citenamefont {Yuh},\ and\ \citenamefont {Levinton}}]{Fredrickson2012NF}%
  \BibitemOpen
  \bibfield  {author} {\bibinfo {author} {\bibfnamefont {E.~D.}\ \bibnamefont
  {Fredrickson}}, \bibinfo {author} {\bibfnamefont {N.~N.}\ \bibnamefont
  {Gorelenkov}}, \bibinfo {author} {\bibfnamefont {E.}~\bibnamefont {Belova}},
  \bibinfo {author} {\bibfnamefont {N.~A.}\ \bibnamefont {Crocker}}, \bibinfo
  {author} {\bibfnamefont {S.}~\bibnamefont {Kubota}}, \bibinfo {author}
  {\bibfnamefont {G.~J.}\ \bibnamefont {Kramer}}, \bibinfo {author}
  {\bibfnamefont {B.}~\bibnamefont {LeBlanc}}, \bibinfo {author} {\bibfnamefont
  {R.~E.}\ \bibnamefont {Bell}}, \bibinfo {author} {\bibfnamefont
  {M.}~\bibnamefont {\Podesta}}, \bibinfo {author} {\bibfnamefont
  {H.}~\bibnamefont {Yuh}}, \ and\ \bibinfo {author} {\bibfnamefont
  {F.}~\bibnamefont {Levinton}},\ }\href
  {http://stacks.iop.org/0029-5515/52/i=4/a=043001} {\bibfield  {journal}
  {\bibinfo  {journal} {Nuclear Fusion}\ }\textbf {\bibinfo {volume} {52}},\
  \bibinfo {pages} {043001} (\bibinfo {year} {2012})}\BibitemShut {NoStop}%
\bibitem [{\citenamefont {Fredrickson}\ \emph {et~al.}(2017)\citenamefont
  {Fredrickson}, \citenamefont {Belova}, \citenamefont {Battaglia},
  \citenamefont {Bell}, \citenamefont {Crocker}, \citenamefont {Darrow},
  \citenamefont {Diallo}, \citenamefont {Gerhardt}, \citenamefont {Gorelenkov},
  \citenamefont {LeBlanc}, \citenamefont {\Podesta},\ and\ \citenamefont {the
  NSTX-U~Team}}]{Fredrickson2017PRL}%
  \BibitemOpen
  \bibfield  {author} {\bibinfo {author} {\bibfnamefont {E.~D.}\ \bibnamefont
  {Fredrickson}}, \bibinfo {author} {\bibfnamefont {E.~V.}\ \bibnamefont
  {Belova}}, \bibinfo {author} {\bibfnamefont {D.~J.}\ \bibnamefont
  {Battaglia}}, \bibinfo {author} {\bibfnamefont {R.~E.}\ \bibnamefont {Bell}},
  \bibinfo {author} {\bibfnamefont {N.~A.}\ \bibnamefont {Crocker}}, \bibinfo
  {author} {\bibfnamefont {D.~S.}\ \bibnamefont {Darrow}}, \bibinfo {author}
  {\bibfnamefont {A.}~\bibnamefont {Diallo}}, \bibinfo {author} {\bibfnamefont
  {S.~P.}\ \bibnamefont {Gerhardt}}, \bibinfo {author} {\bibfnamefont {N.~N.}\
  \bibnamefont {Gorelenkov}}, \bibinfo {author} {\bibfnamefont {B.~P.}\
  \bibnamefont {LeBlanc}}, \bibinfo {author} {\bibfnamefont {M.}~\bibnamefont
  {\Podesta}}, \ and\ \bibinfo {author} {\bibnamefont {the NSTX-U~Team}},\
  }\href {\doibase 10.1103/PhysRevLett.118.265001} {\bibfield  {journal}
  {\bibinfo  {journal} {Phys. Rev. Lett.}\ }\textbf {\bibinfo {volume} {118}},\
  \bibinfo {pages} {265001} (\bibinfo {year} {2017})}\BibitemShut {NoStop}%
\bibitem [{\citenamefont {Gorelenkov}\ \emph {et~al.}(2003)\citenamefont
  {Gorelenkov}, \citenamefont {Fredrickson}, \citenamefont {Belova},
  \citenamefont {Cheng}, \citenamefont {Gates}, \citenamefont {Kaye},\ and\
  \citenamefont {White}}]{Gorelenkov2003NF}%
  \BibitemOpen
  \bibfield  {author} {\bibinfo {author} {\bibfnamefont {N.~N.}\ \bibnamefont
  {Gorelenkov}}, \bibinfo {author} {\bibfnamefont {E.}~\bibnamefont
  {Fredrickson}}, \bibinfo {author} {\bibfnamefont {E.}~\bibnamefont {Belova}},
  \bibinfo {author} {\bibfnamefont {C.~Z.}\ \bibnamefont {Cheng}}, \bibinfo
  {author} {\bibfnamefont {D.}~\bibnamefont {Gates}}, \bibinfo {author}
  {\bibfnamefont {S.}~\bibnamefont {Kaye}}, \ and\ \bibinfo {author}
  {\bibfnamefont {R.}~\bibnamefont {White}},\ }\href
  {http://stacks.iop.org/0029-5515/43/i=4/a=302} {\bibfield  {journal}
  {\bibinfo  {journal} {Nuclear Fusion}\ }\textbf {\bibinfo {volume} {43}},\
  \bibinfo {pages} {228} (\bibinfo {year} {2003})}\BibitemShut {NoStop}%
\bibitem [{\citenamefont {Gorelenkov}\ \emph {et~al.}(2004)\citenamefont
  {Gorelenkov}, \citenamefont {Belova}, \citenamefont {Berk}, \citenamefont
  {Cheng}, \citenamefont {Fredrickson}, \citenamefont {Heidbrink},
  \citenamefont {Kaye},\ and\ \citenamefont {Kramer}}]{Gorelenkov2004POP}%
  \BibitemOpen
  \bibfield  {author} {\bibinfo {author} {\bibfnamefont {N.~N.}\ \bibnamefont
  {Gorelenkov}}, \bibinfo {author} {\bibfnamefont {E.}~\bibnamefont {Belova}},
  \bibinfo {author} {\bibfnamefont {H.~L.}\ \bibnamefont {Berk}}, \bibinfo
  {author} {\bibfnamefont {C.~Z.}\ \bibnamefont {Cheng}}, \bibinfo {author}
  {\bibfnamefont {E.}~\bibnamefont {Fredrickson}}, \bibinfo {author}
  {\bibfnamefont {W.~W.}\ \bibnamefont {Heidbrink}}, \bibinfo {author}
  {\bibfnamefont {S.}~\bibnamefont {Kaye}}, \ and\ \bibinfo {author}
  {\bibfnamefont {G.~J.}\ \bibnamefont {Kramer}},\ }\href {\doibase
  http://dx.doi.org/10.1063/1.1689667} {\bibfield  {journal} {\bibinfo
  {journal} {Physics of Plasmas}\ }\textbf {\bibinfo {volume} {11}},\ \bibinfo
  {pages} {2586} (\bibinfo {year} {2004})}\BibitemShut {NoStop}%
\bibitem [{\citenamefont {Kolesnichenko}\ \emph {et~al.}(2006)\citenamefont
  {Kolesnichenko}, \citenamefont {White},\ and\ \citenamefont
  {Yakovenko}}]{Kolesnichenko2006POP}%
  \BibitemOpen
  \bibfield  {author} {\bibinfo {author} {\bibfnamefont {Y.~I.}\ \bibnamefont
  {Kolesnichenko}}, \bibinfo {author} {\bibfnamefont {R.~B.}\ \bibnamefont
  {White}}, \ and\ \bibinfo {author} {\bibfnamefont {Y.~V.}\ \bibnamefont
  {Yakovenko}},\ }\href {\doibase 10.1063/1.2402129} {\bibfield  {journal}
  {\bibinfo  {journal} {Physics of Plasmas}\ }\textbf {\bibinfo {volume}
  {13}},\ \bibinfo {pages} {122503} (\bibinfo {year} {2006})}\BibitemShut
  {NoStop}%
\bibitem [{\citenamefont {Lestz}\ \emph {et~al.}(2018)\citenamefont {Lestz},
  \citenamefont {Belova},\ and\ \citenamefont {Gorelenkov}}]{Lestz2018POP}%
  \BibitemOpen
  \bibfield  {author} {\bibinfo {author} {\bibfnamefont {J.~B.}\ \bibnamefont
  {Lestz}}, \bibinfo {author} {\bibfnamefont {E.~V.}\ \bibnamefont {Belova}}, \
  and\ \bibinfo {author} {\bibfnamefont {N.~N.}\ \bibnamefont {Gorelenkov}},\
  }\href {\doibase 10.1063/1.4998602} {\bibfield  {journal} {\bibinfo
  {journal} {Physics of Plasmas}\ }\textbf {\bibinfo {volume} {25}},\ \bibinfo
  {pages} {042508} (\bibinfo {year} {2018})}\BibitemShut {NoStop}%
\bibitem [{\citenamefont {Lestz}\ \emph {et~al.}(tion)\citenamefont {Lestz},
  \citenamefont {Belova},\ and\ \citenamefont {Gorelenkov}}]{Lestz2020sim}%
  \BibitemOpen
  \bibfield  {author} {\bibinfo {author} {\bibfnamefont {J.~B.}\ \bibnamefont
  {Lestz}}, \bibinfo {author} {\bibfnamefont {E.~V.}\ \bibnamefont {Belova}}, \
  and\ \bibinfo {author} {\bibfnamefont {N.~N.}\ \bibnamefont {Gorelenkov}},\
  }\href@noop {} {\bibfield  {journal} {\bibinfo  {journal} {Physics of
  Plasmas}\ } (\bibinfo {year} {2020 in preparation})}\BibitemShut {NoStop}%
\bibitem [{\citenamefont {Crocker}\ \emph
  {et~al.}(2018{\natexlab{a}})\citenamefont {Crocker}, \citenamefont {Kubota},
  \citenamefont {Peebles}, \citenamefont {Rhodes}, \citenamefont {Fredrickson},
  \citenamefont {Belova}, \citenamefont {Diallo}, \citenamefont {LeBlanc},\
  and\ \citenamefont {Sabbagh}}]{Crocker2017NF}%
  \BibitemOpen
  \bibfield  {author} {\bibinfo {author} {\bibfnamefont {N.~A.}\ \bibnamefont
  {Crocker}}, \bibinfo {author} {\bibfnamefont {S.}~\bibnamefont {Kubota}},
  \bibinfo {author} {\bibfnamefont {W.~A.}\ \bibnamefont {Peebles}}, \bibinfo
  {author} {\bibfnamefont {T.~L.}\ \bibnamefont {Rhodes}}, \bibinfo {author}
  {\bibfnamefont {E.~D.}\ \bibnamefont {Fredrickson}}, \bibinfo {author}
  {\bibfnamefont {E.}~\bibnamefont {Belova}}, \bibinfo {author} {\bibfnamefont
  {A.}~\bibnamefont {Diallo}}, \bibinfo {author} {\bibfnamefont {B.~P.}\
  \bibnamefont {LeBlanc}}, \ and\ \bibinfo {author} {\bibfnamefont {S.~A.}\
  \bibnamefont {Sabbagh}},\ }\href
  {http://stacks.iop.org/0029-5515/58/i=1/a=016051} {\bibfield  {journal}
  {\bibinfo  {journal} {Nuclear Fusion}\ }\textbf {\bibinfo {volume} {58}},\
  \bibinfo {pages} {016051} (\bibinfo {year} {2018}{\natexlab{a}})}\BibitemShut
  {NoStop}%
\bibitem [{\citenamefont {Fredrickson}\ \emph {et~al.}(2018)\citenamefont
  {Fredrickson}, \citenamefont {V.Belova}, \citenamefont {Gorelenkov},
  \citenamefont {Podestà}, \citenamefont {Bell}, \citenamefont {Crocker},
  \citenamefont {Diallo}, \citenamefont {LeBlanc},\ and\ \citenamefont {the
  NSTX-U~Team}}]{Fredrickson2018NF}%
  \BibitemOpen
  \bibfield  {author} {\bibinfo {author} {\bibfnamefont {E.~D.}\ \bibnamefont
  {Fredrickson}}, \bibinfo {author} {\bibfnamefont {E.}~\bibnamefont
  {V.Belova}}, \bibinfo {author} {\bibfnamefont {N.~N.}\ \bibnamefont
  {Gorelenkov}}, \bibinfo {author} {\bibfnamefont {M.}~\bibnamefont
  {Podestà}}, \bibinfo {author} {\bibfnamefont {R.~E.}\ \bibnamefont {Bell}},
  \bibinfo {author} {\bibfnamefont {N.~A.}\ \bibnamefont {Crocker}}, \bibinfo
  {author} {\bibfnamefont {A.}~\bibnamefont {Diallo}}, \bibinfo {author}
  {\bibfnamefont {B.~P.}\ \bibnamefont {LeBlanc}}, \ and\ \bibinfo {author}
  {\bibnamefont {the NSTX-U~Team}},\ }\href
  {http://stacks.iop.org/0029-5515/58/i=8/a=082022} {\bibfield  {journal}
  {\bibinfo  {journal} {Nuclear Fusion}\ }\textbf {\bibinfo {volume} {58}},\
  \bibinfo {pages} {082022} (\bibinfo {year} {2018})}\BibitemShut {NoStop}%
\bibitem [{\citenamefont {Heidbrink}\ \emph {et~al.}(2006)\citenamefont
  {Heidbrink}, \citenamefont {Fredrickson}, \citenamefont {Gorelenkov},
  \citenamefont {Rhodes},\ and\ \citenamefont {Zeeland}}]{Heidbrink2006NF}%
  \BibitemOpen
  \bibfield  {author} {\bibinfo {author} {\bibfnamefont {W.~W.}\ \bibnamefont
  {Heidbrink}}, \bibinfo {author} {\bibfnamefont {E.~D.}\ \bibnamefont
  {Fredrickson}}, \bibinfo {author} {\bibfnamefont {N.~N.}\ \bibnamefont
  {Gorelenkov}}, \bibinfo {author} {\bibfnamefont {T.~L.}\ \bibnamefont
  {Rhodes}}, \ and\ \bibinfo {author} {\bibfnamefont {M.~A.~V.}\ \bibnamefont
  {Zeeland}},\ }\href {http://stacks.iop.org/0029-5515/46/i=2/a=016} {\bibfield
   {journal} {\bibinfo  {journal} {Nuclear Fusion}\ }\textbf {\bibinfo {volume}
  {46}},\ \bibinfo {pages} {324} (\bibinfo {year} {2006})}\BibitemShut
  {NoStop}%
\bibitem [{\citenamefont {Tang}\ \emph {et~al.}(2018)\citenamefont {Tang},
  \citenamefont {Crocker}, \citenamefont {Carter}, \citenamefont {Thome},
  \citenamefont {Pinsker}, \citenamefont {Pace},\ and\ \citenamefont
  {Heidbrink}}]{Tang2018APS}%
  \BibitemOpen
  \bibfield  {author} {\bibinfo {author} {\bibfnamefont {S.}~\bibnamefont
  {Tang}}, \bibinfo {author} {\bibfnamefont {N.}~\bibnamefont {Crocker}},
  \bibinfo {author} {\bibfnamefont {T.}~\bibnamefont {Carter}}, \bibinfo
  {author} {\bibfnamefont {K.}~\bibnamefont {Thome}}, \bibinfo {author}
  {\bibfnamefont {R.}~\bibnamefont {Pinsker}}, \bibinfo {author} {\bibfnamefont
  {D.}~\bibnamefont {Pace}}, \ and\ \bibinfo {author} {\bibfnamefont
  {W.}~\bibnamefont {Heidbrink}},\ }in\ \href
  {http://meetings.aps.org/Meeting/DPP18/Session/NP11.111} {\emph {\bibinfo
  {booktitle} {$60^{th}$ APS DPP Meeting}}}\ (\bibinfo {address} {Portland,
  OR},\ \bibinfo {year} {2018})\BibitemShut {NoStop}%
\bibitem [{\citenamefont {Crocker}\ \emph
  {et~al.}(2018{\natexlab{b}})\citenamefont {Crocker}, \citenamefont {Barada},
  \citenamefont {Tang}, \citenamefont {Thome}, \citenamefont {Pace},
  \citenamefont {Pinsker}, \citenamefont {Heidbrink}, \citenamefont {Rhodes},\
  and\ \citenamefont {Haye}}]{Crocker2018APS}%
  \BibitemOpen
  \bibfield  {author} {\bibinfo {author} {\bibfnamefont {N.}~\bibnamefont
  {Crocker}}, \bibinfo {author} {\bibfnamefont {K.}~\bibnamefont {Barada}},
  \bibinfo {author} {\bibfnamefont {S.}~\bibnamefont {Tang}}, \bibinfo {author}
  {\bibfnamefont {K.}~\bibnamefont {Thome}}, \bibinfo {author} {\bibfnamefont
  {D.}~\bibnamefont {Pace}}, \bibinfo {author} {\bibfnamefont {R.}~\bibnamefont
  {Pinsker}}, \bibinfo {author} {\bibfnamefont {W.}~\bibnamefont {Heidbrink}},
  \bibinfo {author} {\bibfnamefont {T.}~\bibnamefont {Rhodes}}, \ and\ \bibinfo
  {author} {\bibfnamefont {R.~L.}\ \bibnamefont {Haye}},\ }in\ \href
  {http://meetings.aps.org/Meeting/DPP18/Session/NP11.110} {\emph {\bibinfo
  {booktitle} {$60^{th}$ APS DPP Meeting}}}\ (\bibinfo {address} {Portland,
  OR},\ \bibinfo {year} {2018})\BibitemShut {NoStop}%
\bibitem [{\citenamefont {Duarte}\ \emph
  {et~al.}(2017{\natexlab{a}})\citenamefont {Duarte}, \citenamefont {Berk},
  \citenamefont {Gorelenkov}, \citenamefont {Heidbrink}, \citenamefont
  {Kramer}, \citenamefont {Nazikian}, \citenamefont {Pace}, \citenamefont
  {\Podesta}, \citenamefont {Tobias},\ and\ \citenamefont
  {Zeeland}}]{Duarte2017NF}%
  \BibitemOpen
  \bibfield  {author} {\bibinfo {author} {\bibfnamefont {V.~N.}\ \bibnamefont
  {Duarte}}, \bibinfo {author} {\bibfnamefont {H.~L.}\ \bibnamefont {Berk}},
  \bibinfo {author} {\bibfnamefont {N.~N.}\ \bibnamefont {Gorelenkov}},
  \bibinfo {author} {\bibfnamefont {W.~W.}\ \bibnamefont {Heidbrink}}, \bibinfo
  {author} {\bibfnamefont {G.~J.}\ \bibnamefont {Kramer}}, \bibinfo {author}
  {\bibfnamefont {R.}~\bibnamefont {Nazikian}}, \bibinfo {author}
  {\bibfnamefont {D.~C.}\ \bibnamefont {Pace}}, \bibinfo {author}
  {\bibfnamefont {M.}~\bibnamefont {\Podesta}}, \bibinfo {author}
  {\bibfnamefont {B.~J.}\ \bibnamefont {Tobias}}, \ and\ \bibinfo {author}
  {\bibfnamefont {M.~A.~V.}\ \bibnamefont {Zeeland}},\ }\href
  {http://stacks.iop.org/0029-5515/57/i=5/a=054001} {\bibfield  {journal}
  {\bibinfo  {journal} {Nuclear Fusion}\ }\textbf {\bibinfo {volume} {57}},\
  \bibinfo {pages} {054001} (\bibinfo {year} {2017}{\natexlab{a}})}\BibitemShut
  {NoStop}%
\bibitem [{\citenamefont {Gerhardt}\ \emph
  {et~al.}(2012{\natexlab{a}})\citenamefont {Gerhardt}, \citenamefont {Bell},
  \citenamefont {Bialek}, \citenamefont {Brooks}, \citenamefont {Canik},
  \citenamefont {Chrzanowski}, \citenamefont {Denault}, \citenamefont {Dudek},
  \citenamefont {Gates}, \citenamefont {Gorelenkov}, \citenamefont
  {Guttenfelder}, \citenamefont {Hatcher}, \citenamefont {Hosea}, \citenamefont
  {Kaita}, \citenamefont {Kaye}, \citenamefont {Kessel}, \citenamefont
  {Kolemen}, \citenamefont {Kugel}, \citenamefont {Maingi}, \citenamefont
  {Mardenfeld}, \citenamefont {Mueller}, \citenamefont {Nelson}, \citenamefont
  {Neumeyer}, \citenamefont {Ono}, \citenamefont {Perry}, \citenamefont
  {Ramakrishnan}, \citenamefont {Raman}, \citenamefont {Ren}, \citenamefont
  {Sabbagh}, \citenamefont {Smith}, \citenamefont {Soukhanovskii},
  \citenamefont {Stevenson}, \citenamefont {Strykowsky}, \citenamefont
  {Stutman}, \citenamefont {Taylor}, \citenamefont {Titus}, \citenamefont
  {Tresemer}, \citenamefont {Tritz}, \citenamefont {Viola}, \citenamefont
  {Williams}, \citenamefont {Woolley}, \citenamefont {Yuh}, \citenamefont
  {Zhang}, \citenamefont {Zhai}, \citenamefont {Zolfaghari},\ and\
  \citenamefont {the NSTX~Team}}]{Menard2012NF}%
  \BibitemOpen
  \bibfield  {author} {\bibinfo {author} {\bibfnamefont {J.~E. M.~S.}\
  \bibnamefont {Gerhardt}}, \bibinfo {author} {\bibfnamefont {M.}~\bibnamefont
  {Bell}}, \bibinfo {author} {\bibfnamefont {J.}~\bibnamefont {Bialek}},
  \bibinfo {author} {\bibfnamefont {A.}~\bibnamefont {Brooks}}, \bibinfo
  {author} {\bibfnamefont {J.}~\bibnamefont {Canik}}, \bibinfo {author}
  {\bibfnamefont {J.}~\bibnamefont {Chrzanowski}}, \bibinfo {author}
  {\bibfnamefont {M.}~\bibnamefont {Denault}}, \bibinfo {author} {\bibfnamefont
  {L.}~\bibnamefont {Dudek}}, \bibinfo {author} {\bibfnamefont {D.~A.}\
  \bibnamefont {Gates}}, \bibinfo {author} {\bibfnamefont {N.}~\bibnamefont
  {Gorelenkov}}, \bibinfo {author} {\bibfnamefont {W.}~\bibnamefont
  {Guttenfelder}}, \bibinfo {author} {\bibfnamefont {R.}~\bibnamefont
  {Hatcher}}, \bibinfo {author} {\bibfnamefont {J.}~\bibnamefont {Hosea}},
  \bibinfo {author} {\bibfnamefont {R.}~\bibnamefont {Kaita}}, \bibinfo
  {author} {\bibfnamefont {S.}~\bibnamefont {Kaye}}, \bibinfo {author}
  {\bibfnamefont {C.}~\bibnamefont {Kessel}}, \bibinfo {author} {\bibfnamefont
  {E.}~\bibnamefont {Kolemen}}, \bibinfo {author} {\bibfnamefont
  {H.}~\bibnamefont {Kugel}}, \bibinfo {author} {\bibfnamefont
  {R.}~\bibnamefont {Maingi}}, \bibinfo {author} {\bibfnamefont
  {M.}~\bibnamefont {Mardenfeld}}, \bibinfo {author} {\bibfnamefont
  {D.}~\bibnamefont {Mueller}}, \bibinfo {author} {\bibfnamefont
  {B.}~\bibnamefont {Nelson}}, \bibinfo {author} {\bibfnamefont
  {C.}~\bibnamefont {Neumeyer}}, \bibinfo {author} {\bibfnamefont
  {M.}~\bibnamefont {Ono}}, \bibinfo {author} {\bibfnamefont {E.}~\bibnamefont
  {Perry}}, \bibinfo {author} {\bibfnamefont {R.}~\bibnamefont {Ramakrishnan}},
  \bibinfo {author} {\bibfnamefont {R.}~\bibnamefont {Raman}}, \bibinfo
  {author} {\bibfnamefont {Y.}~\bibnamefont {Ren}}, \bibinfo {author}
  {\bibfnamefont {S.}~\bibnamefont {Sabbagh}}, \bibinfo {author} {\bibfnamefont
  {M.}~\bibnamefont {Smith}}, \bibinfo {author} {\bibfnamefont
  {V.}~\bibnamefont {Soukhanovskii}}, \bibinfo {author} {\bibfnamefont
  {T.}~\bibnamefont {Stevenson}}, \bibinfo {author} {\bibfnamefont
  {R.}~\bibnamefont {Strykowsky}}, \bibinfo {author} {\bibfnamefont
  {D.}~\bibnamefont {Stutman}}, \bibinfo {author} {\bibfnamefont
  {G.}~\bibnamefont {Taylor}}, \bibinfo {author} {\bibfnamefont
  {P.}~\bibnamefont {Titus}}, \bibinfo {author} {\bibfnamefont
  {K.}~\bibnamefont {Tresemer}}, \bibinfo {author} {\bibfnamefont
  {K.}~\bibnamefont {Tritz}}, \bibinfo {author} {\bibfnamefont
  {M.}~\bibnamefont {Viola}}, \bibinfo {author} {\bibfnamefont
  {M.}~\bibnamefont {Williams}}, \bibinfo {author} {\bibfnamefont
  {R.}~\bibnamefont {Woolley}}, \bibinfo {author} {\bibfnamefont
  {H.}~\bibnamefont {Yuh}}, \bibinfo {author} {\bibfnamefont {H.}~\bibnamefont
  {Zhang}}, \bibinfo {author} {\bibfnamefont {Y.}~\bibnamefont {Zhai}},
  \bibinfo {author} {\bibfnamefont {A.}~\bibnamefont {Zolfaghari}}, \ and\
  \bibinfo {author} {\bibnamefont {the NSTX~Team}},\ }\href
  {http://stacks.iop.org/0029-5515/52/i=8/a=083015} {\bibfield  {journal}
  {\bibinfo  {journal} {Nuclear Fusion}\ }\textbf {\bibinfo {volume} {52}},\
  \bibinfo {pages} {083015} (\bibinfo {year} {2012}{\natexlab{a}})}\BibitemShut
  {NoStop}%
\bibitem [{\citenamefont {Ono}\ and\ \citenamefont {Kaita}(2015)}]{Ono2015POP}%
  \BibitemOpen
  \bibfield  {author} {\bibinfo {author} {\bibfnamefont {M.}~\bibnamefont
  {Ono}}\ and\ \bibinfo {author} {\bibfnamefont {R.}~\bibnamefont {Kaita}},\
  }\href {\doibase 10.1063/1.4915073} {\bibfield  {journal} {\bibinfo
  {journal} {Physics of Plasmas}\ }\textbf {\bibinfo {volume} {22}},\ \bibinfo
  {pages} {040501} (\bibinfo {year} {2015})}\BibitemShut {NoStop}%
\bibitem [{\citenamefont {Menard}\ \emph {et~al.}(2011)\citenamefont {Menard},
  \citenamefont {Bromberg}, \citenamefont {Brown}, \citenamefont {Burgess},
  \citenamefont {Dix}, \citenamefont {El-Guebaly}, \citenamefont {Gerrity},
  \citenamefont {Goldston}, \citenamefont {Hawryluk}, \citenamefont {Kastner},
  \citenamefont {Kessel}, \citenamefont {Malang}, \citenamefont {Minervini},
  \citenamefont {Neilson}, \citenamefont {Neumeyer}, \citenamefont {Prager},
  \citenamefont {Sawan}, \citenamefont {Sheffield}, \citenamefont {Sternlieb},
  \citenamefont {Waganer}, \citenamefont {Whyte},\ and\ \citenamefont
  {Zarnstorff}}]{Menard2011NF}%
  \BibitemOpen
  \bibfield  {author} {\bibinfo {author} {\bibfnamefont {J.~E.}\ \bibnamefont
  {Menard}}, \bibinfo {author} {\bibfnamefont {L.}~\bibnamefont {Bromberg}},
  \bibinfo {author} {\bibfnamefont {T.}~\bibnamefont {Brown}}, \bibinfo
  {author} {\bibfnamefont {T.}~\bibnamefont {Burgess}}, \bibinfo {author}
  {\bibfnamefont {D.}~\bibnamefont {Dix}}, \bibinfo {author} {\bibfnamefont
  {L.}~\bibnamefont {El-Guebaly}}, \bibinfo {author} {\bibfnamefont
  {T.}~\bibnamefont {Gerrity}}, \bibinfo {author} {\bibfnamefont {R.~J.}\
  \bibnamefont {Goldston}}, \bibinfo {author} {\bibfnamefont {R.~J.}\
  \bibnamefont {Hawryluk}}, \bibinfo {author} {\bibfnamefont {R.}~\bibnamefont
  {Kastner}}, \bibinfo {author} {\bibfnamefont {C.}~\bibnamefont {Kessel}},
  \bibinfo {author} {\bibfnamefont {S.}~\bibnamefont {Malang}}, \bibinfo
  {author} {\bibfnamefont {J.}~\bibnamefont {Minervini}}, \bibinfo {author}
  {\bibfnamefont {G.~H.}\ \bibnamefont {Neilson}}, \bibinfo {author}
  {\bibfnamefont {C.~L.}\ \bibnamefont {Neumeyer}}, \bibinfo {author}
  {\bibfnamefont {S.}~\bibnamefont {Prager}}, \bibinfo {author} {\bibfnamefont
  {M.}~\bibnamefont {Sawan}}, \bibinfo {author} {\bibfnamefont
  {J.}~\bibnamefont {Sheffield}}, \bibinfo {author} {\bibfnamefont
  {A.}~\bibnamefont {Sternlieb}}, \bibinfo {author} {\bibfnamefont
  {L.}~\bibnamefont {Waganer}}, \bibinfo {author} {\bibfnamefont
  {D.}~\bibnamefont {Whyte}}, \ and\ \bibinfo {author} {\bibfnamefont
  {M.}~\bibnamefont {Zarnstorff}},\ }\href {\doibase
  10.1088/0029-5515/51/10/103014} {\bibfield  {journal} {\bibinfo  {journal}
  {Nuclear Fusion}\ }\textbf {\bibinfo {volume} {51}},\ \bibinfo {pages}
  {103014} (\bibinfo {year} {2011})}\BibitemShut {NoStop}%
\bibitem [{\citenamefont {Menard}\ \emph {et~al.}(2016)\citenamefont {Menard},
  \citenamefont {Brown}, \citenamefont {El-Guebaly}, \citenamefont {Boyer},
  \citenamefont {Canik}, \citenamefont {Colling}, \citenamefont {Raman},
  \citenamefont {Wang}, \citenamefont {Zhai}, \citenamefont {Buxton},
  \citenamefont {Covele}, \citenamefont {D'Angelo}, \citenamefont {Davis},
  \citenamefont {Gerhardt}, \citenamefont {Gryaznevich}, \citenamefont {Harb},
  \citenamefont {Hender}, \citenamefont {Kaye}, \citenamefont {Kingham},
  \citenamefont {Kotschenreuther}, \citenamefont {Mahajan}, \citenamefont
  {Maingi}, \citenamefont {Marriott}, \citenamefont {Meier}, \citenamefont
  {Mynsberge}, \citenamefont {Neumeyer}, \citenamefont {Ono}, \citenamefont
  {Park}, \citenamefont {Sabbagh}, \citenamefont {Soukhanovskii}, \citenamefont
  {Valanju},\ and\ \citenamefont {Woolley}}]{Menard2016NF}%
  \BibitemOpen
  \bibfield  {author} {\bibinfo {author} {\bibfnamefont {J.~E.}\ \bibnamefont
  {Menard}}, \bibinfo {author} {\bibfnamefont {T.}~\bibnamefont {Brown}},
  \bibinfo {author} {\bibfnamefont {L.}~\bibnamefont {El-Guebaly}}, \bibinfo
  {author} {\bibfnamefont {M.}~\bibnamefont {Boyer}}, \bibinfo {author}
  {\bibfnamefont {J.}~\bibnamefont {Canik}}, \bibinfo {author} {\bibfnamefont
  {B.}~\bibnamefont {Colling}}, \bibinfo {author} {\bibfnamefont
  {R.}~\bibnamefont {Raman}}, \bibinfo {author} {\bibfnamefont
  {Z.}~\bibnamefont {Wang}}, \bibinfo {author} {\bibfnamefont {Y.}~\bibnamefont
  {Zhai}}, \bibinfo {author} {\bibfnamefont {P.}~\bibnamefont {Buxton}},
  \bibinfo {author} {\bibfnamefont {B.}~\bibnamefont {Covele}}, \bibinfo
  {author} {\bibfnamefont {C.}~\bibnamefont {D'Angelo}}, \bibinfo {author}
  {\bibfnamefont {A.}~\bibnamefont {Davis}}, \bibinfo {author} {\bibfnamefont
  {S.}~\bibnamefont {Gerhardt}}, \bibinfo {author} {\bibfnamefont
  {M.}~\bibnamefont {Gryaznevich}}, \bibinfo {author} {\bibfnamefont
  {M.}~\bibnamefont {Harb}}, \bibinfo {author} {\bibfnamefont {T.~C.}\
  \bibnamefont {Hender}}, \bibinfo {author} {\bibfnamefont {S.}~\bibnamefont
  {Kaye}}, \bibinfo {author} {\bibfnamefont {D.}~\bibnamefont {Kingham}},
  \bibinfo {author} {\bibfnamefont {M.}~\bibnamefont {Kotschenreuther}},
  \bibinfo {author} {\bibfnamefont {S.}~\bibnamefont {Mahajan}}, \bibinfo
  {author} {\bibfnamefont {R.}~\bibnamefont {Maingi}}, \bibinfo {author}
  {\bibfnamefont {E.}~\bibnamefont {Marriott}}, \bibinfo {author}
  {\bibfnamefont {E.~T.}\ \bibnamefont {Meier}}, \bibinfo {author}
  {\bibfnamefont {L.}~\bibnamefont {Mynsberge}}, \bibinfo {author}
  {\bibfnamefont {C.}~\bibnamefont {Neumeyer}}, \bibinfo {author}
  {\bibfnamefont {M.}~\bibnamefont {Ono}}, \bibinfo {author} {\bibfnamefont
  {J.-K.}\ \bibnamefont {Park}}, \bibinfo {author} {\bibfnamefont {S.~A.}\
  \bibnamefont {Sabbagh}}, \bibinfo {author} {\bibfnamefont {V.}~\bibnamefont
  {Soukhanovskii}}, \bibinfo {author} {\bibfnamefont {P.}~\bibnamefont
  {Valanju}}, \ and\ \bibinfo {author} {\bibfnamefont {R.}~\bibnamefont
  {Woolley}},\ }\href {\doibase 10.1088/0029-5515/56/10/106023} {\bibfield
  {journal} {\bibinfo  {journal} {Nuclear Fusion}\ }\textbf {\bibinfo {volume}
  {56}},\ \bibinfo {pages} {106023} (\bibinfo {year} {2016})}\BibitemShut
  {NoStop}%
\bibitem [{\citenamefont {Menard}(2019)}]{Menard2019RSA}%
  \BibitemOpen
  \bibfield  {author} {\bibinfo {author} {\bibfnamefont {J.~E.}\ \bibnamefont
  {Menard}},\ }\href {\doibase 10.1098/rsta.2017.0440} {\bibfield  {journal}
  {\bibinfo  {journal} {Philosophical Transactions of the Royal Society A}\
  }\textbf {\bibinfo {volume} {377}},\ \bibinfo {pages} {20170440} (\bibinfo
  {year} {2019})}\BibitemShut {NoStop}%
\bibitem [{\citenamefont {Kaye}\ \emph {et~al.}(2007)\citenamefont {Kaye},
  \citenamefont {Levinton}, \citenamefont {Stutman}, \citenamefont {Tritz},
  \citenamefont {Yuh}, \citenamefont {Bell}, \citenamefont {Bell},
  \citenamefont {Domier}, \citenamefont {Gates}, \citenamefont {Horton},
  \citenamefont {Kim}, \citenamefont {LeBlanc}, \citenamefont {Luhmann},
  \citenamefont {Maingi}, \citenamefont {Mazzucato}, \citenamefont {Menard},
  \citenamefont {Mikkelsen}, \citenamefont {Mueller}, \citenamefont {Park},
  \citenamefont {Rewoldt}, \citenamefont {Sabbagh}, \citenamefont {Smith},\
  and\ \citenamefont {Wang}}]{Kaye2007NF}%
  \BibitemOpen
  \bibfield  {author} {\bibinfo {author} {\bibfnamefont {S.}~\bibnamefont
  {Kaye}}, \bibinfo {author} {\bibfnamefont {F.}~\bibnamefont {Levinton}},
  \bibinfo {author} {\bibfnamefont {D.}~\bibnamefont {Stutman}}, \bibinfo
  {author} {\bibfnamefont {K.}~\bibnamefont {Tritz}}, \bibinfo {author}
  {\bibfnamefont {H.}~\bibnamefont {Yuh}}, \bibinfo {author} {\bibfnamefont
  {M.}~\bibnamefont {Bell}}, \bibinfo {author} {\bibfnamefont {R.}~\bibnamefont
  {Bell}}, \bibinfo {author} {\bibfnamefont {C.}~\bibnamefont {Domier}},
  \bibinfo {author} {\bibfnamefont {D.}~\bibnamefont {Gates}}, \bibinfo
  {author} {\bibfnamefont {W.}~\bibnamefont {Horton}}, \bibinfo {author}
  {\bibfnamefont {J.}~\bibnamefont {Kim}}, \bibinfo {author} {\bibfnamefont
  {B.}~\bibnamefont {LeBlanc}}, \bibinfo {author} {\bibfnamefont
  {N.}~\bibnamefont {Luhmann}}, \bibinfo {author} {\bibfnamefont
  {R.}~\bibnamefont {Maingi}}, \bibinfo {author} {\bibfnamefont
  {E.}~\bibnamefont {Mazzucato}}, \bibinfo {author} {\bibfnamefont
  {J.}~\bibnamefont {Menard}}, \bibinfo {author} {\bibfnamefont
  {D.}~\bibnamefont {Mikkelsen}}, \bibinfo {author} {\bibfnamefont
  {D.}~\bibnamefont {Mueller}}, \bibinfo {author} {\bibfnamefont
  {H.}~\bibnamefont {Park}}, \bibinfo {author} {\bibfnamefont {G.}~\bibnamefont
  {Rewoldt}}, \bibinfo {author} {\bibfnamefont {S.}~\bibnamefont {Sabbagh}},
  \bibinfo {author} {\bibfnamefont {D.}~\bibnamefont {Smith}}, \ and\ \bibinfo
  {author} {\bibfnamefont {W.}~\bibnamefont {Wang}},\ }\href {\doibase
  10.1088/0029-5515/47/7/001} {\bibfield  {journal} {\bibinfo  {journal}
  {Nuclear Fusion}\ }\textbf {\bibinfo {volume} {47}},\ \bibinfo {pages} {499}
  (\bibinfo {year} {2007})}\BibitemShut {NoStop}%
\bibitem [{\citenamefont {Valovi{\v{c}}}\ \emph {et~al.}(2009)\citenamefont
  {Valovi{\v{c}}}, \citenamefont {Akers}, \citenamefont {Cunningham},
  \citenamefont {Garzotti}, \citenamefont {Lloyd}, \citenamefont {Muir},
  \citenamefont {Patel}, \citenamefont {Taylor}, \citenamefont {Turnyanskiy},\
  and\ \citenamefont {and}}]{Valovic2009NF}%
  \BibitemOpen
  \bibfield  {author} {\bibinfo {author} {\bibfnamefont {M.}~\bibnamefont
  {Valovi{\v{c}}}}, \bibinfo {author} {\bibfnamefont {R.}~\bibnamefont
  {Akers}}, \bibinfo {author} {\bibfnamefont {G.}~\bibnamefont {Cunningham}},
  \bibinfo {author} {\bibfnamefont {L.}~\bibnamefont {Garzotti}}, \bibinfo
  {author} {\bibfnamefont {B.}~\bibnamefont {Lloyd}}, \bibinfo {author}
  {\bibfnamefont {D.}~\bibnamefont {Muir}}, \bibinfo {author} {\bibfnamefont
  {A.}~\bibnamefont {Patel}}, \bibinfo {author} {\bibfnamefont
  {D.}~\bibnamefont {Taylor}}, \bibinfo {author} {\bibfnamefont
  {M.}~\bibnamefont {Turnyanskiy}}, \ and\ \bibinfo {author} {\bibfnamefont
  {M.~W.}\ \bibnamefont {and}},\ }\href {\doibase
  10.1088/0029-5515/49/7/075016} {\bibfield  {journal} {\bibinfo  {journal}
  {Nuclear Fusion}\ }\textbf {\bibinfo {volume} {49}},\ \bibinfo {pages}
  {075016} (\bibinfo {year} {2009})}\BibitemShut {NoStop}%
\bibitem [{\citenamefont {Valovi{\v{c}}}\ \emph {et~al.}(2011)\citenamefont
  {Valovi{\v{c}}}, \citenamefont {Akers}, \citenamefont {de~Bock},
  \citenamefont {McCone}, \citenamefont {Garzotti}, \citenamefont {Michael},
  \citenamefont {Naylor}, \citenamefont {Patel}, \citenamefont {Roach},
  \citenamefont {Scannell}, \citenamefont {Turnyanskiy}, \citenamefont {Wisse},
  \citenamefont {Guttenfelder},\ and\ \citenamefont {and}}]{Valovic2011NF}%
  \BibitemOpen
  \bibfield  {author} {\bibinfo {author} {\bibfnamefont {M.}~\bibnamefont
  {Valovi{\v{c}}}}, \bibinfo {author} {\bibfnamefont {R.}~\bibnamefont
  {Akers}}, \bibinfo {author} {\bibfnamefont {M.}~\bibnamefont {de~Bock}},
  \bibinfo {author} {\bibfnamefont {J.}~\bibnamefont {McCone}}, \bibinfo
  {author} {\bibfnamefont {L.}~\bibnamefont {Garzotti}}, \bibinfo {author}
  {\bibfnamefont {C.}~\bibnamefont {Michael}}, \bibinfo {author} {\bibfnamefont
  {G.}~\bibnamefont {Naylor}}, \bibinfo {author} {\bibfnamefont
  {A.}~\bibnamefont {Patel}}, \bibinfo {author} {\bibfnamefont {C.~M.}\
  \bibnamefont {Roach}}, \bibinfo {author} {\bibfnamefont {R.}~\bibnamefont
  {Scannell}}, \bibinfo {author} {\bibfnamefont {M.}~\bibnamefont
  {Turnyanskiy}}, \bibinfo {author} {\bibfnamefont {M.}~\bibnamefont {Wisse}},
  \bibinfo {author} {\bibfnamefont {W.}~\bibnamefont {Guttenfelder}}, \ and\
  \bibinfo {author} {\bibfnamefont {J.~C.}\ \bibnamefont {and}},\ }\href
  {\doibase 10.1088/0029-5515/51/7/073045} {\bibfield  {journal} {\bibinfo
  {journal} {Nuclear Fusion}\ }\textbf {\bibinfo {volume} {51}},\ \bibinfo
  {pages} {073045} (\bibinfo {year} {2011})}\BibitemShut {NoStop}%
\bibitem [{\citenamefont {Rewoldt}\ \emph {et~al.}(1996)\citenamefont
  {Rewoldt}, \citenamefont {Tang}, \citenamefont {Kaye},\ and\ \citenamefont
  {Menard}}]{Rewoldt1996POP}%
  \BibitemOpen
  \bibfield  {author} {\bibinfo {author} {\bibfnamefont {G.}~\bibnamefont
  {Rewoldt}}, \bibinfo {author} {\bibfnamefont {W.~M.}\ \bibnamefont {Tang}},
  \bibinfo {author} {\bibfnamefont {S.}~\bibnamefont {Kaye}}, \ and\ \bibinfo
  {author} {\bibfnamefont {J.}~\bibnamefont {Menard}},\ }\href {\doibase
  10.1063/1.871686} {\bibfield  {journal} {\bibinfo  {journal} {Physics of
  Plasmas}\ }\textbf {\bibinfo {volume} {3}},\ \bibinfo {pages} {1667}
  (\bibinfo {year} {1996})}\BibitemShut {NoStop}%
\bibitem [{\citenamefont {Kinsey}\ \emph {et~al.}(2007)\citenamefont {Kinsey},
  \citenamefont {Waltz},\ and\ \citenamefont {Candy}}]{Kinsey2007POP}%
  \BibitemOpen
  \bibfield  {author} {\bibinfo {author} {\bibfnamefont {J.~E.}\ \bibnamefont
  {Kinsey}}, \bibinfo {author} {\bibfnamefont {R.~E.}\ \bibnamefont {Waltz}}, \
  and\ \bibinfo {author} {\bibfnamefont {J.}~\bibnamefont {Candy}},\ }\href
  {\doibase 10.1063/1.2786857} {\bibfield  {journal} {\bibinfo  {journal}
  {Physics of Plasmas}\ }\textbf {\bibinfo {volume} {14}},\ \bibinfo {pages}
  {102306} (\bibinfo {year} {2007})}\BibitemShut {NoStop}%
\bibitem [{\citenamefont {Roach}\ \emph {et~al.}(2009)\citenamefont {Roach},
  \citenamefont {Abel}, \citenamefont {Akers}, \citenamefont {Arter},
  \citenamefont {Barnes}, \citenamefont {Camenen}, \citenamefont {Casson},
  \citenamefont {Colyer}, \citenamefont {Connor}, \citenamefont {Cowley},
  \citenamefont {Dickinson}, \citenamefont {Dorland}, \citenamefont {Field},
  \citenamefont {Guttenfelder}, \citenamefont {Hammett}, \citenamefont
  {Hastie}, \citenamefont {Highcock}, \citenamefont {Loureiro}, \citenamefont
  {Peeters}, \citenamefont {Reshko}, \citenamefont {Saarelma}, \citenamefont
  {Schekochihin}, \citenamefont {Valovic},\ and\ \citenamefont
  {Wilson}}]{Roach2009PPCF}%
  \BibitemOpen
  \bibfield  {author} {\bibinfo {author} {\bibfnamefont {C.~M.}\ \bibnamefont
  {Roach}}, \bibinfo {author} {\bibfnamefont {I.~G.}\ \bibnamefont {Abel}},
  \bibinfo {author} {\bibfnamefont {R.~J.}\ \bibnamefont {Akers}}, \bibinfo
  {author} {\bibfnamefont {W.}~\bibnamefont {Arter}}, \bibinfo {author}
  {\bibfnamefont {M.}~\bibnamefont {Barnes}}, \bibinfo {author} {\bibfnamefont
  {Y.}~\bibnamefont {Camenen}}, \bibinfo {author} {\bibfnamefont {F.~J.}\
  \bibnamefont {Casson}}, \bibinfo {author} {\bibfnamefont {G.}~\bibnamefont
  {Colyer}}, \bibinfo {author} {\bibfnamefont {J.~W.}\ \bibnamefont {Connor}},
  \bibinfo {author} {\bibfnamefont {S.~C.}\ \bibnamefont {Cowley}}, \bibinfo
  {author} {\bibfnamefont {D.}~\bibnamefont {Dickinson}}, \bibinfo {author}
  {\bibfnamefont {W.}~\bibnamefont {Dorland}}, \bibinfo {author} {\bibfnamefont
  {A.~R.}\ \bibnamefont {Field}}, \bibinfo {author} {\bibfnamefont
  {W.}~\bibnamefont {Guttenfelder}}, \bibinfo {author} {\bibfnamefont {G.~W.}\
  \bibnamefont {Hammett}}, \bibinfo {author} {\bibfnamefont {R.~J.}\
  \bibnamefont {Hastie}}, \bibinfo {author} {\bibfnamefont {E.}~\bibnamefont
  {Highcock}}, \bibinfo {author} {\bibfnamefont {N.~F.}\ \bibnamefont
  {Loureiro}}, \bibinfo {author} {\bibfnamefont {A.~G.}\ \bibnamefont
  {Peeters}}, \bibinfo {author} {\bibfnamefont {M.}~\bibnamefont {Reshko}},
  \bibinfo {author} {\bibfnamefont {S.}~\bibnamefont {Saarelma}}, \bibinfo
  {author} {\bibfnamefont {A.~A.}\ \bibnamefont {Schekochihin}}, \bibinfo
  {author} {\bibfnamefont {M.}~\bibnamefont {Valovic}}, \ and\ \bibinfo
  {author} {\bibfnamefont {H.~R.}\ \bibnamefont {Wilson}},\ }\href {\doibase
  10.1088/0741-3335/51/12/124020} {\bibfield  {journal} {\bibinfo  {journal}
  {Plasma Physics and Controlled Fusion}\ }\textbf {\bibinfo {volume} {51}},\
  \bibinfo {pages} {124020} (\bibinfo {year} {2009})}\BibitemShut {NoStop}%
\bibitem [{\citenamefont {Guttenfelder}\ \emph {et~al.}(2013)\citenamefont
  {Guttenfelder}, \citenamefont {Peterson}, \citenamefont {Candy},
  \citenamefont {Kaye}, \citenamefont {Ren}, \citenamefont {Bell},
  \citenamefont {Hammett}, \citenamefont {LeBlanc}, \citenamefont {Mikkelsen},
  \citenamefont {Nevins},\ and\ \citenamefont {Yuh}}]{Guttenfelder2013NF}%
  \BibitemOpen
  \bibfield  {author} {\bibinfo {author} {\bibfnamefont {W.}~\bibnamefont
  {Guttenfelder}}, \bibinfo {author} {\bibfnamefont {J.~L.}\ \bibnamefont
  {Peterson}}, \bibinfo {author} {\bibfnamefont {J.}~\bibnamefont {Candy}},
  \bibinfo {author} {\bibfnamefont {S.~M.}\ \bibnamefont {Kaye}}, \bibinfo
  {author} {\bibfnamefont {Y.}~\bibnamefont {Ren}}, \bibinfo {author}
  {\bibfnamefont {R.~E.}\ \bibnamefont {Bell}}, \bibinfo {author}
  {\bibfnamefont {G.~W.}\ \bibnamefont {Hammett}}, \bibinfo {author}
  {\bibfnamefont {B.~P.}\ \bibnamefont {LeBlanc}}, \bibinfo {author}
  {\bibfnamefont {D.~R.}\ \bibnamefont {Mikkelsen}}, \bibinfo {author}
  {\bibfnamefont {W.~M.}\ \bibnamefont {Nevins}}, \ and\ \bibinfo {author}
  {\bibfnamefont {H.}~\bibnamefont {Yuh}},\ }\href
  {http://stacks.iop.org/0029-5515/53/i=9/a=093022} {\bibfield  {journal}
  {\bibinfo  {journal} {Nuclear Fusion}\ }\textbf {\bibinfo {volume} {53}},\
  \bibinfo {pages} {093022} (\bibinfo {year} {2013})}\BibitemShut {NoStop}%
\bibitem [{\citenamefont {Guttenfelder}\ \emph {et~al.}(2019)\citenamefont
  {Guttenfelder}, \citenamefont {Kaye}, \citenamefont {Kriete}, \citenamefont
  {Bell}, \citenamefont {Diallo}, \citenamefont {LeBlanc}, \citenamefont
  {McKee}, \citenamefont {Podesta}, \citenamefont {Sabbagh},\ and\
  \citenamefont {Smith}}]{Guttenfelder2019NF}%
  \BibitemOpen
  \bibfield  {author} {\bibinfo {author} {\bibfnamefont {W.}~\bibnamefont
  {Guttenfelder}}, \bibinfo {author} {\bibfnamefont {S.~M.}\ \bibnamefont
  {Kaye}}, \bibinfo {author} {\bibfnamefont {D.~M.}\ \bibnamefont {Kriete}},
  \bibinfo {author} {\bibfnamefont {R.~E.}\ \bibnamefont {Bell}}, \bibinfo
  {author} {\bibfnamefont {A.}~\bibnamefont {Diallo}}, \bibinfo {author}
  {\bibfnamefont {B.~P.}\ \bibnamefont {LeBlanc}}, \bibinfo {author}
  {\bibfnamefont {G.~R.}\ \bibnamefont {McKee}}, \bibinfo {author}
  {\bibfnamefont {M.}~\bibnamefont {Podesta}}, \bibinfo {author} {\bibfnamefont
  {S.~A.}\ \bibnamefont {Sabbagh}}, \ and\ \bibinfo {author} {\bibfnamefont
  {D.~R.}\ \bibnamefont {Smith}},\ }\href {\doibase 10.1088/1741-4326/ab0b2c}
  {\bibfield  {journal} {\bibinfo  {journal} {Nuclear Fusion}\ }\textbf
  {\bibinfo {volume} {59}},\ \bibinfo {pages} {056027} (\bibinfo {year}
  {2019})}\BibitemShut {NoStop}%
\bibitem [{\citenamefont {Crocker}\ \emph {et~al.}(2011)\citenamefont
  {Crocker}, \citenamefont {Peebles}, \citenamefont {Kubota}, \citenamefont
  {Zhang}, \citenamefont {Bell}, \citenamefont {Fredrickson}, \citenamefont
  {Gorelenkov}, \citenamefont {LeBlanc}, \citenamefont {Menard}, \citenamefont
  {\Podesta}, \citenamefont {Sabbagh}, \citenamefont {Tritz},\ and\
  \citenamefont {Yuh}}]{Crocker2011PPCF}%
  \BibitemOpen
  \bibfield  {author} {\bibinfo {author} {\bibfnamefont {N.~A.}\ \bibnamefont
  {Crocker}}, \bibinfo {author} {\bibfnamefont {W.~A.}\ \bibnamefont
  {Peebles}}, \bibinfo {author} {\bibfnamefont {S.}~\bibnamefont {Kubota}},
  \bibinfo {author} {\bibfnamefont {J.}~\bibnamefont {Zhang}}, \bibinfo
  {author} {\bibfnamefont {R.~E.}\ \bibnamefont {Bell}}, \bibinfo {author}
  {\bibfnamefont {E.~D.}\ \bibnamefont {Fredrickson}}, \bibinfo {author}
  {\bibfnamefont {N.~N.}\ \bibnamefont {Gorelenkov}}, \bibinfo {author}
  {\bibfnamefont {B.~P.}\ \bibnamefont {LeBlanc}}, \bibinfo {author}
  {\bibfnamefont {J.~E.}\ \bibnamefont {Menard}}, \bibinfo {author}
  {\bibfnamefont {M.}~\bibnamefont {\Podesta}}, \bibinfo {author}
  {\bibfnamefont {S.~A.}\ \bibnamefont {Sabbagh}}, \bibinfo {author}
  {\bibfnamefont {K.}~\bibnamefont {Tritz}}, \ and\ \bibinfo {author}
  {\bibfnamefont {H.}~\bibnamefont {Yuh}},\ }\href
  {http://stacks.iop.org/0741-3335/53/i=10/a=105001} {\bibfield  {journal}
  {\bibinfo  {journal} {Plasma Physics and Controlled Fusion}\ }\textbf
  {\bibinfo {volume} {53}},\ \bibinfo {pages} {105001} (\bibinfo {year}
  {2011})}\BibitemShut {NoStop}%
\bibitem [{\citenamefont {Smith}\ \emph {et~al.}(2008)\citenamefont {Smith},
  \citenamefont {Mazzucato}, \citenamefont {Lee}, \citenamefont {Park},
  \citenamefont {Domier},\ and\ \citenamefont {Luhmann}}]{Smith2008RSI}%
  \BibitemOpen
  \bibfield  {author} {\bibinfo {author} {\bibfnamefont {D.~R.}\ \bibnamefont
  {Smith}}, \bibinfo {author} {\bibfnamefont {E.}~\bibnamefont {Mazzucato}},
  \bibinfo {author} {\bibfnamefont {W.}~\bibnamefont {Lee}}, \bibinfo {author}
  {\bibfnamefont {H.~K.}\ \bibnamefont {Park}}, \bibinfo {author}
  {\bibfnamefont {C.~W.}\ \bibnamefont {Domier}}, \ and\ \bibinfo {author}
  {\bibfnamefont {N.~C.}\ \bibnamefont {Luhmann}},\ }\href {\doibase
  10.1063/1.3039415} {\bibfield  {journal} {\bibinfo  {journal} {Review of
  Scientific Instruments}\ }\textbf {\bibinfo {volume} {79}},\ \bibinfo {pages}
  {123501} (\bibinfo {year} {2008})}\BibitemShut {NoStop}%
\bibitem [{\citenamefont {Barchfeld}\ \emph {et~al.}(2018)\citenamefont
  {Barchfeld}, \citenamefont {Domier}, \citenamefont {Ren}, \citenamefont
  {Ellis}, \citenamefont {Riemenschneider}, \citenamefont {Allen},
  \citenamefont {Kaita}, \citenamefont {Stratton}, \citenamefont {Dannenberg},
  \citenamefont {Zhu},\ and\ \citenamefont {Luhmann}}]{Barchfield2018RSI}%
  \BibitemOpen
  \bibfield  {author} {\bibinfo {author} {\bibfnamefont {R.}~\bibnamefont
  {Barchfeld}}, \bibinfo {author} {\bibfnamefont {C.~W.}\ \bibnamefont
  {Domier}}, \bibinfo {author} {\bibfnamefont {Y.}~\bibnamefont {Ren}},
  \bibinfo {author} {\bibfnamefont {R.}~\bibnamefont {Ellis}}, \bibinfo
  {author} {\bibfnamefont {P.}~\bibnamefont {Riemenschneider}}, \bibinfo
  {author} {\bibfnamefont {N.}~\bibnamefont {Allen}}, \bibinfo {author}
  {\bibfnamefont {R.}~\bibnamefont {Kaita}}, \bibinfo {author} {\bibfnamefont
  {B.}~\bibnamefont {Stratton}}, \bibinfo {author} {\bibfnamefont
  {J.}~\bibnamefont {Dannenberg}}, \bibinfo {author} {\bibfnamefont
  {Y.}~\bibnamefont {Zhu}}, \ and\ \bibinfo {author} {\bibfnamefont {N.~C.}\
  \bibnamefont {Luhmann}},\ }\href {\doibase 10.1063/1.5035410} {\bibfield
  {journal} {\bibinfo  {journal} {Review of Scientific Instruments}\ }\textbf
  {\bibinfo {volume} {89}},\ \bibinfo {pages} {10C114} (\bibinfo {year}
  {2018})}\BibitemShut {NoStop}%
\bibitem [{\citenamefont {Deng}\ \emph {et~al.}(2017)\citenamefont {Deng},
  \citenamefont {Crocker}, \citenamefont {Y.Ren},\ and\ \citenamefont
  {V.Belova}}]{Deng2017APS}%
  \BibitemOpen
  \bibfield  {author} {\bibinfo {author} {\bibfnamefont {Z.}~\bibnamefont
  {Deng}}, \bibinfo {author} {\bibfnamefont {N.~A.}\ \bibnamefont {Crocker}},
  \bibinfo {author} {\bibnamefont {Y.Ren}}, \ and\ \bibinfo {author}
  {\bibfnamefont {E.}~\bibnamefont {V.Belova}},\ }in\ \href
  {http://meetings.aps.org/Meeting/DPP17/Session/JP11.21} {\emph {\bibinfo
  {booktitle} {$59^{th}$ APS DPP Meeting}}}\ (\bibinfo {address} {Milwaukee,
  WI},\ \bibinfo {year} {2017})\BibitemShut {NoStop}%
\bibitem [{\citenamefont {Fredrickson}\ \emph {et~al.}(2014)\citenamefont
  {Fredrickson}, \citenamefont {Gorelenkov}, \citenamefont {\Podesta},
  \citenamefont {Bortolon}, \citenamefont {Gerhardt}, \citenamefont {Bell},
  \citenamefont {Diallo},\ and\ \citenamefont {LeBlanc}}]{Fredrickson2014NF}%
  \BibitemOpen
  \bibfield  {author} {\bibinfo {author} {\bibfnamefont {E.~D.}\ \bibnamefont
  {Fredrickson}}, \bibinfo {author} {\bibfnamefont {N.~N.}\ \bibnamefont
  {Gorelenkov}}, \bibinfo {author} {\bibfnamefont {M.}~\bibnamefont
  {\Podesta}}, \bibinfo {author} {\bibfnamefont {A.}~\bibnamefont {Bortolon}},
  \bibinfo {author} {\bibfnamefont {S.~P.}\ \bibnamefont {Gerhardt}}, \bibinfo
  {author} {\bibfnamefont {R.~E.}\ \bibnamefont {Bell}}, \bibinfo {author}
  {\bibfnamefont {A.}~\bibnamefont {Diallo}}, \ and\ \bibinfo {author}
  {\bibfnamefont {B.}~\bibnamefont {LeBlanc}},\ }\href
  {http://stacks.iop.org/0029-5515/54/i=9/a=093007} {\bibfield  {journal}
  {\bibinfo  {journal} {Nuclear Fusion}\ }\textbf {\bibinfo {volume} {54}},\
  \bibinfo {pages} {093007} (\bibinfo {year} {2014})}\BibitemShut {NoStop}%
\bibitem [{\citenamefont {Goldston}\ \emph {et~al.}(1981)\citenamefont
  {Goldston}, \citenamefont {McCune}, \citenamefont {Towner}, \citenamefont
  {Davis}, \citenamefont {Hawryluk},\ and\ \citenamefont
  {Schmidt}}]{Goldston1982JCP}%
  \BibitemOpen
  \bibfield  {author} {\bibinfo {author} {\bibfnamefont {R.}~\bibnamefont
  {Goldston}}, \bibinfo {author} {\bibfnamefont {D.}~\bibnamefont {McCune}},
  \bibinfo {author} {\bibfnamefont {H.}~\bibnamefont {Towner}}, \bibinfo
  {author} {\bibfnamefont {S.}~\bibnamefont {Davis}}, \bibinfo {author}
  {\bibfnamefont {R.}~\bibnamefont {Hawryluk}}, \ and\ \bibinfo {author}
  {\bibfnamefont {G.}~\bibnamefont {Schmidt}},\ }\href {\doibase
  http://dx.doi.org/10.1016/0021-9991(81)90111-X} {\bibfield  {journal}
  {\bibinfo  {journal} {Journal of Computational Physics}\ }\textbf {\bibinfo
  {volume} {43}},\ \bibinfo {pages} {61 } (\bibinfo {year} {1981})}\BibitemShut
  {NoStop}%
\bibitem [{\citenamefont {Ren}\ \emph {et~al.}(2017)\citenamefont {Ren},
  \citenamefont {Belova}, \citenamefont {Gorelenkov}, \citenamefont
  {Guttenfelder}, \citenamefont {Kaye}, \citenamefont {Mazzucato},
  \citenamefont {Peterson}, \citenamefont {Smith}, \citenamefont {Stutman},
  \citenamefont {Tritz}, \citenamefont {Wang}, \citenamefont {Yuh},
  \citenamefont {Bell}, \citenamefont {Domier},\ and\ \citenamefont
  {LeBlanc}}]{Ren2017NF}%
  \BibitemOpen
  \bibfield  {author} {\bibinfo {author} {\bibfnamefont {Y.}~\bibnamefont
  {Ren}}, \bibinfo {author} {\bibfnamefont {E.}~\bibnamefont {Belova}},
  \bibinfo {author} {\bibfnamefont {N.}~\bibnamefont {Gorelenkov}}, \bibinfo
  {author} {\bibfnamefont {W.}~\bibnamefont {Guttenfelder}}, \bibinfo {author}
  {\bibfnamefont {S.~M.}\ \bibnamefont {Kaye}}, \bibinfo {author}
  {\bibfnamefont {E.}~\bibnamefont {Mazzucato}}, \bibinfo {author}
  {\bibfnamefont {J.~L.}\ \bibnamefont {Peterson}}, \bibinfo {author}
  {\bibfnamefont {D.~R.}\ \bibnamefont {Smith}}, \bibinfo {author}
  {\bibfnamefont {D.}~\bibnamefont {Stutman}}, \bibinfo {author} {\bibfnamefont
  {K.}~\bibnamefont {Tritz}}, \bibinfo {author} {\bibfnamefont {W.~X.}\
  \bibnamefont {Wang}}, \bibinfo {author} {\bibfnamefont {H.}~\bibnamefont
  {Yuh}}, \bibinfo {author} {\bibfnamefont {R.~E.}\ \bibnamefont {Bell}},
  \bibinfo {author} {\bibfnamefont {C.~W.}\ \bibnamefont {Domier}}, \ and\
  \bibinfo {author} {\bibfnamefont {B.~P.}\ \bibnamefont {LeBlanc}},\ }\href
  {http://stacks.iop.org/0029-5515/57/i=7/a=072002} {\bibfield  {journal}
  {\bibinfo  {journal} {Nuclear Fusion}\ }\textbf {\bibinfo {volume} {57}},\
  \bibinfo {pages} {072002} (\bibinfo {year} {2017})}\BibitemShut {NoStop}%
\bibitem [{\citenamefont {Tang}\ \emph {et~al.}(2017)\citenamefont {Tang},
  \citenamefont {Crocker}, \citenamefont {Carter}, \citenamefont {Fredrickson},
  \citenamefont {Gorelenkov},\ and\ \citenamefont
  {Guttenfelder}}]{Tang2017TTF}%
  \BibitemOpen
  \bibfield  {author} {\bibinfo {author} {\bibfnamefont {S.}~\bibnamefont
  {Tang}}, \bibinfo {author} {\bibfnamefont {N.~A.}\ \bibnamefont {Crocker}},
  \bibinfo {author} {\bibfnamefont {T.~A.}\ \bibnamefont {Carter}}, \bibinfo
  {author} {\bibfnamefont {E.~D.}\ \bibnamefont {Fredrickson}}, \bibinfo
  {author} {\bibfnamefont {N.~N.}\ \bibnamefont {Gorelenkov}}, \ and\ \bibinfo
  {author} {\bibfnamefont {W.}~\bibnamefont {Guttenfelder}},\ }in\ \href@noop
  {} {\emph {\bibinfo {booktitle} {2017 US/EU Transport Task Force}}}\
  (\bibinfo {address} {Williamsburg, VA},\ \bibinfo {year} {2017})\BibitemShut
  {NoStop}%
\bibitem [{\citenamefont {Gates}\ \emph {et~al.}(2001)\citenamefont {Gates},
  \citenamefont {Gorelenkov},\ and\ \citenamefont {White}}]{Gates2001PRL}%
  \BibitemOpen
  \bibfield  {author} {\bibinfo {author} {\bibfnamefont {D.~A.}\ \bibnamefont
  {Gates}}, \bibinfo {author} {\bibfnamefont {N.~N.}\ \bibnamefont
  {Gorelenkov}}, \ and\ \bibinfo {author} {\bibfnamefont {R.~B.}\ \bibnamefont
  {White}},\ }\href {\doibase 10.1103/PhysRevLett.87.205003} {\bibfield
  {journal} {\bibinfo  {journal} {Phys. Rev. Lett.}\ }\textbf {\bibinfo
  {volume} {87}},\ \bibinfo {pages} {205003} (\bibinfo {year}
  {2001})}\BibitemShut {NoStop}%
\bibitem [{\citenamefont {Kolesnychenko}\ \emph {et~al.}(2005)\citenamefont
  {Kolesnychenko}, \citenamefont {Lutsenko},\ and\ \citenamefont
  {White}}]{Kolesnychenko2005POP}%
  \BibitemOpen
  \bibfield  {author} {\bibinfo {author} {\bibfnamefont {O.~Y.}\ \bibnamefont
  {Kolesnychenko}}, \bibinfo {author} {\bibfnamefont {V.~V.}\ \bibnamefont
  {Lutsenko}}, \ and\ \bibinfo {author} {\bibfnamefont {R.~B.}\ \bibnamefont
  {White}},\ }\href {\doibase 10.1063/1.2052133} {\bibfield  {journal}
  {\bibinfo  {journal} {Physics of Plasmas}\ }\textbf {\bibinfo {volume}
  {12}},\ \bibinfo {pages} {102101} (\bibinfo {year} {2005})}\BibitemShut
  {NoStop}%
\bibitem [{\citenamefont {Fredrickson}\ \emph {et~al.}(2002)\citenamefont
  {Fredrickson}, \citenamefont {Gorelenkov}, \citenamefont {Cheng},
  \citenamefont {Bell}, \citenamefont {Darrow}, \citenamefont {Gates},
  \citenamefont {Johnson}, \citenamefont {Kaye}, \citenamefont {LeBlanc},
  \citenamefont {McCune}, \citenamefont {Menard}, \citenamefont {Roquemore},\
  and\ \citenamefont {Kubota}}]{Fredrickson2002POP}%
  \BibitemOpen
  \bibfield  {author} {\bibinfo {author} {\bibfnamefont {E.~D.}\ \bibnamefont
  {Fredrickson}}, \bibinfo {author} {\bibfnamefont {N.}~\bibnamefont
  {Gorelenkov}}, \bibinfo {author} {\bibfnamefont {C.~Z.}\ \bibnamefont
  {Cheng}}, \bibinfo {author} {\bibfnamefont {R.}~\bibnamefont {Bell}},
  \bibinfo {author} {\bibfnamefont {D.}~\bibnamefont {Darrow}}, \bibinfo
  {author} {\bibfnamefont {D.}~\bibnamefont {Gates}}, \bibinfo {author}
  {\bibfnamefont {D.}~\bibnamefont {Johnson}}, \bibinfo {author} {\bibfnamefont
  {S.}~\bibnamefont {Kaye}}, \bibinfo {author} {\bibfnamefont {B.}~\bibnamefont
  {LeBlanc}}, \bibinfo {author} {\bibfnamefont {D.}~\bibnamefont {McCune}},
  \bibinfo {author} {\bibfnamefont {J.}~\bibnamefont {Menard}}, \bibinfo
  {author} {\bibfnamefont {L.}~\bibnamefont {Roquemore}}, \ and\ \bibinfo
  {author} {\bibfnamefont {S.}~\bibnamefont {Kubota}},\ }\href {\doibase
  10.1063/1.1464542} {\bibfield  {journal} {\bibinfo  {journal} {Physics of
  Plasmas}\ }\textbf {\bibinfo {volume} {9}},\ \bibinfo {pages} {2069}
  (\bibinfo {year} {2002})}\BibitemShut {NoStop}%
\bibitem [{\citenamefont {Kolesnichenko}\ \emph
  {et~al.}(2018{\natexlab{a}})\citenamefont {Kolesnichenko}, \citenamefont
  {Lutsenko}, \citenamefont {Tyshchenko}, \citenamefont {Weisen}, \citenamefont
  {Yakovenko},\ and\ \citenamefont {Contributors}}]{Kolesnichenko2018NF}%
  \BibitemOpen
  \bibfield  {author} {\bibinfo {author} {\bibfnamefont {Y.}~\bibnamefont
  {Kolesnichenko}}, \bibinfo {author} {\bibfnamefont {V.~V.}\ \bibnamefont
  {Lutsenko}}, \bibinfo {author} {\bibfnamefont {M.~H.}\ \bibnamefont
  {Tyshchenko}}, \bibinfo {author} {\bibfnamefont {H.}~\bibnamefont {Weisen}},
  \bibinfo {author} {\bibfnamefont {Y.}~\bibnamefont {Yakovenko}}, \ and\
  \bibinfo {author} {\bibfnamefont {J.}~\bibnamefont {Contributors}},\ }\href
  {http://stacks.iop.org/0029-5515/58/i=7/a=076012} {\bibfield  {journal}
  {\bibinfo  {journal} {Nuclear Fusion}\ }\textbf {\bibinfo {volume} {58}},\
  \bibinfo {pages} {076012} (\bibinfo {year} {2018}{\natexlab{a}})}\BibitemShut
  {NoStop}%
\bibitem [{\citenamefont {{The NSTX-U Team}}(2013)}]{NSTXU5YP1418}%
  \BibitemOpen
  \bibfield  {author} {\bibinfo {author} {\bibnamefont {{The NSTX-U Team}}},\
  }\href {https://nstx-u.pppl.gov/five-year-plan/five-year-plan-2014-18}
  {\enquote {\bibinfo {title} {{NSTX Upgrade Five Year Plan for FY2014 –
  2018}},}\ } (\bibinfo {year} {2013})\BibitemShut {NoStop}%
\bibitem [{\citenamefont {White}\ and\ \citenamefont
  {Chance}(1984)}]{White1984PF}%
  \BibitemOpen
  \bibfield  {author} {\bibinfo {author} {\bibfnamefont {R.~B.}\ \bibnamefont
  {White}}\ and\ \bibinfo {author} {\bibfnamefont {M.~S.}\ \bibnamefont
  {Chance}},\ }\href {\doibase 10.1063/1.864527} {\bibfield  {journal}
  {\bibinfo  {journal} {The Physics of Fluids}\ }\textbf {\bibinfo {volume}
  {27}},\ \bibinfo {pages} {2455} (\bibinfo {year} {1984})}\BibitemShut
  {NoStop}%
\bibitem [{\citenamefont {Gorelenkov}\ \emph
  {et~al.}(2010{\natexlab{b}})\citenamefont {Gorelenkov}, \citenamefont
  {Stutman}, \citenamefont {Tritz}, \citenamefont {Boozer}, \citenamefont
  {Delgado-Aparicio}, \citenamefont {Fredrickson}, \citenamefont {Kaye},\ and\
  \citenamefont {White}}]{Gorelenkov2010NF}%
  \BibitemOpen
  \bibfield  {author} {\bibinfo {author} {\bibfnamefont {N.~N.}\ \bibnamefont
  {Gorelenkov}}, \bibinfo {author} {\bibfnamefont {D.}~\bibnamefont {Stutman}},
  \bibinfo {author} {\bibfnamefont {K.}~\bibnamefont {Tritz}}, \bibinfo
  {author} {\bibfnamefont {A.}~\bibnamefont {Boozer}}, \bibinfo {author}
  {\bibfnamefont {L.}~\bibnamefont {Delgado-Aparicio}}, \bibinfo {author}
  {\bibfnamefont {E.}~\bibnamefont {Fredrickson}}, \bibinfo {author}
  {\bibfnamefont {S.}~\bibnamefont {Kaye}}, \ and\ \bibinfo {author}
  {\bibfnamefont {R.}~\bibnamefont {White}},\ }\href
  {http://stacks.iop.org/0029-5515/50/i=8/a=084012} {\bibfield  {journal}
  {\bibinfo  {journal} {Nuclear Fusion}\ }\textbf {\bibinfo {volume} {50}},\
  \bibinfo {pages} {084012} (\bibinfo {year} {2010}{\natexlab{b}})}\BibitemShut
  {NoStop}%
\bibitem [{\citenamefont {Kolesnichenko}\ \emph
  {et~al.}(2010{\natexlab{a}})\citenamefont {Kolesnichenko}, \citenamefont
  {Yakovenko},\ and\ \citenamefont {Lutsenko}}]{Kolesnichenko2010PRL}%
  \BibitemOpen
  \bibfield  {author} {\bibinfo {author} {\bibfnamefont {Y.~I.}\ \bibnamefont
  {Kolesnichenko}}, \bibinfo {author} {\bibfnamefont {Y.~V.}\ \bibnamefont
  {Yakovenko}}, \ and\ \bibinfo {author} {\bibfnamefont {V.~V.}\ \bibnamefont
  {Lutsenko}},\ }\href {\doibase 10.1103/PhysRevLett.104.075001} {\bibfield
  {journal} {\bibinfo  {journal} {Phys. Rev. Lett.}\ }\textbf {\bibinfo
  {volume} {104}},\ \bibinfo {pages} {075001} (\bibinfo {year}
  {2010}{\natexlab{a}})}\BibitemShut {NoStop}%
\bibitem [{\citenamefont {Kolesnichenko}\ \emph
  {et~al.}(2010{\natexlab{b}})\citenamefont {Kolesnichenko}, \citenamefont
  {Yakovenko}, \citenamefont {Lutsenko}, \citenamefont {White},\ and\
  \citenamefont {Weller}}]{Kolesnichenko2010NF}%
  \BibitemOpen
  \bibfield  {author} {\bibinfo {author} {\bibfnamefont {Y.}~\bibnamefont
  {Kolesnichenko}}, \bibinfo {author} {\bibfnamefont {Y.}~\bibnamefont
  {Yakovenko}}, \bibinfo {author} {\bibfnamefont {V.~V.}\ \bibnamefont
  {Lutsenko}}, \bibinfo {author} {\bibfnamefont {R.~B.}\ \bibnamefont {White}},
  \ and\ \bibinfo {author} {\bibfnamefont {A.}~\bibnamefont {Weller}},\ }\href
  {http://stacks.iop.org/0029-5515/50/i=8/a=084011} {\bibfield  {journal}
  {\bibinfo  {journal} {Nuclear Fusion}\ }\textbf {\bibinfo {volume} {50}},\
  \bibinfo {pages} {084011} (\bibinfo {year} {2010}{\natexlab{b}})}\BibitemShut
  {NoStop}%
\bibitem [{\citenamefont {Kolesnichenko}\ and\ \citenamefont
  {Tykhyy}(2018)}]{Kolesnichenko2018POP}%
  \BibitemOpen
  \bibfield  {author} {\bibinfo {author} {\bibfnamefont {Y.~I.}\ \bibnamefont
  {Kolesnichenko}}\ and\ \bibinfo {author} {\bibfnamefont {A.~V.}\ \bibnamefont
  {Tykhyy}},\ }\href {\doibase 10.1063/1.5048380} {\bibfield  {journal}
  {\bibinfo  {journal} {Physics of Plasmas}\ }\textbf {\bibinfo {volume}
  {25}},\ \bibinfo {pages} {102507} (\bibinfo {year} {2018})}\BibitemShut
  {NoStop}%
\bibitem [{\citenamefont {Kolesnichenko}\ \emph
  {et~al.}(2018{\natexlab{b}})\citenamefont {Kolesnichenko}, \citenamefont
  {Yakovenko},\ and\ \citenamefont {Tyshchenko}}]{Kolesnichenko2018bPOP}%
  \BibitemOpen
  \bibfield  {author} {\bibinfo {author} {\bibfnamefont {Y.~I.}\ \bibnamefont
  {Kolesnichenko}}, \bibinfo {author} {\bibfnamefont {Y.~V.}\ \bibnamefont
  {Yakovenko}}, \ and\ \bibinfo {author} {\bibfnamefont {M.~H.}\ \bibnamefont
  {Tyshchenko}},\ }\href {\doibase 10.1063/1.5049543} {\bibfield  {journal}
  {\bibinfo  {journal} {Physics of Plasmas}\ }\textbf {\bibinfo {volume}
  {25}},\ \bibinfo {pages} {122508} (\bibinfo {year}
  {2018}{\natexlab{b}})}\BibitemShut {NoStop}%
\bibitem [{\citenamefont {Belova}\ \emph {et~al.}(2015)\citenamefont {Belova},
  \citenamefont {Gorelenkov}, \citenamefont {Fredrickson}, \citenamefont
  {Tritz},\ and\ \citenamefont {Crocker}}]{Belova2015PRL}%
  \BibitemOpen
  \bibfield  {author} {\bibinfo {author} {\bibfnamefont {E.~V.}\ \bibnamefont
  {Belova}}, \bibinfo {author} {\bibfnamefont {N.~N.}\ \bibnamefont
  {Gorelenkov}}, \bibinfo {author} {\bibfnamefont {E.~D.}\ \bibnamefont
  {Fredrickson}}, \bibinfo {author} {\bibfnamefont {K.}~\bibnamefont {Tritz}},
  \ and\ \bibinfo {author} {\bibfnamefont {N.~A.}\ \bibnamefont {Crocker}},\
  }\href {\doibase 10.1103/PhysRevLett.115.015001} {\bibfield  {journal}
  {\bibinfo  {journal} {Phys. Rev. Lett.}\ }\textbf {\bibinfo {volume} {115}},\
  \bibinfo {pages} {015001} (\bibinfo {year} {2015})}\BibitemShut {NoStop}%
\bibitem [{\citenamefont {Belova}\ \emph {et~al.}(2019)\citenamefont {Belova},
  \citenamefont {Fredrickson}, \citenamefont {Lestz},\ and\ \citenamefont
  {Crocker}}]{Belova2019POP}%
  \BibitemOpen
  \bibfield  {author} {\bibinfo {author} {\bibfnamefont {E.~V.}\ \bibnamefont
  {Belova}}, \bibinfo {author} {\bibfnamefont {E.~D.}\ \bibnamefont
  {Fredrickson}}, \bibinfo {author} {\bibfnamefont {J.~B.}\ \bibnamefont
  {Lestz}}, \ and\ \bibinfo {author} {\bibfnamefont {N.~A.}\ \bibnamefont
  {Crocker}},\ }\href {\doibase 10.1063/1.5116357} {\bibfield  {journal}
  {\bibinfo  {journal} {Physics of Plasmas}\ }\textbf {\bibinfo {volume}
  {26}},\ \bibinfo {pages} {092507} (\bibinfo {year} {2019})}\BibitemShut
  {NoStop}%
\bibitem [{\citenamefont {Graves}\ \emph {et~al.}(2012)\citenamefont {Graves},
  \citenamefont {Chapman}, \citenamefont {Coda}, \citenamefont {Lennholm},
  \citenamefont {Albergante},\ and\ \citenamefont {Jucker}}]{Graves2012Nature}%
  \BibitemOpen
  \bibfield  {author} {\bibinfo {author} {\bibfnamefont {J.~P.}\ \bibnamefont
  {Graves}}, \bibinfo {author} {\bibfnamefont {I.~T.}\ \bibnamefont {Chapman}},
  \bibinfo {author} {\bibfnamefont {S.}~\bibnamefont {Coda}}, \bibinfo {author}
  {\bibfnamefont {M.}~\bibnamefont {Lennholm}}, \bibinfo {author}
  {\bibfnamefont {M.}~\bibnamefont {Albergante}}, \ and\ \bibinfo {author}
  {\bibfnamefont {M.}~\bibnamefont {Jucker}},\ }\href
  {https://www.nature.com/articles/ncomms1622} {\bibfield  {journal} {\bibinfo
  {journal} {Nature Communications}\ }\textbf {\bibinfo {volume} {3}} (\bibinfo
  {year} {2012})}\BibitemShut {NoStop}%
\bibitem [{\citenamefont {Bortolon}\ \emph {et~al.}(2013)\citenamefont
  {Bortolon}, \citenamefont {Heidbrink}, \citenamefont {Kramer}, \citenamefont
  {Park}, \citenamefont {Fredrickson}, \citenamefont {Lore},\ and\
  \citenamefont {\Podesta}}]{Bartolon2013PRL}%
  \BibitemOpen
  \bibfield  {author} {\bibinfo {author} {\bibfnamefont {A.}~\bibnamefont
  {Bortolon}}, \bibinfo {author} {\bibfnamefont {W.~W.}\ \bibnamefont
  {Heidbrink}}, \bibinfo {author} {\bibfnamefont {G.~J.}\ \bibnamefont
  {Kramer}}, \bibinfo {author} {\bibfnamefont {J.-K.}\ \bibnamefont {Park}},
  \bibinfo {author} {\bibfnamefont {E.~D.}\ \bibnamefont {Fredrickson}},
  \bibinfo {author} {\bibfnamefont {J.~D.}\ \bibnamefont {Lore}}, \ and\
  \bibinfo {author} {\bibfnamefont {M.}~\bibnamefont {\Podesta}},\ }\href
  {\doibase 10.1103/PhysRevLett.110.265008} {\bibfield  {journal} {\bibinfo
  {journal} {Phys. Rev. Lett.}\ }\textbf {\bibinfo {volume} {110}},\ \bibinfo
  {pages} {265008} (\bibinfo {year} {2013})}\BibitemShut {NoStop}%
\bibitem [{\citenamefont {Velikov}\ \emph {et~al.}(1968)\citenamefont
  {Velikov}, \citenamefont {Kolesnichenko},\ and\ \citenamefont
  {Oraevskii}}]{Belikov1968ZHETF}%
  \BibitemOpen
  \bibfield  {author} {\bibinfo {author} {\bibfnamefont {V.~S.}\ \bibnamefont
  {Velikov}}, \bibinfo {author} {\bibfnamefont {Y.~I.}\ \bibnamefont
  {Kolesnichenko}}, \ and\ \bibinfo {author} {\bibfnamefont {V.~N.}\
  \bibnamefont {Oraevskii}},\ }\href@noop {} {\bibfield  {journal} {\bibinfo
  {journal} {Zh. Eksp. Teor. Fiz.}\ }\textbf {\bibinfo {volume} {55}},\
  \bibinfo {pages} {2210} (\bibinfo {year} {1968})}\BibitemShut {NoStop}%
\bibitem [{\citenamefont {Velikov}\ \emph {et~al.}(1969)\citenamefont
  {Velikov}, \citenamefont {Kolesnichenko},\ and\ \citenamefont
  {Oraevskii}}]{Belikov1969JETP}%
  \BibitemOpen
  \bibfield  {author} {\bibinfo {author} {\bibfnamefont {V.~S.}\ \bibnamefont
  {Velikov}}, \bibinfo {author} {\bibfnamefont {Y.~I.}\ \bibnamefont
  {Kolesnichenko}}, \ and\ \bibinfo {author} {\bibfnamefont {V.~N.}\
  \bibnamefont {Oraevskii}},\ }\href
  {http://www.jetp.ac.ru/cgi-bin/e/index/e/28/6/p1172?a=list} {\bibfield
  {journal} {\bibinfo  {journal} {Journal of Experimental and Theoretical
  Physics}\ }\textbf {\bibinfo {volume} {28}},\ \bibinfo {pages} {1172}
  (\bibinfo {year} {1969})}\BibitemShut {NoStop}%
\bibitem [{\citenamefont {Timofeev}\ and\ \citenamefont
  {Pistunovich}(1970)}]{Timofeevv5}%
  \BibitemOpen
  \bibfield  {author} {\bibinfo {author} {\bibfnamefont {A.~V.}\ \bibnamefont
  {Timofeev}}\ and\ \bibinfo {author} {\bibfnamefont {V.~I.}\ \bibnamefont
  {Pistunovich}},\ }in\ \href@noop {} {\emph {\bibinfo {booktitle} {Reviews of
  Plasma Physics}}},\ Vol.~\bibinfo {volume} {5},\ \bibinfo {editor} {edited
  by\ \bibinfo {editor} {\bibfnamefont {M.~A.}\ \bibnamefont {Leontovich}}}\
  (\bibinfo  {publisher} {Consultants Bureau},\ \bibinfo {address} {New York},\
  \bibinfo {year} {1970})\ pp.\ \bibinfo {pages} {401 -- 445}\BibitemShut
  {NoStop}%
\bibitem [{\citenamefont {Gorelenkov}\ and\ \citenamefont
  {Cheng}(1995{\natexlab{b}})}]{Gorelenkov1995NF}%
  \BibitemOpen
  \bibfield  {author} {\bibinfo {author} {\bibfnamefont {N.~N.}\ \bibnamefont
  {Gorelenkov}}\ and\ \bibinfo {author} {\bibfnamefont {C.~Z.}\ \bibnamefont
  {Cheng}},\ }\href {http://stacks.iop.org/0029-5515/35/i=12/a=I39} {\bibfield
  {journal} {\bibinfo  {journal} {Nuclear Fusion}\ }\textbf {\bibinfo {volume}
  {35}},\ \bibinfo {pages} {1743} (\bibinfo {year}
  {1995}{\natexlab{b}})}\BibitemShut {NoStop}%
\bibitem [{\citenamefont {Gorelenkov}\ and\ \citenamefont
  {Cheng}(2002)}]{Gorelenkov2002bNF}%
  \BibitemOpen
  \bibfield  {author} {\bibinfo {author} {\bibfnamefont {N.~N.}\ \bibnamefont
  {Gorelenkov}}\ and\ \bibinfo {author} {\bibfnamefont {C.~Z.}\ \bibnamefont
  {Cheng}},\ }\href {http://stacks.iop.org/0029-5515/42/i=10/a=307} {\bibfield
  {journal} {\bibinfo  {journal} {Nuclear Fusion}\ }\textbf {\bibinfo {volume}
  {42}},\ \bibinfo {pages} {1216} (\bibinfo {year} {2002})}\BibitemShut
  {NoStop}%
\bibitem [{\citenamefont {Pankin}\ \emph {et~al.}(2004)\citenamefont {Pankin},
  \citenamefont {McCune}, \citenamefont {Andre}, \citenamefont {Bateman},\ and\
  \citenamefont {Kritz}}]{Pankin2004CPC}%
  \BibitemOpen
  \bibfield  {author} {\bibinfo {author} {\bibfnamefont {A.}~\bibnamefont
  {Pankin}}, \bibinfo {author} {\bibfnamefont {D.}~\bibnamefont {McCune}},
  \bibinfo {author} {\bibfnamefont {R.}~\bibnamefont {Andre}}, \bibinfo
  {author} {\bibfnamefont {G.}~\bibnamefont {Bateman}}, \ and\ \bibinfo
  {author} {\bibfnamefont {A.}~\bibnamefont {Kritz}},\ }\href {\doibase
  http://doi.org/10.1016/j.cpc.2003.11.002} {\bibfield  {journal} {\bibinfo
  {journal} {Computer Physics Communications}\ }\textbf {\bibinfo {volume}
  {159}},\ \bibinfo {pages} {157 } (\bibinfo {year} {2004})}\BibitemShut
  {NoStop}%
\bibitem [{\citenamefont {Belova}\ \emph {et~al.}(2003)\citenamefont {Belova},
  \citenamefont {Gorelenkov},\ and\ \citenamefont {Cheng}}]{Belova2003POP}%
  \BibitemOpen
  \bibfield  {author} {\bibinfo {author} {\bibfnamefont {E.~V.}\ \bibnamefont
  {Belova}}, \bibinfo {author} {\bibfnamefont {N.~N.}\ \bibnamefont
  {Gorelenkov}}, \ and\ \bibinfo {author} {\bibfnamefont {C.~Z.}\ \bibnamefont
  {Cheng}},\ }\href {\doibase http://dx.doi.org/10.1063/1.1592155} {\bibfield
  {journal} {\bibinfo  {journal} {Physics of Plasmas}\ }\textbf {\bibinfo
  {volume} {10}},\ \bibinfo {pages} {3240} (\bibinfo {year}
  {2003})}\BibitemShut {NoStop}%
\bibitem [{\citenamefont {Lestz}\ \emph
  {et~al.}(2020{\natexlab{a}})\citenamefont {Lestz}, \citenamefont
  {Gorelenkov}, \citenamefont {Belova}, \citenamefont {Tang},\ and\
  \citenamefont {Crocker}}]{Lestz2020p1}%
  \BibitemOpen
  \bibfield  {author} {\bibinfo {author} {\bibfnamefont {J.~B.}\ \bibnamefont
  {Lestz}}, \bibinfo {author} {\bibfnamefont {N.~N.}\ \bibnamefont
  {Gorelenkov}}, \bibinfo {author} {\bibfnamefont {E.~V.}\ \bibnamefont
  {Belova}}, \bibinfo {author} {\bibfnamefont {S.~X.}\ \bibnamefont {Tang}}, \
  and\ \bibinfo {author} {\bibfnamefont {N.~A.}\ \bibnamefont {Crocker}},\
  }\href {\doibase 10.1063/1.5127551} {\bibfield  {journal} {\bibinfo
  {journal} {Physics of Plasmas}\ }\textbf {\bibinfo {volume} {27}},\ \bibinfo
  {pages} {022513} (\bibinfo {year} {2020}{\natexlab{a}})}\BibitemShut
  {NoStop}%
\bibitem [{\citenamefont {Stix}(1975)}]{Stix1975NF}%
  \BibitemOpen
  \bibfield  {author} {\bibinfo {author} {\bibfnamefont {T.~H.}\ \bibnamefont
  {Stix}},\ }\href {http://stacks.iop.org/0029-5515/15/i=5/a=003} {\bibfield
  {journal} {\bibinfo  {journal} {Nuclear Fusion}\ }\textbf {\bibinfo {volume}
  {15}},\ \bibinfo {pages} {737} (\bibinfo {year} {1975})}\BibitemShut
  {NoStop}%
\bibitem [{\citenamefont {Belikov}\ \emph {et~al.}(2004)\citenamefont
  {Belikov}, \citenamefont {Kolesnichenko},\ and\ \citenamefont
  {White}}]{Belikov2004POP}%
  \BibitemOpen
  \bibfield  {author} {\bibinfo {author} {\bibfnamefont {V.~S.}\ \bibnamefont
  {Belikov}}, \bibinfo {author} {\bibfnamefont {Y.~I.}\ \bibnamefont
  {Kolesnichenko}}, \ and\ \bibinfo {author} {\bibfnamefont {R.~B.}\
  \bibnamefont {White}},\ }\href {\doibase http://dx.doi.org/10.1063/1.1809121}
  {\bibfield  {journal} {\bibinfo  {journal} {Physics of Plasmas}\ }\textbf
  {\bibinfo {volume} {11}},\ \bibinfo {pages} {5409} (\bibinfo {year}
  {2004})}\BibitemShut {NoStop}%
\bibitem [{\citenamefont {Bender}\ and\ \citenamefont
  {Orszag}(1978)}]{BenderOrszagStationaryPhase}%
  \BibitemOpen
  \bibfield  {author} {\bibinfo {author} {\bibfnamefont {C.~M.}\ \bibnamefont
  {Bender}}\ and\ \bibinfo {author} {\bibfnamefont {S.~A.}\ \bibnamefont
  {Orszag}},\ }\enquote {\bibinfo {title} {Advanced mathematical methods for
  scientists and engineers: Asymptotic methods and perturbation theory},}\ \
  (\bibinfo  {publisher} {McGraw-Hill},\ \bibinfo {year} {1978})\ Chap.\
  \bibinfo {chapter} {6.5 Method of Stationary Phase}\BibitemShut {NoStop}%
\bibitem [{\citenamefont {Kuvshinov}(1994)}]{Kuvshinov1994PPCF}%
  \BibitemOpen
  \bibfield  {author} {\bibinfo {author} {\bibfnamefont {B.~N.}\ \bibnamefont
  {Kuvshinov}},\ }\href {http://stacks.iop.org/0741-3335/36/i=5/a=008}
  {\bibfield  {journal} {\bibinfo  {journal} {Plasma Physics and Controlled
  Fusion}\ }\textbf {\bibinfo {volume} {36}},\ \bibinfo {pages} {867} (\bibinfo
  {year} {1994})}\BibitemShut {NoStop}%
\bibitem [{\citenamefont {Lestz}\ \emph
  {et~al.}(2020{\natexlab{b}})\citenamefont {Lestz}, \citenamefont
  {Gorelenkov}, \citenamefont {Belova}, \citenamefont {Tang},\ and\
  \citenamefont {Crocker}}]{Lestz2020p2}%
  \BibitemOpen
  \bibfield  {author} {\bibinfo {author} {\bibfnamefont {J.~B.}\ \bibnamefont
  {Lestz}}, \bibinfo {author} {\bibfnamefont {N.~N.}\ \bibnamefont
  {Gorelenkov}}, \bibinfo {author} {\bibfnamefont {E.~V.}\ \bibnamefont
  {Belova}}, \bibinfo {author} {\bibfnamefont {S.~X.}\ \bibnamefont {Tang}}, \
  and\ \bibinfo {author} {\bibfnamefont {N.~A.}\ \bibnamefont {Crocker}},\
  }\href {\doibase 10.1063/1.5127552} {\bibfield  {journal} {\bibinfo
  {journal} {Physics of Plasmas}\ }\textbf {\bibinfo {volume} {27}},\ \bibinfo
  {pages} {022512} (\bibinfo {year} {2020}{\natexlab{b}})}\BibitemShut
  {NoStop}%
\bibitem [{\citenamefont {Kaye}\ \emph {et~al.}(2019)\citenamefont {Kaye},
  \citenamefont {Battaglia}, \citenamefont {Baver}, \citenamefont {Belova},
  \citenamefont {Berkery}, \citenamefont {Duarte}, \citenamefont {Ferraro},
  \citenamefont {Fredrickson}, \citenamefont {Gorelenkov}, \citenamefont
  {Guttenfelder}, \citenamefont {Hao}, \citenamefont {Heidbrink}, \citenamefont
  {Izacard}, \citenamefont {Kim}, \citenamefont {Krebs}, \citenamefont {Haye},
  \citenamefont {Lestz}, \citenamefont {Liu}, \citenamefont {Morton},
  \citenamefont {Myra}, \citenamefont {Pfefferle}, \citenamefont {\Podesta},
  \citenamefont {Ren}, \citenamefont {Riquezes}, \citenamefont {Sabbagh},
  \citenamefont {Schneller}, \citenamefont {Scotti}, \citenamefont
  {Soukhanovskii}, \citenamefont {Zweben}, \citenamefont {Ahn}, \citenamefont
  {Allain}, \citenamefont {Barchfeld}, \citenamefont {Bedoya}, \citenamefont
  {Bell}, \citenamefont {Bertelli}, \citenamefont {Bhattacharjee},
  \citenamefont {Boyer}, \citenamefont {Brennan}, \citenamefont {Canal},
  \citenamefont {Canik}, \citenamefont {Crocker}, \citenamefont {Darrow},
  \citenamefont {Delgado-Aparicio}, \citenamefont {Diallo}, \citenamefont
  {Domier}, \citenamefont {Ebrahimi}, \citenamefont {Evans}, \citenamefont
  {Fonck}, \citenamefont {Frerichs}, \citenamefont {Gan}, \citenamefont
  {Gerhardt}, \citenamefont {Gray}, \citenamefont {Jarboe}, \citenamefont
  {Jardin}, \citenamefont {Jaworski}, \citenamefont {Kaita}, \citenamefont
  {Koel}, \citenamefont {Kolemen}, \citenamefont {Kriete}, \citenamefont
  {Kubota}, \citenamefont {LeBlanc}, \citenamefont {Levinton}, \citenamefont
  {Luhmann}, \citenamefont {Lunsford}, \citenamefont {Maingi}, \citenamefont
  {Maqueda}, \citenamefont {Menard}, \citenamefont {Mueller}, \citenamefont
  {Myers}, \citenamefont {Ono}, \citenamefont {Park}, \citenamefont {Perkins},
  \citenamefont {Poli}, \citenamefont {Raman}, \citenamefont {Reinke},
  \citenamefont {Rhodes}, \citenamefont {Rowley}, \citenamefont {Russell},
  \citenamefont {Schuster}, \citenamefont {Schmitz}, \citenamefont {Sechrest},
  \citenamefont {Skinner}, \citenamefont {Smith}, \citenamefont
  {Stotzfus-Dueck}, \citenamefont {Stratton}, \citenamefont {Taylor},
  \citenamefont {Tritz}, \citenamefont {Wang}, \citenamefont {Wang},
  \citenamefont {Waters},\ and\ \citenamefont {Wirth}}]{Kaye2019NF}%
  \BibitemOpen
  \bibfield  {author} {\bibinfo {author} {\bibfnamefont {S.~M.}\ \bibnamefont
  {Kaye}}, \bibinfo {author} {\bibfnamefont {D.~J.}\ \bibnamefont {Battaglia}},
  \bibinfo {author} {\bibfnamefont {D.}~\bibnamefont {Baver}}, \bibinfo
  {author} {\bibfnamefont {E.}~\bibnamefont {Belova}}, \bibinfo {author}
  {\bibfnamefont {J.~W.}\ \bibnamefont {Berkery}}, \bibinfo {author}
  {\bibfnamefont {V.~N.}\ \bibnamefont {Duarte}}, \bibinfo {author}
  {\bibfnamefont {N.}~\bibnamefont {Ferraro}}, \bibinfo {author} {\bibfnamefont
  {E.}~\bibnamefont {Fredrickson}}, \bibinfo {author} {\bibfnamefont
  {N.}~\bibnamefont {Gorelenkov}}, \bibinfo {author} {\bibfnamefont
  {W.}~\bibnamefont {Guttenfelder}}, \bibinfo {author} {\bibfnamefont {G.~Z.}\
  \bibnamefont {Hao}}, \bibinfo {author} {\bibfnamefont {W.}~\bibnamefont
  {Heidbrink}}, \bibinfo {author} {\bibfnamefont {O.}~\bibnamefont {Izacard}},
  \bibinfo {author} {\bibfnamefont {D.}~\bibnamefont {Kim}}, \bibinfo {author}
  {\bibfnamefont {I.}~\bibnamefont {Krebs}}, \bibinfo {author} {\bibfnamefont
  {R.~L.}\ \bibnamefont {Haye}}, \bibinfo {author} {\bibfnamefont
  {J.}~\bibnamefont {Lestz}}, \bibinfo {author} {\bibfnamefont
  {D.}~\bibnamefont {Liu}}, \bibinfo {author} {\bibfnamefont {L.~A.}\
  \bibnamefont {Morton}}, \bibinfo {author} {\bibfnamefont {J.}~\bibnamefont
  {Myra}}, \bibinfo {author} {\bibfnamefont {D.}~\bibnamefont {Pfefferle}},
  \bibinfo {author} {\bibfnamefont {M.}~\bibnamefont {\Podesta}}, \bibinfo
  {author} {\bibfnamefont {Y.}~\bibnamefont {Ren}}, \bibinfo {author}
  {\bibfnamefont {J.}~\bibnamefont {Riquezes}}, \bibinfo {author}
  {\bibfnamefont {S.~A.}\ \bibnamefont {Sabbagh}}, \bibinfo {author}
  {\bibfnamefont {M.}~\bibnamefont {Schneller}}, \bibinfo {author}
  {\bibfnamefont {F.}~\bibnamefont {Scotti}}, \bibinfo {author} {\bibfnamefont
  {V.}~\bibnamefont {Soukhanovskii}}, \bibinfo {author} {\bibfnamefont {S.~J.}\
  \bibnamefont {Zweben}}, \bibinfo {author} {\bibfnamefont {J.~W.}\
  \bibnamefont {Ahn}}, \bibinfo {author} {\bibfnamefont {J.~P.}\ \bibnamefont
  {Allain}}, \bibinfo {author} {\bibfnamefont {R.}~\bibnamefont {Barchfeld}},
  \bibinfo {author} {\bibfnamefont {F.}~\bibnamefont {Bedoya}}, \bibinfo
  {author} {\bibfnamefont {R.~E.}\ \bibnamefont {Bell}}, \bibinfo {author}
  {\bibfnamefont {N.}~\bibnamefont {Bertelli}}, \bibinfo {author}
  {\bibfnamefont {A.}~\bibnamefont {Bhattacharjee}}, \bibinfo {author}
  {\bibfnamefont {M.~D.}\ \bibnamefont {Boyer}}, \bibinfo {author}
  {\bibfnamefont {D.}~\bibnamefont {Brennan}}, \bibinfo {author} {\bibfnamefont
  {G.}~\bibnamefont {Canal}}, \bibinfo {author} {\bibfnamefont
  {J.}~\bibnamefont {Canik}}, \bibinfo {author} {\bibfnamefont
  {N.}~\bibnamefont {Crocker}}, \bibinfo {author} {\bibfnamefont
  {D.}~\bibnamefont {Darrow}}, \bibinfo {author} {\bibfnamefont
  {L.}~\bibnamefont {Delgado-Aparicio}}, \bibinfo {author} {\bibfnamefont
  {A.}~\bibnamefont {Diallo}}, \bibinfo {author} {\bibfnamefont
  {C.}~\bibnamefont {Domier}}, \bibinfo {author} {\bibfnamefont
  {F.}~\bibnamefont {Ebrahimi}}, \bibinfo {author} {\bibfnamefont
  {T.}~\bibnamefont {Evans}}, \bibinfo {author} {\bibfnamefont
  {R.}~\bibnamefont {Fonck}}, \bibinfo {author} {\bibfnamefont
  {H.}~\bibnamefont {Frerichs}}, \bibinfo {author} {\bibfnamefont
  {K.}~\bibnamefont {Gan}}, \bibinfo {author} {\bibfnamefont {S.}~\bibnamefont
  {Gerhardt}}, \bibinfo {author} {\bibfnamefont {T.}~\bibnamefont {Gray}},
  \bibinfo {author} {\bibfnamefont {T.}~\bibnamefont {Jarboe}}, \bibinfo
  {author} {\bibfnamefont {S.}~\bibnamefont {Jardin}}, \bibinfo {author}
  {\bibfnamefont {M.~A.}\ \bibnamefont {Jaworski}}, \bibinfo {author}
  {\bibfnamefont {R.}~\bibnamefont {Kaita}}, \bibinfo {author} {\bibfnamefont
  {B.}~\bibnamefont {Koel}}, \bibinfo {author} {\bibfnamefont {E.}~\bibnamefont
  {Kolemen}}, \bibinfo {author} {\bibfnamefont {D.~M.}\ \bibnamefont {Kriete}},
  \bibinfo {author} {\bibfnamefont {S.}~\bibnamefont {Kubota}}, \bibinfo
  {author} {\bibfnamefont {B.~P.}\ \bibnamefont {LeBlanc}}, \bibinfo {author}
  {\bibfnamefont {F.}~\bibnamefont {Levinton}}, \bibinfo {author}
  {\bibfnamefont {N.}~\bibnamefont {Luhmann}}, \bibinfo {author} {\bibfnamefont
  {R.}~\bibnamefont {Lunsford}}, \bibinfo {author} {\bibfnamefont
  {R.}~\bibnamefont {Maingi}}, \bibinfo {author} {\bibfnamefont
  {R.}~\bibnamefont {Maqueda}}, \bibinfo {author} {\bibfnamefont {J.~E.}\
  \bibnamefont {Menard}}, \bibinfo {author} {\bibfnamefont {D.}~\bibnamefont
  {Mueller}}, \bibinfo {author} {\bibfnamefont {C.~E.}\ \bibnamefont {Myers}},
  \bibinfo {author} {\bibfnamefont {M.}~\bibnamefont {Ono}}, \bibinfo {author}
  {\bibfnamefont {J.-K.}\ \bibnamefont {Park}}, \bibinfo {author}
  {\bibfnamefont {R.}~\bibnamefont {Perkins}}, \bibinfo {author} {\bibfnamefont
  {F.}~\bibnamefont {Poli}}, \bibinfo {author} {\bibfnamefont {R.}~\bibnamefont
  {Raman}}, \bibinfo {author} {\bibfnamefont {M.}~\bibnamefont {Reinke}},
  \bibinfo {author} {\bibfnamefont {T.}~\bibnamefont {Rhodes}}, \bibinfo
  {author} {\bibfnamefont {C.}~\bibnamefont {Rowley}}, \bibinfo {author}
  {\bibfnamefont {D.}~\bibnamefont {Russell}}, \bibinfo {author} {\bibfnamefont
  {E.}~\bibnamefont {Schuster}}, \bibinfo {author} {\bibfnamefont
  {O.}~\bibnamefont {Schmitz}}, \bibinfo {author} {\bibfnamefont
  {Y.}~\bibnamefont {Sechrest}}, \bibinfo {author} {\bibfnamefont {C.~H.}\
  \bibnamefont {Skinner}}, \bibinfo {author} {\bibfnamefont {D.~R.}\
  \bibnamefont {Smith}}, \bibinfo {author} {\bibfnamefont {T.}~\bibnamefont
  {Stotzfus-Dueck}}, \bibinfo {author} {\bibfnamefont {B.}~\bibnamefont
  {Stratton}}, \bibinfo {author} {\bibfnamefont {G.}~\bibnamefont {Taylor}},
  \bibinfo {author} {\bibfnamefont {K.}~\bibnamefont {Tritz}}, \bibinfo
  {author} {\bibfnamefont {W.}~\bibnamefont {Wang}}, \bibinfo {author}
  {\bibfnamefont {Z.}~\bibnamefont {Wang}}, \bibinfo {author} {\bibfnamefont
  {I.}~\bibnamefont {Waters}}, \ and\ \bibinfo {author} {\bibfnamefont
  {B.}~\bibnamefont {Wirth}},\ }\href {\doibase 10.1088/1741-4326/ab023a}
  {\bibfield  {journal} {\bibinfo  {journal} {Nuclear Fusion}\ }\textbf
  {\bibinfo {volume} {59}},\ \bibinfo {pages} {112007} (\bibinfo {year}
  {2019})}\BibitemShut {NoStop}%
\bibitem [{\citenamefont {Belova}\ \emph {et~al.}(2020)\citenamefont {Belova},
  \citenamefont {Lestz}, \citenamefont {Crocker},\ and\ \citenamefont
  {Fredrickson}}]{Belova2020IAEA}%
  \BibitemOpen
  \bibfield  {author} {\bibinfo {author} {\bibfnamefont {E.~V.}\ \bibnamefont
  {Belova}}, \bibinfo {author} {\bibfnamefont {J.~B.}\ \bibnamefont {Lestz}},
  \bibinfo {author} {\bibfnamefont {N.~A.}\ \bibnamefont {Crocker}}, \ and\
  \bibinfo {author} {\bibfnamefont {E.~D.}\ \bibnamefont {Fredrickson}},\ }in\
  \href@noop {} {\emph {\bibinfo {booktitle} {28$^{th}$ IAEA Fusion Energy
  Conference}}}\ (\bibinfo {address} {Nice, France},\ \bibinfo {year}
  {2020})\BibitemShut {NoStop}%
\bibitem [{\citenamefont {Belova}\ \emph {et~al.}(1997)\citenamefont {Belova},
  \citenamefont {Denton},\ and\ \citenamefont {Chan}}]{Belova1997JCP}%
  \BibitemOpen
  \bibfield  {author} {\bibinfo {author} {\bibfnamefont {E.~V.}\ \bibnamefont
  {Belova}}, \bibinfo {author} {\bibfnamefont {R.~E.}\ \bibnamefont {Denton}},
  \ and\ \bibinfo {author} {\bibfnamefont {A.~A.}\ \bibnamefont {Chan}},\
  }\href {\doibase http://dx.doi.org/10.1006/jcph.1997.5738} {\bibfield
  {journal} {\bibinfo  {journal} {Journal of Computational Physics}\ }\textbf
  {\bibinfo {volume} {136}},\ \bibinfo {pages} {324 } (\bibinfo {year}
  {1997})}\BibitemShut {NoStop}%
\bibitem [{\citenamefont {Gerhardt}\ \emph
  {et~al.}(2012{\natexlab{b}})\citenamefont {Gerhardt}, \citenamefont {Andre},\
  and\ \citenamefont {Menard}}]{Gerhardt2012NF}%
  \BibitemOpen
  \bibfield  {author} {\bibinfo {author} {\bibfnamefont {S.~P.}\ \bibnamefont
  {Gerhardt}}, \bibinfo {author} {\bibfnamefont {R.}~\bibnamefont {Andre}}, \
  and\ \bibinfo {author} {\bibfnamefont {J.~E.}\ \bibnamefont {Menard}},\
  }\href {http://stacks.iop.org/0029-5515/52/i=8/a=083020} {\bibfield
  {journal} {\bibinfo  {journal} {Nuclear Fusion}\ }\textbf {\bibinfo {volume}
  {52}},\ \bibinfo {pages} {083020} (\bibinfo {year}
  {2012}{\natexlab{b}})}\BibitemShut {NoStop}%
\bibitem [{\citenamefont {Geiser}\ \emph {et~al.}(2016)\citenamefont {Geiser},
  \citenamefont {Crocker}, \citenamefont {Smith},\ and\ \citenamefont
  {Fredrickson}}]{Geiser2016APS}%
  \BibitemOpen
  \bibfield  {author} {\bibinfo {author} {\bibfnamefont {N.}~\bibnamefont
  {Geiser}}, \bibinfo {author} {\bibfnamefont {N.~A.}\ \bibnamefont {Crocker}},
  \bibinfo {author} {\bibfnamefont {H.}~\bibnamefont {Smith}}, \ and\ \bibinfo
  {author} {\bibfnamefont {E.~D.}\ \bibnamefont {Fredrickson}},\ }in\ \href
  {http://meetings.aps.org/Meeting/DPP16/Session/JP10.47} {\emph {\bibinfo
  {booktitle} {$58^{th}$ APS DPP Meeting}}}\ (\bibinfo {address} {San Jose,
  CA},\ \bibinfo {year} {2016})\BibitemShut {NoStop}%
\bibitem [{\citenamefont {Tang}\ \emph {et~al.}(2019)\citenamefont {Tang},
  \citenamefont {Crocker}, \citenamefont {Carter}, \citenamefont {Thome},
  \citenamefont {I.},\ and\ \citenamefont {Heidbrink}}]{Tang2019APS}%
  \BibitemOpen
  \bibfield  {author} {\bibinfo {author} {\bibfnamefont {S.~X.}\ \bibnamefont
  {Tang}}, \bibinfo {author} {\bibfnamefont {N.~A.}\ \bibnamefont {Crocker}},
  \bibinfo {author} {\bibfnamefont {T.~A.}\ \bibnamefont {Carter}}, \bibinfo
  {author} {\bibfnamefont {K.~E.}\ \bibnamefont {Thome}}, \bibinfo {author}
  {\bibfnamefont {R.}~\bibnamefont {I.}}, \ and\ \bibinfo {author}
  {\bibfnamefont {W.~W.}\ \bibnamefont {Heidbrink}},\ }in\ \href
  {http://meetings.aps.org/Meeting/DPP19/Session/BP10.43} {\emph {\bibinfo
  {booktitle} {$61^\text{st}$ APS DPP Meeting}}}\ (\bibinfo {address} {Fort
  Lauderdale, FL},\ \bibinfo {year} {2019})\BibitemShut {NoStop}%
\bibitem [{\citenamefont {Kaufman}(1972)}]{Kaufman1972PF}%
  \BibitemOpen
  \bibfield  {author} {\bibinfo {author} {\bibfnamefont {A.~N.}\ \bibnamefont
  {Kaufman}},\ }\href {\doibase 10.1063/1.1694031} {\bibfield  {journal}
  {\bibinfo  {journal} {The Physics of Fluids}\ }\textbf {\bibinfo {volume}
  {15}},\ \bibinfo {pages} {1063} (\bibinfo {year} {1972})}\BibitemShut
  {NoStop}%
\bibitem [{\citenamefont {Wong}\ and\ \citenamefont
  {Berk}(1999)}]{Wong1999PLA}%
  \BibitemOpen
  \bibfield  {author} {\bibinfo {author} {\bibfnamefont {H.~V.}\ \bibnamefont
  {Wong}}\ and\ \bibinfo {author} {\bibfnamefont {H.~L.}\ \bibnamefont
  {Berk}},\ }\href {\doibase https://doi.org/10.1016/S0375-9601(98)00866-4}
  {\bibfield  {journal} {\bibinfo  {journal} {Physics Letters A}\ }\textbf
  {\bibinfo {volume} {251}},\ \bibinfo {pages} {126 } (\bibinfo {year}
  {1999})}\BibitemShut {NoStop}%
\bibitem [{\citenamefont {Podest{\`{a}}}\ \emph {et~al.}(2018)\citenamefont
  {Podest{\`{a}}}, \citenamefont {Fredrickson},\ and\ \citenamefont
  {Gorelenkova}}]{Podesta2018NF}%
  \BibitemOpen
  \bibfield  {author} {\bibinfo {author} {\bibfnamefont {M.}~\bibnamefont
  {Podest{\`{a}}}}, \bibinfo {author} {\bibfnamefont {E.~D.}\ \bibnamefont
  {Fredrickson}}, \ and\ \bibinfo {author} {\bibfnamefont {M.}~\bibnamefont
  {Gorelenkova}},\ }\href {\doibase 10.1088/1741-4326/aab4ae} {\bibfield
  {journal} {\bibinfo  {journal} {Nuclear Fusion}\ }\textbf {\bibinfo {volume}
  {58}},\ \bibinfo {pages} {082023} (\bibinfo {year} {2018})}\BibitemShut
  {NoStop}%
\bibitem [{\citenamefont {Tataronis}(1975)}]{Tataronis1975JPP}%
  \BibitemOpen
  \bibfield  {author} {\bibinfo {author} {\bibfnamefont {J.~A.}\ \bibnamefont
  {Tataronis}},\ }\href {\doibase 10.1017/S0022377800025897} {\bibfield
  {journal} {\bibinfo  {journal} {Journal of Plasma Physics}\ }\textbf
  {\bibinfo {volume} {13}},\ \bibinfo {pages} {87–105} (\bibinfo {year}
  {1975})}\BibitemShut {NoStop}%
\bibitem [{\citenamefont {Rosenbluth}\ \emph
  {et~al.}(1992{\natexlab{a}})\citenamefont {Rosenbluth}, \citenamefont {Berk},
  \citenamefont {Van~Dam},\ and\ \citenamefont {Lindberg}}]{Rosenbluth1992PRL}%
  \BibitemOpen
  \bibfield  {author} {\bibinfo {author} {\bibfnamefont {M.~N.}\ \bibnamefont
  {Rosenbluth}}, \bibinfo {author} {\bibfnamefont {H.~L.}\ \bibnamefont
  {Berk}}, \bibinfo {author} {\bibfnamefont {J.~W.}\ \bibnamefont {Van~Dam}}, \
  and\ \bibinfo {author} {\bibfnamefont {D.~M.}\ \bibnamefont {Lindberg}},\
  }\href {\doibase 10.1103/PhysRevLett.68.596} {\bibfield  {journal} {\bibinfo
  {journal} {Phys. Rev. Lett.}\ }\textbf {\bibinfo {volume} {68}},\ \bibinfo
  {pages} {596} (\bibinfo {year} {1992}{\natexlab{a}})}\BibitemShut {NoStop}%
\bibitem [{\citenamefont {Rosenbluth}\ \emph
  {et~al.}(1992{\natexlab{b}})\citenamefont {Rosenbluth}, \citenamefont {Berk},
  \citenamefont {Van~Dam},\ and\ \citenamefont {Lindberg}}]{Rosenbluth1992PFB}%
  \BibitemOpen
  \bibfield  {author} {\bibinfo {author} {\bibfnamefont {M.~N.}\ \bibnamefont
  {Rosenbluth}}, \bibinfo {author} {\bibfnamefont {H.~L.}\ \bibnamefont
  {Berk}}, \bibinfo {author} {\bibfnamefont {J.~W.}\ \bibnamefont {Van~Dam}}, \
  and\ \bibinfo {author} {\bibfnamefont {D.~M.}\ \bibnamefont {Lindberg}},\
  }\href {\doibase 10.1063/1.860023} {\bibfield  {journal} {\bibinfo  {journal}
  {Physics of Fluids B: Plasma Physics}\ }\textbf {\bibinfo {volume} {4}},\
  \bibinfo {pages} {2189} (\bibinfo {year} {1992}{\natexlab{b}})}\BibitemShut
  {NoStop}%
\bibitem [{\citenamefont {Berk}\ \emph {et~al.}(1992)\citenamefont {Berk},
  \citenamefont {Van~Dam}, \citenamefont {Guo},\ and\ \citenamefont
  {Lindberg}}]{Berk1992PFB}%
  \BibitemOpen
  \bibfield  {author} {\bibinfo {author} {\bibfnamefont {H.~L.}\ \bibnamefont
  {Berk}}, \bibinfo {author} {\bibfnamefont {J.~W.}\ \bibnamefont {Van~Dam}},
  \bibinfo {author} {\bibfnamefont {Z.}~\bibnamefont {Guo}}, \ and\ \bibinfo
  {author} {\bibfnamefont {D.~M.}\ \bibnamefont {Lindberg}},\ }\href {\doibase
  10.1063/1.860455} {\bibfield  {journal} {\bibinfo  {journal} {Physics of
  Fluids B: Plasma Physics}\ }\textbf {\bibinfo {volume} {4}},\ \bibinfo
  {pages} {1806} (\bibinfo {year} {1992})}\BibitemShut {NoStop}%
\bibitem [{\citenamefont {Grishanov}\ \emph {et~al.}(2001)\citenamefont
  {Grishanov}, \citenamefont {de~Azevedo},\ and\ \citenamefont
  {Neto}}]{Grishanov2001PPCF}%
  \BibitemOpen
  \bibfield  {author} {\bibinfo {author} {\bibfnamefont {N.~I.}\ \bibnamefont
  {Grishanov}}, \bibinfo {author} {\bibfnamefont {C.~A.}\ \bibnamefont
  {de~Azevedo}}, \ and\ \bibinfo {author} {\bibfnamefont {J.~P.}\ \bibnamefont
  {Neto}},\ }\href {\doibase 10.1088/0741-3335/43/8/301} {\bibfield  {journal}
  {\bibinfo  {journal} {Plasma Physics and Controlled Fusion}\ }\textbf
  {\bibinfo {volume} {43}},\ \bibinfo {pages} {1003} (\bibinfo {year}
  {2001})}\BibitemShut {NoStop}%
\bibitem [{\citenamefont {Grishanov}\ \emph {et~al.}(2002)\citenamefont
  {Grishanov}, \citenamefont {Ludwig}, \citenamefont {de~Azevedo},\ and\
  \citenamefont {Neto}}]{Grishanov2002POP}%
  \BibitemOpen
  \bibfield  {author} {\bibinfo {author} {\bibfnamefont {N.~I.}\ \bibnamefont
  {Grishanov}}, \bibinfo {author} {\bibfnamefont {G.~O.}\ \bibnamefont
  {Ludwig}}, \bibinfo {author} {\bibfnamefont {C.~A.}\ \bibnamefont
  {de~Azevedo}}, \ and\ \bibinfo {author} {\bibfnamefont {J.~P.}\ \bibnamefont
  {Neto}},\ }\href {\doibase 10.1063/1.1499953} {\bibfield  {journal} {\bibinfo
   {journal} {Physics of Plasmas}\ }\textbf {\bibinfo {volume} {9}},\ \bibinfo
  {pages} {4089} (\bibinfo {year} {2002})}\BibitemShut {NoStop}%
\bibitem [{\citenamefont {Grishanov}\ \emph {et~al.}(2003)\citenamefont
  {Grishanov}, \citenamefont {Loula}, \citenamefont {de~Azevedo},\ and\
  \citenamefont {Neto}}]{Grishanov2003PPCF}%
  \BibitemOpen
  \bibfield  {author} {\bibinfo {author} {\bibfnamefont {N.~I.}\ \bibnamefont
  {Grishanov}}, \bibinfo {author} {\bibfnamefont {A.~F.~D.}\ \bibnamefont
  {Loula}}, \bibinfo {author} {\bibfnamefont {C.~A.}\ \bibnamefont
  {de~Azevedo}}, \ and\ \bibinfo {author} {\bibfnamefont {J.~P.}\ \bibnamefont
  {Neto}},\ }\href {\doibase 10.1088/0741-3335/45/9/315} {\bibfield  {journal}
  {\bibinfo  {journal} {Plasma Physics and Controlled Fusion}\ }\textbf
  {\bibinfo {volume} {45}},\ \bibinfo {pages} {1791} (\bibinfo {year}
  {2003})}\BibitemShut {NoStop}%
\bibitem [{\citenamefont {Gorelenkov}\ and\ \citenamefont
  {Sharapov}(1992)}]{Gorelenkov1992PS}%
  \BibitemOpen
  \bibfield  {author} {\bibinfo {author} {\bibfnamefont {N.~N.}\ \bibnamefont
  {Gorelenkov}}\ and\ \bibinfo {author} {\bibfnamefont {S.~E.}\ \bibnamefont
  {Sharapov}},\ }\href {\doibase 10.1088/0031-8949/45/2/016} {\bibfield
  {journal} {\bibinfo  {journal} {Physica Scripta}\ }\textbf {\bibinfo {volume}
  {45}},\ \bibinfo {pages} {163} (\bibinfo {year} {1992})}\BibitemShut
  {NoStop}%
\bibitem [{\citenamefont {Coppi}\ \emph {et~al.}(1986)\citenamefont {Coppi},
  \citenamefont {Cowley}, \citenamefont {Kulsrud}, \citenamefont
  {Detragiache},\ and\ \citenamefont {Pegoraro}}]{Coppi1986PF}%
  \BibitemOpen
  \bibfield  {author} {\bibinfo {author} {\bibfnamefont {B.}~\bibnamefont
  {Coppi}}, \bibinfo {author} {\bibfnamefont {S.}~\bibnamefont {Cowley}},
  \bibinfo {author} {\bibfnamefont {R.}~\bibnamefont {Kulsrud}}, \bibinfo
  {author} {\bibfnamefont {P.}~\bibnamefont {Detragiache}}, \ and\ \bibinfo
  {author} {\bibfnamefont {F.}~\bibnamefont {Pegoraro}},\ }\href {\doibase
  10.1063/1.865749} {\bibfield  {journal} {\bibinfo  {journal} {The Physics of
  Fluids}\ }\textbf {\bibinfo {volume} {29}},\ \bibinfo {pages} {4060}
  (\bibinfo {year} {1986})}\BibitemShut {NoStop}%
\bibitem [{\citenamefont {Crocker}\ \emph {et~al.}(2017)\citenamefont
  {Crocker}, \citenamefont {V.Belova}, \citenamefont {White}, \citenamefont
  {Fredrickson}, \citenamefont {Gorelenkov}, \citenamefont {Tritz},
  \citenamefont {Peebles}, \citenamefont {Kubota}, \citenamefont {Diallo},\
  and\ \citenamefont {LeBlanc}}]{Crocker2017IAEA}%
  \BibitemOpen
  \bibfield  {author} {\bibinfo {author} {\bibfnamefont {N.~A.}\ \bibnamefont
  {Crocker}}, \bibinfo {author} {\bibfnamefont {E.}~\bibnamefont {V.Belova}},
  \bibinfo {author} {\bibfnamefont {R.~B.}\ \bibnamefont {White}}, \bibinfo
  {author} {\bibfnamefont {E.~D.}\ \bibnamefont {Fredrickson}}, \bibinfo
  {author} {\bibfnamefont {N.~N.}\ \bibnamefont {Gorelenkov}}, \bibinfo
  {author} {\bibfnamefont {K.}~\bibnamefont {Tritz}}, \bibinfo {author}
  {\bibfnamefont {W.~A.}\ \bibnamefont {Peebles}}, \bibinfo {author}
  {\bibfnamefont {S.}~\bibnamefont {Kubota}}, \bibinfo {author} {\bibfnamefont
  {A.}~\bibnamefont {Diallo}}, \ and\ \bibinfo {author} {\bibfnamefont {B.~P.}\
  \bibnamefont {LeBlanc}},\ }in\ \href
  {https://nucleus.iaea.org/sites/fusionportal/Shared%20Documents/EP%2017th/6.09/I-7.pdf}
  {\emph {\bibinfo {booktitle} {$15^{th}$ IAEA TM on Energetic Particles in
  Magnetic Confinement Systems}}}\ (\bibinfo {address} {Princeton, NJ},\
  \bibinfo {year} {2017})\BibitemShut {NoStop}%
\bibitem [{\citenamefont {Slaby}\ \emph {et~al.}(2016)\citenamefont {Slaby},
  \citenamefont {Könies},\ and\ \citenamefont {Kleiber}}]{Slaby2016POP}%
  \BibitemOpen
  \bibfield  {author} {\bibinfo {author} {\bibfnamefont {C.}~\bibnamefont
  {Slaby}}, \bibinfo {author} {\bibfnamefont {A.}~\bibnamefont {Könies}}, \
  and\ \bibinfo {author} {\bibfnamefont {R.}~\bibnamefont {Kleiber}},\ }\href
  {\doibase 10.1063/1.4961916} {\bibfield  {journal} {\bibinfo  {journal}
  {Physics of Plasmas}\ }\textbf {\bibinfo {volume} {23}},\ \bibinfo {pages}
  {092501} (\bibinfo {year} {2016})}\BibitemShut {NoStop}%
\bibitem [{\citenamefont {\Podesta}\ \emph {et~al.}(2012)\citenamefont
  {\Podesta}, \citenamefont {Bell}, \citenamefont {Bortolon}, \citenamefont
  {Crocker}, \citenamefont {Darrow}, \citenamefont {Diallo}, \citenamefont
  {Fredrickson}, \citenamefont {Fu}, \citenamefont {Gorelenkov}, \citenamefont
  {Heidbrink}, \citenamefont {Kramer}, \citenamefont {Kubota}, \citenamefont
  {LeBlanc}, \citenamefont {Medley},\ and\ \citenamefont
  {Yuh}}]{Podesta2012NF}%
  \BibitemOpen
  \bibfield  {author} {\bibinfo {author} {\bibfnamefont {M.}~\bibnamefont
  {\Podesta}}, \bibinfo {author} {\bibfnamefont {R.~E.}\ \bibnamefont {Bell}},
  \bibinfo {author} {\bibfnamefont {A.}~\bibnamefont {Bortolon}}, \bibinfo
  {author} {\bibfnamefont {N.~A.}\ \bibnamefont {Crocker}}, \bibinfo {author}
  {\bibfnamefont {D.~S.}\ \bibnamefont {Darrow}}, \bibinfo {author}
  {\bibfnamefont {A.}~\bibnamefont {Diallo}}, \bibinfo {author} {\bibfnamefont
  {E.~D.}\ \bibnamefont {Fredrickson}}, \bibinfo {author} {\bibfnamefont
  {G.-Y.}\ \bibnamefont {Fu}}, \bibinfo {author} {\bibfnamefont {N.~N.}\
  \bibnamefont {Gorelenkov}}, \bibinfo {author} {\bibfnamefont {W.~W.}\
  \bibnamefont {Heidbrink}}, \bibinfo {author} {\bibfnamefont {G.~J.}\
  \bibnamefont {Kramer}}, \bibinfo {author} {\bibfnamefont {S.}~\bibnamefont
  {Kubota}}, \bibinfo {author} {\bibfnamefont {B.~P.}\ \bibnamefont {LeBlanc}},
  \bibinfo {author} {\bibfnamefont {S.~S.}\ \bibnamefont {Medley}}, \ and\
  \bibinfo {author} {\bibfnamefont {H.}~\bibnamefont {Yuh}},\ }\href
  {http://stacks.iop.org/0029-5515/52/i=9/a=094001} {\bibfield  {journal}
  {\bibinfo  {journal} {Nuclear Fusion}\ }\textbf {\bibinfo {volume} {52}},\
  \bibinfo {pages} {094001} (\bibinfo {year} {2012})}\BibitemShut {NoStop}%
\bibitem [{\citenamefont {Duarte}\ \emph
  {et~al.}(2017{\natexlab{b}})\citenamefont {Duarte}, \citenamefont {Berk},
  \citenamefont {Gorelenkov}, \citenamefont {Heidbrink}, \citenamefont
  {Kramer}, \citenamefont {Nazikian}, \citenamefont {Pace}, \citenamefont
  {\Podesta},\ and\ \citenamefont {Van~Zeeland}}]{Duarte2017POP}%
  \BibitemOpen
  \bibfield  {author} {\bibinfo {author} {\bibfnamefont {V.~N.}\ \bibnamefont
  {Duarte}}, \bibinfo {author} {\bibfnamefont {H.~L.}\ \bibnamefont {Berk}},
  \bibinfo {author} {\bibfnamefont {N.~N.}\ \bibnamefont {Gorelenkov}},
  \bibinfo {author} {\bibfnamefont {W.~W.}\ \bibnamefont {Heidbrink}}, \bibinfo
  {author} {\bibfnamefont {G.~J.}\ \bibnamefont {Kramer}}, \bibinfo {author}
  {\bibfnamefont {R.}~\bibnamefont {Nazikian}}, \bibinfo {author}
  {\bibfnamefont {D.~C.}\ \bibnamefont {Pace}}, \bibinfo {author}
  {\bibfnamefont {M.}~\bibnamefont {\Podesta}}, \ and\ \bibinfo {author}
  {\bibfnamefont {M.~A.}\ \bibnamefont {Van~Zeeland}},\ }\href {\doibase
  10.1063/1.5007811} {\bibfield  {journal} {\bibinfo  {journal} {Physics of
  Plasmas}\ }\textbf {\bibinfo {volume} {24}},\ \bibinfo {pages} {122508}
  (\bibinfo {year} {2017}{\natexlab{b}})}\BibitemShut {NoStop}%
\bibitem [{\citenamefont {Berk}\ \emph {et~al.}(2001)\citenamefont {Berk},
  \citenamefont {Borba}, \citenamefont {Breizman}, \citenamefont {Pinches},\
  and\ \citenamefont {Sharapov}}]{Berk2001PRL}%
  \BibitemOpen
  \bibfield  {author} {\bibinfo {author} {\bibfnamefont {H.~L.}\ \bibnamefont
  {Berk}}, \bibinfo {author} {\bibfnamefont {D.~N.}\ \bibnamefont {Borba}},
  \bibinfo {author} {\bibfnamefont {B.~N.}\ \bibnamefont {Breizman}}, \bibinfo
  {author} {\bibfnamefont {S.~D.}\ \bibnamefont {Pinches}}, \ and\ \bibinfo
  {author} {\bibfnamefont {S.~E.}\ \bibnamefont {Sharapov}},\ }\href {\doibase
  10.1103/PhysRevLett.87.185002} {\bibfield  {journal} {\bibinfo  {journal}
  {Phys. Rev. Lett.}\ }\textbf {\bibinfo {volume} {87}},\ \bibinfo {pages}
  {185002} (\bibinfo {year} {2001})}\BibitemShut {NoStop}%
\bibitem [{\citenamefont {Breizman}\ \emph {et~al.}(2003)\citenamefont
  {Breizman}, \citenamefont {Berk}, \citenamefont {Pekker}, \citenamefont
  {Pinches},\ and\ \citenamefont {Sharapov}}]{Breizman2003PP}%
  \BibitemOpen
  \bibfield  {author} {\bibinfo {author} {\bibfnamefont {B.~N.}\ \bibnamefont
  {Breizman}}, \bibinfo {author} {\bibfnamefont {H.~L.}\ \bibnamefont {Berk}},
  \bibinfo {author} {\bibfnamefont {M.~S.}\ \bibnamefont {Pekker}}, \bibinfo
  {author} {\bibfnamefont {S.~D.}\ \bibnamefont {Pinches}}, \ and\ \bibinfo
  {author} {\bibfnamefont {S.~E.}\ \bibnamefont {Sharapov}},\ }\href {\doibase
  10.1063/1.1597495} {\bibfield  {journal} {\bibinfo  {journal} {Physics of
  Plasmas}\ }\textbf {\bibinfo {volume} {10}},\ \bibinfo {pages} {3649}
  (\bibinfo {year} {2003})}\BibitemShut {NoStop}%
\bibitem [{\citenamefont {Todo}(2006)}]{Todo2006POP}%
  \BibitemOpen
  \bibfield  {author} {\bibinfo {author} {\bibfnamefont {Y.}~\bibnamefont
  {Todo}},\ }\href {\doibase 10.1063/1.2234296} {\bibfield  {journal} {\bibinfo
   {journal} {Physics of Plasmas}\ }\textbf {\bibinfo {volume} {13}},\ \bibinfo
  {pages} {082503} (\bibinfo {year} {2006})}\BibitemShut {NoStop}%
\bibitem [{\citenamefont {Cheng}\ \emph {et~al.}(1995)\citenamefont {Cheng},
  \citenamefont {Gorelenkov},\ and\ \citenamefont {Hsu}}]{Cheng1995NF}%
  \BibitemOpen
  \bibfield  {author} {\bibinfo {author} {\bibfnamefont {C.~Z.}\ \bibnamefont
  {Cheng}}, \bibinfo {author} {\bibfnamefont {N.~N.}\ \bibnamefont
  {Gorelenkov}}, \ and\ \bibinfo {author} {\bibfnamefont {C.~T.}\ \bibnamefont
  {Hsu}},\ }\href {http://stacks.iop.org/0029-5515/35/i=12/a=I28} {\bibfield
  {journal} {\bibinfo  {journal} {Nuclear Fusion}\ }\textbf {\bibinfo {volume}
  {35}},\ \bibinfo {pages} {1639} (\bibinfo {year} {1995})}\BibitemShut
  {NoStop}%
\bibitem [{\citenamefont {Santoro}\ and\ \citenamefont
  {Chen}(1996)}]{Santoro1996PP}%
  \BibitemOpen
  \bibfield  {author} {\bibinfo {author} {\bibfnamefont {R.~A.}\ \bibnamefont
  {Santoro}}\ and\ \bibinfo {author} {\bibfnamefont {L.}~\bibnamefont {Chen}},\
  }\href {\doibase 10.1063/1.871918} {\bibfield  {journal} {\bibinfo  {journal}
  {Physics of Plasmas}\ }\textbf {\bibinfo {volume} {3}},\ \bibinfo {pages}
  {2349} (\bibinfo {year} {1996})}\BibitemShut {NoStop}%
\bibitem [{\citenamefont {Belova}\ \emph {et~al.}(2000)\citenamefont {Belova},
  \citenamefont {Jardin}, \citenamefont {Ji}, \citenamefont {Yamada},\ and\
  \citenamefont {Kulsrud}}]{Belova2000POP}%
  \BibitemOpen
  \bibfield  {author} {\bibinfo {author} {\bibfnamefont {E.~V.}\ \bibnamefont
  {Belova}}, \bibinfo {author} {\bibfnamefont {S.~C.}\ \bibnamefont {Jardin}},
  \bibinfo {author} {\bibfnamefont {H.}~\bibnamefont {Ji}}, \bibinfo {author}
  {\bibfnamefont {M.}~\bibnamefont {Yamada}}, \ and\ \bibinfo {author}
  {\bibfnamefont {R.}~\bibnamefont {Kulsrud}},\ }\href {\doibase
  http://dx.doi.org/10.1063/1.1318929} {\bibfield  {journal} {\bibinfo
  {journal} {Physics of Plasmas}\ }\textbf {\bibinfo {volume} {7}},\ \bibinfo
  {pages} {4996} (\bibinfo {year} {2000})}\BibitemShut {NoStop}%
\bibitem [{\citenamefont {Tritz}\ \emph {et~al.}(2012)\citenamefont {Tritz},
  \citenamefont {Stutman}, \citenamefont {Finkenthal}, \citenamefont
  {Gorelenkov}, \citenamefont {White}, \citenamefont {Belova}, \citenamefont
  {Fredrickson}, \citenamefont {Kaye},\ and\ \citenamefont
  {Crocker}}]{Tritz2012APS}%
  \BibitemOpen
  \bibfield  {author} {\bibinfo {author} {\bibfnamefont {K.}~\bibnamefont
  {Tritz}}, \bibinfo {author} {\bibfnamefont {D.}~\bibnamefont {Stutman}},
  \bibinfo {author} {\bibfnamefont {M.}~\bibnamefont {Finkenthal}}, \bibinfo
  {author} {\bibfnamefont {N.~N.}\ \bibnamefont {Gorelenkov}}, \bibinfo
  {author} {\bibfnamefont {R.}~\bibnamefont {White}}, \bibinfo {author}
  {\bibfnamefont {E.}~\bibnamefont {Belova}}, \bibinfo {author} {\bibfnamefont
  {E.}~\bibnamefont {Fredrickson}}, \bibinfo {author} {\bibfnamefont
  {S.}~\bibnamefont {Kaye}}, \ and\ \bibinfo {author} {\bibfnamefont
  {N.}~\bibnamefont {Crocker}},\ }in\ \href
  {http://meetings.aps.org/Meeting/DPP12/Session/GO6.4} {\emph {\bibinfo
  {booktitle} {$54^{th}$ APS DPP Meeting}}}\ (\bibinfo {address} {Providence,
  RI},\ \bibinfo {year} {2012})\BibitemShut {NoStop}%
\bibitem [{\citenamefont {Lestz}(2019)}]{Lestz2019APS}%
  \BibitemOpen
  \bibfield  {author} {\bibinfo {author} {\bibfnamefont {J.~B.}\ \bibnamefont
  {Lestz}},\ }in\ \href {http://meetings.aps.org/Meeting/DPP19/Session/TI2.4}
  {\emph {\bibinfo {booktitle} {$61^{st}$ APS DPP Meeting}}}\ (\bibinfo
  {address} {Fort Lauderdale, FL},\ \bibinfo {year} {2019})\BibitemShut
  {NoStop}%
\bibitem [{\citenamefont {Bertelli}\ \emph {et~al.}(2019)\citenamefont
  {Bertelli}, \citenamefont {Ono},\ and\ \citenamefont
  {Jaeger}}]{Bertelli2019NF}%
  \BibitemOpen
  \bibfield  {author} {\bibinfo {author} {\bibfnamefont {N.}~\bibnamefont
  {Bertelli}}, \bibinfo {author} {\bibfnamefont {M.}~\bibnamefont {Ono}}, \
  and\ \bibinfo {author} {\bibfnamefont {E.~F.}\ \bibnamefont {Jaeger}},\
  }\href {\doibase 10.1088/1741-4326/ab1d7f} {\bibfield  {journal} {\bibinfo
  {journal} {Nuclear Fusion}\ }\textbf {\bibinfo {volume} {59}},\ \bibinfo
  {pages} {086006} (\bibinfo {year} {2019})}\BibitemShut {NoStop}%
\end{thebibliography}%

\end{document}